%% file: Thesis.tex
\documentclass[a4paper,9pt,twoside]{ThesisStyle}

\usepackage{amsmath,amssymb,amsthm}      
\usepackage[british,UKenglish,USenglish,english,american]{babel}
\usepackage[utf8]{inputenc}
\usepackage[TS1,T1]{fontenc}
\usepackage{lmodern}
\usepackage{textcomp}
\usepackage[left=2cm,right=2cm,top=1.7cm,bottom=1.7cm,includefoot,includehead,headheight=13.6pt]{geometry}

\usepackage{indentfirst}

\usepackage[mathscr]{euscript}
\usepackage{upgreek}
\DeclareFontFamily{OT1}{pzc}{}
\DeclareFontShape{OT1}{pzc}{m}{it}{<-> s * [1.10] pzcmi7t}{}
\DeclareMathAlphabet{\mathpzc}{OT1}{pzc}{m}{it}
\usepackage{stmaryrd}

\usepackage{calc}
\newlength{\depthofsumsign}
\newcommand{\nprod}[1][1.4]{
    \mathop{%
        \raisebox
            {-#1\depthofsumsign+1\depthofsumsign}
            {\scalebox
                {#1}
                {$\displaystyle\prod$}%
            }
    }
}

\newcommand{\comment}[1]{}

\usepackage[small]{caption}

\usepackage{booktabs}
\newenvironment{changemargin}[2]%
{\begin{list}{}{%
\setlength{\listparindent}{\parindent}%
\setlength{\itemindent}{\parindent}%
\setlength{\leftmargin}{#1}%
\setlength{\rightmargin}{#2}%
}\item }%
{\end{list}}

\usepackage[nottoc, notlof, notlot]{tocbibind}
\usepackage{minitoc}
\dominitoc
\usepackage{aecompl}

\usepackage{ifpdf}

\ifpdf
  \usepackage[pdftex]{graphicx}
  \usepackage{epstopdf}  
  \DeclareGraphicsExtensions{.jpg}
  \usepackage[a4paper,plainpages=false,pdfpagelabels,hyperindex=true]{hyperref}
  \usepackage{color}
  \usepackage[dvipsnames,table]{xcolor}
  \definecolor{darkblue}{rgb}{0.0,0.0,0.4}
  \definecolor{red}{rgb}{0.7,0.0,0.0}
  \definecolor{green}{rgb}{0.0,0.5,0.0}
\else
  \usepackage[dvips]{graphicx}
  \DeclareGraphicsExtensions{.png,.pdf,.ps,.eps}
  \usepackage[a4paper,dvipdfm,hyperindex=true]{hyperref}
  \usepackage{color}
  \usepackage[dvipsnames,table]{xcolor}
  \definecolor{darkblue}{rgb}{0.0,0.0,0.4}
  \definecolor{red}{rgb}{0.7,0.0,0.0}
  \definecolor{green}{rgb}{0.0,0.5,0.0}
\fi

\graphicspath{{.}{images/}}

\usepackage[acronym]{glossaries}

\makeglossaries

\usepackage{tikz}
\usetikzlibrary{shapes,arrows,shadows}

\usepackage{multicol}
\hypersetup
{
bookmarksopen=true,
pdftitle=Florent Leclercq - Bayesian large-scale structure inference and cosmic web analysis,
pdfauthor=Florent Leclercq, 
pdfsubject=Bayesian large-scale structure inference and cosmic web analysis, 
pdfmenubar=true, 
pdfhighlight=/O, 
colorlinks=true, 
pdfpagemode=None, 
pdfpagelayout=SinglePage, 
pdffitwindow=true, 
linkcolor=darkblue, 
citecolor=RoyalBlue, 
urlcolor=darkblue 
}

\usepackage{rotating}                    
\usepackage{natbib}                  
\usepackage{fancyhdr}                    
\let\leftmarkold\leftmark
\let\rightmarkold\rightmark

\fancypagestyle{plain}{
  \fancyhead{}
  \fancyfoot{}
}

\usepackage{algorithm}
\usepackage[noend]{algorithmic}

\makeatletter

\def\cleardoublepage{\clearpage\if@twoside \ifodd\c@page\else%
  \hbox{}%
  \thispagestyle{empty}
  \newpage%
  \if@twocolumn\hbox{}\newpage\fi\fi\fi}

\makeatother
 
{%

\hrulefill
\vspace*{0.5cm}%
\end{minipage}
}

\let\minitocORIG\minitoc
\renewcommand{\minitoc}{\minitocORIG \vspace{1.5em}}

\usepackage{multirow}
{ \begin{list}%
	{$\bullet$}%
	{\setlength{\labelwidth}{25pt}%
	 \setlength{\leftmargin}{30pt}%
	 \setlength{\itemsep}{\parsep}}}%
{ \end{list} }

\renewcommand{\epsilon}{\varepsilon}

\newcommand{\abstract}[1]{ 
\noindent\framebox[\columnwidth][c]{\begin{minipage}{0.98\columnwidth}\small{#1}\normalsize\end{minipage}} 
}

\newenvironment{vcenterpage}
{\newpage\vspace*{\fill}\thispagestyle{empty}}
{\vspace*{\fill}}

\newcommand{\aap}{Astronomy and Astrophysics}
\newcommand{\aj}{Astronomical Journal}
\newcommand{\apj}{Astrophysical Journal}
\newcommand{\apjl}{Astrophysical Journal Letters}
\newcommand{\apjs}{Astrophysical Journal Supplement}
\newcommand{\araa}{Annual review of astronomy and astrophysics}
\newcommand{\jcap}{Journal of Cosmology and Astroparticle Physics}
\newcommand{\mnras}{Monthly Notices of the Royal Astronomical Society}
\newcommand{\na}{New Astronomy}
\newcommand{\nat}{Nature}
\newcommand{\physrep}{Physics Reports}
\newcommand{\prd}{Physical Review D}
\newcommand{\prl}{Physical Review Letters}

\newcommand{\nbsp}{~}

\definecolor{BattleShipGrey}{rgb}{0.52, 0.52, 0.51}
\definecolor{CadmiumGreen}{rgb}{0.0, 0.42, 0.24}
\definecolor{AppleGreen}{rgb}{0.55, 0.71, 0.0}
\definecolor{DarkPastelGreen}{rgb}{0.01, 0.75, 0.24}

\newcommand{\Fuchsia}[1]{\textcolor{Fuchsia}{#1}}
\newcommand{\RoyalBlue}[1]{\textcolor{RoyalBlue}{#1}}

\newcommand{\DarkPastelGreen}[1]{\textcolor{DarkPastelGreen}{#1}}

\newcommand{\RubineRed}[1]{\textcolor{RubineRed}{#1}}
\newcommand{\BattleShipGrey}[1]{\textcolor{BattleShipGrey}{#1}}
\newcommand{\Orange}[1]{\textcolor{Orange}{#1}}
\newcommand{\Gray}[1]{\textcolor{Gray}{#1}}

\newcommand{\draw}[1]{\begin{center}
\colorbox{gray!20}{#1}
\end{center}}

\newcommand{\deriv}[2]{\dfrac{\mathrm{d} #1}{\mathrm{d} #2}}
\newcommand{\pd}[2]{\dfrac{\partial #1}{\partial #2}}

\newcommand{\drm}{\mathrm{d}}
\newcommand{\erm}{\mathrm{e}}

\renewcommand{\i}{\mathrm{i}}

\newcommand{\G}{\mathrm{G}}

\newcommand{\ci}{\perp\!\!\!\perp}
\newcommand{\p}{\mathpzc{P}}
\newcommand{\q}{\mathpzc{Q}}
\newcommand{\rproba}{\mathpzc{R}}

\newcommand{\tproba}{\mathpzc{T}}
\newcommand{\sinc}{\mathrm{sinc}}

\newglossaryentry{LSS} {
  name={LSS},
  description={large-scale structure.},
}

\newglossaryentry{cosmic web} {
  name={cosmic web},
  description={},
}

\newglossaryentry{cosmic web classification} {
  name={cosmic web classification},
  description={},
}

\newglossaryentry{SCDM} {
  name={SCDM},
  description={standard cold dark matter},
}

\newglossaryentry{OCDM} {
  name={OCDM},
  description={open cold dark matter},
}

\newglossaryentry{LCDM} {
  name={$\Lambda$CDM},
  description={lambda cold dark matter},
}

\newglossaryentry{structure formation} {
  name={structure formation},
  description={},
}

\newglossaryentry{dark matter} {
  name={dark matter},
  description={},
}

\newglossaryentry{dark energy} {
  name={dark energy},
  description={},
}

\newglossaryentry{cosmological constant} {
  name={cosmological constant},
  description={},
}

\newglossaryentry{initial conditions} {
  name={initial conditions},
  description={},
}

\newglossaryentry{gravitational evolution} {
  name={gravitational evolution},
  description={},
}

\newglossaryentry{inflation} {
  name={inflation},
  description={},
}

\newglossaryentry{inflaton} {
  name={inflaton},
  description={},
}

\newglossaryentry{equation of state} {
  name={equation of state},
  description={},
}

\newglossaryentry{quantum fluctuations} {
  name={quantum fluctuations},
  description={},
}

\newglossaryentry{statistical homogeneity} {
  name={statistical homogeneity},
  description={},
}

\newglossaryentry{statistical isotropy} {
  name={statistical isotropy},
  description={},
}

\newglossaryentry{Hot Big Bang} {
  name={Hot Big Bang},
  description={},
}

\newglossaryentry{scale factor} {
  name={scale factor},
  description={},
}

\newglossaryentry{quantum fluctuation} {
  name={quantum fluctuation},
  description={},
}

\newglossaryentry{BBN} {
  name={BBN},
  description={Big Bang nucleosynthesis},
}

\newglossaryentry{decoupling} {
  name={decoupling},
  description={},
}

\newglossaryentry{recombination} {
  name={recombination},
  description={},
}

\newglossaryentry{neutrino} {
  name={neutrino},
  description={},
}

\newglossaryentry{BAO} {
  name={BAO},
  description={baryon acoustic oscillations},
}

\newglossaryentry{radiation domination} {
  name={radiation domination},
  description={},
}

\newglossaryentry{matter domination} {
  name={matter domination},
  description={},
}

\newglossaryentry{dark-energy domination} {
  name={dark-energy domination},
  description={},
}

\newglossaryentry{non-linear evolution} {
  name={non-linear evolution},
  description={},
}

\newglossaryentry{pdf} {
  name={pdf},
  description={probability distribution function},
}

\newglossaryentry{CDM} {
  name={CDM},
  description={cold dark matter},
}

\newglossaryentry{HDM} {
  name={HDM},
  description={hot dark matter},
}

\newglossaryentry{quantum field theory} {
  name={quantum field theory},
  description={},
}

\newglossaryentry{general relativity} {
  name={general relativity},
  description={},
}

\newglossaryentry{bias} {
  name={bias},
  description={},
}

\newglossaryentry{Alcock-Paczynski effect} {
  name={Alcock-Paczynski effect},
  description={},
}

\newglossaryentry{paradigms of science} {
  name={paradigms of science},
  description={},
}

\newglossaryentry{exascale computers} {
  name={exascale computers},
  description={},
}

\newglossaryentry{SDSS} {
  name={SDSS},
  description={Sloan Digital Sky Survey},
}

\newglossaryentry{LPT} {
  name={LPT},
  description={Lagrangian perturbation theory},
}

\newglossaryentry{expansion} {
  name={expansion},
  description={},
}

\newglossaryentry{linear evolution} {
  name={linear evolution},
  description={},
}

\newglossaryentry{potential well} {
  name={potential well},
  description={},
}

\newglossaryentry{EPT} {
 name={EPT},
 description={Eulerian perturbation theory},
}

\newglossaryentry{non-linear approximation} {
 name={non-linear approximation},
 description={},
}

\newglossaryentry{cosmic time} {
 name={cosmic time},
 description={},
}

\newglossaryentry{conformal time} {
 name={conformal time},
 description={},
}

\newglossaryentry{redshift} {
 name={redshift},
 description={},
}

\newglossaryentry{Einstein's equations} {
 name={Einstein's equations},
 description={},
}

\newglossaryentry{conformal expansion rate} {
 name={conformal expansion rate},
 description={},
}

\newglossaryentry{Hubble parameter} {
 name={Hubble parameter},
 description={},
}

\newglossaryentry{gravitational constant} {
 name={gravitational constant},
 description={},
}

\newglossaryentry{Friedmann's equations} {
 name={Friedmann's equations},
 description={},
}

\newglossaryentry{cosmological parameters} {
 name={cosmological parameters},
 description={},
}

\newglossaryentry{homogeneous Universe} {
 name={homogeneous Universe},
 description={},
}

\newglossaryentry{scalar field} {
 name={scalar field},
 description={},
}

\newglossaryentry{density contrast} {
 name={density contrast},
 description={},
}

\newglossaryentry{gravitational potential} {
 name={gravitational potential},
 description={},
}

\newglossaryentry{density field} {
 name={density field},
 description={},
}

\newglossaryentry{velocity field} {
 name={velocity field},
 description={},
}

\newglossaryentry{sample average} {
 name={sample average},
 description={},
}

\newglossaryentry{ensemble average} {
 name={ensemble average},
 description={},
}

\newglossaryentry{ergodicity} {
 name={ergodicity},
 description={},
}

\newglossaryentry{galaxy survey} {
 name={galaxy survey},
 description={},
}

\newglossaryentry{redshift-space distortions} {
 name={redshift-space distortions},
 description={},
}

\newglossaryentry{grf} {
 name={grf},
 description={Gaussian random field},
}

\newglossaryentry{non-Gaussianity} {
 name={non-Gaussianity},
 description={},
}

\newglossaryentry{CMB} {
 name={CMB},
 description={cosmic microwave background},
}

\newglossaryentry{log-normal distribution} {
 name={log-normal distribution},
 description={},
}

\newglossaryentry{one-point distribution} {
 name={one-point distribution},
 description={},
}

\newglossaryentry{moment} {
 name={moment},
 description={},
}

\newglossaryentry{Wick's theorem} {
 name={Wick's theorem},
 description={},
}

\newglossaryentry{marginal pdf} {
 name={marginal pdf},
 description={},
}

\newglossaryentry{conditional pdf} {
 name={conditional pdf},
 description={},
}

\newglossaryentry{conditional independence} {
 name={conditional independence},
 description={},
}

\newglossaryentry{N-body simulation} {
 name={$N$-body simulation},
 description={},
}

\newglossaryentry{two-point correlation function} {
 name={two-point correlation function},
 description={},
}

\newglossaryentry{power spectrum} {
 name={power spectrum},
 description={},
}

\newglossaryentry{Dirac delta distribution} {
 name={Dirac delta distribution},
 description={},
}

\newglossaryentry{Kronecker symbol} {
 name={Kronecker symbol},
 description={},
}

\newglossaryentry{conditional density contrast} {
 name={conditional density contrast},
 description={},
}

\newglossaryentry{high-order correlation function} {
 name={high-order correlation function},
 description={},
}

\newglossaryentry{bispectrum} {
 name={bispectrum},
 description={},
}

\newglossaryentry{reduced bispectrum} {
 name={reduced bispectrum},
 description={},
}

\newglossaryentry{trispectrum} {
 name={trispectrum},
 description={},
}

\newglossaryentry{model comparison} {
 name={model comparison},
 description={},
}

\newglossaryentry{dark matter particles} {
 name={dark matter particles},
 description={},
}

\newglossaryentry{equation of motion} {
 name={equation of motion},
 description={},
}

\newglossaryentry{Poisson equation} {
 name={Poisson equation},
 description={},
}

\newglossaryentry{Vlasov equation} {
 name={Vlasov equation},
 description={},
}

\newglossaryentry{Euler's equation} {
 name={Euler's equation},
 description={},
}

\newglossaryentry{continuity equation} {
 name={continuity equation},
 description={},
}

\newglossaryentry{Newtonian gravity} {
 name={Newtonian gravity},
 description={},
}

\newglossaryentry{fluid} {
 name={fluid},
 description={},
}

\newglossaryentry{Hubble flow} {
 name={Hubble flow},
 description={},
}

\newglossaryentry{comoving coordinates} {
 name={comoving coordinates},
 description={},
}

\newglossaryentry{peculiar velocity} {
 name={peculiar velocity},
 description={},
}

\newglossaryentry{peculiar velocity flow} {
 name={peculiar velocity flow},
 description={},
}

\newglossaryentry{momentum} {
 name={momentum},
 description={},
}

\newglossaryentry{Vlasov-Poisson system} {
 name={Vlasov-Poisson system},
 description={},
}

\newglossaryentry{stress tensor} {
 name={stress tensor},
 description={},
}

\newglossaryentry{velocity dispersion} {
 name={velocity dispersion},
 description={},
}

\newglossaryentry{viscosity} {
 name={viscosity},
 description={},
}

\newglossaryentry{single-stream approximation} {
 name={single-stream approximation},
 description={},
}

\newglossaryentry{shell-crossing} {
 name={shell-crossing},
 description={},
}

\newglossaryentry{linear regime} {
 name={linear regime},
 description={},
}

\newglossaryentry{vorticity} {
 name={vorticity},
 description={},
}

\newglossaryentry{linear growth factor} {
 name={linear growth factor},
 description={},
}

\newglossaryentry{second-order growth factor} {
 name={second-order growth factor},
 description={},
}

\newglossaryentry{growing mode} {
 name={growing mode},
 description={},
}

\newglossaryentry{decaying mode} {
 name={decaying mode},
 description={},
}

\newglossaryentry{Einstein-de Sitter universe} {
 name={Einstein-de Sitter universe},
 description={},
}

\newglossaryentry{tidal tensor} {
 name={tidal tensor},
 description={},
}

\newglossaryentry{ZA} {
 name={ZA},
 description={Zel'dovich approximation},
}

\newglossaryentry{divergence of the Lagrangian displacement field} {
 name={divergence of the Lagrangian displacement field},
 description={},
}

\newglossaryentry{eigenvalue} {
 name={eigenvalue},
 description={},
}

\newglossaryentry{pancake} {
 name={pancake},
 description={},
}

\newglossaryentry{void} {
 name={void},
 description={},
}

\newglossaryentry{sheet} {
 name={sheet},
 description={},
}

\newglossaryentry{filament} {
 name={filament},
 description={},
}

\newglossaryentry{cluster} {
 name={cluster},
 description={},
}

\newglossaryentry{halo} {
 name={halo},
 description={},
}

\newglossaryentry{2LPT} {
 name={2LPT},
 description={second-order Lagrangian perturbation theory},
}

\newglossaryentry{non-local} {
 name={non-local},
 description={},
}

\newglossaryentry{local} {
 name={local},
 description={},
}

\newglossaryentry{tidal effects} {
 name={tidal effects},
 description={},
}

\newglossaryentry{free-particle approximation} {
 name={free-particle approximation},
 description={},
}

\newglossaryentry{Burgers' equation} {
 name={Burgers' equation},
 description={},
}

\newglossaryentry{Schrodinger equation} {
 name={Schr\"odinger equation},
 description={},
}

\newglossaryentry{adhesion approximation} {
 name={adhesion approximation},
 description={},
}

\newglossaryentry{velocity potential} {
 name={velocity potential},
 description={},
}

\newglossaryentry{gravitational field} {
 name={gravitational field},
 description={},
}

\newglossaryentry{non-magnetic approximation} {
 name={non-magnetic approximation},
 description={},
}

\newglossaryentry{frozen flow approximation} {
 name={frozen flow approximation},
 description={},
}

\newglossaryentry{linear potential approximation} {
 name={linear potential approximation},
 description={},
}

\newglossaryentry{local tidal approximation} {
 name={local tidal approximation},
 description={},
}

\newglossaryentry{diffusion equation} {
 name={diffusion equation},
 description={},
}

\newglossaryentry{characteristic function} {
 name={characteristic function},
 description={},
}

\newglossaryentry{Kac's theorem} {
 name={Kac's theorem},
 description={},
}

\newglossaryentry{Schur-complement} {
 name={Schur-complement},
 description={},
}

\newglossaryentry{Lagrangian potential} {
 name={Lagrangian potential},
 description={},
}

\newglossaryentry{full gravity} {
 name={full gravity},
 description={},
}

\newglossaryentry{particle realization} {
 name={particle realization},
 description={},
}

\newglossaryentry{VFF} {
 name={VFF},
 description={volume filling fraction},
}

\newglossaryentry{MFF} {
 name={MFF},
 description={mass filling fraction},
}

\newglossaryentry{reconstruction} {
 name={reconstruction},
 description={},
}

\newglossaryentry{non-linear regime} {
 name={non-linear regime},
 description={},
}

\newglossaryentry{mildly non-linear regime} {
 name={mildly non-linear regime},
 description={},
}

\newglossaryentry{displacement field} {
 name={displacement field},
 description={},
}

\newglossaryentry{Gadget-2} {
 name={Gadget-2},
 description={},
}

\newglossaryentry{2LPTic} {
 name={2LPTic},
 description={},
}

\newglossaryentry{N-GenIC} {
 name={N-GenIC},
 description={},
}

\newglossaryentry{CiC} {
 name={CiC},
 description={cloud-in-cell},
}

\newglossaryentry{NGP} {
 name={NGP},
 description={nearest grid point},
}

\newglossaryentry{TSC} {
 name={TSC},
 description={triangular shaped cloud},
}

\newglossaryentry{Cosmic Emulator} {
 name={Cosmic Emulator},
 description={},
}

\newglossaryentry{WMAP-7} {
 name={WMAP-7},
 description={},
}

\newglossaryentry{ZARM} {
 name={ZARM},
 description={Zel'dovich approximation remapped},
}

\newglossaryentry{2LPTRM} {
 name={2LPTRM},
 description={second-order Lagrangian perturbation theory remapped},
}

\newglossaryentry{remapping} {
 name={remapping},
 description={},
}

\newglossaryentry{skewness} {
 name={skewness},
 description={},
}

\newglossaryentry{excess kurtosis} {
 name={excess kurtosis},
 description={},
}

\newglossaryentry{aliasing} {
 name={aliasing},
 description={},
}

\newglossaryentry{Nyquist wavenumber} {
 name={Nyquist wavenumber},
 description={},
}

\newglossaryentry{cross-correlation} {
 name={cross-correlation},
 description={},
}

\newglossaryentry{cosmic variance} {
 name={cosmic variance},
 description={},
}

\newglossaryentry{phase} {
 name={phase},
 description={},
}

\newglossaryentry{three-point correlation function} {
 name={three-point correlation function},
 description={},
}

\newglossaryentry{shot noise} {
 name={shot noise},
 description={},
}

\newglossaryentry{3LPT} {
 name={3LPT},
 description={third-order Lagrangian perturbation theory},
}

\newglossaryentry{vector part} {
 name={vector part},
 description={},
}

\newglossaryentry{SC} {
 name={SC},
 description={spherical collapse},
}

\newglossaryentry{ALPT} {
 name={ALPT},
 description={Augmented Lagrangian perturbation theory},
}

\newglossaryentry{muscle} {
 name={MUSCLE},
 description={Multiscale spherical collapse},
}

\newglossaryentry{local Lagrangian approximations} {
 name={local Lagrangian approximations},
 description={},
}

\newglossaryentry{Gaussian kernel} {
 name={Gaussian kernel},
 description={},
}

\newglossaryentry{void-in-cloud} {
 name={void-in-cloud},
 description={},
}

\newglossaryentry{void-in-void} {
 name={void-in-void},
 description={},
}

\newglossaryentry{scalar part} {
 name={scalar part},
 description={},
}

\newglossaryentry{Helmholtz decomposition} {
 name={Helmholtz decomposition},
 description={},
}

\newglossaryentry{structure type} {
 name={structure type},
 description={},
}

\newglossaryentry{large-scale structure inference} {
 name={large-scale structure inference},
 description={},
}

\newglossaryentry{data assimilation} {
 name={data assimilation},
 description={},
}

\newglossaryentry{probability theory} {
 name={probability theory},
 description={},
}

\newglossaryentry{Bayesian statistics} {
 name={Bayesian statistics},
 description={},
}

\newglossaryentry{extended logic} {
 name={extended logic},
 description={},
}

\newglossaryentry{no-free lunch theorem} {
 name={no-free lunch theorem},
 description={},
}

\newglossaryentry{MCMC} {
 name={MCMC},
 description={Markov Chain Monte Carlo},
}

\newglossaryentry{prior} {
 name={prior},
 description={},
}

\newglossaryentry{posterior} {
 name={posterior},
 description={},
}

\newglossaryentry{likelihood} {
 name={likelihood},
 description={},
}

\newglossaryentry{evidence} {
 name={evidence},
 description={},
}

\newglossaryentry{prior choice} {
 name={prior choice},
 description={},
}

\newglossaryentry{parameter inference} {
 name={parameter inference},
 description={},
}

\newglossaryentry{borg} {
 name={BORG},
 description={Bayesian Origin Reconstruction from Galaxies},
}

\newglossaryentry{ares} {
 name={ARES},
 description={Algorithm for REconstruction and Sampling},
}

\newglossaryentry{hades} {
 name={HADES},
 description={HAmiltonian Density Estimation and Sampling},
}

\newglossaryentry{cola} {
 name={COLA},
 description={COmoving Lagrangian Acceleration},
}

\newglossaryentry{vide} {
 name={VIDE},
 description={Void IDentification and Examination toolkit},
}

\newglossaryentry{diva} {
 name={DIVA},
 description={DynamIcal Void Analysis},
}

\newglossaryentry{origami} {
 name={ORIGAMI},
 description={Order-ReversIng Gravity, Apprehended Mangling Indices},
}

\newglossaryentry{frequentist statistics} {
 name={frequentist statistics},
 description={},
}

\newglossaryentry{plausible reasoning} {
 name={plausible reasoning},
 description={},
}

\newglossaryentry{inference} {
 name={inference},
 description={},
}

\newglossaryentry{HMC} {
 name={HMC},
 description={Hamiltonian Monte Carlo},
}

\newglossaryentry{definition of probability} {
 name={probability (definition)},
 description={},
}

\newglossaryentry{estimator} {
 name={estimator},
 description={},
}

\newglossaryentry{hypothesis testing} {
 name={hypothesis testing},
 description={},
}

\newglossaryentry{Cox-Jaynes theorem} {
 name={Cox-Jaynes theorem},
 description={},
}

\newglossaryentry{Cox's desiderata} {
 name={Cox's desiderata},
 description={},
}

\newglossaryentry{plausibility} {
 name={plausibility},
 description={},
}

\newglossaryentry{maximum-entropy} {
 name={maximum-entropy},
 description={},
}

\newglossaryentry{information theory} {
 name={information theory},
 description={},
}

\newglossaryentry{machine learning} {
 name={machine learning},
 description={},
}

\newglossaryentry{forward modeling} {
 name={forward modeling},
 description={},
}

\newglossaryentry{inverse problem} {
 name={inverse problem},
 description={},
}

\newglossaryentry{Bayes' theorem} {
 name={Bayes' theorem},
 description={},
}

\newglossaryentry{data} {
 name={data},
 description={},
}

\newglossaryentry{photometric redshift} {
 name={photometric redshift},
 description={},
}

\newglossaryentry{noise} {
 name={noise},
 description={},
}

\newglossaryentry{mask} {
 name={mask},
 description={},
}

\newglossaryentry{selection effects} {
 name={selection effects},
 description={},
}

\newglossaryentry{survey geometry} {
 name={survey geometry},
 description={},
}

\newglossaryentry{Cromwell's rule} {
 name={Cromwell's rule},
 description={},
}

\newglossaryentry{Bernstein-von Mises theorem} {
 name={Bernstein-von Mises theorem},
 description={},
}

\newglossaryentry{constrained likelihood} {
 name={constrained likelihood},
 description={},
}

\newglossaryentry{Jeffreys' priors} {
 name={Jeffreys' priors},
 description={},
}

\newglossaryentry{flat prior} {
 name={flat prior},
 description={},
}

\newglossaryentry{proper prior} {
 name={proper prior},
 description={},
}

\newglossaryentry{nuisance parameters} {
 name={nuisance parameters},
 description={},
}

\newglossaryentry{exploration of the posterior} {
 name={exploration of the posterior},
 description={},
}

\newglossaryentry{high-dimensional parameter space} {
 name={high-dimensional parameter space},
 description={},
}

\newglossaryentry{sampling} {
 name={sampling},
 description={},
}

\newglossaryentry{sample} {
 name={sample},
 description={},
}

\newglossaryentry{systematic uncertainty} {
 name={systematic uncertainty},
 description={},
}

\newglossaryentry{statistical uncertainty} {
 name={statistical uncertainty},
 description={},
}

\newglossaryentry{ABC} {
 name={ABC},
 description={Approximate Bayesian Computation},
}

\newglossaryentry{likelihood-free methods} {
 name={likelihood-free methods},
 description={},
}

\newglossaryentry{Occam's razor} {
 name={Occam's razor},
 description={},
}

\newglossaryentry{non-committal prior} {
 name={non-committal prior},
 description={},
}

\newglossaryentry{Bayes factor} {
 name={Bayes factor},
 description={},
}

\newglossaryentry{Jeffreys' scale} {
 name={Jeffreys' scale},
 description={},
}

\newglossaryentry{nested model} {
 name={nested model},
 description={},
}

\newglossaryentry{Savage-Dickey ratio} {
 name={Savage-Dickey ratio},
 description={},
}

\newglossaryentry{transition probability} {
 name={transition probability},
 description={},
}

\newglossaryentry{invariant distribution} {
 name={invariant distribution},
 description={},
}

\newglossaryentry{CosmoMC} {
 name={CosmoMC},
 description={},
}

\newglossaryentry{burn-in} {
 name={burn-in},
 description={},
}

\newglossaryentry{MH} {
 name={MH},
 description={Metropolis-Hastings},
}

\newglossaryentry{acceptance rate} {
 name={acceptance rate},
 description={},
}

\newglossaryentry{proposal distribution} {
 name={proposal distribution},
 description={},
}

\newglossaryentry{Hastings ratio} {
 name={Hastings ratio},
 description={},
}

\newglossaryentry{Metropolis ratio} {
 name={Metropolis ratio},
 description={},
}

\newglossaryentry{detailed balance} {
 name={detailed balance},
 description={},
}

\newglossaryentry{reversibility} {
 name={reversibility},
 description={},
}

\newglossaryentry{symplecticity} {
 name={symplecticity},
 description={},
}

\newglossaryentry{auto-correlation function (Markov chain)} {
 name={auto-correlation function (Markov chain)},
 description={},
}

\newglossaryentry{Gibbs sampling} {
 name={Gibbs sampling},
 description={},
}

\newglossaryentry{classical mechanics} {
 name={classical mechanics},
 description={},
}

\newglossaryentry{Hamilton's equations} {
 name={Hamilton's equations},
 description={},
}

\newglossaryentry{Liouville's theorem} {
 name={Liouville's theorem},
 description={},
}

\newglossaryentry{canonical distribution} {
 name={canonical distribution},
 description={},
}

\newglossaryentry{Boltzmann constant} {
 name={Boltzmann constant},
 description={},
}

\newglossaryentry{partition function} {
 name={partition function},
 description={},
}

\newglossaryentry{statistical mechanics} {
 name={statistical mechanics},
 description={},
}

\newglossaryentry{kick} {
 name={kick},
 description={},
}

\newglossaryentry{drift} {
 name={drift},
 description={},
}

\newglossaryentry{mass matrix} {
 name={mass matrix},
 description={},
}

\newglossaryentry{Metropolis update} {
 name={Metropolis update},
 description={},
}

\newglossaryentry{prior volume} {
 name={prior volume},
 description={},
}

\newglossaryentry{posterior odds} {
 name={posterior odds},
 description={},
}

\newglossaryentry{data model} {
 name={data model},
 description={},
}

\newglossaryentry{galaxy formation} {
 name={galaxy formation},
 description={},
}

\newglossaryentry{lightcone} {
 name={lightcone},
 description={},
}

\newglossaryentry{curse of dimensionality} {
 name={curse of dimensionality},
 description={},
}

\newglossaryentry{machine epsilon} {
 name={machine epsilon},
 description={},
}

\newglossaryentry{mock catalog} {
 name={mock catalog},
 description={},
}

\newglossaryentry{high-dimensional function} {
 name={high-dimensional function},
 description={},
}

\newglossaryentry{formation history} {
 name={formation history},
 description={},
}

\newglossaryentry{Euler's method} {
 name={Euler's method},
 description={},
}

\newglossaryentry{noise parameter} {
 name={noise parameter},
 description={},
}

\newglossaryentry{principal component analysis} {
 name={principal component analysis},
 description={},
}

\newglossaryentry{large-scale structure likelihood} {
 name={large-scale structure likelihood},
 description={},
}

\newglossaryentry{physical density prior} {
 name={physical density prior},
 description={},
}

\newglossaryentry{posterior mean} {
 name={posterior mean},
 description={},
}

\newglossaryentry{posterior standard deviation} {
 name={posterior standard deviation},
 description={},
}

\newglossaryentry{final conditions} {
 name={final conditions},
 description={},
}

\newglossaryentry{declination} {
 name={declination},
 description={},
}

\newglossaryentry{right ascension} {
 name={right ascension},
 description={},
}

\newglossaryentry{Poisson likelihood} {
 name={Poisson likelihood},
 description={},
}

\newglossaryentry{Poisson process} {
 name={Poisson process},
 description={},
}

\newglossaryentry{Poisson intensity field} {
 name={Poisson intensity field},
 description={},
}

\newglossaryentry{luminosity} {
 name={luminosity},
 description={},
}

\newglossaryentry{Schechter luminosity functions} {
 name={Schechter luminosity functions},
 description={},
}

\newglossaryentry{survey response operator} {
 name={survey response operator},
 description={},
}

\newglossaryentry{Gamma distribution} {
 name={$\Gamma$-distribution},
 description={},
}

\newglossaryentry{KDK} {
 name={KDK},
 description={kick drift kick},
}

\newglossaryentry{leapfrog} {
 name={leapfrog},
 description={},
}

\newglossaryentry{periodic boundary conditions} {
 name={periodic boundary conditions},
 description={},
}

\newglossaryentry{chrono-cosmography} {
 name={chrono-cosmography},
 description={},
}

\newglossaryentry{spectroscopic redshift} {
 name={spectroscopic redshift},
 description={},
}

\newglossaryentry{Lagrangian transport} {
 name={Lagrangian transport},
 description={},
}

\newglossaryentry{information content} {
 name={information content},
 description={},
}

\newglossaryentry{uncertainty quantification} {
 name={uncertainty quantification},
 description={},
}

\newglossaryentry{non-linear filtering} {
 name={non-linear filtering},
 description={},
}

\newglossaryentry{constrained simulation} {
 name={constrained simulation},
 description={},
}

\newglossaryentry{dark matter void} {
 name={dark matter void},
 description={},
}

\newglossaryentry{PM} {
 name={PM},
 description={particle-mesh},
}

\newglossaryentry{mode coupling} {
 name={mode coupling},
 description={},
}

\newglossaryentry{remapping function} {
 name={remapping function},
 description={},
}

\newglossaryentry{transfer function} {
 name={transfer function},
 description={},
}

\newglossaryentry{decision theory} {
 name={decision theory},
 description={},
}

\newglossaryentry{cosmostatistics} {
 name={cosmostatistics},
 description={},
}

\newglossaryentry{renormalized perturbation theory} {
 name={renormalized perturbation theory},
 description={},
}

\newglossaryentry{path integral formalism} {
 name={path integral formalism},
 description={},
}

\newglossaryentry{renormalization group flow} {
 name={renormalization group flow},
 description={},
}

\newglossaryentry{Gaussianization} {
 name={Gaussianization},
 description={},
}

\newglossaryentry{cdf} {
 name={cdf},
 description={cumulative distribution function},
}

\newglossaryentry{21 cm surveys} {
 name={21 cm surveys},
 description={},
}

\newglossaryentry{Lyman-alpha forest} {
 description={Lyman-$\alpha$ forest},
 name={},
}

\newglossaryentry{reionization} {
 name={reionization},
 description={},
}

\newglossaryentry{sparsity} {
 name={sparsity},
 description={},
}

\newglossaryentry{compensation} {
 name={compensation},
 description={},
}

\newglossaryentry{density profile} {
 name={density profile},
 description={},
}

\newglossaryentry{galaxy void} {
 name={galaxy void},
 description={},
}

\newglossaryentry{number function} {
 name={number function},
 description={},
}

\newglossaryentry{ellipticity distribution} {
 name={ellipticity distribution},
 description={},
}

\newglossaryentry{void hierarchy} {
 name={void hierarchy},
 description={},
}

\newglossaryentry{weak gravitational lensing} {
 name={weak gravitational lensing},
 description={},
}

\newglossaryentry{integrated Sachs-Wolfe effect} {
 name={integrated Sachs-Wolfe effect},
 description={},
}

\newglossaryentry{Rees-Sciama effect} {
 name={Rees-Sciama effect},
 description={},
}

\newglossaryentry{Blackwell-Rao estimator} {
 name={Blackwell-Rao estimator},
 description={},
}

\newglossaryentry{velocity profile} {
 name={velocity profile},
 description={},
}

\newglossaryentry{watershed transform} {
 name={watershed transform},
 description={},
}

\newglossaryentry{Voronoi tessellation} {
 name={Voronoi tessellation},
 description={},
}

\newglossaryentry{tidal field} {
 name={tidal field},
 description={},
}

\newglossaryentry{Sunyaev-Zel'dovich effect} {
 name={Sunyaev-Zel'dovich effect},
 description={},
}

\newglossaryentry{mass resolution} {
 name={mass resolution},
 description={},
}

\newglossaryentry{T-web} {
 name={T-web},
 description={},
}

\newglossaryentry{V-web} {
 name={V-web},
 description={},
}

\newglossaryentry{velocity shear field} {
 name={velocity shear field},
 description={},
}

\newglossaryentry{Kullback-Leibler divergence} {
 name={Kullback-Leibler divergence},
 description={},
}

\newglossaryentry{entropy} {
 name={entropy},
 description={},
}

\newglossaryentry{mesh assignment} {
 name={mesh assignment},
 description={},
}

\newglossaryentry{utility function} {
 name={utility function},
 description={},
}

\newglossaryentry{gain function} {
 name={gain function},
 description={},
}

\newglossaryentry{fair game} {
 name={fair game},
 description={},
}

\newglossaryentry{risk aversion} {
 name={risk aversion},
 description={},
}

\newglossaryentry{speculative map} {
 name={speculative map},
 description={},
}

\newglossaryentry{conservative map} {
 name={conservative map},
 description={},
}

\newglossaryentry{Green function} {
 name={Green function},
 description={},
}

\newglossaryentry{Fourier transform} {
 name={Fourier transform},
 description={},
}

\newglossaryentry{interpolation} {
 name={interpolation},
 description={},
}

\newglossaryentry{assignment function} {
 name={assignment function},
 description={},
}

\newglossaryentry{shape function} {
 name={shape function},
 description={},
}

\newglossaryentry{Nyquist-Shannon sampling theorem} {
 name={Nyquist-Shannon sampling theorem},
 description={},
}

\newglossaryentry{FDA} {
 name={FDA},
 description={finite difference approximation},
}

\newglossaryentry{Box-Muller method} {
 name={Box-M\"uller method},
 description={},
}

\newglossaryentry{white noise} {
 name={white noise},
 description={},
}

\newglossaryentry{Cholesky decomposition} {
 name={Cholesky decomposition},
 description={},
}

\newglossaryentry{low-pass filter} {
 name={low-pass filter},
 description={},
}

\newglossaryentry{Hubble radius} {
 name={Hubble radius},
 description={},
}

\newglossaryentry{phase space} {
 name={phase space},
 description={},
}

\newglossaryentry{zobov} {
 name={ZOBOV},
 description={},
}

\newglossaryentry{WDM} {
 name={WDM},
 description={warm dark matter},
}

\newglossaryentry{isocurvature perturbations} {
 name={isocurvature perturbations},
 description={},
}

\newglossaryentry{CMB lensing} {
 name={CMB lensing},
 description={},
}

\newglossaryentry{Wiener filter} {
 name={Wiener filter},
 description={},
}

\newcommand{\borg}{\glslink{borg}{\textsc{borg}}}
\newcommand{\ares}{\glslink{ares}{\textsc{ares}}}
\newcommand{\hades}{\glslink{hades}{\textsc{hades}}}
\newcommand{\cola}{\glslink{cola}{\textsc{cola}}}
\newcommand{\diva}{\glslink{diva}{\textsc{diva}}}
\newcommand{\origami}{\glslink{origami}{\textsc{origami}}}
\newcommand{\vide}{\glslink{vide}{\textsc{vide}}}
\newcommand{\LCDM}{\glslink{LCDM}{$\Lambda$CDM}}

\begin{document}

\frontmatter


\begin{titlepage}
\vspace*{-2.5cm}

\begin{changemargin}{-0.5cm}{-0.5cm}
\includegraphics[height=140px]{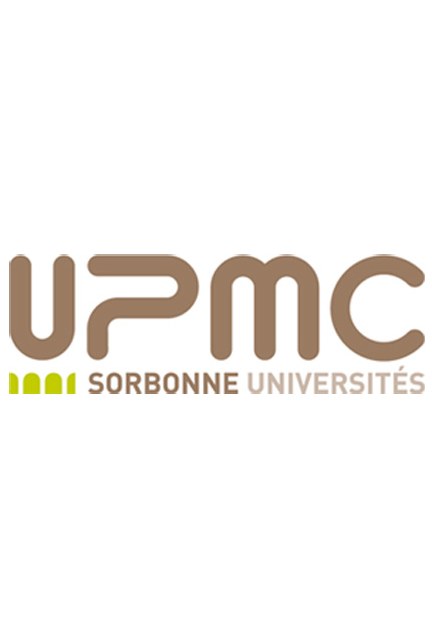} \hfill \includegraphics[height=120px]{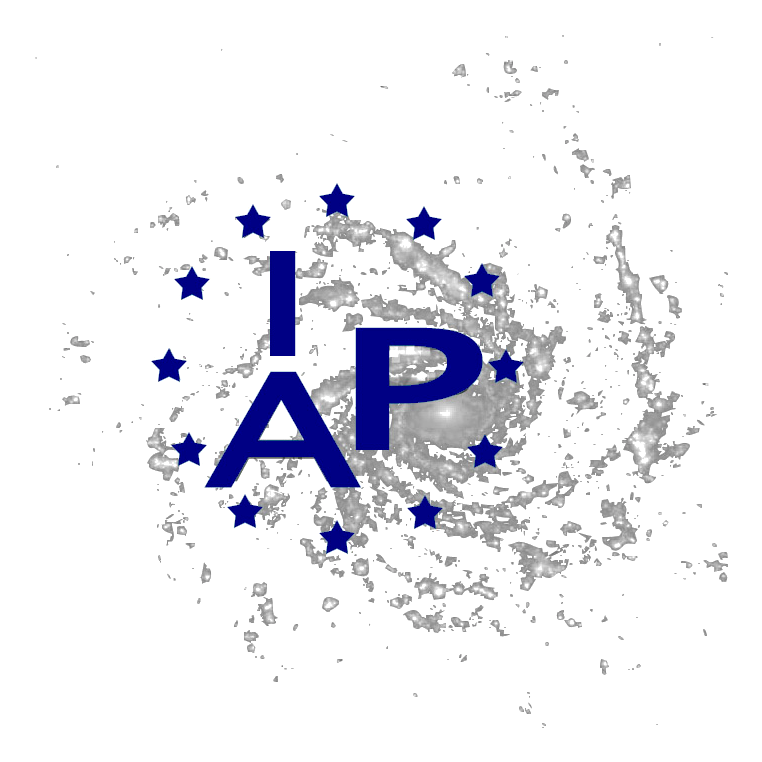} \hfill \includegraphics[height=140px]{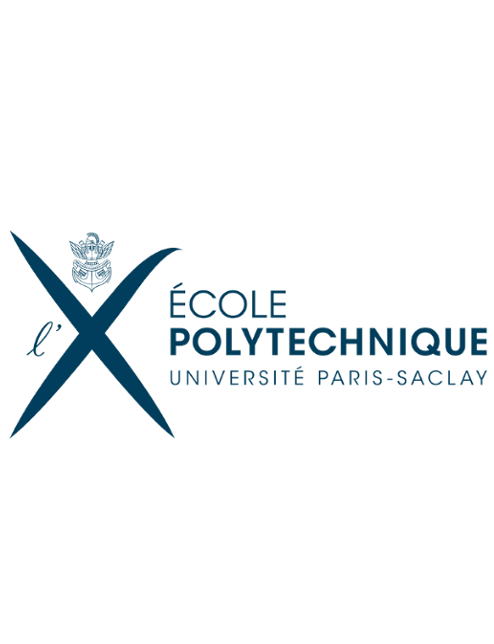}
\end{changemargin}

\begin{center}
\vspace*{0.5cm}
{\LARGE \textsc{Université Pierre et Marie Curie}} \\*
\vspace*{0.5cm}
{\Large \textsc{École doctorale d'Astronomie et d'Astrophysique d'Île-de-France (ED127)}} \\*
\vspace*{0.5cm}
{\LARGE \textsc{Institut d'Astrophysique de Paris}} \\

\vspace*{1.5cm}
{\huge \sffamily \textbf{Bayesian large-scale structure inference\\ and cosmic web analysis}} \\
\vspace*{1.0cm}
{\Large \textsc{Thèse de doctorat}} \\
\vspace*{0.5cm}
{\large \textsc{Discipline : Cosmologie}} \\
\vspace*{0.5cm}
{\large présentée et soutenue publiquement le 24 septembre 2015} \\
\vspace*{0.5cm}
{par} \\
\vspace*{0.5cm}
{\huge Florent \textsc{Leclercq}} \\
\vspace*{2.5cm}
{\large devant un jury composé de :} \\
\vspace*{0.5cm}
\begin{changemargin}{-0.7cm}{-0.7cm}
\large{\begin{tabular}{lll}
    Bertrand \textsc{Laforge} & Professeur, Université Pierre et Marie Curie & \textit{Président du jury}\\
	Benjamin \textsc{Wandelt} & Professeur, Université Pierre et Marie Curie & \textit{Directeur de thèse}\\
	Oliver \textsc{Hahn} & Professeur, Université Nice Sophia Antipolis & \textit{Rapporteur}\\
	Alan \textsc{Heavens} & Professeur, Imperial College London & \textit{Rapporteur}\\
	Ofer \textsc{Lahav} & Professeur, University College London & \textit{Examinateur}\\
	Will \textsc{Percival} & Professeur, University of Portsmouth & \textit{Examinateur}\\
	Rien \textsc{van de Weygaert} & Professeur, University of Groningen & \textit{Invité}\\
	Mat\'ias \textsc{Zaldarriaga} & Professeur, Institute for Advanced Study, Princeton & \textit{Invité}
\end{tabular}}
\end{changemargin}
\end{center}

\end{titlepage}
\sloppy

\titlepage


\cleardoublepage

\begin{vcenterpage}

\noindent Because all things balance – as on a wheel – and we cannot see nine-tenths of what is real,\\ our claims of self-reliance are pieced together by unpanned gold.
\smallskip

\defcitealias{Reeve1995}{Franklin D'Olier}
\noindent --- \citetalias{Reeve1995} \citet{Reeve1995}, \textit{Coasting}

\bigskip
\bigskip

\noindent There's gold, and it's haunting and haunting;\\
\indent It's luring me on as of old;\\
\noindent Yet it isn't the gold that I'm wanting\\
\indent So much as just finding the gold.\\
\noindent It's the great, big, broad land 'way up yonder,\\
\indent It's the forests where silence has lease;\\
\noindent It's the beauty that thrills me with wonder,\\
\indent It's the stillness that fills me with peace.
\smallskip

\defcitealias{Service1907}{Robert William}
\noindent --- \citetalias{Service1907} \citet{Service1907}, \textit{The Spell of the Yukon}


\end{vcenterpage}

\cleardoublepage

\include{Preface/PrefaceContent}

\tableofcontents

\mainmatter

\include{Intro/IntroContent}
\include{Chapter1/Chapter1Content}
\include{Chapter2/Chapter2Content}
\include{Chapter3/Chapter3Content}
\include{Chapter4/Chapter4Content}
\include{Chapter5/Chapter5Content}
\include{Chapter6/Chapter6Content}
\include{Chapter7/Chapter7Content}
\include{Chapter8/Chapter8Content}
\include{Chapter9/Chapter9Content}
\include{Chapter10/Chapter10Content}

\include{Conclusion/ConclusionContent}

\clearpage
\appendix

\include{AppendixA/AppendixAContent}
\include{AppendixB/AppendixBContent}

\include{AppendixC/AppendixCContent}

\bibliography{Thesis}

\newpage
\begin{multicols}{2}
\printglossary[title=Index]
\addstarredchapter{Index}
\end{multicols}



\cleardoublepage
\begin{vcenterpage}
\noindent\rule[2pt]{\textwidth}{0.5pt}
\begin{center}
{\large\textbf{Bayesian large-scale structure inference and cosmic web analysis\\}}
\end{center}
\noindent Surveys of the cosmic large-scale structure carry opportunities for building and testing cosmological theories about the origin and evolution of the Universe. This endeavor requires appropriate data assimilation tools, for establishing the contact between survey catalogs and models of structure formation.

\noindent In this thesis, we present an innovative statistical approach for the \textit{ab initio} simultaneous analysis of the formation history and morphology of the cosmic web: the \textsc{borg} algorithm infers the primordial density fluctuations and produces physical reconstructions of the dark matter distribution that underlies observed galaxies, by assimilating the survey data into a cosmological structure formation model. The method, based on Bayesian probability theory, provides accurate means of uncertainty quantification.

\noindent We demonstrate the application of \textsc{borg} to the Sloan Digital Sky Survey data and describe the primordial and late-time large-scale structure in the observed volume. We show how the approach has led to the first quantitative inference of the cosmological initial conditions and of the formation history of the observed structures. We then use these results for several cosmographic projects aiming at analyzing and classifying the large-scale structure. In particular, we build an enhanced catalog of cosmic voids probed at the level of the dark matter
 distribution, deeper than with the galaxies. We present detailed probabilistic maps of the dynamic cosmic web, and offer a general solution to the problem of classifying structures in the presence of uncertainty.

\noindent The results described in this thesis constitute accurate chrono-cosmography of the inhomogeneous cosmic structure.

\noindent\rule[2pt]{\textwidth}{0.5pt}

\begin{center}
{\large\textbf{Inférence bayésienne et analyse des grandes structures de l'Univers\\}}
\end{center}
\noindent Les observations de la structure à grande échelle de l'Univers sont précieuses pour établir et tester des théories cosmologiques sur son origine et son évolution. Cette démarche requiert des outils appropriés d'assimilation des données, afin d'établir le contact entre les catalogues de galaxies et les modèles de formation des structures.

\noindent Dans cette thèse, une nouvelle approche pour l'analyse \textit{ab initio} et simultanée de la formation et de la morphologie de la toile cosmique est présentée : l'algorithme \textsc{borg} infère les fluctuations de densité primordiales et produit des reconstructions physiques de la distribution de matière noire, en assimilant les relevés de galaxies dans un modèle cosmologique de formation des structures. La méthode, basée sur la théorie bayésienne des probabilités, fournit un moyen de quantifier précisément les incertitudes.

\noindent On présente l'application de \textsc{borg} aux données du Sloan Digital Sky Survey et on décrit la structure de l'Univers dans le volume considéré. On démontre que cette approche a mené à la première inférence quantitative des conditions initiales et du scénario de formation des structures observées. On utilise ces résultats pour plusieurs projets cosmographiques visant à analyser et classifier la toile cosmique. En particulier, on construit un catalogue de vides, décrits au niveau de la matière noire et non des galaxies. On présente des cartes probabilistes détaillées de la dynamique de la toile cosmique et on propose une solution générale pour la classification des structures en présence d'incertitude.

\noindent Les résultats de cette thèse constituent une précise description chrono-cosmographique des inhomogénéités de la structure cosmique.

\noindent\rule[2pt]{\textwidth}{0.5pt}
\end{vcenterpage}

\end{document}

%% file: Preface/PrefaceContent.tex
\chapter*{Preface}

\defcitealias{London1903}{Jack}
\begin{flushright}
\begin{minipage}[c]{0.6\textwidth}
\rule{\columnwidth}{0.4pt}

``But especially he loved to run in the dim twilight of the summer midnights, listening to the subdued and sleepy murmurs of the forest, reading signs and sounds as a man may read a book, and seeking for the mysterious something that called -- called, waking or sleeping, at all times, for him to come.''\\
--- \citetalias{London1903} \citet{London1903}, \textit{The Call of the Wild}

\vspace{-5pt}\rule{\columnwidth}{0.4pt}
\end{minipage}
\end{flushright}

This PhD thesis is written as completion of my work at the Institut d'Astrophysique de Paris from 2012 to 2015. Three years ago, I started this project with heartfelt enthusiasm, but it turned out to be much more than expected -- an incredibly rewarding journey. This thesis is the report of this long process. It expresses my vision -- that I have had incredible trouble in organizing linearly -- of the final scientific products.

Unfortunately, it cannot describe the thought process and the strange mechanism by which something out in there in the Universe, be it a galaxy cluster, a scientific tool, or a concept -- something that a month ago was a stranger -- becomes intimate. Neither does it capture my feelings during the long days -- and nights -- spent in front of a black board, a paper or a computer, alone or in the lab: the sadness and tiredness with failed endeavors, the bittersweet taste of learning I was wrong, the hope for good results, the joy for successes, and the wonder at the elegance of the cosmos.

As a consequence, this preface takes the occasion to describe how this thesis came into being.

\section*{Stars and physical sciences}

I have always thought that modern physics is fascinating. Not only does it manipulate extraordinary ideas and concepts (quantum mechanics, relativity), it also deals with important societal issues (energy, natural resources). In this respect, I am fascinated by how far the physical sciences, in just a few hundred years, have taken us in our understanding of nature. Given this incredible evolution, it is amazing to realize that looking at the sky -- the amusement during the warm summers in ancient Greece -- still gathers so much attention. This strange mixture of tradition and modernity may be the reason why, as far as I remember, I have always had a particular attraction for astrophysics.

For most of the history of humanity, cosmology was part of religion or metaphysics. Only recently did it become a science, its peculiarity being the uniqueness of its object of interest -- the Universe as a whole. The idea that the entire Universe can be treated as a physical system was one of the most striking revelation of my life as a student. It is commonly predicted that early 21st century cosmology is on the verge of a revolution. In upcoming years, surprising or unexpected results may or may not be found, but I believe that cosmology will stay one of humanity's greatest intellectual endeavors, and certainly the one that has produced the deepest description of the natural world as we find it.

\section*{Models and beliefs}

Contrary to my long-held passion for the Universe, my interest in probability theory came in a rather fortuitous manner. But faced with the immodest, enthralling questions of cosmology, one soon realizes that there is no absolute truth, only beliefs. Colleagues showed me that this viewpoint makes all the problems of modern cosmology appear in a very different light. Then, in a quick succession, reading about probability theory, which truly is the ``logic of science'', in the words of \citet{Jaynes2003}, made me realize that in the much larger and permanent world of plausible reasoning, i.e. rational thinking in the presence of uncertainty, the current problems of physics appear only as details: what matters is the road, not the destination. 

Few, if any, of our ordinary-life beliefs are certain to the degree that we cannot imagine them being overthrown by
sufficient contradictory evidence. Similarly, typical commonsense inferences rely on applying rules that are general, but not universal. Therefore, deduction does not entirely characterize commonsense reasoning. This thesis exploits theories of inductive and abductive logic, to draw and assess the strength of conclusions from uncertain rules and partial evidence.

Probability theory, when seen as an extension of ordinary logic incorporates the description of randomness but also statistical inference and becomes a field of logical unity and simplicity. It allows us to solve problems of great complexity, and reproduces many aspects of human cognitive activity, often in disturbing detail. In doing so, it captures something about how our minds operate when we form inductive judgments, of which we may not be consciously aware. This aspect takes a very particular meaning when we deal with the Universe.

\section*{Why bother about this thesis?}

This thesis focuses on methodology for the analysis of the large-scale structure of the Universe. I should say from the beginning that the methods presented do not have the same degree of maturity as standard techniques for the analysis of galaxy surveys. So why should the reader bother?
\begin{enumerate}
\item I believe that new solutions to complex problems involving both data and uncertainty are needed to exploit the full potential of future, but also of existing surveys. However many data sets we record and analyze, if we use the same old models without questioning them, we will always miss the same crucially important feature that the experiment was competent to find, of which we may not be aware. If we want to detect any phenomenon, we must have a data model that at least allows the possibility that it may exist.
\item Innovative methods also allow the possibility to crosscheck the validity of cosmological analyses that are widely accepted. A false premise or a confirmation bias, when built into a model that is never questioned, cannot be removed by any amount of new data. Only a fresh look can get rid of that.
\end{enumerate}

In the hope to make progress in the analysis of the cosmic large-scale structure, this thesis tries to develop a ``healthy disrespect for tradition and authority, which have retarded progress throughout the 20th century'' \citep{Jaynes2003}.

\section*{How to read this thesis}

The thesis is divided into four parts. Part \ref{part:I} is a preparatory discussion on the analytical and numerical description of the large-scale structure. The heart of this thesis is part \ref{part:II} on Bayesian large-scale structure inference and part \ref{part:IV} on cosmic web analysis. The transitional part \ref{part:III} focuses on the non-linear regime of structure formation.

It is my hope that at least some chapters are written in a sufficiently engaging textbook style for graduate students. This should be the case in particular for chapter \ref{chap:theory} on structure formation, chapter \ref{chap:stats} on Bayesian statistics and appendix \ref{apx:simulations} on numerical simulations. There, except in section \ref{sec:homogeneous}, I have struggled to avoid the dreadful sentence ``It can be shown that...'' as much as possible. I have tried to give references to the original literature whenever it is possible, but I certainly did not attempt a true bibliography as can be found in excellent review papers on the large-scale structure and on probability theory. Chapter \ref{chap:BORG} describes the \textsc{borg} algorithm, which is the basis this entire thesis. It gathers information scattered through published journal papers, and hence is also intended for reference use. The rest of the thesis directly draws from the research papers that have been published during my PhD work: chapter \ref{chap:lpt} from \citet{Leclercq2013} and its addendum, \citet{Leclercq2015ADDENDUM}, chapter \ref{chap:BORGSDSS} from \citet{Jasche2015BORGSDSS}, chapter \ref{chap:remapping} from \citet{Leclercq2013}, chapter \ref{chap:filtering} from \citet{Leclercq2015DMVOIDS} and \citet{Leclercq2015ST}, chapter \ref{chap:dmvoids} from \citet{Leclercq2015DMVOIDS}, chapter \ref{chap:ts} from \citet{Leclercq2015ST}, and chapter \ref{chap:decision} from \citet{Leclercq2015DT}. There, the style becomes more succinct and the aim is rather to describe specific projects, give a guide to the literature and report on the results obtained.

Considering that this thesis is a rare occasion to include whatever I want in a research work, I decided to tackle the difficult task of choosing epigraphs. The various quotes spread throughout the thesis may or may not have something to do with the main text. Some are inspirational, thought-provoking, some are openly provocative, some can just be considered as Easter eggs, and some are just there for free.

The online version of thesis will be revised to correct for any mistakes, typographical and otherwise, found after it goes to press and archiving. I will try to maintain a list of corrections on my website, currently hosted at \href{http://www2.iap.fr/users/leclercq/}{\texttt{http://www2.iap.fr/users/leclercq/}}. Please feel free to send me any comments at \texttt{florent.leclercq@polytechnique.org}.

\comment{
I try to give references to the original literature whenever it is possible, but I certainly did not attempt a true bibliography as can be found in excellent review papers on the large-scale structure and on probability theory. In these situations, I may have suffered from that puzzling peculiarity of the human mind of tending to favor my own work. Contrary maybe to what some people would demand, I do not apologize for that in any way.

The choice of topics is certainly not encyclopedic. 

The online version of thesis will be revised to correct for any mistakes, typographical and otherwise, found after it goes to press. I will try to maintain a list of corrections on my website, currently hosted at \href{http://www2.iap.fr/users/leclercq/}{\texttt{http://www2.iap.fr/users/leclercq/}}.

Please feel free to send me any comments at \texttt{florent.leclercq@polytechnique.org}.

So the various quotes may or may not have something to do with the may text. Some are inspirational, thought-provoking, some are openly provocative, some may just be considered as Easter eggs, and some are just here for free. Like this one:
}

\newpage

\section*{Cosmology is a journey}
\defcitealias{Coelho1988}{Paulo}
\defcitealias{Hugo1856}{Victor}

\begin{flushright}
\begin{minipage}[c]{0.6\textwidth}
\rule{\columnwidth}{0.4pt}

``Bien lire l'Univers, c'est bien lire la vie.''\\
--- \citetalias{Hugo1856} \citet{Hugo1856}, \textit{Les Contemplations}

\vspace{-5pt}\rule{\columnwidth}{0.4pt}
\end{minipage}
\end{flushright}

Though physical cosmology is celebrating its first century, it is no relic of the past. We live unique and very exciting times, when we expect to see a qualitative leap in our knowledge of the Universe within a lifetime. I consider myself incredibly fortunate to be part of this adventure.

In my experience, the loneliness felt by some researchers is easily overcome in our field by a simple thought: a cosmologist's quest is the quest of all humanity. This is why, I believe, cosmology resonates with people all around the world well beyond professional scientists, in different places and cultures. It touches everybody intellectually, but also emotionally and spiritually, without prejudice. As probability theory says something about how our mind works, physical cosmology tells us how we can think of ourselves as a species.

Before moving to the traditionally must-read acknowledgement section, I would like to quote \citetalias{Coelho1988} \citeauthor{Coelho1988}'s prologue to \textit{The Alchemist} \citetext{\citeyear{Coelho1988}}. When Narcissus falls into the lake and dies, the lake weeps, and declares: ``I weep for Narcissus, but I never noticed that Narcissus was beautiful. I weep because, each time he knelt beside my banks, I could see, in the depths of his eyes, my own beauty reflected.'' When we look into the deep Universe, the Universe also may be looking deeply into us.

\section*{Acknowledgments}

\defcitealias{Fitzgerald1925}{Francis Scott}
\begin{flushright}
\begin{minipage}[c]{0.6\textwidth}
\rule{\columnwidth}{0.4pt}

``There are all kinds of love in this world, but never the same love twice.''\\
--- \citetalias{Fitzgerald1925} \citet{Fitzgerald1925}, \textit{The Great Gatsby}

\vspace{-5pt}\rule{\columnwidth}{0.4pt}
\end{minipage}
\end{flushright}

What a long road this has been. It is difficult to believe that the end has finally arrived. That I could write this thesis, and much more importantly, that I became who I am to write it, is due in no small part to the support of a great many people. It is not a process that started three years ago, but way before that. Along the road, I met thousands of wonderful people. I wish I could mention by name anyone who has helped, in one way or another, but the list is very, very long. I hope that I have been as kind with you as you have been with me, and I apologize for not writing down your name.

\vspace{20pt}

I first wish to express my gratitude to my supervisor, Benjamin Wandelt, who was abundantly helpful and offered invaluable assistance, support and guidance. I cannot emphasize enough how delightful it has been to work with you, not only for your knowledge, dedication and modesty, but also for your interest in the human being that lies behind the scientist. Your expertise, understanding, and patience added considerably to my experience during the preparation of this thesis. Our meetings were always rich in ideas, jokes and laughs; I learned so much from you, not only scientifically, but also on how to be successful as a researcher. This is something very precious that I intend to keep forever.

I would like to address very special thanks to Jens Jasche for the opportunity to work within the larger effort surrounding Bayesian large-scale structure inference -- such a stimulating endeavor -- and for his assistance at all levels during this research project. I am very grateful for the time that you gave selflessly to provide me with direction and technical assistance, especially during my first two years. I greatly enjoyed our many scientific discussions and the way you influenced my scientific approaches. You have been a friendly, creative and supportive advisor.

I am also grateful to several researchers with whom I have had the chance to collaborate in the last three years: Jacopo Chevallard, Héctor Gil-Mar\'in, Nico Hamaus, Guilhem Lavaux, Emilio Romano-D\'iaz, Paul Sutter, and Alice Pisani. I greatly benefited from the experience of each of you, and it has been a pleasure to work on common projects. To Nico, thanks for the organization of the weekly happy hour and for your unfailing contribution to the good atmosphere. To Paul, thanks for a very nice workshop in Ohio and for a memorable void barbecue at the occasion of the ``Breaking of the Fellowship''. À Guilhem, merci de m'avoir fait partager ton expérience, tout en me rappelant qu'il ne faut pas trop parler de science comme si c'était un sujet sérieux ! À Alice, merci pour ton sourire, pour nos discussions et tout ce qu'on a partagé.

Two people, with whom I have not worked directly, deserve however a very special mention: Joseph Silk and Mat\'ias Zaldarriaga. Joe is one of the scientists I admire most, not only for his immense knowledge in astrophysics and cosmology, but also for his kindness and simplicity. Many thanks for sharing valuable insights into your vision of science and research. Mat\'ias followed my work at various stages during my PhD project. I am grateful for your invitations in Princeton, for your encouragements and for many useful discussions and suggestions. This year, applying for postdoctoral positions has been both stimulating and challenging -- yet another experience of what it means to grow up. In this respect, I want to thank my recommendation letter writers, Ben Wandelt, Jens Jasche, Joe Silk and Mat\'ias Zaldarriaga, who very kindly accepted to support my applications. I am very obliged to Will Percival for his trust and for the perspective of thought-provoking exchanges, stimulating discussions and a fruitful collaboration. 

I am indebted to my defense committee, Oliver Hahn, Alan Heavens, Bertrand Laforge, Ofer Lahav, Will Percival, Rien van de Weygaert and Mat\'ias Zaldarriaga for the honor to judge my work. I am extremely proud to be able to gather such a prestigious jury. Additional thanks go to the rapporteurs, Oliver Hahn and Alan Heavens, for having spent part of their summer reviewing the manuscript and for their positive judgment of my work. I am very obliged to Alan Heavens for the honor to feature one of the cosmic web maps presented in this thesis as the front cover of the proceedings of IAU Symposium 306 ``Statistical challenges in 21st century cosmology''. 

Participating in various international conferences and summer schools has always been a fruitful and enjoyable experience. I want to thank the organizers of the ICTP summer school on cosmology and workshop on large-scale structure (2012), the Varenna summer school (2013), the Les Houches summer school (2013), the Rencontres de Moriond (2014, Cosmology session), the IAU symposia 306 and 308 in Lisbon and Tallinn (2014), the CCAPP workshop on cosmic voids (2014), COSMO 2014 in Chicago, the MPA-EXC workshop on the dynamic Universe (2014, Garching), the ICTP workshop on cosmological structures (2015), the ESO-MPA-EXC large-scale structure conference (2014, Garching), and the Rencontres du Vietnam in Quy Nhon (2015, Cosmology session). Best greetings, in particular, to the Les Houches' cosmologists group; thanks also to the organizers of the student conferences I attended: Elbereth 2012, 2013, 2014, and the SCGSC 2013. On various occasions, I have had the chance to have friendly and interesting discussions (even if sometimes short) within the cosmology community. In particular, my work benefited from interactions with Niayesh Afshordi, Raul Angulo, Stephen Bailey, Robert Cahn, Olivier Doré, Torsten En{\ss}lin, Luigi Guzzo, Oliver Hahn, Jean-Christophe Hamilton, Alan Heavens, Shirley Ho, Mike Hudson, Eiichiro Komatsu, Ofer Lahav, Mark Neyrinck, Nelson Padilla, Bruce Partridge, Will Percival, David Schlegel, Uro\v{s} Seljak, Sergei Shandarin, Ravi Sheth, Svetlin Tassev, and Rien van de Weygaert (among many others). At this point, it also seems needed to acknowledge the decisive contribution of a familiar $\sim 20$ Mpc/$h$ void (at coordinates $x \approx -100$, $y \approx 200$ in the slice that I usually show), which very nicely makes my point during presentations. 

Cette thèse n'aurait pas été la même si elle n'avait pas été préparée dans l'excellent environnement scientifique de l'IAP. Pour cette raison, je suis reconnaissant à tous ses chercheurs, et en particuliar, pour d'intéressantes conversations, à Francis Bernardeau, Luc Blanchet, François Bouchet, Jean-François Cardoso, Stéphane Charlot, Yohan Dubois, Florence Durret, Silvia Galli, Valérie de Lapparent, Matt Lehnert, Gary Mamon, Henry McCracken, Jean-Philippe Uzan, Sébastien Peirani, Patrick Peter, Cyril Pitrou, et Sébastien Renaux-Petel. Merci à Valérie de Lapparent pour son aide, utile pour la rédaction de l'introduction de cette thèse, concernant l'historique des relevés de galaxies et plusieurs questions observationnelles. Thanks to Matt Lehnert for sharing his experience of the politics of research. Une pensée amicale aux étudiants en thèse que j'ai eu l'opportunité de côtoyer pendant trois ans: Sylvain et Guillaume (docteurs en 2013); Vincent, Vincent, Maxime, Jean, Hayley, Guillaume, Manuel et Flavien (docteurs en 2014); Nicolas, Hélène, Sandrine, Charlotte, Vivien et Clément (vous me devancez de quelques jours, félicitations !); et ceux pour qui ce n'est pas encore fini (bon courage !): Pierre, Thomas, Alba, Clotilde, Jean-Baptiste, Julia, Laura, Mélanie, Rebekka, Caterina, Erwan, Federico, Nicolas, Sébastien, Tilman. Amitiés également à quelques uns des labos voisins: Fabien, Julian, Benjamin, Julien, Mathilde, Tico, Anaïs, François, Linc, Agnès, Marta.  Pour la gestion sans faille des serveurs Horizon, dont mon travail a beaucoup bénéficié, merci à Stéphane Rouberol et Christophe Pichon. Enfin, merci aux services informatique et administratif de l'IAP, en particulier à Isabelle Guillerme et Olivia Leroy pour la gestion de mes missions.

Je voudrais exprimer ma reconnaissance à l'École polytechnique pour une formation hors pair. Tout particulièrement, je dois à certains professeurs de m'avoir ouvert les yeux sur la relativité, l'astrophysique et la cosmologie, et d'avoir partagé avec moi leur vision de la recherche : Francis Bernardeau, David Langlois et Martin Lemoine. Merci à Fabio Iocco d'avoir guidé mes premiers pas dans le monde de la recherche, et à tous les professeurs de la première promotion du M2 Physique des Hautes Énergies à l'X, en particulier aux directeurs Ignatios Antoniadis et Jean-Claude Brient, de m'avoir engagé sur la bonne voie pour commencer cette thèse. Pour leur suivi pendant ces trois ans, merci à l'ED 127 et à la commission des thèses de physique, en particulier, respectivement, à Florence Durret et Benoît Semelin.

\vspace{20pt}

Bisous à mes amis de longue date (je ne suis pas sûr que vous pensiez trouver votre nom ici !): les BIP Antoine (comment vont les caribous ?), Sébastien, Nico, Vincent, Arnaud et Nicopathe; Alex et Élise -- plus un p'tit gars dont je ne connais pas encore le prénom au moment d'écrire ces lignes; Thomas et Agathe; Amandine, Arnaud et Aymeric; Anaëlle, Justin, Kéan et Aaron; Manuela, Thibaut, Marine, Clotilde, Benji, Julien, Samuel, Stéphane, et tous les autres. Merci pour les soirées, les vacances, les jeux, et pour me rappeler que s'évader n'est pas uniquement faire de la cosmologie. 

« \textit{Vivre d'orgies est ma seule espérance, le seul bonheur que j'aie pu conquérir. C'est sur les flots qu'j'ai passé mon enfance, c'est sur les flots qu'un forban doit mourir.} » Une pensée très particulière pour les corsaires de la Ciotat et les bretteurs de l'X: Cap'taine Daniel; mes différents partenaires Kornichon, Sasha, Seva, et Daniel le Savant; Emmanuel le Gaucher, Guillaume la Montagne, Charles Ventre-Sec, Dino l'Italien, Lauriane Joli-c{\oe}ur, Céline Fleur d'épine, Benoît la Grenouille; et tous les corsaires ciotadens. Merci pour les combats, l'ambiance et les chants. May your swords stay sharp!

Enfin, il conviendrait sûrement d'exprimer ma reconnaissance à ma famille, mais ce que j'aurais à leur dire va bien au-delà de ce qu'il est possible d'écrire ici.

My final thanks go to you, the reader. If you are reading this thesis linearly, you have already read 9 pages and only have 228 to go. I hope the galaxies shine brightly over you through the rest of your life!

\begin{flushright}
Florent Leclercq\\
Paris, September 2015
\end{flushright}

%% file: Intro/IntroContent.tex
\chapter*{Introduction}
\label{chap:intro}
\renewcommand{\leftmark}{Introduction}
\renewcommand{\rightmark}{Introduction}
\addstarredchapter{Introduction}

\defcitealias{Lewis1955}{Clive Staples}
\begin{flushright}
\begin{minipage}[c]{0.6\textwidth}
\rule{\columnwidth}{0.4pt}

``Make your choice, adventurous Stranger,\\
Strike the bell and bide the danger,\\
Or wonder, till it drives you mad,\\
What would have followed if you had.''\\
--- \citetalias{Lewis1955} \citet{Lewis1955}, \textit{The Chronicles of Narnia, The Magician's Nephew}

\vspace{-5pt}\rule{\columnwidth}{0.4pt}
\end{minipage}
\end{flushright}

\section*{Large-scale structure surveys during the age of precision cosmology}

Understanding the \glslink{LSS}{structure of the Universe at the largest scales} is one of the main goals of cosmology. The existence of such a structure has been suggested by early observational projects aimed at mapping the distribution of galaxies, which resulted in a number of discoveries of individual elements -- filamentary bridges between superclusters, and large voids -- on scales of tens of megaparsecs \citep{Gregory1978,Gregory1981,Kirshner1981,Zeldovich1982}. In 1986, the results of the Center for Astrophysics redshift survey marked a milestone, with the discovery of bubble-like structures separated by sheets on which galaxies tend to lie \citep{deLapparent1986}. These results renewed interest for large-scale structure cartography, leading to new galaxy catalogs up to a depth of $\sim 400$ Mpc \citep{Geller1989,Shectman1996,Vettolani1997,Schuecker1991}. In spite of their incompleteness, these maps conclusively confirmed the existence of a large-scale organization of galaxies into a \glslink{void hierarchy}{hierarchical structure}, the \textit{\gls{cosmic web}}. At the turn of the century, massive surveys, aimed at obtaining the spectra of hundreds of thousands of galaxies \citetext{e.g. the \href{http://www.2dfgrs.net/}{2dFGRS}, \citealp{Colless2003}; the \href{http://www.sdss.org/}{SDSS}, \citealp{Strauss2002}; \href{http://wigglez.swin.edu.au/site/}{WiggleZ}, \citealp{Drinkwater2010} or the \href{http://www.ipac.caltech.edu/2mass/}{2MASS} redshift survey, \citealp{Huchra2012}}, mapped large volumes of the nearby Universe. They allowed to largely increase the completeness of observations and to obtain large enough samples for statistical analyses. Other observational programs \citetext{e.g. \href{http://deep.ps.uci.edu/}{DEEP2}, \citealp{Davis2003,Davis2007}; \href{http://cesam.oamp.fr/vvdsproject/}{VVDS}, \citealp{LeFevre2005,LeFevre2013}; \href{http://www.exp-astro.phys.ethz.ch/zCOSMOS/}{zCOSMOS}, \citealp{Lilly2007}; \href{http://www.gama-survey.org/}{GAMA}, \citealp{Driver2009}; \href{http://vipers.inaf.it/}{VIPERS}, \citealp{Guzzo2014}} focused on targeting galaxies in a smaller area on the sky, but at higher redshift.

In the coming decade, ongoing or planned cosmological programs will measure the distribution of galaxies at an unprecedented level. These include wide photometric surveys \citetext{\href{http://www.darkenergysurvey.org/}{DES}, \citealp{TheDarkEnergySurveyCollaboration2005}; \href{http://subarutelescope.org/Projects/HSC/}{HSC}, \citealp{Miyazaki2012}; \href{http://jpas.astro.ufsc.br/}{J-PAS}, \citealp{Benitez2015}; \href{http://www.lsst.org/lsst/}{LSST}, \citealp{LSSTScienceCollaboration2009,LSSTScienceCollaboration2012}}, deep spectroscopic surveys \citetext{\href{https://www.sdss3.org/future/eboss.php}{eBOSS}; \href{http://hetdex.org/}{HETDEX}, \citealp{Hill2008}; the \href{http://sumire.ipmu.jp/pfs/intro.html}{Subaru Prime Focus Spectrograph}, \citealp{Takada2014}; \href{http://desi.lbl.gov/}{DESI}, \citealp{Schlegel2011,Abdalla2012,Levi2013}}, and the \href{http://www.euclid-ec.org/}{Euclid} \citep{Laureijs2011,Amendola2013} and \href{http://wfirst.gsfc.nasa.gov/}{WFIRST} \citep{Green2012,Spergel2013} satellites.

How do we compare this avalanche of data to cosmological models? The standard picture of \gls{LSS} formation, developed over the last three decades, relies of the \glslink{gravitational evolution}{gravitational self-evolution} of a set of \glslink{initial conditions}{initial density fluctuations}, giving rise to the complex structures observed in \glslink{galaxy survey}{galaxy surveys}. Extracting the wealth of information that surveys contain thus requires a quantitative understanding of both the generation of the initial seed perturbations and of the dynamics of gravitational instability.

\section*{Early Universe physics and generation of the initial conditions of the Universe}

\Gls{inflation} and the \gls{Hot Big Bang} scenario provide an observationally well-supported physical model for the \gls{initial conditions}. The \glslink{inflation}{inflationary paradigm} \citep[see e.g.][for a review]{Baumann2011} is generally favored over other theories for the origin of seed perturbations, since it also provides explanations for some shortcomings of the standard \gls{Hot Big Bang} picture, e.g. the statistical \glslink{statistical homogeneity}{homogeneity} and \glslink{statistical isotropy}{isotropy} of the Universe, and the horizon problem \citep{Guth1981,Linde1982,Albrecht1982}. According to this model, during the \glslink{inflation}{inflationary era}, the \gls{equation of state} of the Universe is governed by a potential-dominated quantum \gls{scalar field} with negative pressure, the so-called \textit{\gls{inflaton} field}. This quantum field drives an exponential growth of the cosmic \gls{scale factor}. What is remarkable with \gls{inflation} is that the accelerated \gls{expansion} in the very early Universe can magnify the vacuum \glslink{quantum fluctuation}{quantum fluctuations} of the \gls{inflaton} into macroscopic cosmological perturbations. This model naturally provides us with a \glslink{statistical homogeneity}{statistically homogeneous} and \glslink{statistical isotropy}{isotropic} \gls{density field} with small, very nearly \glslink{grf}{Gaussian-distributed}, and nearly scale-invariant density perturbations \citep{Guth1982,Hawking1982,Starobinsky1982,Bardeen1983}.

Phenomena such as \glslink{BBN}{primordial nucleosynthesis} \citep{Alpher1948}, \gls{decoupling} and \gls{recombination}, free-streaming of \glslink{neutrino}{neutrinos}, \glslink{BAO}{acoustic oscillations of the photon-baryon plasma}, and transition from \glslink{radiation domination}{radiation} to \gls{matter domination}, come next. They are predicted by the \gls{Hot Big Bang} model, which remains a cornerstone of our understanding of the past and present Universe \citep[see e.g.][]{Kolb1990,Peacock1999}. They change the post-\glslink{inflation}{inflationary} \gls{density field} into what we call the ``\gls{initial conditions}'' for \gls{gravitational evolution}. Then, during the matter and \glslink{dark-energy domination}{dark-energy dominated eras}, self-gravity and the \glslink{expansion}{expansion of the Universe} modify these \gls{initial conditions} into an evolved \gls{density field}, at first through \glslink{linear evolution}{linear transfer} and then through \glslink{non-linear evolution}{non-linear structure formation}.

Due to their quantum origin, the process of generating seed perturbations is stochastic \citep[see e.g.][section 2.3]{Baumann2011}. Therefore, a \glslink{pdf}{probability distribution function} is the most fundamental characterization of the \glslink{LSS}{large-scale structure} of the Universe. As a consequence, it is now standard to describe in a probabilistic way the generation of the initial density fluctuations by the above-mentioned early Universe processes.

\section*{Large-scale structure evolution and galaxy formation}

According to the current picture of cosmic \gls{structure formation}, all presently observed structures have their origins in primordial seed fluctuations. \citet{Zeldovich1983} recognized the central role played by gravitational instability. \citet{Peebles1982a,Peebles1984} realized that baryonic models of \gls{structure formation} are insufficient to explain observed galaxies morphology and distribution, and consequently proposed the introduction of \glslink{CDM}{cold dark matter}. The ensuing controversy between the ``top-down'' (in which large structures form first, then fragment; as is the case when \glslink{HDM}{hot dark matter}, such as \glslink{neutrino}{neutrinos}, dominates) and ``bottom-up'' (in which small structures such as galaxies form first, then aggregate; as is the case when \glslink{CDM}{cold dark matter} dominates) \gls{structure formation} scenarios was subsequently settled in favor of the latter \citep{Bond1982,Melott1983,Blumenthal1984}. Therefore, it is currently believed that \gls{structure formation} is mostly governed by the gravitational aggregation of a \gls{dark matter} \gls{fluid}. As proposed by \citet{Rees1977,Silk1977,White1978}, luminous objects such as galaxies form via condensation and cooling of baryonic matter in gravitational \glslink{potential well}{potential wells} shaped by the \gls{dark matter} structure.

\section*{Physical processes and information content}

The detailed appearance of the presently observed galaxy distribution contains a record of its origin and \gls{formation history}. Large-scale \gls{structure formation} therefore encodes information on a wide range of processes involving very different physics, ranging from \gls{quantum field theory} and \gls{general relativity}, to the dynamics of collisionless \gls{dark matter} and the hydrodynamics and radiation transfer processes involved in \gls{galaxy formation}. The next generation of \glslink{galaxy survey}{galaxy surveys} is therefore expected to provide insights into many fundamental physics questions: What is the Universe made of? What is the microphysics of \gls{dark matter}? How does \gls{dark energy} behave? What is the mass of \glslink{neutrino}{neutrinos}? Is \gls{general relativity} complete or does it require modifications? What were the \glslink{initial conditions}{conditions in the early Universe}?

All \gls{LSS} observations are informative in some ways about these questions, but due to an incomplete understanding of the dark matter-galaxy connection \citetext{the ``\gls{bias} problem'': see in particular the ``peak-background split model'', \citealp{Bardeen1986,Cole1989}, and the ``\gls{halo} model'', \citealp{Seljak2000,Peacock2000,Cooray2002}} and observational effects \citetext{the \gls{Alcock-Paczynski effect}, \citealp{Alcock1979}; \gls{redshift-space distortions}, \citealp{Kaiser1987,Peacock2001,Hawkins2003,Guzzo2008}; non-trivial \glslink{selection effects}{selection functions}; see e.g. \citealp{Percival2014}, for a review}, the message is encoded and sometimes hard to extricate. Hence, crucial to the aim of answering the above questions is identifying where is the \gls{information content} and developing efficient tools to extract it. 

The usual strategy is to look at the shape and length scales imprinted in the galaxy \gls{power spectrum}, such as the \glslink{BAO}{baryon acoustic oscillation} scale \citep{Percival2001,Cole2005,Eisenstein2005,Percival2010}. However, at small scales and at late times, non-linear dynamics shifts the \gls{information content} away from the \glslink{two-point correlation function}{two-point function} to the \glslink{high-order correlation function}{higher-order correlators}. One of the main goals of this thesis is to access the untapped information in late-time, non-linear modes. The number of modes accessible for cosmological analyses grows like $k^3$, where $k$ is the largest usable wavenumber. In the case of \glslink{BAO}{BAOs}, a technique known as ``\gls{reconstruction}'' has been designed to correct for the effects of non-linearities, and has been shown to improve distance measurements \citep{Eisenstein2007,Padmanabhan2012}. Hence, our strategy is twofold: pushing down the smallest scale that can be both modeled and resolved; and inferring the \gls{initial conditions} that give rise to the observed \gls{LSS}. The reward for undertaking this project is a potentially vast gain of information for the determination of model parameters.

\section*{The scientific method and the process of data assimilation}

Generally, contact between theory and observations cannot be established directly. Historically, scientific progress relied on experimental assessment (the ``\glslink{paradigms of science}{first paradigm}'' of science) and theoretical modeling (the ``\glslink{paradigms of science}{second paradigm}''). In the last few decades, with growing complexity of the real-world processes to be described, testable predictions of theories often had to be obtained through numerical simulations of phenomena. Additionally, even elaborate experiments do not allow for direct comparison to the results of simulations, due to the fact that there exists no ideal observation in reality, as they are subject to a variety of uncertainties. One has to model the response functions of devices and treat their outputs, a step we refer to as signal processing (for example, the representation of a real-world signal and the application of fast \glslink{Fourier transform}{Fourier transforms} require its computer representation to be discrete both in configuration and in frequency space). Numerical simulations and data processing constitute the ``\glslink{paradigms of science}{third paradigm}'' of science. Their outputs are compared to judge and evaluate current models. These results can be used to perform inferences, i.e. update our knowledge on theoretical parameters, test hypotheses and compare competing models. They can also be used to optimize the design of new experiments. These last two steps close loops that go from theory to data, and from data to theory, as illustrated in figure \ref{tb:scientific_process}. Scientific progress in any of the physical sciences crucially depends on these steps.

\pgfdeclarelayer{background}
\pgfdeclarelayer{foreground}
\pgfsetlayers{background,main,foreground}

\begin{figure}[h]
\begin{center}
\begin{tikzpicture}

	\tikzstyle{data}=[draw, fill=LimeGreen!20, text width=10em, text centered, minimum height=3em,drop shadow]
	\tikzstyle{observations}=[draw, fill=Orange!20, text width=10em, text centered, minimum height=3em,drop shadow]
	\tikzstyle{theory}=[draw, fill=Fuchsia!20, text width=10em, text centered, minimum height=3em,drop shadow]

	\def\blockdist{2.3}
	\def\edgedist{2.5}

    \node (evaluation) [data] {Evaluation};
    \path (evaluation.north)+(-\blockdist,0.5*\blockdist) node (simulation) [theory] {Simulation};
    \path (evaluation.north)+(\blockdist,0.5*\blockdist) node (data) [observations] {Data processing};
    \path (simulation.north)+(0.,0.5*\blockdist) node (theory) [theory] {Theoretical\\ modeling};
    \path (data.north)+(0.,0.5*\blockdist) node (observations) [observations] {Observations, Experiments};    
    \path (simulation.west)+(-\blockdist,0.) node (inference) [theory] {Inference};
    \path (data.east)+(\blockdist,0.) node (design) [observations] {Experimental Design};

	\path [draw, line width=0.7pt, arrows={-latex}] (evaluation.east) -- (design.south) ;
	\path [draw, line width=0.7pt, arrows={-latex}] (evaluation.west) -- (inference.south) ;
	\path [draw, line width=0.7pt, arrows={-latex}] (inference.north) -- (theory.west) ;
	\path [draw, line width=0.7pt, arrows={-latex}] (design.north) -- (observations.east) ;
	\path [draw, line width=0.7pt, arrows={-latex}] (theory.south) -- (simulation.north) ;
	\path [draw, line width=0.7pt, arrows={-latex}] (observations.south) -- (data.north) ;
	\path [draw, line width=0.7pt, arrows={-latex}] (simulation.south) -- (evaluation.north) ;
	\path [draw, line width=0.7pt, arrows={-latex}] (data.south) -- (evaluation.north) ;

    \begin{pgfonlayer}{background}
        \path (inference.west |- inference.north)+(-0.3,0.3) node (a) {};
        \path (design.south -| design.east)+(+0.3,-0.3) node (c) {};
        \path [fill=yellow!20,rounded corners, draw=black!50, dashed] (a) rectangle (c);           
    \end{pgfonlayer}
\end{tikzpicture}
\end{center}
\caption{Diagram illustrating steps in the scientific method. Progress in physical sciences depends on each of these steps: experimental assessment (the \glslink{paradigms of science}{first paradigm}), theoretical modeling (the \glslink{paradigms of science}{second paradigm}), computational studies (simulation of phenomena and data processing -- the \glslink{paradigms of science}{third paradigm}). The outputs of simulations and data processing are compared to
judge and evaluate current models. These results are used to infer theoretical parameters and to design new experiments. The yellow rectangle shows the emergence of a \glslink{paradigms of science}{fourth research paradigm}: data-intensive scientific discovery, where extremely large data sets captured by instruments and generated by complex simulations are used.\label{tb:scientific_process}}
\end{figure}

Several authors are now describing the emergence of a so-called ``\glslink{paradigms of science}{fourth paradigm}'' of research: data-intensive scientific discovery \citep{MicrosoftResearch2009,Szalay2014}. Scientific insights are wrested from extremely vast data sets. This transition from hypothesis-driven to data-driven research is made possible by new technologies for gathering, processing, manipulating, analyzing, mining, and displaying data. For example, \gls{exascale computers}, expected around 2018, will be of the order of processing power of the human brain at the neural level. 

This thesis falls within the context of this emerging \glslink{paradigms of science}{fourth paradigm}. Its field is \gls{cosmostatistics}, the discipline that uses stochastic quantities as seeds of structure to make the connection between cosmological models and observations. This area is at the interface between theory and observational data \citep[see][]{Leclercq2014Varenna}:
\begin{itemize}
\item It consists of predicting cosmological observables from stochastic quantities as seeds of structure in the Universe (\textit{from theory to data}). Theoretical hypotheses are used to model, predict and anticipate results.
\item It uses the departures from \glslink{statistical homogeneity}{homogeneity} and \glslink{statistical isotropy}{isotropy}, observed in astronomical surveys, to distinguish between cosmological models (\textit{from data to theory}). Data sets are used to infer parameters of the theoretical models and to compare their relative suitability.
\end{itemize}

More specifically, this work focuses on the process of \textit{\gls{data assimilation}} for the \glslink{LSS}{large-scale structure}, i.e. the process by which observations are incorporated into a numerical model of a real system. Data assimilation is a field of statistics, widely understood and used outside of astrophysics (e.g. in meteorology, geophysics, oceanography and climate sciences). The general mechanism consists of iteratively correcting the state of a prediction based on a physical model, using successive observations. In this work, we borrow ideas from these sciences and apply them to \glslink{LSS}{large-scale structure} data analysis. For all quantities of interest, we do not only provide a best-guess estimate, but fully account for all credible regions by a detailed quantification of the probability density.

\section*{Goal and structure of this thesis}

The ambition of this work is to describe progress towards enriching the standard for the analysis of \glslink{galaxy survey}{galaxy surveys}. The central ingredient is the recently proposed {\borg} \citep[Bayesian Origin Reconstruction from Galaxies,][]{Jasche2013BORG,Jasche2015BORGSDSS} algorithm. {\borg} is a Bayesian \gls{data assimilation} code, which provides a fully probabilistic, physical model of the \glslink{non-linear evolution}{non-linearly evolved} \gls{density field} as probed by \gls{LSS} surveys.

The goal of this thesis is to demonstrate that Bayesian \gls{large-scale structure inference} with the {\borg} algorithm has moved beyond the proof-of-concept stage. In particular, it describes the first application to real cosmological data from the \glslink{SDSS}{Sloan Digital Sky Survey}, and shows how these results can be used for \gls{cosmic web classification} and analysis.

This thesis is organized as follows. Part \ref{part:I} focuses on the analytical and numerical description of the morphology and growth of the \gls{LSS}. Chapter \ref{chap:theory} is an introduction to the theory of \gls{structure formation}, as relevant for this thesis. As \glslink{LPT}{Lagrangian perturbation theory} is a key ingredient of the {\borg} algorithm, the reliability of its numerical predictions is investigated in chapter \ref{chap:lpt}.

Part \ref{part:II} introduces Bayesian \gls{large-scale structure inference}. In chapter \ref{chap:stats}, we present the framework of \glslink{Bayesian statistics}{Bayesian} \gls{probability theory}. Chapter \ref{chap:BORG} reviews the latest version of the {\borg} algorithm and its implementation. Chapter \ref{chap:BORGSDSS} presents the application of {\borg} to the Sloan Digital Sky Survey data. These results quantify the distribution of \gls{initial conditions} as well as the possible \gls{formation history} of the observed structures.

As \gls{LSS} surveys contain a wealth of \glslink{information content}{information} that cannot be trivially extracted due to the \glslink{non-linear evolution}{non-linear dynamical evolution} of the \gls{density field}, part \ref{part:III} discusses methods designed to improve upon standard techniques by including \glslink{non-Gaussianity}{non-Gaussian} and non-linear \glslink{data model}{data models} for the description of late-time \gls{structure formation}. Chapter \ref{chap:remapping} presents a computationally fast and flexible model of \glslink{mildly non-linear regime}{mildly non-linear} \glslink{density field}{density fields} via the technique of \gls{remapping} \gls{LPT}. In chapter \ref{chap:filtering}, we introduce the concept of \gls{non-linear filtering}, designed to improve \gls{LSS} \glslink{sample}{samples} at non-linear scales. 

Finally, part \ref{part:IV} exploits the {\borg} \gls{SDSS} analysis for different cosmographic projects aiming at characterizing and analyzing the \gls{cosmic web}. Chapter \ref{chap:dmvoids} demonstrates that the inference of \glslink{dark matter void}{voids in the dark matter distribution} is possible, and that, in addition, our method yields a drastic reduction of \gls{statistical uncertainty} in \gls{void} catalogs. In chapter \ref{chap:ts}, we describe a probabilistic \glslink{cosmic web classification}{classification} of the dynamic \gls{cosmic web} into four distinct components: \glslink{void}{voids}, \glslink{sheet}{sheets}, \glslink{filament}{filaments}, and \glslink{cluster}{clusters}. Subsequently, chapter \ref{chap:decision} introduces a new \glslink{decision theory}{decision criterion} for labeling different regions, on the basis of \gls{posterior} probabilities and the strength of data constraints. 

In \hyperref[chap:ccl]{the last chapter}, we summarize our results and give our conclusions. Prospective applications and possible directions for future investigations are also mentioned.

The appendices provide complementary material: a mathematical exposition of \glslink{grf}{Gaussian random fields} (appendix \ref{apx:complements GRFs}), a review of the \glslink{PM}{particle-mesh} technique for dark matter \glslink{N-body simulation}{simulations} (appendix \ref{apx:simulations}), and a description of the \gls{cosmic web} analysis algorithms used in this thesis (appendix \ref{apx:classification}).

%% file: Chapter1/Chapter1Content.tex
\part{Morphology and dynamics of the large-scale structure}
\label{part:I}
\renewcommand{\leftmark}{\leftmarkold}
\renewcommand{\rightmark}{\rightmarkold}

\chapter{Cosmological perturbations and structure formation}
\label{chap:theory}
\minitoc

\begin{flushright}
\begin{minipage}[c]{0.6\textwidth}
\rule{\columnwidth}{0.4pt}

``For the mind wants to discover by reasoning what exists in the infinity of space that lies out there, beyond the ramparts of this world -- that region into which the intellect longs to peer and into which the free projection of the mind does actually extend its flight.''\\
--- \citeauthor{Lucretius55BC}, \textit{De Rerum Natura}

\vspace{-5pt}\rule{\columnwidth}{0.4pt}
\end{minipage}
\end{flushright}

\abstract{\section*{Abstract} This chapter provides an overview of the current paradigm of cosmic \gls{structure formation}, as relevant for this thesis. It also reviews standard tools for large-scale structure analysis.}

This chapter is organized as follows. In section \ref{sec:homogeneous}, key equations of \glslink{general relativity}{general relativistic} Friedmann-Lemaître cosmological models are briefly reviewed, followed by a discussion of the statistical description of cosmological fields in section \ref{sec:stats} and of the dynamics of \glslink{gravitational evolution}{gravitational instability} in section \ref{sec:Dynamics}. In section \ref{sec:EPT} and \ref{sec:LPT} we describe cosmological perturbation theory in \glslink{EPT}{Eulerian} and \glslink{LPT}{Lagrangian} descriptions. Finally, section \ref{sec:NL-approxs} deals with various \glslink{non-linear approximation}{non-linear approximations} to \glslink{gravitational evolution}{gravitational instability}.

\section{The homogeneous Universe}
\label{sec:homogeneous}

This section provides an overview of the standard picture of cosmology, describing the \glslink{homogeneous Universe}{homogeneous} evolution of the Universe. In particular, we reproduce some very standard equations around which perturbation theory will be implemented in the following. A demonstration can be found in any introduction to cosmology, see for example \citet{Peebles1980,Kolb1990,Liddle2000,Bernardeau2002,Lesgourgues2004,Trodden2004,Langlois2005,Langlois2010}.

Let $a$ be the cosmic \gls{scale factor}, normalized to unity at the present time: $a_0=1$. We denote by $t$ the \gls{cosmic time}, by $\tau$ the \gls{conformal time}, defined by $\drm t = a(\tau) \, \drm \tau$, and by $z$ the \gls{redshift}, defined by $a=1/(1+z)$. In the following, a dot denotes a differentiation with respect to $t$ and a prime a differentiation with respect to $\tau$. \gls{Friedmann's equations}, describing the dynamics of the Universe, are derived from \gls{Einstein's equations} of \gls{general relativity}. In \gls{conformal time}, they read:
\begin{eqnarray}
\mathcal{H}^2 & = & \dfrac{8\pi\G}{3}a^2 \rho - k , \label{eq:Friedmann1-conformal-time} \\
\mathrm{and} \quad \mathcal{H}' & = & - \dfrac{4\pi\G}{3}a^2 (\rho + 3P) ,
\label{eq:Friedmann2-conformal-time}
\end{eqnarray}
where $\mathcal{H} \equiv a'/a = a H$ is the \gls{conformal expansion rate}, $H \equiv \dot{a}/a$ is the \gls{Hubble parameter}, $\rho$ the total energy density and $P$ the pressure. $k$ is the reduced curvature parameter, taking one of the following values: $-1$ for an open universe, $0$ for a flat universe, $1$ for a closed universe. $\G$ denotes the \gls{gravitational constant}, and we adopt units such that $\mathrm{c} = 1$.

As a direct consequence, \gls{Friedmann's equations} immediately determine the evolution of the energy density, described as:
\begin{equation}
\label{eq:energy-density-conservation}
\rho' = - 3\mathcal{H}(\rho + P) .
\end{equation}

Throughout this thesis, we will particularly focus on the eras of \gls{matter domination} and \gls{dark-energy domination} within the standard {\LCDM} paradigm. Hence, we will consider that the content of the Universe is limited to two components: matter (mostly \glslink{CDM}{cold dark matter}) and \gls{dark energy} in the form of a \gls{cosmological constant} $\Lambda$. We denote by $\rho_\mathrm{m}$ and $\rho_\Lambda \equiv \Lambda/8\pi\G$ their respective energy densities. Introducing their respective \glslink{equation of state}{equations of state}, $w_i \equiv P_i/\rho_i$, we have $w \approx 0$ for \glslink{CDM}{cold dark matter} and $w = -1$ for the \gls{cosmological constant}. For this cosmology, equation \eqref{eq:Friedmann2-conformal-time} reads
\begin{equation}
\label{eq:Friedmann-conformal-time-2}
\mathcal{H}' = -\dfrac{4\pi\G}{3} a^2 \rho_\mathrm{m} + \dfrac{8\pi\G}{3} a^2 \rho_\Lambda .
\end{equation}

It is convenient to introduce the dimensionless \gls{cosmological parameters} as the ratio of density to critical density, $\rho_\mathrm{crit}(t)\nbsp\equiv\nbsp3H^2(t)/8\pi\G$, which corresponds to the total energy density in a flat universe: $\Omega_\mathrm{m}(t) \equiv 8\pi\G \rho_\mathrm{m}(t) / 3H^2(t)$ and $\Omega_\Lambda(t) \equiv 8\pi\G \rho_\Lambda / 3H^2(t) = \Lambda / 3H^2(t)$. Their expression in terms of \gls{conformal time} is given by
\begin{eqnarray}
\Omega_\mathrm{m}(\tau) \mathcal{H}^2(\tau) & = & \dfrac{8\pi\G}{3} \rho_\mathrm{m}(\tau) a^2(\tau), \\
\Omega_\Lambda(\tau) \mathcal{H}^2(\tau) & = & \dfrac{8\pi\G}{3} \rho_\Lambda a^2(\tau) \equiv \dfrac{\Lambda}{3} a^2(\tau) .
\end{eqnarray}
Note that $\Omega_\mathrm{m}(\tau)$ and $\Omega_\Lambda(\tau)$ are time-dependent. Inserting these two expressions in equation \eqref{eq:Friedmann-conformal-time-2} yields the following form of the second \glslink{Friedmann's equations}{Friedmann equation},
\begin{equation}
\label{eq:Friedmann-2}
\mathcal{H}'(\tau) = \left( -\dfrac{\Omega_\mathrm{m}(\tau)}{2} + \Omega_\Lambda(\tau) \right) \mathcal{H}^2(\tau) ,
\end{equation}
and the first one reads
\begin{equation}
\mathcal{H}^2 = \dfrac{8\pi\G}{3} a^2 \rho_\mathrm{m} + \dfrac{8\pi\G}{3} a^2 \rho_\Lambda - k = \Omega_\mathrm{m} \mathcal{H}^2 + \Omega_\Lambda \mathcal{H}^2 - k ,
\end{equation}
which yields
\begin{equation}
\label{eq:Friedmann-1}
k = (\Omega_{\mathrm{tot}}(\tau) - 1) \mathcal{H}^2(\tau) ,
\end{equation}
where $\Omega_{\mathrm{tot}}(\tau) \equiv \Omega_\mathrm{m}(\tau) + \Omega_\Lambda(\tau)$. In the following, we will note $\Omega_\mathrm{m}^{(0)}=\Omega_\mathrm{m}(a=1)$ and $\Omega_\Lambda^{(0)} = \Omega_\Lambda(a=1)$.

\section{Statistical description of cosmological fields}
\label{sec:stats}

In this section, we consider some cosmic \gls{scalar field} $\lambda(\textbf{x})$ whose statistical properties are to be described. It denotes either the cosmological \gls{density contrast}, $\delta(\textbf{x})$, the \gls{gravitational potential}, $\Phi(\textbf{x})$ (see section \ref{sec:Dynamics}), or any other field of interest derived from vectorial fields (e.g. the \glslink{velocity field}{velocity divergence field}), polarization fields, etc.

As discussed in the \hyperref[chap:intro]{introduction}, values of $\lambda(\textbf{x})$ have to be treated as stochastic variables. For an arbitrary number $n$ of spatial positions $\textbf{x}_i$, one can define the \textit{joint multivariate \glslink{pdf}{probability distribution function}} to have $\lambda(\textbf{x}_1)$ between $\lambda_1$ and $\lambda_1+\drm \lambda_1$, $\lambda(\textbf{x}_2)$ between $\lambda_2$ and $\lambda_2+\drm \lambda_2$, etc. This \gls{pdf} is written
\begin{equation}
\p(\lambda_1, \lambda_2, ..., \lambda_n) \, \drm \lambda_1 \drm \lambda_2 \, ... \, \drm \lambda_n .
\end{equation}

\subsection{Average and ergodicity}

\paragraph{Average.}The word ``average'' (and in the following, the corresponding $\left\langle \right\rangle $ symbols) may have two different meanings. First, one can average by taking many realizations drawn from the distribution, all of them produced in the same way (e.g. by $N$-body simulations). This is the \textit{\gls{ensemble average}}, defined to be for any quantity $X(\lambda_1, \lambda_2, ..., \lambda_n)$:
\begin{equation}
\left\langle X \right\rangle \equiv \int X(\lambda_1, \lambda_2, ..., \lambda_n)\p(\lambda_1, \lambda_2, ..., \lambda_n) \, \drm \lambda_1 \drm \lambda_2 \, ... \, \drm \lambda_n ,
\end{equation}
where $\p(\lambda_1, \lambda_2, ..., \lambda_n)$ is the joint multivariate \gls{pdf}.

One can also average by considering the quantity of interest at different locations within the same realization of the distribution. This is the \textit{\gls{sample average}}. For some volume $V$ in the Universe, the \gls{sample average} over $V$ of a quantity $X$ is defined to be:
\begin{equation}
\bar{X} \equiv \dfrac{1}{V}\int_V X(\textbf{x}) \, \drm^3 \textbf{x} .
\end{equation}

\paragraph{Ergodicity.}If the \gls{ensemble average} of any quantity coincides with the \gls{sample average} of the same quantity, the system is said to be \textit{ergodic}. In cosmology, the hypothesis of \gls{ergodicity} is often adopted, at least if the considered catalogue is large enough. For instance, if \gls{ergodicity} holds, the mean density of the Universe is given by
\begin{equation}
\left\langle \rho(\textbf{x}) \right\rangle = \bar{\rho} \equiv \dfrac{1}{V} \int_V \rho(\textbf{x}) \, \drm^3 \textbf{x} ,
\end{equation}
in the limit where $V \rightarrow \infty$. The term of \gls{ergodicity} historically refers to time processes, not to spatial ones. If the above property is fulfilled in cosmology, one says that the system is a \textit{fair sample} of the Universe.\footnote{In the case of the \gls{LSS}, care should be taken with deep surveys. Indeed, as data lie on the surface of the relativistic \gls{lightcone}, we cannot have access to a fair sample of the Universe at the present time. Rigorously, \gls{ergodicity} is not verified.}

\subsection{Statistical homogeneity and isotropy}

A random field is said to be \textit{\glslink{statistical homogeneity}{statistically homogeneous}} if all joint multivariate \glslink{pdf}{pdfs} $\p(\lambda(\textbf{x}_1), \lambda(\textbf{x}_2), ..., \lambda(\textbf{x}_n))$ are invariant under translations of the coordinates $\textbf{x}_1$, $\textbf{x}_2$, ..., $\textbf{x}_n$ in space. Thus probabilities depend only on relative positions, but not on locations. Note that \gls{statistical homogeneity} is a weaker assumption than homogeneity, which would mean that $\lambda(\textbf{x})$ takes the same value everywhere in space.

Similarly, a random field is said to be \textit{\glslink{statistical isotropy}{statistically isotropic}} if all $\p(\lambda(\textbf{x}_1), \lambda(\textbf{x}_2), ..., \lambda(\textbf{x}_n))$ are invariant under spatial rotations. 

From now on, cosmic fields will be considered \glslink{statistical homogeneity}{statistically homogeneous} and \glslink{statistical isotropy}{isotropic}, as a consequence of the cosmological principle that underlies most \glslink{inflation}{inflationary calculations} \citep[see][]{Guth1981,Linde1982,Albrecht1982,Linde1995}, and of standard \gls{gravitational evolution} \citep[e.g.][]{Peebles1980}. Of course, the validity of this assumption has to be checked against observational data. It is also important to note that a lot of the information from \glslink{galaxy survey}{galaxy surveys} comes from effects that distort the observed signal away from this ideal. In particular, observational effects such as the \gls{Alcock-Paczynski effect} \citep{Alcock1979} and \gls{redshift-space distortions} \citep{Kaiser1987} in \glslink{galaxy survey}{galaxy surveys} introduce significant deviations from \gls{statistical homogeneity} and \glslink{statistical isotropy}{isotropy}.

\subsection{Gaussian and log-normal random fields in cosmostatistics}

\draw{This section draws from subsection 2.2 of \citet{Leclercq2014Varenna}.}

\glslink{grf}{Gaussian random fields} are ubiquitous in \gls{cosmostatistics} \citep[see][for reviews]{Lahav2004,Wandelt2013,Leclercq2014Varenna}. Indeed, as mentioned in the \hyperref[chap:intro]{introduction}, \glslink{inflation}{inflationary models} predict the \glslink{initial conditions}{initial density perturbations} to arise from a large number of independent \glslink{quantum fluctuation}{quantum fluctuations}, and therefore to be very nearly \glslink{grf}{Gaussian-distributed}. Even in models which are said to produce ``large'' \glslink{non-Gaussianity}{non-Gaussianities}, deviations from Gaussianity are strongly constrained by observational tests \citep[see][for the latest results]{Planck2013PNG,Planck2015PNG}. \glslink{grf}{Grfs} are essential for the analysis of the \glslink{CMB}{cosmic microwave background}, but the large scale distribution of galaxies can also be approximately modeled as a \gls{grf}, at least on very large scales, where \gls{gravitational evolution} is still well-described by linear perturbation theory (see sections \ref{sec:EPT} and \ref{sec:LPT}). The \gls{log-normal distribution} is convenient for modeling the statistical behavior of evolved \glslink{density field}{density fields}, partially accounting for \glslink{non-linear evolution}{non-linear gravitational effects} at the level of the \gls{one-point distribution}.

In the following, we summarize some results about finite-dimensional \glslink{grf}{Gaussian} and \glslink{log-normal distribution}{log-normal} random fields. Without loss of generality, infinite-dimensional fields can be discretized. If the field is sufficiently regular and the discretization scale is small enough, no information will be lost. In practice, throughout this thesis, any field that we want to describe is already discretized on a grid of particles or voxels. Let us denote the values of the considered cosmic \gls{scalar field} $\lambda(\textbf{x})$ at \glslink{comoving coordinates}{comoving positions} $\textbf{x}_i$ as $\lambda_i \equiv \lambda(\textbf{x}_i)$ for $i$ from $1$ to any arbitrary integer $n$.  

\subsubsection{Gaussian random fields}
\label{sec:Gaussian random fields}

The $n$-dimensional vector $\lambda = \left( \lambda_i \right)_{1 \leq i \leq n}$ is a \glslink{grf}{Gaussian random field} (we will often say ``is Gaussian'' in the following) with mean $\mu \equiv \left( \mu_i \right)_{1 \leq i \leq n}$ and covariance matrix $C \equiv \left( C_{ij} \right)_{1 \leq i \leq n, 1 \leq j \leq n}$ if its joint multivariate \gls{pdf} is a multivariate Gaussian:\footnote{Here we use the common terminology in physics and refer to this \gls{pdf} as a ``multivariate Gaussian''. It is called a ``multivariate normal'' distribution in statistics. Note that it is possible to generalize this definition to the case where $C$ is only a positive semi-definite Hermitian matrix, using the notion of characteristic function (see appendix \ref{apx:complements GRFs}).}
\begin{equation}
\label{eq:Gaussian}
\p(\lambda | \mu,C) = \frac{1}{\sqrt{\vert 2\pi C\vert}} \exp\left(-\frac{1}{2}(\lambda-\mu)^* C^{-1}(\lambda-\mu) \right) = \frac{1}{\sqrt{\vert 2\pi C\vert}} \exp\left(-\frac{1}{2} \sum_{i=1}^n \sum_{j=1}^n (\lambda_i-\mu_i) C_{ij}^{-1}(\lambda_j-\mu_j) \right).
\end{equation}
where $z^*$ denotes the conjugate transpose of $z$, vertical bars indicate the determinant of the surrounded matrix and $C$ is assumed to be a positive-definite Hermitian matrix (and therefore invertible). In practical cases, $\mu$ is often taken to be zero. As can be seen from this definition, a \gls{grf} is completely specified by its mean $\mu$ and its covariance matrix $C$.

It is interesting to note that for the \gls{density contrast} $\delta(\textbf{x})$, the Gaussian assumption has to break down at later epochs of \gls{structure formation} since it predicts density amplitudes to be symmetrically distributed among positive and negative values, but weak and strong energy conditions require $\delta(\textbf{x}) \geq -1$. Even in the \gls{initial conditions}, Gaussianity can not be exact due to the existence of this lower bound. The Gaussian assumption is therefore strictly speaking only valid in the limit of infinitesimally small density fluctuations, $\vert \delta(\textbf{x}) \vert \ll 1$.

\subsubsection{Moments of Gaussian random fields, Wick's theorem}
\label{sec:Moments_GRF_Wick_theorem}

From equation \eqref{eq:Gaussian} it is easy to check that the mean $\left\langle \lambda \right\rangle$ is really $\mu$ and the covariance matrix is really $\left\langle (\lambda-\mu)^*(\lambda-\mu) \right\rangle = C$, just by evaluating the Gaussian integrals:
\begin{eqnarray}
\left\langle \lambda_i \right\rangle & = & \int \lambda_i \, \p(\lambda|\mu,C) \, \drm \lambda_i = \mu_i , \\
\left\langle (\lambda_i-\mu_i)^*(\lambda_j-\mu_j) \right\rangle & = & \int (\lambda_i-\mu_i)^*(\lambda_j-\mu_j) \, \p(\lambda|\mu,C) \, \drm \lambda_i \drm \lambda_j = C_{ij} . \label{eq:covariance-GRF}
\label{eq:covariance-GRF}
\end{eqnarray}
We now want to compute \glslink{high-order correlation function}{higher-order} \glslink{moment}{moments} of a \gls{grf}. Let us focus on central \glslink{moment}{moments}, or equivalently, let us assume in the following that the mean is $\mu = 0$. Here we omit the star denoting conjugate transpose for simplicity. Any odd \glslink{moment}{moments}, e.g. the third $\left\langle \lambda_i \lambda_j \lambda_k \right\rangle$, the fifth $\left\langle \lambda_i \lambda_j \lambda_k \lambda_l \lambda_m \right\rangle$, etc. are found to be zero by symmetry of the Gaussian \gls{pdf}. The \glslink{high-order correlation function}{higher-order} even ones (e.g. the fourth, the sixth, etc.) can be evaluated through the application of \gls{Wick's theorem}, an elegant method of reducing \glslink{high-order correlation function}{high-order statistics} to a combinatorics problem. 

\gls{Wick's theorem} states that \glslink{high-order correlation function}{high-order even} \glslink{moment}{moments} of a \gls{grf} are computed by connecting up all possible pairs of the field (Wick contractions) and writing down the covariance matrix for each pair using equation \eqref{eq:covariance-GRF}. For instance,
\begin{eqnarray}
\left\langle \lambda_i \lambda_j \lambda_k \lambda_l \right\rangle & = & \left\langle \lambda_i \lambda_j \right\rangle \left\langle \lambda_k \lambda_l \right\rangle + \left\langle \lambda_i \lambda_k \right\rangle \left\langle \lambda_j \lambda_l \right\rangle + \left\langle \lambda_i \lambda_l \right\rangle \left\langle \lambda_j \lambda_k \right\rangle \nonumber \\
& = & C_{ij}C_{kl} + C_{ik}C_{jl} + C_{il}C_{jk} .
\end{eqnarray}
The number of terms generated in this fashion for the $n$-th order \gls{moment} is $\prod_{i=1}^{n/2} (2i-1)$.

\subsubsection{Marginals and conditionals of Gaussian random fields}

Let us the split the \gls{grf} up into two parts $x = \llbracket 1,m \rrbracket$ and $y = \llbracket m+1,n \rrbracket$ ($m<n$), so that
\begin{equation}
\lambda=
\begin{pmatrix}
\lambda_{x} \\
\lambda_{y}
\end{pmatrix}, \quad
\mu=
\begin{pmatrix}
\mu_{x} \\
\mu_{y}
\end{pmatrix} \quad \mathrm{and} \quad C=
\begin{pmatrix}C_{xx}&C_{xy}\\ C_{yx}&C_{yy}
\end{pmatrix}.
\label{eq:split-GRF}
\end{equation}
$C_{xy}=\left(C_{yx}\right)^*$ since $C$ is Hermitian.

Easy computation of \glslink{marginal pdf}{marginal} and \glslink{conditional pdf}{conditional pdfs} is a very convenient property of \glslink{grf}{grfs}. First of all, \glslink{marginal pdf}{marginal} and \glslink{conditional pdf}{conditional} densities of \glslink{grf}{grfs} are multivariate Gaussians. Therefore, all we need to calculate are their means and covariances. 
For the \glslink{marginal pdf}{marginal} \glslink{pdf}{pdfs}, the results are
\begin{eqnarray}
\left\langle \lambda_x \right\rangle & = & \mu_{x}, \label{eq:GRF-marginals-1}\\
\left\langle (\lambda_x-\mu_x)^*(\lambda_x-\mu_x) \right\rangle & = & C_{xx}, \label{eq:GRF-marginals-2}\\
\left\langle \lambda_y \right\rangle & = & \mu_{y}, \label{eq:GRF-marginals-3}\\
\left\langle (\lambda_y-\mu_y)^*(\lambda_y-\mu_y) \right\rangle & = & C_{yy}. \label{eq:GRF-marginals-4}
\end{eqnarray}
These expressions simply mean that the \glslink{marginal pdf}{marginal} means and \glslink{marginal pdf}{marginal} covariances are just the corresponding parts of the joint mean and covariance, as defined by equation \eqref{eq:split-GRF}. 

Less trivially, here are the parameters of the \glslink{conditional pdf}{conditional} densities:
\begin{eqnarray}
\mu_{x|y} \equiv \left\langle \lambda_x|\lambda_y \right\rangle & = & \mu_x + C_{xy}C^{-1}_{yy}(\lambda_y-\mu_{y}), \label{eq:GRF-conditionals-1}\\
C_{x|y} \equiv \left\langle (\lambda_x-\mu_x)^*(\lambda_x-\mu_x) | \lambda_y \right\rangle & = & C_{xx}-C_{xy}C^{-1}_{yy}C_{yx}, \label{eq:GRF-conditionals-2}\\
\mu_{y|x} \equiv \left\langle \lambda_y|\lambda_x \right\rangle & = & \mu_y + C_{yx}C^{-1}_{xx}(x-\mu_{x}), \label{eq:GRF-conditionals-3}\\
C_{y|x} \equiv \left\langle (\lambda_y-\mu_y)^*(\lambda_y-\mu_y) | \lambda_x \right\rangle & = & C_{yy}-C_{yx}C^{-1}_{xx}C_{xy}. \label{eq:GRF-conditionals-4}
\end{eqnarray}

A demonstration of these formulae can be found in appendix \ref{apx:complements GRFs}. From these expressions, it is easy to see that for \glslink{grf}{grfs}, lack of covariance ($C_{xy}=0$) implies \glslink{conditional independence}{independence}, i.e. $\p(x,y)=\p(x)\p(y)$. This is most certainly not the case for general random fields. Similarly, if $\lambda_1, ..., \lambda_n$ are jointly Gaussian, then each $\lambda_i$ is Gaussian-distributed, but not conversely.

A particular case is the optimal de-noising of a data set $d$, modeled as the sum of some signal $s$ and a stochastic noise contribution $n$: $d = s+n$. We model all three fields $d$, $s$, $n$ as \glslink{grf}{grfs}. Assuming a vanishing signal mean, an unbiased measurement (i.e. $\mu_d = \mu_s = \mu_n =0$), and lack of covariance between signal and noise (i.e. $C_{sn}=C_{ns}=0$, which implies $C_{sd}=C_{ss}$ and $C_{dd}=C_{ss}+C_{nn}$), equations \eqref{eq:GRF-conditionals-1} and \eqref{eq:GRF-conditionals-2} yield the famous \gls{Wiener filter} equations:
\begin{eqnarray}
\mu_{s|d} & = & C_{ss} \left( C_{ss} + C_{nn} \right)^{-1} d, \label{eq:Wiener-filter-1}\\
C_{s|d} & = & C_{ss} - C_{ss} \left( C_{ss} + C_{nn} \right)^{-1} C_{ss} = \left( C_{nn}^{-1} + C_{ss}^{-1} \right)^{-1} .\label{eq:Wiener-filter-2}
\end{eqnarray}

\subsubsection{Log-normal random fields}
\label{sec:Lognormal-fields}

In the case where $\delta(\textbf{x})$ is the \gls{density contrast}, \gls{gravitational evolution} will yield very high positive \gls{density contrast} amplitudes. In order to prevent negative mass ($\delta(\textbf{x}) < -1$) while preserving $\left\langle \delta(\textbf{x}) \right\rangle = 0$, the resultant \gls{pdf} must be strongly skewed \citep[e.g.][]{Peacock1999}. In the absence of an exact \gls{pdf} for the \gls{density field} in \glslink{non-linear regime}{non-linear regimes}, solution to dynamical equations, one can describe the statistical properties of the evolved matter distribution by phenomenological probability distributions. A common choice is the \gls{log-normal distribution}, which approximates well the \glslink{one-point distribution}{one-point behavior} observed in \glslink{galaxy survey}{galaxy observations} and \glslink{N-body simulation}{$N$-body simulations} \citep[e.g.][]{Hubble1934,Peebles1980,Coles1991,Gaztanaga1993,Colombi1994,Kayo2001,Neyrinck2009}. This is the model adopted as a \gls{prior} for the \gls{density field} in the {\hades} algorithm \citep[][see also table \ref{tb:LSS_algos}]{JascheKitaura2010,Jasche2012}. The assumption is that the log-density, $\ln(1+\delta)$, instead of the \gls{density contrast} $\delta$, obeys Gaussian statistics.

If $\lambda$ is a $n$-dimensional vector having multivariate Gaussian distribution with mean $\mu$ and covariance matrix $C$, then $\xi$, defined by its components $\xi_i=\exp(\lambda_i)$, has a multivariate \gls{log-normal distribution} given by
\begin{equation}
\p(\xi|\mu,C) = \frac{1}{\sqrt{\vert 2\pi C\vert}} \exp\left(-\frac{1}{2} \sum_{i=1}^n \sum_{j=1}^n \left(\ln(\xi_i)-\mu_i\right)^* C_{ij}^{-1} \left(\ln(\xi_j)-\mu_j\right) \right) \prod_{k} \frac{1}{\xi_k} .
\end{equation}
The mean of $\xi$ is $\nu$ defined by
\begin{equation}
\nu_i \equiv \left\langle \xi_i \right\rangle = \exp\left( \mu_i + \frac{1}{2}C_{ii} \right) ,
\end{equation}
and its covariance matrix is $D$ defined by
\begin{equation}
D_{ij} \equiv \left\langle(\xi_i-\mu_i)^*(\xi_j-\mu_j)\right\rangle = \exp\left(\mu_i + \mu_j + \frac{1}{2}(C_{ii}+C_{jj}) \right) \left( \exp(C_{ij}) -1 \right).
\end{equation}

In cosmology, we assume that $\lambda = \ln(1+\delta)$ is a \gls{grf} with mean $\mu$ and covariance matrix $C$. Then $\delta = \exp(\lambda) -1$ follows a \gls{log-normal distribution}, given by
\begin{equation}
\p(\delta|\mu,C) = \frac{1}{\sqrt{\vert 2\pi C\vert}} \exp\left(-\frac{1}{2} \sum_{i=1}^n \sum_{j=1}^n \left(\ln(1+\delta_i)-\mu_i\right)^* C_{ij}^{-1} \left(\ln(1+\delta_j)-\mu_j\right) \right) \prod_{k} \frac{1}{1+\delta_k} .
\end{equation}
To ensure that $\left\langle \delta \right\rangle$ vanishes everywhere, i.e. that
\begin{equation}
\nu_i = \left\langle 1+ \delta_i \right\rangle = \exp\left(\mu_i + \frac{1}{2} C_{ii}\right) = 1,
\end{equation}
one has to impose the following choice for $\mu$:
\begin{equation}
\mu_i = -\frac{1}{2} C_{ii} =  -\frac{1}{2} C_{00} = \mu_0.
\end{equation}
We have used that $C_{ii} = C_{00}$, since the correlation function depends only on distance (assuming \gls{statistical homogeneity} and \glslink{statistical isotropy}{isotropy}). Hence, the mean for the lognormal distribution is the same throughout the entire field.

For further discussion on the \glslink{log-normal distribution}{log-normal} behavior of \glslink{density field}{density fields}, see chapters \ref{chap:lpt} and \ref{chap:remapping}.

\subsection{Correlation functions and power spectra}

\subsubsection{Two-point correlation function and power spectrum}
\label{sec:power-spectrum}

\paragraph{Definitions.}The \gls{two-point correlation function} is defined in configuration space as the joint \gls{ensemble average} of the field at two different locations:
\begin{equation}
\xi(r) = \left\langle \lambda^\ast(\textbf{x})\lambda(\textbf{x}+\textbf{r}) \right\rangle.
\end{equation}
It depends only on the norm of $\textbf{r}$ if \gls{statistical isotropy} and \glslink{statistical homogeneity}{homogeneity} hold.

The \gls{scalar field} $\lambda(\textbf{x})$ is usually written in terms of its Fourier components,
\begin{equation}
\lambda(\textbf{x}) = \frac{1}{(2\pi)^{3/2}} \int \lambda(\textbf{k}) \exp(\i \textbf{k} \cdot \textbf{x}) \, \drm^3 \textbf{k},
\end{equation}
or, equivalently,
\begin{equation}
\lambda(\textbf{k}) = \frac{1}{(2\pi)^{3/2}} \int \lambda(\textbf{x}) \exp(-\i \textbf{k} \cdot \textbf{x}) \, \drm^3 \textbf{x}.
\end{equation}

\noindent The quantities $\lambda(\textbf{k})$ are complex random variables. If $\lambda(\textbf{x})$ is real, one has $\lambda(-\textbf{k}) = \lambda^\ast(\textbf{k})$ which means that half of the Fourier space contains redundant information.

The computation of the \glslink{two-point correlation function}{two-point correlator} for $\lambda(\textbf{k})$ in Fourier space gives:
\begin{eqnarray}
\left\langle \lambda^\ast(\textbf{k})\lambda(\textbf{k}') \right\rangle & = & \dfrac{1}{(2\pi)^{3/2}} \dfrac{1}{(2\pi)^{3/2}} \iint \left\langle \lambda^\ast(\textbf{x})\lambda(\textbf{x}+\textbf{r}) \right\rangle \exp(\i (\textbf{k}-\textbf{k}') \cdot \textbf{x} - \i \textbf{k}' \cdot \textbf{r} ) \, \drm^3 \textbf{x} \, \drm^3 \textbf{r} \\
& = & \dfrac{1}{(2\pi)^{3}} \iint \xi(r) \exp(\i (\textbf{k}-\textbf{k}') \cdot \textbf{x} - \i \textbf{k}' \cdot \textbf{r} ) \, \drm^3 \textbf{x} \, \drm^3 \textbf{r} \\
& = & \dfrac{1}{(2\pi)^{3}} \, \updelta_\mathrm{D}(\textbf{k}-\textbf{k}') \int \xi(r) \exp(\i \textbf{k} \cdot \textbf{r}) \, \drm^3 \textbf{r} \\
& \equiv & \dfrac{1}{(2\pi)^{3/2}} \, \updelta_{\mathrm{D}}(\textbf{k}-\textbf{k}') \, P(k), \label{eq:two-pt-correlator-FS}
\end{eqnarray}
where $\updelta_{\mathrm{D}}$ is a \gls{Dirac delta distribution} and 
\begin{equation}
P(k) \equiv \frac{1}{(2\pi)^{3/2}} \int \xi(r) \exp(\i \textbf{k} \cdot \textbf{r}) \, \drm^3 \textbf{r}
\end{equation}
is defined to be the \textit{\gls{power spectrum}} of the field $\lambda(\textbf{x})$ (this relation is known as the Wiener-Khinchin theorem). Because of \gls{statistical homogeneity} and \glslink{statistical isotropy}{isotropy}, it depends only on the norm of $\textbf{k}$. The inverse relation between the \gls{two-point correlation function}, $\xi(r)$, and the \gls{power spectrum}, $P(k)$, reads
\begin{equation}
\xi(r) = \dfrac{1}{(2\pi)^{3/2}} \int P(k) \exp(- \i \textbf{k} \cdot \textbf{r}) \, \drm^3 \textbf{k}.
\end{equation}
In spherical coordinates, using
\begin{equation}
\int_\Omega \exp\left(-\i kr \cos \theta \right) \drm \Omega = \int_{\theta=0}^{\pi} \int_{\varphi=0}^{2\pi} \exp\left(-\i kr \cos \theta \right) \sin \theta \, \drm \theta \, \drm \varphi = 4\pi \frac{\sin(kr)}{kr} ,
\end{equation}
we obtain the one-dimensional relations between $\xi(r)$ and $P(k)$,
\begin{eqnarray}
P(k) & = & \frac{2}{\sqrt{\pi}} \int_{0}^{\infty} \xi(r) \, j_0(kr) \, r^2 \, \drm r ,\\
\xi(r) & = & \frac{2}{\sqrt{\pi}} \int_{0}^{\infty} P(k) \, j_0(kr) \, k^2 \, \drm k ,
\end{eqnarray}
where $j_0$ is the zero-th order spherical Bessel function,
\begin{equation}
j_0(x) \equiv \frac{\sin(x)}{x} .
\end{equation}

\paragraph{Two-point probability function and two-point correlation function.}The following physical interpretation of the \gls{two-point correlation function} establishes a link between the \gls{ensemble average} and the \gls{sample average}. Indeed, correlation functions are directly related to multivariate \glslink{pdf}{probability functions} (in fact, they are sometimes defined from them). Here we exemplify this fact for the \gls{density contrast} at position $\textbf{x}$, $\delta(\textbf{x}) \equiv \rho(\textbf{x})/\bar{\rho}-1$.

Let us consider two infinitesimal volumes $\drm V_1$ and $\drm V_2$ inside the volume $V$. Let $n_1$ and $n_2$ be the particle densities at locations $\textbf{x}_1$ and $\textbf{x}_2$ and $n \equiv N/V$ the average numerical density. Then the \glslink{density contrast}{density contrasts} are $\delta(\textbf{x}_1) = n_1/(n \, \drm V_1) -1$ and $\delta(\textbf{x}_2) = n_2/(n \, \drm V_2) -1$ and the \gls{two-point correlation function} reads
\begin{equation}
\xi(x_{12}) = \left\langle \delta(\textbf{x}_1)\delta(\textbf{x}_2) \right\rangle = \dfrac{\drm N_{12}}{n^2 \, \drm V_1 \,  \drm V_2}-1,
\end{equation}
where $x_{12} \equiv \vert \textbf{x}_2-\textbf{x}_1 \vert$ and $\drm N_{12} = \left\langle n_1 n_2 \right\rangle$ is the average number of \textit{pairs} in the volumes $\drm V_1$ and $\drm V_2$ (i.e. the product of the number of particles in one volume times the number in the other volume). One can then rewrite
\begin{equation}
\drm N_{12} = \left\langle n_1 n_2 \right\rangle = n^2 \, (1+\xi(x_{12})) \, \drm V_1 \, \drm V_2.
\end{equation}

The physical interpretation of the \gls{two-point correlation function} is that it measures the excess over uniform probability that two particles at volume elements $\drm V_1$ and $\drm V_2$ are separated by a distance $x_{12}$. If particle positions are drawn from uniform distributions (i.e. if there is no clustering), then $\drm N_{12}$ is independent of the separation. In this case, the average number of pairs is the product of the average number of particles in the two volumes, $\left\langle n_1 n_2 \right\rangle = \left\langle n_1 \right\rangle \! \left\langle n_2 \right\rangle = n^2 \, \drm V_1 \, \drm V_2$ and the correlation $\xi$ vanishes. Particles are said to be uncorrelated. Conversely, if $\xi$ is non-zero, particle distributions are said to be correlated (if $\xi>0$) or anti-correlated (if $\xi<0$).

It is sometimes easier to derive the correlation function as the average density of particles at a distance $r$ from another particle, i.e. by choosing the volume element $\drm V_1$ such as $n \, \drm V_1 = 1$. Then the number of pairs is given by the number of particles in volume $\drm V_2$:
\begin{equation}
\drm N_2 = n \, (1+\xi(r)) \, \drm V_2.
\end{equation}
Hence, one can evaluate the correlation function as follows:
\begin{equation}
\xi(r) = \dfrac{\drm N(r)}{n \, \drm V} -1 = \dfrac{\left\langle n(r) \right\rangle}{n} -1,
\end{equation}
i.e. as the average number of particles at distance $r$ from any given particle, divided by the expected number of particles at the same distance in a uniform distribution, minus one. As $\drm N_2$ is linked to the \glslink{conditional pdf}{conditional probability} that there is a particle in $\drm V_2$ given that there is one in $\drm V_1$, the previous expression is sometimes referred to as the \textit{\gls{conditional density contrast}}.

\paragraph{Two-point correlation function and power spectrum of Gaussian fields.}If $\lambda(\textbf{x})$ is a real \gls{grf} of mean 0 and covariance matrix $C$, then equation \eqref{eq:covariance-GRF} means that its \gls{two-point correlation function} in configuration space is directly given by the covariance matrix: $\left\langle \lambda(\textbf{x}_i) \lambda(\textbf{x}_j) \right\rangle = C_{ij}$.

Additionally, if $\lambda(\textbf{x})$ is also \glslink{statistical homogeneity}{statistically homogeneous}, equation \eqref{eq:two-pt-correlator-FS} implies that $\lambda(\textbf{k})$ has independent Fourier modes and that its covariance matrix in Fourier space is diagonal and contains the \gls{power spectrum} coefficients $P(k)/(2\pi)^{3/2}$. Finally, according to \gls{Wick's theorem} (section \ref{sec:Moments_GRF_Wick_theorem}), one can write for any integer $p$:
\begin{eqnarray}
\left\langle \lambda(\textbf{k}_1) \, ... \, \lambda(\textbf{k}_{2p+1}) \right\rangle & = & 0 ,\\
\left\langle \lambda(\textbf{k}_1) \, ... \, \lambda(\textbf{k}_{2p}) \right\rangle & = & \sum_{\mathrm{all\nbsp pair\nbsp associations}} \quad \prod_{p\nbsp\mathrm{pairs}\nbsp(i,j)} \left\langle \lambda(\textbf{k}_i) \lambda(\textbf{k}_j) \right\rangle \nonumber \\
& = & \sum_{\mathrm{all\nbsp pair\nbsp associations}} \quad \prod_{p\nbsp\mathrm{pairs}\nbsp(i,j)} \updelta_\mathrm{D}(\textbf{k}_i-\textbf{k}_j) \frac{P(k_i)}{(2\pi)^{3/2}} .
\end{eqnarray}
Hence, for \glslink{grf}{grfs}, all statistical properties are included in \glslink{two-point correlation function}{two-point correlations}. More specifically, all statistical properties of random variables $\lambda(\textbf{k})$ are conclusively determined by the shape of the \gls{power spectrum} $P(k)$.

\subsubsection{Higher-order correlation functions}
\label{sec:higher-order-correlators}

\paragraph{Higher-order correlation functions in configuration space.}It is possible to define \glslink{high-order correlation function}{higher-order correlation functions}, as the \textit{connected part} (denoted by a subscript c) of the joint \gls{ensemble average} of the field $\lambda(\textbf{x})$ in an arbitrary number of locations. This can be formally written as
\begin{eqnarray}
\xi_n(\textbf{x}_1, \textbf{x}_2, ..., \textbf{x}_n) & = & \langle \lambda(\textbf{x}_1) \lambda(\textbf{x}_2) \, ... \, \lambda(\textbf{x}_n) \rangle_\mathrm{c} \\
& \equiv & \langle \lambda(\textbf{x}_1) \lambda(\textbf{x}_2) \, ... \, \lambda(\textbf{x}_n) \rangle - \sum_{\mathcal{S} \in \mathcal{P}(\left\lbrace \textbf{x}_1, \textbf{x}_2, ..., \textbf{x}_n \right\rbrace)} \prod_{s_i \in \mathcal{S}} \xi_{\# s_i}(\textbf{x}_{s_i(1)}, \textbf{x}_{s_i(2)}, ..., \textbf{x}_{s_i(\# s_i)}) , \nonumber
\end{eqnarray}
where the sum is made over the proper partitions (any partition except the set itself) of $\left\lbrace \textbf{x}_1, \textbf{x}_2, ..., \textbf{x}_n \right\rbrace$ and $s_i$ is a subset of $\left\lbrace \textbf{x}_1, \textbf{x}_2, ..., \textbf{x}_n \right\rbrace$ contained in partition $\mathcal{S}$. When the average of $\lambda(\textbf{x})$ is zero, only partitions that contain no singlets contribute. The decomposition in connected and non-connected parts of the joint \gls{ensemble average} of the field can be easily visualized in a diagrammatic way \citep[see e.g.][]{Bernardeau2002}.

For \glslink{grf}{grfs}, as a consequence of \gls{Wick's theorem} (section \ref{sec:Moments_GRF_Wick_theorem}), all connected correlations functions are zero except $\xi_2$. 

\paragraph{Higher-order correlators in Fourier space.}This definition in configuration space can be extended to Fourier space. By \gls{statistical isotropy} of the field, the $n$-th Fourier-space correlator does not depend on the direction of the $\textbf{k}$-vectors. By \gls{statistical homogeneity}, the $\textbf{k}$-vectors have to sum up to zero. We can thus define $P_n(\textbf{k}_1, \textbf{k}_2, ..., \textbf{k}_n)$ by
\begin{equation}
\left\langle \lambda(\textbf{k}_1) \lambda(\textbf{k}_2) \, ... \, \lambda(\textbf{k}_n) \right\rangle_\mathrm{c} \equiv \updelta_\mathrm{D}(\textbf{k}_1+\textbf{k}_2+ ... +\textbf{k}_n) \, P_n(\textbf{k}_1, \textbf{k}_2, ..., \textbf{k}_n) .
\end{equation}
The \gls{Dirac delta distribution} $\updelta_\mathrm{D}$ ensures that $\textbf{k}$-vector configurations form closed polygons: $\sum_i \textbf{k}_i = \textbf{0}$. 

Let us now examine the lowest-order connected \glslink{moment}{moments}.

\paragraph{Bispectrum.}After the \gls{power spectrum}, the second statistic of interest is the \gls{bispectrum} $B$, for $n=3$, defined by
\begin{equation}
\left\langle \lambda(\textbf{k}_1) \lambda(\textbf{k}_2) \lambda(\textbf{k}_3) \right\rangle = \left\langle \lambda(\textbf{k}_1) \lambda(\textbf{k}_2) \lambda(\textbf{k}_3) \right\rangle_\mathrm{c} \equiv \updelta_\mathrm{D}(\textbf{k}_1+\textbf{k}_2+\textbf{k}_3) \,  B(\textbf{k}_1, \textbf{k}_2, \textbf{k}_3) .
\label{eq:def-bispectrum}
\end{equation}

\paragraph{Reduced bispectrum.}It is convenient to define the \gls{reduced bispectrum} $Q(\textbf{k}_1, \textbf{k}_2, \textbf{k}_3)$ as
\begin{equation}
Q(\textbf{k}_1, \textbf{k}_2, \textbf{k}_3) \equiv \dfrac{B(\textbf{k}_1, \textbf{k}_2, \textbf{k}_3)}{P(k_1)P(k_2) + P(k_1)P(k_3) + P(k_2)P(k_3)} ,
\end{equation}
which takes away most of the dependence on scale and cosmology. The \gls{reduced bispectrum} is useful for \glslink{model comparison}{comparing different models}, because its weak dependence on cosmology is known to break degeneracies between \gls{cosmological parameters} and to isolate the effects of gravity \citep[see][for an example]{Gil-Marin2011}.

\paragraph{Trispectrum.}The \gls{trispectrum} is the following correlator, for $n=4$. It is defined as
\begin{equation}
\left\langle \lambda(\textbf{k}_1) \lambda(\textbf{k}_2) \lambda(\textbf{k}_3) \lambda(\textbf{k}_4) \right\rangle_\mathrm{c} \equiv \updelta_\mathrm{D}(\textbf{k}_1+\textbf{k}_2+\textbf{k}_3+\textbf{k}_4) \, T(\textbf{k}_1, \textbf{k}_2, \textbf{k}_3, \textbf{k}_4) .
\end{equation}

\section{Dynamics of gravitational instability}
\label{sec:Dynamics}

The standard picture for the formation of the \gls{LSS} as seen in \glslink{galaxy survey}{galaxy surveys} is the \glslink{gravitational evolution}{gravitational amplification} of \glslink{initial conditions}{primordial density fluctuations}. The dynamics of this process is mostly governed by gravitational interactions of collisionless (or at least, weakly-interacting) \gls{dark matter particles} in an expanding universe.

For scales much smaller than the \gls{Hubble radius}, \glslink{general relativity}{relativistic effects} (such as the curvature of the Universe or the apparent distance-redshift relation) are believed to be subdominant or negligible \citep[e.g.][and references therein]{Kolb1990} and as we will show, the \gls{expansion} of the background can be factored out by a redefinition of variables. Although the microscopic nature of \gls{dark matter particles} remains unknown, candidates have to pass several tests in order to be viable \citep{Taoso2008}. In particular, particles which are relativistic at the time of \gls{structure formation} lead to a large damping of small-scale fluctuations \citep{Silk1968,Bond1983}, incompatible with observed structures. The standard theory thus requires \gls{dark matter particles} to be \glslink{CDM}{cold} during \gls{structure formation}, i.e. non-relativistic well before the \glslink{matter domination}{matter-dominated era} \citep{Peebles1982,Blumenthal1984,Davis1985}. For these two reasons, at scales much smaller than the \gls{Hubble radius} the \glslink{equation of motion}{equations of motion} can be well approximated by \gls{Newtonian gravity}.

In addition, all dark matter particle candidates are extremely light compared to the mass of typical astrophysical objects such as stars or galaxies, with an expected number density of a least $10^{50}$ particles per $\mathrm{Mpc}^3$. Therefore, discreteness effects are negligible and collisionless \gls{dark matter} can be well described in the \gls{fluid} limit approximation.

In this section, we present the dynamics of \glslink{gravitational evolution}{gravitational instability} in the framework of \gls{Newtonian gravity} within a flat, expanding background and in the fluid limit approximation. It is of course possible to do a detailed relativistic treatment of \gls{structure formation} dynamics and cosmological perturbation theory \citep{Bardeen1980,Mukhanov1992} and to derive the \glslink{Newtonian gravity}{Newtonian limit} from \gls{general relativity} \citep[see e.g.][]{Peebles1980}.

\subsection{The Vlasov-Poisson system}
\label{sec:The Vlasov-Poisson system}

\paragraph{The cosmological Poisson equation.} Let us consider a large number of particles that interact only gravitationally in an \glslink{expansion}{expanding universe}. For a particle of velocity $\textbf{v}$ at position  $\textbf{r}$, the action of all other particles can be treated as a smooth gravitational potential induced by the local mass density $\rho(\textbf{r})$,
\begin{equation}
\phi(\textbf{r}) = \G \int \dfrac{\rho(\textbf{r}'-\textbf{r})}{\vert \textbf{r}' - \textbf{r} \vert} \, \drm^3 \textbf{r}' ,
\end{equation}
and the \gls{equation of motion} reads
\begin{equation}
\deriv{\textbf{v}}{t} = - \nabla_\textbf{r} \phi = \G \int \dfrac{\rho(\textbf{r}'-\textbf{r}) \, (\textbf{r}'-\textbf{r})}{\vert \textbf{r}' - \textbf{r} \vert^3} \, \drm^3 \textbf{r}'.
\label{eq:velocity-derivative}
\end{equation}
Examining \glslink{gravitational evolution}{gravitational instabilities} in the context of an expanding universe requires to consider the departure from the \glslink{homogeneous Universe}{homogeneous} \gls{Hubble flow}. It is natural to describe the positions of particles in terms of their \gls{comoving coordinates} $\textbf{x}$ such that the physical coordinates are $\textbf{r}=a\textbf{x}$ and of the \gls{conformal time} $\tau$, defined by $\drm t = a(\tau) \, \drm \tau$. Hereafter, when there is no ambiguity, we will denote $\nabla \equiv \nabla_\textbf{x}$ and $\Delta \equiv \Delta_\textbf{x}$. The Jacobian of the spatial coordinate transformation is $\vert J \vert = a^3$ so that the right-hand side of the previous equation becomes
\begin{eqnarray}
\G \int \dfrac{\rho(\textbf{r}'-\textbf{r}) \, (\textbf{r}'-\textbf{r})}{\vert \textbf{r}' - \textbf{r} \vert^3} \, \drm^3 \textbf{r}' & = & \G \int \dfrac{\rho(\textbf{x}'-\textbf{x}) \, a \, (\textbf{x}'-\textbf{x})}{a^3\vert \textbf{x}'-\textbf{x} \vert^3} \, a^3 \, \drm^3 \textbf{x}' \\
& = & \G a \bar{\rho} \int \dfrac{(\textbf{x}'-\textbf{x})}{\vert \textbf{x}' - \textbf{x} \vert^3} \, \drm^3 \textbf{x}' + \G a \bar{\rho} \int \delta(\textbf{x}'-\textbf{x}) \dfrac{(\textbf{x}'-\textbf{x})}{\vert \textbf{x}' - \textbf{x} \vert^3} \, \drm^3 \textbf{x}' ,\label{eq:velocity-derivative-2}
\end{eqnarray}
where we have introduced the \gls{density contrast} $\delta(\textbf{x})$, defined as
\begin{equation}
\rho(\textbf{x},t) \equiv \bar{\rho}(t) \left[ 1 + \delta(\textbf{x},t) \right] ,
\end{equation}
where $\bar{\rho}(t) \propto 1/a^3$ (consequence of equation \eqref{eq:energy-density-conservation} with $w=0$).

\glslink{velocity field}{Velocities} of particles are $\textbf{v} = \dot{a}\textbf{x} + a \, \drm \textbf{x}/\drm t$, permitting us to define \glslink{peculiar velocity}{peculiar velocities} as the difference between total velocities and the \gls{Hubble flow}:
\begin{equation}
\textbf{u} \equiv a \, \deriv{\textbf{x}}{t} = \textbf{v} - \dot{a}\textbf{x}.
\end{equation}
$\drm \textbf{v} / \drm t$ is written in terms of $\textbf{u}$ as
\begin{eqnarray}
\deriv{\textbf{v}}{t} & = & \deriv{\textbf{u}}{t} + \ddot{a} \textbf{x} + \dot{a} \, \deriv{\textbf{x}}{t} \\
& = & \deriv{\textbf{u}}{t} + \ddot{a} \textbf{x} + \dfrac{\dot{a}}{a} \, \textbf{u}. \label{eq:velocity-derivative-3}
\end{eqnarray}
By the use of the second \glslink{Friedmann's equations}{Friedmann equation} for the \glslink{homogeneous Universe}{homogeneous} background (equation \eqref{eq:Friedmann2-conformal-time}),
\begin{equation}
\ddot{a} = -\frac{4\pi \G}{3} a \bar{\rho},
\end{equation}
and Gauss's theorem,
\begin{equation}
\frac{4\pi}{3}\textbf{x} = -\int \frac{(\textbf{x}'-\textbf{x})}{\vert \textbf{x}' - \textbf{x} \vert^3} \drm^3 \textbf{x}',
\end{equation}
the term $\ddot{a} \textbf{x}$ is equal to 
\begin{equation}
\G a \bar{\rho} \int \dfrac{(\textbf{x}'-\textbf{x})}{\vert \textbf{x}' - \textbf{x} \vert^3} \, \drm^3 \textbf{x}' \equiv -\frac{1}{a} \nabla_\textbf{x} \upphi, 
\end{equation}
which leaves for the \gls{peculiar velocity} the following \gls{equation of motion} (see equations \eqref{eq:velocity-derivative}, \eqref{eq:velocity-derivative-2} and \eqref{eq:velocity-derivative-3}):
\begin{equation}
\deriv{\textbf{u}}{t} + \frac{\dot{a}}{a} \, \textbf{u} = \G a \bar{\rho} \int \delta(\textbf{x}'-\textbf{x}) \dfrac{(\textbf{x}'-\textbf{x})}{\vert \textbf{x}' - \textbf{x} \vert^3} \, \drm^3 \textbf{x}' \equiv - \dfrac{1}{a} \nabla_\textbf{x} \Phi .
\label{eq:equation-of-motion-intermediate}
\end{equation}
Here we have defined the cosmological \gls{gravitational potential} $\Phi$ such that $\phi \equiv \upphi + \Phi$ with, for the background,
\begin{equation}
\upphi(\textbf{x}) = \frac{4\pi \G}{3} a^2 \bar{\rho} \left( \frac{1}{2} \vert\textbf{x}\vert^2 \right) = -\mathcal{H}' \left( \frac{1}{2} \vert\textbf{x}\vert^2 \right), \quad \mathrm{satisfying} \quad \Delta \upphi = 4 \pi \G a^2 \bar{\rho} .
\end{equation}
Using the overall \gls{Poisson equation}, $\Delta_\textbf{r} \phi = \Delta \phi / a^2= 4\pi\G \bar{\rho}(1+\delta)$, we find that $\Phi$ follows a cosmological \gls{Poisson equation} sourced only by \glslink{density contrast}{density fluctuations}, as expected:
\begin{equation}
\label{eq:Poisson}
\Delta \Phi = 4\pi \G a^2 \bar{\rho} \delta = \dfrac{3}{2} \Omega_\mathrm{m}(\tau) \mathcal{H}^2(\tau) \delta.
\end{equation}
The second equality comes from the first \glslink{Friedmann's equations}{Friedmann equation} in a flat universe (equation \eqref{eq:Friedmann1-conformal-time} with $k=0$).

\paragraph{The Vlasov equation.} Looking at equation \eqref{eq:equation-of-motion-intermediate}, the \gls{momentum} of a single particle of mass $m$ is identified as:
\begin{equation}
\textbf{p} = m a \textbf{u},
\end{equation}
and the \gls{equation of motion} reads:
\begin{equation}
\label{eq:equation-of-motion-p}
\deriv{\textbf{p}}{t} = - m \nabla_\textbf{x} \Phi = -ma \nabla_\textbf{r} \Phi \quad \mathrm{or} \quad \deriv{\textbf{p}}{\tau} = - ma \nabla_\textbf{x} \Phi .
\end{equation}
Let us now define the particle number density in \gls{phase space} by $f(\textbf{x},\textbf{p},\tau)$. \glslink{phase space}{Phase-space} conservation and \gls{Liouville's theorem} imply the \gls{Vlasov equation} (collisionless version of the Boltzmann equation):
\begin{equation}
\label{eq:Vlasov}
\deriv{f}{\tau} = \pd{f}{\tau} + \dfrac{\textbf{p}}{ma} \cdot \nabla f - ma \nabla \Phi \cdot \pd{f}{\textbf{p}} = 0 .
\end{equation}

Given equations \eqref{eq:Poisson} and \eqref{eq:Vlasov}, the \gls{Vlasov-Poisson system} is closed.

\subsection{Fluid dynamics approach, evolution equations in phase space}
\label{sec:fluid-dynamics-approach}

The \gls{Vlasov equation} is very difficult to solve, since it is a partial differential equation involving seven variables, with non-linearity induced by the dependence of the \glslink{gravitational potential}{potential} $\Phi$ on the \glslink{density field}{density} through the \gls{Poisson equation}. Its complicated structure prevents the analytic analysis of \gls{dark matter} dynamics. In practice, we are usually not interested in solving the full \glslink{phase space}{phase-space} dynamics, but only the evolution of the spatial distribution. It is therefore convenient to take \gls{momentum} \glslink{moment}{moments} of the distribution function. This yields a \gls{fluid} dynamics approach for the motion of collisionless dark matter. The zeroth-order \gls{momentum}, by construction, relates the \glslink{phase space}{phase-space} density to the \gls{density field},
\begin{equation}
\int f(\textbf{x},\textbf{p},\tau) \, \drm^3 \textbf{p} \equiv \rho(\textbf{x},\tau) .
\end{equation}
The next order \glslink{moment}{moments},
\begin{equation}
\int \dfrac{\textbf{p}}{ma} f(\textbf{x},\textbf{p},\tau) \, \drm^3 \textbf{p} \equiv \rho(\textbf{x},\tau) \textbf{u}(\textbf{x},\tau) ,
\end{equation}
\begin{equation}
\int \dfrac{\textbf{p}_i \textbf{p}_j}{m^2 a^2} f(\textbf{x},\textbf{p},\tau) \, \drm^3 \textbf{p} \equiv \rho(\textbf{x},\tau) \textbf{u}_i(\textbf{x},\tau) \textbf{u}_j(\textbf{x},\tau) + \sigma_{ij}(\textbf{x},\tau) ,
\label{eq:def-stress-tensor}
\end{equation}
define the \textit{\gls{peculiar velocity flow}} $\textbf{u}(\textbf{x},\tau)$ (average local velocity of particles in a region of space; for simplification, we use the same notation as the \gls{peculiar velocity} of a single particle) and the \textit{\gls{stress tensor}} $\sigma_{ij}(\textbf{x},\tau)$ which can be related to the \textit{\gls{velocity dispersion} tensor}, $v_{ij}(\textbf{x},\tau)$, by $\sigma_{ij}(\textbf{x},\tau)~\equiv~\rho(\textbf{x},\tau) v_{ij}(\textbf{x},\tau)$.

By taking successive \gls{momentum} \glslink{moment}{moments} of the \gls{Vlasov equation} and integrating out \glslink{phase space}{phase-space} information, a hierarchy of equations that couple successive \glslink{moment}{moments} of the distribution function can be constructed. The zeroth \gls{moment} of the \gls{Vlasov equation} gives the \gls{continuity equation},
\begin{equation}
\label{eq:continuity}
\pd{\delta(\textbf{x},\tau)}{\tau} + \nabla \cdot \left\lbrace \left[ 1 + \delta(\textbf{x},\tau) \right] \textbf{u}(\textbf{x},\tau) \right\rbrace = 0,
\end{equation}
which describes the conservation of mass. Taking the first \gls{moment} and subtracting $\bar{\rho}\, \textbf{u}(\textbf{x},\tau)$ times the \gls{continuity equation} yields the \glslink{Euler's equation}{Euler equation},
\begin{equation}
\label{eq:Euler}
\pd{\textbf{u}_i(\textbf{x},\tau)}{\tau} + \mathcal{H}(\tau) \textbf{u}_i(\textbf{x},\tau) + \textbf{u}_j(\textbf{x},\tau) \cdot \nabla_j \textbf{u}_i(\textbf{x},\tau) = - \nabla_i \Phi(\textbf{x},\tau) - \dfrac{1}{\rho(\textbf{x},\tau)} \nabla_j(\sigma_{ij}(\textbf{x},\tau)),
\end{equation}
which describes the conservation of \gls{momentum}. This equation is very similar to that of hydrodynamics, apart from an additional term which accounts for the \gls{expansion} of the Universe and the fact that, contrary to perfect fluids, auto-gravitating systems may have an anisotropic \gls{stress tensor}.

The infinite sequence of \gls{momentum} \glslink{moment}{moments} of the \gls{Vlasov equation} is usually truncated at this point and completed by fluid dynamics assumptions to close the system. Specifically, one postulates an Ansatz for the \gls{stress tensor}, namely the \gls{equation of state} of the cosmological fluid. For example, if the fluid is locally thermalized, the \gls{velocity dispersion} becomes isotropic and proportional to the pressure \citep[e.g.][]{Bernardeau2002}:
\begin{equation}
\sigma_{ij} = - P \updelta_{\mathrm{K}}^{ij},
\end{equation}
where $\updelta_{\mathrm{K}}^{ab}$ is a \gls{Kronecker symbol}. Standard fluid dynamics also prescribes, with the addition of a \glslink{viscosity}{viscous} \gls{stress tensor}, the following equation \citep[e.g.][]{Bernardeau2002}:
\begin{equation}
\sigma_{ij} = -P \updelta_{\mathrm{K}}^{ij} + \zeta (\nabla \cdot \textbf{u}) \updelta_{\mathrm{K}}^{ij} + \mu \left[ \nabla_i \textbf{u}_j + \nabla_j \textbf{u}_i - \dfrac{2}{3} (\nabla \cdot \textbf{u}) \updelta_{\mathrm{K}}^{ij} \right] ,
\end{equation}
where $\zeta$ is the coefficient of bulk \gls{viscosity} and $\mu$ is the coefficient of shear \gls{viscosity}.

\subsection{The single-stream approximation}
\label{sec:single-stream-approx}

At the early stages of cosmological \glslink{gravitational evolution}{gravitational instability}, it is possible to further simplify and to rely on a different hypothesis, namely the \textit{\gls{single-stream approximation}}. At sufficiently large scales, gravity-induced cosmic flows will dominate over the \gls{velocity dispersion} due to \glslink{peculiar velocity}{peculiar motions}. The \gls{single-stream approximation} consists in assuming that for \gls{CDM}, \gls{velocity dispersion} and pressure are negligible, i.e. $\sigma_{ij} = 0$, and that all particles have identical \glslink{peculiar velocity}{peculiar velocities}. Hence, the density in \gls{phase space} can be written
\begin{equation}
f(\textbf{x},\textbf{p},\tau) = \rho(\textbf{x},\tau) \, \updelta_{\mathrm{D}} \! \left[ \textbf{p}-ma\textbf{u}(\textbf{x}) \right] .
\end{equation}
Note, that from its definition, equation \eqref{eq:def-stress-tensor}, the \gls{stress tensor} characterizes the deviation of particle motions from a single coherent flow.

The \gls{single-stream approximation} only works at the beginning of gravitational \gls{structure formation}, when structures had no time to collapse and virialize. Because of non-linearity in the \gls{Vlasov-Poisson system}, later stages will involve the superposition of three or more streams in \gls{phase space}, indicating the break down of the approximation at increasingly larger scales. The breakdown of $\sigma_{ij} \approx 0$, describing the generation of \gls{velocity dispersion} or \glslink{stress tensor}{anisotropic stress} due to the multiple-stream regime, is generically known as \textit{\gls{shell-crossing}}. Beyond that point, the density in \gls{phase space} exhibits no simple form, generally preventing further analytic analysis. This issue will be discussed further in chapters \ref{chap:lpt} and \ref{chap:remapping}.

The \gls{single-stream approximation} yields the following system of equations:
\begin{eqnarray}
\label{eq:continuity-single-stream}
\pd{\delta(\textbf{x},\tau)}{\tau} + \nabla \cdot \left\lbrace \left[ 1 + \delta(\textbf{x},\tau) \right] \textbf{u}(\textbf{x},\tau) \right\rbrace & = & 0, \\
\label{eq:Euler-single-stream}
\pd{\textbf{u}_i(\textbf{x},\tau)}{\tau} + \mathcal{H}(\tau) \textbf{u}_i(\textbf{x},\tau) + \textbf{u}_j(\textbf{x},\tau) \cdot \nabla_j \textbf{u}_i(\textbf{x},\tau) & = & - \nabla_i \Phi(\textbf{x},\tau), \\
\label{eq:Poisson-single-stream}
\Delta \Phi(\textbf{x},\tau) & = & 4\pi \G a^2(\tau) \bar{\rho}(\tau) \delta(\textbf{x},\tau) .
\end{eqnarray}
It is a non-linear, closed system of equations involving the local \gls{density contrast}, the local \gls{velocity field} and the local \gls{gravitational potential}.

There exists no general analytic solution to the \gls{fluid} dynamics of collisionless self-gravitating \gls{CDM}, even in the single-stream regime. However, literature provides several different analytic perturbative expansion techniques to yield approximate solutions for the \gls{dark matter} dynamics, which we briefly review below (sections \ref{sec:EPT} and \ref{sec:LPT}).

\section{Eulerian perturbation theory}
\label{sec:EPT}

\subsection{Eulerian linear perturbation theory}
\label{sec:EPT-linear}

\glslink{EPT}{}As mentioned above, at large scales and during the early stages of \gls{gravitational evolution}, we expect the matter distribution to be smooth and to follow a single velocity stream. In this regime, it is therefore possible to linearize equation \eqref{eq:continuity-single-stream} and \eqref{eq:Euler-single-stream}, assuming that fluctuations of density are small compared to the \glslink{homogeneous Universe}{homogeneous} contribution and that gradients of \glslink{velocity field}{velocity fields} are small compared to the \gls{Hubble parameter},
\begin{eqnarray}
\vert \delta(\textbf{x},\tau) \vert & \ll & 1, \\
\vert \nabla_j \textbf{u}_i(\textbf{x},\tau) \vert & \ll & \mathcal{H}(\tau) .
\end{eqnarray}
We obtain the \gls{equation of motion} in the \textit{\gls{linear regime}},
\begin{eqnarray}
\label{eq:continuity-linear}
\pd{\delta(\textbf{x},\tau)}{\tau} + \theta(\textbf{x},\tau) & = & 0, \\
\label{eq:Euler-linear}
\pd{\textbf{u}(\textbf{x},\tau)}{\tau} + \mathcal{H}(\tau) \textbf{u}(\textbf{x},\tau) & = & - \nabla \Phi(\textbf{x},\tau) ,
\end{eqnarray}
where $\theta(\textbf{x},\tau) \equiv \nabla \cdot \textbf{u}(\textbf{x},\tau)$ is the divergence of the \gls{velocity field}. The \gls{velocity field}, as any vector field, is completely described by its divergence $\theta(\textbf{x},\tau)$ and its curl, referred to as the \gls{vorticity}, $\textbf{w}(\textbf{x},\tau)~\equiv~\nabla~\times~\textbf{u}(\textbf{x},\tau)$, whose equations of motion follow from taking the divergence and the curl of equation \eqref{eq:Euler-single-stream} and using the \gls{Poisson equation}:
\begin{eqnarray}
\label{eq:divergence-linear}
\pd{\theta(\textbf{x},\tau)}{\tau} + \mathcal{H}(\tau)\theta(\textbf{x},\tau) + 4\pi \G a^2(\tau) \bar{\rho}(\tau) \delta(\textbf{x},\tau) & = & 0, \\
\label{eq:vorticity-linear}
\pd{\textbf{w}(\textbf{x},\tau)}{\tau} + \mathcal{H}(\tau)\textbf{w}(\textbf{x},\tau) & = & 0.
\end{eqnarray}
Since \gls{vorticity} is not sourced in the \gls{linear regime}, any initial \gls{vorticity} rapidly decays due to the \glslink{expansion}{expansion of the Universe}. Indeed, its evolution immediately follows from equation \eqref{eq:vorticity-linear}: $\textbf{w}(\tau) \propto 1/a(\tau)$. In the \gls{non-linear regime}, the emergence of \glslink{stress tensor}{anisotropic stress} in the right-hand side of \gls{Euler's equation} can lead to \gls{vorticity} generation \citep{Pichon1999}.

The \gls{density contrast} evolution follows by replacing equation \eqref{eq:continuity-linear} and its time derivative in equation \eqref{eq:divergence-linear}:
\begin{equation}
\pd{^2\delta(\textbf{x},\tau)}{\tau^2} + \mathcal{H}(\tau) \pd{\delta(\textbf{x},\tau)}{\tau} = 4\pi \G a^2(\tau) \bar{\rho}(\tau) \delta(\textbf{x},\tau) = \dfrac{3}{2} \Omega_\mathrm{m}(\tau) \mathcal{H}^2(\tau) \delta(\textbf{x},\tau) .
\end{equation}

\subsection{The growth of fluctuations in linear theory}

This linear equation allows us to look for different fluctuation modes, decoupling spatial and time contributions by writing $\delta(\textbf{x},\tau) = D_1(\tau) \, \delta(\textbf{x},0)$, where some ``initial'' reference time is labeled as $0$ and where $D_1(\tau)$ is called the \textit{\gls{linear growth factor}}. The time dependence of the fluctuation growth rate satisfies
\begin{equation}
\label{eq:evolution-growth-factor}
\deriv{^2 D_1(\tau)}{\tau^2} + \mathcal{H}(\tau) \deriv{D_1(\tau)}{\tau} = \dfrac{3}{2} \Omega_\mathrm{m}(\tau) \mathcal{H}^2(\tau) D_1(\tau),
\end{equation}
regardless of $\textbf{x}$ (or of the Fourier mode $\textbf{k}$): in the \gls{linear regime}, the growth of fluctuations is scale-independent. This equation, together with \gls{Friedmann's equations}, equations \eqref{eq:Friedmann-2} and \eqref{eq:Friedmann-1}, determines the growth of \glslink{density contrast}{density perturbations} in the \gls{linear regime} as a function of \gls{cosmological parameters}. There are two independent solutions, the \glslink{growing mode}{fastest growing mode} $D_1^{(+)}(\tau)$ and the \glslink{decaying mode}{slowest growing mode} $D_1^{(-)}(\tau)$. The evolution of the \gls{density contrast} is then given by:
\begin{equation}
\label{eq:density-field-linearEPT}
\delta(\textbf{x},\tau) = D_1^{(+)}(\tau)\delta_+(\textbf{x}) + D_1^{(-)}(\tau)\delta_-(\textbf{x}),
\end{equation}
where $\delta_+(\textbf{x})$ and $\delta_-(\textbf{x})$ are two functions of position only describing the initial \gls{density field} configuration.

In terms of the \gls{scale factor} and using \gls{Friedmann's equations}, equation \eqref{eq:evolution-growth-factor} can be rewritten as
\begin{equation}
\label{eq:evolution-growth-factor-a}
a^2 \deriv{^2 D_1}{a^2} + \left( \Omega_\Lambda(a) - \dfrac{\Omega_\mathrm{m}(a)}{2} + 2 \right) a \deriv{D_1}{a} = \dfrac{3}{2} \Omega_\mathrm{m}(a) D_1 ,
\end{equation}
where the \gls{cosmological parameters} $\Omega_\Lambda(a)$ and $\Omega_\mathrm{m}(a)$ now depend on the \gls{scale factor} \citep[for more details on this derivation and a generalization to time-varying \gls{dark energy} models, see][]{Percival2005b}.

Using the linearized \gls{continuity equation}, equation \eqref{eq:continuity-linear}, the velocity divergence is given by
\begin{equation}
\label{eq:velocity-field-linearEPT}
\theta(\textbf{x},\tau) = -\mathcal{H}(\tau) \left[ f(\Omega_i) \delta_+(\textbf{x},\tau) + g(\Omega_i) \delta_-(\textbf{x},\tau) \right] .
\end{equation}
It does not depend directly on the \gls{linear growth factor} of each mode, but on its logarithmic derivative, the exponent in the momentary power law relating $D$ to $a$,
\begin{equation}
f(\Omega_i) \equiv \dfrac{1}{\mathcal{H}(\tau)} \deriv{\ln D_1^{(+)}}{\tau} = \deriv{\ln D_1^{(+)}}{\ln a}, \quad g(\Omega_i) \equiv \dfrac{1}{\mathcal{H}(\tau)} \deriv{\ln D_1^{(-)}}{\tau} = \deriv{\ln D_1^{(-)}}{\ln a} ,
\end{equation}
with $\delta_\pm(\textbf{x},\tau) \equiv D_1^{(\pm)}(\tau) \delta_\pm(\textbf{x})$.

We now review some cosmological models for which analytic expressions exist.
\begin{enumerate}
\item For a \glslink{SCDM}{standard cold dark matter} (\gls{SCDM}) model, i.e. a particular case of an \gls{Einstein-de Sitter universe} \citep{Einstein1932} where dark matter is \glslink{CDM}{cold}, the \gls{cosmological parameters} are time-independent: $\Omega_\mathrm{m}(a) = 1$ and $\Omega_\Lambda(a) = 0$. Using equation \eqref{eq:evolution-growth-factor-a}, the evolution of the \gls{linear growth factor} satisfies
\begin{equation}
a^2 \deriv{^2 D_1}{a^2} + \dfrac{3}{2} a \deriv{D_1}{a} = \dfrac{3}{2} D_1 .
\end{equation}
Two independent solutions are
\begin{equation}
\label{eq:solution-EdS}
D_1^{(+)} \propto a, \quad f(\Omega_\mathrm{m} = 1, \Omega_\Lambda = 0) = 1, \quad D_1^{(-)} \propto a^{-3/2}, \quad g(\Omega_\mathrm{m} = 1, \Omega_\Lambda = 0) = - \dfrac{3}{2} ,
\end{equation}
thus density fluctuations grow as the \gls{scale factor}, $\delta \propto a$, once the \gls{decaying mode} has vanished.
\item For an \glslink{OCDM}{open cold dark matter} (\gls{OCDM}) model, the \gls{cosmological parameters} are $\Omega_\mathrm{m}(a) < 1$ and $\Omega_\Lambda(a) = 0$. The solutions of equation \eqref{eq:evolution-growth-factor-a} are \citep{Groth1975}, with $x \equiv a(\tau)(1/\Omega_\mathrm{m}^{(0)} - 1)$,
\begin{equation}
D_1^{(+)} = 1 + \dfrac{3}{x} + 3 \, \dfrac{(1+x)^{1/2}}{x^{3/2}} \ln \left[ (1+x)^{1/2} - x^{1/2} \right], \quad D_1^{(-)} = \dfrac{(1+x)^{1/2}}{x^{3/2}} .
\end{equation}
The dimensionless parameter $g$ is calculated to be
\begin{equation}
g(\Omega_\mathrm{m}, \Omega_\Lambda = 0) = - \dfrac{\Omega_\mathrm{m}}{2} - 1 ,
\end{equation}
and the dimensionless parameter $f$ can be approximated by \citep{Peebles1976,Peebles1980}
\begin{equation}
f(\Omega_\mathrm{m}, \Omega_\Lambda = 0) \approx \Omega_\mathrm{m}^{3/5} .
\end{equation}
As $\Omega_\mathrm{m} \rightarrow 0$ ($a \rightarrow \infty$ and $x \rightarrow \infty$), $D_1^{(+)} \rightarrow 1$ and $D_1^{(-)} \sim x^{-1}$: perturbations cease to grow.
\item For a universe with \glslink{CDM}{cold dark matter} and a \gls{cosmological constant}, $\Omega_\mathrm{m}(a) < 1$ and $0 < \Omega_\Lambda(a) \leq 1$ ({\LCDM} model), allowing the possibility of a curvature term ($\Omega_\mathrm{tot}(a) = \Omega_\mathrm{m}(a) + \Omega_\Lambda(a) \neq 1$), the first \glslink{Friedmann's equations}{Friedmann equation}, equation \eqref{eq:Friedmann1-conformal-time}, allows to write the \gls{Hubble parameter} as
\begin{equation}
H(a) = H_0 \sqrt{\Omega_\Lambda^{(0)} + (1 -\Omega_\Lambda^{(0)} -\Omega_\mathrm{m}^{(0)}) a^{-2} + \Omega_\mathrm{m}^{(0)} a^{-3}} .
\end{equation}
It can be checked that this expression is a solution of equation \eqref{eq:evolution-growth-factor-a}. The \gls{decaying mode} is then
\begin{equation}
\label{eq:decaying-growth-factor-LCDM}
D_1^{(-)} \propto H(a) = \dfrac{\mathcal{H}(a)}{a}.
\end{equation}
Using this particular solution and the variation of parameters method, the other solution is found to be \citep{Heath1977,Carroll1992,Bernardeau2002}
\begin{equation}
D_1^{(+)} \propto a^3 H^3(a) \int_0^a \dfrac{\drm \tilde{a}}{\tilde{a}^3 H^3(\tilde{a})} .
\end{equation}
Due to equations \eqref{eq:Friedmann-2} and \eqref{eq:decaying-growth-factor-LCDM}, one finds for arbitrary $\Omega_\mathrm{m}$ and $\Omega_\Lambda$,
\begin{equation}
g(\Omega_\mathrm{m}, \Omega_\Lambda) = \Omega_\Lambda -\dfrac{\Omega_\mathrm{m}}{2} - 1 ,
\end{equation}
and $f$ can be approximated by \citep{Lahav1991}
\begin{equation}
f(\Omega_\mathrm{m}, \Omega_\Lambda) \approx \left[ \dfrac{\Omega_\mathrm{m}^{(0)}a^{-3}}{\Omega_\mathrm{m}^{(0)}a^{-3} + (1 -\Omega_\Lambda^{(0)} -\Omega_\mathrm{m}^{(0)})a^{-2} + \Omega_\Lambda^{(0)} } \right]^{3/5}
\end{equation}
or \citep{Lightman1990,Carroll1992}
\begin{equation}
f(\Omega_\mathrm{m}, \Omega_\Lambda) \approx \left[ \dfrac{\Omega_\mathrm{m}^{(0)}a^{-3}}{\Omega_\mathrm{m}^{(0)}a^{-3} + (1 -\Omega_\Lambda^{(0)} -\Omega_\mathrm{m}^{(0)})a^{-2} + \Omega_\Lambda^{(0)} } \right]^{4/7} .
\end{equation}
For flat Universe, $\Omega_\mathrm{m} + \Omega_\Lambda = 1$, we have \citep{Bouchet1995,Bernardeau2002}
\begin{equation}
f(\Omega_\mathrm{m}, \Omega_\Lambda = 1 - \Omega_\mathrm{m}) \approx \Omega_\mathrm{m}^{5/9} .
\end{equation}
\end{enumerate}

In the case of the \gls{Einstein-de Sitter universe}, an interpretation of the \glslink{growing mode}{growing} and \glslink{decaying mode}{decaying modes} can be easily given. Referring to the solution for the \glslink{linear growth factor}{growth factor}, equation \eqref{eq:solution-EdS}, the initial \gls{density field} (equation \eqref{eq:density-field-linearEPT}) and the initial \gls{velocity field} (equation \eqref{eq:velocity-field-linearEPT}) are written
\begin{eqnarray}
\delta_\mathrm{init}(\textbf{x}) & \equiv & \delta(\textbf{x},0) = \delta_+(\textbf{x}) + \delta_-(\textbf{x}) ,\\
\theta_\mathrm{init}(\textbf{x}) & \equiv & \theta(\textbf{x},0) = - \mathcal{H}(0) \left[ \delta_+(\textbf{x}) - \dfrac{3}{2} \delta_-(\textbf{x}) \right] ,
\end{eqnarray}
if we assume that $D_+$ and $D_-$ are normalized to unity at the initial time. These relations can be inverted to give
\begin{eqnarray}
\delta_+(\textbf{x}) & = & \dfrac{3}{5} \left( \delta_\mathrm{init}(\textbf{x}) - \dfrac{2}{3} \dfrac{\theta_\mathrm{init}(\textbf{x})}{\mathcal{H}(0)} \right), \\
\delta_-(\textbf{x}) & = & \dfrac{2}{5} \left( \delta_\mathrm{init}(\textbf{x}) + \dfrac{\theta_\mathrm{init}(\textbf{x})}{\mathcal{H}(0)} \right).
\end{eqnarray}
From these expressions, the interpretation of the modes become clear. The sign is significant: recall that for a \gls{growing mode} alone we would expect $\theta_\mathrm{init} = - \mathcal{H}(0) \delta_\mathrm{init}$ and for a \gls{decaying mode} alone, $\theta_\mathrm{init} = 3/2 \, \mathcal{H}(0) \delta_\mathrm{init}$. A pure \gls{growing mode} corresponds to the case where the \glslink{density field}{density} and \glslink{velocity field}{velocity fields} are initially ``in phase'', in the sense that the \gls{velocity field} converges towards the \glslink{potential well}{potential wells} defined by the \gls{density field}. A pure \gls{decaying mode} corresponds to the case where the \glslink{density field}{density} and \glslink{velocity field}{velocity fields} are initially ``opposite in phase'', the \gls{velocity field} being such that particles escape \glslink{potential well}{potential wells}.

\subsection{Eulerian perturbation theory at higher order}
\label{sec:EPT-higher-order}

At higher order, \glslink{EPT}{Eulerian perturbation theory} can be implemented by expanding the density and velocity fields,
\begin{eqnarray}
\delta(\textbf{x},\tau) & = & \sum_{n=1}^{\infty} \delta^{(n)}(\textbf{x},\tau) = \delta^{(1)}(\textbf{x},\tau) + \delta^{(2)}(\textbf{x},\tau) + ... \, , \\
\theta(\textbf{x},\tau) & = & \sum_{n=1}^{\infty} \theta^{(n)}(\textbf{x},\tau) = \theta^{(1)}(\textbf{x},\tau) + \theta^{(2)}(\textbf{x},\tau) + ... \, ,
\end{eqnarray}
where $\delta^{(1)}(\textbf{x},\tau)$ and $\theta^{(1)}(\textbf{x},\tau)$ are the linear order solution studied in the previous section. Focusing only on the \gls{growing mode}, the first-order \gls{density field} reads,
\begin{equation}
\delta^{(1)}(\textbf{x},\tau) = D_1(\tau)\delta_\mathrm{init}(\textbf{x}) ,
\end{equation}
with $D_1(\tau) \equiv D_1^{(+)}(\tau)$ and $\delta_\mathrm{init}(\textbf{x}) = \delta_+(\textbf{x})$. $\delta^{(2)}(\textbf{x},\tau)$ describes to leading order the \gls{non-local} evolution of the \gls{density field} due to gravitational interactions. It is found to be proportional to the \textit{\gls{second-order growth factor}}, $D_2(\tau)$, which satisfies the differential equation \citep[equation 43 in][]{Bouchet1995}
\begin{equation}
\label{eq:evolution-second-growth-factor-a}
a^2 \deriv{^2 D_2}{a^2} + \left( \Omega_\Lambda(a) - \dfrac{\Omega_\mathrm{m}(a)}{2} + 2 \right) a \deriv{D_2}{a} = \dfrac{3}{2} \Omega_\mathrm{m}(a) \left[D_2 - (D_1^{(+)})^2 \right].
\end{equation}
In the codes implemented for this thesis, we use the fitting function
\begin{equation}
\label{eq:fitting_D2}
D_2(\tau) \approx - \dfrac{3}{7} D_1^2(\tau) \Omega_\mathrm{m}^{-1/143},
\end{equation}
valid for a flat {\LCDM} model \citep{Bouchet1995}. Depending on the \gls{cosmological parameters}, different expressions can be found in the literature \citep[see e.g.][]{Bernardeau2002}, but $D_2(\tau)$ always stays of the order of $D_1^2(\tau)$ as expected in perturbation theory.

A detailed presentation of non-linear Eulerian perturbation theory involves new types of objects (kernels, propagators, vertices) and is beyond the scope of this thesis. For an existing review, see e.g. \citet{Bernardeau2002}.

\section{Lagrangian perturbation theory}
\label{sec:LPT}

\subsection{Lagrangian fluid approach for cold dark matter}

As we have seen (section \ref{sec:fluid-dynamics-approach}), our approach is based on the assumption that \gls{CDM} is well described by a \gls{fluid}. A way of looking at \gls{fluid} motion is to focus on specific locations in space through which the \gls{fluid} flows as time passes. It is then possible to study dynamics of \glslink{density field}{density} and \glslink{velocity field}{velocity fields} in this context, which constitutes the Eulerian point of view. We have developed Eulerian perturbation theory in section \ref{sec:EPT}.

Alternatively, in \gls{fluid} dynamics, one can choose to describe the field by following the trajectories of particles or \gls{fluid} elements. This is the so-called Lagrangian description. The goal of this paragraph is to apply this description to the cosmological \gls{fluid} and to build \textit{\glslink{LPT}{Lagrangian perturbation theory}} in this framework.

\paragraph{Mapping from Lagrangian to Eulerian coordinates.}In Lagrangian description, the object of interest is not the position of particles but the \textit{\gls{displacement field}} $\Psi(\textbf{q})$ which maps the initial \glslink{comoving coordinates}{comoving particle position} $\textbf{q}$ into its final \glslink{comoving coordinates}{comoving} Eulerian position \textbf{x}, \citep[e.g.][]{Buchert1989,Bouchet1995,Bernardeau2002}:
\begin{equation}
\label{eq:Lagrangian-Eulerian-mapping}
\textbf{x}(\textbf{q},\tau) \equiv \textbf{q} + \boldsymbol{\Psi}(\textbf{q},\tau) .
\end{equation}
Let $J(\textbf{q},\tau)$ be the Jacobian of the transformation between Lagrangian and Eulerian coordinates, 
\begin{equation}
J(\textbf{q},\tau) \equiv \left| \pd{\textbf{x}}{\textbf{q}} \right| = \left| \det \mathscr{D} \right| = \left| \det ( \mathscr{I} + \mathscr{R} ) \right|,
\end{equation}
where the deformation tensor $\mathscr{D}$ can be written as the identity tensor $\mathscr{I}$ plus the shear of the displacement,\footnote{$\mathscr{R}$ is mathematically a tensor. It is sometimes referred to as the \gls{tidal tensor} and noted $\mathscr{T}$. We will avoid this nomenclature and notation here, so as not to introduce confusion with the Hessian of the gravitational potential $\mathscr{T} \equiv \partial^2 \Phi/\partial\textbf{x}^2$ (see section \ref{sec:apx-tweb}).} $\mathscr{R}\nbsp \equiv\nbsp \partial \boldsymbol{\Psi} / \partial \textbf{q}$. The Jacobian can be obtained by requiring that the Lagrangian mass element be conserved in the relationship between \gls{density contrast} and trajectories:
\begin{equation}
\rho(\textbf{x},\tau) \, \drm^3 \textbf{x} = \rho(\textbf{q}) \, \drm^3 \textbf{q} \quad \Rightarrow \quad \bar{\rho}(\tau) \left[ 1+\delta(\textbf{x},\tau) \right] \drm^3 \textbf{x} = \bar{\rho}(\tau) \drm^3 \textbf{q} ,
\end{equation}
Hence,
\begin{equation}
\label{eq:Jacobian}
J(\textbf{q},\tau) = \dfrac{1}{1 + \delta(\textbf{x},\tau )} \quad \mathrm{or} \quad \delta(\textbf{x},\tau) = J^{-1}(\textbf{q},\tau) - 1 .
\end{equation}
Note that this result (without the absolute value for $J$) is valid as long as no \gls{shell-crossing} occurs. At the first crossing of trajectories, \gls{fluid} elements with different initial positions $\textbf{q}$ end up at the same Eulerian position $\textbf{x}$ through the mapping in equation \eqref{eq:Lagrangian-Eulerian-mapping}. The Jacobian vanishes and one expects a singularity, namely a collapse to infinite
density. At this point, the description of dynamics in terms of a mapping does not hold anymore, the correct description involves a summation over all possible streams.

\paragraph{Equation of motion in Lagrangian coordinates.}The \gls{equation of motion} for a \gls{fluid} element, equation \eqref{eq:equation-of-motion-intermediate}, reads in \gls{conformal time},
\begin{equation}
\label{eq:motion_Lagrangian_coords}
\pd{\textbf{u}}{\tau} + \mathcal{H}(\tau) \textbf{u} = -\nabla_\textbf{x} \Phi ,
\end{equation}
where $\Phi$ is the cosmological \gls{gravitational potential} and $\nabla_\textbf{x}$ is the gradient operator in Eulerian \gls{comoving coordinates} $\textbf{x}$. Taking the divergence of this equation, noting that $\textbf{u} = \drm \textbf{x} / \drm \tau = \partial \boldsymbol{\Psi} / \partial \tau$, using equation \eqref{eq:Jacobian} and the \gls{Poisson equation}, equation \eqref{eq:Poisson}, and multiplying by the Jacobian, we obtain
\begin{equation}
\label{eq:equation-motion-Lagrangian}
J(\textbf{q},\tau) \, \nabla_\textbf{x} \cdot \left[ \pd{^2 \boldsymbol{\Psi}}{\tau^2} + \mathcal{H}(\tau) \pd{\boldsymbol{\Psi}}{\tau} \right] = \dfrac{3}{2} \Omega_\mathrm{m}(\tau) \mathcal{H}^2(\tau) \left[ J(\textbf{q},\tau) - 1 \right] .
\end{equation}
This equation shows the principal difficulty of the Lagrangian approach: the gradient operator has to be taken with reference to the Eulerian variable $\textbf{x}$, which depends on $\textbf{q}$ according to equation \eqref{eq:Lagrangian-Eulerian-mapping}. Equation \eqref{eq:equation-motion-Lagrangian} can be rewritten in terms of Lagrangian coordinates only by using $(\nabla_\textbf{x})_i = \left[ \updelta_{\mathrm{K}}^{ij} + \boldsymbol{\Psi}_{i,j} \right]^{-1} (\nabla_\textbf{q})_j$, where $\boldsymbol{\Psi}_{i,j} \equiv \partial \boldsymbol{\Psi}_i / \partial \textbf{q}_j = \mathscr{R}_{ij}$ are the shears of the displacement. The resulting non-linear differential equation for $\boldsymbol{\Psi}(\textbf{q},\tau)$ is then solved perturbatively, expanding about its linear solution.

\subsection{The Zel'dovich approximation}
\label{sec:ZA}

\paragraph{Displacement field in the Zel'dovich approximation.}In Lagrangian approach, \glslink{non-linear evolution}{non-linearities of the dynamics} are encoded in the relation between $\textbf{q}$ and $\textbf{x}$ (equation \eqref{eq:Lagrangian-Eulerian-mapping}) and in the relation between the \gls{displacement field} and the local density (equation \eqref{eq:Jacobian}). The \glslink{ZA}{Zel'dovich approximation} \citep[][hereafter \gls{ZA}]{Zeldovich1970,Shandarin1989} is first order \glslink{LPT}{Lagrangian perturbation theory}. It consists of taking the linear solution of equation \eqref{eq:equation-motion-Lagrangian} for the \gls{displacement field} while keeping the general equation with the Jacobian, equation \eqref{eq:Jacobian}, to reconstruct the \gls{density field}.
At linear order in the \gls{displacement field}, the relation between the gradients in Eulerian and Lagrangian coordinates is $J(\textbf{q},\tau) \nabla_\textbf{x} \approx \nabla_\textbf{q}$, and the first-order Jacobian is $J(\textbf{q},\tau) \approx 1 + \nabla_\textbf{q} \cdot \boldsymbol{\Psi}$. The equation to solve becomes
\begin{equation}
\label{eq:equation-motion-Lagrangian-ZA}
\nabla_\textbf{q} \cdot \left[ \pd{^2 \boldsymbol{\Psi}}{\tau^2} + \mathcal{H}(\tau) \pd{\boldsymbol{\Psi}}{\tau} \right] = \dfrac{3}{2} \Omega_\mathrm{m}(\tau) \mathcal{H}^2(\tau) \left( \nabla_\textbf{q} \cdot \boldsymbol{\Psi} \right) .
\end{equation}
The addition of any divergence-free \gls{displacement field} to a solution of the previous equation will also be a solution. In the following, we remove this indeterminacy by assuming that the movement is potential, i.e. $\nabla_\textbf{q} \times \boldsymbol{\Psi} = 0$. Introducing the \gls{divergence of the Lagrangian displacement field}, $\psi \equiv \nabla_\textbf{q} \cdot \boldsymbol{\Psi}$, one has to solve,
\begin{equation}
\label{eq:eqdiff-ZA}
\psi '' + \mathcal{H}(\tau) \psi ' = \dfrac{3}{2} \Omega_\mathrm{m}(\tau) \mathcal{H}^2(\tau) \psi .
\end{equation}
Therefore, the linear solution of equation \eqref{eq:equation-motion-Lagrangian} is separable into a product of a temporal and a spatial contribution. It can be written as $\boldsymbol{\Psi}^{(1)}(\textbf{q},\tau)$ such that
\begin{equation}
\label{eq:divergence-Psi1}
\psi^{(1)}(\textbf{q},\tau) \equiv \nabla_\textbf{q} \cdot \boldsymbol{\Psi}^{(1)}(\textbf{q},\tau) = -D_1(\tau) \, \delta(\textbf{q}) ,
\end{equation}
where $D_1(\tau)$ denotes the \gls{linear growth factor} studied in section \ref{sec:EPT-linear} and $\delta(\textbf{q})$ describes the \gls{growing mode} of the initial \gls{density contrast} field in Lagrangian coordinates. This can be checked in equation \eqref{eq:eqdiff-ZA} using the differential equation satisfied by the \glslink{linear growth factor}{growth factor}, equation \eqref{eq:evolution-growth-factor}. The above choice for the spatial contribution permits to recover the linear Eulerian behaviour, since initially $\delta(\textbf{x}) \approx D_1(\tau) \delta(\textbf{q}) \approx (1+\psi)^{-1} -1  \approx - \psi$. 

Note that the evolution of \gls{fluid} elements at linear order is \textit{\gls{local} evolution}, i.e. it does not depend on the behavior of the rest of \gls{fluid} elements. We have assumed that at linear order, the \gls{displacement field} is entirely determined by its divergence, i.e. that \gls{vorticity} vanishes. As we have already noted from equation \eqref{eq:vorticity-linear}, in the \gls{linear regime}, any initial \gls{vorticity} decays away due to the \glslink{expansion}{expansion of the Universe}. Thus, one might consider that the solutions will apply anyway, even if \gls{vorticity} is initially present, because at later times it will have negligible effect. Similarly, we have neglected the effect of the \gls{decaying mode} in equation \eqref{eq:density-field-linearEPT}.

\paragraph{Shell-crossing in the Zel'dovich approximation.\label{sec:Shell-crossing in the Zel'dovich approximation}}Since the \gls{displacement field} in the \gls{ZA} is curl-free, it is convenient to introduce the potential from which it derives, $\phi^{(1)}(\textbf{q})$, such that $\boldsymbol{\Psi}^{(1)}(\textbf{q},\tau) = -D_1(\tau) \nabla_\textbf{q} \phi^{(1)}(\textbf{q})$. At linear order in the \gls{displacement field}, its shear $\mathscr{R} \equiv \partial \boldsymbol{\Psi}^{(1)}/\partial \textbf{q}$ is equal to $-D_1(\tau) \mathrm{H}(\phi^{(1)}(\textbf{q}))$. Let $\lambda_1(\textbf{q}) \leq \lambda_2(\textbf{q}) \leq \lambda_3(\textbf{q})$ be the local \glslink{eigenvalue}{eigenvalues} of the Hessian of the Zel'dovich potential $\phi^{(1)}(\textbf{q})$. At \gls{conformal time} $\tau$, these values have grown of a factor $-D_1(\tau)$ to give the \glslink{eigenvalue}{eigenvalues} of the shear of the displacement $\mathscr{R}$. Using equation \eqref{eq:Jacobian}, the \gls{density contrast} may then be written as \citep[e.g.][]{Bouchet1995,Bernardeau2002}
\begin{equation}
1 + \delta(\textbf{x},\tau ) = \dfrac{1}{\left[ 1 - \lambda_1(\textbf{q}) D_1(\tau) \right] \left[ 1 - \lambda_2(\textbf{q}) D_1(\tau) \right] \left[ 1 - \lambda_3(\textbf{q}) D_1(\tau) \right]} .
\end{equation}
This equation allows an interpretation of what happens at \gls{shell-crossing} in the \gls{ZA}. If all \glslink{eigenvalue}{eigenvalues} $\lambda_i$ are negative, this is a developing underdense region, eventually reaching $\delta = -1$.  If $\lambda_3$ only is positive, when $\lambda_3 D_1(\tau) \rightarrow 1$, the \gls{ZA} leads to a planar collapse to infinite density along the axis of $\lambda_3$ and the formation of a two-dimensional ``cosmic \gls{pancake}''. In the case when two \glslink{eigenvalue}{eigenvalues} are positive, $\lambda_2, \lambda_3 > 0$, there is collapse to a \gls{filament}. The case $\lambda_1, \lambda_2, \lambda_3 > 0$ leads to gravitational collapse along all directions (\glslink{SC}{spherical collapse} if $\lambda_1 \approx \lambda_2 \approx \lambda_3$). This picture of gravitational \gls{structure formation} leads to a \gls{cosmic web classification} algorithm, which labels different regions either as \glslink{void}{voids}, \glslink{sheet}{sheets}, \glslink{filament}{filaments}, or \glslink{halo}{halos} \citep[see][an section \ref{sec:apx-tweb}]{Hahn2007a,Lavaux2010}.

\subsection{Second-order Lagrangian perturbation theory}
\label{sec:2LPT}

\paragraph{Displacement field in second-order Lagrangian perturbation theory.} The \glslink{ZA}{Zel'dovich approximation} being \gls{local}, it fails at sufficiently non-linear stages when particles are forming gravitationally bound structures instead of following straight lines. Already \glslink{2LPT}{second-order Lagrangian perturbation theory} (hereafter \gls{2LPT}) provides a remarkable improvement over the \gls{ZA} in describing the global properties of \glslink{density field}{density} and \glslink{velocity field}{velocity fields} \citep{Melott1995}. The solution of equation \eqref{eq:equation-motion-Lagrangian} up to second order takes into account the fact that \glslink{gravitational evolution}{gravitational instability} is \textit{\gls{non-local}}, i.e. it includes the correction to the \gls{ZA} displacement due to gravitational \gls{tidal effects}. It reads
\begin{equation}
\label{eq:mapping-2LPT}
\textbf{x}(\tau) = \textbf{q} + \boldsymbol{\Psi}(\textbf{q},\tau) = \textbf{q} + \boldsymbol{\Psi}^{(1)}(\textbf{q},\tau) + \boldsymbol{\Psi}^{(2)}(\textbf{q},\tau) , \quad \mathrm{or} \quad \boldsymbol{\Psi}(\textbf{q},\tau) = \boldsymbol{\Psi}^{(1)}(\textbf{q},\tau) + \boldsymbol{\Psi}^{(2)}(\textbf{q},\tau),
\end{equation}
where the divergence of the first order solution is the same as in the \gls{ZA} (equation \eqref{eq:divergence-Psi1}),
\begin{equation}
\label{eq:2LPT-Psi1}
\psi^{(1)}(\textbf{q},\tau) = \nabla_\textbf{q} \cdot \boldsymbol{\Psi}^{(1)}(\textbf{q},\tau) = -D_1(\tau) \, \delta(\textbf{q}) ,
\end{equation}
and the divergence of the second order solution describes the \gls{tidal effects},
\begin{equation}
\label{eq:2LPT-Psi2}
\psi^{(2)}(\textbf{q},\tau) = \nabla_\textbf{q} \cdot \boldsymbol{\Psi}^{(2)}(\textbf{q},\tau) = \dfrac{1}{2} \dfrac{D_2(\tau)}{D_1^2(\tau)} \sum_{i \neq j} \left[ \boldsymbol{\Psi}^{(1)}_{i,i}\boldsymbol{\Psi}^{(1)}_{j,j} - \boldsymbol{\Psi}^{(1)}_{i,j}\boldsymbol{\Psi}^{(1)}_{j,i} \right] ,
\end{equation}
where $\boldsymbol{\Psi}^{(1)}_{k,l} \equiv \partial \boldsymbol{\Psi}^{(1)}_k / \partial \textbf{q}_l$ and $D_2(\tau)$ denotes the \gls{second-order growth factor}, defined in section \ref{sec:EPT-higher-order}.

\paragraph{Lagrangian potentials.} Since Lagrangian solutions up to second order are irrotational \citetext{\citealp{Melott1995,Buchert1994,Bernardeau2002}; this is assuming that \gls{initial conditions} are only in the \gls{growing mode}, in the same spirit as neglecting completely the decaying \gls{vorticity}}, it is convenient to define the \glslink{Lagrangian potential}{Lagrangian potentials} $\phi^{(1)}$ and $\phi^{(2)}$ from which $\boldsymbol{\Psi}^{(1)}$ and $\boldsymbol{\Psi}^{(2)}$ derive, so that in \gls{2LPT},
\begin{equation}
\label{eq:Psi_2LPT}
\boldsymbol{\Psi}^{(1)}(\textbf{q},\tau) = - D_1(\tau) \nabla_\textbf{q} \phi^{(1)}(\textbf{q}) \quad \mathrm{and} \quad \boldsymbol{\Psi}^{(2)}(\textbf{q},\tau) = D_2(\tau) \nabla_\textbf{q} \phi^{(2)}(\textbf{q}) .
\end{equation}
Since $\boldsymbol{\Psi}^{(1)}$ is of order $D_1(\tau)$ (equation \eqref{eq:2LPT-Psi1}) and $\boldsymbol{\Psi}^{(2)}(\tau)$ is of order $D_2(\tau)$ (equation \eqref{eq:2LPT-Psi2}), the above potentials are time-independent. They satisfy \glslink{Poisson equation}{Poisson-like equations} \citep{Buchert1994},
\begin{eqnarray}
\Delta_\textbf{q} \phi^{(1)}(\textbf{q}) & = & \delta(\textbf{q}) ,\label{eq:Poisson_phi1}\\
\Delta_\textbf{q} \phi^{(2)}(\textbf{q}) & = & \sum_{i>j} \left[ \phi^{(1)}_{,ii}(\textbf{q})\phi^{(1)}_{,jj}(\textbf{q}) - (\phi^{(1)}_{,ij}(\textbf{q}))^2 \right] .\label{eq:Poisson_phi2}
\end{eqnarray}
The mapping from Eulerian to Lagrangian, equation \eqref{eq:mapping-2LPT}, thus reads
\begin{equation}
\textbf{x}(\tau) = \textbf{q} - D_1(\tau) \nabla_\textbf{q} \phi^{(1)}(\textbf{q}) + D_2(\tau) \nabla_\textbf{q} \phi^{(2)}(\textbf{q}) .
\label{eq:mapping_LPT_code}
\end{equation}

\paragraph{Velocity field in second-order Lagrangian perturbation theory.} Taking the derivative of the previous equation yields for the \gls{velocity field},
\begin{equation}
\textbf{u} = - f_1(\tau) D_1(\tau) \mathcal{H}(\tau) \nabla_\textbf{q} \phi^{(1)}(\textbf{q}) + f_2(\tau)  D_2(\tau) \mathcal{H}(\tau) \nabla_\textbf{q} \phi^{(2)}(\textbf{q}) .
\label{eq:velocities_LPT_code}
\end{equation}
which involves the logarithmic derivatives of the growth factors, $f_i \equiv \drm \ln D_i / \drm \ln a$, well approximated in a flat {\LCDM} model by \citep{Bouchet1995}
\begin{equation}
\label{eq:fitting_f2}
f_1 \approx \Omega_\mathrm{m}^{5/9} \quad \mathrm{and} \quad f_2 \approx 2 \, \Omega_\mathrm{m}^{6/11} \approx 2 \, f_1^{54/55}.
\end{equation}
Other expressions for different cosmologies can be found in \citet{Bouchet1995,Bernardeau2002}.

\section{Non-linear approximations to gravitational instability}
\label{sec:NL-approxs}

When fluctuations become strongly \glslink{non-linear evolution}{non-linear} in the \gls{density field}, \glslink{EPT}{Eulerian perturbation theory} breaks down. \glslink{LPT}{Lagrangian perturbation theory} is often more successful, since the Lagrangian picture is intrinsically non-linear in the \gls{density field} (see e.g. equation \eqref{eq:equation-motion-Lagrangian}). A small perturbation in the Lagrangian \gls{displacement field} carries a considerable amount of non-linear information about the corresponding Eulerian \glslink{density field}{density} and \glslink{velocity field}{velocity fields}. However, at some point, computers are required to study \glslink{gravitational evolution}{gravitational instability} (in particular through \glslink{N-body simulation}{$N$-body simulations}), the important drawback being that the treatment becomes numerical instead of analytical. We will adopt this approach in this thesis. However, several non-linear approximations to the equations of motion have been suggested in the literature to allow the extrapolation of analytical calculations in the \gls{non-linear regime}. We now briefly review some of them.

Non-linear approximations consist of replacing one of the equations of the dynamics (\glslink{Poisson equation}{Poisson} -- equation \eqref{eq:Poisson} --, \glslink{continuity equation}{continuity} -- equation \eqref{eq:continuity} -- or \glslink{Euler's equation}{Euler} -- equation \eqref{eq:Euler}) by a different assumption.\footnote{In this section, we have come back to a Eulerian description of the cosmological \gls{fluid}.} In general, the \gls{Poisson equation} is replaced \citep{Munshi1994}. These modified dynamics are often \gls{local}, in the sense described above for the \gls{ZA}, in order to provide a simpler way of calculating the evolution of fluctuations than the full \gls{non-local} dynamics.

\subsection{The Zel'dovich approximation as a non-linear approximation}

As we have seen in section \ref{sec:Dynamics}, in Eulerian dynamics, non-linearity is encoded in the \gls{Poisson equation}, equation \eqref{eq:Poisson}, $\Delta \Phi = 4\pi\G a^2\bar{\rho}\delta$. The goal of this paragraph is to see what replaces the \gls{Poisson equation} in the Eulerian description of the \gls{ZA}. From this point of view, the \gls{ZA} is the original \glslink{non-linear approximation}{non-linear Eulerian approximation}, and it remains one of the most famous.

If we restrict our attention to potential movements, the \glslink{peculiar velocity flow}{peculiar velocity field} $\textbf{u}$ is irrotational. It can be written as the gradient of a \gls{velocity potential},
\begin{equation}
\label{eq:velocity-potential}
\textbf{u} = - \dfrac{\nabla_\textbf{x} V}{a} .
\end{equation}
As discussed before, the main reason to restrict to this case is the decay of \glslink{vorticity}{vortical perturbations}.

It is then possible to postulate various forms for the \gls{velocity potential} $V$. The \gls{ZA} corresponds to the Ansatz (\citealp{Munshi1994,Hui1996}; appendix B in \citealp{Scoccimarro1997})
\begin{equation}
V = \dfrac{2fa}{3\Omega_\mathrm{m} \mathcal{H}} \Phi ,
\end{equation}
where $\Phi$ is the cosmological \gls{gravitational potential} and $f$ is the logarithmic derivative of the \gls{linear growth factor}. The Zel'dovich approximation is therefore equivalent to the replacement of the \gls{Poisson equation} by
\begin{equation}
\label{eq:velocity-ZA-nonlinear}
\textbf{u} = -\dfrac{2f}{3\Omega_\mathrm{m}\mathcal{H}} \nabla \Phi .
\end{equation}
This can be explicitly checked as follows. Combining equations \eqref{eq:motion_Lagrangian_coords} and \eqref{eq:velocity-ZA-nonlinear}, one gets
\begin{equation}
\pd{\textbf{u}}{\tau} + \mathcal{H} \textbf{u} = \frac{3 \Omega_\mathrm{m} \mathcal{H}}{2f} \textbf{u} .
\end{equation}
Then, noting that $\nabla_\textbf{q} \cdot \textbf{u} = \psi'$, the differential equation for $\psi$ is
\begin{equation}
\psi'' + \mathcal{H} \psi' = \frac{3 \Omega_\mathrm{m} \mathcal{H}}{2f} \psi',
\end{equation}
Using the time evolution of $D_1$ (equation \eqref{eq:evolution-growth-factor}) and the identity $D_1' = \mathcal{H} f D_1$, one can check that the Zel'dovich solution, $\psi = -D_1 \delta(\textbf{q})$ indeed verifies the above equation.

Equation \eqref{eq:velocity-ZA-nonlinear} means that at linear order, particles just go straight (in \gls{comoving coordinates}) in the direction set by their initial velocity. In the \glslink{ZA}{Zel'dovich approximation}, the proportionality between \gls{velocity field} and \gls{gravitational field} always holds (not just to first order in $\boldsymbol{\Psi}$).

Note that during the matter era, $a \propto t^{2/3}$ and thus $\mathcal{H} \equiv \dot{a} = 2a/(3t)$, which means that an equivalent form for the \gls{ZA} Ansatz is
\begin{equation}
V=\frac{f}{\Omega_\mathrm{m}} \Phi t \approx \Phi t .
\end{equation}

The \gls{ZA} is a \gls{local} approximation that represents exactly the true dynamics in one-dimensional collapse \citep{Buchert1989,Yoshisato2006}. It is also possible to formulate \gls{local} approximations that besides describing correctly planar collapse like the \gls{ZA}, are suited for cylindrical or \glslink{SC}{spherical collapse} (leading to the formation of cosmic \glslink{filament}{filaments} and \glslink{halo}{halos}, in addition to cosmic \glslink{pancake}{pancakes}). These approximations, namely the \glslink{non-magnetic approximation}{``non-magnetic'' approximation} \citep[NMA,][]{Bertschinger1994} and the \glslink{local tidal approximation}{``local tidal'' approximation} \citep[LTA,][]{Hui1996}, are not straightforward to implement for the calculation of statistical properties of \glslink{density field}{density} and \glslink{velocity field}{velocity fields}.

\subsection{Other velocity potential approximations}

Some other possibilities for the \gls{velocity potential} can be found in literature \citep{Coles1993,Munshi1994}. The \glslink{frozen flow approximation}{frozen flow} (FF) approximation postulates
\begin{equation}
\label{eq:FF}
V = \Phi^{(1)} t ,
\end{equation}
where $\Phi^{(1)}$ is the first-order solution (the linear approximation) for the \gls{gravitational potential}. It satisfies the \gls{Poisson equation} in the \gls{linear regime},
\begin{equation}
\label{eq:FF-Poisson}
\Delta \Phi^{(1)} = \dfrac{3}{2} \Omega_\mathrm{m}(\tau) \mathcal{H}^2(\tau) \delta_1(\textbf{x},\tau) ,
\end{equation}
where $\delta_1(\textbf{x},\tau) = D_1(\tau) \, \delta_1(\textbf{x})$ is the linearly extrapolated \gls{density field}. In FF, the \gls{Poisson equation} is replaced by the analog of equation \eqref{eq:velocity-ZA-nonlinear}, substituting equation \eqref{eq:FF},
\begin{equation}
\textbf{u} = -\dfrac{2f}{3 \Omega_\mathrm{m} \mathcal{H}} \nabla \Phi^{(1)} ,
\end{equation}
or, by taking the divergence and using equation \eqref{eq:FF-Poisson},
\begin{equation}
\theta(\textbf{x},\tau) = - \mathcal{H}(\tau) f \, \delta_1(\textbf{x},\tau) .
\end{equation}
The physical meaning of this approximation is that the \gls{velocity field} is assumed to remain linear while the \gls{density field} is allowed to explore the \gls{non-linear regime}.

In the \glslink{linear potential approximation}{linear potential} (LP) approximation, the \gls{gravitational potential} is instead assumed to remain the same as in the \gls{linear regime}; therefore, the \gls{Poisson equation} is replaced by
\begin{equation}
\Phi = \Phi^{(1)}, \quad \Delta \Phi = \dfrac{3}{2} \Omega_\mathrm{m}(\tau) \mathcal{H}^2(\tau) \delta_1(\textbf{x},\tau) .
\end{equation}
The idea is that since $\Phi \propto \delta/k^2$ in Fourier space, the \gls{gravitational potential} is dominated by the long-wavelength modes more than the \gls{density field}, and therefore it ought to obey \glslink{linear regime}{linear perturbation theory} to a better approximation.

\subsection{The adhesion approximation}

All the above approximations (\gls{ZA}, NMA, LTA, FF, LP) are \gls{local}, which means that we neglect the self-gravity of inhomogeneities. A significant problem of the \gls{ZA}, and of subsequent variations, is the fact that after \gls{shell-crossing}, matter continues to flow throughout the newly-formed structure, which should instead be gravitationally bound. This phenomenon washes out cosmic structures on small scales.

A possible phenomenological solution is to add a \gls{viscosity} term to the \glslink{single-stream approximation}{single-stream} \glslink{Euler's equation}{Euler equation}, equation \eqref{eq:Euler-single-stream}, which then becomes \gls{Burgers' equation},
\begin{equation}
\pd{\textbf{u}_i(\textbf{x},\tau)}{\tau} + \mathcal{H}(\tau) \textbf{u}_i(\textbf{x},\tau) + \textbf{u}_j(\textbf{x},\tau) \cdot \nabla_j \textbf{u}_i(\textbf{x},\tau) = - \nabla_i \Phi(\textbf{x},\tau) + v \, \Delta \textbf{u}_i(\textbf{x},\tau) .
\end{equation}
This is the so-called \textit{\gls{adhesion approximation}} \citep{Kofman1988,Gurbatov1989,Kofman1992,Valageas2011,Hidding2012}. For a potential flow, it can be reduced to a linear \gls{diffusion equation}, and therefore solved exactly. Surprisingly, in the \gls{adhesion approximation}, the dynamical equations describing the evolution of the self-gravitating cosmological \gls{fluid} can be written in the form of a \glslink{Schrodinger equation}{Schr\"odinger equation} coupled to a \gls{Poisson equation} describing \gls{Newtonian gravity} \citep{Short2006a}. The dynamics can therefore be studied with the tools of wave mechanics. An alternative to the \glslink{adhesion approximation}{adhesion model} is the \gls{free-particle approximation} (FPA), in which the artificial \gls{viscosity} term in \gls{Burgers' equation} is replaced by a non-linear term known as the \textit{quantum pressure}. This also leads to a free-particle \glslink{Schrodinger equation}{Schr\"odinger equation} \citep{Short2006a,Short2006b}. 

Comparisons of the \gls{adhesion approximation} to full-gravitational numerical simulations show an improvement over the \gls{ZA} at small scales, even if the fragmentation of structures into dense clumps is still underestimated \citep{Weinberg1990}. At weakly non-linear scales, the \gls{adhesion approximation} is essentially equal to the \gls{ZA}.

%% file: Chapter2/Chapter2Content.tex
\chapter{Numerical diagnostics of Lagrangian perturbation theory}
\label{chap:lpt}
\minitoc

\begin{flushright}
\begin{minipage}[c]{0.6\textwidth}
\rule{\columnwidth}{0.4pt}

``\textbf{Hector Barbossa}: The world used to be a bigger place.\\
\textbf{Jack Sparrow}: The world's still the same. There's just... less in it.''\\
--- \citet{POTC2007}

\vspace{-5pt}\rule{\columnwidth}{0.4pt}
\end{minipage}
\end{flushright}

\abstract{\section*{Abstract}
This chapter is intended as a guide on the approximation error in using \glslink{LPT}{Lagrangian perturbation theory} instead of \glslink{full gravity}{fully non-linear gravity} in large-scale structure analysis. We compare various properties of \glslink{particle realization}{particle realizations} produced by \gls{LPT} and by \glslink{N-body simulation}{\glslink{N-body simulation}{$N$-body simulations}}. In doing so, we characterize differences and similarities, as a function of scale, resolution and \gls{redshift}.}

The goal of this chapter is to characterize the accuracy of \glslink{LPT}{Lagrangian perturbation theory} in terms of a set of numerical diagnostics. It is organized as follows. In section \ref{sec:Correlation functions of the density field}, we look at the correlations functions of the \gls{density field}, which usually are the final observable in cosmological \glslink{galaxy survey}{surveys}. As the \gls{displacement field} plays a central role in \gls{LPT}, we study its statistics in section \ref{sec:Statistics of the Lagrangian displacement field}. In particular, we illustrate that in some regimes, when the perturbative parameter is large, \gls{2LPT} performs worse than the \gls{ZA}. We examine the decomposition of the \gls{displacement field} in a scalar and rotational part and review various approximations based on its \glslink{divergence of the Lagrangian displacement field}{divergence}. Finally, in section \ref{sec:Comparison of structure types in LPT and $N$-body dynamics}, we compare \gls{cosmic web} elements (\glslink{void}{voids}, \glslink{sheet}{sheets}, \glslink{filament}{filaments}, and \glslink{cluster}{clusters}) as predicted by \gls{LPT} and by \glslink{N-body simulation}{non-linear simulations} of the \gls{LSS}.\footnote{In the following, we will often write ``\gls{full gravity}'', even if, strictly speaking, \glslink{N-body simulation}{$N$-body simulations} also involve some degree of approximation.}

Corresponding \gls{LPT} and \glslink{N-body simulation}{$N$-body simulations} used in this chapter have been run from the same \gls{initial conditions}, generated at \gls{redshift} $z=63$ using \glslink{2LPT}{second-order Lagrangian perturbation theory}. The \glslink{N-body simulation}{$N$-body simulations} have been run with the \textsc{\gls{Gadget-2}} cosmological code \citep{Springel2001,Springel2005}. Evolutions of the \glslink{ZA}{Zel'dovich approximation} were performed with \textsc{\gls{N-GenIC}} \citep{Springel2005}, and of \glslink{2LPT}{second-order Lagrangian perturbation} theory with \textsc{\gls{2LPTic}} \citep{Crocce2006b}. To ensure sufficient statistical significance, we used eight realizations of the same cosmology, changing the seed used to generate respective \gls{initial conditions}. All computations are done after binning the \gls{dark matter particles} with a \glslink{CiC}{Cloud-in-Cell} (\gls{CiC}) method (see section \ref{sec:apx-Density assignments}). The simulations contain $512^3$ \glslink{dark matter particles}{particles} in a $1024$ Mpc/$h$ cubic box with \gls{periodic boundary conditions}. We checked that with this setup, the \gls{power spectrum} agrees with the non-linear \gls{power spectrum} of simulations run with higher \gls{mass resolution}, provided by \textsc{\gls{Cosmic Emulator}} tools \citep{Heitmann2009,Heitmann2010,Lawrence2010} (deviations are at most sub-percent level for $k~\lesssim~1~(\mathrm{Mpc}/h)^{-1}$). Therefore, at the scales of interest of this work, $k~\leq~0.4~(\mathrm{Mpc}/h)^{-1}$ (corresponding to the \glslink{linear regime}{linear} and \gls{mildly non-linear regime} at \gls{redshift} zero), the clustering of dark matter is correctly reproduced by our set of simulations.

The \gls{cosmological parameters} used are \gls{WMAP-7} fiducial values \citep{Komatsu2011},
\begin{equation}
\label{eq:comological-parameters}
\Omega_\Lambda = 0.728, \Omega_\mathrm{m}~=~0.2715, \Omega_\mathrm{b}~=~0.0455, \sigma_8 = 0.810, h = 0.704, n_{\mathrm{s}} = 0.967 .
\end{equation}
Thus, each particle carries a mass of $6.03\times10^{11}~\mathrm{M}_\odot/h$.

\section{Correlation functions of the density field}
\label{sec:Correlation functions of the density field}

\draw{This section draws from section III in \citet{Leclercq2013}.}

In this section, we analyze the correlation functions of the \glslink{density field}{density contrast field}, $\delta$, in \gls{LPT} and \glslink{N-body simulation}{$N$-body fields}. 

\textit{Note.} All plots presented in this section contain lines labeled as ``\gls{ZARM}'' and ``\gls{2LPTRM}'' which correspond to remapped fields based on the \gls{ZA} and on \gls{2LPT}, respectively. They are ignored in this chapter, which focuses on diagnostics of \gls{LPT}. For a description of the \gls{remapping} procedure and for comments on these approximations in comparison to the \gls{ZA}, \gls{2LPT} and \glslink{N-body simulation}{$N$-body dynamics}, the reader is referred to chapter \ref{chap:remapping}.

\subsection{One-point statistics}
\label{sec:eulerian-one-point-stats}

\begin{figure}
\begin{center}
\includegraphics[width=0.5\columnwidth]{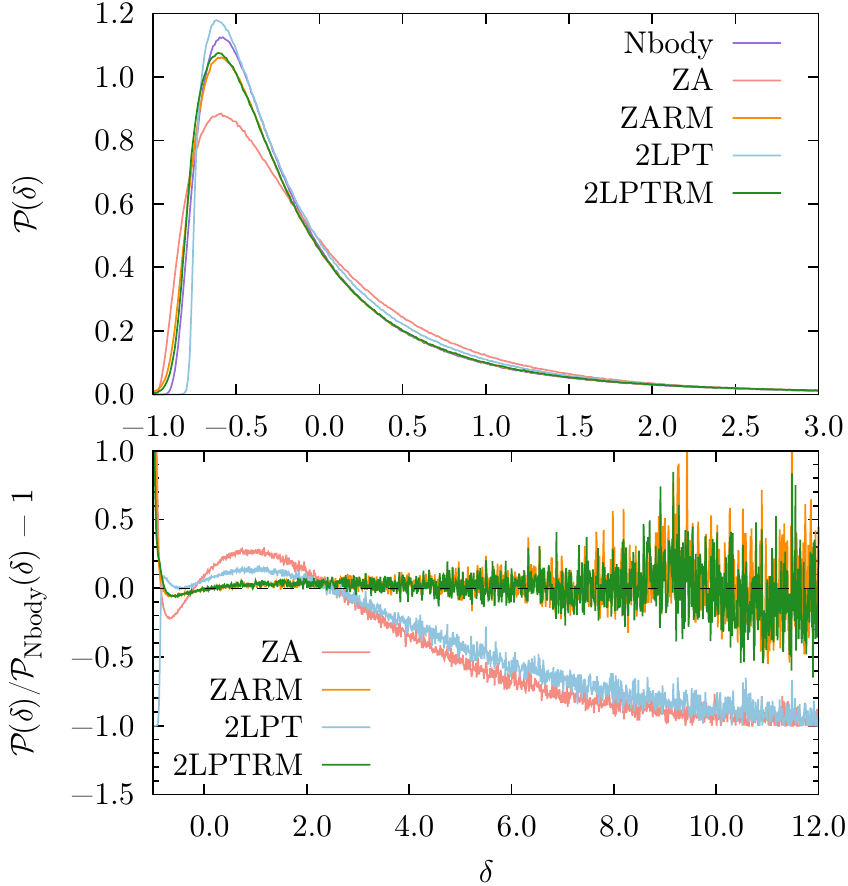}
\caption{\textit{Upper panel}. \Gls{redshift}-zero probability distribution function for the \gls{density contrast} $\delta$, computed from eight 1024~Mpc/$h$-box simulations of $512^3$ particles. The particle distribution is determined using: a full \glslink{N-body simulation}{$N$-body simulation} (purple curve), the Zel'dovich approximation, alone (\gls{ZA}, light red curve) and after \gls{remapping} (\gls{ZARM}, orange curve), second-order Lagrangian perturbation theory, alone (\gls{2LPT}, light blue curve) and after \gls{remapping} (\gls{2LPTRM}, green curve). \textit{Lower panel}. Relative deviations of the same \glslink{pdf}{pdfs} with reference to \glslink{N-body simulation}{$N$-body simulation} results. Note that, contrary to standard \gls{LPT} approaches, remapped fields follow the \gls{one-point distribution} of full \glslink{N-body simulation}{$N$-body dynamics} in an unbiased way, especially in the high density regime.}
\label{fig:pdf}
\end{center}
\end{figure}

Figure \ref{fig:pdf} shows the \gls{pdf} for the \gls{density contrast}, $\mathcal{P}_\delta$, at \gls{redshift} zero, for \glslink{N-body simulation}{$N$-body simulations}, and for \gls{ZA} and \gls{2LPT} \glslink{density field}{density fields}. All \glslink{pdf}{pdfs} are \glslink{non-Gaussianity}{non-Gaussian} with a substantial \gls{skewness}, are tied down to $0$ at $\delta=-1$ with a large tail in the high-density values. As discussed in section \ref{sec:Lognormal-fields}, the late-time \gls{pdf} for \glslink{density field}{density fields} is approximately \glslink{log-normal distribution}{log-normal}. However, already at the level of \glslink{one-point distribution}{one-point statistics}, the detailed behaviors of \gls{LPT} and \glslink{N-body simulation}{$N$-body simulations} disagree: the peak of the \gls{pdf} is shifted and the tails differ. In particular, \gls{LPT} largely underpredicts the number of voxels in the high-density regime. This effect is more severe for the \gls{ZA} than for \gls{2LPT}. This comes from the fact that \gls{2LPT} captures some of \gls{non-local} effects involved in the formation of the densest \glslink{halo}{halos}.

The \glslink{one-point distribution}{one-point pdf} of the density is further analyzed in section \ref{sec:psi_vs_delta}, in comparison to the \glslink{one-point distribution}{one-point pdf} of the Lagrangian \gls{displacement field}.

\subsection{Two-point statistics}
\label{sec:eulerian-two-point-stats}

\subsubsection{Power spectrum}
\label{sec:eulerian-two-point-stats-power-spectrum}

\begin{figure}
\begin{center}
\includegraphics[width=0.5\columnwidth]{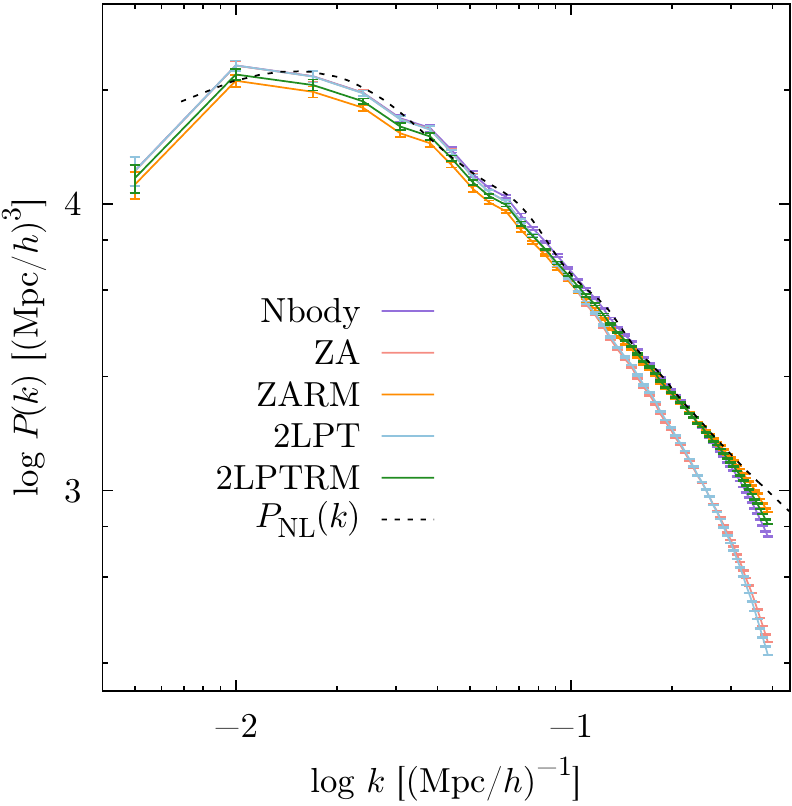}
\caption{\Gls{redshift}-zero dark matter \glslink{power spectrum}{power spectra} in a 1024~Mpc/$h$ simulation, with \glslink{density field}{density fields} computed with a mesh size of 8~Mpc/$h$. The particle distribution is determined using: a full \glslink{N-body simulation}{$N$-body simulation} (purple curve), the Zel'dovich approximation, alone (\gls{ZA}, light red curve) and after \gls{remapping} (\gls{ZARM}, orange curve), second-order Lagrangian perturbation theory, alone (\gls{2LPT}, light blue curve) and after \gls{remapping} (\gls{2LPTRM}, green curve). The dashed black curve represents $P_{\mathrm{NL}}(k)$, the theoretical \gls{power spectrum} expected at $z=0$. Both \gls{ZARM} and \gls{2LPTRM} show increased power in the \gls{mildly non-linear regime} compared to \gls{ZA} and \gls{2LPT} (at scales corresponding to $k~\gtrsim~0.1$~(Mpc/$h$)$^{-1}$ for this \gls{redshift}), indicating an improvement of \glslink{two-point correlation function}{two-point statistics} with the \gls{remapping} procedure.}
\label{fig:PS}
\end{center}
\end{figure}

\begin{figure*}
\begin{center}
\includegraphics[width=\textwidth]{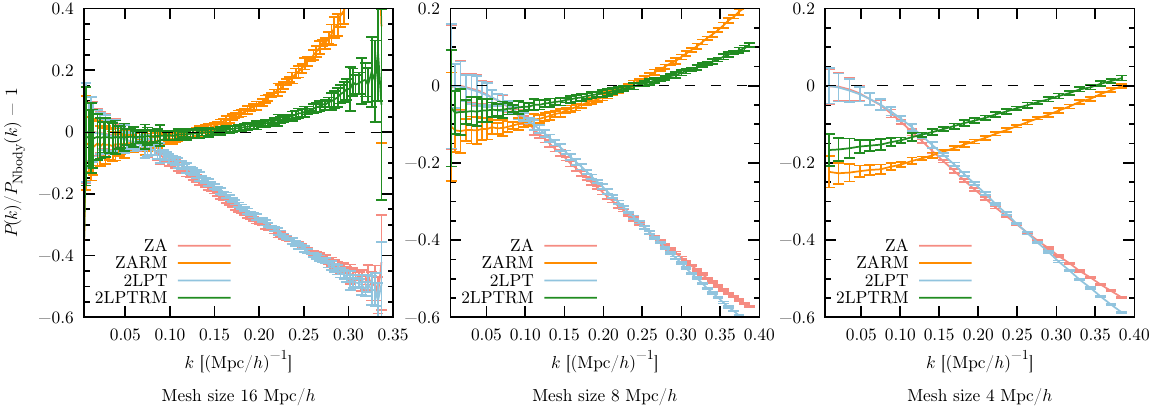}
\caption{\textit{Power spectrum: mesh size-dependence}. Relative deviations for the \glslink{power spectrum}{power spectra} of various particle distributions, with reference to the \gls{density field} computed with a full \glslink{N-body simulation}{$N$-body simulation}. The particle distribution is determined using: the Zel'dovich approximation, alone (\gls{ZA}, light red curve) and after \gls{remapping} (\gls{ZARM}, orange curve), second-order Lagrangian perturbation theory, alone (\gls{2LPT}, light blue curve) and after \gls{remapping} (\gls{2LPTRM}, green curve). The computation is done on different meshes: 16 Mpc/$h$ ($64^3$-voxel grid, left panel), 8 Mpc/$h$ ($128^3$-voxel grid, central panel) and 4 Mpc/$h$ ($256^3$-voxel grid, right panel). All results are shown at \gls{redshift} $z=0$. \gls{LPT} fields exhibit more small-scale correlations after \gls{remapping} and their \glslink{power spectrum}{power spectra} get closer to the shape of the full non-linear \gls{power spectrum}.}
\label{fig:PS-deviations-mesh}
\end{center}
\end{figure*}

\begin{figure*}
\begin{center}
\includegraphics[width=\textwidth]{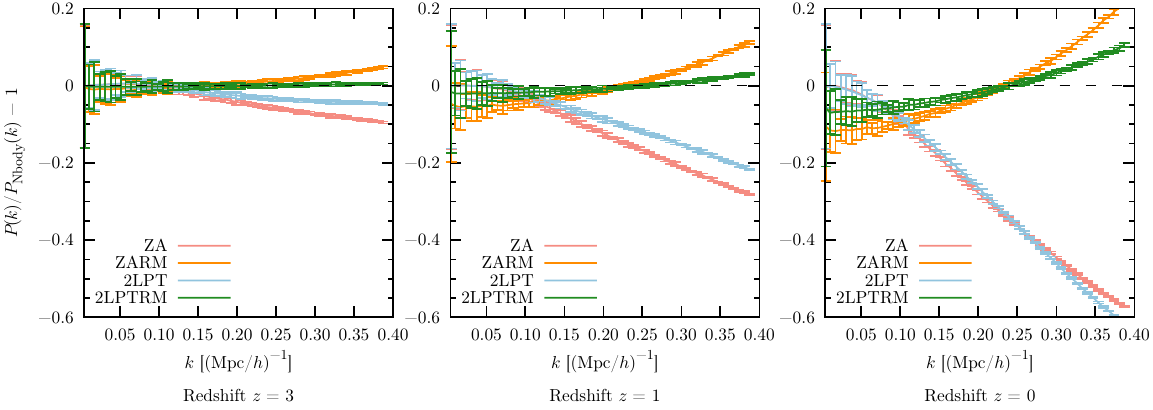}
\caption{\textit{Power spectrum: \gls{redshift}-dependence}. Relative deviations for the \glslink{power spectrum}{power spectra} of various particle distributions (see the caption of figure \ref{fig:PS-deviations-mesh}), with reference to the \gls{density field} computed with a full \glslink{N-body simulation}{$N$-body simulation}. The computation is done on a 8 Mpc/$h$ mesh ($128^3$-voxel grid). Results at different \glslink{redshift}{redshifts} are shown: $z=3$ (right panel), $z=1$ (central panel) and $z=0$ (left panel). The \gls{remapping} procedure is increasingly successful with increasing \gls{redshift}.}
\label{fig:PS-deviations-redshift}
\end{center}
\end{figure*}

We measured the \gls{power spectrum} of dark matter \glslink{density field}{density fields}, as defined by equation \eqref{eq:two-pt-correlator-FS}. \glslink{dark matter particles}{Dark matter particles} have been displaced according to each prescription and assigned to cells with a \gls{CiC} scheme, for different mesh sizes. \glslink{power spectrum}{Power spectra} were measured from theses meshes, with a correction for \gls{aliasing} effects \citep{Jing2005}. \Gls{redshift}-zero results computed on a 8~Mpc/$h$ mesh are presented in figure \ref{fig:PS}. There, the dashed line corresponds to the theoretical, non-linear \gls{power spectrum} expected, computed with \textsc{\gls{Cosmic Emulator}} tools \citep{Heitmann2009,Heitmann2010,Lawrence2010}. A deviation of full \glslink{N-body simulation}{$N$-body simulations} from this theoretical prediction can be observed at small scales. This discrepancy is a gridding artifact, completely due to the finite mesh size used for the analysis. As a rule of thumb, a maximum threshold in $k$ for trusting the simulation data is set by a quarter of the \gls{Nyquist wavenumber}, defined by $k_{\mathrm{N}} \equiv 2 \pi/L \times N_{\mathrm{p}}^{1/3}/2$, where $L$ is the size of the box and $N_{\mathrm{p}}$ is the number of cells in the Lagrangian grid on which particles are placed in the \gls{initial conditions}; which makes for our analysis ($L=1024$~Mpc/$h$, $N_{\mathrm{p}} = 512^3$), $k_\mathrm{N}/4~\approx~0.39$~(Mpc/$h$)$^{-1}$. At this scale, it has been observed that the \gls{power spectrum} starts to deviate at the percent-level with respect to higher resolution simulations \citep{Heitmann2010}. The relative deviations of various \glslink{power spectrum}{power spectra} with reference to \gls{full gravity} are presented in figures \ref{fig:PS-deviations-mesh} and \ref{fig:PS-deviations-redshift}. In all the plots, the error bars represent the dispersion of the mean among eight independent realizations.

Generally, \gls{LPT} correctly predicts the largest scales, when $k \rightarrow 0$ (the smallest wavelength mode accessible here is set by the box size: $k_\mathrm{min}=2\pi/L$ with $L=1024$ Mpc/$h$, giving $k_\mathrm{min} \approx 0.006~(\mathrm{Mpc}/h)^{-1}$), as these are in the \gls{linear regime}. These are affected by \gls{cosmic variance}, but the effect is not visible in our plots, as corresponding \gls{LPT} and \glslink{N-body simulation}{$N$-body fields} start from the same \gls{initial conditions}. Differences arise in the \glslink{mildly non-linear regime}{mildly non-linear} and \gls{non-linear regime}, where \gls{LPT} predicts too little power. Indeed, as \gls{LPT} only captures part of the non-linearity of the \gls{Vlasov-Poisson system}, presented in section \ref{sec:The Vlasov-Poisson system}, the clustering of \gls{dark matter particles} is underestimated. 

The discrepancy between \gls{LPT} and \glslink{N-body simulation}{$N$-body} \glslink{power spectrum}{power spectra} depends both on the target resolution (see figure \ref{fig:PS-deviations-mesh}) and on the desired \gls{redshift} (see figure \ref{fig:PS-deviations-redshift}). For example, at a resolution of $8$ Mpc/$h$ and at a comoving wavelength of $k=0.40~(\mathrm{Mpc}/h)^{-1}$, \gls{2LPT} only lacks $5$\% power at $z=3$ but more than $50$\% at $z=0$. At fixed \gls{redshift}, the lack of small scale power in \gls{LPT} weakly depends on the mesh size.

\subsubsection{Fourier-space cross-correlation coefficient}
\label{sec:eulerian-two-point-stats-cross-correlation}

\begin{figure}
\begin{center}
\includegraphics[width=0.5\columnwidth]{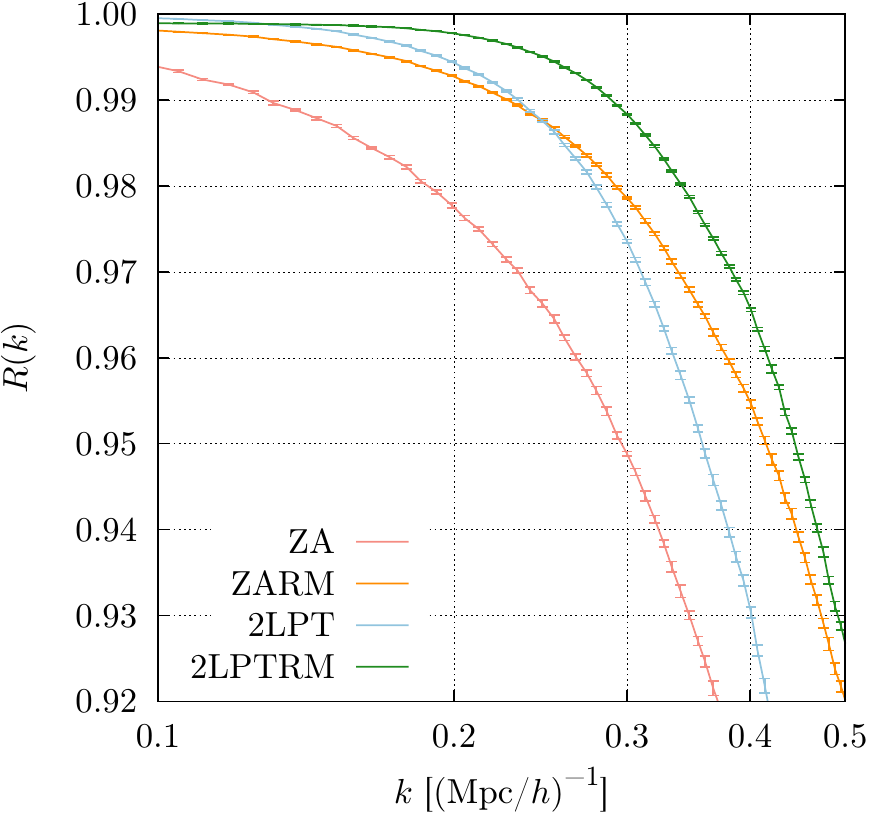}
\caption{Fourier-space \gls{cross-correlation} coefficient between various approximately-evolved \glslink{density field}{density fields} and the particle distribution as evolved with full \glslink{N-body simulation}{$N$-body dynamics}, all at \gls{redshift} zero. The binning of \glslink{density field}{density fields} is done on a 8~Mpc/$h$ mesh ($128^3$-voxel grid). At small scales, $k~\geq~0.2$~(Mpc/$h$)$^{-1}$, the \glslink{cross-correlation}{cross-correlations} with respect to the \glslink{N-body simulation}{$N$-body}-evolved field are notably better after \gls{remapping} than with \gls{LPT} alone.}
\label{fig:crosscorr}
\end{center}
\end{figure}

The Fourier space \gls{cross-correlation} coefficient between two \glslink{density field}{density fields} $\delta$ and $\delta'$ is defined as the cross-power spectrum of $\delta$ and $\delta'$, normalized by the auto-power spectra of the same fields:
\begin{equation}
R(k) \equiv \frac{P_{\delta \times \delta'}(k)}{\sqrt{P_\delta(k) P_{\delta'}(k)}} \equiv \frac{\left\langle \delta^\ast(\textbf{k})\delta'(\textbf{k}) \right\rangle}{\sqrt{\left\langle \delta^\ast(\textbf{k})\delta(\textbf{k}) \right\rangle \left\langle \delta'^\ast(\textbf{k})\delta'(\textbf{k}) \right\rangle}} .
\end{equation}
It is a dimensionless coefficient, in modulus between $0$ and $1$, representing the agreement, at the level of \glslink{two-point correlation function}{two-point statistics}, between the \textit{phases} of $\delta$ and $\delta'$ (as the overall power has been divided out). Here we choose as a reference the \gls{density field} predicted by \glslink{N-body simulation}{$N$-body simulations}, $\delta' = \delta_\mathrm{Nbody}$, and compare with approximate \glslink{density field}{density fields} generated from the same \gls{initial conditions} with \gls{LPT}. In this fashion, we characterize the \gls{phase} accuracy of the \gls{ZA} and \gls{2LPT}.

In figure \ref{fig:crosscorr} we present the Fourier-space cross-correlation coefficient between the \gls{redshift}-zero \gls{density field} in the \glslink{N-body simulation}{$N$-body simulation} and other \glslink{density field}{density fields}. At this point, it is useful to recall that an approximation well-correlated with the non-linear \gls{density field} can be used in a variety of cosmological applications, such as the \gls{reconstruction} of the non-linear \gls{power spectrum} \citep{Tassev2012b}. As pointed out by \citet{Neyrinck2013}, the \gls{cross-correlation} between \gls{2LPT} and full gravitational dynamics is higher at small $k$ than the cross-correlation between the \gls{ZA} and the full dynamics, meaning that the position of structures is more correct when additional physics (\gls{non-local} \gls{tidal effects}) is taken into account.

\subsection{Three-point statistics}
\label{sec:eulerian-three-point-stats}

\begin{figure}
\begin{center}
\includegraphics[width=0.5\columnwidth]{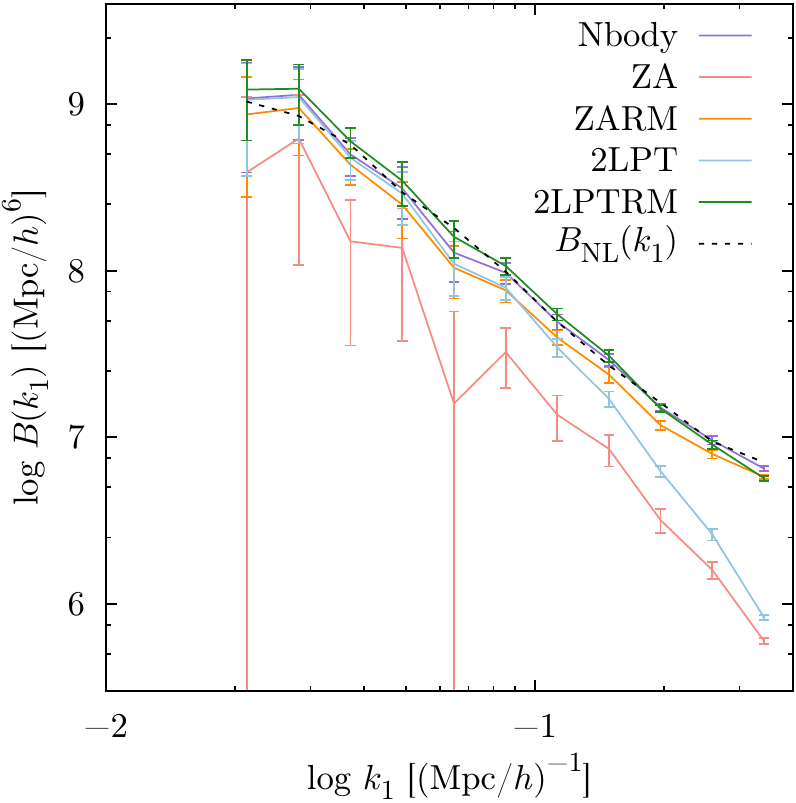}
\caption{\Gls{redshift}-zero dark matter \glslink{bispectrum}{bispectra} for equilateral triangle shape, in 1024~Mpc/$h$ simulations, with \glslink{density field}{density fields} computed on mesh of 8~Mpc/$h$ size. The particle distribution is determined using: a full \glslink{N-body simulation}{$N$-body simulation} (purple curve), the Zel'dovich approximation, alone (\gls{ZA}, light red curve) and after \gls{remapping} (\gls{ZARM}, orange curve), second-order Lagrangian perturbation theory, alone (\gls{2LPT}, light blue curve) and after \gls{remapping} (\gls{2LPTRM}, green curve). The dashed line, $B_{\mathrm{NL}}(k)$, corresponds to theoretical predictions for the \gls{bispectrum}, found using the fitting formula of \citep{Gil-Marin2012}. Note that both \gls{ZARM} and \gls{2LPTRM} show increased \gls{bispectrum} in the \gls{mildly non-linear regime} compared to \gls{ZA} and \gls{2LPT}, indicating an improvement of \glslink{three-point correlation function}{three-point statistics} with the \gls{remapping} procedure.}
\label{fig:BS}
\end{center}
\end{figure}

In this section, we analyze the accuracy of \gls{LPT} beyond \glslink{two-point correlation function}{second-order statistics}, by studying the \gls{three-point correlation function} of the \gls{density field} in Fourier space, i.e. the \gls{bispectrum}, defined by equation \eqref{eq:def-bispectrum}. The importance of \glslink{three-point correlation function}{three-point statistics} relies in their ability to test the shape of structures. Some of the natural applications are to test gravity \citep{Shiratra2007,Gil-Marin2011}, to break degeneracies due to the galaxy \gls{bias} \citep{Matarrese1997,Verde1998,Scoccimarro2001,Verde2002} or to test the existence of primordial \glslink{non-Gaussianity}{non-Gaussianities} in the initial matter \gls{density field} \citep{Sefusatti2007,Jeong2009}.

\begin{figure*}
\begin{center}
\includegraphics[width=\textwidth]{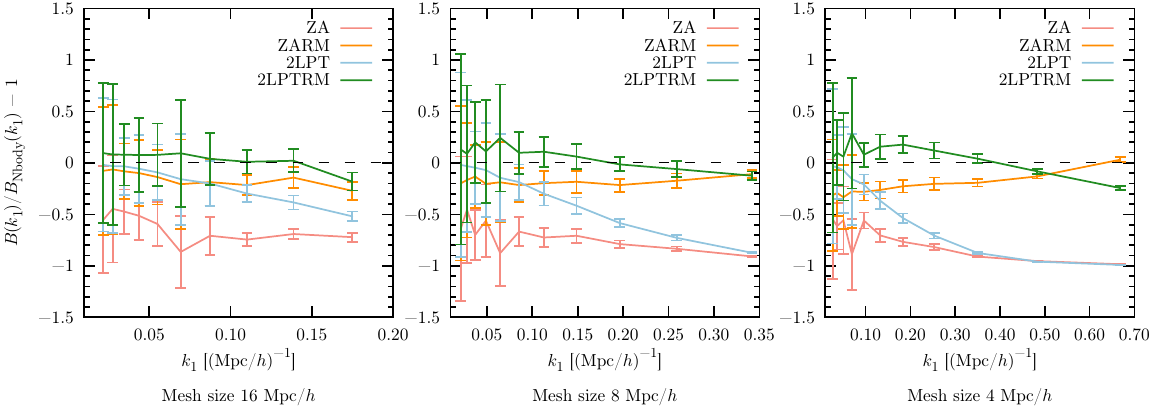}
\caption{\textit{Bispectrum: mesh size-dependence}. Relative deviations for the \glslink{bispectrum}{bispectra} $B(k_1)$ of various particle distributions, with reference to the prediction from a full \glslink{N-body simulation}{$N$-body simulation}, $B_{\mathrm{Nbody}}(k_1)$. The particle distribution is determined using: the Zel'dovich approximation, alone (\gls{ZA}, light red curve) and after \gls{remapping} (\gls{ZARM}, orange curve), second-order Lagrangian perturbation theory, alone (\gls{2LPT}, light blue curve) and after \gls{remapping} (\gls{2LPTRM}, green curve). The computation of \glslink{bispectrum}{bispectra} is done for equilateral triangles and on different meshes: 16 Mpc/$h$ ($64^3$-voxel grid, left panel), 8 Mpc/$h$ ($128^3$-voxel grid, central panel) and 4 Mpc/$h$ ($256^3$-voxel grid, right panel). All results are shown at \gls{redshift} $z=0$. \gls{LPT} fields exhibit more small-scale \glslink{three-point correlation function}{three-point correlations} after \gls{remapping} and their \glslink{bispectrum}{bispectra} get closer to the shape of the full non-linear \gls{bispectrum}.}
\label{fig:BS-deviations-mesh}
\end{center}
\end{figure*}

\begin{figure*}
\begin{center}
\includegraphics[width=\textwidth]{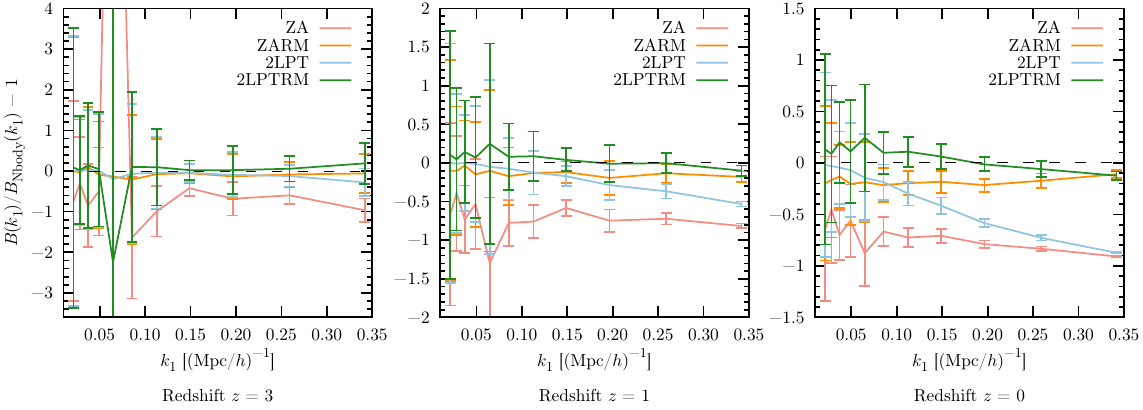}
\caption{\textit{Bispectrum: \gls{redshift}-dependence}. Relative deviations for the \glslink{bispectrum}{bispectra} $B(k_1)$ of various particle distributions (see the caption of figure \ref{fig:BS-deviations-mesh}), with reference to a full \glslink{N-body simulation}{$N$-body simulation}, $B_{\mathrm{Nbody}}(k_1)$. The computation of \glslink{bispectrum}{bispectra} is done on a 8~Mpc/$h$ mesh ($128^3$-voxel grid) and for equilateral triangles. Results at different \glslink{redshift}{redshifts} are shown: $z=3$ (right panel), $z=1$ (central panel) and $z=0$ (left panel).}
\label{fig:BS-deviations-redshift}
\end{center}
\end{figure*}

\begin{figure*}
\begin{center}
\includegraphics[width=\textwidth]{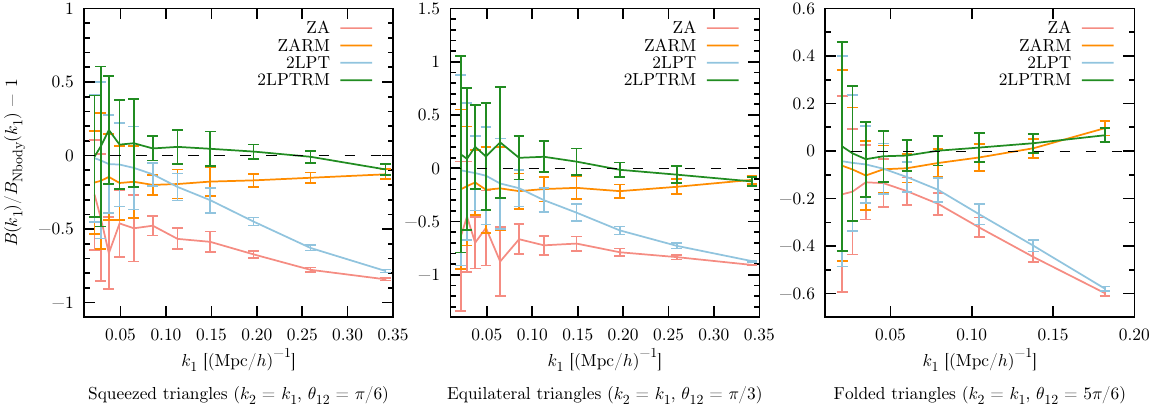}
\caption{\textit{Bispectrum: scale-dependence for different triangle shapes}. Relative deviations for the \glslink{bispectrum}{bispectra} $B(k_1)$ of various particle distributions (see the caption of figure \ref{fig:BS-deviations-mesh}), with reference to a full \glslink{N-body simulation}{$N$-body simulation}, $B_{\mathrm{Nbody}}(k_1)$. The computation is done on a 8~Mpc/$h$ mesh ($128^3$-voxel grid) and results are shown at \gls{redshift} $z=0$ for various triangle shapes as defined above.}
\label{fig:BS-deviations-triangle}
\end{center}
\end{figure*}

\begin{figure*}
\begin{center}
\includegraphics[width=\textwidth]{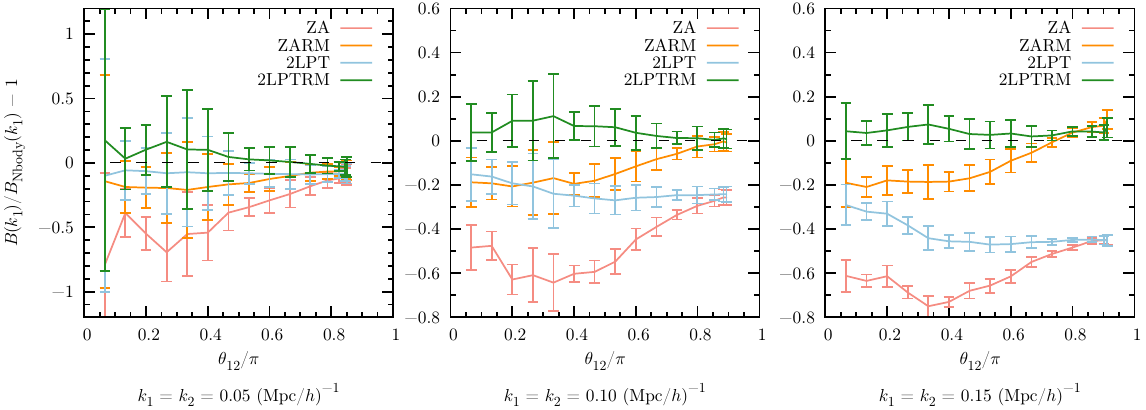}
\caption{\textit{Bispectrum: triangle shape-dependence}. Relative deviations for the \glslink{bispectrum}{bispectra} $B(k_1)$ of various particle distributions (see the caption of figure \ref{fig:BS-deviations-mesh}), with reference to a full \glslink{N-body simulation}{$N$-body simulation}, $B_{\mathrm{Nbody}}(k_1)$. The computation is done on a 8~Mpc/$h$ mesh ($128^3$-voxel grid) and results are shown at \gls{redshift} $z=0$. The dependence on the angle of the triangle $\theta_{12} = (\textbf{k}_1,\textbf{k}_2)$ is shown for different scales: $k_1=k_2=0.05~(\mathrm{Mpc}/h)^{-1}$ (corresponding to 125~Mpc/$h$), $k_1=k_2=0.10~(\mathrm{Mpc}/h)^{-1}$ (corresponding to 63~Mpc/$h$), $k_1=k_2=0.15~(\mathrm{Mpc}/h)^{-1}$ (corresponding to 42~Mpc/$h$).}
\label{fig:BS-deviations-theta}
\end{center}
\end{figure*}

As for the \gls{power spectrum}, we construct the dark matter \gls{density contrast} field, by \glslink{mesh assignment}{assigning} \glslink{dark matter particles}{particles} to the grid using a \gls{CiC} scheme. We then deconvolve the \gls{CiC} kernel to correct for corresponding smoothing effects. The algorithm used to compute the \gls{bispectrum} $B(\textbf{k}_1,\textbf{k}_2,\textbf{k}_3)$ from this $\delta(\textbf{k})$ field consists of randomly drawing $k$-vectors from a specified bin, namely $\Delta k$, and randomly orientating the $(\textbf{k}_1,\textbf{k}_2,\textbf{k}_3)$ triangle in space. We chose the number of random triangles to depend on the number of fundamental triangle per bin, that scales as $k_1 k_2 k_3 \Delta k^3$ \citep{Scoccimarro1997}, where $\Delta k$ is the chosen $k$-binning: given $k_i$ we allow triangles whose $i$-side lies between $k_i-\Delta k/2$ and $k_i+\Delta k/2$. In this paper we always set $\Delta k = k_{\mathrm{min}} = 2\pi/L$, where $L$ is the size of the box. For the equilateral case, at scales of $k \approx 0.1$~(Mpc/$h$)$^{-1}$ we generate $\sim 1.7 \times 10^6$ random triangles. We have verified that increasing the number of triangles beyond this value does not have any effect on the measurement. The rule of thumb presented in section \ref{sec:eulerian-two-point-stats-power-spectrum} for the smallest scale to trust applies for the \gls{bispectrum} as well. Also, as a lower limit in $k$, we have observed that for scales larger than $\sim 3\,k_{\mathrm{min}}$, effects of \gls{cosmic variance} start to be important and considerable deviations with respect to linear theory can be observed. For this reason, we limit the largest scale for our \gls{bispectrum} analysis to $3\,k_{\mathrm{min}} \approx 1.8 \times 10^{-2}~(\mathrm{Mpc}/h)^{-1}$.

Error bars in \gls{bispectrum} plots represent the dispersion of the mean among eight independent realizations, all of them with the same \gls{cosmological parameters}. It has been tested \citep{Gil-Marin2012}, that this \gls{estimator} for the error is in good agreement with theoretical predictions based on the Gaussianity of \gls{initial conditions} \citep{Scoccimarro1998b}.

The subtracted \gls{shot noise} is always assumed to be Poissonian:
\begin{equation}
B_{\mathrm{SN}}(\textbf{k}_1,\textbf{k}_2,\textbf{k}_3) = \frac{1}{\bar{n}} \left[ P(k_1)+P(k_2)+P(k_3) \right] + \frac{1}{\bar{n}^2} ,
\end{equation}
\citep[see e.g.][and references therein]{Peebles1980}, where $\bar{n}$ is the number density of particles in the box.

A triangle shape is defined by the relative length of vectors $\textbf{k}_1$ and $\textbf{k}_2$ and the inner angle $\theta_{12}$, in such a way that $\textbf{k}_1+\textbf{k}_2+\textbf{k}_3 = 0$ and $\textbf{k}_1~\cdot~\textbf{k}_2~=~k_1 k_2 \cos(\pi-\theta_{12})$. In figure \ref{fig:BS}, we plot the \gls{redshift}-zero \gls{bispectrum}, computed on a 8~Mpc/$h$ mesh, of the different \glslink{density field}{density fields} for equilateral triangles ($\theta_{12} = \pi/3$ and $k_2/k_1 = 1$). There, the dashed line corresponds to theoretical predictions for the non-linear \gls{bispectrum}, found using the fitting formula of \citet{Gil-Marin2012}. The relative deviations of various \glslink{bispectrum}{bispectra} with reference to full \glslink{N-body simulation}{$N$-body simulations} are shown in figures \ref{fig:BS-deviations-mesh}, \ref{fig:BS-deviations-redshift}, \ref{fig:BS-deviations-triangle} and \ref{fig:BS-deviations-theta}.

The main result is that \gls{LPT} predicts less \glslink{three-point correlation function}{three-point correlation} than \gls{full gravity}. This is true even at large scales for the \gls{ZA}: as it is \gls{local}, it generally fails to predict the shape of structures. \gls{2LPT} agrees with \glslink{N-body simulation}{$N$-body simulations} at large scales, with differences starting to appear only in the \gls{mildly non-linear regime}, $k \gtrsim 0.1$ (Mpc/$h$)$^{-1}$ at $z=0$.

\section{Statistics of the Lagrangian displacement field}
\label{sec:Statistics of the Lagrangian displacement field}

\subsection{Lagrangian $\psi$ versus Eulerian $\delta$: one-point statistics}
\label{sec:psi_vs_delta}

\draw{This section draws from \citet{Leclercq2015ADDENDUM}, addendum to \citet{Leclercq2013}.}

\begin{figure}
\begin{center}
\includegraphics[width=0.5\columnwidth]{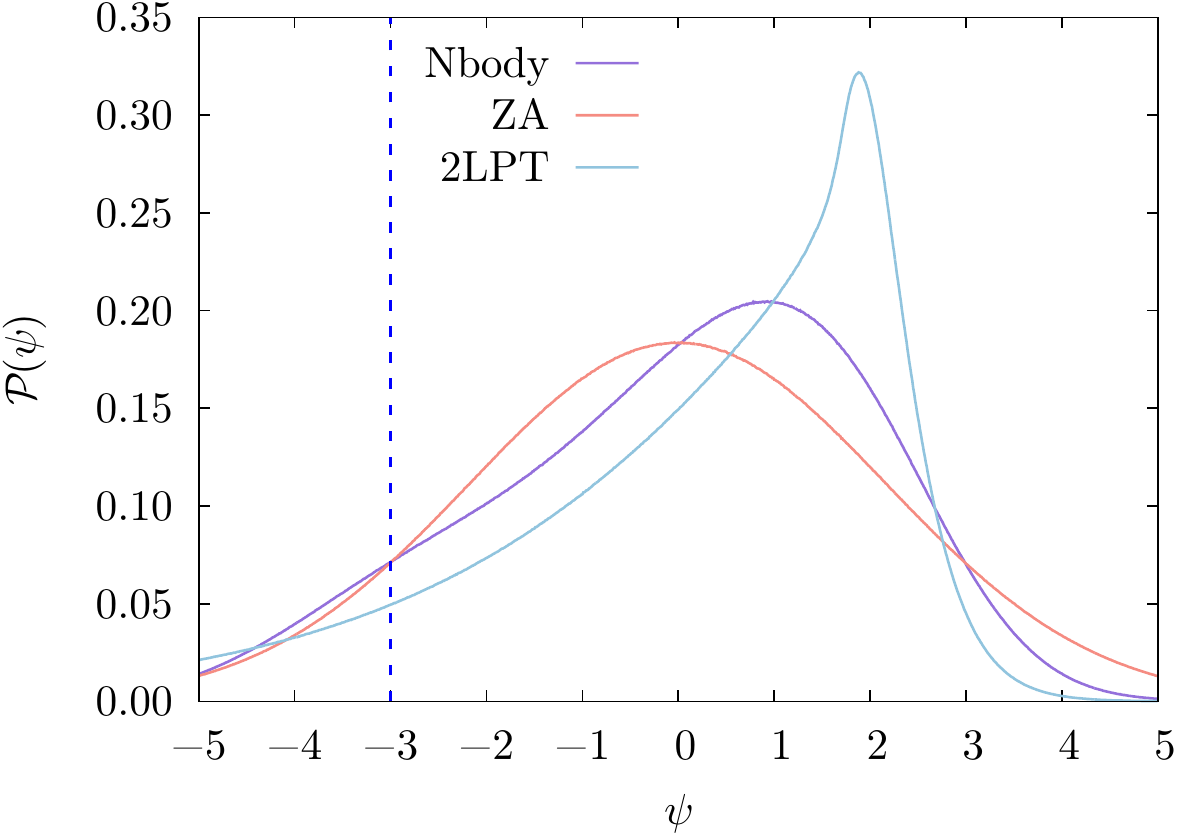}
\caption{\Gls{redshift}-zero probability distribution function for the \glslink{divergence of the Lagrangian displacement field}{divergence of the displacement field} $\psi$, computed from eight 1024~Mpc/$h$-box simulations of $512^3$ particles. A quantitative analysis of the deviation from Gaussianity of these \glslink{pdf}{pdfs} is given in table \ref{tb:NGparam}. The particle distribution is determined using: a full \glslink{N-body simulation}{$N$-body simulation} (purple curve), the Zel'dovich approximation (\gls{ZA}, light red curve) and second-order Lagrangian perturbation theory (\gls{2LPT}, light blue curve). The vertical line at $\psi =-3$ represents the collapse barrier about which $\psi$ values bob around after gravitational collapse. A bump at this value is visible with \gls{full gravity}, but \gls{LPT} is unable to reproduce this feature. This regime corresponds to virialized, overdense \glslink{cluster}{clusters}.}
\label{fig:divpsi_distrib}
\end{center}
\end{figure}

\begin{table}\centering
\begin{tabular}{lcc}
\hline\hline
Model & $\mathcal{P}_\delta$ & $\mathcal{P}_\psi$ \\
\hline
\multicolumn{1}{c}{} & \multicolumn{2}{c}{Skewness $\gamma_1$} \\
ZA & $2.36 \pm 0.01$ & $-0.0067 \pm 0.0001$ \\
2LPT & $2.83 \pm 0.01$ & $-1.5750 \pm 0.0002$ \\
$N$-body & $5.14 \pm 0.05$ & $-0.4274 \pm 0.0001$ \\
\hline
\multicolumn{1}{c}{} & \multicolumn{2}{c}{Excess kurtosis $\gamma_2$} \\
ZA & $9.95 \pm 0.09$ & $-2.2154 \times 10^{-6} \pm 0.0003$ \\
2LPT & $13.91 \pm 0.15$ & $3.544 \pm 0.0011$ \\
$N$-body & $62.60 \pm 2.75$ & $-0.2778 \pm 0.0004$ \\
\hline\hline
\end{tabular}
\caption{Non-Gaussianity parameters (the \gls{skewness} $\gamma_1$ and the \gls{excess kurtosis} $\gamma_2$) of the \gls{redshift}-zero probability distribution functions $\mathcal{P}_\delta$ and $\mathcal{P}_\psi$ of the \gls{density contrast} $\delta$ and the \glslink{divergence of the Lagrangian displacement field}{divergence of the displacement field} $\psi$, respectively. The confidence intervals given correspond to the 1-$\sigma$ standard deviations among eight realizations. In all cases, $\gamma_1$ and $\gamma_2$ are reduced when measured from $\psi$ instead of $\delta$.}
\label{tb:NGparam}
\end{table}

\begin{figure*}
\begin{center}
\includegraphics[width=0.85\textwidth]{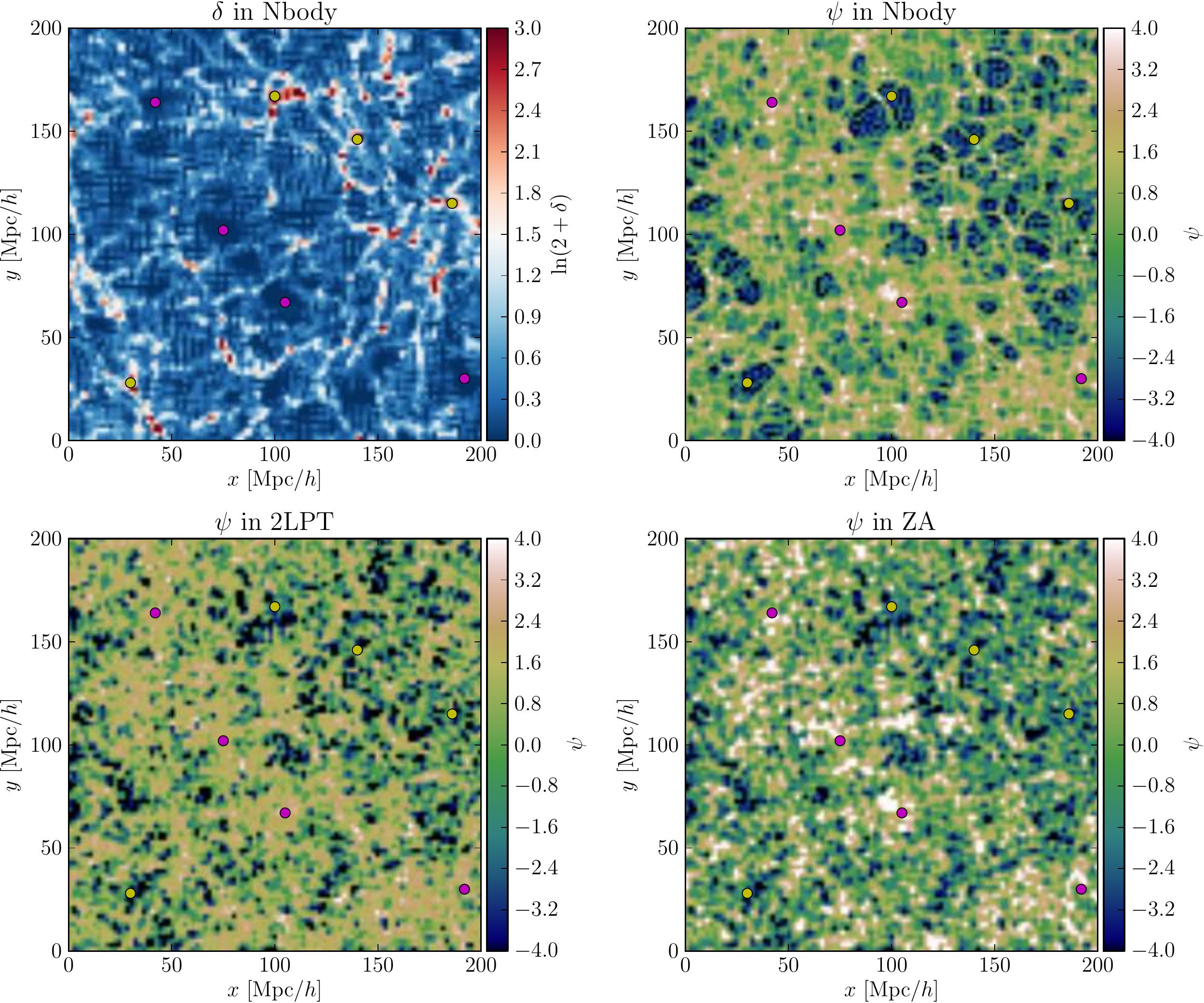}
\caption{Slices of the \glslink{divergence of the Lagrangian displacement field}{divergence of the displacement field}, $\psi$, on a Lagrangian sheet of $512^2$ particles from a $512^3$-particle simulation of box size 1024 Mpc/$h$, run to \gls{redshift} zero. For clarity we show only a 200~Mpc/$h$ region. Each pixel corresponds to a particle. The particle distribution is determined using respectively a full \glslink{N-body simulation}{$N$-body simulation}, the Zel'dovich approximation (\gls{ZA}) and second-order Lagrangian perturbation theory (\gls{2LPT}). In the upper left panel, the \gls{density contrast} $\delta$ in the \glslink{N-body simulation}{$N$-body simulation} is shown, after binning on a $512^3$-voxel grid. To guide the eye, some \glslink{cluster}{clusters} and \glslink{void}{voids} are identified by yellow and purple dots, respectively. The ``lakes'', Lagrangian regions that have collapsed to form \glslink{halo}{halos}, are only visible in the \glslink{N-body simulation}{$N$-body simulation}, while the ``mountains'', Lagrangian regions corresponding to cosmic \glslink{void}{voids}, are well reproduced by \gls{LPT}.}
\label{fig:slices_divpsi}
\end{center}
\end{figure*}

\begin{figure*}
\begin{center}
\includegraphics[width=0.35\textwidth]{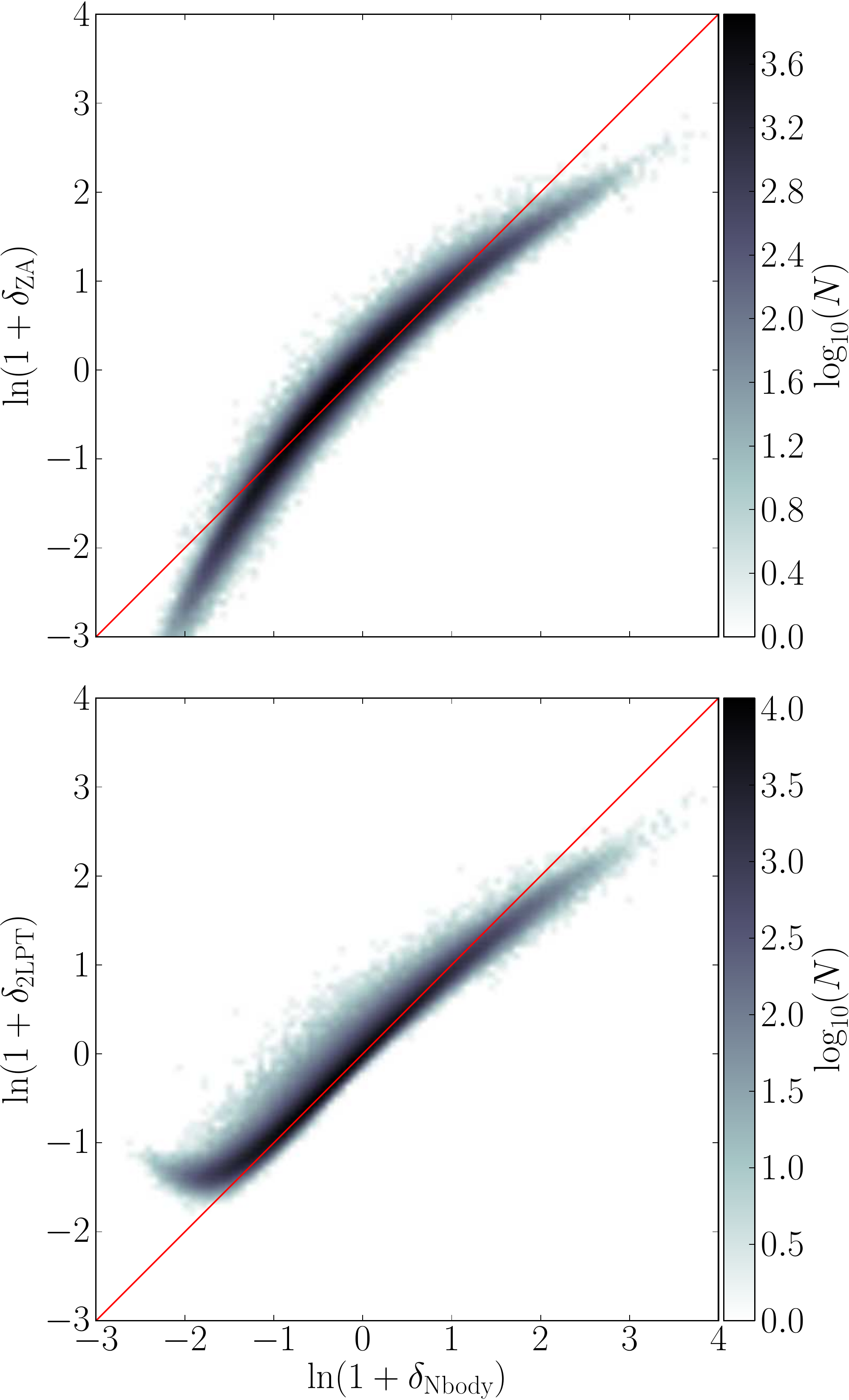} \quad \includegraphics[width=0.35\textwidth]{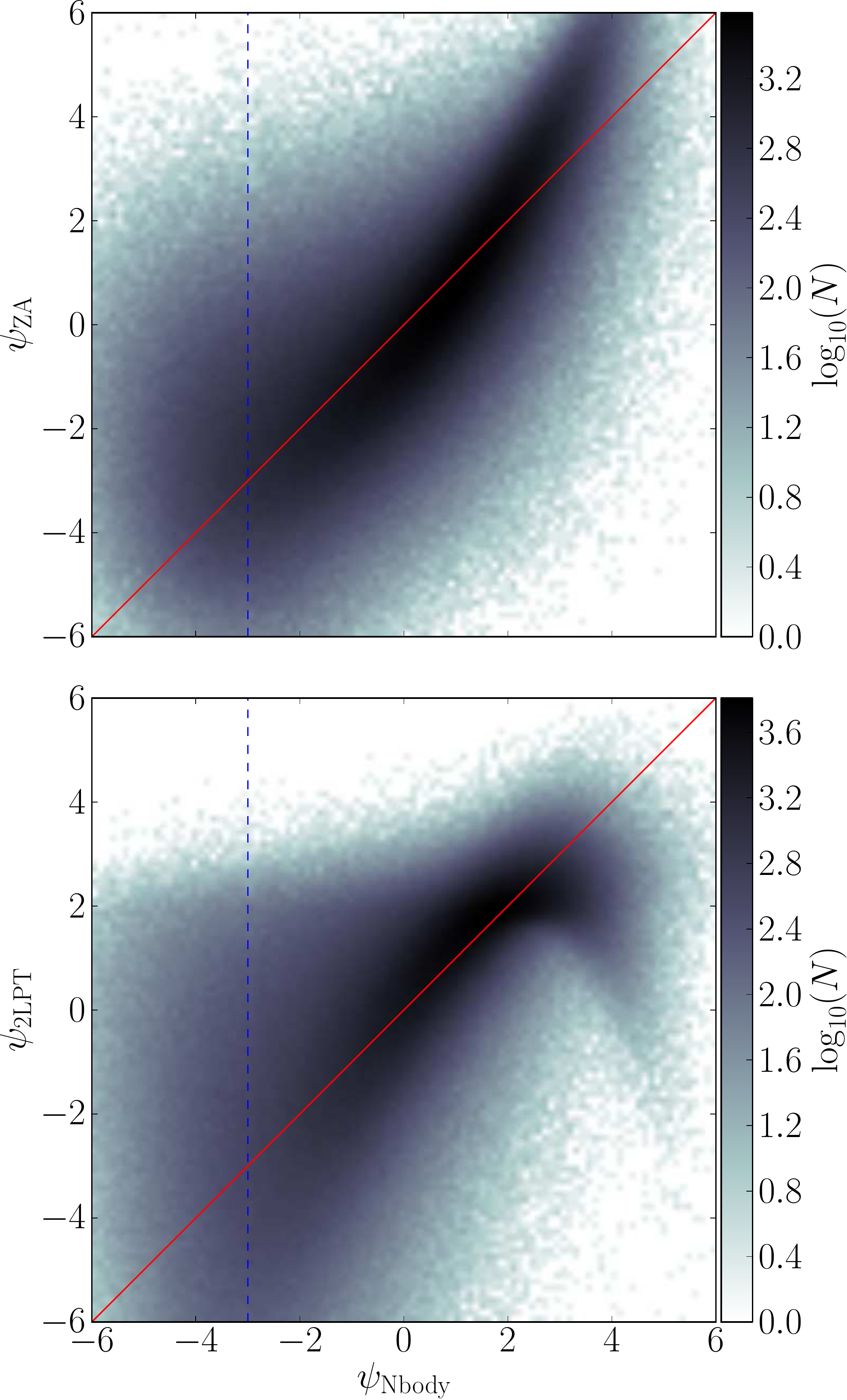}
\caption{\emph{Left panel}. Two-dimensional histograms comparing particle densities evolved with full \glslink{N-body simulation}{$N$-body dynamics} (the $x$-axis) to densities in the \gls{LPT}-evolved particle distributions (the $y$-axis). The red lines show the ideal $y = x$ locus. A turn-up at low densities is visible with \gls{2LPT}, meaning that some overdense regions are predicted where there should be deep \glslink{void}{voids}. \emph{Right panel}. Same plot for the \glslink{divergence of the Lagrangian displacement field}{divergence of the displacement field} $\psi$. Negative $\psi$ corresponds to overdensities and positive $\psi$ correspond to underdensities. The dotted blue line shows the collapse barrier at $\psi = -3$ where particle get clustered in \gls{full gravity}. The scatter is bigger with $\psi$ than with $\delta$, in particular in overdensities, since with \gls{LPT}, particles do not cluster. The turn-up at low densities with \gls{2LPT}, observed with the \gls{density contrast}, is also visible with the \glslink{divergence of the Lagrangian displacement field}{divergence of the displacement field}.}
\label{fig:scatters}
\end{center}
\end{figure*}

As noted by previous authors \citep[see in particular][]{Neyrinck2013}, in the Lagrangian representation of the \gls{LSS}, it is natural to use the \glslink{divergence of the Lagrangian displacement field}{divergence of the displacement field} $\psi$ instead of the \glslink{density contrast}{Eulerian density contrast} $\delta$. This section comments the \glslink{one-point distribution}{one-point statistics} of $\psi$ in \gls{LPT} and \gls{full gravity} and comparatively analyzes key features of $\psi$ and $\delta$.

As seen in section \ref{sec:LPT}, in the Lagrangian frame, the quantity of interest is not the position, but the \gls{displacement field} $\boldsymbol{\Psi}(\textbf{q})$ which maps the initial \glslink{comoving coordinates}{comoving} particle position $\textbf{q}$ to its final \glslink{comoving coordinates}{comoving} Eulerian position \textbf{x} (see e.g. \citealp{Bouchet1995} or \citealp{Bernardeau2002} for overviews),
\begin{equation}
\textbf{x} \equiv \textbf{q} + \boldsymbol{\Psi}(\textbf{q}) .
\end{equation}
It is important to note that, though $\boldsymbol{\Psi}(\textbf{q})$ is \textit{a priori} a full three-dimensional vector field, it is curl-free up to second order in \gls{LPT} (appendix D in \citealp{Bernardeau1994} or \citealp{Bernardeau2002} for a review). In this thesis, we do not consider perturbative contributions beyond \gls{2LPT}.

Let $\psi(\textbf{q}) \equiv \nabla_{\textbf{q}} \cdot \boldsymbol{\Psi}(\textbf{q})$ denote the \glslink{divergence of the Lagrangian displacement field}{divergence of the displacement field}, where $\nabla_{\textbf{q}}$ is the divergence operator in Lagrangian coordinates. $\psi$ quantifies the angle-averaged spatial-stretching of the Lagrangian \gls{dark matter} ``sheet'' in \gls{comoving coordinates} \citep{Neyrinck2013}. Let $\mathcal{P}_{\psi,\mathrm{LPT}}$ and $\mathcal{P}_{\psi,\mathrm{Nbody}}$ be the \glslink{one-point distribution}{one-point probability distribution functions} for the \glslink{divergence of the Lagrangian displacement field}{divergence of the displacement field} in \gls{LPT} and in full $N$-body fields, respectively. We denote by $\mathcal{P}_\delta$ the corresponding \glslink{pdf}{pdfs} for the \glslink{density contrast}{Eulerian density contrast}.

In figure \ref{fig:divpsi_distrib}, we show the \glslink{pdf}{pdfs} of $\psi$ for the \gls{ZA}, \gls{2LPT} and full $N$-body gravity. The most important feature of $\psi$ is that, whatever the model for \gls{structure formation}, the \gls{pdf} exhibits reduced \gls{non-Gaussianity} compared to the \gls{pdf} for the \gls{density contrast} $\delta$ (see the upper panel of figure \ref{fig:pdf} for comparison). The main reason is that $\mathcal{P}_{\delta}$, unlike $\mathcal{P}_\psi$, is tied down to zero at $\delta = -1$. It is highly \glslink{non-Gaussianity}{non-Gaussian} in the \gls{final conditions}, both in \glslink{N-body simulation}{$N$-body simulations} and in approximations to the true dynamics. For a quantitative analysis, we looked at the first and second-order \gls{non-Gaussianity} statistics: the skewness $\gamma_1$ and the \gls{excess kurtosis} $\gamma_2$,
\begin{equation}
\gamma_1 \equiv \frac{\mu_3}{\sigma^3} \quad \mathrm{and} \quad \gamma_2 \equiv \frac{\mu_4}{\sigma^4}-3,
\end{equation}
where $\mu_n$ is the $n$-th \gls{moment} about the mean and $\sigma$ is the standard deviation. We estimated $\gamma_1$ and $\gamma_2$ at \gls{redshift} zero in our simulations, in the \glslink{one-point distribution}{one-point statistics} of the \gls{density contrast} $\delta$ and of the \glslink{divergence of the Lagrangian displacement field}{divergence of the displacement field} $\psi$. The results are shown in table \ref{tb:NGparam}. In all cases, we found that both $\gamma_1$ and $\gamma_2$ are much smaller when measured from $\mathcal{P}_\psi$ instead of $\mathcal{P}_\delta$.

At linear order in \glslink{LPT}{Lagrangian perturbation theory} (the \glslink{ZA}{Zel'dovich approximation}), the \glslink{divergence of the Lagrangian displacement field}{divergence of the displacement field} is proportional to the \gls{density contrast} in the \gls{initial conditions}, $\delta(\mathbf{q})$, scaling with the negative \glslink{linear growth factor}{growth factor}, $-D_1(\tau)$:
\begin{equation}
\label{eq:mapping-ZA}
\psi^{(1)}(\textbf{q},\tau) = \nabla_\textbf{q} \cdot \boldsymbol{\Psi}^{(1)}(\textbf{q},\tau) = -D_1(\tau) \, \delta(\textbf{q}) .
\end{equation}
Since we take Gaussian \gls{initial conditions}, the \gls{pdf} for $\psi$ is Gaussian at any time with the \gls{ZA}. In \gls{full gravity}, \gls{non-linear evolution} slightly breaks Gaussianity. $\mathcal{P}_{\psi,\mathrm{Nbody}}$ is slightly skewed towards negative values while its mode gets shifted around $\psi \approx 1$. Taking into account \gls{non-local} effects, \gls{2LPT} tries to get closer to the shape observed in \glslink{N-body simulation}{$N$-body simulations}, but the correct \gls{skewness} is overshot and the \gls{pdf} is exceedingly peaked.

Figure \ref{fig:slices_divpsi} shows a slice of the \glslink{divergence of the Lagrangian displacement field}{divergence of the displacement field}, measured at \gls{redshift} zero for particles occupying a flat $512^2$-pixel Lagrangian sheet from one of our simulations. For comparison, see also the figures in \citet{Mohayaee2006,Pueblas2009} and \citet{Neyrinck2013}. We used the color scheme of the latter paper, suggesting a topographical analogy when working in Lagrangian coordinates. As structures take shape, $\psi$ departs from its initial value; it takes positive values in underdensities and negative values in overdensities. The shape of \glslink{void}{voids} (the ``mountains'') is found to be reasonably similar in \gls{LPT} and in the \glslink{N-body simulation}{$N$-body simulation}. For this reason, the influence of late-time non-linear effects in \glslink{void}{voids} is milder as compared to overdense structures, which makes them easier to relate to the \gls{initial conditions}. However, in overdense regions where $\psi$ decreases, it is not allowed to take arbitrary values: where gravitational collapse occurs, ``lakes'' form and $\psi$ gets stuck around a collapse barrier, $\psi \approx -3$. As expected, these ``lakes'', corresponding to virialized \glslink{cluster}{clusters}, can only be found in \glslink{N-body simulation}{$N$-body simulations}, since \gls{LPT} fails to accurately describe the highly non-linear physics involved. A small bump at $\psi=-3$ is visible in $\mathcal{P}_{\psi,\mathrm{Nbody}}$ (see figure \ref{fig:divpsi_distrib}). We checked that this bump gets more visible in higher mass-resolution simulations (200 Mpc/$h$ box for $256^3$ particles), where matter is more clustered. This means that part of the information about gravitational clustering can be found in the \glslink{one-point distribution}{one-point statistics} of $\psi$. Of course, the complete description of \glslink{halo}{halos} requires to precisely account for the shape of the ``lakes'', which can only be done via \glslink{high-order correlation function}{higher-order correlation functions}. More generally, it is possible to use Lagrangian information in order to classify structures of the \gls{cosmic web}. In particular, {\diva} \citep{Lavaux2010} uses the shear of the \gls{displacement field} and {\origami} \citep{Falck2012} the number of \glslink{phase space}{phase-space} folds. While these techniques cannot be straightforwardly used for the analysis of \glslink{galaxy survey}{galaxy surveys}, where we lack Lagrangian information, recently proposed techniques for physical inference of the \gls{initial conditions} \citep[chapters \ref{chap:BORG} and \ref{chap:BORGSDSS}][]{Jasche2013BORG,Jasche2015BORGSDSS} should allow their use with observational data.

Figure \ref{fig:scatters} shows two-dimensional histograms comparing \glslink{N-body simulation}{$N$-body simulations} to the \gls{LPT} realizations for the \gls{density contrast} $\delta$ and the \glslink{divergence of the Lagrangian displacement field}{divergence of the displacement field} $\psi$. At this point, it is useful to note that a good mapping exists in the case where the relation shown is monotonic and the scatter is narrow. As pointed out by \citet{Sahni1996} and \citet{Neyrinck2013}, matter in the substructure of \gls{2LPT}-\glslink{void}{voids} has incorrect statistical properties: there are overdense particles in the low density region of the \gls{2LPT} $\delta$-scatter plot. This degeneracy is also visible in the $\psi > 0$ region of the \gls{2LPT} $\psi$-scatter plot. On average, the scatter is bigger with $\psi$ than with $\delta$, in particular in overdensities ($\psi <0$), since with \gls{LPT}, particles do not cluster: $\psi$ takes any value between 2 and $-3$ where it should remain around $-3$.

Summing up our discussions in this paragraph, we analyzed the relative merits of the \glslink{divergence of the Lagrangian displacement field}{Lagrangian divergence of the displacement field} $\psi$, and the \glslink{density contrast}{Eulerian density contrast} $\delta$ at the level of \glslink{one-point distribution}{one-point statistics}. The important differences are the following:

\begin{enumerate}
\item $\boldsymbol{\Psi}$ being irrotational up to order two, its divergence $\psi$ contains nearly all information on the \gls{displacement field} in one dimension, instead of three. The collapse barrier at $\psi=-3$ is visible in $\mathcal{P}_\psi$ for \glslink{N-body simulation}{$N$-body simulations} but not for \gls{LPT}. A part of the information about non-linear gravitational clustering is therefore encoded in the \glslink{one-point distribution}{one-point statistics} of $\psi$.
\item $\psi$ exhibits much fewer gravitationally-induced \glslink{non-Gaussianity}{non-Gaussian} features than $\delta$ in the \gls{final conditions} (figure \ref{fig:divpsi_distrib} and table \ref{tb:NGparam}).
\item However, the values of $\psi$ are more scattered than the values of $\delta$ with respect to the true dynamics (figure \ref{fig:scatters}), meaning that an unambiguous mapping is more difficult.
\end{enumerate}

\subsection{Perturbative and non-perturbative prescriptions for $\psi$}

Even if $\psi$ does not contain all the information about the vector \gls{displacement field} $\boldsymbol{\Psi}$, knowledge of its evolution allows for methods to produce approximate \glslink{particle realization}{particle realizations} at the desired \gls{redshift}, for the variety of cosmological applications described in the \hyperref[chap:intro]{introduction} of this thesis. These methods include, but are not limited to, the \gls{ZA} and \gls{2LPT}. On the contrary, \gls{3LPT} involves a non-zero \glslink{vector part}{rotational component} and comes at the expense of significantly greater complexity, for an agreement with \gls{full gravity} that does not improve substantially \citep{Buchert1994,Bouchet1995,Sahni1996}. Since we have adopted the approximation that the \gls{displacement field} is potential, we stop our analysis of \gls{LPT} at second order. However, we will describe various non-perturbative schemes.

Importantly, $\psi$-based methods are essentially as fast as producing \gls{initial conditions} for \glslink{N-body simulation}{$N$-body simulations}. Their implementation can be decomposed in several steps:
\begin{enumerate}
\item Generation of a voxel-wise initial-\gls{density field} $\delta$. It is typically a \gls{grf}, given a prescription for the linear \gls{power spectrum} (see section \ref{sec:apx-Setting up initial conditions}), but it can also include primordial \glslink{non-Gaussianity}{non-Gaussianities}.
\item Estimation of $\psi$ from $\delta$ at the desired \gls{redshift}.
\item Generation of the final vector \gls{displacement field} $\boldsymbol{\Psi}$ from $\psi$ with an inverse-divergence operator.
\item Application of $\boldsymbol{\Psi}$ to the \glslink{dark matter particles}{particles} of a regular Lagrangian lattice to get their final positions.
\end{enumerate}
In practice, steps 1 and 3 are performed in Fourier space, using fast \glslink{Fourier transform}{Fourier transforms} to translate between configuration space and Fourier space when necessary. In the remainder of this paragraph, we review various prescriptions that have been proposed in the literature to estimate $\psi(\textbf{q},\tau)$ from $\delta(\textbf{q})$ (step 2).

\paragraph{The Zel'dovich approximation.} The first scheme, already studied in section \ref{sec:ZA}, is the \gls{ZA} (equation \eqref{eq:divergence-Psi1}),
\begin{equation}
\psi_\mathrm{ZA}(\textbf{q},\tau) = -D_1(\tau) \, \delta(\textbf{q}) \equiv - \delta_\mathrm{L}(\textbf{q},\tau) .
\end{equation} The \gls{ZA} allows to separate prescriptions for $\psi$ into two classes: \textit{\gls{local}} Lagrangian approximations, where $\psi$ depends only on its linear value, $\psi_\mathrm{L}(\textbf{q},\tau) \equiv - \delta_\mathrm{L}(\textbf{q},\tau)$ and \textit{\gls{non-local}} ones (e.g. higher-order \gls{LPT}) where $\psi$ depends on derivatives of $\psi_\mathrm{L}$ as well (which means that the behavior of a Lagrangian particle depends on its neighbours).

\paragraph{Second-order Lagrangian perturbation theory.} In \gls{2LPT}, the \gls{non-local} prescription for $\psi$ is (see equation \eqref{eq:Psi_2LPT})
\begin{equation}
\psi_\mathrm{2LPT}(\textbf{q},\tau) = -D_1(\tau) \Delta_\textbf{q} \phi^{(1)}(\textbf{q}) + D_2(\tau) \Delta_\textbf{q} \phi^{(2)}(\textbf{q}) ,
\end{equation} where the \glslink{Lagrangian potential}{Lagrangian potentials} follow the \glslink{Poisson equation}{Poisson-like equations} \eqref{eq:Poisson_phi1} and \eqref{eq:Poisson_phi2}. As pointed out by \citet{Neyrinck2013}, since \gls{2LPT} is a second-order scheme, $\psi_\mathrm{2LPT}$ is roughly parabolic in the local $\delta_\mathrm{L}$, which yields, using $D_2(\tau) \approx -\frac{3}{7} D_1^2(\tau)$ \citep{Bouchet1995},
\begin{equation}
\psi_\mathrm{2LPT}(\textbf{q},\tau) \approx \psi_\mathrm{2LPT,parab}(\textbf{q},\tau) \equiv -\delta_\mathrm{L}(\textbf{q},\tau) + \frac{1}{7}\left( \delta_\mathrm{L}(\textbf{q},\tau) \right)^2.
\end{equation}

\paragraph{The spherical collapse approximation.} \citet{Bernardeau1994} provides a simple formula for the time-evolution (collapse or expansion) of a spherical Lagrangian volume element, independent of \gls{cosmological parameters}:
\begin{equation}
\label{eq:SC1}
V(\textbf{q},\tau) = V(\textbf{q}) \left( 1 - \frac{2}{3} \delta_\mathrm{L}(\textbf{q},\tau) \right)^{3/2} .
\end{equation}
Building upon this result, \citet{Mohayaee2006,Lavaux2008} and \citet{Neyrinck2013} derived a prescription for the \glslink{divergence of the Lagrangian displacement field}{divergence of the displacement field}. Considering the isotropic stretch of a Lagrangian mass element that occupies a cube of side length $1+\psi/3$ (giving $\nabla_\textbf{q} \cdot \boldsymbol{\Psi} = \psi$), mass conservation imposes
\begin{equation}
\label{eq:SC2}
\frac{V(\textbf{q},\tau)}{V(\textbf{q})}= \frac{1}{1+\delta} = \left(1+\frac{\psi}{3}\right)^3 .
\end{equation}
Equations \eqref{eq:SC1} and \eqref{eq:SC2} yield
\begin{equation}
\label{eq:SC3}
\psi =  3 \left( \sqrt{1-\dfrac{2}{3} \delta_\mathrm{L}} -1 \right) .
\end{equation}
However, there exists no solution for $\delta_\mathrm{L} > 3/2$. \citet{Neyrinck2013} proposes to fix $\psi=-3$ in such volume elements. This corresponds to the ideal case of a Lagrangian patch contracting to a single point ($\nabla_\textbf{q} \cdot \textbf{x} = 0$). The final prescription for the \glslink{SC}{spherical collapse} (\gls{SC}) approximation is then
\begin{equation}
\psi_\mathrm{SC}(\textbf{q},\tau) = \left\{
\begin{array}{ll}
      3 \left( \sqrt{1-\dfrac{2}{3} \delta_\mathrm{L}(\textbf{q},\tau)} -1 \right) & \mathrm{if~} \delta_\mathrm{L} < 3/2, \\
      -3 & \mathrm{if~}  \delta_\mathrm{L} \geq 3/2 .
\end{array} 
\right.
\label{eq:psi_SC}
\end{equation}
One possible concern with this formula is that, in \gls{full gravity}, there are roughly as many \glslink{dark matter particles}{particles} with $\psi > -3$ as with $\psi < -3$ \citep[see e.g. trajectories in $\psi$ as a function of the \gls{scale factor} $a$, figure 7 in][]{Neyrinck2013}. Yet, this remains more correct than what happens with \gls{LPT}, where $\psi$ can take any negative value, indicating severe unphysical \glslink{shell-crossing}{over-crossing} of particles in collapsed structures.

Compared to \gls{LPT}, the \gls{SC} approximation gives reduced \glslink{shell-crossing}{stream-crossing}, better small-scale flows and \glslink{one-point distribution}{one-point pdf} correspondence to the results of \gls{full gravity}. However, a significant drawback is its incorrect treatment of large-scale flows, leading to a negative offset in the large-scale \gls{power spectrum} \citep[figure 14 in][]{Neyrinck2013}.\footnote{An empirical correction may be added to the SC formula to fix this issue: multiplying $\delta_\mathrm{L}$ in equation \eqref{eq:psi_SC} by a factor such that the large-scale \gls{power spectrum} of SC realizations agrees with that of \gls{LPT} realizations \citep{Neyrinck2013}. See also \hyperref[sec:MUSCLE]{the paragraph on \glslink{muscle}{\textsc{muscle}}}.} \gls{LPT} realizations, on the other hand, give more accurate large-scale \glslink{power spectrum}{power spectra}, as well as improved \gls{cross-correlation} to the \gls{density field} evolved with \gls{full gravity}.

\paragraph{Local Lagrangian approximations.} The SC approximation belongs to a more general family of \glslink{local Lagrangian approximations}{``local Lagrangian'' approximations} investigated by \citet{Protogeros1997}, parameterized by $1 \leq \alpha \leq 3$, the effective number of axes along which the considered volume element undergoes gravitational collapse. The corresponding density is given by
\begin{equation}
\delta_\alpha(\psi) = \left( 1 + \frac{\psi}{\alpha} \right)^{-\alpha} - 1.
\end{equation}
Here, $\psi$ is the actual non-linear displacement-divergence of a volume element, not necessarily related to the linearly evolved $\psi_\mathrm{L}$.
From equations \eqref{eq:SC1} and \eqref{eq:SC2}, we get
\begin{equation}
\delta = \left( 1 - \frac{2}{3} \delta_\mathrm{L} \right)^{-3/2} -1= \left( 1 + \frac{2}{3} \psi_\mathrm{L}\right)^{-3/2} -1, 
\end{equation}
therefore the \glslink{SC}{spherical collapse} approximation corresponds to the case $\alpha=3/2$ for $\psi = \psi_\mathrm{L}$. The cubic mass-element approximation that would follow directly from using equation \eqref{eq:SC2} without equation \eqref{eq:SC1} corresponds to the case $\alpha=3$ for the full $\psi$. \citet{Neyrinck2013} shows that the $\delta$--$\psi$ relation closely follows $\delta_3(\psi)$ for $\psi<0$, whereas for $\psi>0$ the result is between $\delta_3(\psi)$ and $\delta_{3/2}(\psi)$, when accounting for the anisotropy of gravitational expansion. 

\paragraph{Augmented Lagrangian Perturbation Theory.\label{sec:ALPT}} As discussed before, \gls{LPT} correctly describes large scales and SC more accurately captures small, collapsed structures. \citet{Kitaura2013ALPT} proposed a recipe to interpolate between the \gls{LPT} displacement on large scales and the \gls{SC} displacement on small scales, calling it Augmented Lagrangian Perturbation Theory (\gls{ALPT}). It reads
\begin{equation}
\psi_\mathrm{ALPT}(\textbf{q},\tau) = (K_{R_\mathrm{s}} * \psi_\mathrm{2LPT})(\textbf{q},\tau) + \left[(1-K_{R_\mathrm{s}}) * \psi_\mathrm{SC}\right]\!(\textbf{q},\tau),
\end{equation}
or, in Lagrangian Fourier space,\footnote{We denote by $\boldsymbol{\upkappa}$ a Fourier mode on the Lagrangian grid, $\kappa$ its norm.}
\begin{equation}
\psi_\mathrm{ALPT}(\boldsymbol{\upkappa},\tau) = K_{R_\mathrm{s}}(\kappa) \, \psi_\mathrm{2LPT}(\boldsymbol{\upkappa},\tau) + \left[1-K_{R_\mathrm{s}}(\kappa)\right] \psi_\mathrm{SC}(\boldsymbol{\upkappa},\tau) .
\end{equation}
This method introduces a free parameter, $R_\mathrm{s}$, the width of the \gls{Gaussian kernel} used in the above equations to filter between large and small displacements, $K_{R_\mathrm{s}}(k) \propto \exp(-k^2/2 \times (R_\mathrm{s}/2\pi)^2 )$. In numerical experiments, \citet{Kitaura2013ALPT} empirically found that the range $R_\mathrm{s} = 4-5$\nbsp Mpc/$h$ yields the best density \gls{cross-correlation} to \gls{full gravity}.

\paragraph{Multi-scale spherical collapse evolution.\label{sec:MUSCLE}} \citet{Neyrinck2015} argued that the major deficiency in the SC approximation is its treatment of the \gls{void-in-cloud} process \citep[in the terminology originally introduced by][]{Sheth2004}, i.e. of small underdensities in larger-scale overdensities. Such regions should eventually collapse, which is not accounted for in SC. To overcome this problem, he proposes to use the SC prescription as a function of the initial \gls{density contrast} on multiple Gaussian-smoothed scales, thus including the \gls{void-in-cloud} process. The resulting parameter-free scheme, \glslink{muscle}{\textsc{muscle}} (MUltiscale Spherical-CoLlapse Evolution), mathematically reads
\begin{equation}
\psi_\mathrm{MUSCLE}(\textbf{q},\tau) = \left\{
\begin{array}{ll}
	3 \left( \sqrt{1-\dfrac{2}{3} \delta_\mathrm{L}(\textbf{q},\tau)} -1 \right) & \mathrm{if~} \delta_\mathrm{L} < 3/2 \mathrm{~and~} \forall R_\mathrm{s} \geq R_\mathrm{i}, K_{R_\mathrm{s}} * \delta_\mathrm{L} < 3/2, \\
    -3 & \mathrm{otherwise} ,
\end{array}
\right.
\end{equation}
where $R_\mathrm{i}$ is the resolution of the \glslink{initial conditions}{initial density field} $\delta(\textbf{q}$), and $K_{R_\mathrm{s}} * \delta_\mathrm{L}$ is the linearly extrapolated initial \gls{density field}, smoothed using a \gls{Gaussian kernel} of width $R_\mathrm{s}$. In practice, a finite number of scales $r >R_\mathrm{i}$ have to be tried (for example $r = 2^n R_\mathrm{i}$ for integers $0 \leq n \leq n_\mathrm{max}$ such that $2^{n_\mathrm{max}}R_\mathrm{i} \leq L$ and $2^{n_\mathrm{max}+1}R_\mathrm{i}>L$).

\citet{Neyrinck2015} checked that \glslink{muscle}{\textsc{muscle}} corrects the problems of \gls{SC} at large scales and outperforms the \gls{ZA} and \gls{2LPT} in terms of the density \gls{cross-correlation} to \gls{full gravity}.

\subsection{Non-linear evolution of {$\psi$} and generation of a vector part}

Beyond the approximations presented in the previous section, \citet{Chan2014} analyzed the \gls{non-linear evolution} of $\boldsymbol{\Psi}$ in \gls{full gravity}, splitting it into its \glslink{scalar part}{scalar} and \glslink{vector part}{vector parts} (the so-called ``\gls{Helmholtz decomposition}''):
\begin{equation}
\boldsymbol{\Psi}(\textbf{q}) = \nabla_\textbf{q} \phi(\textbf{q}) + \nabla_\textbf{q} \times \textbf{A}(\textbf{q}),
\end{equation}
with
\begin{eqnarray}
\Delta_\textbf{q} \phi & = & \nabla_\textbf{q} \cdot \boldsymbol{\Psi}(\textbf{q}),\\
\Delta_\textbf{q} \textbf{A}(\textbf{q}) & = & - \nabla_\textbf{q} \times \boldsymbol{\Psi}(\textbf{q}) .
\end{eqnarray}

Looking at \glslink{two-point correlation function}{two-point statistics} of $\boldsymbol{\Psi}$, he found that \gls{shell-crossing} leads to a suppression of small-scale power in the \gls{scalar part}, and, subdominantly, to the generation of a \glslink{vector part}{vector contribution}. Even at late-time and non-linear scales, the \gls{scalar part} of the \gls{displacement field} remains the dominant contribution. The rotational component is much smaller and does not have a coherent large-scale component. Therefore, the potential approximation is still good even when \gls{shell-crossing} is non-negligible.

However, as pointed out by \citet{Neyrinck2015}, even if we neglect the \glslink{vector part}{rotational component}, there is still a long way to go before we can perfectly predict $\psi$. Variants of \gls{LPT}, such as \gls{ALPT} (primarily motivated by the agreement in the scatter plot of final versus initial $\psi$ -- see figure 6 in \citealp{Neyrinck2013} and figure 10 in \citealp{Chan2014}) or the inclusion of a suppression factor in the \glslink{Lagrangian potential}{LPT displacement potential} (\citealp{Chan2014} -- designed for fitting the non-linear \gls{power spectrum} of $\boldsymbol{\Psi}$) extract information from \glslink{N-body simulation}{simulations} by taking the average of some statistics. Since \gls{shell-crossing} is a highly non-linear process, it may not be surprising that such approaches yield limited success compared to standard \gls{LPT} for some other statistics (such as the density \gls{power spectrum} or \gls{phase} accuracy). This suggests that a more detailed understanding and modeling of the small-scale physics beyond the simple phenomenological approach is required for improvement in $\psi$-based schemes, which would substantially increase the accuracy of \glslink{particle realization}{particle realizations}.

\section{Comparison of structure types in LPT and $N$-body dynamics}
\label{sec:Comparison of structure types in LPT and $N$-body dynamics}

\begin{figure*}
\begin{center}
\includegraphics[width=0.5\textwidth]{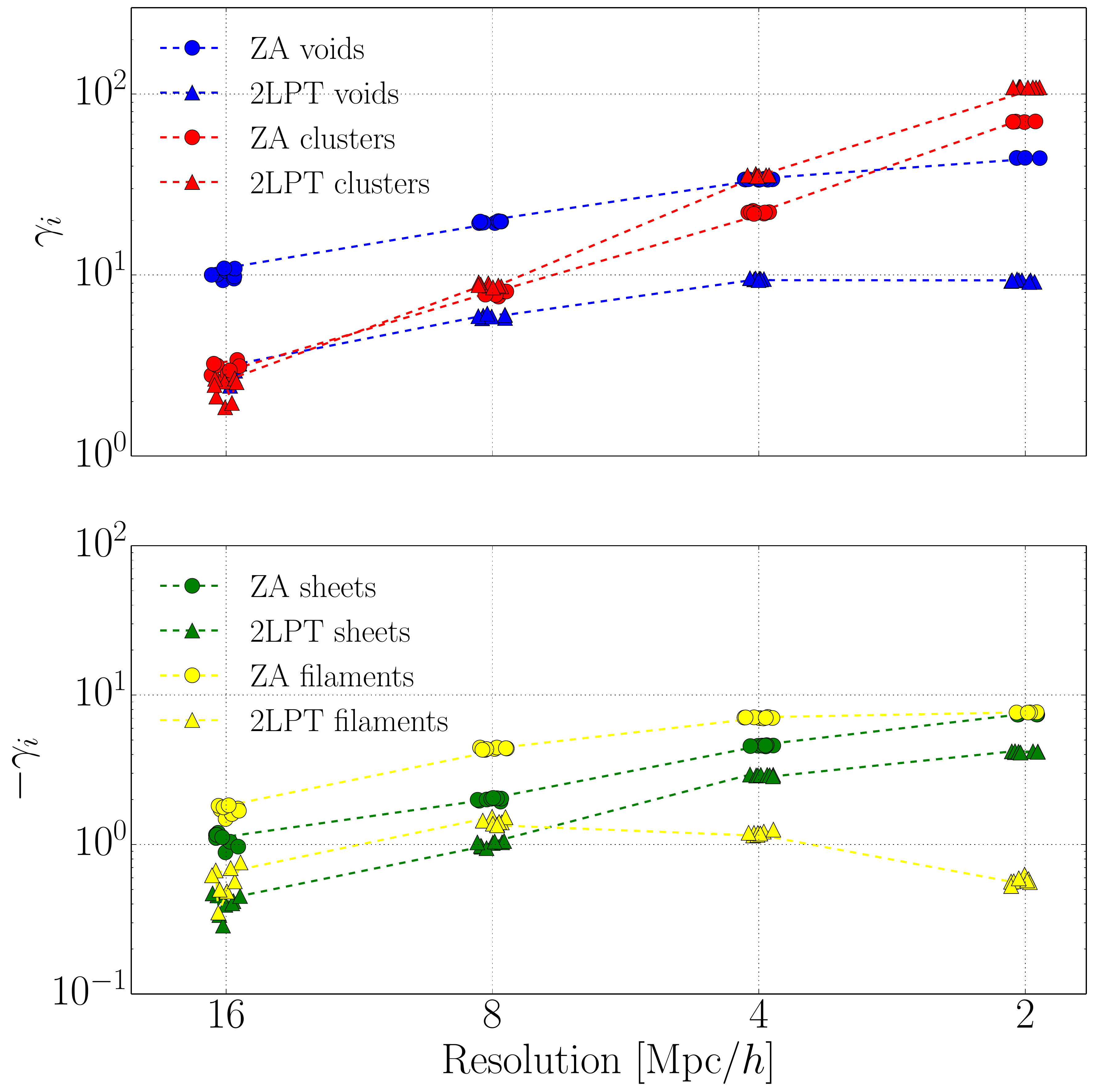}
\end{center}
\caption{Relative \glslink{VFF}{volume fraction} of \glslink{void}{voids}, \glslink{sheet}{sheets}, \glslink{filament}{filaments} and \glslink{cluster}{clusters} predicted by \gls{LPT}, compared to \glslink{N-body simulation}{$N$-body simulations}, as a function of the resolution used for the definition of the \glslink{density field}{density fields}. The points are sightly randomized on the $x$-axis for clarity. The \glslink{estimator}{estimators} $\gamma_i$ are defined by eq \eqref{eq:gamma}. Eight realizations of the \gls{ZA} (circles) and \gls{2LPT} (triangles) are compared to the corresponding \glslink{N-body simulation}{$N$-body realization}, for various resolutions. The volume fraction of incorrectly predicted structures in \gls{LPT} generally increases with increasing resolution.}
\label{fig:structure_type_analysis}
\end{figure*}

\draw{This section draws from section II.B. in \citet{Leclercq2013}.}

In this section, we perform a study of differences in \glslink{structure type}{structure types} in \glslink{density field}{density fields} predicted by \gls{LPT} and \glslink{N-body simulation}{$N$-body simulations}. We employ the \glslink{cosmic web classification}{web-type classification algorithm} proposed by \citet{Hahn2007a}, which relies on estimating the \glslink{eigenvalue}{eigenvalues} of the Hessian of the \gls{gravitational potential} (see section \ref{sec:apx-tweb}). This algorithm dissects the voxels into four different web types (\glslink{void}{voids}, \glslink{sheet}{sheets}, \glslink{filament}{filaments} and \glslink{cluster}{clusters}). Due to the different representations of the \gls{non-linear regime} of \gls{structure formation}, we expect differences in \glslink{structure type}{structure types} in \gls{LPT} and \glslink{N-body simulation}{$N$-body simulations}. In particular, overdense \glslink{cluster}{clusters} are objects in the strongly \gls{non-linear regime}, far beyond \gls{shell-crossing}, where predictions of \gls{LPT} fail, while underdense \glslink{void}{voids} are believed to be better apprehended \citep[e.g.][]{Bernardeau2002}.

As an indicator of the mismatch between the volume occupied by different \glslink{structure type}{structure types} in \gls{LPT} and \glslink{N-body simulation}{$N$-body dynamics}, we use the quantities $\gamma_i$ defined by
\begin{equation}
\label{eq:gamma}
\gamma_i \equiv \frac{N_{i}^{\mathrm{LPT}}-N_{i}^{\mathrm{Nbody}}}{N_{i}^{\mathrm{Nbody}}},
\end{equation}
where $i$ indexes one of the four \glslink{structure type}{structure types} ($\mathrm{T}_0=$ \gls{void}, $\mathrm{T}_1=$ \gls{sheet}, $\mathrm{T}_2=$ \gls{filament}, $\mathrm{T}_3=$ \gls{cluster}), and $N_{i}^{\mathrm{LPT}}$ and $N_{i}^{\mathrm{Nbody}}$ are the numbers of voxels flagged as belonging to a structure of type $\mathrm{T}_i$, in corresponding \gls{LPT} and in \glslink{N-body simulation}{$N$-body realizations}, respectively. At fixed resolution, corresponding \glslink{particle realization}{realizations} have the same total number of voxels $N_\mathrm{tot}$, so we also have
\begin{equation}
\gamma_i = \frac{\mathrm{VFF}_i^\mathrm{LPT}}{\mathrm{VFF}_i^\mathrm{Nbody}} -1 ,
\end{equation}
where the \glslink{VFF}{volume filling fraction} of structure type $\mathrm{T}_i$ is defined by $\mathrm{VFF}_i \equiv N_i/N_\mathrm{tot}$.

In figure \ref{fig:structure_type_analysis}, we plot $\gamma_i$ as a function of the voxel size used to define the \glslink{density field}{density fields}. $\gamma_i$ is positive for \glslink{cluster}{clusters} and \glslink{void}{voids}, and negative for \glslink{sheet}{sheets} and \glslink{filament}{filaments}, meaning that too large \gls{cluster} and \gls{void} regions are predicted in \gls{LPT}, at the detriment of \glslink{sheet}{sheets} and \glslink{filament}{filaments}. More specifically, \gls{LPT} predicts fuzzier \glslink{halo}{halos} than \glslink{N-body simulation}{$N$-body dynamics}, and incorrectly predicts the surroundings of \glslink{void}{voids} as part of them. This result indicates that even though \gls{LPT} and \glslink{N-body simulation}{$N$-body fields} look visually similar, there are crucial differences in the representation of \glslink{structure type}{structure types}. As demonstrated by figure \ref{fig:structure_type_analysis}, this mismatch increases with increasing resolution. This effect is of general interest when employing \gls{LPT} in \gls{LSS} data analysis.

%% file: Chapter3/Chapter3Content.tex
\part{Bayesian large-scale structure inference}
\label{part:II}

\chapter{Bayesian cosmostatistics}
\label{chap:stats}
\minitoc

\defcitealias{Jaynes2003}{Edwin Thompson}
\begin{flushright}
\begin{minipage}[c]{0.6\textwidth}
\rule{\columnwidth}{0.4pt}

``A previous acquaintance with probability and statistics is not necessary; indeed, a certain amount of innocence in this area may be desirable, because there will be less to unlearn.''\\
--- \citetalias{Jaynes2003} \citet{Jaynes2003}, \textit{Probability Theory: The Logic of Science}

\vspace{-5pt}\rule{\columnwidth}{0.4pt}
\end{minipage}
\end{flushright}

\abstract{\section*{Abstract}
In this chapter, essential concepts of \gls{Bayesian statistics} in the context of cosmological data analysis are presented. We discuss motivations for seeing probabilistic calculations as an \glslink{extended logic}{extension of ordinary logic} and justify the use of a \gls{prior} in an experimental learning process by referring to the ``\gls{no-free lunch theorem}''. This chapters also reviews \gls{parameter inference}, \gls{model comparison}, and contains a brief introduction to the subject of \glslink{MCMC}{Markov Chain Monte Carlo} methods.
}

\glslink{cosmostatistics}{}This chapter aims at introducing the necessary background in Bayesian \gls{probability theory} for presenting the {\borg} algorithm in chapter \ref{chap:BORG} and applications in the following chapters. A much more complete picture can be found in the reference book of \citet{Gelman2013}. For introductions to \gls{Bayesian statistics} in a cosmological context, see \citet{Hobson2010} and the reviews or lecture notes by \cite{Trotta2008,Heavens2009,Verde2010,Leclercq2014Varenna}.

This chapter is organized as follows. Section \ref{sec:Proba_intro} is a general introduction on \gls{plausible reasoning}. Basic concepts and definitions used in \gls{Bayesian statistics} are presented in section \ref{sec:Inverse problems and the mechanism of experimental learning}. In section \ref{sec:Bayesian data analysis problems}, we discuss standard statistical \gls{inference} problems. Finally, section \ref{sec:Markov Chain Monte Carlo techniques for parameter inference} is summarizes the basics of \glslink{MCMC}{Markov Chain Monte Carlo} methods.

\section{Introduction: plausible reasoning}
\label{sec:Proba_intro}

When discussing statistical data analysis, two different points of view are traditionally reviewed and opposed: the \glslink{frequentist statistics}{frequentist} \citep[see e.g.][]{Kendall1968} and the \glslink{Bayesian statistics}{Bayesian approaches}. In this author's experience, arguments for or against each of them are generally on the level of a philosophical or ideological position, at least among cosmologists in 2015. Before criticizing this controversy, somewhat dated to the 20th century, and stating that more recent scientific work suppresses the need to appeal to such arguments, we report the most common statements encountered. 

\subsection{On the definition of probability}
\label{sec:On the definition of probability}

\glslink{frequentist statistics}{Frequentist} and \gls{Bayesian statistics} differ in the epistemological interpretation of probability and their consequences for \glslink{hypothesis testing}{testing hypotheses} and \glslink{model comparison}{comparing models}. First and foremost, the methods differ on the understanding of the concept of the probability $\p(A)$ of an event $A$. In \gls{frequentist statistics}, one defines the probability $\p(A)$ as the relative frequency with which the event $A$ occurs in repeated experiments, i.e. the number of times the event occurs over the total number of trials, in the limit of a infinite series of equiprobable repetitions. As forcefully argued for example by \citet{Trotta2008}, this \gls{definition of probability} has several shortcomings. Besides being useless in real life (as it assumes an infinite repetition of experiments with nominally identical test conditions, requirement that is never met in most practical cases), it cannot handle unrepeatable situations, which have a particular importance in cosmology, as we have exactly one \gls{sample} of the Universe. More importantly, this definition is surprisingly circular, in the sense that it assumes that repeated trials are equiprobable, whereas it is the very notion of probability that is being defined in the first place.

On the other hand, in \gls{Bayesian statistics}, the probability $\p(A)$ represents the degree of belief that any reasonable person (or machine) shall attribute to the occurrence of event $A$ under consideration of all available information. This definition implies that in Bayesian theory, probabilities are used to \glslink{uncertainty quantification}{quantify uncertainties} independently of their origin, and therefore applies to any event. In other words, probabilities represent a state of knowledge in presence of partial information. This is the intuitive concept of probability as introduced by Laplace, Bayes, Bernoulli, Gau\ss, Metropolis, Jeffreys, etc. \citep[see][]{Jaynes2003}.

\subsection{On parameter determination}

Translated to the \glslink{parameter inference}{measurement of a parameter} in an experiment, the \glslink{definition of probability}{definitions of probabilities} given in the previous section yield differences in the questions addressed by \glslink{frequentist statistics}{frequentist} and Bayesian statistical analyses.

In the \glslink{frequentist statistics}{frequentist} point of view, statements are of the form: ``the measured value $x$ occurs with probability $\p(x)$ if the measurand $X$ has the true value $\mathpzc{X}$''. This means that the only questions that can be answered are of the form: ``given the true value $\mathpzc{X}$ of the measurand $X$, what is the probability distribution of the measured values $x$?''. It also implies that statistical analyses are about building \textit{\glslink{estimator}{estimators}}, $\hat{X}$, of the truth, $\mathpzc{X}$.

In contrast, \gls{Bayesian statistics} allows statements of the form: ``given the measured value $x$, the measurand $X$ has the true value $\mathpzc{X}$ with probability $\q$''. Therefore, one can also answer the question: ``given the observed measured value $x$, what is the probability that the true value of $X$ is $\mathpzc{X}$?'', which arguably is the only natural thing to demand from data analysis. For this reason, \gls{Bayesian statistics} offers a principled approach to the question underlying every measurement problem, of how to \textit{\glslink{inference}{infer}} the true value of the measurand given all available information, including observations.

In summary, in the context of parameter determination, the fundamental difference between the two approaches is that \glslink{frequentist statistics}{frequentist} statistics assumes the measurement to be uncertain and the measurand known, while \gls{Bayesian statistics} assumes the observation to be known and the measurand uncertain. Similar considerations can be formulated regarding the problems of \gls{hypothesis testing} and \gls{model comparison}.

\subsection{Probability theory as extended logic}
\label{sec:Probability theory as extended logic}

As outlined in the seminal work of \citet{Cox1946}, popularized and generalized by Jaynes \citep[in particular in his inspirational posthumous book,][]{Jaynes2003},\footnote{At this point, the influence of \citet{Shannon1948} and \citet{Polya1954a,Polya1954b} should also be emphasized.} neither the \glslink{Bayesian statistics}{Bayesian} nor the \glslink{frequentist statistics}{frequentist approach} is universally applicable. It is possible to adopt a more general viewpoint that can simply be referred to as ``\gls{probability theory}'', which encompasses both approaches. This framework automatically includes all \glslink{Bayesian statistics}{Bayesian} and \glslink{frequentist statistics}{frequentist} calculations, but also contains concepts that do not fit into either category (for example, the principle of maximum entropy, which can be applied in the absence of a particular model, when very little is known beyond the raw data).

In the author's view, this approach is a breakthrough that remains shockingly unknown in astrophysics. As we believe that a conceptual understanding of these concepts are of interest for the purpose of this thesis, we now qualitatively describe the salient features of this way of thinking.

The \gls{Cox-Jaynes theorem} (1946) states that there is only a single set of rules for doing \gls{plausible reasoning} which is consistent with a set of axioms that is in qualitative correspondence with common sense. These axioms, or \glslink{Cox's desiderata}{desiderata}, are \citep[][section 1.7]{Jaynes2003}:
\begin{enumerate}
\item \textit{Degrees of \gls{plausibility} are represented by real numbers.} We denote by $w(A|B)$ the real number assigned to the \gls{plausibility} of some proposition $A$, given some other proposition $B$. 
\item \textit{\glslink{plausible reasoning}{Plausible reasoning} qualitatively agrees with human common sense with respect to the ``direction'' in which reasoning is to go.} Formally, we introduce a continuity assumption: $w(A)$ changes only infinitesimally if $A$ changes infinitesimally. In addition, if some old information $C$ gets updated to $C'$ in such a way that the \gls{plausibility} of $A$ is increased, but the \gls{plausibility} of $A$ given $B$ is unchanged, i.e. $w(A|C') > w(A|C)$ and $w(B|AC')=w(B|AC)$, we demand that the \gls{plausibility} that $A$ is false decrease, i.e. $w(\bar{A}|C')<w(\bar{A}|C)$, and that the \gls{plausibility} of $A$ and $B$ can only increase, i.e. $w(AB|C') \geq w(AB|C)$.
\item \textit{\glslink{plausible reasoning}{Plausible reasoning} is performed consistently.} This is requiring the three common colloquial meanings of the word ``consistent'':
\begin{enumerate}
\item \textit{If a conclusion can be reached in more than one way, then every possible way must lead to the same result.}
\item \textit{Consistent \gls{plausible reasoning} always takes into account all of the evidence it has relevant to a question. It does not arbitrarily ignore some of the available information, basing its conclusion on what remains. In other words, it is completely non-ideological.}
\item \textit{Equivalent states of knowledge (up to the labeling of propositions) are represented by equal \gls{plausibility} assignments.}
\end{enumerate}
\end{enumerate}
The \gls{Cox-Jaynes theorem} demonstrates that the only consistent system to manipulate numerical ``\glslink{plausibility}{plausibilities}'' that respect these rules is isomorphic to \gls{probability theory},\footnote{Formally, the theorem states that there exists an isomorphism $f$ such that for any two propositions $A$, $B$, we have $f \circ w(A|B) = \p(A|B)$.} and shows that this system consistently \glslink{extended logic}{extends} the two-valued Boolean algebra $\left\lbrace 0,1 \right\rbrace$ to the continuum $\left[ 0,1 \right]$. This paradigm therefore introduces a ``\glslink{extended logic}{logical}'' interpretation of probabilities that can be deduced without any reference to frequencies.

In this perspective, statistical techniques that use \gls{Bayes' theorem} or the \gls{maximum-entropy} inference rule are fully as valid as any based on the \glslink{frequentist statistics}{frequentist} interpretation of probability. In fact, they are the \textit{unique} consistent generalization of logical deduction in the presence of uncertainty. As demonstrated by Jaynes, their introduction enables to broaden the scope of \gls{probability theory} so that it includes various seemingly unrelated fields, such as \glslink{information theory}{communication theory} of the \gls{maximum-entropy} interpretation of thermodynamics. They also provides a rational basis to the mechanism of logical induction and therefore to \gls{machine learning}.

\section{Inverse problems and the mechanism of experimental learning}
\label{sec:Inverse problems and the mechanism of experimental learning}

\draw{This section draws from section 3 in \citet{Leclercq2014Varenna}.}

The ``\gls{plausible reasoning}'' framework described in section \ref{sec:Proba_intro} can be formulated mathematically by introducing the concept of \glslink{conditional pdf}{conditional probability} $\p(A|B)$, which describes the probability that event $A$ will occur given whatever information $B$ is given on the right side of the vertical conditioning bar. To \glslink{conditional pdf}{conditional probabilities} applies the following famous identity, which allows to go from \gls{forward modeling} to the \gls{inverse problem}, by noting that if one knows how $x$ arises from $y$, then one can use $x$ to constrain $y$:
\begin{equation}
\p(y|x)\p(x) = \p(x|y)\p(y) = \p(x,y).
\end{equation}
This observation forms the basis of \gls{Bayesian statistics}.

\subsection{What is Bayesian analysis?}

\glslink{Bayesian statistics}{Bayesian analysis} is a general method for updating the probability estimate for a theory in light of new data. It is based on \gls{Bayes' theorem},
\begin{equation}
\label{eq:Bayes}
\p(\theta	|d) = \frac{\p(d|\theta)\p(\theta)}{\p(d)} .
\end{equation}
In the previous formula, $\theta$ represents the set of model parameters for a particular theory and $d$ the \gls{data} (before it is known), written as a vector. Therefore, 
\begin{itemize}
\item $\p(d|\theta)$ is the probability of the data \textit{before it is known}, given the theory. It is called the \textit{\gls{likelihood}};
\item $\p(\theta)$ is the probability of the theory in the absence of data. It is called the \gls{prior} probability distribution function or simply the \textit{\gls{prior}};
\item $\p(\theta|d)$ is the probability of the theory	after the data is known. It is called the \gls{posterior} probability distribution function or simply the \textit{\gls{posterior}};
\item $\p(d)$ is the probability of the data \textit{before it is known}, without any assumption about the theory. It is called the \textit{\gls{evidence}}.
\end{itemize}

A simple way to summarize \glslink{Bayesian statistics}{Bayesian analysis} can be formulated by the following:
\begin{center}
\textit{Whatever is uncertain gets a \gls{pdf}.}
\end{center}
This statement can be a little disturbing at first (e.g. the value of $\Omega_\mathrm{m}$ is a constant of nature, certainly not a random number of an experiment). What it means is that in \gls{Bayesian statistics}, \glslink{pdf}{pdfs} are used to \glslink{uncertainty quantification}{quantify uncertainty} of all kinds, not just what is usually referred to as ``randomness'' in the outcome of an experiment. In other words, the \gls{pdf} for an uncertain parameter can be thought as a ``belief distribution function'', quantifying the degree of truth that one attributes to the possible values for some parameter (see the discussion in section \ref{sec:On the definition of probability}). Certainty can be represented by a \glslink{Dirac delta distribution}{Dirac distribution}, e.g. if the data determine the parameters completely.

The inputs of a \glslink{Bayesian statistics}{Bayesian analysis} are of two sorts:
\begin{itemize}
\item the \textit{\gls{prior}}: it includes modeling assumptions, both theoretical and experimental. Specifying a \gls{prior} is a systematic way of quantifying what one assumes true about a theory before looking at the data.
\item the \textit{\gls{data}}: in cosmology, these can include the temperature in pixels of a \gls{CMB} map, galaxy \glslink{redshift}{redshifts}, \glslink{photometric redshift}{photometric redshifts} \glslink{pdf}{pdfs}, etc. Details of the survey specifications have also to be accounted for at this point: \gls{noise}, \gls{mask}, \gls{survey geometry}, \gls{selection effects}, \glslink{bias}{biases}, etc.
\end{itemize}

A key point is that the output of a \glslink{Bayesian statistics}{Bayesian analysis} is a \gls{pdf}, the \textit{\gls{posterior} density}. Therefore, contrary to \gls{frequentist statistics}, the output of the analysis is not an \gls{estimator} for the parameters. The word ``\gls{estimator}'' has a precise meaning in \gls{frequentist statistics}: it is a function of the \gls{data} which returns a number that is meant to be close to the parameter it is designed to estimate; or the left and right ends of a confidence interval, etc. The outcome of a \glslink{Bayesian statistics}{Bayesian analysis} is the \gls{posterior} \gls{pdf}, a \gls{pdf} whose values give a quantitative measure of the relative degree of rational belief in different parameter values given the combination of \gls{prior} information and the \gls{data}. 

\subsection{Prior choice}
\label{sec:Prior choice}

The \gls{prior choice} is a key ingredient of \gls{Bayesian statistics}. It is sometimes considered problematic, since there is no unique prescription for selecting the \gls{prior}. Here we argue that \gls{prior} specification is not a limitation of \gls{Bayesian statistics} and does not undermine objectivity as sometimes misstated.

The guiding principle is that there can be no \gls{inference} without assumptions, that there does not exist an ``external truth'', but that science is building predictive models in certain axiomatic frameworks. In this regard, stating a \gls{prior} in Bayesian \gls{probability theory} becomes a systematic way to quantify one's assumptions and state of knowledge about the problem in question before the data is examined. While it is true that such probability assignment does not describe something that could be measured in a physical experiment, it is completely objective in the sense that it is independent of the ``personal feelings'' of the user. Anyone who has the same information, but comes to a different conclusion, is necessarily violating one of \gls{Cox's desiderata} (see the discussion in section \ref{sec:Probability theory as extended logic}).

\glslink{Bayes' theorem}{Bayes' theorem} gives an unequivocal procedure to update even different degrees of beliefs. As long as the \gls{prior} has a support that is non-zero in regions where the \gls{likelihood} is large (\gls{Cromwell's rule}), the repeated application of the theorem will converge to a unique \gls{posterior} distribution (\gls{Bernstein-von Mises theorem}). Generally, objectivity is assured in \gls{Bayesian statistics} by the fact that, if the \gls{likelihood} is more informative than the \gls{prior}, the \gls{posterior} converges to a common function.

Specifying \glslink{prior}{priors} exposes assumptions to falsification and scientific criticism. This is a positive feature of Bayesian \gls{probability theory}, because \glslink{frequentist statistics}{frequentists} also have to make assumptions that may be more difficult to find within the analysis. An important theorem \citep{Wolpert1997} states that there is ``\glslink{no-free lunch theorem}{no-free lunch}'' for optimization problems: when searching for the local extremum of a target function (the \gls{likelihood} in our case) in a finite space, the average performance of algorithms (that do not resample points) across all possible problems is identical. An important implication is that no universally good algorithm exists \citep{Ho2002}; \gls{prior} information should always be used to match procedures to problems.

In many situations, domain knowledge is highly relevant and should be included in the analysis. For example, when trying to estimate a mass $m$ from some data, one should certainly enforce it to be a positive quantity by setting a \gls{prior} such that $\p(m)~=~0$ for $m~<~0$. \glslink{frequentist statistics}{Frequentist techniques} based on the \gls{likelihood} can give estimates and confidence intervals that include negative values. Taken at face value, this result is meaningless, unless special care is taken (e.g. the so-called ``\gls{constrained likelihood}'' methods). The use of \gls{Bayes' theorem} ensures that meaningless results are excluded from the beginning and that one knows how to place bets on values of the parameter given the actual data set at hand.

As discussed in the \hyperref[chap:intro]{introduction}, in cosmology, the current state-of-the-art is that previous data (COBE, WMAP, Planck, \gls{SDSS}~etc.) allowed to establish an extremely solid theoretical footing: the so-called {\LCDM} model. Even when trying to detect deviations from this model in the most recent data, it is absolutely well-founded to use it as \gls{prior} knowledge about the physical behaviour of the Universe. Therefore, using less informative \glslink{prior}{priors} would be refusing to ``climb on the shoulder of giants''.

It can happen that the data are not informative enough to override the \gls{prior} (e.g. for \glslink{sparsity}{sparsely sampled} \gls{data} or very \gls{high-dimensional parameter space}), in which case care must be given in assessing how much of the final (first level, see section \ref{sec:First level analysis: Bayesian parameter inference}) inference depends on the \gls{prior choice}. A good way to perform such a check is to simulate data using the \gls{posterior} and see if it agrees with the observed data. This can be thought of as ``calculating doubt'' \citep{Starkman2008,March2011} to quantify the degree of belief in a model given observational data in the absence of explicit alternative models. Note that even in the case where the inference strongly depends on \gls{prior} knowledge, information has been gained on the constraining power (or lack thereof) of the data.

For \glslink{model comparison}{model selection} questions (second level analysis, see section \ref{sec:Second level analysis: Bayesian model comparison}), the impact of the \gls{prior choice} is much stronger, since it is precisely the available \gls{prior volume} that matters in determining the penalty that more complex models should incur. Hence, care should be taken in assessing how much the outcome would change for physically reasonable changes in the \gls{prior}.

There exists a vast literature about quantitative prescriptions for \gls{prior choice} that we cannot summarize here. An important topic concerns the determination of ``ignorance priors'' or ``\gls{Jeffreys' priors}'': a systematic way to \glslink{uncertainty quantification}{quantify a maximum level of uncertainty} and to reflect a state of indifference with respect to symmetries of the problem considered. While the ignorance prior is unphysical (nothing is ever completely uncertain) it can be viewed as a convenient approximation to the problem of carefully constructing an accurate representation of weak \gls{prior} information, which can be very challenging -- especially in \glslink{high-dimensional parameter space}{high dimensional parameter spaces}. 

For example, it can be shown that, if one is wholly uncertain about the position of the \gls{pdf}, a ``\gls{flat prior}'' should be chosen. In this case, the \gls{prior} is taken to be constant (within some minimum and maximum value of the parameters so as to be \glslink{proper prior}{proper}, i.e. normalizable to unity). In this fashion, equal probability is assigned to equal states of knowledge. However, note that a flat \gls{prior} on a parameter $\theta$ does not necessarily correspond to a flat \gls{prior} on a non-linear function of that parameter, $\varphi(\theta)$. Since $\p(\varphi)~=~\p(\theta)~\times~|\mathrm{d}\theta/\mathrm{d}\varphi|$, a non-informative \glslink{flat prior}{(flat) prior} on $\theta$ can be strongly informative about $\varphi$. Analogously, if one is entirely uncertain about the width of the \gls{pdf}, i.e. about the scale of the inferred quantity $\theta$, it can be shown that the appropriate \gls{prior} is $\p(\theta) \propto 1/\theta$, which gives the same probability in logarithmic bins, i.e. the same weight to all orders of magnitude. 

\section{Bayesian data analysis problems}
\label{sec:Bayesian data analysis problems}

\draw{This section draws from section 3 in \citet{Leclercq2014Varenna}.}

Bayesian data analysis problems can be typically classified as: \gls{parameter inference}, \gls{model comparison}, \gls{hypothesis testing}. For example, cosmological questions of these three types, related to the large-scale structure, would be respectively
\begin{itemize}
\item What is the value of $w$, the \gls{equation of state} of \gls{dark energy}?
\item Is \gls{structure formation} driven by \gls{general relativity} or by massive gravity?
\item Are large-scale structure observations consistent with the hypothesis of a spatially flat universe?
\end{itemize}

In this section; we describe the methodology for questions of the first two types. \glslink{hypothesis testing}{Hypothesis testing}, i.e. \gls{inference} within an uncertain model, in the absence of an explicit alternative, can be treated in a similar manner.

\subsection{First level analysis: Bayesian parameter inference}
\label{sec:First level analysis: Bayesian parameter inference}

The general problem of Bayesian \gls{parameter inference} can be stated as follows. Given a physical model $\mathcal{M}$,\footnote{In this section, we make explicit the choice of a model $\mathcal{M}$ by writing it on the right-hand side of the conditioning symbol of all \glslink{pdf}{pdfs}.} a set of hypotheses is specified in the form of a vector of parameters, $\theta$. Together with the model, \glslink{prior}{priors} for each parameter must be specified: $\p(\theta|\mathcal{M})$. The next step is to construct the \gls{likelihood} function for the measurement, with a probabilistic, generative model of the data: $\p(d|\theta,\mathcal{M})$. The \gls{likelihood} reflects \glslink{data model}{how the data are obtained}: for example, a measurement with Gaussian \gls{noise} will be represented by a normal distribution.

Once the \gls{prior} is specified and the data is incorporated in the \gls{likelihood} function, one immediately obtains the \gls{posterior} distribution for the model parameters, integrating all the information known to date, by using \gls{Bayes' theorem} (equation \eqref{eq:Bayes}):
\begin{equation}
\label{eq:First_level_inference}
\p(\theta|d,\mathcal{M}) \propto \p(d|\theta,\mathcal{M}) \p(\theta|\mathcal{M}) .
\end{equation}
Note that the normalizing constant $\p(d|\mathcal{M})$ (the Bayesian \gls{evidence}) is irrelevant for \gls{parameter inference} (but fundamental for \gls{model comparison}, see section \ref{sec:Second level analysis: Bayesian model comparison}).

Usually, the set of parameters $\theta$ can be divided in some physically interesting quantities $\varphi$ and a set of \gls{nuisance parameters} $\psi$. The \gls{posterior} obtained by equation \eqref{eq:First_level_inference} is the joint \gls{posterior} for $\theta = (\varphi,\psi)$. The \glslink{marginal pdf}{marginal} \gls{posterior} for the parameters of interest is written as (marginalizing over the \gls{nuisance parameters})
\begin{equation}
\p(\varphi|d,\mathcal{M}) \propto \int \p(d|\varphi, \psi, \mathcal{M}) \p(\varphi,\psi|\mathcal{M}) \, \mathrm{d}\psi .
\end{equation}
This \gls{pdf} is the final inference on $\varphi$ from the joint \gls{posterior}. The following step, to apprehend and exploit this information, is to explore the \gls{posterior}. It is the subject of the next section.

\subsection{Exploration of the posterior}
\label{sec:Exploration of the posterior}

\begin{figure}
\begin{center}
\includegraphics[width=0.3\textwidth]{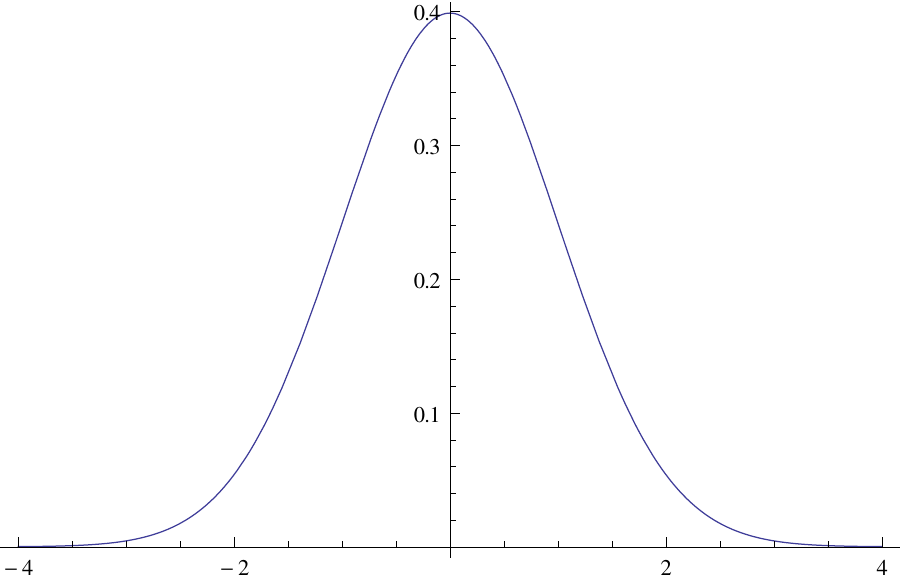} \hspace{0.25cm}\includegraphics[width=0.25\textwidth]{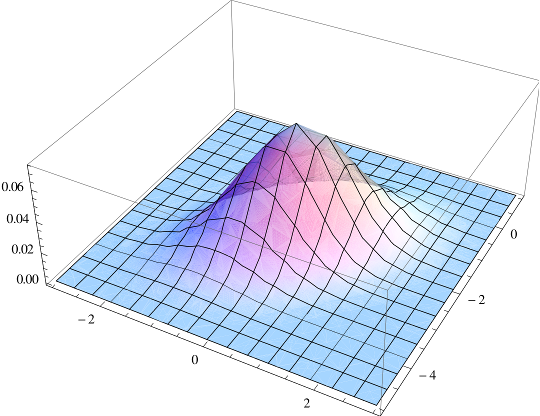}\hspace{0.5cm} \includegraphics[width=0.25\textwidth]{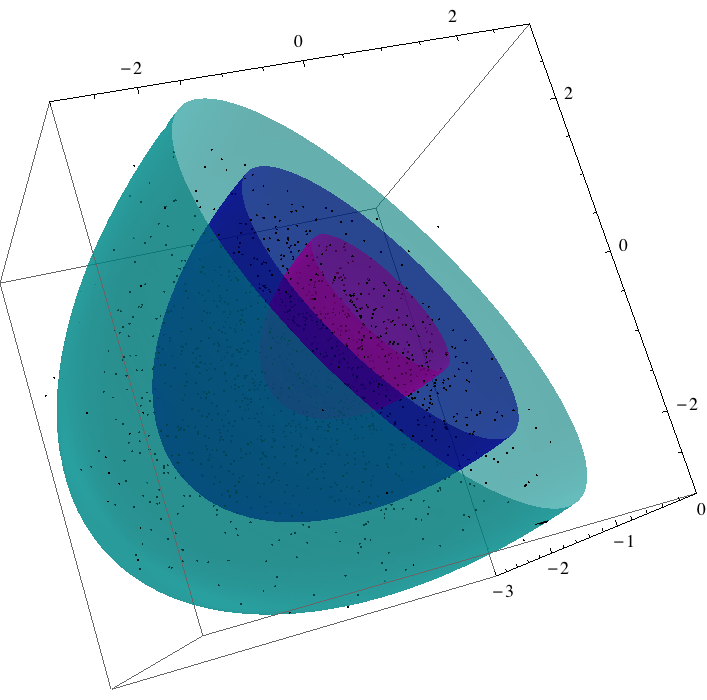}
\end{center}
\caption{Example visualizations of \gls{posterior} densities in low-dimensional parameter spaces (from left to right: one, two and three).\label{fig:posterior_visualization}}
\end{figure}

The result of \gls{parameter inference} is contained in the \gls{posterior} \gls{pdf}, which is the actual output of the statistical analysis. Since this \gls{pdf} cannot always be easily represented, convenient \glslink{exploration of the posterior}{communication} of the \gls{posterior} information can take different forms:

\begin{itemize}
\item a direct visualization, which is only possible if the parameter space has sufficiently small dimension (see figure \ref{fig:posterior_visualization}).
\item the computation of statistical summaries of the \gls{posterior}, e.g. the mean, the median, or the mode of the distribution of each parameter, \glslink{marginal pdf}{marginalizing} over all others, its standard deviation; the \glslink{posterior mean}{means} and \glslink{posterior standard deviation}{covariance} matrices of some groups of parameters, etc. It is also common to present the inference by plotting two-dimensional subsets of parameters, with the other components \glslink{marginal pdf}{marginalized} over (this is especially useful when the \gls{posterior} is multi-modal or with heavy tails).
\end{itemize}

For typical problems in cosmology, the \glslink{exploration of the posterior}{exploration of a posterior} density meets practical challenges, depending on the \glslink{high-dimensional parameter space}{dimension} $D$ of the parameter space. Due to the computational time requirements, direct integration and mapping of the \gls{posterior} density is almost never a smart idea, except for $D<4$. Besides, computing statistical summaries by \glslink{marginal pdf}{marginalization} means integrating out the other parameters. This is rarely possible analytically (\glslink{grf}{Gaussian random fields} being one notable exception), and even numerical direct integration is basically hopeless for $D>5$.

In this thesis, we will be looking at cases where $D$ is of the order of $10^7$: the density in each voxel of the map to infer is a parameter of the analysis. This means that direct evaluation of the \gls{posterior} is impossible and one has to rely on a numerical approximation: \gls{sampling} the \gls{posterior} distribution.

\begin{figure}
\begin{center}
\includegraphics[width=0.6\textwidth]{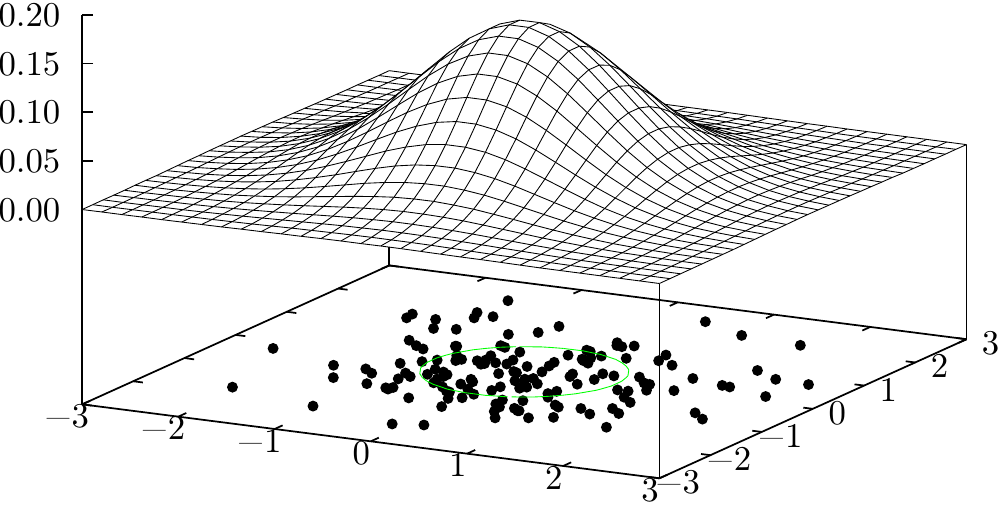}
\end{center}
\caption{Example of a sampled representation of a \gls{posterior} distribution in two dimensions. A set of \glslink{sample}{samples} is constructed in such a way that at any point, the \gls{posterior} probability is proportional to the local density of \glslink{sample}{samples} in parameter space.\label{fig:posterior_sample}}
\end{figure}

The idea is to approximate the \gls{posterior} by a set of \glslink{sample}{samples} drawn from the real \gls{posterior} distribution. In this fashion, one replaces the real \gls{posterior} distribution, $\p(\theta|d)$, by the sum of $N$ \glslink{Dirac delta distribution}{Dirac delta distributions}, $\p_N(\theta|d)$:
\begin{equation}
\p(\theta|d) \approx \p_N(\theta|d) = \frac{1}{N} \sum_{i=1}^{N} \updelta_{\mathrm{D}}(\theta - \theta_i) .
\end{equation}
A sampled representation of the \gls{posterior} is constructed in such a way that at any point, the \gls{posterior} probability is proportional to the local density of \glslink{sample}{samples} in parameter space (see figure \ref{fig:posterior_sample}).

An intuitive way to think about these \glslink{sample}{samples} is to consider each of them as a ``possible version of the truth''. The variation between different \glslink{sample}{samples} \glslink{uncertainty quantification}{quantifies the uncertainty}. At this point, it is worth stressing again that an advantage of Bayesian approach is that it deals with uncertainty independently of its origin, i.e. there is no fundamental distinction between ``\gls{statistical uncertainty}'' coming from the stochastic nature of the experiment and ``\gls{systematic uncertainty}'', deriving from deterministic effects that are only partially known. 

The advantage of a \gls{sampling} approach is that \glslink{marginal pdf}{marginalization} over some parameters becomes trivial: one just has to histogram. Specifically, it is sufficient to count the number of \glslink{sample}{samples} falling within different bins of some subset of parameters, simply ignoring the values of the others parameters. Integration to get means and variances is also much simpler, since the problem is limited to the computation of discrete sums. More generally, the expectation value of any function of the parameters, $f(\theta)$ is
\begin{equation}
\left\langle f(\theta) \right\rangle = \int f(\theta) \p(\theta) \mathrm{d}\theta \approx \frac{1}{N} \sum_{i=1}^{N} f(\theta_i) .
\end{equation}
We make an extensive use of this last property in part \ref{part:IV} of this thesis, when exploiting the {\borg} \gls{SDSS} analysis for \gls{cosmic web classification}.

How can one get a sampled representation of the \gls{posterior}? The ideal case would be to have an infinitely powerful computer. Then, a na\"ive but straightforward \gls{sampling} algorithm would be the following: simulate data from the generative model (draw $\theta$ from the \gls{prior}, then data from the \gls{likelihood} knowing $\theta$) and check that the real data agree with the simulated data. If it is the case, keep $\theta$ as one \gls{sample}, otherwise try again. This is correct in principle, but hugely inefficient, particularly in \glslink{high-dimensional parameter space}{high dimensions} where it can become prohibitively expensive to evaluate the \gls{posterior} \gls{pdf}. Fortunately, a battery of powerful methods exists for approximating and \gls{sampling} from probability distributions. Interestingly, \gls{sampling} algorithms exist that do not evaluate the \gls{posterior} \gls{pdf} (except perhaps occasionally, to maintain high numerical precision).

One class of approaches is Approximate Bayesian Computation (\gls{ABC}) sometimes also known as \glslink{likelihood-free methods}{``likelihood-free'' methods} \citetext{see \citealp{Marin2011} for an overview, or \citealp{Cameron2012,Weyant2013,Lin2015} for applications to astrophysics}. The general principle is similar to the na\"ive approach described above, but \gls{ABC} makes it practical by using an approximate forward model, the outcomes $\tilde{d}$ of which are compared with the observed data $d$. The candidate \gls{sample} $\tilde{d}$ is accepted with tolerance $\varepsilon >0$ if $\rho(\tilde{d},d) \leq \varepsilon$, where the distance measure $\rho$ determines the allowed level of discrepancy between $\tilde{d}$ and $d$ based on a given metric.

Another important class of standard techniques to \glslink{sampling}{sample} the \gls{posterior} is to use \glslink{MCMC}{Markov Chain Monte Carlo}, which is the subject of section \ref{sec:Markov Chain Monte Carlo techniques for parameter inference}.

\subsection{Second level analysis: Bayesian model comparison}
\label{sec:Second level analysis: Bayesian model comparison}

In the case where there are several competing theoretical models, second level inference (or Bayesian \gls{model comparison}) provides a systematic way of evaluating their relative probability in light of the data and any \gls{prior} information available. It does not replace \gls{parameter inference}, but rather extends the assessment of hypotheses to the space of theoretical models.

This allows quantitatively to address everyday questions in cosmology -- Is the Universe flat or should one allow a non-zero curvature parameter? Are the primordial perturbations Gaussian or \glslink{non-Gaussianity}{non-Gaussian}? Are there \glslink{isocurvature perturbations}{isocurvature modes}? Are the perturbations strictly scale-invariant ($n_\mathrm{s} = 1$) or should the spectrum be allowed to deviate from scale-invariance? Is there evidence for a deviation from \gls{general relativity}? Is the \gls{equation of state} of \gls{dark energy} equal to $-1$? 

In many of the situations above, Bayesian \gls{model comparison} offers a way of balancing complexity and goodness of fit: it is obvious that a model with more free parameters will always fit the data better, but it should also be ``penalized'' for being more complex and hence, less predictive. The notion of predictiveness really is central to Bayesian \gls{model comparison} in a very specific way: the \gls{evidence} is actually the prior predictive \gls{pdf}, the \gls{pdf} over all data sets predicted for the experiment before data are taken. Since predictiveness is a criterion for good science everyone can agree on, it is only natural to compare models based on how well they predicted the data set before it was obtained. This criterion arises automatically in the \glslink{Bayesian statistics}{Bayesian framework}.

The guiding scientific principle is known as \gls{Occam's razor}: the simplest model compatible with the available information ought to be preferred. We now understand this principle as a consequence of using predictiveness as the criterion. A model that is so vague (e.g. has so many parameters) that it can predict a large range of possible outcomes will predict any data set with smaller probability than a model that is highly specific and therefore has to commit to predicting only a small range of possible data sets. It is clear that the specific model should be preferred if the data falls within the narrow range of its prediction. Conversely we default to the broader more general model only if the data are incompatible with the specific model. Therefore, Bayesian \gls{model comparison} offers formal statistical grounds for selecting models based on an evaluation whether the data truly favor the extra complexity of one model compared to another. 

Contrary to \glslink{frequentist statistics}{frequentists} goodness-of-fit tests, second level inference always requires an alternative explanation for comparison (finding that the data are unlikely within a theory does not mean that the theory itself is improbable, unless compared with an alternative). The \gls{prior} specification is crucial for \glslink{model comparison}{model selection} issues: since it is the range of values that parameters can take that controls the sharpness of \gls{Occam's razor}, the \gls{prior} should exactly reflect the available parameter space under the model before obtaining the data.

The evaluation of model $\mathcal{M}$'s performance given the data is quantified by $\p(\mathcal{M}|d)$. Using \gls{Bayes' theorem} to invert the order of conditioning, we see that it is proportional to the product of the \gls{prior} probability for the model itself, $\p(\mathcal{M})$, and of the Bayesian \gls{evidence} already encountered in first level inference, $\p(d|\mathcal{M})$:
\begin{equation}
\p(\mathcal{M}|d) \propto \p(\mathcal{M})\, \p(d|\mathcal{M}) .
\end{equation}
Usually, \gls{prior} probabilities for the models are taken as all equal to $1/N_\mathrm{m}$ if one considers $N_\mathrm{m}$ different models (this choice is said to be \textit{\glslink{non-committal prior}{non-committal}}). When comparing two competing models denoted by $\mathcal{M}_1$ and $\mathcal{M}_2$, one is interested in the ratio of the \gls{posterior} probabilities, or \textit{\gls{posterior odds}}, given by
\begin{equation}
\mathcal{P}_{12} \equiv \frac{\p(\mathcal{M}_1|d)}{\p(\mathcal{M}_2|d)} = \frac{\p(\mathcal{M}_1)}{\p(\mathcal{M}_2)} \frac{\p(d|\mathcal{M}_1)}{\p(d|\mathcal{M}_2)} .
\end{equation}
With \glslink{non-committal prior}{non-committal priors} on the models, $\p(\mathcal{M}_1) = \p(\mathcal{M}_2)$, the ratio simplifies to the ratio of \glslink{evidence}{evidences}, called the \textit{\gls{Bayes factor}},
\begin{equation}
\mathcal{B}_{12} \equiv \frac{\p(d|\mathcal{M}_1)}{\p(d|\mathcal{M}_2)} .
\end{equation}

The \gls{Bayes factor} is the relevant quantity to update our state of belief in two competing models in light of the data, regardless of the relative \gls{prior} probabilities we assign to them: a value of $\mathcal{B}_{12}$ greater than one means that the data support model $\mathcal{M}_1$ over model $\mathcal{M}_2$. Note that, generally, the \gls{Bayes factor} is very different from the ratio of \glslink{likelihood}{likelihoods}: a more complicated model will always yield higher \gls{likelihood} values, whereas the \gls{evidence} will favor a simpler model if the fit is nearly as good, through the smaller \gls{prior volume}.

\glslink{posterior odds}{Posterior odds} (or directly the \gls{Bayes factor} in case of \glslink{non-committal prior}{non-committal priors}) are usually interpreted against the \gls{Jeffreys' scale} for the strength of evidence. For two competing models $\mathcal{M}_1$ and $\mathcal{M}_2$ with \glslink{non-committal prior}{non-committal priors} ($\p(\mathcal{M}_1) = \p(\mathcal{M}_2) = 1/2$) and exhausting the model space ($\p(\mathcal{M}_1|d) + \p(\mathcal{M}_2|d) = 1$), the relevant quantity is the logarithm or the \gls{Bayes factor}, $\ln \mathcal{B}_{12}$ for which thresholds at values of 1.0, 2.5 and 5.0 are set (corresponding to \glslink{posterior odds}{odds} of about 3:1, 12:1 and 150:1, representing weak, moderate and strong evidence, respectively). The use of a logarithm in this empirical scale quantifies the principle that the evidence for a model only accumulates slowly with new informative data: rising up one level in the evidence strength requires about one order of magnitude more support. 

The computation of the Bayesian \gls{evidence} is generally technically challenging. For this reason, simplifying assumptions often have to be introduced \citep[see][for the Gaussian \gls{likelihood} approximation within a \glslink{model comparison}{model selection} context]{Heavens2007}. Another important particular situation is when $\mathcal{M}_2$ is a simpler model, described by fewer ($n'<n$) parameters than $\mathcal{M}_1$. $\mathcal{M}_2$ is said to be \textit{\glslink{nested model}{nested}} in model $\mathcal{M}_1$ if the $n'$ parameters of $\mathcal{M}_2$ are also parameters of $\mathcal{M}_1$. $\mathcal{M}_1$ has $p \equiv n-n'$ extra parameters that are fixed to fiducial values in $\mathcal{M}_2$. For simplicity, let us assume that there is only one extra parameter $\zeta$ in model $\mathcal{M}_1$, fixed to 0 in $\mathcal{M}_2$ ($\zeta$ describes the continuous deformation from one model to the other). Let us denote the set of other parameters by $\theta$. Under these hypotheses, the \gls{evidence} for $\mathcal{M}_1$ is $\p(d|\mathcal{M}_1) \equiv \p(d|\mathcal{M}_{\theta,\zeta})$ and the \gls{evidence} for $\mathcal{M}_2$ is $\p(d|\mathcal{M}_2)~\equiv~\p(d|\mathcal{M}_{\theta,\zeta=0})~=~\p(d|\zeta=0, \mathcal{M}_{\theta,\zeta})$. We also assume \glslink{non-committal prior}{non-committal priors} for $\mathcal{M}_1$ and $\mathcal{M}_2$.

If the \gls{prior} for the additional parameter $\zeta$ is independent of the other parameters (which makes the joint \gls{prior} separable: $\p(\zeta,\theta|\mathcal{M}_{\theta,\zeta}) = \p(\zeta|\mathcal{M}_{\theta,\zeta}) \p(\theta|\mathcal{M}_{\theta,\zeta=0})$), it can be shown that the \gls{Bayes factor} takes a simple form, the \gls{Savage-Dickey ratio} \citep{Dickey1971,Verdinelli1995}
\begin{equation}
\mathcal{B}_{12} = \frac{\p(d|\mathcal{M}_{\theta,\zeta})}{\p(d|\mathcal{M}_{\theta,\zeta=0})} = \frac{\p(\zeta=0|\mathcal{M}_{\theta,\zeta})}{\p(\zeta=0|d,\mathcal{M}_{\theta,\zeta})} ,
\end{equation}
that is, the ratio of the \glslink{marginal pdf}{marginal} \gls{prior} and the \glslink{marginal pdf}{marginal} \gls{posterior} of the larger model $\mathcal{M}_{1}$, where the additional parameter $\zeta$ is held at its fiducial value. The \gls{Bayes factor} favors the ``larger'' model only if the data decreases the \gls{posterior} \gls{pdf} at the fiducial value compared to the \gls{prior}. Operationally, if $n-n'$ is small, one can easily compute the \gls{Savage-Dickey ratio} given \glslink{sample}{samples} from the \gls{posterior} and \gls{prior} of $\mathcal{M}_{1}$ by simply estimating the \glslink{marginal pdf}{marginal} densities at the fiducial value.

\section{Markov Chain Monte Carlo techniques for parameter inference}
\label{sec:Markov Chain Monte Carlo techniques for parameter inference}

\draw{This section draws from section 3 in \citet{Leclercq2014Varenna}.}

\subsection{Markov Chains}
\label{sec:Markov Chains}

The purpose of Markov Chain Monte Carlo (\gls{MCMC}) \gls{sampling} is to construct a sequence of points in parameter space (a so-called ``chain''), whose density is proportional to the \gls{pdf} that we want to \glslink{sampling}{sample}.

A sequence $\left\lbrace \theta_0, \theta_1, \theta_2, ..., \theta_n, ...\right\rbrace $ of random elements of some set (the ``state space'') is called a \textit{Markov Chain} if the \glslink{conditional pdf}{conditional distribution} of $\theta_{n+1}$ given all the previous elements $\theta_1$, ... $\theta_n$ depends only on $\theta_n$ (the \textit{Markov property}). It is said to have \textit{stationary \gls{transition probability}} if, additionally, this distribution does not depend on $n$. This is the main kind of Markov chains of interest for \gls{MCMC}.

Such stationary chains are completely characterized by the \glslink{marginal pdf}{marginal distribution} for the first element $\theta_0$ (the \textit{initial distribution}) and the \glslink{conditional pdf}{conditional distribution} of $\theta_{n+1}$ given $\theta_n$, called the \textit{\gls{transition probability} distribution}. 

Let us denote by $\p(\theta)$ the target \gls{pdf} and by $\tproba(\theta'|\theta)$ the \glslink{transition probability}{transition pdf}. When designing a \gls{MCMC} method, we want to construct a chain with the following properties.
\begin{enumerate}
\item The desired distribution $\p(\theta)$ should be an \textit{\gls{invariant distribution}} of the chain, namely the probability of the next state being $\theta$ must satisfy the general balance property,
\begin{equation}
\label{eq:MCMC_invariance}
\p(\theta) = \int \tproba(\theta|\theta') \, \p(\theta') \, \drm \theta' .
\end{equation}
Formally, an \gls{invariant distribution} is a fixed point of the \gls{transition probability} operator, i.e. an eigenvector  with eigenvalue 1.
\item The chain should be \textit{\glslink{ergodicity}{ergodic}} (or \textit{irreducible}) which means that it is possible to go from every state to every state (not necessarily in one move).
\end{enumerate}

Property 1 ensures the existence of an invariant distribution, and property 2 its uniqueness: it is the target pdf $\p(\theta)$. Therefore, the crucial property of such Markov chains is that, after some steps depending on the initial position (the so-called ``\gls{burn-in}'' phase), they reach a state where successive elements of the chain are drawn from the high-density regions of the target distribution, in our case the \gls{posterior} of a Bayesian \gls{parameter inference}: the probability to draw $\theta$ as the $n$-th element of the chain, $\p^{(n)}(\theta)$, satisfies
\begin{equation}
\p^{(n)}(\theta) \rightarrow \p(\theta) \mathrm{~as~} n \rightarrow \infty, \mathrm{~for~any~} \theta_0.
\end{equation}
Exploiting this property, \gls{MCMC} algorithms use Markovian processes to move from one state to another in parameter space; then, given a set of random \glslink{sample}{samples}, they reconstruct the probability heuristically. Several \gls{MCMC} algorithms exist and the relevant choice is highly dependent on the problem addressed and on the \gls{posterior} distribution to be explored (see the discussion of the \glslink{no-free lunch theorem}{``no-free lunch'' theorem} in section \ref{sec:Prior choice}), but the basic principle is always similar to that of the popular \textsc{\gls{CosmoMC}} code \citep{Lewis2002}: perform a random walk in parameter space, constrained by the \gls{posterior} probability distribution.

Many useful \glslink{transition probability}{transition probabilities} satisfy the \gls{detailed balance} property,
\begin{equation}
\label{eq:detailed_balance_T}
\tproba(\theta|\theta') \, \p(\theta') = \tproba(\theta'|\theta) \, \p(\theta) .
\end{equation}
While general balance expresses the ``balance of flow'' into and out of any state $\theta$, detailed balance expresses the ``balance of flow'' between every pair of states: the flow from $\theta$ to $\theta'$ is the flow from $\theta'$ to $\theta$. Markov chains that satisfy \gls{detailed balance} are also called \textit{\glslink{reversibility}{reversible Markov chains}}. The reason why the \gls{detailed balance} property is of interest is that it is a sufficient (but not necessary) condition for the invariance of the distribution $\p$ under the \glslink{transition probability}{transition pdf} $\tproba$ (equation \eqref{eq:MCMC_invariance}), which can be easily checked:
\begin{equation}
\int \tproba(\theta|\theta') \, \p(\theta') \, \drm \theta' = \int \tproba(\theta'|\theta) \, \p(\theta) \, \drm \theta' = \p(\theta) \int \tproba(\theta'|\theta) \, \drm \theta' = \p(\theta) .
\end{equation}

\subsection{The Metropolis-Hastings algorithm}
\label{sec:The Metropolis-Hastings algorithm}

\begin{figure}
\begin{center}
\includegraphics[width=0.32\textwidth]{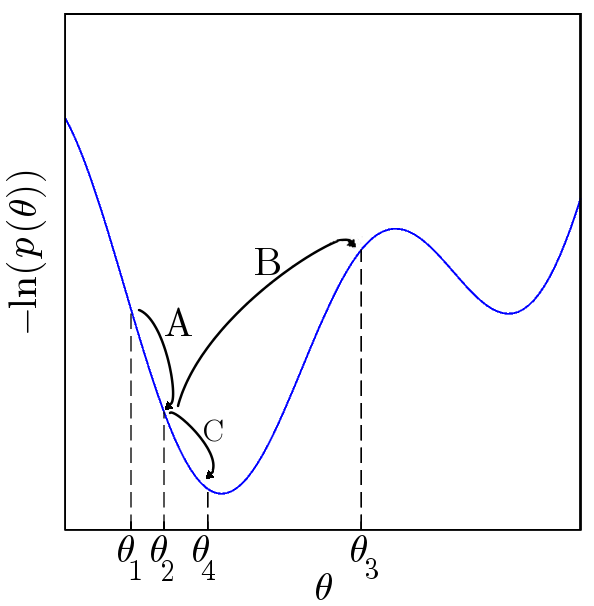} \includegraphics[width=0.32\textwidth]{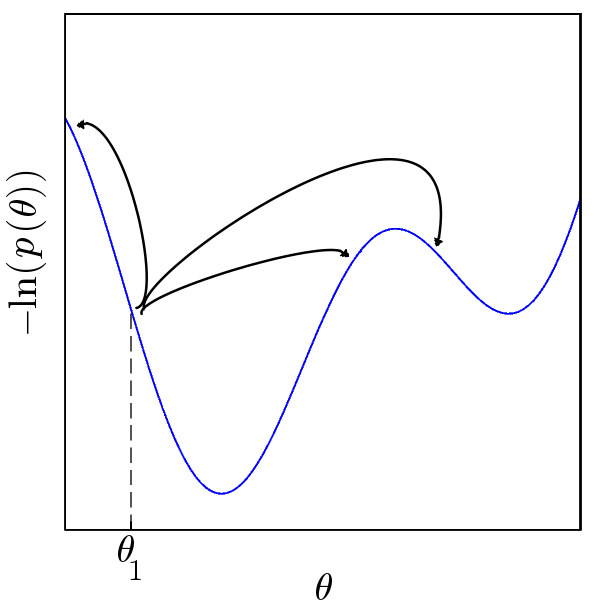} \includegraphics[width=0.32\textwidth]{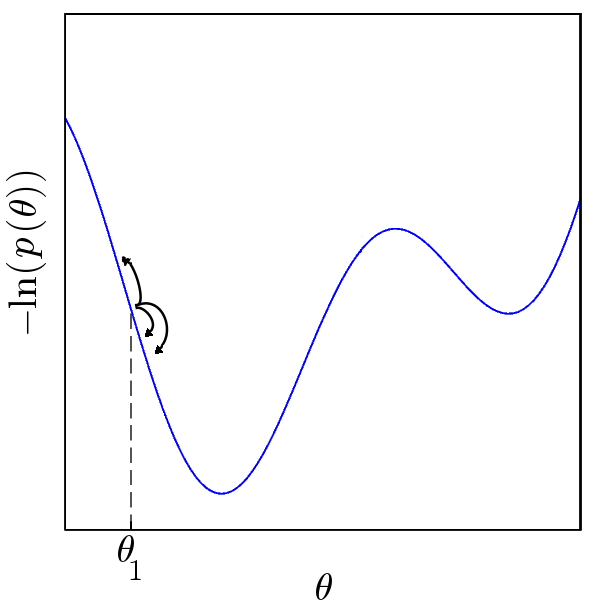}
\end{center}
\caption{\textit{Left panel}. An example of Markov chain constructed by the \glslink{MH}{Metropolis-Hastings} algorithm: starting at $\theta_1$, $\theta_2$ is proposed and accepted (step A), $\theta_3$ is proposed and refused (step B), $\theta_4$ is proposed and accepted (step C). The resulting chain is $\left\lbrace \theta_1, \theta_2, \theta_2, \theta_4, ...\right\rbrace$. \textit{Central panel}. An example of what happens with too broad a jump size: the chain lacks mobility because all the proposals are unlikely. \textit{Right panel}. An example of what happens with too narrow a jump size: the chain \glslink{sampling}{samples} the parameter space very slowly.\label{fig:MH_visualization}}
\end{figure}

A popular version of \gls{MCMC} is called the \glslink{MH}{Metropolis-Hastings} (\gls{MH}) algorithm, which works as follows. Initially, one chooses an arbitrary point $\theta_0$ to be the first \gls{sample}, and specifies a distribution $\q(\theta'|\theta)$ which proposes a candidate $\theta'$ for the next \gls{sample} value, given the previous \gls{sample} value $\theta$ ($\q$ is called the \glslink{proposal distribution}{proposal density} or jumping distribution). At each step, one draws a realization $\theta'$ from $\q(\theta'|\theta)$ and calculates the \textit{\gls{Hastings ratio}}:
\begin{equation}
\label{eq:Hastings_ratio}
r(\theta,\theta') \equiv \frac{\p(\theta')}{\p(\theta)} \frac{\q(\theta|\theta')}{\q(\theta'|\theta)} .
\end{equation}
The proposed move to $\theta'$ is \glslink{acceptance rate}{accepted} with probability $a(\theta,\theta') \equiv \min\left[ 1; r(\theta,\theta') \right] = \tproba(\theta'|\theta)$\glslink{transition probability}{}. In case it is accepted, $\theta'$ becomes the new state of the chain, otherwise the chain stays at $\theta$. A graphical illustration of the \gls{MH} algorithm is shown in figure \ref{fig:MH_visualization}. Note that each step only depends on the previous one and is also independent of the number of previous steps, therefore the ensemble of \glslink{sample}{samples} of the target distribution, constructed by the algorithm, is a stationary Markov chain.

The probability that the next state is $\theta'$ is the sum of the probability that the current state is $\theta'$ and the update leads to rejection -- which happens that a probability that we note $\rproba(\theta')$ -- and of the probability that the current state is some $\theta$ and a move from $\theta$ to $\theta'$ is proposed and accepted. This is formally written
\begin{equation}
\p(\theta') = \int \p(\theta) \tproba(\theta'|\theta) \, \drm \theta = \p(\theta') \, \rproba(\theta') + \int \p(\theta) \, \q(\theta'|\theta) \, \drm \theta .
\end{equation}
The probability to depart from $\theta'$ to any $\theta$ is $\int \q(\theta|\theta') \, \drm \theta = 1 - \rproba(\theta')$. 

The special case of a symmetric \gls{proposal distribution}, i.e. $\q(\theta|\theta') = \q(\theta'|\theta)$ for all $\theta$ and $\theta'$ is called the \textit{\gls{Metropolis update}}. Then the \gls{Hastings ratio} simplifies to
\begin{equation}
\label{eq:Metropolis_ratio}
r(\theta,\theta')=\frac{\p(\theta')}{\p(\theta)}
\end{equation}
and is called the \textit{\gls{Metropolis ratio}}. Given this result, the \gls{detailed balance} condition, equation \eqref{eq:detailed_balance_T} reads
\begin{equation}
\p(\theta') \min \left[ 1 \, ; \frac{\p(\theta)}{\p(\theta')} \right]  = \p(\theta) \min \left[ 1 \, ; \frac{\p(\theta')}{\p(\theta)} \right],
\end{equation}
which is easily seen to be true.

\begin{figure}
\begin{center}
\begin{tabular}{ccc}
suitable step size & step size too large & step size too small \\
\includegraphics[width=0.32\textwidth]{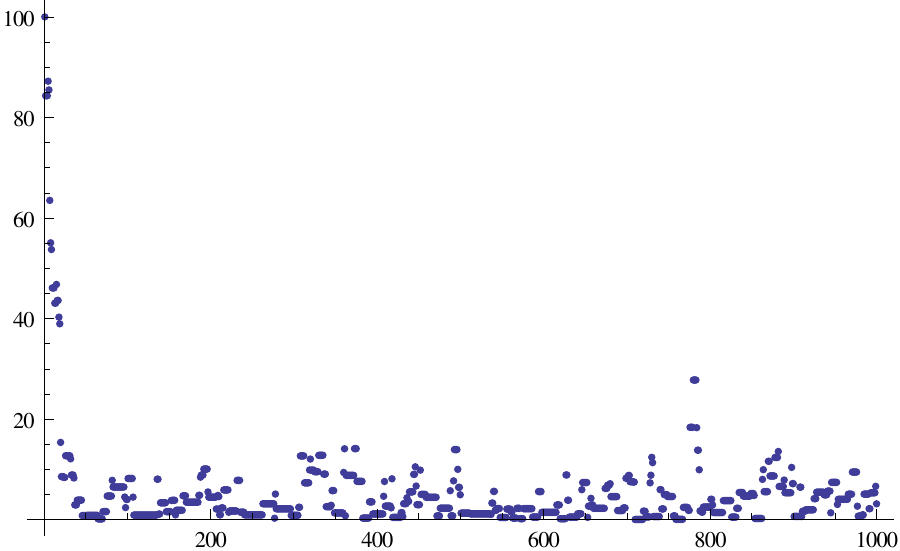} & \includegraphics[width=0.32\textwidth]{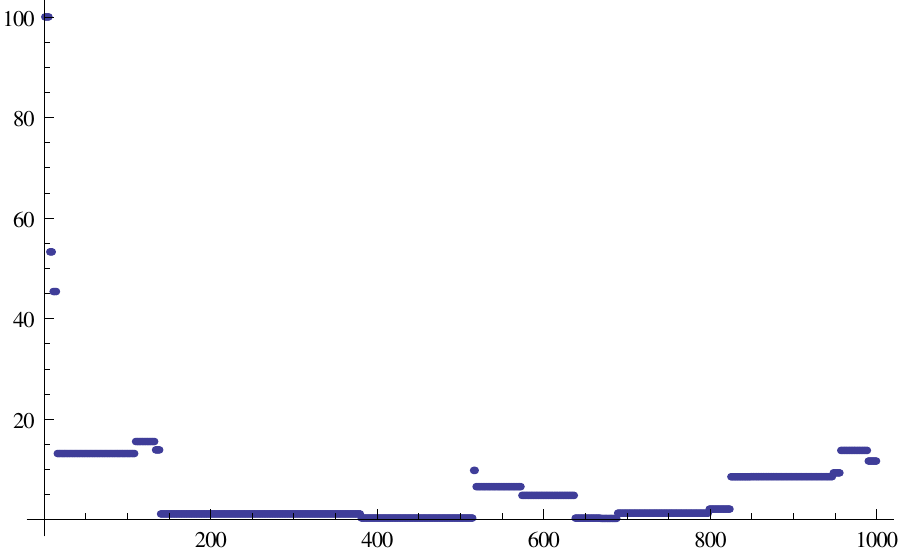} & \includegraphics[width=0.32\textwidth]{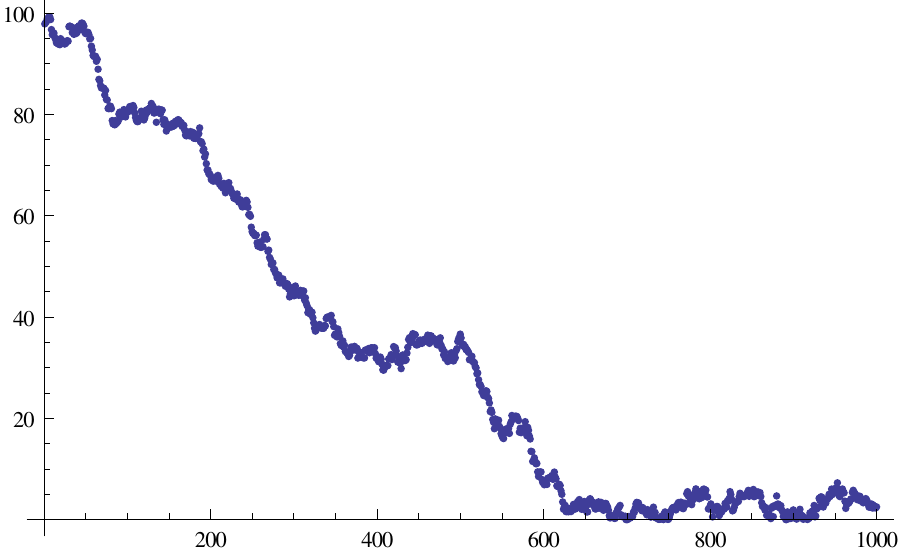}
\end{tabular}
\end{center}
\caption{Example of Markov chains constructed by the \glslink{MH}{Metropolis-Hastings} algorithm, \gls{sampling} the same target distribution but with varying \gls{proposal distribution} (step size). The plots show the value of the sampled parameter as a function of the position in the chain. The ideal behavior with a suitable step size is shown in the left panel. On the central panel, the step size is too large: the maximum \gls{likelihood} region is not well sampled. On the right panel, the step size is too small: the \gls{burn-in} phase is very long and the \gls{sampling} is slow. Note that this phenomena are easily diagnosed using the \glslink{auto-correlation function (Markov chain)}{auto-correlation function of the chain}, equation \eqref{eq:auto-correlation-MC}. \label{fig:MH_visualization_2}}
\end{figure}

In many cases, the \gls{MH} algorithm will be inefficient if the \gls{proposal distribution} is sub-optimal. It is often hard to find good \glslink{proposal distribution}{proposal distributions} if the parameter space has \glslink{high-dimensional parameter space}{high dimension} (e.g. larger than 10). Typically, the chain moves very slowly, either due to a tiny step size, either because only a tiny fraction of proposals are accepted. The initial \gls{burn-in} phase can be very long, i.e. the chain takes some time to reach high \gls{likelihood} regions, where the initial position chosen has no influence on the statistics of the chain. Even in the stationary state, sufficient \gls{sampling} of the \gls{likelihood} surface can take a very large number of steps. In the central and left panels of figure \ref{fig:MH_visualization}, we illustrate what happens with too broad a jump size (the chain lacks mobility and all proposals are unlikely) or too narrow (the chain moves slowly to \glslink{sampling}{sample} all the parameter space). Note that the step-size issues can be diagnosed using the lagged \glslink{auto-correlation function (Markov chain)}{auto-correlation function of the chain},
\begin{equation}
\label{eq:auto-correlation-MC}
\xi(\Delta) = \int \theta(t) \theta(t+\Delta) \, \mathrm{d}t .
\end{equation}
A convergence criterion using different chains or sections of chains is proposed in \cite{Gelman1992}. Possible solutions to the issues mentioned involve an adaptive step size or refinements of the standard \glslink{MH}{Metropolis-Hastings} procedure.

In some particular cases, the \glslink{proposal distribution}{proposal density} itself satisfies the \gls{detailed balance} property,
\begin{equation}
\label{eq:detailed_balance_Q}
\q(\theta|\theta') \, \p(\theta') = \q(\theta'|\theta) \, \p(\theta) ,
\end{equation}
which implies that the \gls{Hastings ratio} is always unity, i.e. that proposed states are always accepted ($\q$ is \glslink{transition probability}{}$\tproba$ and $\rproba$ is zero). For example, \gls{Gibbs sampling} is a particular case of a generalized \gls{MH} algorithm, alternating between different proposals \citetext{see e.g. \citealp{Wandelt2004} for a cosmological example}. It is particularly helpful when the joint probability distribution is difficult to \glslink{sampling}{sample} directly, but the \glslink{conditional pdf}{conditional distribution} of some parameters given the others is known. It uses a block scheme of individual \textit{\glslink{Gibbs sampling}{Gibbs updates}} to \glslink{sampling}{sample} an instance from the distribution of each variable in turn, \glslink{conditional pdf}{conditional} on the current values of the other variables. Formally, the \glslink{proposal distribution}{proposal} for a single \glslink{Gibbs sampling}{Gibbs update} is from a \glslink{conditional pdf}{conditional distribution} of the target \gls{pdf}: $\q(\theta'|\theta) \equiv \p(\theta'|f(\theta))$ where $f(\theta)$ is $\theta$ with some components omitted. $\theta'$ is an update of these missing components, keeping the others at the values they had in $\theta$. Therefore, $f(\theta')=f(\theta)$, and we have
\begin{equation}
\q(\theta'|\theta) \equiv \p(\theta'|f(\theta)) = \p(\theta'|f(\theta')) = \p(\theta'),
\end{equation}
which trivially implies the \gls{detailed balance} property (equation \eqref{eq:detailed_balance_Q}) and ensures an \gls{acceptance rate} of unity.

\subsection{Hamiltonian Monte Carlo}
\label{sec:Hamiltonian Monte Carlo}

A very efficient \gls{MCMC} algorithm for \glslink{high-dimensional parameter space}{high-dimensional problems} such as those encountered in cosmology is \glslink{HMC}{Hamiltonian Monte Carlo} \citep[\gls{HMC}, originally introduced under the name of hybrid Monte Carlo,][]{Duane1987}. A detailed overview is provided by \cite{Neal2011}. 

The general idea of \gls{HMC} is to use concepts borrowed from \gls{classical mechanics} to solve statistical problems. As it is a core ingredient in the {\borg} code, we now discuss the most important features of \gls{HMC}. We start by reviewing physical properties of Hamiltonian dynamics. The system is described by the Hamiltonian $H(\boldsymbol{\uptheta},\textbf{p})$, a function of the $D$-dimensional position vector $\boldsymbol{\uptheta}$ and of the $D$-dimensional \gls{momentum} vector $\textbf{p}$.\footnote{In this section we use boldface notations for all vectors, to strengthen the link between physics and statistics.} Its time evolution is described by \gls{Hamilton's equations},
\begin{eqnarray}
\deriv{\boldsymbol{\uptheta}}{t} & = & \pd{H}{\textbf{p}}, \\
\deriv{\textbf{p}}{t} & = & - \pd{H}{\boldsymbol{\uptheta}}.
\end{eqnarray}
For any time interval of duration $s$, these equations define a mapping $T_s$ from the state at any time $t$ to the state at time $t+s$. The first important property of Hamiltonian dynamics is time \gls{reversibility}, which means for any $s$, that the mapping $T_s$ has an inverse. It is easy to check that this inverse is $T_{-s}$. 

A second property of the dynamics is that it conserves the Hamiltonian during the evolution, which can be checked explicitly:
\begin{equation}
\deriv{H}{t} = \pd{H}{\boldsymbol{\uptheta}} \deriv{\boldsymbol{\uptheta}}{t} + \pd{H}{\textbf{p}} \deriv{\textbf{p}}{t} = \pd{H}{\boldsymbol{\uptheta}} \pd{H}{\textbf{p}} - \pd{H}{\textbf{p}}\pd{H}{\boldsymbol{\uptheta}} = 0.
\end{equation}

In $2D$ dimensions, using $\textbf{z} = (\boldsymbol{\uptheta},\textbf{p})$ and the matrix
\begin{equation}
\textbf{J} = \begin{pmatrix}
\textbf{0}_D & \textbf{I}_D \\
-\textbf{I}_D & \textbf{0}
\end{pmatrix} ,
\end{equation}
one can rewrite \gls{Hamilton's equations} as
\begin{equation}
\deriv{\textbf{z}}{t} = \textbf{J} \cdot \nabla H .
\end{equation}
The third important property is that Hamiltonian dynamics is \textit{\glslink{symplecticity}{symplectic}}, which means that the Jacobian matrix $\textbf{B}_s$ of the mapping $T_s$ satisfies
\begin{equation}
\textbf{B}_s^\intercal \, \textbf{J}^{-1} \, \textbf{B}_s = \textbf{J}^{-1} .
\end{equation}
This property implies volume conservation in $(\boldsymbol{\uptheta},\textbf{p})$ \gls{phase space} (a result also known as \gls{Liouville's theorem}), since $\mathrm{det}(\textbf{B}_s)^2$ must be one.

Crucially, \gls{reversibility} and \gls{symplecticity} are properties that can be maintained exactly, even when Hamiltonian dynamics is approximated by numerical integrators (see section \ref{sec:The leapfrog scheme integrator}).

The link between probabilities and Hamiltonian dynamics is established via the concept of \textit{\gls{canonical distribution}} from \gls{statistical mechanics}. Given the energy distribution $E(\textbf{x})$ for possibles states $\textbf{x}$ of the physical system, the \gls{canonical distribution} over states $\textbf{x}$ has \gls{pdf}
\begin{equation}
\p(\textbf{x}) = \frac{1}{Z} \exp \left( \frac{-E(\textbf{x})}{k_\mathrm{B}T}\right)
\end{equation}
where $k_\mathrm{B}$ is the \gls{Boltzmann constant}, $T$ the temperature of the system, and the \textit{\gls{partition function}} $Z$ is the normalization constant needed to ensure $\int \p(\textbf{x}) \, \drm \textbf{x} = 1$. In Hamiltonian dynamics, $H$ is an energy function for the joint state of positions $\boldsymbol{\uptheta}$ and \glslink{momentum}{momenta} $\textbf{p}$, and hence defines a joint \gls{pdf} as
\begin{equation}
\p(\boldsymbol{\uptheta},\textbf{p}) = \frac{1}{Z} \exp \left( \frac{-H(\boldsymbol{\uptheta},\textbf{p})}{k_\mathrm{B}T}\right)
\end{equation} 
Viewing this the opposite way, if we are interested in some joint distribution with probability $\p(\boldsymbol{\uptheta},\textbf{p})$, we can obtain it as a \gls{canonical distribution} with temperature $k_\mathrm{B}T = 1$, by setting $H(\boldsymbol{\uptheta},\textbf{p}) = - \ln \p(\boldsymbol{\uptheta},\textbf{p}) - \ln Z$, where $Z$ is any convenient positive constant (we choose $Z=1$ in the following for simplicity). 

We are now ready to discuss the \glslink{HMC}{Hamiltonian Monte Carlo} algorithm. \gls{HMC} interprets the negative logarithm of the \gls{pdf} to \glslink{sampling}{sample} as a physical potential, $\psi(\boldsymbol{\uptheta}) = -\ln \p(\boldsymbol{\uptheta})$ and introduces auxiliary variables: ``conjugate \glslink{momentum}{momenta}'' $p_i$ for all the different parameters. Using these new variables as \gls{nuisance parameters}, one can formulate a Hamiltonian describing the dynamics in the multi-dimensional \gls{phase space}. Such a Hamiltonian is given as: 
\begin{equation}
\label{eq:HMC_hamiltonian}
H(\boldsymbol{\uptheta},\textbf{p}) = \frac{1}{2} \textbf{p}^\intercal \, \textbf{M}^{-1} \, \textbf{p} + \psi(\boldsymbol{\uptheta}) = - \ln \p(\boldsymbol{\uptheta},\textbf{p}) ,
\end{equation}
where the kinetic term, $K(\textbf{p}) \equiv \frac{1}{2}\textbf{p}^\intercal \, \textbf{M}^{-1} \, \textbf{p}$ involves $\textbf{M}$, a symmetric positive definite ``\gls{mass matrix}'' whose choice can strongly impact the performance of the sampler. Masses characterize the inertia of parameters when moving through the parameter space. Consequently, too large masses will result in slow \glslink{exploration of the posterior}{exploration} efficiency, while too light masses will result in large rejection rates (see also figure \ref{fig:MH_visualization_2}).

Each iteration of the \gls{HMC} algorithm works as follows. One draws a realization of the \glslink{momentum}{momenta} from the distribution defined by the kinetic energy term, i.e. a multi-dimensional Gaussian with a covariance matrix $\textbf{M}$, then moves the positions $\boldsymbol{\uptheta}$ using a Hamiltonian integrator in parameter space, respecting \glslink{symplecticity}{symplectic} symmetry. In other words, we first ``\gls{kick} the system'' then follow its deterministic dynamical evolution in \gls{phase space} according to \gls{Hamilton's equations}, which read
\begin{eqnarray}
\deriv{\boldsymbol{\uptheta}}{t} & = & \textbf{M}^{-1} \, \textbf{p} , \\
\deriv{\textbf{p}}{t} & = & - \pd{\psi(\boldsymbol{\uptheta})}{\boldsymbol{\uptheta}} . \label{eq:Hamiltonian_force}
\end{eqnarray}

If the integrator is \glslink{reversibility}{reversible}, then the \glslink{proposal distribution}{proposal} is symmetric, and the \glslink{acceptance rate}{acceptance probability} for the new point $(\boldsymbol{\uptheta}',\textbf{p}')$ follows the \glslink{Metropolis update}{Metropolis rule} (see equation \eqref{eq:Metropolis_ratio}):
\begin{equation}
a(\boldsymbol{\uptheta}',\textbf{p}',\boldsymbol{\uptheta},\textbf{p}) = \min\left[ 1 \, ; \frac{\p(\boldsymbol{\uptheta}',\textbf{p}')}{\p(\boldsymbol{\uptheta},\textbf{p})} \right] = \min \left[ 1 \, ; \exp(- H(\boldsymbol{\uptheta}',\textbf{p}') + H(\boldsymbol{\uptheta},\textbf{p})) \right] .
\end{equation}
Using the results of sections \ref{sec:Markov Chains} and \ref{sec:The Metropolis-Hastings algorithm}, this proves that detailed balance is verified and that \gls{HMC} leaves the \gls{canonical distribution} invariant.

In exact Hamiltonian dynamics, the energy is conserved, and therefore, ideally, this procedure always provides an \gls{acceptance rate} of unity. In practice, numerical errors can lead to a somewhat lower \gls{acceptance rate} but \gls{HMC} remains computationally much cheaper than standard \gls{MH} techniques in which proposals are often refused. In the end, we discard the \glslink{momentum}{momenta} and yield the target parameters by \glslink{marginal pdf}{marginalization}:
\begin{equation}
\p(\boldsymbol{\uptheta}) = \int \p(\boldsymbol{\uptheta},\textbf{p}) \, \mathrm{d}\textbf{p} .
\end{equation}

Applications of \gls{HMC} in cosmology include: the determination of \gls{cosmological parameters} \citetext{\citealp{Hajian2007}; in combination with \textsc{Pico}, \citealp{Fendt2007}}, \gls{CMB} \gls{power spectrum} \gls{inference} \citep{Taylor2008} and \glslink{Bayesian statistics}{Bayesian approach} to \gls{non-Gaussianity} analysis \citep{Elsner2010}, \glslink{log-normal distribution}{log-normal} density reconstruction \citetext{\citealp{JascheKitaura2010}; including from \gls{photometric redshift} \glslink{galaxy survey}{surveys}, \citealp{Jasche2012}}, dynamical, non-linear \gls{reconstruction} of the \gls{initial conditions} from \glslink{galaxy survey}{galaxy surveys} \citep{Jasche2013BORG}, joint \gls{power spectrum} and \gls{bias} model inference \citep{Jasche2013BIAS}, inference of \gls{CMB lensing} \citep{Anderes2015}.

%% file: Chapter4/Chapter4Content.tex
\chapter{Physical large-scale structure inference with the BORG algorithm}
\label{chap:BORG}
\minitoc

\begin{flushright}
\begin{minipage}[c]{0.6\textwidth}
\rule{\columnwidth}{0.4pt}

``We are the Borg. Lower your shields and surrender your ships. We will add your biological and technological distinctiveness to our own. Your culture will adapt to service us. Resistance is futile.''\\
--- \citet{StarTrek1996}

\vspace{-5pt}\rule{\columnwidth}{0.4pt}
\end{minipage}
\end{flushright}

\abstract{\section*{Abstract}
This chapter describes the development and implementation of the {\borg} algorithm, which aims at physical \gls{large-scale structure inference} in the \glslink{linear regime}{linear} and \gls{mildly non-linear regime}. It describes the \gls{data model}, which jointly accounts for the shape of three-dimensional matter field and its \gls{formation history}. Based on an efficient implementation of the \glslink{HMC}{Hamiltonian Monte Carlo} algorithm, {\borg} \glslink{sample}{samples} the joint \gls{posterior} of the several millions parameters involved, which allows for thorough \gls{uncertainty quantification}.
}

This chapter presents {\borg} (Bayesian Origin Reconstruction from Galaxies), a \gls{data assimilation} method for probabilistic, physical \gls{large-scale structure inference}. In section \ref{sec:The challenge: the curse of dimensionality}, the main challenge faced, namely the \gls{curse of dimensionality}, is discussed. In section \ref{sec:The BORG data model}, we describe the latest formulation of {\borg} \gls{data model}, initially introduced by \citet{Jasche2013BORG} and updated by \citet{Jasche2015BORGSDSS}. Section \ref{sec:Sampling procedure and numerical implementation} gives considerations about the \gls{sampling} procedure and the numerical implementation of the algorithm. Finally, in section \ref{sec:Testing BORG}, we report on a test of the {\borg} algorithm using a \glslink{mock catalog}{synthetic catalog} of tracers. 

\section{The challenge: the curse of dimensionality}
\label{sec:The challenge: the curse of dimensionality}

Statistical analyses of \glslink{LSS}{large-scale structure} surveys require to go from the few parameters describing the \gls{homogeneous Universe} to a point-by-point characterization of the inhomogeneous Universe. The latter description typically involves \glslink{high-dimensional parameter space}{tens of millions of parameters}: the density in each voxel of the discretized \glslink{galaxy survey}{survey} volume.

``\glslink{curse of dimensionality}{Curse of dimensionality}'' phenomena \citep{Bellman1961} are the significant obstacle in this \glslink{high-dimensional parameter space}{high-dimensional} data analysis problem. They refer to the difficulties caused by the exponential increase in volume associated with adding extra dimensions to a mathematical space. In the following, we discuss the basic aspects of the \glslink{high-dimensional parameter space}{high-dimensional} situation. In particular, we outline three aspects of the \gls{curse of dimensionality} phenomena.

\subsection{Sparse sampling}

The first and most obvious aspect is the exponential increase of \gls{sparsity} given a fixed amount of \glslink{sample}{sampling points}. Reciprocally, the number of points drawn from a uniform distribution, needed for \gls{sampling} at a constant density a region in parameter space, increases exponentially with its dimension.

\begin{figure}
\begin{center}
\includegraphics[width=0.3\columnwidth]{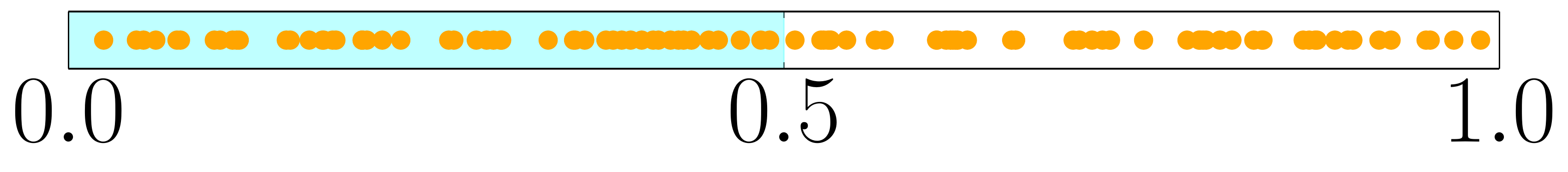}
\includegraphics[width=0.3\columnwidth]{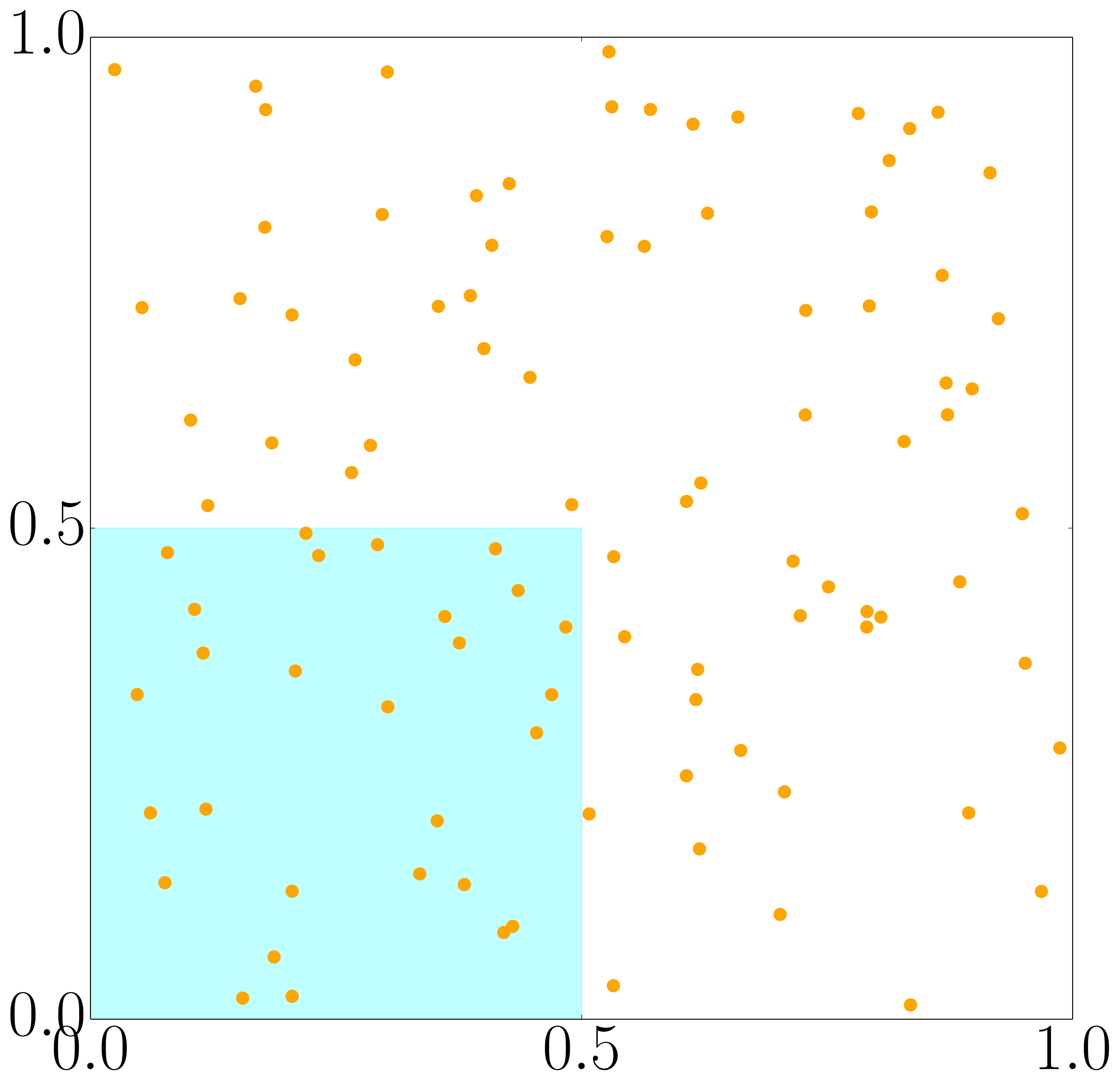}
\includegraphics[width=0.3\columnwidth]{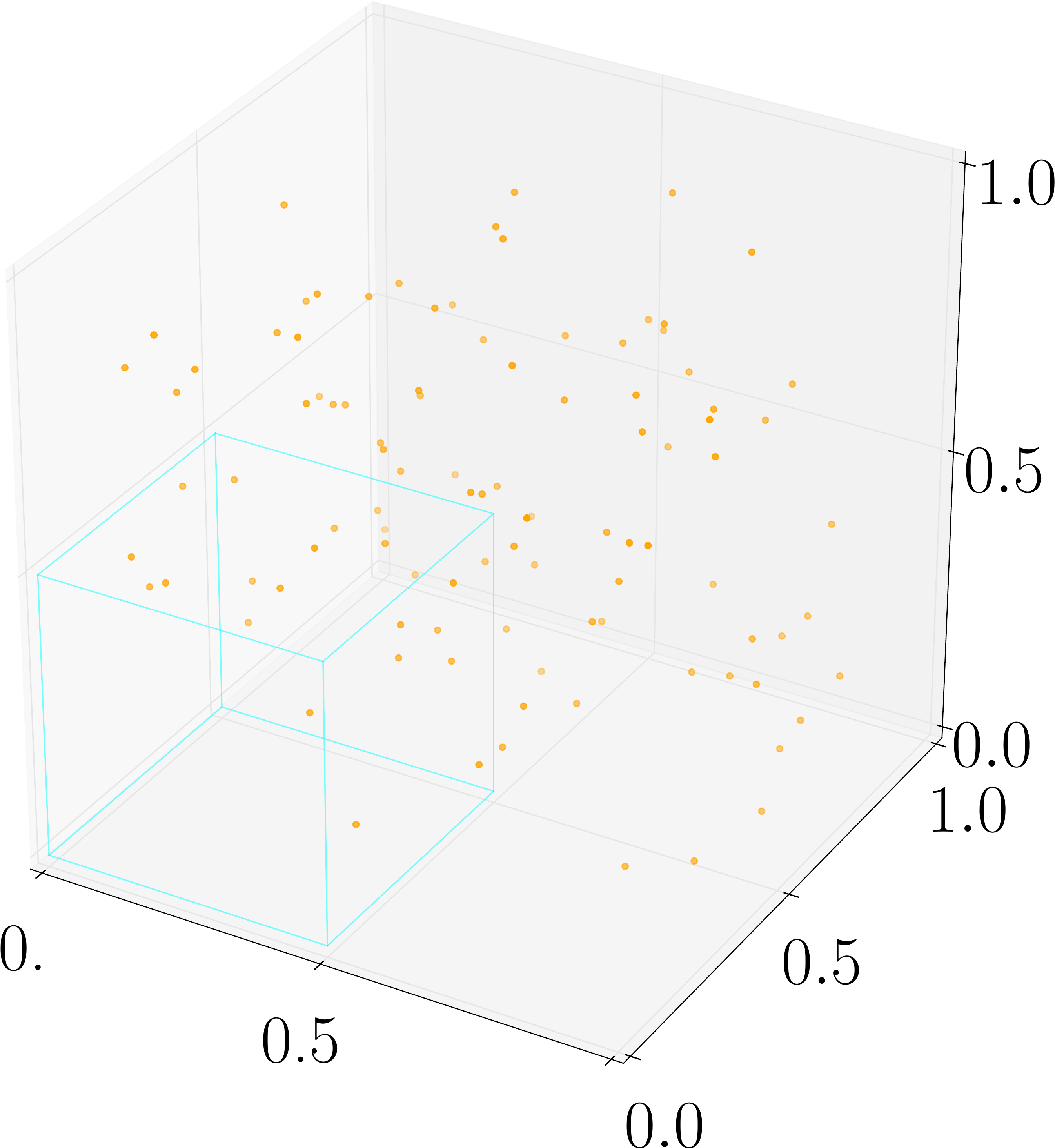}
\end{center}
\caption{Illustration of the \gls{curse of dimensionality} in one, two and three dimensions. We draw an original \gls{sample} of 100 random points uniformly distributed in $\left[0;1\right]$, then progressively add a second and third coordinate, also uniformly drawn in $\left[0;1\right]$. The \gls{sparsity} of the data (here illustrated by the number of \glslink{sample}{samples} in the $\left[0;\frac{1}{2}\right]$ hypercube, in cyan) increases exponentially with the number of dimensions.\label{fig:curse}}
\end{figure}

We illustrate this phenomenon in figure \ref{fig:curse} with 100 points randomly drawn in $\left[0;1\right]^D$ for $D=1,2,3$. The number of \glslink{sample}{samples} that fall in some fixed region in parameter space exponentially decreases with the dimensionality of the problem. For example, the probability $\p_D$ for a random point to be in the $\left[0; \frac{1}{2}\right]^D$ hyperquadrant (shown in cyan in figure \ref{fig:curse}) is $2^{-D}$. Difficulties to represent such probabilities numerically (table \ref{tb:curse}) arise well before $D=10^7$, as we now discuss.

\begin{table}\centering
\begin{tabular}{lll}
\hline\hline
Dimension $D$ & $\p_D = 2^{-D}$ & Numerical representation\\
\hline
$1$ & $2^{-1}$ & $0.5$\\
$10$ & $2^{-10}$ & $9.77 \times 10^{-4}$\\
$100$ & $2^{-100}$ & $7.89 \times 10^{-31}$\\
$1000$ & $2^{-1000}$ & $9.33 \times 10^{-302}$\\
$10000$ & $2^{-10000}$ & $0.$\\
\hline\hline
\end{tabular}
\caption{Probability for a \gls{sample} uniformly drawn in $\left[ 0;1 \right]^D$ to be in $\left[ 0;\frac{1}{2} \right]^D$, as a function of the dimension $D$. The mathematical result, $2^{-D}$ (second column) is compared to its double-precision computer representation (third column). For $D \geq 1075$, $\p_D$ is below the minimum positive subnormal double representable.}
\label{tb:curse}
\end{table}

\begin{figure}
\begin{center}
\includegraphics[width=0.6\columnwidth]{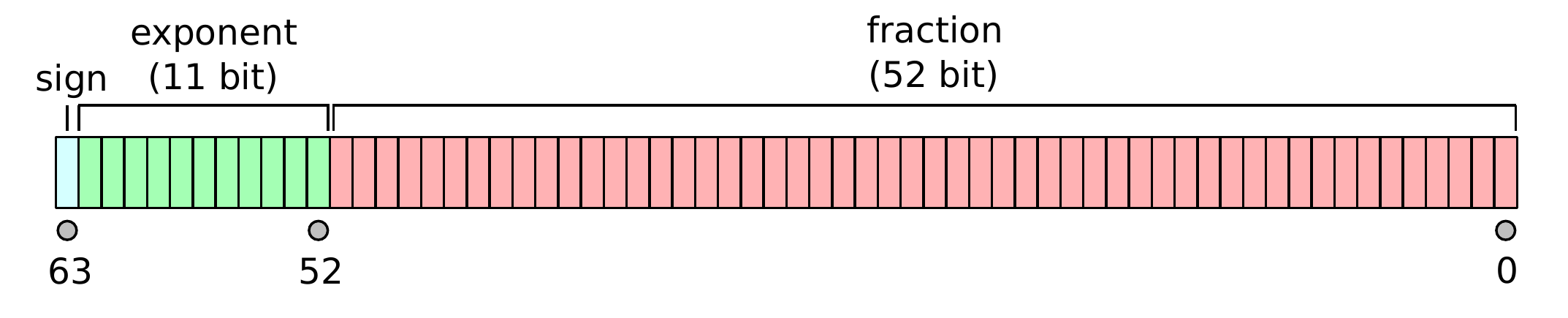}
\end{center}
\caption{Computer representation of double-precision binary floating-point numbers. One bit is used to store the sign, $11$ to store the exponent, and $52$ bits to store the fractional part. This representation on a finite number of bits implies the existence of both a minimal and a maximal positive representable number.\label{fig:binary64}}
\end{figure}

In standard double-precision binary floating-point format (the IEEE 754 ``\texttt{binary64}'' norm), numbers are represented in base $b=2$. The bits are laid out as follows (figure \ref{fig:binary64}): $1$ sign bit, $11$ bits for the exponent width, and $p=52$ bits for the significand precision. The real value assigned by the machine to a set of \texttt{binary64} digits is
\begin{equation}
(-1)^{\mathrm{sign}} \left( 1+ \sum_{i=1}^{52} b_{52-i} 2^{-i} \right) \times 2^{e-1023} ,
\end{equation}
where $1 \leq e \leq 2046$ is the ``biased exponent'' encoded in the $11$ exponent bits and $b_i$ are the values of the significand bits.

This representation implies that the maximum relative rounding error when rounding a number to the nearest representable one (the ``\gls{machine epsilon}'') is $b^{-(p-1)} = 2^{-52}$. Therefore, the maximum positive double is $\texttt{max\_double} \equiv (1+(1-2^{-52})) \times 2^{1023} \approx 1.798 \times 10^{308}$ and the minimum positive double is $\texttt{min\_normal\_double} \equiv 2^{-1022} \approx 2.225 \times 10^{-308}$.

In a normal floating-point value, there are no leading zeros in the significand; instead leading zeros are moved to the exponent. By using leading zeros in the significand, it is possible to represent ``subnormal numbers'', i.e. numbers where this representation would result in an exponent that is too small for the allowed number of bits. The smallest subnormal number representable with the \texttt{binary64} norm is $\texttt{min\_subnormal\_double} \equiv 2^{-52} \times 2^{-1022} \approx 4.941 \times 10^{-324}$.

Coming back to the representation of $\p_D$ is a large number of dimensions, the discussion above implies that $\p_D$ is exactly zero, at computer precision, for $D \geq 1075$. More generally, typical probabilities are often below $\texttt{min\_subnormal\_double}$ for $D \gtrsim 1000$, which means that their computer representations as doubles is impossible. Representing such numbers requires more than 64 bits. This number of dimensions is well below that of the problem that we want to tackle, $D \approx 10^7$.

\subsection{Shape of high-dimensional pdfs}

Generally, \glslink{high-dimensional function}{high-dimensional functions} can have more complex features than low-dimensional functions (there is more ``space'' for that), and hence can be harder to characterize.

Since it is not possible to store arbitrarily small positive numbers, numerical representations of \glslink{high-dimensional function}{high-dimensional pdfs} will tend to have narrow support and very peaked features. This can also cause difficulties, as \glslink{pdf}{pdfs} have to be normalized to unity: if the support is sufficiently small, the value of the \gls{pdf} at its peaks can easily be above the maximum double $\texttt{max\_double}$, which will cause computer crashes. 

\subsection{Algorithms in high dimensions}

It is important to note that \gls{curse of dimensionality} phenomena are generally not an intrinsic problem of \glslink{high-dimensional parameter space}{high-dimensional} problems, but a joint problem of the data set and the algorithm used. In particular, a dramatic increase of computational time (both to get one \gls{sample} and to reach the required number of \glslink{sample}{samples}) is common. The \gls{curse of dimensionality} often means that the number of samples available is small compared to the dimension of the space, which can lead to issues such as overfitting the data or getting poor classification or clustering when searching for specific patterns \citep{Verleysen2005}.

For most \gls{MCMC} algorithms, the slow convergence, due a high rejection rate, is the most significant obstacle. In particular, for many interesting problems (typically non-linear and where components are not independently distributed), traditional \gls{sampling} techniques that perform a random walk in parameter space, like the \glslink{MH}{Metropolis-Hastings algorithm} (see section \ref{sec:The Metropolis-Hastings algorithm}) will unequivocally fail in $D \approx 10^7$.\footnote{At least, unless the \gls{proposal distribution} approximates extremely well the target distribution -- which would imply to have already solved the problem!} However, gradients of \glslink{pdf}{pdfs} carry capital information, as they indicate the direction to high-density regions, permitting fast travel through a large volume in parameter space.

One way forward is to reduce the dimensionality of the problem, which is actually an entire research field. For example, \gls{principal component analysis} converts a set of correlated variables to a set of linearly uncorrelated ``principal components''. Unfortunately, due to the highly \glslink{non-linear evolution}{non-linear} and complex physics involved in \gls{structure formation} (see chapter \ref{chap:theory}), no obvious reduction of the problem size exists in our case. Under the assumption of an \glslink{initial conditions}{initial} \gls{grf} with independent density amplitudes in Fourier space, we cannot make any further dimension reduction, and we have to deal with all $D \approx 10^7$ dimensions. Dimensionality can only be reduced by coarsening the resolution and discarding information. 

As we will demonstrate in the rest of this chapter, \glslink{HMC}{Hamiltonian Monte Carlo} (see section \ref{sec:Hamiltonian Monte Carlo}) beats the \gls{curse of dimensionality} for the problem of physical \gls{large-scale structure inference}. In particular, the approximate conservation of the Hamiltonian enables us to keep a high \gls{acceptance rate}, and the use of gradients of the \gls{posterior} \gls{pdf} ($\partial \psi(\boldsymbol{\uptheta})/\partial \boldsymbol{\uptheta}$ in \gls{Hamilton's equations}) allows efficient search for high density of probability regions.

\section{The BORG data model}
\label{sec:The BORG data model}

\begin{table*}
\defcitealias{Jasche2010b}{J+10b}
\defcitealias{JascheKitaura2010}{JK10}
\defcitealias{Jasche2012}{JW12}
\defcitealias{Jasche2013BORG}{JW13a}
\defcitealias{Jasche2013BIAS}{JW13b}
\defcitealias{Jasche2015BORGSDSS}{JLW15}
\defcitealias{Lavaux2016BORG2MPP}{LJ16}
\begin{center}
\begin{tabular}{|m{1.05cm}|m{1.3cm}|m{1.3cm}|m{1.3cm}|m{1.1cm}|m{1.3cm}|m{1.3cm}|m{1.3cm}|m{1.3cm}|m{1.3cm}|}
  \hline
  \footnotesize{Code} & \footnotesize{Density field model} & \footnotesize{Response operator} & \footnotesize{Multi-survey} & \footnotesize{$P(k)$} & \footnotesize{Photo-$z$} & \footnotesize{Galaxy bias model} & \footnotesize{$b$} & \footnotesize{$\tilde{N}$} & \footnotesize{RSD}\\
  \hline
  {\ares} & \cellcolor{green!25}\footnotesize{Gaussian \citepalias{Jasche2010b}} & \cellcolor{green!25}\footnotesize{\citepalias{Jasche2010b}} & \cellcolor{green!25}\footnotesize{\citepalias{Jasche2013BIAS}} & \cellcolor{green!25}\footnotesize{\citepalias{Jasche2010b}} &  & \cellcolor{green!25}\footnotesize{linear \footnotesize{\citepalias{Jasche2010b}}; $M$-dep., linear \citepalias{Jasche2013BIAS}} & \cellcolor{green!25}\footnotesize{sampled \citepalias{Jasche2013BIAS}} & \cellcolor{green!25}\footnotesize{\citepalias{Jasche2013BIAS}} & \cellcolor{blue!20}\footnotesize{(J+in prep.)}\\
  \hline
  {\hades} & \cellcolor{green!25}\footnotesize{Log-normal \citepalias{JascheKitaura2010}} & \cellcolor{green!25}\footnotesize{\citepalias{JascheKitaura2010}} & \cellcolor{blue!20}\footnotesize{(J+in prep.)} & \cellcolor{blue!20}\footnotesize{(J+in prep.)} & \cellcolor{green!25}\footnotesize{\citepalias{Jasche2012}} & \cellcolor{green!25}\footnotesize{linear \citepalias{JascheKitaura2010}} &  & \cellcolor{blue!20}\footnotesize{(J+in prep.)} & \\
  \hline
  {\borg} & \cellcolor{green!25}\footnotesize{2LPT \citepalias{Jasche2013BORG}} & \cellcolor{green!25}\footnotesize{\citepalias{Jasche2013BORG}} & \cellcolor{green!25}\footnotesize{\citepalias{Jasche2015BORGSDSS}} &  &  & \cellcolor{green!25}\footnotesize{linear \citepalias{Jasche2013BORG}; $M$-dep., power-law \citepalias{Jasche2015BORGSDSS}} & \cellcolor{blue!20}\footnotesize{calibrated with {\ares} \footnotesize{\citepalias{Lavaux2016BORG2MPP}}}; sampled \footnotesize{(J+in prep.)} & \cellcolor{green!25}\footnotesize{\citepalias{Jasche2015BORGSDSS}} & \\
  \hline
\end{tabular}
\end{center}
\caption{Current status of Bayesian large-scale structure analysis codes {\ares}, {\hades} and {\borg}. Green cells 
correspond to features implemented in the \gls{data model} and tested, as reported in the corresponding papers. Blue cells correspond to features which will be described in upcoming publications. The column correspond respectively to: the model used to describe the \gls{prior} \gls{density field}; treatment of the \gls{survey response operator} (survey \gls{mask} and \gls{selection effects}); treatment of multiple, independent \glslink{galaxy survey}{surveys} (or sub-samples of the same survey); \gls{power spectrum} \gls{sampling}; \glslink{photometric redshift}{photometric redshifts} \gls{sampling}; galaxy \gls{bias} model ($M$-dep. stands for \gls{luminosity}-dependent bias); treatment of \glslink{noise parameter}{bias parameters}; \gls{sampling} of \gls{noise} levels; treatment of \glslink{peculiar velocity}{peculiar velocities} and \gls{redshift-space distortions}. The references are \citetalias{Jasche2010b} $=$ \citet{Jasche2010b}; \citetalias{JascheKitaura2010} $=$ \citet{JascheKitaura2010}; \citetalias{Jasche2012} $=$ \citet{Jasche2012}; \citetalias{Jasche2013BORG} $=$ \citet{Jasche2013BORG}; \citetalias{Jasche2013BIAS} $=$ \citet{Jasche2013BIAS}; \citetalias{Jasche2015BORGSDSS} $=$ \citet{Jasche2015BORGSDSS}; \citetalias{Lavaux2016BORG2MPP} $=$ \citet{Lavaux2016BORG2MPP}.}\label{tb:LSS_algos}
\end{table*}

In this section, we discuss the {\borg} \gls{data model}, i.e. the set of assumptions concerning the generation of observed \glslink{LSS}{large-scale structure} \gls{data}. In other words, we write down a probabilistic data-generating process.

This model was initially introduced by \citet{Jasche2013BORG}. In \citet{Jasche2015BORGSDSS}, we updated the \gls{data model} and modified to the original formulation of the {\borg} \gls{sampling} scheme to introduce the improvements presented in \citet{Jasche2013BIAS}. These improvements permit to account for \gls{luminosity}-dependent galaxy \gls{bias} and to perform automatic \glslink{noise parameter}{noise level} calibration.

{\borg} is the successor of {\ares} \citep[Algorithm for REconstruction and Sampling,][]{Jasche2010b,Jasche2013BIAS} and {\hades} \citep[HAmiltonian Density Estimation and Sampling][]{JascheKitaura2010,Jasche2012}. In table \ref{tb:LSS_algos}, we summarize the different aspects covered by the {\ares}, {\hades}, and {\borg} data models. Contrary to {\ares} and {\hades}, which use phenomenological models to describe the \gls{density field}, {\borg} involves a physical \gls{structure formation} model (see table \ref{tb:LSS_algos}). \gls{LSS} observations are merged with actual dynamics. Therefore, even if it is the least advanced algorithm in terms of the aspects covered by the \gls{data model}, its physical modeling is the most sophisticated. 

In the following, $x$ labels one of the $D$ voxels of the discretized domain, $\BattleShipGrey{\delta}^{\BattleShipGrey{\mathrm{i}}}$ and $\BattleShipGrey{\delta}^{\BattleShipGrey{\mathrm{f}}}$ are realizations of the \glslink{initial conditions}{initial} (at $a=10^{-3}$) and \glslink{final conditions}{final} (at $a=1$) \gls{density contrast}, respectively, expressed as $D$-dimensional vectors. For improved clarity, we use colors in equations to distinguish the different quantities that are involved in the \gls{data model}.

\subsection{The physical density prior}

In contrast to earlier algorithms (see table \ref{tb:LSS_algos}) {\borg} includes a \gls{physical density prior} i.e. involves a model for \gls{structure formation}. This makes the \gls{prior} (expressed in terms of the \glslink{final conditions}{final} \gls{density contrast}) highly \glslink{non-Gaussianity}{non-Gaussian} and non-linear. Writing down this \gls{prior} is the subject of the present section.

\subsubsection{The initial Gaussian prior}
\label{sec:The initial Gaussian prior}

As discussed in the \hyperref[chap:intro]{introduction} and in chapter \ref{chap:theory}, it is commonly admitted that the \gls{density contrast} early in the \glslink{matter domination}{matter era} obeys \glslink{grf}{Gaussian statistics}. Consistently with the discussion of section \ref{sec:Prior choice}, this is the \gls{prior} that we adopt.

Explicitly, in Fourier space, the \gls{prior} for the \glslink{initial conditions}{initial} \gls{density contrast} is a \glslink{grf}{multivariate Gaussian process} with zero mean and diagonal covariance matrix $\DarkPastelGreen{\hat{S}}$ (see equation \eqref{eq:Gaussian}):
\begin{equation}
\p(\BattleShipGrey{\hat{\delta}}^\mathrm{\BattleShipGrey{i}} | \DarkPastelGreen{\hat{S}}) = \frac{1}{\sqrt{\left|2\pi \DarkPastelGreen{\hat{S}} \right|}} \exp\left( - \frac{1}{2} \sum_{k,k'} \BattleShipGrey{\hat{\delta}}^\mathrm{\BattleShipGrey{i}}_k \DarkPastelGreen{\hat{S}}_{kk'}^{-1} \BattleShipGrey{\hat{\delta}}^\mathrm{\BattleShipGrey{i}}_{k'} \right) .
\end{equation}
where we explicitly noted by a hat the Fourier-space quantities.

The elements in matrix $\DarkPastelGreen{\hat{S}}$ are fixed parameters in {\borg}. They characterize the variance of the \glslink{initial conditions}{initial} \gls{density field} and therefore contain a cosmological dependence. We further assume that the covariance matrix $\DarkPastelGreen{\hat{S}}$ is diagonal in Fourier space (this is assuming \gls{statistical homogeneity} of the \glslink{initial conditions}{initial} \gls{density contrast}, as seen in section \ref{sec:power-spectrum}). The diagonal coefficients are $\sqrt{\DarkPastelGreen{P}(k)/(2\pi)^{3/2}}$, where $\DarkPastelGreen{P}(k)$ are the initial \glslink{power spectrum}{power spectra} coefficients for the adopted fiducial \gls{cosmological parameters}. They are chosen to follow the prescription of \citet{Eisenstein1998,Eisenstein1999}, including \glslink{BAO}{baryonic wiggles}.

Alternatively, using the configuration space representation yields
\begin{equation}
\label{eq:borg_prior_initial}
\p(\BattleShipGrey{\delta}^\mathrm{\BattleShipGrey{i}} | \DarkPastelGreen{S}) = \frac{1}{\sqrt{|2\pi \DarkPastelGreen{S}|}} \exp\left( - \frac{1}{2} \sum_{x,x'} \BattleShipGrey{\delta}^\mathrm{\BattleShipGrey{i}}_x \DarkPastelGreen{S}_{xx'}^{-1} \BattleShipGrey{\delta}^\mathrm{\BattleShipGrey{i}}_{x'} \right) .
\end{equation}

\subsubsection{Translating to the final density field}
\label{sec:Translating to the final density field}

Following \citet{Jasche2013BORG}, we now show that the problem of physical inference of \glslink{final conditions}{final} \glslink{density field}{density fields} can be recast into the problem of inferring the corresponding \gls{initial conditions}, given the \gls{structure formation} model.

As seen before, it is straightforward to express a \gls{prior} in the \gls{initial conditions}, $\p(\BattleShipGrey{\delta}^\mathrm{\BattleShipGrey{i}})$. Given this, we can obtain a \gls{prior} distribution for the \glslink{final conditions}{final} \gls{density contrast} at \gls{scale factor} $a$ by using the standard formula for \glslink{conditional pdf}{conditional probabilities}:
\begin{eqnarray}
\p(\BattleShipGrey{\delta}^\mathrm{\BattleShipGrey{f}}) & = & \int \p(\BattleShipGrey{\delta}^\mathrm{\BattleShipGrey{f}}, \BattleShipGrey{\delta}^\mathrm{\BattleShipGrey{i}}) \, \drm \BattleShipGrey{\delta}^\mathrm{\BattleShipGrey{i}} \\
& = & \int \p(\BattleShipGrey{\delta}^\mathrm{\BattleShipGrey{f}} | \BattleShipGrey{\delta}^\mathrm{\BattleShipGrey{i}}) \, \p(\BattleShipGrey{\delta}^\mathrm{\BattleShipGrey{i}}) \, \drm \BattleShipGrey{\delta}^\mathrm{\BattleShipGrey{i}} .
\end{eqnarray}
For a deterministic model of \gls{structure formation} $\BattleShipGrey{\delta}^\mathrm{\BattleShipGrey{i}} \mapsto \mathcal{G}\left(\BattleShipGrey{\delta}^\mathrm{\BattleShipGrey{i}}, a\right)$, the \glslink{conditional pdf}{conditional probability} is given by \glslink{Dirac delta distribution}{Dirac delta distributions}:
\begin{equation}
\p(\BattleShipGrey{\delta}^\mathrm{\BattleShipGrey{f}} | \BattleShipGrey{\delta}^\mathrm{\BattleShipGrey{i}}) = \prod_x \updelta_\mathrm{D} \left( \BattleShipGrey{\delta}^\mathrm{\BattleShipGrey{f}}_x - \left[\mathcal{G}(\BattleShipGrey{\delta}^\mathrm{\BattleShipGrey{i}}, a)\right]_x \right) .
\end{equation}
Therefore, given a model $\mathcal{G}$ for \gls{structure formation}, a \gls{prior} distribution for the late-time \gls{density field} can be obtained by a two-step \gls{sampling} process:
\begin{enumerate}
\item drawing an \glslink{initial conditions}{initial condition} \glslink{particle realization}{realization} from the \gls{prior} $\p(\BattleShipGrey{\delta}^\mathrm{\BattleShipGrey{i}})$;
\item propagating the \glslink{initial conditions}{initial} state \glslink{forward modeling}{forward} in time with $\mathcal{G}$ (this step is entirely deterministic).
\end{enumerate}
This process amounts to drawing \glslink{sample}{samples} from the joint \gls{prior} distribution of \glslink{initial conditions}{initial} and \gls{final conditions}:
\begin{equation}
\p(\BattleShipGrey{\delta}^\mathrm{\BattleShipGrey{f}} , \BattleShipGrey{\delta}^\mathrm{\BattleShipGrey{i}}) = \p(\BattleShipGrey{\delta}^\mathrm{\BattleShipGrey{i}}) \prod_x \updelta_\mathrm{D} \left( \BattleShipGrey{\delta}^\mathrm{\BattleShipGrey{f}}_x - \left[\mathcal{G}(\BattleShipGrey{\delta}^\mathrm{\BattleShipGrey{i}}, a)\right]_x \right) .
\label{eq:joint_prior_borg}
\end{equation}
\glslink{marginal pdf}{Marginalization} over \glslink{initial conditions}{initial} density \glslink{particle realization}{realizations} then yields \glslink{sample}{samples} of the \glslink{physical density prior}{non-Gaussian prior} for \glslink{final conditions}{final} \glslink{density field}{density fields}. In practice, as \gls{initial conditions} are also interesting for a variety of cosmological applications, we do not discard them and we always store them, whenever we draw a \gls{sample} from the prior.

\subsubsection{The structure formation model}

Ideally, the \gls{structure formation} model should be \glslink{full gravity}{fully non-linear gravity}. For reasons of computational feasibility, in {\borg}, $\mathcal{G}$ is obtained from \glslink{2LPT}{second-order Lagrangian perturbation theory} and the \glslink{CiC}{cloud-in-cell} scheme. More specifically, the \glslink{initial conditions}{initial} \gls{density field} is populated by \gls{dark matter particles} that are evolved according to the equations for \gls{2LPT} displacements given in section \ref{sec:2LPT}. In the \glslink{final conditions}{final state}, these \glslink{dark matter particles}{particles} are assigned to the grid using a \gls{CiC} scheme, yielding the \glslink{final conditions}{final} \gls{density contrast} $\BattleShipGrey{\delta}^\mathrm{\BattleShipGrey{f}}$. The reader is referred to appendix \ref{apx:simulations} for details on the numerical implementation of \gls{2LPT} and \gls{CiC}.

Using equations \eqref{eq:borg_prior_initial} and \eqref{eq:joint_prior_borg}, the joint \glslink{physical density prior}{physical prior} for \glslink{initial conditions}{initial} and late-time \glslink{density field}{density fields} is found to be
\begin{equation}
\label{eq:borg_prior_full}
\p(\BattleShipGrey{\delta}^\mathrm{\BattleShipGrey{f}} , \BattleShipGrey{\delta}^\mathrm{\BattleShipGrey{i}} | \DarkPastelGreen{S}) = \frac{1}{\sqrt{|2\pi \DarkPastelGreen{S}|}} \exp\left( - \frac{1}{2} \sum_{x,x'} \BattleShipGrey{\delta}^\mathrm{\BattleShipGrey{i}}_x \DarkPastelGreen{S}_{xx'}^{-1} \BattleShipGrey{\delta}^\mathrm{\BattleShipGrey{i}}_{x'} \right) \prod_x \updelta_\mathrm{D} \left( \BattleShipGrey{\delta}^\mathrm{\BattleShipGrey{f}}_x - \left[\mathcal{G}(\BattleShipGrey{\delta}^\mathrm{\BattleShipGrey{i}}, a)\right]_x \right) .
\end{equation}

Note that the first part (corresponding to the \gls{initial conditions}) is more easily handled in Fourier space, while the second part (corresponding to the propagation from \glslink{initial conditions}{initial} to \gls{final conditions}) involves \glslink{density field}{density fields} in configuration space.

\subsection{The large-scale structure likelihood}

This section discusses the {\borg} \glslink{large-scale structure likelihood}{likelihood}, $\p(\RubineRed{d}|\BattleShipGrey{\delta}^\mathrm{\BattleShipGrey{i}})$. The \gls{data} $\RubineRed{d}$ used by {\borg} are galaxy (or matter tracer) \glslink{number function}{number counts} in each voxel of the discretized domain. To compute it, the position of galaxies is translated from spherical to Cartesian coordinates using the following coordinate transform:
\begin{eqnarray}
x & = & d_\mathrm{com}(z) \cos(\lambda) \cos(\eta) ,\\
y & = & d_\mathrm{com}(z) \cos(\lambda) \sin(\eta) ,\\
z & = & d_\mathrm{com}(z) \sin(\lambda) ,
\end{eqnarray}
with $\lambda$ being the \gls{declination}, $\eta$ the \gls{right ascension} and $d_\mathrm{com}(z)$ the radial \glslink{comoving coordinates}{comoving distance} to \gls{redshift} $z$ for the fiducial cosmology. Galaxies are then binned using the \glslink{NGP}{Nearest Grid Point} (\gls{NGP}) \glslink{mesh assignment}{assignment scheme} to get voxel-wise galaxy \glslink{number function}{number counts}.

\subsubsection{Splitting the galaxy distribution}

In order to account for the \gls{luminosity}-dependence of \gls{selection effects} and galaxy \glslink{bias}{biases}, we split the \gls{data} into several bins of \glslink{luminosity}{absolute magnitude}. In the following, $\ell$ labels one of these bins, and $\RubineRed{N^\ell}$ is the \gls{data} set containing the \glslink{number function}{number counts} of galaxies in the \gls{luminosity} bin $\ell$ and in voxel $x$, $\RubineRed{N}^{\RubineRed{\ell}}_x$.

{\borg} treats different \glslink{luminosity}{magnitude} bins as independent \gls{data} sets. Each of them is assigned a \gls{likelihood} function, $\p(\RubineRed{N^\ell}|\BattleShipGrey{\delta}^\mathrm{\BattleShipGrey{i}})$. Since it is fair to assume that galaxies in different \gls{luminosity} bins are independent and identically distributed, once the density field is given, the final \gls{likelihood} of the total \gls{data} set $\RubineRed{d} = \{ \RubineRed{N^\ell} \}$ is obtained by multiplying these \gls{likelihood} functions,
\begin{equation}
\label{eq:likelihood_splitting}
\p(\RubineRed{d}|\BattleShipGrey{\delta}^\mathrm{\BattleShipGrey{i}}) = \prod_\ell \p(\RubineRed{N^\ell}|\BattleShipGrey{\delta}^\mathrm{\BattleShipGrey{i}}) .
\end{equation}

\subsubsection{The galaxy distribution as an inhomogeneous Poisson process}

Galaxies are tracers of the mass distribution. The \gls{statistical uncertainty} due to the discrete nature of their distribution is often modeled as a \gls{Poisson process} \citep{Layzer1956,Peebles1980,Martinez2002}. Before {\borg}, \glslink{Poisson likelihood}{Poissonian likelihoods} have been successfully applied to perform reconstructions of the matter density by \citet{Kitaura2010,JascheKitaura2010,Jasche2010b}. Adopting this picture, we write
\begin{equation}
\label{eq:Poisson_likelihood}
\p(\RubineRed{N^\ell} | \lambda(\BattleShipGrey{\delta}^\mathrm{\BattleShipGrey{i}}) ) = \nprod_x \frac{\exp\left( - \lambda_x^\ell(\BattleShipGrey{\delta}^\mathrm{\BattleShipGrey{i}}) \right) \left( \lambda_x^\ell(\BattleShipGrey{\delta}^\mathrm{\BattleShipGrey{i}}) \right)^{\RubineRed{N}^{\RubineRed{\ell}}_x}}{\RubineRed{N}^{\RubineRed{\ell}}_x !} .
\end{equation}
The \gls{Poisson intensity field}, $\lambda^\ell(\BattleShipGrey{\delta}^\mathrm{\BattleShipGrey{i}})$, characterizes the expected number of galaxies in voxel $x$ given the \glslink{initial conditions}{initial} \gls{density contrast} $\BattleShipGrey{\delta}^\mathrm{\BattleShipGrey{i}}$. As it depends on the position, it is an \textit{inhomogeneous} \gls{Poisson process}.

Real galaxy \glslink{sample}{samples} can have a sub- or super-Poissonian behavior (i.e. be under- or over-dispersed), depending on \gls{local} and \gls{non-local} properties \citep{Mo1996,Somerville2001,Casas-Miranda2002}. These effects are neglected here, but in the context of large-scale structure \glslink{reconstruction}{reconstructions}, deviations from Poissonity have been introduced in the likelihood by \citet{Kitaura2012a,Ata2015}.

\subsubsection{The Poisson intensity field}

The expected number of galaxies in a voxel depends -- of course -- on the underlying large-scale structure, but also on galaxy \gls{bias}, \gls{redshift-space distortions}, dynamical processes along the observer's backwards \gls{lightcone}, \gls{selection effects}, and instrumental \gls{noise}. All these effects should in principle be taken into account in the \gls{Poisson intensity field}. In the following, we detail, step by step, how to go from $\BattleShipGrey{\delta}^\mathrm{\BattleShipGrey{i}}$ to $\lambda(\BattleShipGrey{\delta}^\mathrm{\BattleShipGrey{i}})$ in the {\borg} \glslink{large-scale structure likelihood}{likelihood}.

\begin{enumerate}
\item \textsf{\textbf{\small{Structure formation.}}} The first step is to translate \glslink{initial conditions}{initial} to evolved dark matter overdensity: \begin{equation}
\BattleShipGrey{\delta}^\mathrm{\BattleShipGrey{i}} \mapsto \mathcal{G}(\BattleShipGrey{\delta}^\mathrm{\BattleShipGrey{i}},a).
\end{equation}
As discussed before, for this step {\borg} relies on \gls{2LPT} instead of \glslink{full gravity}{fully non-linear gravitational dynamics}, meaning that there exists some degree of approximation in the \glslink{large-scale structure inference}{inference} process. Accurate quantification this level of approximation is unfortunately not currently possible, as it would require the fully non-linear inference process for reference, which so far is not computationally tractable.
\item \textsf{\textbf{\small{Lightcone effects.}}} Along with step 1, we could account for lightcone effects so that the distant structures are less evolved than the closest ones. This is exploiting the dependence of $\mathcal{G}$ on $a$ to build the dark matter density on the \gls{lightcone}. For simplicity, this is not currently implemented in {\borg}; rather, we run \gls{2LPT} up to $a=1$ everywhere. In the following we simplify the notations and we write $\BattleShipGrey{\delta}^\mathrm{\BattleShipGrey{f}} \equiv \mathcal{G}(\BattleShipGrey{\delta}^\mathrm{\BattleShipGrey{i}}) \equiv \mathcal{G}(\BattleShipGrey{\delta}^\mathrm{\BattleShipGrey{i}},a=1)$.
\item \textsf{\textbf{\small{Redshift-space distortions.}}} At this point, the \gls{data model} could also include a treatment of \gls{redshift-space distortions} \citep[see][]{Heavens1995,Tadros1999,Percival2004,Percival2005a,Percival2009}. Though not explicitly included in the present {\borg} \gls{data model}, we find empirically that \gls{redshift-space distortions} are mitigated by the \gls{prior} preference for \glslink{statistical homogeneity}{homogeneity} and \glslink{statistical isotropy}{isotropy} (see chapter \ref{chap:BORGSDSS}): {\borg} interprets deviations from isotropy as noise, and fits an isotropic distribution to the data.
\item \textsf{\textbf{\small{Galaxy bias.}}} The following step is to get the galaxy density $\rho_\mathrm{g}$ given the dark matter density $\rho$. This is making assumptions for \glslink{bias}{physical biasing} in \gls{galaxy formation}. Various \glslink{large-scale structure inference}{LSS inference} algorithms assume a linear \gls{bias} model. In order to be well defined, a \gls{Poisson likelihood} requires intensities of the inhomogeneous \gls{Poisson process} to be strictly positive. Since a linear \gls{bias} model does not guarantee a positive \gls{density field} and corresponding \glslink{Poisson intensity field}{Poisson intensity}, it is not applicable to the present case. For this reason, we assume a phenomenological power-law to account for galaxy \glslink{bias}{biasing}:
\begin{equation}
\rho_g \propto \Fuchsia{\beta} \rho^{\Fuchsia{\alpha}} .
\end{equation}
In \gls{luminosity} bin $\ell$ and in terms of the dark matter overdensity, this is step written 
\begin{equation}
\BattleShipGrey{\delta}^\mathrm{\BattleShipGrey{f}} \mapsto \Fuchsia{\beta^\ell} (1+\BattleShipGrey{\delta}^\mathrm{\BattleShipGrey{f}})^{\Fuchsia{\alpha^\ell}} \propto \rho_g^\ell.
\end{equation}
Note that coefficients $\Fuchsia{\alpha^\ell}$ and $\Fuchsia{\beta^\ell}$ depend on $\ell$, which means that the \gls{data model} accounts for \textit{\gls{luminosity}-dependent} galaxy \glslink{bias}{biases}. Parameters $\Fuchsia{\beta^\ell}$ are automatically calibrated during the generation of the Markov Chain (see section \ref{sec:Calibration of the noise level}). For simplicity, parameters $\Fuchsia{\alpha^\ell}$ are kept at fixed, fiducial values. In the {\borg} analysis of the \gls{SDSS} (chapter \ref{chap:BORGSDSS}), these values are determined using a standard model for \gls{luminosity}-dependent galaxy \gls{bias}. In their analysis of the 2M++ catalog \citep{Lavaux2011}, \citet{Lavaux2016BORG2MPP} show that it is possible to calibrate these values with a preliminary {\ares} inference, for subsequent use in {\borg}.
\item \textsf{\textbf{\small{Mean number of galaxies.}}} To get the expected number of galaxies from the unnormalized galaxy density, the quantity $\Fuchsia{\beta^\ell} (1+\BattleShipGrey{\delta}^\mathrm{\BattleShipGrey{f}})^{\Fuchsia{\alpha^\ell}}$ has to be multiplied by the mean number of galaxies in bin $\ell$, $\bar{N}^\ell$. This step is therefore simply:
\begin{equation}
\Fuchsia{\beta^\ell} (1+\BattleShipGrey{\delta}^\mathrm{\BattleShipGrey{f}})^{\Fuchsia{\alpha^\ell}} \mapsto \bar{N}^\ell \Fuchsia{\beta^\ell} (1+\BattleShipGrey{\delta}^\mathrm{\BattleShipGrey{f}})^{\Fuchsia{\alpha^\ell}} .
\end{equation}
\item \textsf{\textbf{\small{Observational effects.}}} The last step is to put in the \gls{luminosity}-dependent \gls{selection effects} and the survey \gls{mask}. For this, we multiply with the linear \gls{survey response operator} $\RoyalBlue{R}^{\RoyalBlue{\ell}}_x$, a voxel-wise three-dimensional function that incorporates \glslink{survey geometry}{survey geometries} and \gls{selection effects}:
\begin{equation}
\bar{N}^\ell \Fuchsia{\beta^\ell} (1+\BattleShipGrey{\delta}^\mathrm{\BattleShipGrey{f}}_x)^{\Fuchsia{\alpha^\ell}} \mapsto \RoyalBlue{R}^{\RoyalBlue{\ell}}_x \bar{N}^\ell \Fuchsia{\beta^\ell} (1+\BattleShipGrey{\delta}^\mathrm{\BattleShipGrey{f}}_x)^{\Fuchsia{\alpha^\ell}} .
\end{equation}
\end{enumerate}

\begin{figure}
\begin{center}
\includegraphics[width=\textwidth]{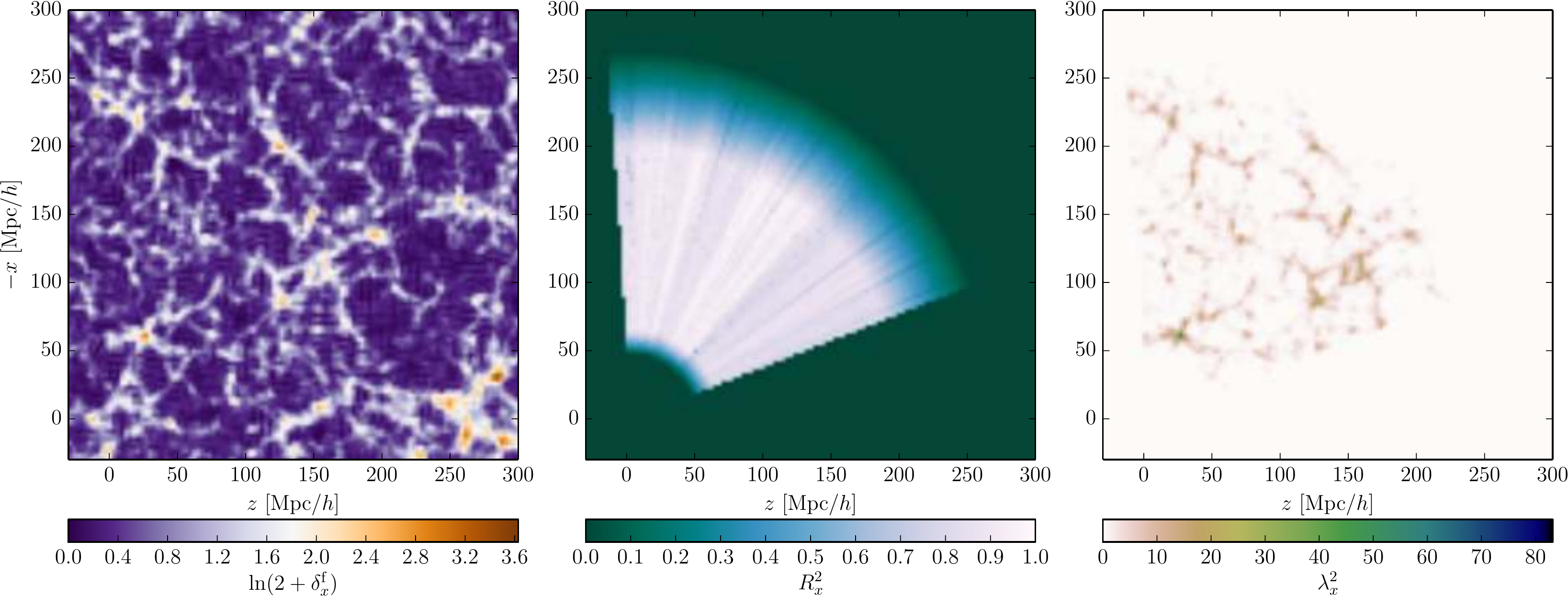} 
\caption{Slices through the box used in the {\borg} \gls{SDSS} analysis (see chapter \ref{chap:BORGSDSS}). \textit{Left panel}. Density in one \gls{sample} (for clarity, the quantity shown is $\ln(2+\BattleShipGrey{\delta}^\mathrm{\BattleShipGrey{f}}_x)$). \textit{Middle panel}. \glslink{survey response operator}{Survey response operator} $\RoyalBlue{R}^{\RoyalBlue{2}}_x$ in the $\ell=2$ \gls{luminosity} bin, corresponding to absolute $r$-band \glslink{luminosity}{magnitudes} in the range $-19.67~<~M^2_{{0.1}_r}~<~-19.00$. \textit{Right panel}. \gls{Poisson intensity field} $\lambda^2_x$ for this \gls{sample} and \gls{luminosity} bin, computed with equation \eqref{eq:Poisson_intensity}. The \gls{bias} and \glslink{noise parameter}{noise parameters} are respectively $\Fuchsia{\alpha^2}=1.30822$ and $\Orange{\widetilde{N}^2}=1.39989$ (see table \ref{tb:table_borg_sdss_lum_bins}).}
\label{fig:borg_poisson_intensity}
\end{center}
\end{figure}

Eventually, the \gls{Poisson intensity field} is given by
\begin{equation}
\lambda^\ell_x(\BattleShipGrey{\delta}^\mathrm{\BattleShipGrey{i}}) = \RoyalBlue{R}^{\RoyalBlue{\ell}}_x \bar{N}^\ell \Fuchsia{\beta^\ell} \left(1+\left[\mathcal{G}(\BattleShipGrey{\delta}^\mathrm{\BattleShipGrey{i}})\right]_x\right)^{\Fuchsia{\alpha^\ell}} .
\end{equation}
We note that $\bar{N}^\ell$ and $\Fuchsia{\beta^\ell}$ are degenerate, in the sense that only the product $\bar{N}^\ell \Fuchsia{\beta^\ell}$ matters. We define $\Orange{\widetilde{N}^\ell} \equiv \bar{N}^\ell \Fuchsia{\beta^\ell}$, so that
\begin{equation}
\label{eq:Poisson_intensity}
\lambda^\ell_x(\BattleShipGrey{\delta}^\mathrm{\BattleShipGrey{i}}) = \RoyalBlue{R}^{\RoyalBlue{\ell}}_x \Orange{\widetilde{N}^\ell} \left(1+\left[\mathcal{G}(\BattleShipGrey{\delta}^\mathrm{\BattleShipGrey{i}})\right]_x\right)^{\Fuchsia{\alpha^\ell}}.
\end{equation}
$\Orange{\widetilde{N}^\ell}$ represents the overall \glslink{noise parameter}{noise level} in bin $\ell$. With the improved {\borg} \gls{data model} \citep{Jasche2015BORGSDSS}, we automatically calibrate this parameter (see section \ref{sec:Calibration of the noise level}). In figure \ref{fig:borg_poisson_intensity}, we illustrate the construction of the \gls{Poisson intensity field} for the $\ell=2$ bin of the \gls{SDSS} analysis. We show the dark matter density, $\BattleShipGrey{\delta}^\mathrm{\BattleShipGrey{f}}_x$, the \gls{survey response operator} $\RoyalBlue{R}^{\RoyalBlue{2}}_x$ and the \glslink{Poisson intensity field}{Poisson intensity} $\lambda^2_x$.

\subsubsection{The comprehensive large-scale structure likelihood}

Noting $\RubineRed{d} \equiv \{\RubineRed{N^\ell}\}$ the total \gls{data} set, i.e. all available galaxy \glslink{number function}{number counts}, and $\Orange{\widetilde{N}} \equiv \{\Orange{\widetilde{N}^\ell}\}$ the set of \glslink{noise parameter}{noise parameters} in each bin, we obtain the final expression for the \glslink{large-scale structure likelihood}{LSS likelihood} using equations \eqref{eq:likelihood_splitting}, \eqref{eq:Poisson_likelihood} and \eqref{eq:Poisson_intensity}. It reads
\begin{equation}
\label{eq:borg_likelihood}
\p(\RubineRed{d}|\BattleShipGrey{\delta}^\mathrm{\BattleShipGrey{i}}, \Orange{\widetilde{N}}) = \nprod_{x,\ell} \frac{\exp\left( - \RoyalBlue{R}^{\RoyalBlue{\ell}}_x \Orange{\widetilde{N}^\ell} (1+\left[\mathcal{G}(\BattleShipGrey{\delta}^\mathrm{\BattleShipGrey{i}})\right]_x)^{\Fuchsia{\alpha^\ell}} \right) \left( \RoyalBlue{R}^{\RoyalBlue{\ell}}_x \Orange{\widetilde{N}^\ell} (1+\left[\mathcal{G}(\BattleShipGrey{\delta}^\mathrm{\BattleShipGrey{i}})\right]_x)^{\Fuchsia{\alpha^\ell}} \right)^{\RubineRed{N}^{\RubineRed{\ell}}_x}}{\RubineRed{N}^{\RubineRed{\ell}}_x !}
\end{equation}
In this equation, we omitted on the right side of the conditioning bar the sets $\{\RoyalBlue{R}^{\RoyalBlue{\ell}}_x\}$ and $\{\Fuchsia{\alpha^\ell}\}$ (one can consider that all probabilities inferred by {\borg} are \glslink{conditional pdf}{conditional} on these). However, we now write explicitly $\Orange{\widetilde{N}}$, as this will be of importance later.

\subsection{The posterior distribution}

As usual in \gls{Bayesian statistics}, the \gls{posterior} distribution is obtained, up to a normalization constant, by the use of \glslink{Bayes' theorem}{Bayes' formula},
\begin{equation}
\p(\BattleShipGrey{\delta}^\mathrm{\BattleShipGrey{i}}|\RubineRed{d}, \DarkPastelGreen{S}, \Orange{\widetilde{N}}) \propto \p(\BattleShipGrey{\delta}^\mathrm{\BattleShipGrey{i}}|\DarkPastelGreen{S}, \Orange{\widetilde{N}}) \, \p(\RubineRed{d}|\BattleShipGrey{\delta}^\mathrm{\BattleShipGrey{i}},\DarkPastelGreen{S}, \Orange{\widetilde{N}}) = \p(\BattleShipGrey{\delta}^\mathrm{\BattleShipGrey{i}}|\DarkPastelGreen{S}) \, \p(\RubineRed{d}|\BattleShipGrey{\delta}^\mathrm{\BattleShipGrey{i}}, \Orange{\widetilde{N}}).
\end{equation}

Substituting equations \eqref{eq:borg_prior_initial} and \eqref{eq:borg_likelihood} allows to write down the full problem solved by {\borg} for the density distribution:
\begin{equation}
\label{eq:borg_posterior}
\p(\BattleShipGrey{\delta}^\mathrm{\BattleShipGrey{i}}|\RubineRed{d}, \DarkPastelGreen{S}, \Orange{\widetilde{N}}) \propto \frac{1}{\sqrt{|2\pi \DarkPastelGreen{S}|}} \exp\left( - \frac{1}{2} \sum_{x,x'} \BattleShipGrey{\delta}^\mathrm{\BattleShipGrey{i}}_x \DarkPastelGreen{S}_{xx'}^{-1} \BattleShipGrey{\delta}^\mathrm{\BattleShipGrey{i}}_{x'} \right) \nprod_{x,\ell} \frac{\exp\left( - \RoyalBlue{R}^{\RoyalBlue{\ell}}_x \Orange{\widetilde{N}^\ell} (1+\left[\mathcal{G}(\BattleShipGrey{\delta}^\mathrm{\BattleShipGrey{i}})\right]_x)^{\Fuchsia{\alpha^\ell}} \right) \left( \RoyalBlue{R}^{\RoyalBlue{\ell}}_x \Orange{\widetilde{N}^\ell} (1+\left[\mathcal{G}(\BattleShipGrey{\delta}^\mathrm{\BattleShipGrey{i}})\right]_x)^{\Fuchsia{\alpha^\ell}} \right)^{\RubineRed{N}^{\RubineRed{\ell}}_x}}{\RubineRed{N}^{\RubineRed{\ell}}_x !} .
\end{equation}
It is simpler to express the {\borg} \gls{posterior} in terms of the \gls{initial conditions}, but recall that one gets the \gls{final conditions} (and in fact the entire \gls{LSS} \glslink{formation history}{history}, as demonstrated in chapter \ref{chap:BORGSDSS}) automatically and entirely deterministically via the \gls{structure formation} model $\mathcal{G}$ (see section \ref{sec:Translating to the final density field}):
\begin{equation}
\p(\BattleShipGrey{\delta}^\mathrm{\BattleShipGrey{f}},\BattleShipGrey{\delta}^\mathrm{\BattleShipGrey{i}}|\RubineRed{d}, \DarkPastelGreen{S}, \Orange{\widetilde{N}}) = \p(\BattleShipGrey{\delta}^\mathrm{\BattleShipGrey{i}}|\RubineRed{d}, \DarkPastelGreen{S}, \Orange{\widetilde{N}}) \prod_x \updelta_\mathrm{D} \left( \BattleShipGrey{\delta}^\mathrm{\BattleShipGrey{f}}_x - \left[\mathcal{G}(\BattleShipGrey{\delta}^\mathrm{\BattleShipGrey{i}})\right]_x \right) .
\end{equation}

\subsection{The $\Gamma$-distribution for noise sampling}
\label{sec:The Gamma-distribution for noise sampling}

\draw{This section draws from appendix A of \citet{Jasche2015BORGSDSS}.}

We aim at automatically calibrating, during the \gls{sampling} procedure, the \glslink{noise parameter}{noise level} of each \gls{luminosity} bin, given the \gls{data} and the current density \gls{sample}. This requires to write down the \glslink{conditional pdf}{conditional probability} $\p(\Orange{\widetilde{N}^\ell} | \RubineRed{N^\ell}, \BattleShipGrey{\delta}^\mathrm{\BattleShipGrey{f}})$, which we do in this section.

According to \glslink{Bayes' theorem}{Bayes' formula}, we can write
\begin{equation}
\label{eq:Bayes_noise_sampling}
\p(\Orange{\widetilde{N}^\ell} | \RubineRed{N^\ell}, \BattleShipGrey{\delta}^\mathrm{\BattleShipGrey{f}}) \propto \p(\Orange{\widetilde{N}^\ell}) \, \p(\RubineRed{N^\ell} | \Orange{\widetilde{N}^\ell}, \BattleShipGrey{\delta}^\mathrm{\BattleShipGrey{f}}),
\end{equation}
where we have assumed the \gls{conditional independence} $\p(\Orange{\widetilde{N}^\ell} | \BattleShipGrey{\delta}^\mathrm{\BattleShipGrey{f}}) = \p(\Orange{\widetilde{N}^\ell})$. In the absence of any further information on the parameter $\Orange{\widetilde{N}^\ell}$, we follow the maximum agnostic approach pursued by \citet{Jasche2013BIAS} by setting the \gls{prior} distribution $\Orange{\widetilde{N}^\ell}$ constant. By using the \gls{Poisson likelihood} for $\p(\RubineRed{N^\ell} | \Orange{\widetilde{N}^\ell}, \BattleShipGrey{\delta}^\mathrm{\BattleShipGrey{f}})$ (equations \eqref{eq:Poisson_likelihood} and \eqref{eq:Poisson_intensity}) into equation \eqref{eq:Bayes_noise_sampling}, we obtain the \glslink{conditional pdf}{conditional} \gls{posterior} for the \gls{noise parameter} $\Orange{\widetilde{N}^\ell}$ as:
\begin{equation}
\p(\Orange{\widetilde{N}^\ell} | \RubineRed{N^\ell}, \BattleShipGrey{\delta}^\mathrm{\BattleShipGrey{f}}) \propto \exp\left( - \Orange{\widetilde{N}^\ell} A_\ell \right) \times \left( \Orange{\widetilde{N}^\ell} \right)^{B_\ell}, 
\end{equation}
where $A_\ell \equiv \sum\limits_x \RoyalBlue{R}^{\RoyalBlue{\ell}}_x (1+\BattleShipGrey{\delta}^\mathrm{\BattleShipGrey{f}}_x)^{\Fuchsia{\alpha^\ell}}$ and $B_\ell \equiv \sum\limits_x \RubineRed{N}^{\RubineRed{\ell}}_x$. By choosing $\RubineRed{k_\ell} \equiv B_\ell + 1$ and $\theta_\ell \equiv 1/A_\ell$, we yield a properly normalized \glslink{Gamma distribution}{$\Gamma$-distribution} for the \gls{noise parameter} $\Orange{\widetilde{N}^\ell}$, given as:
\begin{equation}
\p(\Orange{\widetilde{N}^\ell} | \RubineRed{N^\ell}, \BattleShipGrey{\delta}^\mathrm{\BattleShipGrey{f}}) = \Gamma \! \left[ \RubineRed{k_\ell}, \theta_\ell \right] \!\left( \Orange{\widetilde{N}^\ell} \right) = \frac{\left(\Orange{\widetilde{N}^\ell}\right)^{\RubineRed{k_\ell} -1} \exp\left( - \frac{\Orange{\widetilde{N}^\ell}}{\theta_\ell} \right)}{\theta_\ell^{\RubineRed{k_\ell}} \Gamma(\RubineRed{k_\ell})} .
\end{equation}
with shape parameter
\begin{equation}
\RubineRed{k_\ell} \equiv 1 + \sum_x \RubineRed{N}^{\RubineRed{\ell}}_x,
\end{equation}
and scale parameter
\begin{equation}
\theta_\ell \equiv \frac{1}{\sum\limits_x \RoyalBlue{R}^{\RoyalBlue{\ell}}_x (1+\BattleShipGrey{\delta}^\mathrm{\BattleShipGrey{f}}_x)^{\Fuchsia{\alpha^\ell}}} .
\end{equation}

\section{Sampling procedure and numerical implementation}
\label{sec:Sampling procedure and numerical implementation}

\subsection{Calibration of the noise level}
\label{sec:Calibration of the noise level}

\draw{This section draws from section 3.2. in \citet{Jasche2015BORGSDSS}.}

Following the approach described in \citet{Jasche2013BIAS}, \glslink{density field}{density fields} and \glslink{noise parameter}{noise level parameters} can be jointly inferred by introducing an additional \gls{sampling} block to the original implementation of the {\borg} algorithm. The additional \gls{sampling} block is designed to provide random \glslink{sample}{samples} of the \glslink{noise parameter}{noise parameters} $\Orange{\widetilde{N}^\ell}$ given the galaxy \gls{data} set $\RubineRed{N^\ell}$ and the current \glslink{final conditions}{final} density \gls{sample} $\BattleShipGrey{\delta}^\mathrm{\BattleShipGrey{f}}$. 

As indicated by figure \ref{fig:borg_block_sampling}, in a first step, the algorithm infers \glslink{density field}{density fields}, then \glslink{conditional pdf}{conditionally} \glslink{sampling}{samples} the \glslink{noise parameter}{noise parameters}. Iteration of this procedure yields Markovian \glslink{sample}{samples} from the joint target distribution. 

\begin{figure}
\begin{center}
\begin{tikzpicture}
	\pgfdeclarelayer{background}
	\pgfdeclarelayer{foreground}
	\pgfsetlayers{background,main,foreground}

	\tikzstyle{sampling}=[draw, text width=8em, text centered, rounded corners, minimum height=2.5em, minimum width=8em]
	\tikzstyle{data}=[draw, fill=RubineRed!20, text width=5em, text centered, minimum height=2.5em,drop shadow]
	\tikzstyle{Ntilde}=[draw, fill=Orange!20, text width=5em, text centered, minimum height=2.5em,drop shadow]
	\tikzstyle{delta}=[draw, fill=black!20, text width=5em, text centered, minimum height=2.5em,drop shadow]

	\def\blockdist{2.3}
	\def\edgedist{2.5}

    \node (datatop) [data] {data $\RubineRed{d} = \{ \RubineRed{N^\ell} \}$};
    \path (datatop.south)+(0,-1.0) node (deltasampling) [sampling] {density sampling\linebreak $\p(\BattleShipGrey{\delta}^\mathrm{\BattleShipGrey{f}},\BattleShipGrey{\delta}^\mathrm{\BattleShipGrey{i}}|\RubineRed{d}, \DarkPastelGreen{S}, \Orange{\widetilde{N}})$};
    \path (datatop.east)+(\blockdist,-\blockdist-1.0) node (Ntildesamples) [Ntilde] {new $\Orange{\widetilde{N}}$ sample};
    \path (datatop.west)+(-\blockdist,-\blockdist-1.0) node (deltasamples) [delta] {new $\delta$ sample};
    \path (datatop.south)+(0,-\blockdist-\blockdist) node (Ntildesampling) [sampling] {noise sampling\linebreak $\p(\Orange{\widetilde{N}^\ell} | \RubineRed{N^\ell}, \BattleShipGrey{\delta}^\mathrm{\BattleShipGrey{f}})$};
    \path (datatop.south)+(0,-\blockdist-\blockdist-1.6) node (databottom) [data] {data $\RubineRed{d} = \{ \RubineRed{N^\ell} \}$};

	\path [draw, line width=0.7pt, arrows={-latex}] (datatop.south) -- (deltasampling.north) {} (deltasampling) ;
	\path [draw, line width=0.7pt, arrows={-latex}] (databottom.north) -- (Ntildesampling.south) {} (Ntildesampling) ;
	\path [draw, line width=0.7pt, arrows={-latex}] (Ntildesamples.north) -- (deltasampling.east);
	\path [draw, line width=0.7pt, arrows={-latex}] (deltasampling.west) -- (deltasamples.north);
	\path [draw, line width=0.7pt, arrows={-latex}] (deltasamples.south) -- (Ntildesampling.west);
	\path [draw, line width=0.7pt, arrows={-latex}] (Ntildesampling.east) -- (Ntildesamples.south);

\end{tikzpicture}
\end{center}
\caption{Flow chart depicting the multi-step iterative block \gls{sampling} procedure. In the first step, {\borg} generates random realizations of the \glslink{initial conditions}{initial} and \glslink{final conditions}{final} \glslink{density field}{density fields} \glslink{conditional pdf}{conditional} on the galaxy \glslink{sample}{samples} $\RubineRed{d}$ and on the \glslink{noise parameter}{noise levels} $\{\Orange{\widetilde{N}^\ell}\}$. In a subsequent step, the \glslink{noise parameter}{noise parameters} $\Orange{\widetilde{N}^\ell}$ are \glslink{sampling}{sampled} \glslink{conditional pdf}{conditional} on the previous density \glslink{particle realization}{realizations}.\label{fig:borg_block_sampling}}
\end{figure}

As demonstrated in section \ref{sec:The Gamma-distribution for noise sampling}, the \gls{posterior} distributions of \glslink{noise parameter}{noise parameters} $\Orange{\widetilde{N}^\ell}$ are \glslink{Gamma distribution}{$\Gamma$-distributions}. In the new \gls{sampling} block, random variates of the \glslink{Gamma distribution}{$\Gamma$-distribution} are generated by standard routines provided by the GNU scientific library \citep{GSL}.

\subsection{Hamiltonian Monte Carlo and equations of motion for the LSS density}
\label{sec:Hamiltonian Monte Carlo and equations of motion for the LSS density}

\glslink{sampling}{Sampling} of the \gls{posterior} distribution for \glslink{density field}{density fields} is achieved via \glslink{HMC}{Hamiltonian Monte Carlo}. As described in section \ref{sec:Hamiltonian Monte Carlo}, \gls{HMC} permits to explore the non-linear \gls{posterior} by following Hamiltonian dynamics in the \gls{high-dimensional parameter space}. Omitting normalization constants, the Hamiltonian potential $\psi(\BattleShipGrey{\delta}^\mathrm{\BattleShipGrey{i}})$ can be written as:
\begin{eqnarray}
\psi(\BattleShipGrey{\delta}^\mathrm{\BattleShipGrey{i}}) & = & -\ln \p(\BattleShipGrey{\delta}^\mathrm{\BattleShipGrey{i}}|\RubineRed{d}, \DarkPastelGreen{S}, \Orange{\widetilde{N}}) - \ln Z\\
& = & \psi_\mathrm{prior}(\BattleShipGrey{\delta}^\mathrm{\BattleShipGrey{i}}) + \psi_\mathrm{likelihood}(\BattleShipGrey{\delta}^\mathrm{\BattleShipGrey{i}}) ,
\end{eqnarray}
with the ``\gls{prior} potential'' $\psi_\mathrm{prior}(\BattleShipGrey{\delta}^\mathrm{\BattleShipGrey{i}})$ given as
\begin{equation}
\psi_\mathrm{prior}(\BattleShipGrey{\delta}^\mathrm{\BattleShipGrey{i}}) = \frac{1}{2} \sum_{x,x'} \BattleShipGrey{\delta}^\mathrm{\BattleShipGrey{i}}_x \DarkPastelGreen{S}_{xx'}^{-1} \BattleShipGrey{\delta}^\mathrm{\BattleShipGrey{i}}_{x'} ,
\end{equation}
and the ``\gls{likelihood} potential'' $\psi_\mathrm{likelihood}(\BattleShipGrey{\delta}^\mathrm{\BattleShipGrey{i}})$ given as
\begin{equation}
\psi_\mathrm{likelihood}(\BattleShipGrey{\delta}^\mathrm{\BattleShipGrey{i}}) = \sum_{x,\ell} \RoyalBlue{R}^{\RoyalBlue{\ell}}_x \Orange{\widetilde{N}^\ell} \left(1+\left[\mathcal{G}(\BattleShipGrey{\delta}^\mathrm{\BattleShipGrey{i}})\right]_x\right)^{\Fuchsia{\alpha^\ell}} - \RubineRed{N}^{\RubineRed{\ell}}_x \ln \left( \RoyalBlue{R}^{\RoyalBlue{\ell}}_x \Orange{\widetilde{N}^\ell} \left(1+\left[\mathcal{G}(\BattleShipGrey{\delta}^\mathrm{\BattleShipGrey{i}})\right]_x \right)^{\Fuchsia{\alpha^\ell}} \right) .
\end{equation}

Given the above definitions of the potential $\psi(\BattleShipGrey{\delta}^\mathrm{\BattleShipGrey{i}})$, one can obtain the required Hamiltonian force (see equation \eqref{eq:Hamiltonian_force}) by differentiating with respect to $\BattleShipGrey{\delta}^\mathrm{\BattleShipGrey{i}}_x$:
\begin{equation}
\pd{\psi(\BattleShipGrey{\delta}^\mathrm{\BattleShipGrey{i}})}{\BattleShipGrey{\delta}^\mathrm{\BattleShipGrey{i}}_x} = \pd{\psi_\mathrm{prior}(\BattleShipGrey{\delta}^\mathrm{\BattleShipGrey{i}})}{\BattleShipGrey{\delta}^\mathrm{\BattleShipGrey{i}}_x} + \pd{\psi_\mathrm{likelihood}(\BattleShipGrey{\delta}^\mathrm{\BattleShipGrey{i}})}{\BattleShipGrey{\delta}^\mathrm{\BattleShipGrey{i}}_x} .
\end{equation}
The \gls{prior} term is given by
\begin{equation}
\pd{\psi_\mathrm{prior}(\BattleShipGrey{\delta}^\mathrm{\BattleShipGrey{i}})}{\BattleShipGrey{\delta}^\mathrm{\BattleShipGrey{i}}_x} = \sum_{x'} \DarkPastelGreen{S}_{xx'}^{-1} \BattleShipGrey{\delta}^\mathrm{\BattleShipGrey{i}}_{x'}
\end{equation}

The \gls{likelihood} term cannot be obtained trivially. However, the choice of \gls{2LPT} and a \gls{CiC} kernel to model $\mathcal{G}(\BattleShipGrey{\delta}^\mathrm{\BattleShipGrey{i}})$ makes possible to derive this term analytically. This is of crucial importance, because a numerical estimation of gradients is very expensive. A detailed computation can be found in appendix D of \citet{Jasche2013BORG}. The result is
\begin{equation}
\pd{\psi_\mathrm{likelihood}(\BattleShipGrey{\delta}^\mathrm{\BattleShipGrey{i}})}{\BattleShipGrey{\delta}^\mathrm{\BattleShipGrey{i}}_x} = - D_1 J_x + D_2 \sum_{a>b} \left( \uptau_x^{aabb} + \uptau_x^{bbaa} - 2\uptau_x^{abab} \right) ,
\end{equation}
where $D_1$ and $D_2$ are the \glslink{linear growth factor}{first} and \glslink{second-order growth factor}{second-order growth factors} at the desired time ($a=1$), and $J_x$ and $\uptau_x^{abcd}$ are a vector and a tensor that depend on $\RoyalBlue{R}^{\RoyalBlue{\ell}}_x$, $\Orange{\widetilde{N}^\ell}$, $\Fuchsia{\alpha^\ell}$, $\RubineRed{N}^{\RubineRed{\ell}}_x$. 

Finally, the \glslink{equation of motion}{equations of motion} for the Hamiltonian system can be written as 
\begin{eqnarray}
\deriv{\BattleShipGrey{\delta}^\mathrm{\BattleShipGrey{i}}_x}{t} & = & \sum_{x'} M_{xx'}^{-1} \, \BattleShipGrey{p}_{x'} , \label{eq:Hamilton_borg_1}\\
\deriv{\BattleShipGrey{p}_x}{t} & = & - \sum_{x'} \DarkPastelGreen{S}_{xx'}^{-1} \BattleShipGrey{\delta}^\mathrm{\BattleShipGrey{i}}_{x'} + D_1 J_x(\BattleShipGrey{\delta}^\mathrm{\BattleShipGrey{i}}) - D_2 \sum_{a>b} \left( \uptau_x^{aabb}(\BattleShipGrey{\delta}^\mathrm{\BattleShipGrey{i}}) + \uptau_x^{bbaa}(\BattleShipGrey{\delta}^\mathrm{\BattleShipGrey{i}}) - 2\uptau_x^{abab}(\BattleShipGrey{\delta}^\mathrm{\BattleShipGrey{i}}) \right) \label{eq:Hamilton_borg_2}
\end{eqnarray}

\subsection{The mass matrix}

As mentioned in section \ref{sec:Hamiltonian Monte Carlo}, the \gls{HMC} algorithm possesses a large number of tunable parameters contained in the \gls{mass matrix} $M$, whose choice can strongly impact the efficiency of the sampler. As shown in \citet[][section 5.2 and appendix F]{Jasche2013BORG}, a good approach to obtain suitable masses is to perform a stability analysis of the numerical \gls{leapfrog} scheme (see section \ref{sec:The leapfrog scheme integrator}) implemented as integrator. This results in the following prescription:
\begin{equation}
\label{eq:mass_matrix}
M_{xx'} \equiv \DarkPastelGreen{S}^{-1}_{xx'} - \updelta_\mathrm{K}^{xx'} D_1 \pd{J_x(\BattleShipGrey{\delta}^\mathrm{\BattleShipGrey{i}})}{\BattleShipGrey{\delta}^{\mathrm{\BattleShipGrey{i}}}_x} \left( \xi_x \right) ,
\end{equation}
where $\updelta_\mathrm{K}$ is a \glslink{Kronecker symbol}{Kronecker delta symbol} and $\xi_x$ is assumed to be the mean \glslink{initial conditions}{initial} \gls{density contrast} in high probability regions, i.e. once the sampler has moved beyond the \gls{burn-in} phase.

Due to the \glslink{high-dimensional parameter space}{high-dimensionality} of the problem, inverting $M$ and storing $M^{-1}$ is computationally impractical. Therefore, a diagonal \gls{mass matrix} is constructed from equation \eqref{eq:mass_matrix}.

\subsection{The leapfrog scheme integrator}
\label{sec:The leapfrog scheme integrator}

For computer implementation, \gls{Hamilton's equations}, \eqref{eq:Hamilton_borg_1} and \eqref{eq:Hamilton_borg_2}, must be approximated by discretizing time, using some small stepsize, $\varepsilon$. Several choices of integrator, such as the popular Euler's method, are possible (see section \ref{sec:apx_Time_integrators}).

As discussed in section \ref{sec:Hamiltonian Monte Carlo}, it is essential that the adopted scheme respect \gls{reversibility} and \gls{symplecticity}, to ensure incompressibility in \gls{phase space}. Additionally, achieving high \glslink{acceptance rate}{acceptance rates} require the numerical integration scheme to be very accurate in order to conserve the Hamiltonian. For these reasons, the integrator adopted for implementing {\borg} is the \gls{leapfrog} scheme \citep[e.g.][]{Birdsall1985}, which relies on a sequence of ``\glslink{kick}{}\glslink{drift}{}\glslink{KDK}{kick--drift--kick}'' operations that work as follows (see also figure \ref{fig:KDK}):
\begin{eqnarray}
\BattleShipGrey{p}_x \left(t+\frac{\varepsilon}{2}\right) & = & \BattleShipGrey{p}_x(t) - \frac{\varepsilon}{2} \, \pd{\psi(\BattleShipGrey{\delta}^\mathrm{\BattleShipGrey{i}})}{\BattleShipGrey{\delta}^\mathrm{\BattleShipGrey{i}}_x} \left( \BattleShipGrey{\delta}^\mathrm{\BattleShipGrey{i}}_x \left( t \right) \right) , \\
\BattleShipGrey{\delta}^\mathrm{\BattleShipGrey{i}}_x \left( t+\varepsilon \right) & = & \BattleShipGrey{\delta}^\mathrm{\BattleShipGrey{i}}_x \left( t \right) + \varepsilon \, \frac{\BattleShipGrey{p}_x\left(t+\frac{\varepsilon}{2}\right)}{m_x} ,\\
\BattleShipGrey{p}_x \left( t+\varepsilon \right) & = & \BattleShipGrey{p}_x \left(t+\frac{\varepsilon}{2}\right) - \frac{\varepsilon}{2} \, \pd{\psi(\BattleShipGrey{\delta}^\mathrm{\BattleShipGrey{i}})}{\BattleShipGrey{\delta}^\mathrm{\BattleShipGrey{i}}_x} \left( \BattleShipGrey{\delta}^\mathrm{\BattleShipGrey{i}}_x \left( t+\varepsilon \right) \right) ,
\end{eqnarray}
where $m_x$ is the element of the diagonal \gls{mass matrix} at position $x$.

The \glslink{equation of motion}{equations of motion} are integrated by making $n$ such steps with a finite step size $\varepsilon$. In order to prevent resonant trajectories, time steps are slightly randomized ($\varepsilon$ is randomly drawn from a uniform distribution).

\section{Testing BORG}
\label{sec:Testing BORG}

Demonstrating of the performance of the {\borg} algorithm is the subject of sections 6 and 7 in \citet{Jasche2013BORG}. As these results are relevant to set the {\borg} \gls{SDSS} analysis on firm statistical grounds, in the following, we briefly report on the original test using \glslink{mock catalog}{mock observations}. 

\subsection{Generating mock observations}

The first step is to generate an \glslink{initial conditions}{initial} \glslink{grf}{Gaussian random field} (see section \ref{sec:apx-Density assignments}). This was done on a three-dimensional Cartesian grid of $128^3$ voxels covering a \glslink{comoving coordinates}{comoving} cubic box of length 750~Mpc/$h$ with \gls{periodic boundary conditions}. The Fourier-space covariance matrix includes an \citet{Eisenstein1998,Eisenstein1999} cosmological \gls{power spectrum} with \glslink{BAO}{baryonic wiggles}. The \gls{cosmological parameters} are fixed at fiducial values,
\begin{equation}
\Omega_\Lambda = 0.78, \Omega_\mathrm{m} = 0.22, \Omega_\mathrm{b} = 0.04, \sigma_8 = 0.807, h = 0.702, n_{\mathrm{s}} = 0.961.
\end{equation}

The \glslink{grf}{Gaussian} \gls{initial conditions} are populated by a Lagrangian lattice of $256^3$ \glslink{dark matter particles}{particles}, that are propagated \glslink{forward modeling}{forward} in time using the same implementation of \glslink{2LPT}{second-order Lagrangian perturbation theory} as used in {\borg}. The \glslink{final conditions}{final} \gls{density field} is constructed from the resultant \glslink{particle realization}{particle distribution} using the \glslink{CiC}{cloud-in-cell} scheme. Note that it is crucial to use the \gls{2LPT} model for \gls{structure formation} at this point, instead of, for example, a full \glslink{N-body simulation}{$N$-body simulation}, in order to demonstrate that {\borg} correctly infers the input field. Only in this fashion can we demonstrate that the {\borg} complicated statistical machinery works, and compare the input and output without differences due to additional physics.

An \glslink{mock catalog}{artificial tracer catalog} is then generated by simulating an inhomogeneous \gls{Poisson process} characterized by equations \eqref{eq:Poisson_likelihood} and \eqref{eq:Poisson_intensity} (see also figure \ref{fig:borg_poisson_intensity} for an illustration). For the purpose of the test run, the problem is simplified to only one \gls{luminosity} bin ($\ell=0$), the mean number of galaxies $\bar{N}^0$ is fixed, and the tracers are supposed to be \glslink{bias}{unbiased} (which amounts to fixing $\Fuchsia{\alpha^0}=1$, $\Fuchsia{\beta^0}=1$). However, the \gls{survey response operator} $\RoyalBlue{R}^{\RoyalBlue{0}}_x$ involves a highly-structured survey \gls{mask} (mimicking the \glslink{survey geometry}{geometry} of the \glslink{SDSS}{Sloan Digital Sky Survey} data release 7) and realistic \glslink{selection effects}{selection functions} (based on standard \gls{Schechter luminosity functions}), in order to demonstrate the possibility of doing \gls{large-scale structure inference} from real \gls{data} sets.

\subsection{Convergence and correlations of the Markov Chain}
\label{sec:Convergence and correlations of the Markov Chain}

As mentioned in section \ref{sec:Hamiltonian Monte Carlo}, \gls{HMC} is designed to have the target distribution as its stationary distribution. Therefore, the \gls{sampling} process provides \glslink{sample}{samples} of the \gls{posterior} distribution (equation \eqref{eq:borg_posterior}) after an initial \gls{burn-in} phase. \citet{Jasche2013BORG} showed that during this phase, of the order of 600 \glslink{sample}{samples}, the \gls{power spectrum} converges at all scales towards the true power in the \glslink{initial conditions}{initial} \gls{density field}. The absence of any power excess or deficiency demonstrates the correct treatment of the \glslink{survey response operator}{response operator}. The analysis also showed that \gls{burn-in} also manifests itself in the \gls{acceptance rate}, which has a dip around after 100 \glslink{sample}{samples}, then increases and asymptotes at a constant value of around 84\%. 

Generally, successive \glslink{sample}{samples} of the chain will be correlated to previous \glslink{sample}{samples}. The correlation length of the chain determines the amount of independent \glslink{sample}{samples} that can be drawn from the total chain. \citet{Jasche2013BORG} estimated the correlation length to about 200 \glslink{sample}{samples} and obtained a total of 15,000 \glslink{sample}{samples}; which amounts to around 72 independent \glslink{sample}{samples} after \gls{burn-in}.

These statistical tests demonstrate that exploring the \glslink{large-scale structure likelihood}{large-scale structure posterior} is numerically feasible despite the \glslink{high-dimensional parameter space}{high dimensionality} of the problem.

\subsection{Large-scale structure inference}

\begin{figure}
\begin{center}
\includegraphics[width=\textwidth]{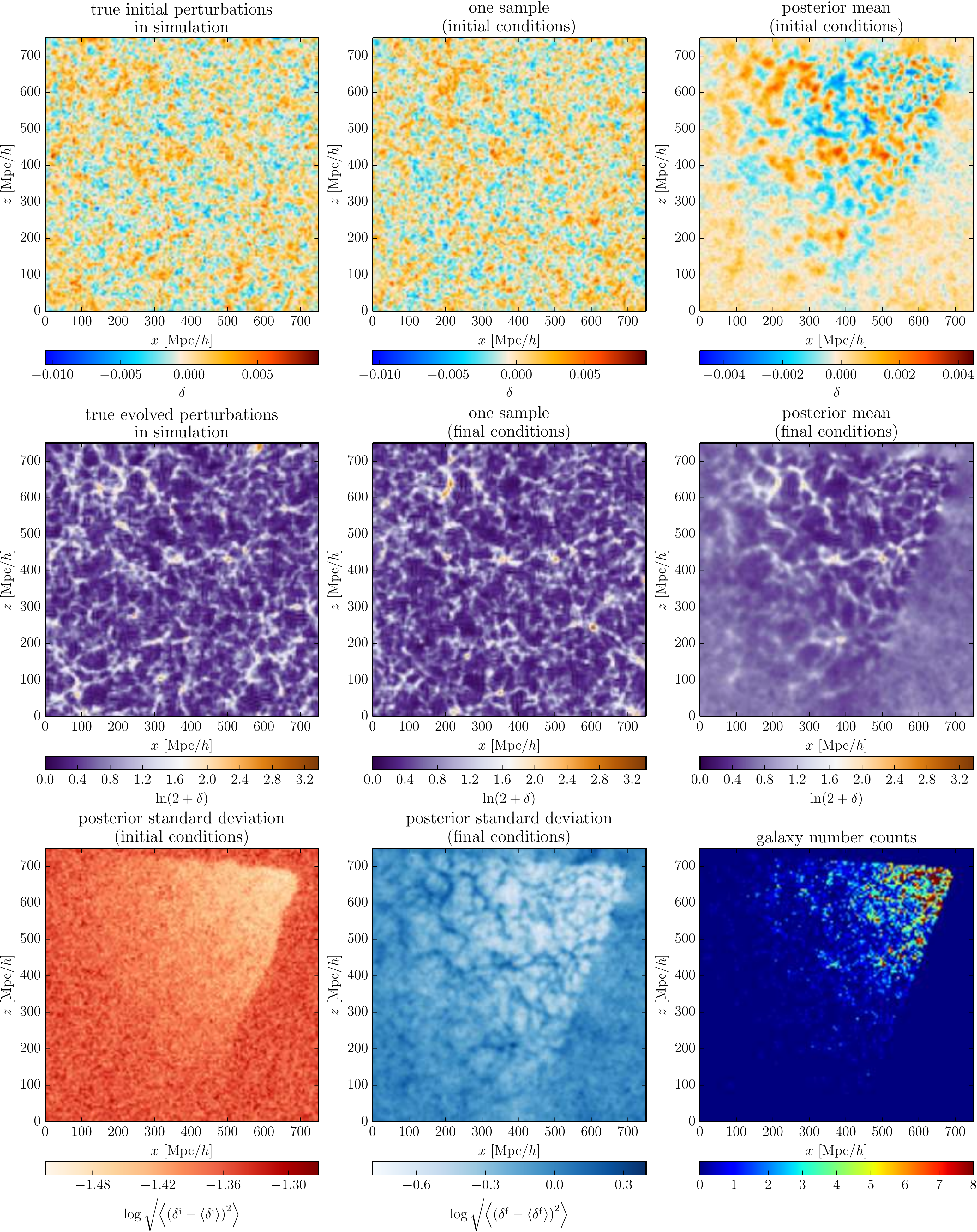} 
\caption{Slices through the box used for testing {\borg} on a \glslink{mock catalog}{synthetic data set}. Various quantities (indicated above the panels) are shown. The comparison between panels illustrates the performance of {\borg} at inferring \glslink{density field}{density fields} and demonstrates its capability of \glslink{uncertainty quantification}{quantifying uncertainties}. This figure shows results originally obtained by \citet{Jasche2013BORG}, courtesy of Jens Jasche.}
\label{fig:borg_test}
\end{center}
\end{figure}

This section discusses the \glslink{large-scale structure inference}{large-scale structure} inferred via the application of {\borg} to the \glslink{mock catalog}{synthetic data set}. Figure \ref{fig:borg_test} shows slices through various three-dimensional quantities: the true \glslink{initial conditions}{initial} \gls{density field}, one \gls{sample} of \gls{initial conditions}, the \gls{posterior mean} for the \glslink{initial conditions}{initial} \gls{density field}; the same quantities for \glslink{final conditions}{final} \glslink{density field}{density fields}; the \gls{posterior standard deviation} in the \glslink{initial conditions}{initial} and \gls{final conditions}; and the \glslink{mock catalog}{mock data set}.

Comparison of \glslink{initial conditions}{initial} and \glslink{final conditions}{final} \glslink{density field}{density fields} permits to check the correspondence between structures with growing statistical complexity. Furthermore, comparison of \glslink{final conditions}{final} \glslink{density field}{density fields} to the \gls{data} demonstrates the accuracy of the inference of the underlying dark matter \gls{density field}. In particular, one can see that the algorithm extrapolates unobserved \glslink{filament}{filaments} between \glslink{cluster}{clusters}, based on the physical picture of \gls{structure formation} provided by \gls{2LPT}. At high \gls{redshift} or near the \glslink{survey geometry}{survey boundaries}, complex structures appear continuous, which proves that the algorithm augments unobserved or poorly constrained regions with statistically correct information, consistently with the \gls{structure formation} model. Therefore, each individual \gls{sample} is a physical dark matter \glslink{particle realization}{realization}, to the level of accuracy of \gls{2LPT}.

The variation between \glslink{sampling}{samples} quantifies joint and correlated uncertainties. This is illustrated in figure \ref{fig:borg_test} by unobserved regions in the \glslink{posterior mean}{posterior means}, where the values in different \glslink{sample}{samples} average to cosmic mean density, and by the \glslink{posterior standard deviation}{posterior standard deviations}. Therefore, contrary to other \gls{reconstruction} approaches found in the literature, {\borg} possesses a demonstrated capability of \glslink{uncertainty quantification}{quantifying uncertainty} of inferred maps, locally and globally. These uncertainties can then be propagated to any derived quantity, as we demonstrate for example with \gls{cosmic web} \glslink{structure type}{types} in chapter \ref{chap:ts}.

Finally, \cite{Jasche2013BORG} demonstrated that the inferred \glslink{initial conditions}{initial} \gls{density contrast} follows Gaussian \glslink{one-point distribution}{one-point} statistics, that inferred \glslink{density field}{density fields} \glslink{cross-correlation}{cross-correlate} with the true solution as expected (i.e. $R(k) \equiv P_{\delta_\mathrm{inferred} \times \delta_\mathrm{true}} / \sqrt{P_{\delta_\mathrm{inferred}}P_{\delta_\mathrm{true}}} \rightarrow 1$ as $k \rightarrow 0$), and that {\borg} also infers the underlying \gls{velocity field} in detail.

\section{Future extensions of BORG}

The method described in this chapter forms the basis of a sophisticated, but also extensible, physical \gls{large-scale structure inference} framework. In particular, natural extensions of the {\borg} algorithm would enable automatic calibration of \gls{bias} parameters (the exponents $\Fuchsia{\alpha^\ell}$ in previous sections) and of the covariance matrix of initial fluctuations (the matrix $\DarkPastelGreen{S}$). This would allow precise inference of the early-time matter \gls{power spectrum} from \glslink{bias}{biased} catalogs of tracers. As noted in the \hyperref[chap:intro]{introduction}, this endeavor could yield a vast gain of information for the determination of \gls{cosmological parameters}, in comparison to state-of-the-art techniques.

Let us consider a set of \glslink{comoving coordinates}{comoving} wavenumbers $\{ k_n \}$ and let us denote by $\DarkPastelGreen{P} \equiv \{ \DarkPastelGreen{P}(k_n) \}$ the set of corresponding \gls{power spectrum} coefficients. Since direct sampling from $\p(\DarkPastelGreen{P} | \RubineRed{d})$ is impossible, or at least difficult, \citet{Jasche2010b} proposed to explore the full multi-dimensional joint \gls{posterior} of \glslink{power spectrum}{power spectra} coefficients and \glslink{density field}{density fluctuations}, $\p(\BattleShipGrey{\delta}^\mathrm{\BattleShipGrey{f}}, \DarkPastelGreen{P} | \RubineRed{d})$. They employ a two-steps \gls{Gibbs sampling} scheme, a method previously applied to \gls{CMB} data analysis \citep{Wandelt2004,Eriksen2004,Jewell2004}:
\begin{eqnarray}
\BattleShipGrey{\delta}^\mathrm{\BattleShipGrey{f}} & \curvearrowleft & \p(\BattleShipGrey{\delta}^\mathrm{\BattleShipGrey{f}} | \DarkPastelGreen{P},\RubineRed{d}), \\
\DarkPastelGreen{P} & \curvearrowleft & \p(\DarkPastelGreen{P} | \BattleShipGrey{\delta}^\mathrm{\BattleShipGrey{f}}, \RubineRed{d}),
\end{eqnarray}
where the arrow denotes a random draw from the \gls{pdf} on its right. The {\ares} code is an implementation of this scheme. It assumes the \gls{conditional independence} $\p(\DarkPastelGreen{P} | \BattleShipGrey{\delta}^\mathrm{\BattleShipGrey{f}}, \RubineRed{d}) = \p(\DarkPastelGreen{P} | \BattleShipGrey{\delta}^\mathrm{\BattleShipGrey{f}})$, which yields an inverse-Gamma distribution for \gls{power spectrum} coefficients, and a \glslink{grf}{Gaussian} \gls{prior} for $\BattleShipGrey{\delta}^\mathrm{\BattleShipGrey{f}}$ \citetext{i.e. a \glslink{Wiener filter}{Wiener} posterior for $\p(\BattleShipGrey{\delta}^\mathrm{\BattleShipGrey{f}} | \DarkPastelGreen{P},\RubineRed{d})$; see \citealp{Jasche2010b}}. In \citet{Jasche2013BIAS}, updates and improvements of {\ares} are introduced, in order to account for uncertainties arising from galaxy \glslink{bias}{biases} and normalizations of the galaxy density (i.e. \glslink{noise parameter}{noise levels}). 

\begin{figure}
\begin{center}
\begin{tikzpicture}
	\pgfdeclarelayer{background}
	\pgfdeclarelayer{foreground}
	\pgfsetlayers{background,main,foreground}

	\tikzstyle{sampling}=[draw, text width=8em, text centered, rounded corners, minimum height=2.5em, minimum width=8em]
	\tikzstyle{powersampling}=[draw, text width=12em, text centered, rounded corners, minimum height=2.5em, minimum width=8em]
	\tikzstyle{data}=[draw, fill=RubineRed!20, text width=5em, text centered, minimum height=2.5em,drop shadow]
	\tikzstyle{Ntilde}=[draw, fill=Orange!20, text width=5em, text centered, minimum height=2.5em,drop shadow]
	\tikzstyle{delta}=[draw, fill=black!20, text width=5em, text centered, minimum height=2.5em,drop shadow]
	\tikzstyle{alpha}=[draw, fill=Fuchsia!20, text width=5em, text centered, minimum height=2.5em,drop shadow]
	\tikzstyle{S}=[draw, fill=DarkPastelGreen!20, text width=5em, text centered, minimum height=2.5em,drop shadow]

	\def\blockdist{2.3}
	\def\edgedist{2.5}

    \node (datatop) [data] {data $\RubineRed{d} = \{ \RubineRed{N^\ell} \}$};
    \path (datatop.south)+(0,-1.0) node (deltasampling) [sampling] {density sampling\linebreak $\p(\BattleShipGrey{\delta}^\mathrm{\BattleShipGrey{f}},\BattleShipGrey{\delta}^\mathrm{\BattleShipGrey{i}}|\RubineRed{d}, \DarkPastelGreen{S}, \Orange{\widetilde{N}}, \Fuchsia{\alpha})$};
    \path (deltasampling.west)+(-\blockdist,0.0) node (deltasamples) [delta] {new $\delta$ sample};
    \path (deltasamples.south)+(0.0,-1.5) node (biassampling) [sampling] {bias sampling\linebreak 
$\p(\Fuchsia{\alpha^\ell} | \RubineRed{d}, \BattleShipGrey{\delta}^\mathrm{\BattleShipGrey{f}}, \Orange{\widetilde{N}^\ell})$};
    \path (biassampling.west)+(-1.7,0.0) node (dataleft) [data] {data $\RubineRed{d} = \{ \RubineRed{N^\ell} \}$};
    \path (biassampling.south)+(0.0,-1.5) node (alphasamples) [alpha] {new $\alpha$ sample};
    \path (deltasampling.east)+(\blockdist,0.0) node (Ntildesamples) [Ntilde] {new $\Orange{\widetilde{N}}$ sample};
    \path (Ntildesamples.south)+(0.0,-1.5) node (Ntildesampling) [sampling] {noise sampling\linebreak $\p(\Orange{\widetilde{N}^\ell} | \RubineRed{N^\ell}, \BattleShipGrey{\delta}^\mathrm{\BattleShipGrey{f}})$};
    \path (Ntildesampling.east)+(1.7,0.0) node (dataright) [data] {data $\RubineRed{d} = \{ \RubineRed{N^\ell} \}$};
    \path (Ntildesampling.south)+(0.0,-1.45) node (Ssamples) [S] {new $\DarkPastelGreen{S}$ sample};
    \path (deltasampling.south)+(0.0,-\blockdist-1.2) node (powersampling) [powersampling] {power spectrum sampling\linebreak 
$\p(\DarkPastelGreen{S} | \BattleShipGrey{\delta}^\mathrm{\BattleShipGrey{i}}, \Orange{\widetilde{N}}, \Fuchsia{\alpha})$};

	\path [draw, line width=0.7pt, arrows={-latex}] (datatop.south) -- (deltasampling.north) {} (deltasampling) ;
	\path [draw, line width=0.7pt, arrows={-latex}] (dataleft.east) -- (biassampling.west) {} (biassampling) ;
	\path [draw, line width=0.7pt, arrows={-latex}] (dataright.west) -- (Ntildesampling.east) {} (Ntildesampling) ;
	\path [draw, line width=0.7pt, arrows={-latex}] (deltasampling.west) -- (deltasamples.east) {} (deltasamples) ;
	\path [draw, line width=0.7pt, arrows={-latex}] (deltasamples.south) -- (biassampling.north) {} (biassampling) ;
	\path [draw, line width=0.7pt, arrows={-latex}] (biassampling.south) -- (alphasamples.north) {} (alphasamples) ;
	\path [draw, line width=0.7pt, arrows={-latex}] (alphasamples.east) -- (powersampling.west) {} (powersampling) ;
	\path [draw, line width=0.7pt, arrows={-latex}] (powersampling.east) -- (Ssamples.west) {} (Ssamples) ;
	\path [draw, line width=0.7pt, arrows={-latex}] (Ssamples.north) -- (Ntildesampling.south) {} (Ntildesampling) ;
	\path [draw, line width=0.7pt, arrows={-latex}] (Ntildesampling.north) -- (Ntildesamples.south) {} (Ntildesamples) ;
	\path [draw, line width=0.7pt, arrows={-latex}] (Ntildesamples.west) -- (deltasampling.east) {} (deltasampling) ;
\end{tikzpicture}
\end{center}
\caption{Flow chart depicting the multi-step iterative block \gls{sampling} procedure for a natural extension of the {\borg} algorithm. In the first step, {\borg} generates random realizations of \glslink{initial conditions}{initial} and \glslink{final conditions}{final} \glslink{density field}{density fields} \glslink{conditional pdf}{conditional} on the galaxy samples $\RubineRed{d}$, on the covariance matrix of \glslink{initial conditions}{initial fluctuations}, $\DarkPastelGreen{S}$, on the \glslink{noise parameter}{noise levels} $\{ \Orange{\widetilde{N}^\ell} \}$ and on the \gls{bias} parameters $\{ \Fuchsia{\alpha^\ell} \}$. In subsequent steps, the bias parameters, the covariance matrix and the \glslink{noise parameter}{noise parameters} are sampled conditional on respective previous \glslink{sample}{samples} and on the data when necessary. Iterations of this procedure yield \glslink{sample}{samples} from the full joint \gls{posterior} distribution, $\p(\BattleShipGrey{\delta}^\mathrm{\BattleShipGrey{f}},\BattleShipGrey{\delta}^\mathrm{\BattleShipGrey{i}}, \DarkPastelGreen{S}, \Orange{\widetilde{N}}, \Fuchsia{\alpha} |\RubineRed{d})$.\label{fig:borg_block_sampling_ideal}}
\end{figure}

Following these ideas, an extended {\borg} algorithm should perform iterative block sampling according to the scheme given in figure \ref{fig:borg_block_sampling_ideal} \citetext{for reference, see also figure \ref{fig:borg_block_sampling} for the current {\borg} algorithm, and figure 1 in \citealp{Jasche2013BIAS}, for the {\ares} algorithm}. In comparison to the \glslink{conditional pdf}{conditional} \gls{posterior} expressions written down by \citet{Jasche2010b} and \citet{Jasche2013BIAS}, this procedure would involve the expression of $\p(\Fuchsia{\alpha^\ell} | \RubineRed{d}, \BattleShipGrey{\delta}^\mathrm{\BattleShipGrey{f}}, \Orange{\widetilde{N}^\ell})$ in terms of the {\borg} power-law \gls{bias} model (instead of the linear \gls{bias} model of {\ares}) and of $\p(\DarkPastelGreen{S} | \BattleShipGrey{\delta}^\mathrm{\BattleShipGrey{i}}, \Orange{\widetilde{N}}, \Fuchsia{\alpha})$ in terms of \glslink{initial conditions}{initial} (instead of \glslink{final conditions}{final}) \glslink{density field}{density fields}.\footnote{As noted in section \ref{sec:The initial Gaussian prior}, the Fourier-space representation of $\DarkPastelGreen{S}$ is a diagonal matrix containing the coefficients $\sqrt{\DarkPastelGreen{P}(k)/(2\pi)^{3/2}}$.} In {\ares}, density sampling is by far the most expensive step. It can be done by constructing the \glslink{Wiener filter}{Wiener-filtered} map (which requires inversions of large matrices, see equations \eqref{eq:Wiener-filter-1} and \eqref{eq:Wiener-filter-2}) and augmenting missing fluctuations from the \gls{prior} \citep{Jasche2010b}, by means of \gls{HMC} \citep{Jasche2013BIAS}, or by using an auxiliary messenger field, which removes the need for matrix inversion \citetext{\citealp{Jasche2015}; see also \citealp{Elsner2013}}. For the {\borg} data model, involving a structure formation model instead of a \glslink{grf}{Gaussian} prior for the galaxy density, \gls{HMC} is the state-of-the-art technique.

An upcoming improvement of {\borg} will involve the joint sampling of \glslink{density field}{density} $\BattleShipGrey{\delta}^\mathrm{\BattleShipGrey{i}}$, \glslink{noise parameter}{noise levels} $\Orange{\widetilde{N}^\ell}$ and \gls{bias} parameters $\Fuchsia{\alpha^\ell}$. Unfortunately, computational time issues mean that joint, physical inference of \glslink{density field}{density} and \glslink{power spectrum}{power spectra} is still out of reach. Correlation lengths are of the order of 200 \glslink{sample}{samples} for {\borg} \glslink{density field}{density fields} \citep{Jasche2013BORG} and 100 \glslink{sample}{samples} for {\ares} \gls{power spectrum} coefficients \citep{Jasche2013BIAS}.\footnote{See \citealp{Jasche2013BIAS,Jewell2009}, for the discussion of a method designed to reduce the otherwise prohibitively long correlation length of {\ares} chains.} Preliminary tests indicate that the correlation length for the joint inference process is of the order of a few hundred \glslink{sample}{samples}. However, even with a correlation length of 100 \glslink{sample}{samples}, accurate characterization of \glslink{power spectrum}{power spectra} and corresponding uncertainties require, at least, about 40,000 \glslink{sample}{samples}. With the current performance of the {\borg} sampler (discussed in sections \ref{sec:Convergence and correlations of the Markov Chain} and \ref{sec:The BORG SDSS analysis}), such a run would take several years on a typical computer. For this reason, this thesis focuses on sampling the matter \gls{density field} for a fixed \gls{power spectrum} of \glslink{initial conditions}{primordial fluctuations}, rather than sampling this as well. Algorithmic and methodological innovations that would render such a run possible are currently being discussed but will require a considerable additional implementation effort and are outside the scope of this thesis.

%% file: Chapter5/Chapter5Content.tex
\chapter{Past and present cosmic structure in the Sloan Digital Sky Survey}
\label{chap:BORGSDSS}
\minitoc

\defcitealias{Pratchett1990}{Terry}
\begin{flushright}
\begin{minipage}[c]{0.6\textwidth}
\rule{\columnwidth}{0.4pt}

``Map-making had never been a precise art on the Discworld. People tended to start off with good intentions and then get so carried away with the spouting whales, monsters, waves, and other twiddly bits of cartographic furniture that they often forgot to put the boring mountains and rivers in at all.''\\
--- \citetalias{Pratchett1990} \citet{Pratchett1990}, \textit{Moving Pictures}

\vspace{-5pt}\rule{\columnwidth}{0.4pt}
\end{minipage}
\end{flushright}

\abstract{\section*{Abstract}
We present a \gls{chrono-cosmography} project, aiming at the \glslink{large-scale structure inference}{inference} of the four dimensional \gls{formation history} of the observed \glslink{LSS}{large-scale structure} from its origin to the present epoch. To do so, we perform a full-scale Bayesian analysis of the northern galactic cap of the Sloan Digital Sky Survey (\gls{SDSS}) Data Release 7 main galaxy sample, relying on a fully probabilistic, physical model of the non-linearly evolved \gls{density field}. Besides inferring \gls{initial conditions} from observations, our methodology naturally and accurately reconstructs non-linear features at the present epoch, such as \glslink{sheet}{walls} and \glslink{filament}{filaments}, corresponding to \glslink{high-order correlation function}{high-order correlation functions} generated by late-time \gls{structure formation}. Our \gls{inference} framework self-consistently accounts for typical observational \glslink{systematic uncertainty}{systematic} and \glslink{statistical uncertainty}{statistical uncertainties} such as \gls{noise}, \gls{survey geometry} and \gls{selection effects}. We further account for \gls{luminosity} dependent galaxy \glslink{bias}{biases} and automatic \gls{noise} calibration within a fully Bayesian approach. As a result, this analysis provides highly-detailed and accurate \glslink{reconstruction}{reconstructions} of the present \gls{density field} on scales larger than $\sim~3$ Mpc$/h$, constrained by \gls{SDSS} observations. This approach also leads to the first quantitative \gls{inference} of plausible \glslink{formation history}{formation histories} of the dynamic large scale structure underlying the observed galaxy distribution. The results described in this chapter constitute the first full Bayesian non-linear analysis of the cosmic large scale structure with the demonstrated capability of \gls{uncertainty quantification}. Some of these results have been made publicly available along with the corresponding paper. The level of detail of inferred results and the high degree of control on observational uncertainties pave the path towards high precision \gls{chrono-cosmography}, the subject of simultaneously studying the dynamics and the morphology of the inhomogeneous Universe.
}

\draw{This chapter is adapted from its corresponding publication, \citet{Jasche2015BORGSDSS}.}
\medskip

This chapter describes the {\borg} analysis of the Sloan Digital Sky Survey Data Release 7 main galaxy sample. It is structured as follows. In section \ref{sec:galaxy_sample}, we give a brief overview about the \gls{SDSS} \gls{data} set used in the analysis. In section \ref{sec:The BORG SDSS analysis}, we demonstrate the application of the {\borg} \gls{inference} algorithm to observations and discuss the general performance of the \glslink{HMC}{Hamiltonian Monte Carlo} sampler. Section \ref{sec:inference_results} describes the \gls{inference} results obtained in the course of this work. In particular, we present results on inferred 3D \glslink{initial conditions}{initial} and \glslink{final conditions}{final} \glslink{density field}{density} as well as \glslink{velocity field}{velocity fields} and show the ability of our method to provide accurate \gls{uncertainty quantification} for any finally inferred quantity. Further, we also demonstrate the ability of our methodology to perform \gls{chrono-cosmography}, by accurately inferring plausible 4D \glslink{formation history}{formation histories} for the observed \gls{LSS} from its origins to the present epoch. In section \ref{sec:Summary and conclusions BORGSDSS}, we conclude by summarizing and discussing the results obtained in the course of this project.

\section{The SDSS galaxy sample}
\label{sec:galaxy_sample}

In this work, we follow a similar procedure as described in \citet{Jasche2010a}, by applying the {\borg} algorithm to the \gls{SDSS} main galaxy sample. Specifically, we employ the \texttt{Sample dr72} of the New York University Value Added Catalogue\footnote{\href{http://sdss.physics.nyu.edu/vagc/}{http://sdss.physics.nyu.edu/vagc/}} (NYU-VAGC). This is an updated version of the catalogue originally constructed by \citet{Blanton2005} and is based on the final \gls{data} release \citep[DR7;][]{Abazajian2009} of the Sloan Digital Sky Survey \citep[\gls{SDSS};][]{York2000}. Based on \texttt{Sample dr72}, we construct a flux-limited galaxy sample with \glslink{spectroscopic redshift}{spectroscopically measured redshifts} in the range $0.001<z<0.4$, $r$-band Petrosian \glslink{luminosity}{apparent magnitude} $r\leq 17.6$ after correction for Galactic extinction, and $r$-band \glslink{luminosity}{absolute magnitude} $-21<M_{^{0.1}r}<-17$. Absolute $r$-band \glslink{luminosity}{magnitudes} are corrected to their $z=0.1$ values using the $K$-correction code of \citet{Blanton2003a,Blanton2007} and the \gls{luminosity} evolution model described in \citet{Blanton2003}. We also restrict our analysis to the main contiguous region of the \gls{SDSS} in the northern Galactic cap, excluding the three survey strips in the southern cap (about 10 per cent of the full survey area). The NYU-VAGC provides required information on the incompleteness in our \glslink{spectroscopic redshift}{spectroscopic sample}. This includes a \gls{mask}, indicating which areas of the sky have been targeted and which not. The \gls{mask} defines the effective area of the \glslink{galaxy survey}{survey} on the sky, which is 6437 deg$^2$ for the sample we use here. This \glslink{galaxy survey}{survey} area is divided into a large number of smaller subareas, called {\it polygons}, for each of which the NYU-VAGC lists a \glslink{spectroscopic redshift}{spectroscopic} completeness, defined as the fraction of photometrically identified target galaxies in the polygon for which usable spectra were obtained. Throughout our sample the average completeness is $0.92$. To account for radial \glslink{selection effects}{selection functions}, defined as the fraction of galaxies in the \glslink{luminosity}{absolute magnitude} range considered here, that are within the \glslink{luminosity}{apparent magnitude} range of the sample at a given \gls{redshift}, we use a standard \glslink{Schechter luminosity functions}{luminosity function} proposed by \citet{Schechter1976} with $r$-band parameters $\alpha = -1.05$, $M_{*} - 5\log_{10}(h) = -20.44$ \citep{Blanton2003b}.

Our analysis accounts for \gls{luminosity} dependent galaxy \glslink{bias}{biases}, by following the approach described in section \ref{sec:The BORG data model}. In order to do so, we subdivide our galaxy sample into six equidistant bins in absolute $r$-band \glslink{luminosity}{magnitude} in the range $-21<M_{^{0.1}r}<-17$, resulting in a total of $372,198$ main sample galaxies to be used in the analysis. As described in section \ref{sec:The BORG data model}, splitting the galaxy sample permits us to treat each of these sub-samples as an individual \gls{data} set, with its respective \gls{selection effects}, \glslink{bias}{biases} and \glslink{noise parameter}{noise levels}.

\section{The BORG SDSS analysis}
\label{sec:The BORG SDSS analysis}

We performed the analysis of the \gls{SDSS} main galaxy sample on a cubic Cartesian domain with a side length of $750$ Mpc$/h$ consisting of $256^3$ equidistant grid nodes, resulting in \glslink{high-dimensional parameter space}{$\sim~1.6\times~10^7$ inference parameters}. Thus, the \gls{inference} procedure provides data-constrained realizations for \glslink{initial conditions}{initial} and \glslink{final conditions}{final} \glslink{density field}{density fields} at a grid resolution of about $\sim~3$ Mpc$/h$. For the analysis, we assume a standard {\LCDM} cosmology with the set of \gls{cosmological parameters}
\begin{equation}
\Omega_\Lambda = 0.728, \Omega_\mathrm{m} = 0.272, \Omega_\mathrm{b} = 0.045, \sigma_8 = 0.807, h = 0.702, n_{\mathrm{s}} = 0.961.
\label{eq:cosmo-BORGSDSS}
\end{equation}
The cosmological \gls{power spectrum} for \glslink{initial conditions}{initial} \glslink{density field}{density fields} is calculated according to the prescription provided by \citet{Eisenstein1998,Eisenstein1999}. In order to sufficiently resolve the \glslink{final conditions}{final} \gls{density field}, the \gls{2LPT} model is evaluated with $512^3$ \glslink{dark matter particles}{particles}, by oversampling \gls{initial conditions} by a factor of eight.

We adjusted the parameters $\Fuchsia{\alpha^\ell}$ of the assumed power-law \gls{bias} model during the initial 1000 \glslink{sample}{sampling steps}, but kept them fixed afterwards. For the purpose of this work, the power-law indices $\Fuchsia{\alpha^\ell}$ of the \gls{bias} relations are determined by requiring them to resemble the linear \gls{luminosity} dependent \gls{bias} when expanded in a Taylor series to linear order as:
\begin{equation}
(1+\Gray{\delta}^\mathrm{\Gray{f}})^{\Fuchsia{\alpha^\ell}} = 1 + \Fuchsia{\alpha^\ell} \Gray{\delta}^\mathrm{\Gray{f}} + \mathcal{O}\left(\left(\Gray{\delta}^\mathrm{\Gray{f}}\right)^2\right) . 
\end{equation}
In particular, we assume the functional shape of the \gls{luminosity} dependent \gls{bias} parameter $\Fuchsia{\alpha^\ell}$ to follow a standard model for the linear \gls{luminosity} dependent \gls{bias} in terms of absolute $r$-band \glslink{luminosity}{magnitudes} $M_{^{0.1}r}$, as given by:
\begin{equation}
\Fuchsia{\alpha^\ell}=b(M_{^{0.1}r}^\ell)=b_{*}\left( a + b \times 10^{0.4\left(M_{*}-M_{^{0.1}r}^\ell \right)} + c \times \left(M_{^{0.1}r}^\ell-M_{*}\right)\right) \, ,
\label{eq:rel_bias}
\end{equation}
with the fitting parameters $a=0.895$, $b=0.150$, $c=-0.040$ and $M_{*}=-20.40$ \citep[see e.g.][for details]{Norberg2001,Tegmark2004}. The parameter $b_{*}$ was adjusted during the initial \gls{burn-in} phase and was finally set to a fixed value of $b_{*}=1.44$, such that the sampler recovers the correct shape of the assumed \glslink{initial conditions}{initial} \gls{power spectrum}.

As described in sections \ref{sec:The Gamma-distribution for noise sampling} and \ref{sec:Calibration of the noise level}, contrary to \gls{bias} exponents, corresponding \glslink{noise parameter}{noise parameters} $\Orange{\widetilde{N}^\ell}$ are sampled and explored throughout the entire Markov chain. Inferred \glslink{posterior mean}{ensemble means} and \glslink{posterior standard deviation}{standard deviations} for the $\Orange{\widetilde{N}^\ell}$ along with chosen power-law parameters $\Fuchsia{\alpha^\ell}$ are provided in table \ref{tb:table_borg_sdss_lum_bins}.

\begin{table}
\center
\renewcommand{\arraystretch}{1.5}
    \begin{tabular}{ccc}
    \hline
      $M_{^{0.1}r}^\ell$ & $\Fuchsia{\alpha^\ell}$ & $\Orange{\widetilde{N}^\ell}$  \\
      \hline
      $-21.00< M_{^{0.1}r}^0 <-20.33$ & 1.58029 & 4.67438 $\times 10^{-2}\, \pm 3.51298\times 10^{-4}$\\
      $-20.33< M_{^{0.1}r}^1 < -19.67$ & 1.41519 & 9.54428 $\times 10^{-2}\, \pm 5.77786\times 10^{-4}$\\
      $  -19.67< M_{^{0.1}r}^2 <-19.00$ & 1.30822 & 1.39989 $\times 10^{-1}\, \pm 1.21087\times 10^{-3}$\\
      $ -19.00< M_{^{0.1}r}^3 <-18.33$ & 1.23272 & 1.74284 $\times 10^{-1}\, \pm 1.89168\times 10^{-3}$\\
      $-18.33< M_{^{0.1}r}^4 <-17.67$  & 1.17424 & 2.19634 $\times 10^{-1}\, \pm 3.42586\times 10^{-3}$\\
      $-17.67 < M_{^{0.1}r}^5 <-17.00$  & 1.12497 & 2.86236 $\times 10^{-1}\, \pm 5.57014\times 10^{-3}$\\
      \hline
    \end{tabular}
     \caption{Bias and \glslink{noise parameter}{noise parameters}, as described in the text, for six galaxy sub-samples, subdivided by their absolute $r$-band \glslink{luminosity}{magnitudes}.}
    \label{tb:table_borg_sdss_lum_bins}
\end{table}

\begin{figure}
\begin{center}
\includegraphics[width=\columnwidth]{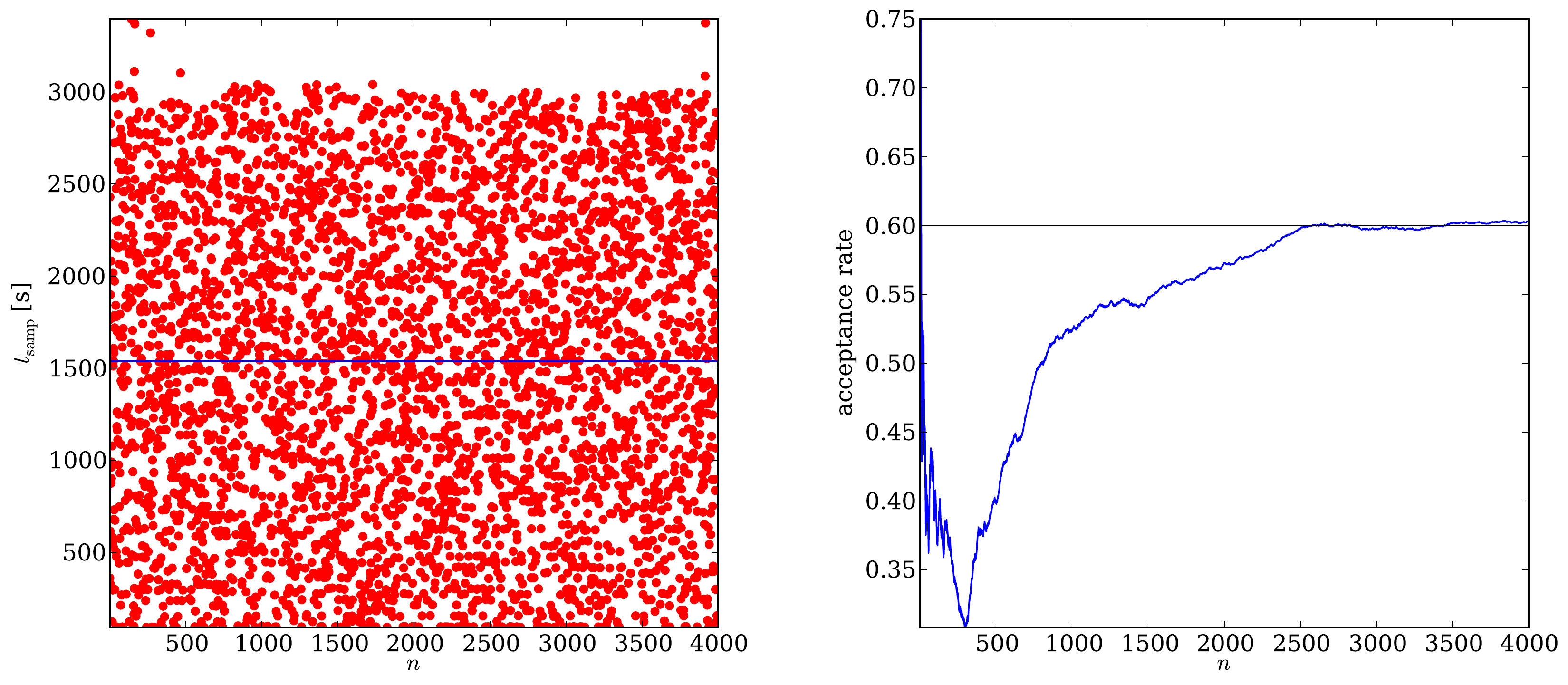}
\caption{Diagnotics of the Markov chain: scatter plot of \gls{sample} generation times (left panel) and Markov \glslink{acceptance rate}{acceptance rates} during the initial \gls{burn-in} phase (right panel). As shown by the left panel, times to generate individual \glslink{sample}{samples} range from zero to about 3000 seconds. The average execution time per \gls{sample} generation is about 1500 seconds on \(16\) cores. Initially, \glslink{acceptance rate}{acceptance rates} drop during \gls{burn-in} but rise again to reach an asymptotic value of about 60 percent.\label{fig:markov_properties}}
\end{center}
\end{figure}

The entire analysis yielded $12,000$ realizations for \glslink{initial conditions}{initial} and \glslink{final conditions}{final} \glslink{density field}{density fields}. The generation of a single Markov \gls{sample} requires an operation count equivalent to about $\sim~200$ \gls{2LPT} model evaluations. Typical generation times for data-constrained realizations are shown in the left panel of figure \ref{fig:markov_properties}. On average the sampler requires about 1500 seconds to generate a single \gls{density field} realization on $16$ cores. The total analysis consumed several months of computing time and produced on the order of $\sim~3$ TB of information represented by the set of Markov \glslink{sample}{samples}.

\begin{figure}
\begin{center}
\includegraphics[width=0.5\columnwidth]{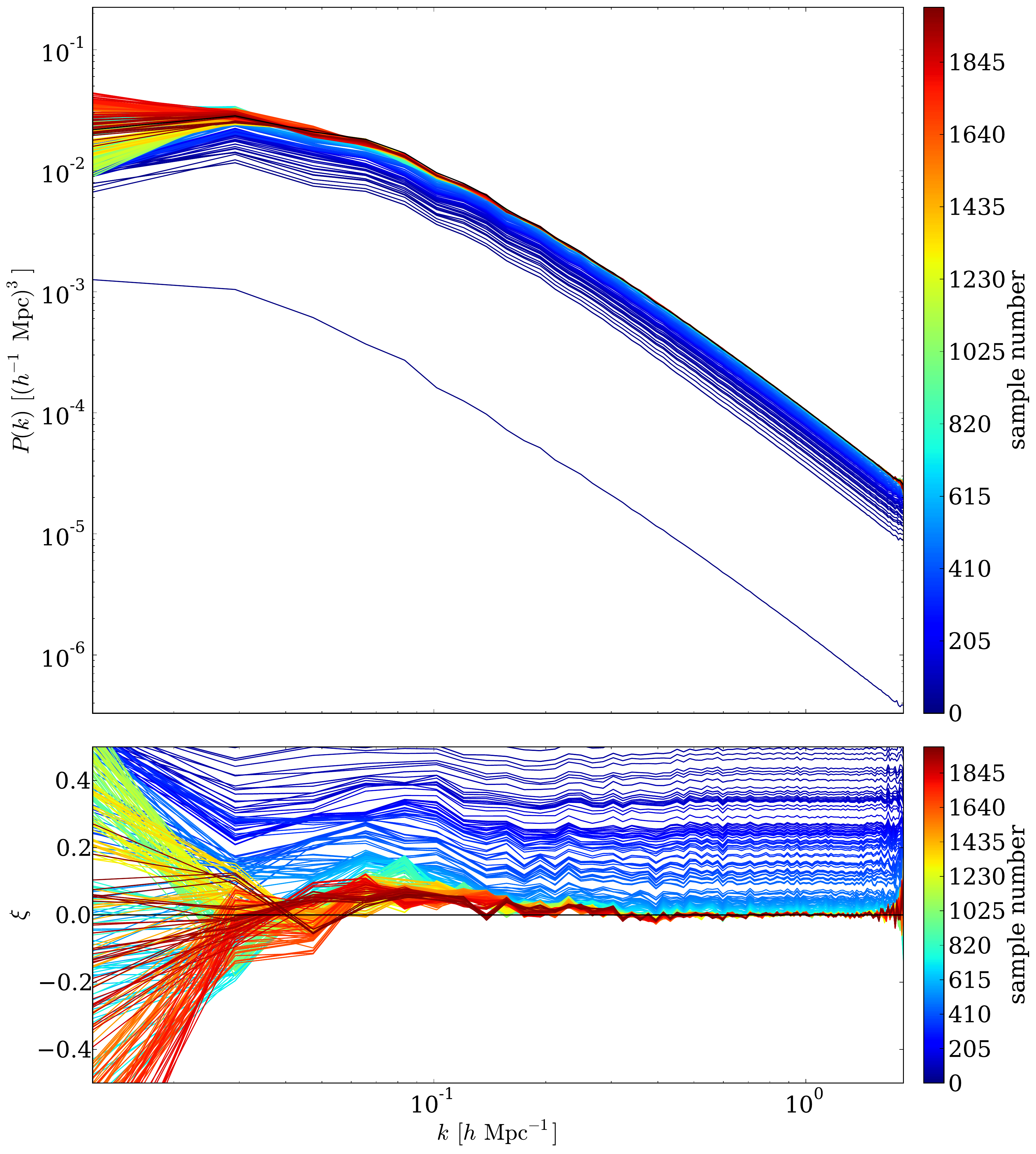}
\caption{\glslink{burn-in}{Burn-in} \glslink{power spectrum}{power spectra} measured from the first $2000$ \glslink{sample}{samples} of the Markov chain colored corresponding to their \gls{sample} number as indicated by the colorbar. The black line represents a fiducial reference \gls{power spectrum} for the cosmology assumed in this work. Subsequent \glslink{power spectrum}{power spectra} approach the fiducial cosmological \gls{power spectrum} homogeneously throughout all scales in Fourier space.\label{fig:burnin_specs}}
\end{center}
\end{figure}

The numerical efficiency of any \glslink{MCMC}{Markov Chain Monte Carlo} algorithm, particularly in \glslink{high-dimensional parameter space}{high dimensions}, is crucially determined by the average \gls{acceptance rate}. As demonstrated by the right panel of figure \ref{fig:markov_properties}, after an initial \gls{burn-in} period, the \gls{acceptance rate} asymptotes at a value of about 60 percent, rendering our analysis numerically feasible. As a simple consistency check, we follow a standard procedure to determine the initial \gls{burn-in} behavior of the sampler via a simple experiment \citep[see e.g.][for more details]{Eriksen2004,JascheKitaura2010,Jasche2013BORG}. The sampler is initialized with an overdispersed state, far remote from the target region in parameter space, by scaling normal random amplitudes of the \glslink{initial conditions}{initial} \gls{density field} at a cosmic \gls{scale factor} of $a=10^{-3}$ by a constant factor of 0.01. In the course of the initial \gls{burn-in} phase, the Markov chain should then drift towards preferred regions in parameter space. As demonstrated by figure \ref{fig:burnin_specs}, this drift is manifested by a sequence of \gls{posterior} \glslink{power spectrum}{power spectra} measured from subsequent \glslink{initial conditions}{initial} \gls{density field} realizations. It can be clearly seen that the chain approaches the target region  within the first $2000$ \glslink{sample}{sampling steps}. The sequence of \glslink{power spectrum}{power spectra} shows a homogeneous drift of all modes with no indication of any particular hysteresis or bias across different scales in Fourier space. As improper treatment of survey \glslink{systematic uncertainty}{systematics}, uncertainties and galaxy \gls{bias} typically result in obvious erroneous features in \glslink{power spectrum}{power spectra}, figure \ref{fig:burnin_specs} clearly demonstrates that these effects have been accurately accounted for by the algorithm.

\section{Inference results}
\label{sec:inference_results}

This section describes \glslink{large-scale structure inference}{inference} results obtained by our Bayesian analysis of the \gls{SDSS} main galaxy sample.

\subsection{Inferred 3D density fields}
\label{sec:inf_density_field}

A major goal of this work is to provide inferred 3D \glslink{initial conditions}{initial} and \glslink{final conditions}{final} \glslink{density field}{density fields} along with corresponding \gls{uncertainty quantification} in a \glslink{high-dimensional parameter space}{$\sim~1.6\times 10^7$ dimensional parameter space}. To do this, the {\borg} algorithm provides a sampled \gls{LSS} \gls{posterior} distribution in terms of an ensemble of data-constrained \glslink{sample}{samples}, via an efficient implementation of a \glslink{MCMC}{Markov Chain Monte Carlo} algorithm. It should be remarked that, past the initial \gls{burn-in} phase, all individual \glslink{sample}{samples} reflect physically meaningful \glslink{density field}{density fields}, limited only by the validity of the employed \gls{2LPT} model. In particular, the present analysis correctly accounts for \gls{selection effects}, \glslink{survey geometry}{survey geometries}, \gls{luminosity} dependent galaxy \glslink{bias}{biases} and automatically calibrates the \glslink{noise parameter}{noise levels} of the six \gls{luminosity} bins as described above. As can be seen in figure \ref{fig:burnin_specs}, past the initial \gls{burn-in} phase, individual \glslink{sample}{samples} possess physically correct power throughout all ranges in Fourier space, and do not show any sign of attenuation due to survey characteristics such as \gls{survey geometry}, \gls{selection effects} or galaxy \glslink{bias}{biases}.

\begin{figure}
\begin{center}
\includegraphics[width=\columnwidth]{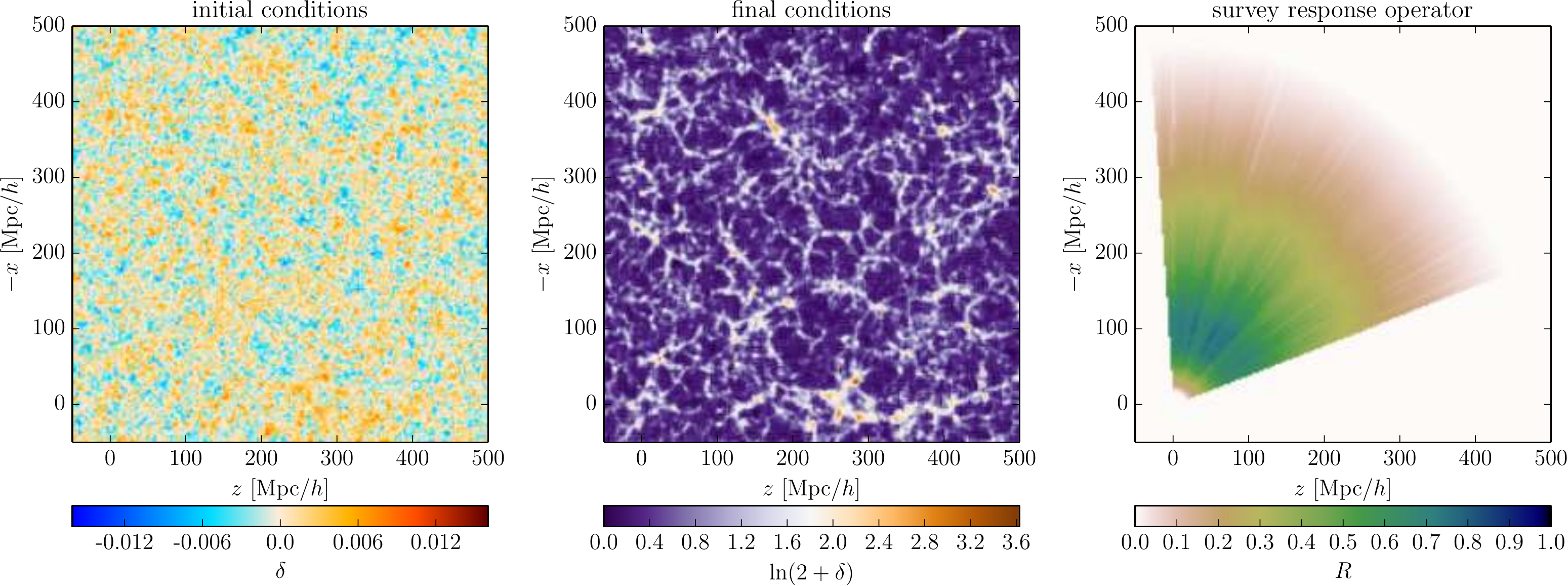}
\caption{Slices through the \glslink{initial conditions}{initial} (left panel) and corresponding \glslink{final conditions}{final} (middle panel) \glslink{density field}{density fields} of the $5000$th \gls{sample}. The right panel shows a corresponding slice through the combined \gls{survey response operator} $R$ for the six \glslink{luminosity}{absolute magnitude} bins considered in this work.  As can be seen, unobserved and observed regions in the inferred \glslink{initial conditions}{initial} and \glslink{final conditions}{final} \glslink{density field}{density fields} do not appear visually distinct, demonstrating the fact that individual data-constrained realizations constitute physically meaningful \glslink{density field}{density fields}. It also shows that the sampler naturally extends observed large scale structures beyond the \glslink{survey geometry}{survey boundaries} in a physically and statistically fully consistent fashion.\label{fig:sample_slices}}
\end{center}
\end{figure} 

To further illustrate that individual \glslink{sample}{samples} qualify for physically meaningful \glslink{density field}{density fields}, in figure \ref{fig:sample_slices} we show slices through data-constrained realizations of the \glslink{initial conditions}{initial} and \glslink{final conditions}{final} \glslink{density field}{density fields} of the $5000$th \gls{sample} as well as the corresponding slice through the combined \gls{survey response operator} $R$, averaged over the six \gls{luminosity} bins. It can be seen that the algorithm correctly augments unobserved regions with statistically correct information. Note that unobserved and observed regions in the inferred \glslink{final conditions}{final} \glslink{density field}{density fields} do not appear visually distinct, a consequence of the excellent approximation of \gls{2LPT} not just to the first but also \glslink{high-order correlation function}{higher-order moments}  \citep{Moutarde1991,Buchert1994,Bouchet1995,Scoccimarro2000,Scoccimarro2002}. 
Figure \ref{fig:sample_slices} therefore clearly reflects the fact that our sampler naturally extends observed large scale structures beyond the \glslink{survey geometry}{survey boundaries} in a physically and statistically fully consistent fashion.
This is a great advantage over previous methods relying on \glslink{grf}{Gaussian} or \glslink{log-normal distribution}{log-normal} models specifying the statistics of the \gls{density field} correctly only to \glslink{two-point correlation function}{two-point statistics} by assuming a cosmological \gls{power spectrum}. The interested reader may want to qualitatively compare with figure 2 in \citet{Jasche2010a}, where a \glslink{log-normal distribution}{log-normal model}, unable to represent \glslink{filament}{filamentary structures}, was employed.

\begin{figure}
\begin{center}
\includegraphics[width=\columnwidth]{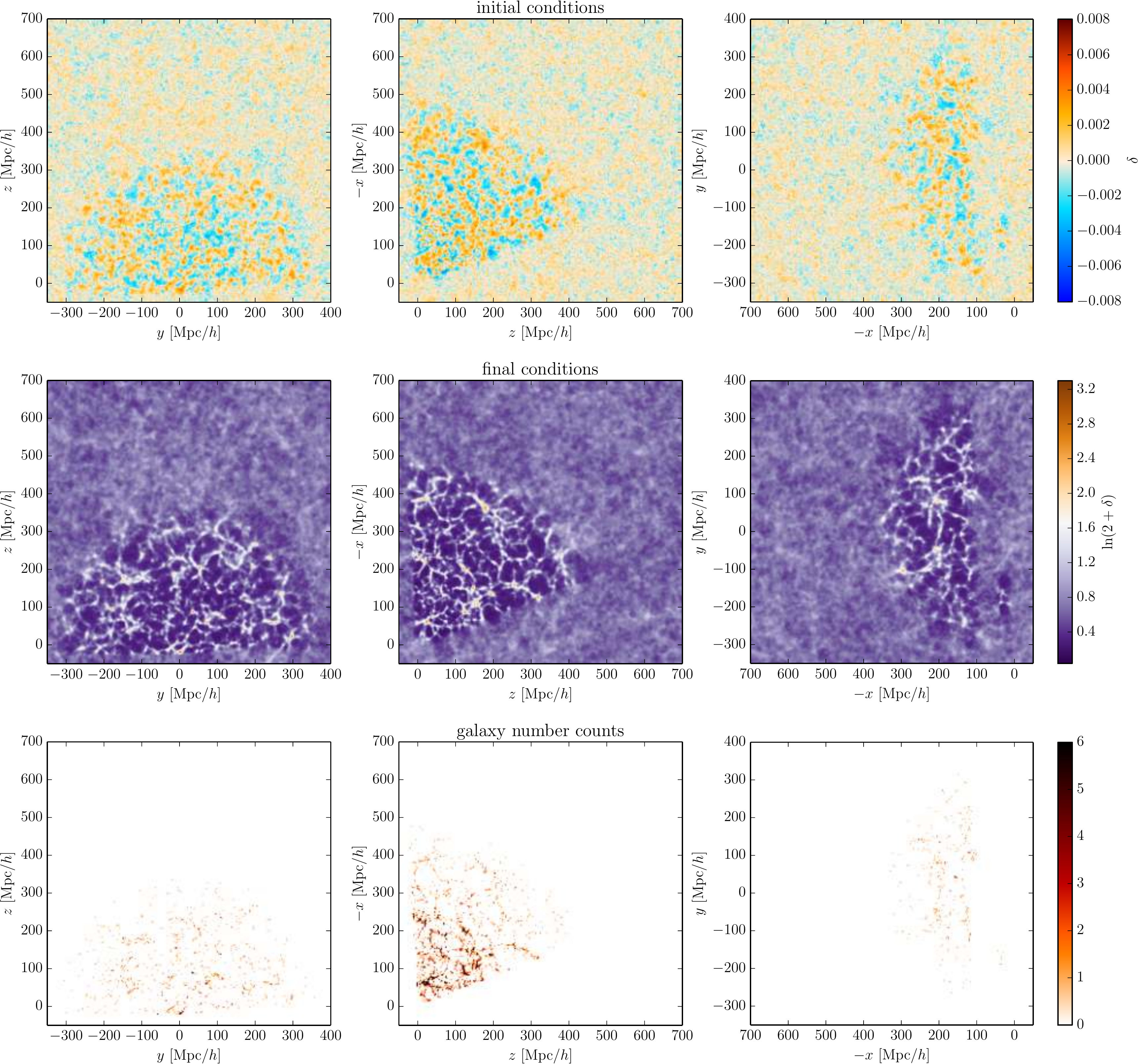}
\caption{Three slices from different directions through the three dimensional ensemble \glslink{posterior mean}{posterior means} for the \glslink{initial conditions}{initial} (upper panels) and \glslink{final conditions}{final} \glslink{density field}{density fields} (middle panels) estimated from $12,000$ \glslink{sample}{samples}. The lower panels depict corresponding slices through the galaxy \glslink{number function}{number counts} field of the \gls{SDSS} main sample.\label{fig:mean_var_dens}}
\end{center}
\end{figure}

The ensemble of the $12,000$ inferred data-constrained \glslink{initial conditions}{initial} and \glslink{final conditions}{final} \glslink{density field}{density fields} permits us to provide any desired statistical summary, such as \glslink{posterior mean}{mean} and \glslink{posterior standard deviation}{variance}, for full 3D fields. In figure \ref{fig:mean_var_dens}, we show slices through the \glslink{posterior mean}{ensemble mean} \glslink{initial conditions}{initial} and \glslink{final conditions}{final} \glslink{density field}{density fields}, to be used in subsequent analyses. The plot shows the correct anticipated behavior for inferred \gls{posterior mean} \glslink{final conditions}{final} \glslink{density field}{density fields}, since observed regions represent \gls{data} constraints, while unobserved regions approach cosmic mean density. This behavior is also present in corresponding \glslink{initial conditions}{initial} \glslink{density field}{density fields}. In particular, the \glslink{posterior mean}{ensemble mean} \glslink{final conditions}{final} \gls{density field} shows a highly detailed \gls{LSS} in regions where \gls{data} constraints are available, and approaches cosmic mean density in regions where \gls{data} are uninformative on average \citep[see also][for comparison]{Jasche2010a}. Analogously, these results translate to the \glslink{posterior mean}{ensemble mean} \glslink{initial conditions}{initial} \gls{density field}. Comparing the \glslink{posterior mean}{ensemble mean} \glslink{final conditions}{final} \gls{density field} to the galaxy number densities, depicted in the lower panels of figure \ref{fig:mean_var_dens}, demonstrates the performance of the method in regions only poorly sampled by galaxies. In particular, comparing the right middle and right lower panel of figure \ref{fig:mean_var_dens} reveals the capability of our algorithm to recover highly detailed structures even in \gls{noise} dominated regions \citep[for a discussion see chapter \ref{chap:BORG} and][]{Jasche2013BORG}. By comparing \glslink{posterior mean}{ensemble mean} \glslink{initial conditions}{initial} and \glslink{final conditions}{final} \glslink{density field}{density fields}, upper and middle panels in figure \ref{fig:mean_var_dens}, one can also see correspondences between structures in the present Universe and their origins at a \gls{scale factor} of $a=10^{-3}$.

\begin{figure}
\begin{center}
\includegraphics[width=\columnwidth]{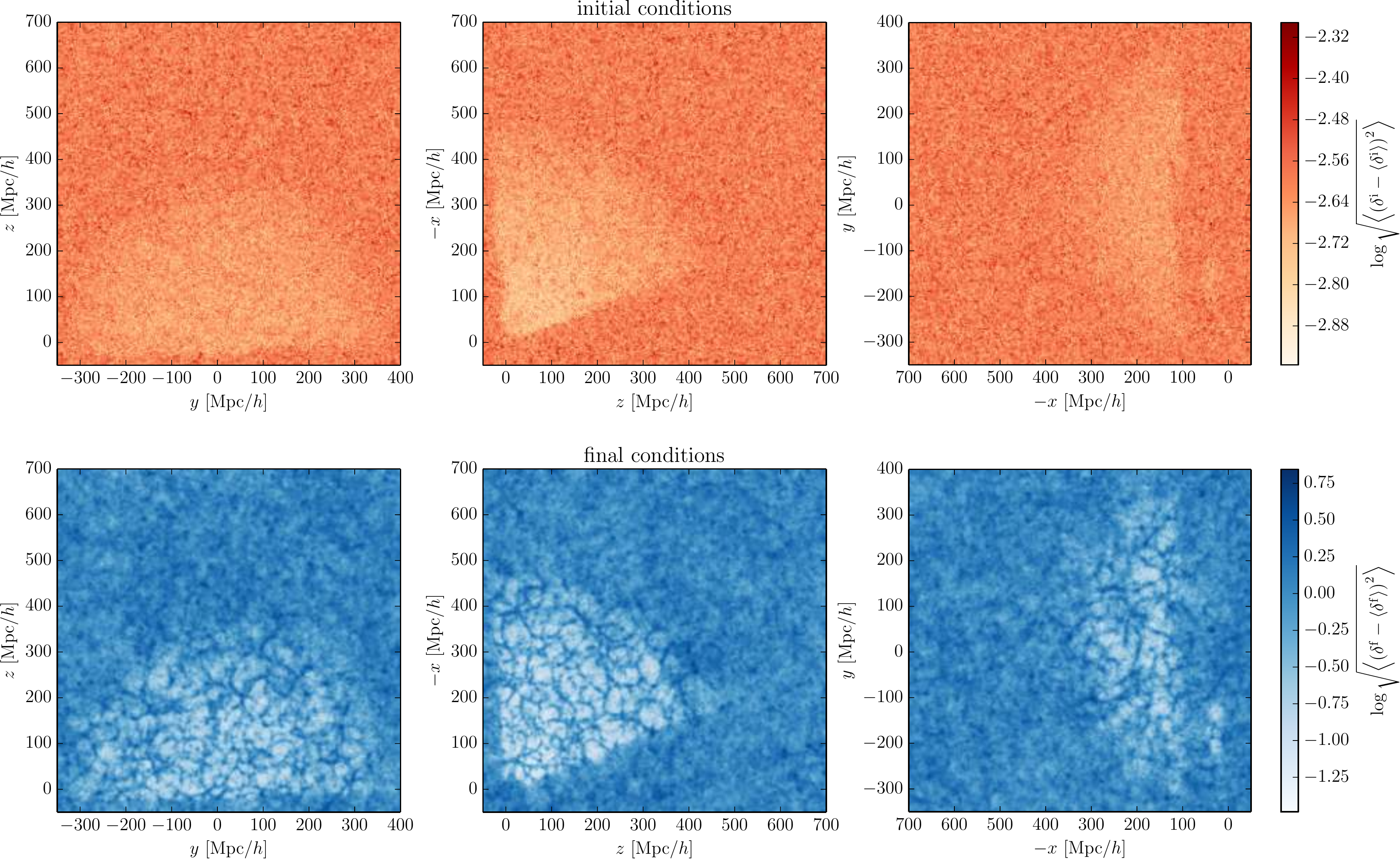}
\caption{Three slices from different directions through the three dimensional voxel-wise \gls{posterior standard deviation} for the \glslink{initial conditions}{initial} (upper panels) and \glslink{final conditions}{final} \glslink{density field}{density fields} (lower panels) estimated from $12,000$ \glslink{sample}{samples}. It can be seen that regions covered by observations show on average lower variance than unobserved regions. Also note, that voxel-wise \glslink{posterior standard deviation}{standard deviations} for the \glslink{final conditions}{final} \glslink{density field}{density fields} are highly structured, reflecting the signal-dependence of the inhomogeneous \gls{shot noise} of the galaxy distribution. In contrast, voxel-wise \glslink{posterior standard deviation}{standard deviations} in the \gls{initial conditions} are more homogeneously distributed, manifesting the flow of information between \gls{data} and \gls{initial conditions} as discussed in the text.\label{fig:var_dens}}
\end{center}
\end{figure}

The ensemble of data-constrained realizations also permits to provide corresponding \gls{uncertainty quantification}. In figure \ref{fig:var_dens} we plot voxel-wise \glslink{posterior standard deviation}{standard deviations} for \glslink{initial conditions}{initial} and \glslink{final conditions}{final} \glslink{density field}{density fields} estimated from $12,000$ \glslink{sample}{samples}. It can be seen that regions covered by \gls{data} exhibit on average lower variances than unobserved regions, as expected. Note that for non-linear \gls{inference} problems, signal and \gls{noise} are typically correlated. This is particularly true for inhomogeneous \glslink{Poisson process}{point processes}, such as discrete galaxy distributions tracing an underlying \gls{density field}. In figure \ref{fig:var_dens}, the correlation between signal and \gls{noise} is clearly visible for \glslink{posterior standard deviation}{standard deviation} estimates of \glslink{final conditions}{final} \glslink{density field}{density fields}. In particular high density regions also correspond to high variance regions, as is expected for \glslink{Poisson likelihood}{Poissonian likelihoods} since signal-to-noise ratios scale as the square root of the number of observed galaxies \citep[also see][for a similar discussion]{Jasche2010a}. Also note that voxel-wise \glslink{posterior standard deviation}{standard deviations} for \glslink{final conditions}{final} \glslink{density field}{density fields} are highly structured, while \glslink{posterior standard deviation}{standard deviations} of \gls{initial conditions} appear to be more homogeneous. This is related to the fact that our algorithm naturally and correctly translates information of the observations \glslink{non-local}{non-locally} to the \gls{initial conditions} via \gls{Lagrangian transport}, as discussed below in section \ref{sec:cosmic_history}.

As mentioned in the introduction, results for the \glslink{posterior mean}{ensemble mean} \glslink{final conditions}{final} \gls{density field} and corresponding voxel-wise \glslink{posterior standard deviation}{standard deviations} have been published as as supplementary material to the article \citep{Jasche2015BORGSDSS}.\footnote{These data can be accessed at \href{http://iopscience.iop.org/1475-7516/2015/01/036}{http://iopscience.iop.org/1475-7516/2015/01/036}.}

\subsection{Inference of 3D velocity fields}
\label{sec:3d_velocity_fields}

In addition to \glslink{initial conditions}{initial} and \glslink{final conditions}{final} \glslink{density field}{density fields}, the analysis further provides information on the dynamics of the large scale structure as mediated by the employed \gls{2LPT} model. Indeed, the {\borg} algorithm shows excellent performance in recovering large scale modes, typically poorly constrained by \glslink{mask}{masked} galaxy observations \citep{Jasche2013BORG}.

\begin{figure}
\begin{center}
\includegraphics[width=0.6\columnwidth]{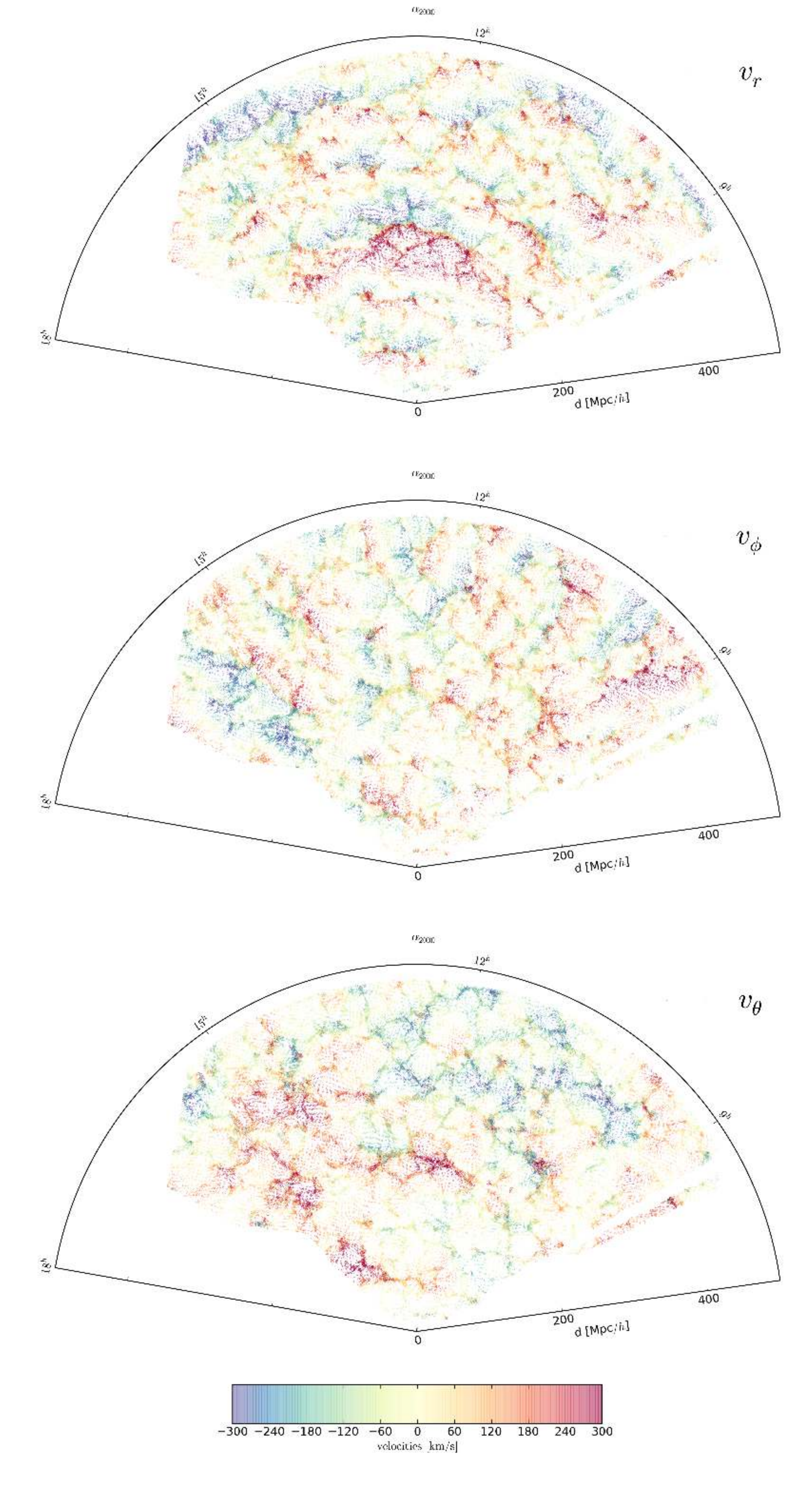}
\caption{Slices through the 3D \glslink{velocity field}{velocity fields}, derived from the $5000$th \gls{sample}, for the radial (upper panel), polar (middle panel) and the azimuthal (lower panel) velocity components. The plot shows \gls{2LPT} \glslink{dark matter particles}{particles} in a 4 Mpc$/h$ thick slice around the celestial equator for the observed domain, colored by their respective velocity components.\label{fig:velocity_field}}
\end{center}
\end{figure}

This a crucial feature when deriving 3D \glslink{velocity field}{velocity fields}, which are predominantly governed by the largest scales. In this fashion, we can derive 3D \glslink{velocity field}{velocity fields} from our \glslink{large-scale structure inference}{inference results}. Note that these \glslink{velocity field}{velocity fields} are derived \textit{a posteriori} and are only predictions of the \gls{2LPT} model given inferred \glslink{initial conditions}{initial} \glslink{density field}{density fields}, since currently the algorithm does not exploit velocity information contained in the \gls{data}. However, since inferred \gls{2LPT} displacement vectors are constrained by observations, and since \gls{2LPT} displacement vectors and velocities differ only by constant prefactors given a fixed cosmology, inferred velocities are considered to be accurate. For this reason, exploitation of velocity information contained in the \gls{data} itself, being the subject of a future publication, is not expected to crucially change present results. To demonstrate the capability of recovering 3D \glslink{velocity field}{velocity fields}, in figure \ref{fig:velocity_field} we show the three components of the \gls{velocity field} for the $5000$th \gls{sample} in spherical coordinates. More precisely, figure \ref{fig:velocity_field} shows the corresponding \gls{2LPT} \glslink{particle realization}{particle distribution} evolved to \gls{redshift} $z=0$ in a $4$ Mpc$/h$ slice around the celestial equator. \glslink{dark matter particles}{Particles} are colored by their radial (upper panel), polar (middle panel) and azimuthal (lower panel) velocity components. To translate between Cartesian and spherical coordinates we used the standard coordinate transform,
\begin{eqnarray}
x & = & d_\mathrm{com} \cos(\lambda) \cos(\eta) \\
y & = & d_\mathrm{com} \cos(\lambda) \sin(\eta) \\
z & = & d_\mathrm{com} \sin(\lambda) ,
\end{eqnarray}
where $\lambda$ is the \gls{declination}, $\eta$ is the \gls{right ascension} and $d_\mathrm{com}$ is the radial \glslink{comoving coordinates}{comoving distance}.

\subsection{Inference of LSS formation histories}
\label{sec:cosmic_history}

\begin{figure}
\begin{center}
\includegraphics[width=\columnwidth]{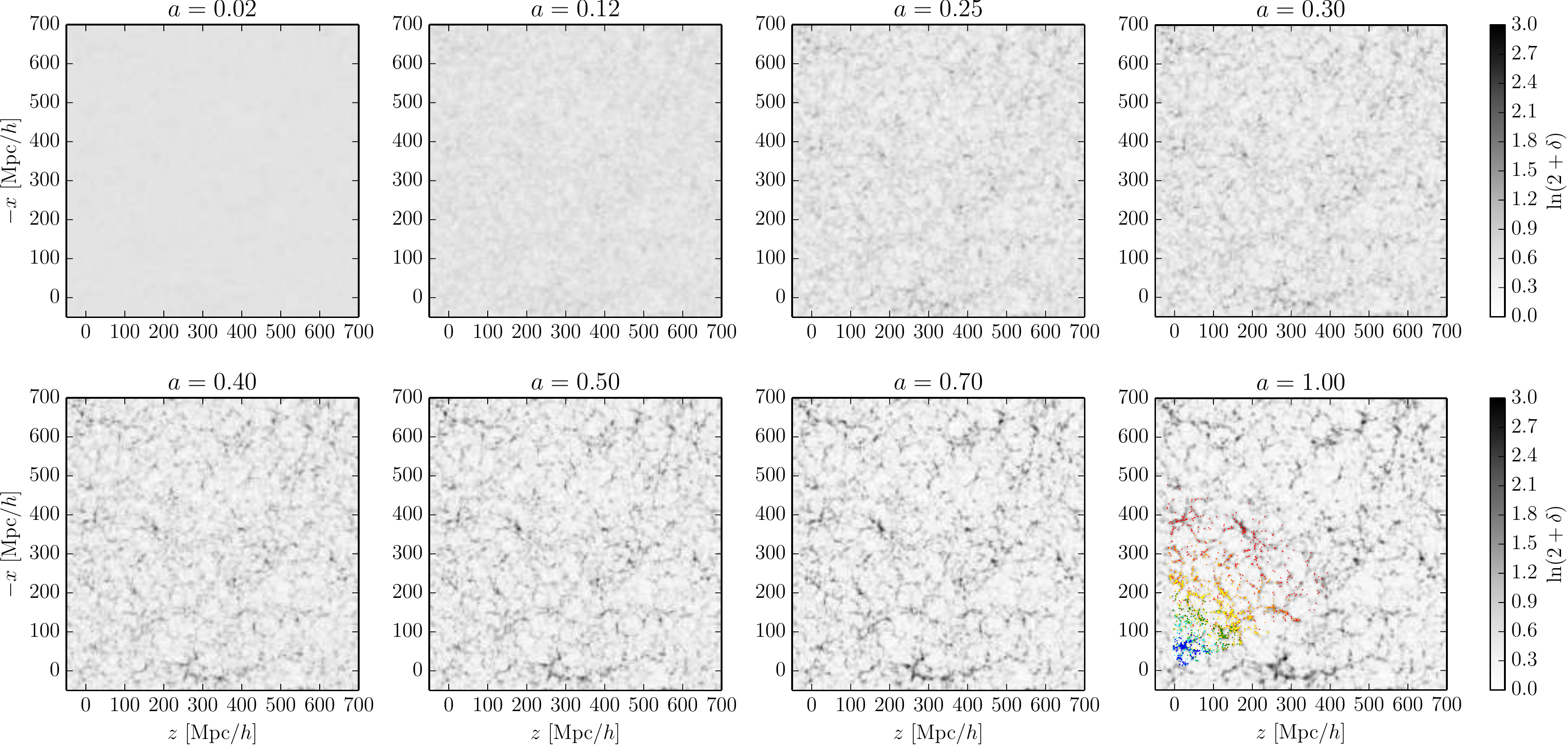}
\caption{Slices through the inferred three dimensional \gls{density field} of the $5000$th \gls{sample} at different stages of its evolution, as indicated by the cosmic \gls{scale factor} in the respective panels. The plot describes a possible \glslink{formation history}{formation scenario} for the \gls{LSS} in the observed domain starting at a \gls{scale factor} of $a=0.02$ to the present epoch $a=1.0$. In the lower right panel, we overplotted the inferred present \gls{density field} with the observed galaxies in the respective six \glslink{luminosity}{absolute magnitude} ranges \( -21.00< M_{^{0.1}r} <-20.33\) (red dots), \(-20.33< M_{^{0.1}r} < -19.67\) (orange dots), \(  -19.67< M_{^{0.1}r} <-19.00\) (yellow dots), \( -19.00< M_{^{0.1}r} <-18.33\) (green dots) , \(-18.33< M_{^{0.1}r} <-17.67\) (cyan dots) and , \(-17.67 <M_{^{0.1}r} <-17.00\) (blue dots). As can be clearly seen, observed galaxies trace the recovered three dimensional \gls{density field}. Besides measurements of three dimensional \glslink{initial conditions}{initial} and \glslink{final conditions}{final} \glslink{density field}{density fields}, this plot demonstrates that our algorithm also provides plausible four dimensional \glslink{formation history}{formation histories}, describing the evolution of the presently observed \gls{LSS}.\label{fig:cosmic_evolution}}
\end{center}
\end{figure}

As described in chapter \ref{chap:BORG}, the {\borg} algorithm employs a \gls{2LPT} model to connect \gls{initial conditions} to present \gls{SDSS} observations in a fully probabilistic approach. Besides inferred 3D \glslink{initial conditions}{initial} and \glslink{final conditions}{final} \glslink{density field}{density fields}, our algorithm therefore also provides full four dimensional \glslink{formation history}{formation histories} for the observed \gls{LSS} as mediated by the \gls{2LPT} model. As an example, in figure \ref{fig:cosmic_evolution} we depict the \gls{LSS} \gls{formation history} for the $5000$th Markov \gls{sample} ranging from a \gls{scale factor} of $a=0.02$ to the present epoch at $a=1.00$. Initially, the \gls{density field} seems to obey close to \glslink{grf}{Gaussian statistics} and corresponding amplitudes are low. In the course of cosmic history, amplitudes grow and \glslink{high-order correlation function}{higher-order statistics} such as \glslink{three-point correlation function}{three-point statistics} are generated, as indicated by the appearance of \glslink{filament}{filamentary structures}. The final panel of figure \ref{fig:cosmic_evolution}, at a cosmic \gls{scale factor} of $a=1.00$, shows the inferred \glslink{final conditions}{final} \gls{density field} overplotted by \gls{SDSS} galaxies for the six bins in \glslink{luminosity}{absolute magnitude}, as described previously. Observed galaxies nicely trace the underlying \gls{density field}. This clearly demonstrates that our algorithm infers plausible \glslink{formation history}{formation histories} for large scale structures observed by the \gls{SDSS} survey. By exploring the corresponding \gls{LSS} \gls{posterior} distribution, the {\borg} algorithm naturally generates an ensemble of such data-constrained \gls{LSS} \glslink{formation history}{formation histories}, permitting to accurately quantify the 4D dynamical state of our Universe and corresponding observational uncertainties inherent to \glslink{galaxy survey}{galaxy surveys}. Detailed and quantitative analysis of these cosmic \glslink{formation history}{formation histories} will be the subject of forthcoming publications (see also chapter \ref{chap:ts}). 

The {\borg} algorithm also provides a statistically valid framework for propagating observational \glslink{systematic uncertainty}{systematics} and uncertainties from observations to any finally inferred result. This is of particular importance, since detailed treatment of \glslink{survey geometry}{survey geometries} and \gls{selection effects} is a crucial issue if inferred results are to be used for thorough scientific analyses. These effects generally vary greatly across the observed domain and will result in erroneous artifacts if not accounted for properly. Since large scale \gls{structure formation} is a \gls{non-local} process, exact information propagation is complex, as it requires to translate uncertainties and \glslink{systematic uncertainty}{systematics} from observations to the inferred \gls{initial conditions}. Consequently, the \gls{information content} of observed \gls{data} has to be distributed differently in \glslink{initial conditions}{initial} and \glslink{final conditions}{final} \glslink{density field}{density fields}, even though the  total amount of information is conserved. Following \gls{2LPT} \glslink{dark matter particles}{particles} from high density regions, and corresponding high signal-to-noise regions in the \gls{data}, backward in time, demonstrates that the same amount of information contained in the \gls{data} will be distributed over a larger region in the \gls{initial conditions}. Analogously, for underdense regions, such as \glslink{void}{voids}, the \gls{information content} of the \gls{data} will amass in a smaller volume at the \glslink{initial conditions}{initial} state. This means that the signal-to-noise ratio for a given \glslink{comoving coordinates}{comoving} Eulerian volume is a function of time along inferred cosmic \glslink{formation history}{histories} \citep{Jasche2013BORG}. This fact manifests itself in the different behaviour of voxel-wise \glslink{posterior standard deviation}{standard deviations} for \glslink{final conditions}{final} and \gls{initial conditions}, as presented in figure \ref{fig:var_dens}. While the signal-to-noise ratio is highly clustered in \gls{final conditions}, the same amount of observational information is distributed more evenly over the entire volume in corresponding \gls{initial conditions}.

\begin{figure}
\begin{center}
\includegraphics[width=\columnwidth]{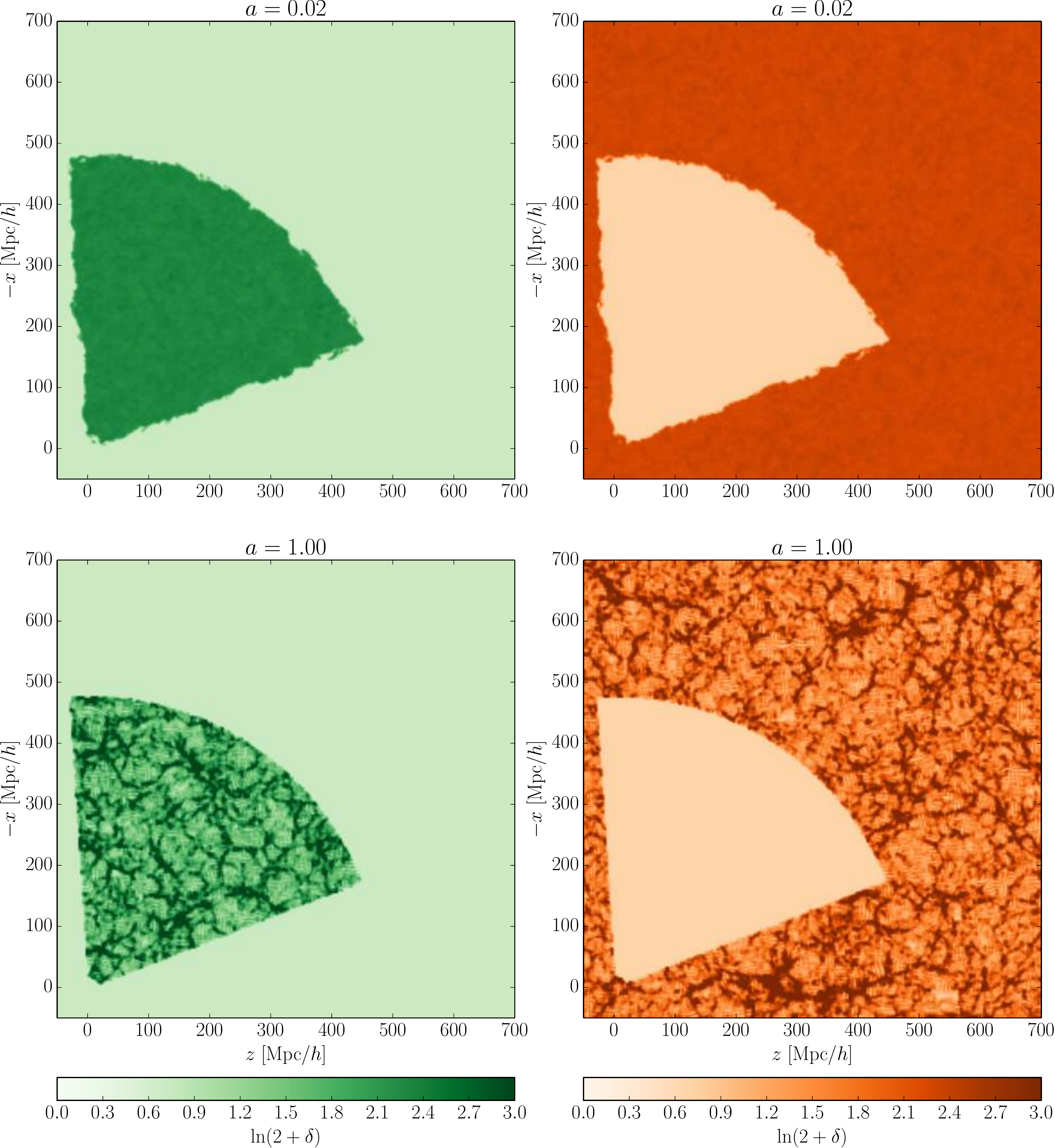}
\caption{Slices through the distribution of \glslink{dark matter particles}{particles} in the $5000$th \gls{sample}, which are located inside (left panels) and outside (right panels) the observed domain at the time of observation, at two time snapshots as indicated above the panels. It can be seen that \glslink{dark matter particles}{particles} located within the observed region at the present time may originate from regions outside the corresponding \glslink{comoving coordinates}{comoving} Eulerian volume at an earlier epoch and vice versa. As discussed in the text, this plot demonstrates the \gls{non-local} \glslink{Lagrangian transport}{transport} of information, which provides accurate \glslink{large-scale structure inference}{inference of the cosmic large scale structure} beyond \glslink{survey geometry}{survey boundaries} within a rigorous probabilistic approach.\label{fig:inside_out}}
\end{center}
\end{figure}

\glslink{non-local}{Non-local} propagation of observational information across \glslink{survey geometry}{survey boundaries}, together with cosmological correlations in the \glslink{initial conditions}{initial} \gls{density field}, is also the reason why our method is able to extrapolate the cosmic \gls{LSS} beyond \glslink{survey geometry}{survey boundaries}, as discussed in section \ref{sec:inf_density_field} above and demonstrated by figure \ref{fig:mean_var_dens}. To further demonstrate this fact, in figure \ref{fig:inside_out}, we show the \gls{density field} of the $5000$th \gls{sample} traced by \glslink{dark matter particles}{particles} from inside and outside the observed domain at the present epoch. At the present epoch, the set of \glslink{dark matter particles}{particles} can be sub-divided into two sets for \glslink{dark matter particles}{particles} inside and outside the observed domain. The boundary between these two sets of \glslink{dark matter particles}{particles} is the sharp outline of the \gls{SDSS} \gls{survey geometry}. When tracing these \glslink{dark matter particles}{particles} back to an earlier epoch at a \gls{scale factor} of $a=0.02$, it can clearly be seen that this sharp boundary starts to frazzle. \glslink{dark matter particles}{Particles} within the observed domain at the \glslink{final conditions}{final} state may originate from regions outside the corresponding Eulerian volume at the \glslink{initial conditions}{initial} state, and vice versa. Information from within the observed domain \glslink{non-local}{non-locally} influences the large scale structure outside the observed domain, thus increasing the region influenced by \gls{data} beyond the \glslink{survey geometry}{survey boundaries}. Figure \ref{fig:inside_out} therefore demonstrates the ability of our algorithm to correctly account for information propagation via \gls{Lagrangian transport} within a fully probabilistic approach. The ability to provide 4D dynamic \glslink{formation history}{formation histories} for \gls{SDSS} \gls{data} together with accurate \gls{uncertainty quantification} paves the path towards high precision \gls{chrono-cosmography}, permitting us to study the inhomogeneous evolution of our Universe. Detailed and quantitative analysis of the various aspects of the results obtained in this chapter are discussed in part \ref{part:IV} of this thesis and will be the subject of future publications.

\section{Summary and conclusions}
\label{sec:Summary and conclusions BORGSDSS}

This chapter discusses a fully Bayesian \glslink{chrono-cosmography}{chrono-cosmographic} analysis of the 3D cosmological large scale structure underlying the \gls{SDSS} main galaxy sample \citep{Abazajian2009}. We presented a \gls{data} application of the recently proposed {\borg} algorithm \citep[see chapter \ref{chap:BORG} and][]{Jasche2013BORG}, which permits to simultaneously infer \glslink{initial conditions}{initial} and present non-linear 3D \glslink{density field}{density fields} from galaxy observations within a fully probabilistic approach. As discussed in chapter \ref{chap:BORG}, the algorithm incorporates a \glslink{2LPT}{second-order Lagrangian perturbation} model to connect observations to \gls{initial conditions} and to perform dynamical \gls{large-scale structure inference} from \glslink{galaxy survey}{galaxy redshift surveys}.

Besides correctly accounting for usual \glslink{statistical uncertainty}{statistical} and \glslink{systematic uncertainty}{systematic uncertainties}, such as \gls{noise}, \glslink{survey geometry}{survey geometries} and \gls{selection effects}, this methodology also physically treats gravitational \gls{structure formation} in the \glslink{linear regime}{linear} and \gls{mildly non-linear regime} and captures \glslink{high-order correlation function}{higher-order statistics} present in non-linear \glslink{density field}{density fields} \citep[see e.g.][]{Moutarde1991,Buchert1994,Bouchet1995,Scoccimarro2000,Scoccimarro2002}. The {\borg} algorithm explores a \glslink{high-dimensional function}{high-dimensional posterior distribution} via an efficient implementation of a \glslink{HMC}{Hamiltonian Monte Carlo} sampler and therefore provides naturally and fully self-consistently accurate \gls{uncertainty quantification} for any finally inferred quantity. 

In the paper corresponding to this work \citep{Jasche2015BORGSDSS}, we upgraded the original \gls{sampling} procedure described in \citet{Jasche2013BORG} to account for automatic \gls{noise} calibration and \gls{luminosity} dependent galaxy \glslink{bias}{biases} (see sections \ref{sec:The Gamma-distribution for noise sampling} and \ref{sec:Calibration of the noise level}). To do so, we followed the philosophy described in \citet{Jasche2013BIAS} and splitted the main galaxy sample into six \glslink{luminosity}{absolute magnitude} bins in the range $-21<M_{^{0.1}r}<-17$. The Bayesian analysis treats each of this six galaxy sub-samples as an individual \gls{data} set with its individual \glslink{statistical uncertainty}{statistical} and \glslink{systematic uncertainty}{systematic uncertainties}. As described in sections \ref{sec:The Gamma-distribution for noise sampling} and \ref{sec:Calibration of the noise level}, the original algorithm described in \citet{Jasche2013BORG} has been augmented by a power-law \gls{bias} model and an additional \gls{sampling} procedure to jointly infer corresponding \glslink{noise parameter}{noise levels} for the respective galaxy samples. 

As discussed in section \ref{sec:The BORG SDSS analysis}, we applied this modified version of the {\borg} algorithm to the \gls{SDSS} DR7 main galaxy samples and generated about 12,000 full three dimensional data-constrained \gls{initial conditions} in the course of this work. The \glslink{initial conditions}{initial} \gls{density field}, at a \gls{scale factor} of $a=10^{-3}$, has been inferred on a \glslink{comoving coordinates}{comoving} Cartesian equidistant grid, of side length $750$ Mpc$/h$ and $256^3$ grid nodes. This amounts to a target resolution of about $\sim~3~\mathrm{Mpc}/h$ for respective volume elements. Density amplitudes at these Lagrangian grid nodes correspond to about \glslink{high-dimensional parameter space}{$\sim ~10^7$ parameters} to be constrained by our \gls{inference} procedure. Typically, the generation of individual data-constrained realizations involves an equivalent of $\sim ~200$ \gls{2LPT} evaluations and requires on the order of $1500$ seconds on $16$ cores. Despite the complexity of the problem, we demonstrated that our sampler can explore \glslink{high-dimensional parameter space}{multi-million dimensional parameter spaces} via efficient \glslink{MCMC}{Markov Chain Monte Carlo} algorithms with an asymptotic \gls{acceptance rate} of about 60 percent, rendering our numerical \gls{inference} framework numerically feasible.

To test the performance of the sampler, we followed a standard approach for testing the initial \gls{burn-in} behavior via experiments \citep[see e.g.][]{Eriksen2004,JascheKitaura2010,Jasche2013BORG}. We initialized the sampler with a \glslink{grf}{Gaussian random field} scaled by a factor of $0.01$, to start from an over-dispersed state. During an initial \gls{burn-in} period the sampler performed a systematic drift towards the target region in parameter space. We examined the initial \gls{burn-in} behavior by following the sequence of \textit{a posteriori} \glslink{power spectrum}{power spectra}, measured from the first $2500$ \glslink{sample}{samples}, and showed that subsequent \glslink{sample}{samples} homogeneously approach the target spectrum throughout all regions in Fourier space without any sign of hysteresis. This indicates the efficiency of the sampler to rapidly explore all scales of the \gls{inference} problem. The absence of any particular \gls{bias} or erroneous power throughout all scales in Fourier space, further demonstrates the fact that \gls{survey geometry}, \gls{selection effects}, galaxy \glslink{bias}{biasing} and observational \gls{noise} have been accurately accounted for in this analysis. These \textit{a posteriori} \glslink{power spectrum}{power spectra} also indicate that individual data-constrained realizations possess the correct physical power in all regions in Fourier space, and can therefore be considered as physically meaningful \glslink{density field}{density fields}. This fact has been further demonstrated in section \ref{sec:inf_density_field} by showing slices through an arbitrary data-constrained realization. These results clearly demonstrate the power of our Bayesian methodology to correctly treat the ill-posed \gls{inverse problem} of inferring signals from incomplete observations, by augmenting unobserved regions with statistically and physically meaningful information. In particular, constrained and unconstrained regions in the \glslink{sample}{samples} are visually indistinguishable, demonstrating a major improvement over previous approaches, typically relying on \glslink{grf}{Gaussian} or \glslink{log-normal distribution}{log-normal statistics}, incapable of representing the \glslink{filament}{filamentary structure} of the \gls{cosmic web} \citep[see e.g.][]{JascheKitaura2010}. It should be remarked that this fact not only demonstrates the ability to access \glslink{high-order correlation function}{high-order statistics} in finally inferred quantities such as 3D \glslink{density field}{density} maps, but also reflects the control of \glslink{high-order correlation function}{high-order statistics} in \gls{uncertainty quantification} far beyond standard normal statistics.

The ensemble of $12,000$ full 3D data-constrained \glslink{sample}{samples} permits us to estimate any desired statistical summary. In particular, in section \ref{sec:inf_density_field}, we showed \glslink{posterior mean}{ensemble mean} \glslink{density field}{density fields} for \glslink{final conditions}{final} and \gls{initial conditions}. A particularly interesting aspect is the fact that the algorithm manages to infer highly-detailed large scale structures even in regimes only poorly covered by observations \citep[for further comments see chapter \ref{chap:BORG} and][]{Jasche2013BORG}. To demonstrate the possibility of \gls{uncertainty quantification}, we also calculated the ensemble voxel-wise \gls{posterior standard deviation}, which reflects the degree of \gls{statistical uncertainty} at every volume element in the \gls{inference} domain. As discussed in section \ref{sec:inf_density_field}, these results clearly reflect the signal-dependence of \gls{noise} for any inhomogeneous \glslink{Poisson process}{point processes}, such as discrete \glslink{Poisson process}{Poissonian} galaxy distribution. As expected, high signal regions correspond to high variance regions. These results further demonstrate the ability to accurately translate uncertainties in the \gls{final conditions} to \glslink{initial conditions}{initial} \glslink{density field}{density fields}, as demonstrated by the plots of voxel-wise \glslink{posterior standard deviation}{standard deviations} for corresponding \glslink{initial conditions}{initial} \glslink{density field}{density fields}. However, note that voxel-wise \glslink{posterior standard deviation}{standard deviations} are just an approximation to the full joint and correlated uncertainty that otherwise can by correctly \glslink{uncertainty quantification}{quantified} by considering the entire set of data-constrained realizations. Besides 3D \glslink{initial conditions}{initial} and \glslink{final conditions}{final} \glslink{density field}{density fields}, the methodology also provides information on cosmic dynamics, as mediated by the \gls{2LPT} model. In section \ref{sec:3d_velocity_fields}, we showed a \gls{velocity field} realization in one \gls{sample}. In particular, we showed the radial, polar and azimuthal velocity components in a 4 Mpc$/h$ thick slice around the celestial equator for the observed domain. These velocities are not primarily constrained by observations, but are derived from the \gls{2LPT} model. However, since \gls{2LPT} displacement vectors are data-constrained, and since displacement vectors and velocities differ only by constant factors independent of the \gls{inference} process, derived velocities are considered to be accurate.

As pointed out frequently, the {\borg} algorithm employs \gls{2LPT} as a dynamical model to connect \gls{initial conditions} to present observations of \gls{SDSS} galaxies. As a consequence, the algorithm not only provides 3D \glslink{density field}{density} and \glslink{velocity field}{velocity fields} but also infers plausible 4D \glslink{formation history}{formation histories} for the observed \gls{LSS}. In section \ref{sec:cosmic_history}, we illustrated this feature with an individual \gls{sample}. We followed its cosmic evolution from a \glslink{initial conditions}{initial} \gls{scale factor} of $a=0.02$ to the present epoch at $a=1.00$. As could be seen, the \glslink{initial conditions}{initial} \gls{density field} appears homogeneous and obeys \glslink{grf}{Gaussian statistics}. In the course of \gls{structure formation} \glslink{cluster}{clusters}, \glslink{filament}{filaments} and \glslink{void}{voids} are formed. To demonstrate that this \gls{formation history} correctly recovers the observed large scale structure, we plotted the observed galaxies, for the six \gls{luminosity} bins, on top of the \glslink{final conditions}{final} \gls{density field}. These results clearly demonstrate the ability of our algorithm to infer plausible large scale structure \glslink{formation history}{formation histories} compatible with observations. Additionally, since the {\borg} algorithm is a full Bayesian \gls{inference} framework, it not only provides a single 4D history, but an ensemble of such data-constrained \glslink{formation history}{formation histories} and thus accurate means to \glslink{uncertainty quantification}{quantify corresponding observational uncertainties}. In particular, our methodology correctly accounts for the \gls{non-local} \glslink{Lagrangian transport}{transport} of observational information between present observations and corresponding inferred \gls{initial conditions}. As discussed in section \ref{sec:cosmic_history}, the \gls{information content} in \glslink{initial conditions}{initial} and \gls{final conditions} has to be conserved but can be distributed differently. High-density regions in the \gls{final conditions}, typically coinciding with high signal-to-noise regions in the \gls{data}, form by clustering of matter which was originally distributed over a larger Eulerian volume in the \gls{initial conditions}. For this reason, the observational information associated to a \gls{cluster} in the \glslink{final conditions}{final} \gls{density field} will be distributed over a larger volume in the corresponding \glslink{initial conditions}{initial} \gls{density field}. Conversely, the \gls{information content} of \glslink{void}{voids} in the \gls{final conditions} will be confined to a smaller volume in the \gls{initial conditions}. This fact is also reflected by the analysis of voxel-wise \glslink{posterior standard deviation}{standard deviations} presented in section \ref{sec:inf_density_field}. While the signal-to-noise ratio is highly clustered in the \gls{final conditions}, the same amount of observational information is distributed more homogeneously over the entire volume in corresponding \gls{initial conditions}. As discussed in section \ref{sec:cosmic_history}, \glslink{dark matter particles}{particles} within the observed domain at the \glslink{final conditions}{final} state may originate from regions outside the corresponding \glslink{comoving coordinates}{comoving} Eulerian volume in the \gls{initial conditions} and vice versa \citep[also see chapter \ref{chap:BORG} and][]{Jasche2013BORG}. This \gls{non-local} \glslink{Lagrangian transport}{translation of information along Lagrangian trajectories} is also the reason for the ability of our methodology to extrapolate beyond the \glslink{survey geometry}{survey boundaries} of the \gls{SDSS} and infer the \gls{LSS} there within a fully probabilistic and rigorous approach. In particular, the high degree of control on \glslink{statistical uncertainty}{statistical uncertainties} permit us to perform accurate \glslink{inference}{inferences} on the nature of \gls{initial conditions} and \glslink{formation history}{formation histories} for the observed \gls{LSS} in these regions. For these reasons we believe that inferred \glslink{final conditions}{final} \glslink{posterior mean}{ensemble mean} fields and corresponding voxel-wise \glslink{posterior standard deviation}{standard deviations} as a means of \gls{uncertainty quantification}, may be of interest to the scientific community. These data products have been published as supplementary material along with the article, and are accessible at \href{http://iopscience.iop.org/1475-7516/2015/01/036}{http://iopscience.iop.org/1475-7516/2015/01/036}.

In summary, this chapter describes an application of the previously proposed {\borg} algorithm to the \gls{SDSS} DR7 main galaxy sample. As demonstrated, our methodology produces a rich variety of scientific results, various aspects of which are objects of detailed and quantitative analyses in subsequent chapters of this thesis and forthcoming publications. Besides pure three dimensional \glslink{reconstruction}{reconstructions} of the present \gls{density field}, the algorithm provides detailed information on corresponding \gls{initial conditions}, large scale dynamics and \glslink{formation history}{formation histories} for the observed \gls{LSS}. Together with a thorough \glslink{uncertainty quantification}{quantification of joint and correlated observational uncertainties}, these results mark the first steps towards high precision \gls{chrono-cosmography}, the subject of analyzing the four dimensional state of our Universe. 

%% file: Chapter6/Chapter6Content.tex
\part{The non-linear regime of structure formation}
\label{part:III}

\chapter{Remapping Lagrangian perturbation theory}
\label{chap:remapping}
\minitoc

\begin{flushright}
\begin{minipage}[c]{0.6\textwidth}
\rule{\columnwidth}{0.4pt}

``Everyone thinks of changing the world, but no one thinks of changing himself.''\\
--- Leo Tolstoy\\
Quoted in \citet{Bryan1999}, \textit{The Artist's Way at Work: Riding the Dragon}

\vspace{-5pt}\rule{\columnwidth}{0.4pt}
\end{minipage}
\end{flushright}

\abstract{\section*{Abstract}

On the smallest scales, three-dimensional \glslink{LSS}{large-scale structure} \glslink{galaxy survey}{surveys} contain a wealth of \glslink{information content}{cosmological information} which cannot be trivially extracted due to the \glslink{non-linear evolution}{non-linear dynamical evolution} of the \gls{density field}. \glslink{LPT}{Lagrangian perturbation theory} is widely applied to the generation of \glslink{mock catalog}{mock halo catalogs} and data analysis. In this chapter, we propose a method designed to improve the correspondence between these \glslink{density field}{density fields}, in the \gls{mildly non-linear regime}. We develop a computationally fast and flexible tool for a variety of cosmological applications. Our method is based on a \gls{remapping} of the approximately-evolved \gls{density field}, using information extracted from \glslink{N-body simulation}{$N$-body simulations}. The \gls{remapping} procedure consists of replacing the \gls{one-point distribution} of the \gls{density contrast} by one which better accounts for the \glslink{full gravity}{full gravitational dynamics}. As a result, we obtain a physically more pertinent \gls{density field} on a point-by-point basis, while also improving \glslink{high-order correlation function}{higher-order statistics} predicted by \gls{LPT}. We quantify the approximation error in the \gls{power spectrum} and in the \gls{bispectrum} as a function of scale and \gls{redshift}. Our \gls{remapping} procedures improves \glslink{one-point distribution}{one-}, \glslink{two-point correlation function}{two-} and \glslink{three-point correlation function}{three-point statistics} at scales down to 8~Mpc/$h$.}

\draw{This chapter is adapted from its corresponding publication, \citet{Leclercq2013}.}

\section{Introduction}
\label{sec:Remapping_intro}

At present, observations of the three-dimensional \gls{LSS} are major sources of \glslink{information content}{information} on the origin and evolution of the Universe. According to the current paradigm of cosmological \gls{structure formation}, the presently observed structures formed via gravitational clustering of \glslink{CDM}{cold} \gls{dark matter particles} and condensation of baryonic matter in gravitational \glslink{potential well}{potential wells}. Consequently, the large-scale matter distribution retains a memory of its \gls{formation history}, enabling us to study the \glslink{homogeneous Universe}{homogeneous} as well as the inhomogeneous evolution of our Universe.

Due to \glslink{non-linear evolution}{non-linearities} involved in the formation process, at present there exists just limited analytic understanding of \gls{structure formation} in terms of perturbative expansions in \glslink{EPT}{Eulerian} or \glslink{LPT}{Lagrangian} representations. Both of these approaches rely on a truncated sequence of \gls{momentum} \glslink{moment}{moments} of the \gls{Vlasov-Poisson system}, completed by \gls{fluid} dynamic assumptions \citep[see chapter \ref{chap:theory} or e.g.][and references therein]{Bernardeau2002}. For this reason, the validity of these approaches ceases, once the evolution of the \gls{LSS} enters the \glslink{shell-crossing}{multi-stream} regime \citep[see e.g.][]{Pueblas2009}.

Nevertheless, \glslink{EPT}{Eulerian} and \glslink{LPT}{Lagrangian} approximations have been successfully applied to the analysis of three-dimensional \glslink{density field}{density fields} in regimes where they are still applicable, either at large scales or in the early Universe. Particularly, \gls{LPT} captures significant \glslink{mode coupling}{mode-coupling} information that is encoded beyond \glslink{linear evolution}{linear theory}, such as large-scale flows and free-streaming, yielding three-dimensional matter distributions approximating those of \glslink{full gravity}{full scale numerical simulations} with reasonable accuracy \citep{Moutarde1991,Buchert1994,Bouchet1995,Scoccimarro1998b,Scoccimarro2000,Scoccimarro2002,Yoshisato2006}. Especially, \glslink{2LPT}{second-order Lagrangian perturbation theory} has been widely applied in data analysis and for fast generation of \glslink{mock catalog}{galaxy mock catalogs} (e.g. \textsc{PTHalos}: \citealp{Scoccimarro2002,Manera2013}; \textsc{Pinocchio}: \citealp{Monaco2002a,Monaco2002b,Taffoni2002,Heisenberg2011,Monaco2013}) that can be useful to estimate error bounds when analyzing observations.

Modern cosmological data analysis has an increasing demand for analytic and computationally inexpensive models providing accurate representations of the \gls{mildly non-linear regime} of \gls{structure formation}. Over the years, various \glslink{non-linear approximation}{non-linear approximations} and attempts to extend the validity of \gls{LPT} have been proposed (see section \ref{sec:NL-approxs}). These include the \glslink{SC}{spherical collapse model} \citep{Gunn1972,Bernardeau1994}, the truncated Zel'dovich approximation \citep{Melott1994} and models with various forms for the \gls{velocity potential} \citep{Coles1993,Munshi1994} or the addition of a \gls{viscosity} term in the \glslink{Euler's equation}{Euler equation} \citep[the \glslink{adhesion approximation}{adhesion} model,][]{Gurbatov1989}. Analytical techniques to improve the convergence and behavior of \glslink{EPT}{standard perturbation theory}, successfully employed in \gls{quantum field theory} and \glslink{statistical mechanics}{statistical physics}, have also been applied in the context of gravitational clustering. These include \gls{renormalized perturbation theory} \citep{Crocce2006a}, the \gls{path integral formalism} \citep{Valageas2007}, and the \gls{renormalization group flow} \citep{Matarrese1997}. More recently, \citet{Tassev2012a,Tassev2012b} constructed a physical picture of the matter distribution in the \gls{mildly non-linear regime}, and developed a method yielding improvements over \gls{LPT} \citep{Tassev2013}, in particular at the scales relevant for \glslink{BAO}{baryon acoustic peak} \gls{reconstruction} \citep{Tassev2012c}.

In this chapter, we propose a numerically efficient method designed to improve the correspondence between approximate models and \glslink{full gravity}{full numerical simulations} of gravitational large-scale \gls{structure formation}. Generally, it can be applied to any approximate model of gravitational instability, but it is especially targeted to improving \glslink{LPT}{Lagrangian} methods. We illustrate both these methods on fields evolved with \gls{LPT}: at order one, the \glslink{ZA}{Zel'dovich approximation} \citep{Zeldovich1970,Shandarin1989} and \glslink{2LPT}{second-order Lagrangian perturbation theory}.

\gls{LPT} and \glslink{N-body simulation}{$N$-body} \glslink{density field}{density fields} are visually similar, which suggests that the properties of \gls{LPT} could be improved by one-to-one mapping in voxel space, following a similar line of thoughts as the ``\gls{Gaussianization}'' idea originally proposed by \citet{Weinberg1992} and inspired existing techniques, widely used in cosmology \citep{Melott1993,Croft1998,Narayanan1998,Croft1999,Feng2000,Neyrinck2011,Yu2011,Yu2012,Neyrinck2013b}. The method described in this chapter is based on a \textit{\gls{remapping}} of the approximately evolved particle distribution using information extracted from \glslink{N-body simulation}{$N$-body simulations}. It basically consists of replacing the \gls{one-point distribution} of the approximately evolved distribution by one which better accounts for the \glslink{full gravity}{full gravitational system}. In this fashion, we adjust the \gls{one-point distribution} to construct a physically more reasonable representation of the three-dimensional matter distribution, while retaining or improving \glslink{high-order correlation function}{higher order statistics}, described already reasonably well by the \gls{ZA} \citep{Zeldovich1970,Doroshkevich1970a,Shandarin1989,Buchert1989,Moutarde1991,Yoshisato2006} and by \gls{2LPT} \citep{Moutarde1991,Buchert1994,Bouchet1995,Scoccimarro1998b,Scoccimarro2000,Scoccimarro2002}.

Major problems with naive approaches to \gls{remapping} \gls{LPT} \glslink{density field}{density fields} arise from minor deviations in \glslink{structure type}{structure types} represented by \gls{LPT} models and \gls{full gravity}. For this reason, in chapter \ref{chap:lpt}, we discussed the different representations of \glslink{cluster}{clusters}, \glslink{void}{voids}, \glslink{sheet}{sheets}, and \glslink{filament}{filaments}, predicted by \gls{LPT} and \glslink{N-body simulation}{$N$-body simulations}. Besides being of general interest for \gls{LSS} data analysis, the insights gained from this comparison will allow us to improve the \gls{remapping} procedure.

Implementing and testing the accuracy and the regime of validity of our method is essential, and is subject of the present chapter. Our study quantifies the approximation error as a function of scale in terms of a set of statistical diagnostics. From cosmographic measurements, $\sigma_8$ is known to be of order unity, which means that gravity becomes highly \glslink{non-linear evolution}{non-linear} at some scale around 8~Mpc/$h$. Our method is expected to break down due to \gls{shell-crossing} in \gls{LPT}, at some scale larger than 8~Mpc/$h$. Achieving a resolution of 16~Mpc/$h$ would already constitute substantial improvement with respect to existing methods, since \glslink{non-linear evolution}{non-linearities} begin to affect even large-scale cosmographic measurements such as the determination of the \glslink{BAO}{baryon acoustic oscillations} scale from \glslink{galaxy survey}{galaxy surveys} \citep[about 125 Mpc/$h$, e.g.][]{Eisenstein2005}. However, we explore the validity of the improvement at 8~Mpc/$h$ down to 4~Mpc/$h$, to see to what extent we can push the limit for possible data analysis applications into the \gls{non-linear regime}. Recall that in three-dimensional \gls{LSS} \glslink{galaxy survey}{surveys}, the number of modes usable for cosmological exploitation scales as the cube of the largest wavenumber, $k^3$, meaning that even minor improvements in the \gls{mildly non-linear regime} would give access to much more \glslink{information content}{cosmological information} from existing and upcoming observations.

As will be demonstrated, this method can be used to generate realizations of \glslink{density field}{density fields} much faster than \glslink{N-body simulation}{$N$-body simulations}. Even though approximate, these fast realizations of \glslink{mock catalog}{mock density fields} may be sufficient to model the salient features of the \glslink{non-linear evolution}{non-linear} \gls{density field} for certain applications.

This chapter is structured as follows. In section \ref{par:remapping}, we describe the \gls{remapping} procedure for the \gls{density contrast} of \glslink{final conditions}{present-day} \glslink{density field}{density fields}, analyze the obstacles to straightforward application and present an improved method. In section \ref{par:results}, we apply the procedure to cosmological models using data from numerical simulations, we study the statistics of remapped fields and quantify the approximation error. We discuss our results and give our conclusions in section \ref{par:conclusion}.

The setup of \gls{LPT} and \glslink{N-body simulation}{$N$-body simulations} used in this chapter are described at the beginning of chapter \ref{chap:lpt}.

\section{Method}
\label{par:remapping}

In this section, we discuss the \gls{remapping} procedure and apply it to cosmological \glslink{density field}{density fields} evolved with \gls{LPT}. Naively following the approach of \citet{Weinberg1992} for \glslink{final conditions}{present-day} \glslink{density field}{density fields} yields the procedure described in section \ref{par:remapping-procedure}. This approach is not entirely satisfactory and we analyze the reasons for its shortcomings in section \ref{par:comparison-structure-types}. In consequence, we propose a improvement of the \gls{remapping} procedure in section \ref{par:improvement-remapping-procedure}. The properties of the \gls{remapping function} are examined in section \ref{par:remapping-function}.

\subsection{Remapping procedure}
\label{par:remapping-procedure}

\begin{figure*}
\begin{center}
\includegraphics[width=0.90\textwidth]{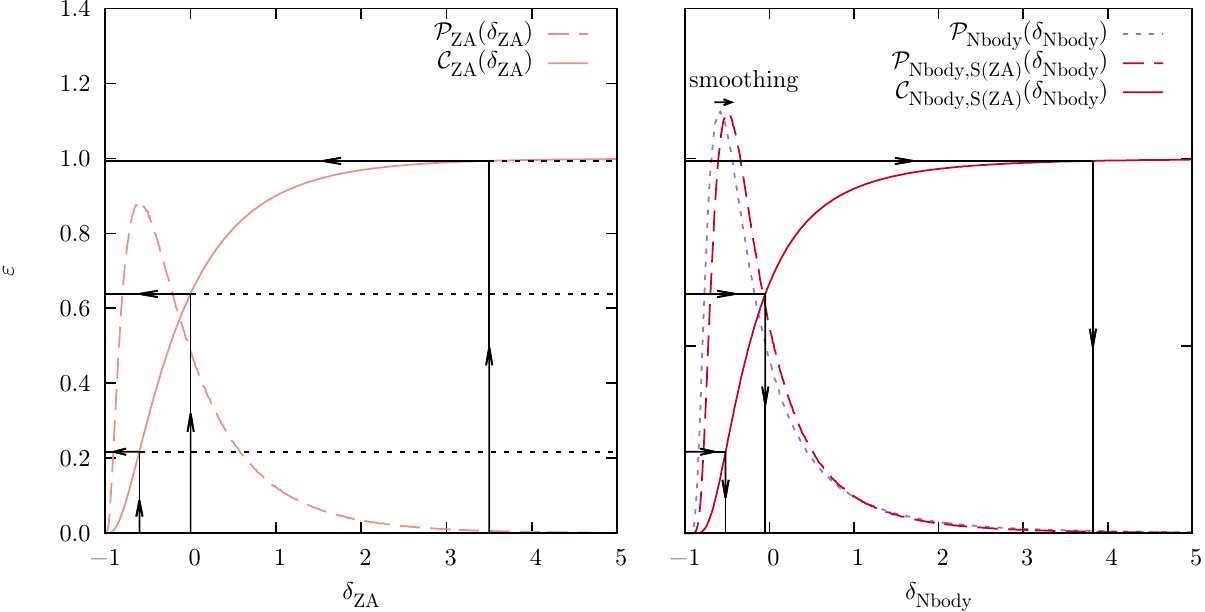} \\
\includegraphics[width=0.90\textwidth]{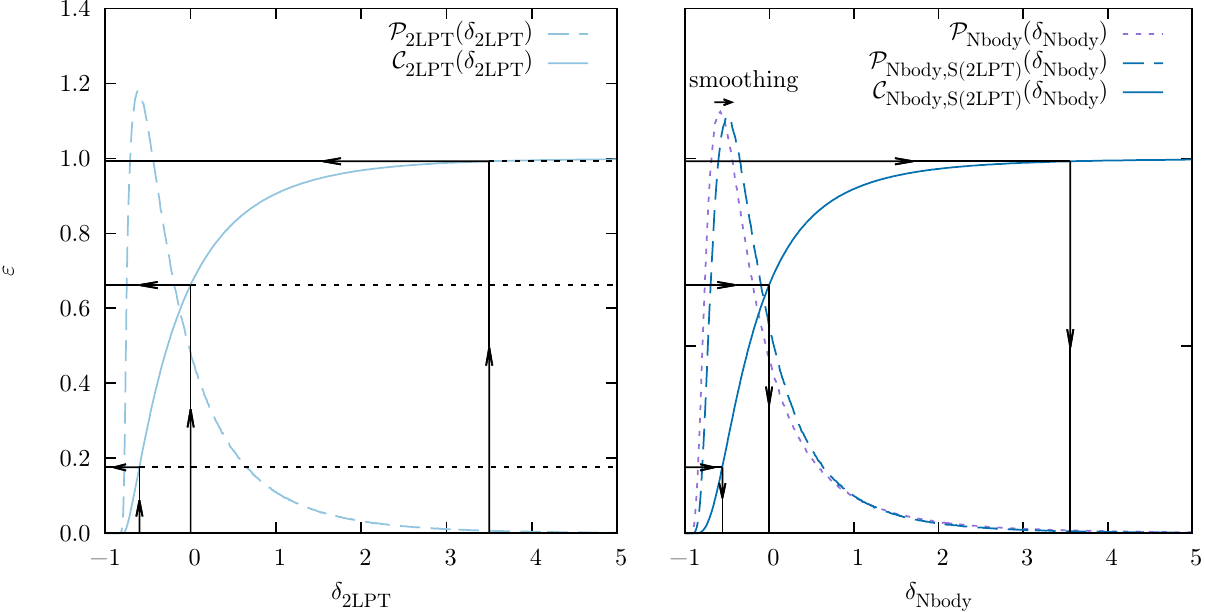}
\caption{A graphical illustration of the improved \gls{remapping} procedure at \gls{redshift} zero, for the \gls{ZA} (upper panel) and \gls{2LPT} (lower panel). On the right panels, the dotted purple curves are the \glslink{pdf}{probability distribution function} for the \gls{density contrast} in \glslink{full gravity}{full $N$-body dynamics}. The first step of the procedure is to smooth the \glslink{N-body simulation}{$N$-body} \gls{density field} using the \gls{transfer function} given by equation \eqref{eq:transfer-function}, which yields slightly different \glslink{pdf}{pdfs} (dashed dark red and dark blue curves on the right panels). On the left panels, the dashed curves are the \glslink{pdf}{pdfs} for the \gls{density contrast} in the \glslink{final conditions}{final} \gls{density field}, evolved from the same \gls{initial conditions} with \gls{LPT} (\gls{ZA}: red line and \gls{2LPT}: blue line). The \glslink{pdf}{pdfs} are computed on a 8~Mpc/$h$ mesh ($128^3$-voxel grid) using the full statistics from eight \glslink{particle realization}{realizations} of $512^3$-particles in a 1024~Mpc/$h$ box with \gls{periodic boundary conditions}. The solid curves are their respective integrals, the cumulative distribution functions. The second step is \gls{remapping}, which assigns a voxel with \gls{density contrast} $\delta_\mathrm{LPT}$ and fractional rank $\varepsilon$ the value of $\delta_\mathrm{Nbody}$ that would have the same fractional rank in the smoothed $N$-body distribution (equation \eqref{eq:remapping-eulerian-improved}). This \gls{remapping} is illustrated for \gls{2LPT} with three sample points: $\delta_\mathrm{2LPT}=-0.60$ maps to $\delta_\mathrm{Nbody}=-0.56$, $\delta_\mathrm{2LPT}=0.00$ maps to $\delta_\mathrm{Nbody}=-0.01$, and $\delta_\mathrm{2LPT}=3.50$ maps to $\delta_\mathrm{Nbody}=3.56$. The \gls{remapping} procedure imposes the \gls{one-point distribution} of the smoothed $N$-body field while maintaining the rank order of the \gls{LPT}-evolved \glslink{density field}{density fields}. The last step is an increase of small-scale power in the remapped distribution using the reciprocal \gls{transfer function}, equation \eqref{eq:transfer-function-reciprocal}.}
\label{fig:remapping_procedure_eulerian}
\end{center}
\end{figure*}

In this section, we describe the \gls{remapping} algorithm used to go from a low-\gls{redshift} realization of a \gls{density field} evolved with \gls{LPT} to one evolved with \glslink{full gravity}{full $N$-body gravitational dynamics}. Note that both fields obey the same \gls{initial conditions} but are evolved by different physical models.

\glslink{density field}{Density fields} are defined on Cartesian grids of cubic voxels. Linear \gls{gravitational evolution} exactly maintains the relative amplitude of fluctuations in different voxels. Due to \gls{mode coupling}, positive and negative fluctuations grow at different rates in the \gls{non-linear regime}, but even \gls{non-linear evolution} tends to preserve the \textit{rank order} of the voxels, sorted by density.

The \glslink{one-point distribution}{one-point} \glslink{pdf}{probability distribution functions} and the \glslink{cdf}{cumulative distribution functions} (\gls{cdf}) of the \glslink{final conditions}{final} \glslink{density field}{density fields}, evolved with either \gls{LPT} or \glslink{full gravity}{full $N$-body gravitational dynamics}, exhibit similar, but not identical shapes. This result suggests a way to improve the approximation with information extracted from the \glslink{N-body simulation}{$N$-body simulation}: maintain the rank order of the voxels, but reassign densities so that the two \glslink{cdf}{cdfs} match. The method therefore resembles the ``\gls{Gaussianization}'' procedure proposed by \citet{Weinberg1992}, an attempt to reconstruct the \gls{initial conditions} of a \gls{density field} from its \glslink{final conditions}{final} \gls{cdf}.

Let $\p_{\mathrm{LPT}}$ and $\p_{\mathrm{Nbody}}$ denote the \glslink{pdf}{probability distribution functions} for the \gls{density contrast} in the \gls{LPT} and in the \glslink{full gravity}{full $N$-body} \glslink{density field}{density fields}, respectively. Let $\mathpzc{C}_{\mathrm{LPT}}$ and $\mathpzc{C}_{\mathrm{Nbody}}$ be their integrals, the cumulative distribution functions. $\mathpzc{C}_{\mathrm{LPT}}(\delta_\mathrm{LPT})$ is the \textit{fractional rank} for $\delta_\mathrm{LPT}$ i.e. the probability that the \gls{density contrast} at a given voxel is smaller than $\delta_\mathrm{LPT}$, $\p_{\mathrm{LPT}}(\delta \leq \delta_\mathrm{LPT})$, and the analogous for the \glslink{N-body simulation}{$N$-body} field. The \gls{remapping} procedure works as follows. A voxel with rank order $\delta_\mathrm{LPT}$ is assigned a new density $\delta_\mathrm{Nbody}$ such that
\begin{equation}
\label{eq:remapping-eulerian}
\mathpzc{C}_{\mathrm{LPT}}(\delta_\mathrm{LPT}) = \mathpzc{C}_{\mathrm{Nbody}}(\delta_\mathrm{Nbody})
\end{equation}
(also see figure \ref{fig:remapping_procedure_eulerian} for a schematic outline of this method). The left panels of figure \ref{fig:remapping_procedure_eulerian} show $\p_{\mathrm{LPT}}$ (dashed curves) and the corresponding cumulative distributions, $\mathpzc{C}_{\mathrm{LPT}}$ (solid curves). On the right panels, the dotted curves represent the \gls{pdf} of the corresponding \glslink{N-body simulation}{$N$-body} \glslink{particle realization}{realization}, $\p_{\mathrm{Nbody}}$. Remapping assigns to a voxel with \gls{density contrast} $\delta_\mathrm{LPT}$ and fractional rank $\varepsilon~=~\mathpzc{C}_{\mathrm{LPT}}(\delta_\mathrm{LPT})$ the value of $\delta_\mathrm{Nbody}$ that would have the same fractional rank in the \glslink{N-body simulation}{$N$-body} distribution.

Since $\mathpzc{C}_{\mathrm{Nbody}}$ contains exactly the same information as $\p_{\mathrm{Nbody}}$, the \gls{remapping} procedure imposes the \gls{one-point distribution} taken from the \glslink{N-body simulation}{$N$-body-evolved} \gls{density field} while maintaining the rank order of the \gls{LPT}-evolved \gls{density field}. In other words, only the weight of underdensities and overdensities is modified, while their locations remain unchanged. In this fashion, we seek to adjust the \gls{density field} while maintaining \glslink{high-order correlation function}{higher-order statistics} provided by \gls{LPT} with reasonable accuracy. We checked numerically that mass is always conserved in this procedure.

\subsection{Comparison of structure types in LPT and in $N$-body dynamics}
\label{par:comparison-structure-types}

\begin{figure}
\begin{center}
\includegraphics[width=0.5\textwidth]{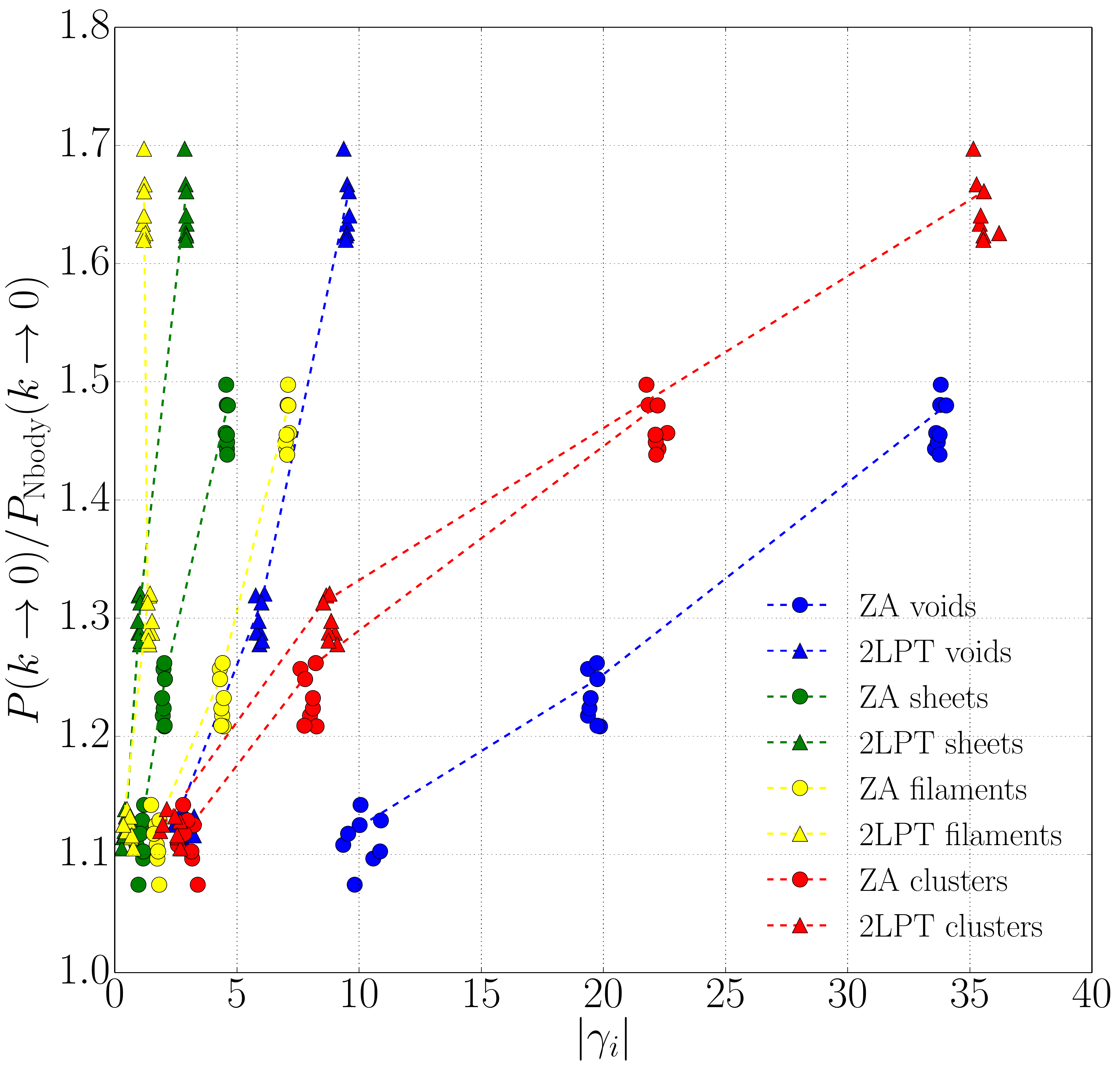}
\caption{The large-scale bias in the \gls{power spectrum} of \glslink{density field}{density fields}, remapped with the procedure described in section \ref{par:remapping-procedure}, as a function of the mismatch between \glslink{structure type}{structure types}. The \glslink{estimator}{estimators} $\gamma_i$ are defined by equation \eqref{eq:gamma}. Eight realizations of the \gls{ZA} (circles) and \gls{2LPT} (triangles) are compared to the corresponding \glslink{N-body simulation}{$N$-body} \glslink{particle realization}{realization}, for various resolutions (from bottom to top: 16~Mpc/$h$, 8~Mpc/$h$, 4~Mpc/$h$). The large-scale bias in the \glslink{power spectrum}{power spectra} of remapped fields is strongly correlated to the \glslink{VFF}{volume fraction} of structures incorrectly predicted by \gls{LPT}.}
\label{fig:bias_analysis}
\end{center}
\end{figure}

We implemented the \gls{remapping} procedure described in the previous section and checked that we are able to modify \gls{LPT} \glslink{density field}{density fields} so as to correctly reproduce the \gls{one-point distribution} of a \glslink{full gravity}{full $N$-body simulation}. However, we experience a large-scale bias in the \gls{power spectrum}, namely the amplitude of the spectrum of the remapped fields is slightly too high. Generally, a non-linear transformation in voxel space can change the variance of a field. This is consistent with the findings of \citet{Weinberg1992}, who found a similar effect in his \glslink{reconstruction}{reconstructions} of the \gls{initial conditions}, and who addressed the issue by rescaling the \gls{density field} in Fourier space. However, such an approach will generally destroy the remapped \gls{one-point distribution}, and may even further lead to Gibbs ringing effects which will make the remapped field unphysical.

The bias indicates a stronger clustering in the remapped \gls{density field}. Since \gls{remapping} is a \gls{local} operation in voxel space, this large-scale bias means that erroneous \gls{remapping} of small-scale structures affects the clustering at larger scales. To identify the cause of this bias, we performed a study of differences in \glslink{structure type}{structure types} (\glslink{void}{voids}, \glslink{sheet}{sheets}, \glslink{filament}{filaments}, and \glslink{cluster}{clusters}) in \glslink{density field}{density fields} predicted by \gls{LPT} and \glslink{N-body simulation}{$N$-body simulations}. With this analysis, we wanted to understand the effect of \gls{remapping} in different density and dynamical regimes of the \gls{LSS}. The results are presented in section \ref{sec:Comparison of structure types in LPT and $N$-body dynamics}. We identified, in particular, a mismatch between the volume occupied by different \glslink{structure type}{structure types} in \gls{LPT} and \glslink{N-body simulation}{$N$-body dynamics}, quantified by the parameters $\gamma_i$ defined by equation \eqref{eq:gamma}.

In figure \ref{fig:bias_analysis}, we plot the large-scale bias observed in remapped fields obtained with the procedure of section \ref{par:remapping-procedure} as a function of $\gamma_i$, for various resolutions. A strong correlation is observed between the bias and the mismatch in the volume occupied by different \glslink{structure type}{structure types}. The difference in the prediction of the volume of extended objects is the cause of the bias: in \glslink{cluster}{clusters} and in \glslink{void}{voids}, \gls{remapping} enhances a larger volume than should be enhanced, which yields on average a stronger clustering in the box.

\subsection{Improvement of the remapping procedure}
\label{par:improvement-remapping-procedure}

In the previous section, we noted that because too much volume of \gls{LPT} \glslink{density field}{density fields} is mapped to the tails of the \glslink{N-body simulation}{$N$-body} \gls{one-point distribution} (dense \glslink{cluster}{clusters} and deep \glslink{void}{voids}), the average \glslink{two-point correlation function}{two-point correlations} are larger after \gls{remapping}. Given this result, minor modifications of the procedure described in section \ref{par:remapping-procedure} can solve the problem. We now propose to improve the \gls{remapping} procedure of \citet{Weinberg1992} for \glslink{final conditions}{present-day} \glslink{density field}{density fields}, in a similar fashion as in \citet{Narayanan1998}, combining it with \gls{transfer function} techniques to deal with the \glslink{mildly non-linear regime}{mildly non-linear} modes as in \citet{Tassev2012a,Tassev2012b}. Our method works as follows.

\begin{enumerate}
\item We degrade the \glslink{N-body simulation}{$N$-body} \gls{density field} to the same degree of smoothness as the \gls{LPT} \gls{density field}, by multiplying the Fourier modes of the \gls{density field} by the \gls{transfer function}
\begin{equation}
\label{eq:transfer-function}
T(k) \equiv \sqrt{\frac{P_{\mathrm{LPT}}(k)}{P_{\mathrm{Nbody}}(k)}} .
\end{equation}
This steps yields a new \gls{density field}, noted Nbody,S(LPT), whose \gls{power spectrum} matches that of the \gls{LPT} \gls{density field}.
\item We \glslink{remapping}{remap} the \gls{LPT} \gls{density field} in the fashion described in section \ref{par:remapping-procedure}, but using as a reference the \gls{cdf} of the smoothed \gls{density field}, $\mathpzc{C}_{\mathrm{Nbody,S(LPT)}}$, instead of the \glslink{full gravity}{full $N$-body} \gls{density field} (see figure \ref{fig:remapping_procedure_eulerian}). The \gls{remapping} condition, equation \eqref{eq:remapping-eulerian}, now reads
\begin{equation}
\label{eq:remapping-eulerian-improved}
\mathpzc{C}_{\mathrm{LPT}}(\delta_\mathrm{LPT}) = \mathpzc{C}_{\mathrm{Nbody,S(LPT)}}(\delta_\mathrm{Nbody}) .
\end{equation}
\item We increase the power of small scales modes in the remapped distribution to the same value as in a \glslink{full gravity}{full $N$-body simulation}, using the reciprocal of the \gls{transfer function} \eqref{eq:transfer-function}, namely
\begin{equation}
\label{eq:transfer-function-reciprocal}
T^{-1}(k) = \sqrt{\frac{P_{\mathrm{Nbody}}(k)}{P_{\mathrm{LPT}}(k)}} .
\end{equation}
\end{enumerate}

This procedure cures the large-scale bias issue experienced with the simple implementation of the \gls{remapping} described in section \ref{par:remapping-procedure}, without requiring any prior knowledge on the corresponding \glslink{N-body simulation}{$N$-body simulation}. As we will demonstrate in section \ref{par:results}, it yields improvement of \glslink{one-point distribution}{one-}, \glslink{two-point correlation function}{two-} and \glslink{three-point correlation function}{three-point statistics} of \gls{LPT}.

\subsection{Remapping function and transfer function}
\label{par:remapping-function}

Since $\mathpzc{C}_{\mathrm{LPT}}$ and $\mathpzc{C}_{\mathrm{Nbody,S(LPT)}}$ are monotonically increasing functions, there is no ambiguity in the choice of $\delta_\mathrm{Nbody}$, and this procedure defines a \textit{\gls{remapping function}} $f$ such that

\begin{equation}
\label{eq:remapping-function-eulerian}
\delta_\mathrm{LPT} \mapsto \delta_\mathrm{Nbody} = \mathpzc{C}_{\mathrm{Nbody,S(LPT)}}^{-1}(\mathpzc{C}_{\mathrm{LPT}}(\delta_\mathrm{LPT})) \equiv f(\delta_\mathrm{LPT}) .
\end{equation}

Establishing a \gls{remapping function} $f$ requires knowledge of both \gls{LPT} and \glslink{N-body simulation}{$N$-body} \gls{density field} statistics. Ideally, several realizations with different \gls{initial conditions} should be combined in order to compute a precise \gls{remapping function}. Indeed, a limited amount of available \glslink{N-body simulation}{$N$-body simulations} results in a lack of statistics and hence uncertainties for the \gls{remapping} procedure in the high density regime. However, this effect is irrelevant from a practical point of view, since these high density events are very unlikely and affect only a negligible number of voxels. As a consequence this uncertainty will only affect to sub-percent level the usual statistical summaries of the \gls{density field}. Note that in any case, if desired, the accuracy of the \gls{remapping function} in the high density regime can be trivially enhanced by enlarging the size or number of \glslink{N-body simulation}{$N$-body simulations} used for its construction. For the analysis presented in this chapter, the \glslink{remapping function}{remapping functions} have been computed using the full statistics from eight \glslink{particle realization}{realizations} of $512^3$ particles in a 1024 Mpc/$h$ box.

Note that once the relationship between the statistical behavior of the \gls{LPT} fields and the \glslink{full gravity}{full non-linear field} is known, this procedure can be used on \gls{LPT} realizations without the need of evolving corresponding \glslink{N-body simulation}{$N$-body simulations}. More specifically, the \gls{remapping function}~$f$ (equation \eqref{eq:remapping-function-eulerian}) and the \gls{transfer function}~$T$ (equation \eqref{eq:transfer-function}) can be tabulated and stored, then used for the fast construction of a large number of \glslink{LSS}{large-scale structure} \glslink{density field}{density fields}. Since producing \gls{LPT} realizations is computationally faster than running \glslink{full gravity}{full gravitational} \glslink{N-body simulation}{simulations} by a factor of several hundreds, our method can be used to produce a large set of \glslink{N-body simulation}{$N$-body}-like realizations in a short time.

\begin{figure*}
\begin{center}
\includegraphics[width=0.49\textwidth]{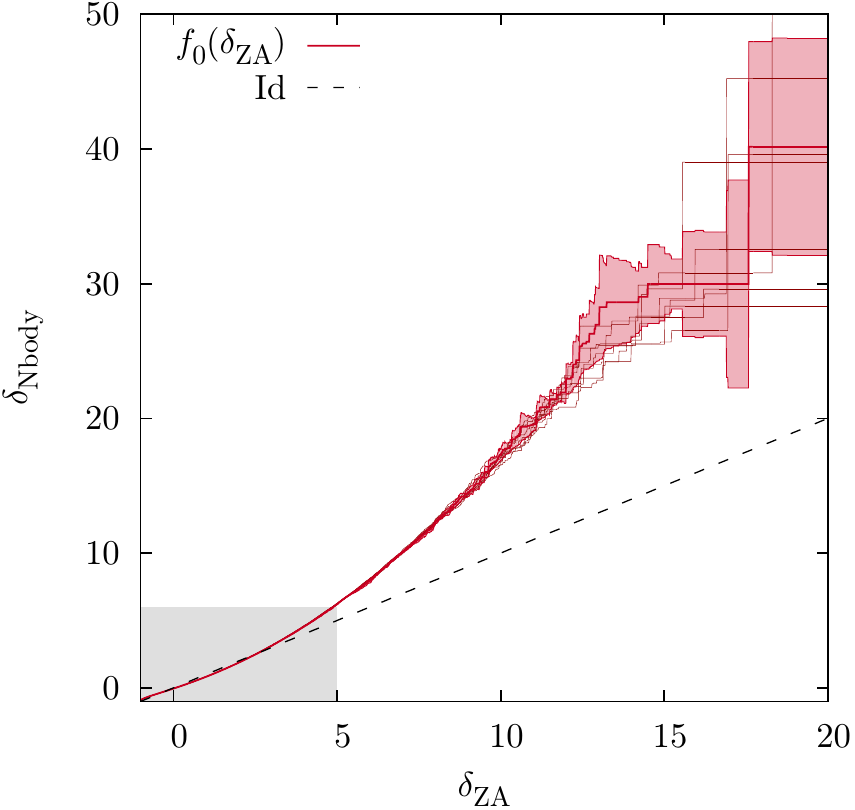}
\includegraphics[width=0.49\textwidth]{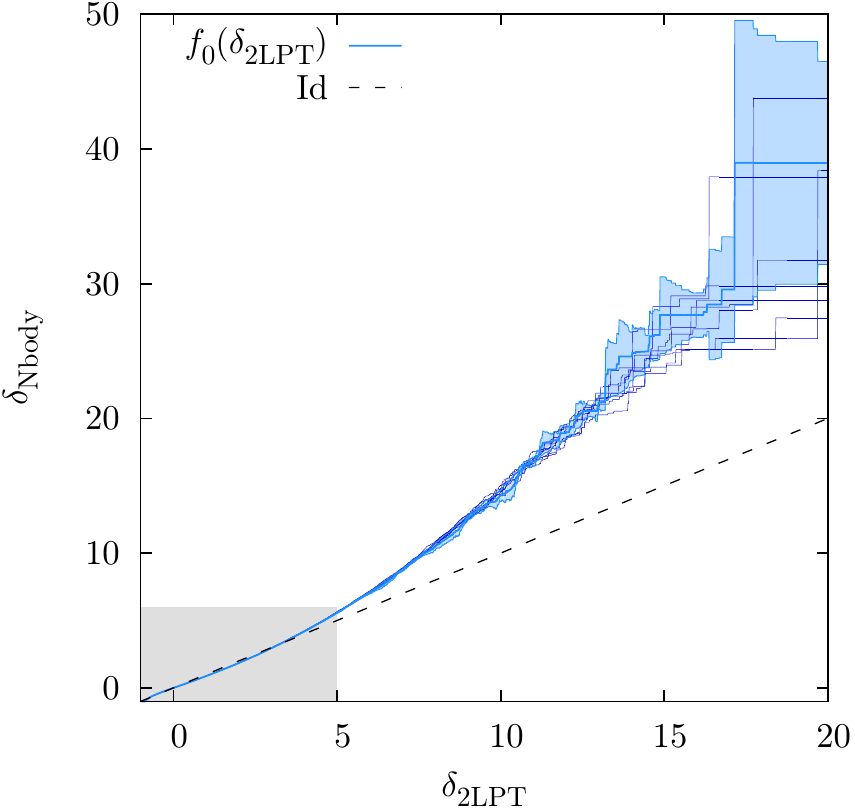} \\
\includegraphics[width=0.48\textwidth]{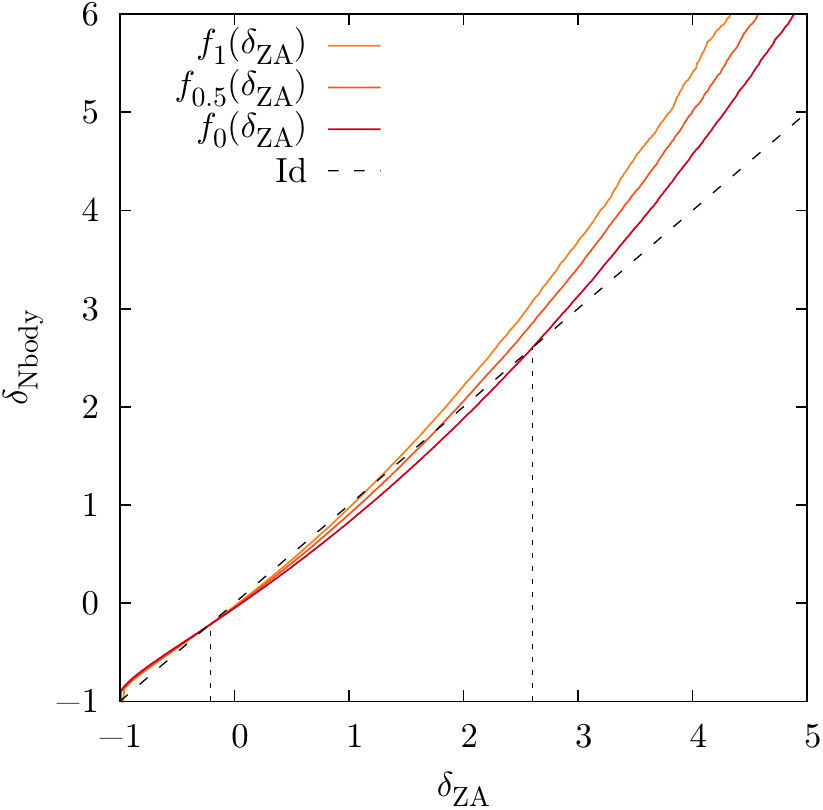}
\includegraphics[width=0.48\textwidth]{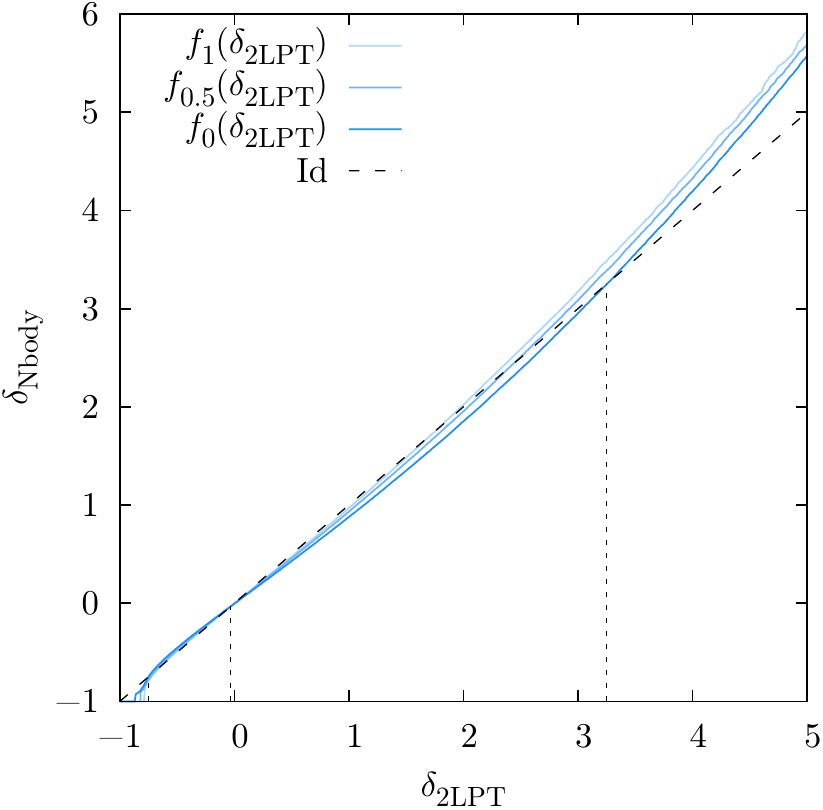} 
\end{center}
\caption{The \glslink{remapping function}{remapping functions} from \gls{LPT} to smoothed \glslink{N-body simulation}{$N$-body} \glslink{density field}{density fields}, for the \gls{ZA} (left panel) and \gls{2LPT} (right panel), all computed on a 8~Mpc/$h$ mesh. The precise \gls{redshift}-zero \glslink{remapping function}{remapping functions} $f_0$ (red and blue solid curves) have been computed using the full statistics from eight realizations (darker red and blue solid curves). The error bars shown are the 1-$\sigma$ dispersion among the eight runs with reference to the full \gls{remapping function}. The lower plots show the detail of the shaded area, in a density range containing most of the voxels. The \gls{redshift}-dependence of the \gls{remapping function} $f_z$ is shown for $z=1$, $z=0.5$ and $z=0$. The dashed line shows the identity function. Critical values of $\delta_{\mathrm{LPT}}$ for which \gls{remapping} does not change the local density are identified.}
\label{fig:rE_function}
\end{figure*}

Some \glslink{remapping function}{remapping functions} are presented in figure \ref{fig:rE_function}. In each panel, the solid curves represent the \gls{remapping function} $f_z$ at \gls{redshift} $z$, computed with the \gls{LPT} and \glslink{N-body simulation}{$N$-body simulations}. The dashed black line shows the identity function. We checked that the \gls{remapping function} converges to the identity function with increasing \gls{redshift}, as expected. Critical values where the \gls{remapping function} crosses the identity function are identified. Between these critical values, \gls{remapping} either increases or decreases the local density.

The \glslink{pdf}{pdfs} for the \gls{density contrast} are evaluated on a grid after a \gls{CiC} \glslink{mesh assignment}{assignment of particles}. This means that the \gls{remapping function} \textit{a priori} depends on the size of voxels. The problem of choosing a voxel size for the computation of the \gls{remapping function} is linked to the more general problem of choosing a mesh for the \gls{CiC} evaluation of density. Choosing too coarse a binning will result in an underestimation of the clustering of particles, whereas choosing too fine a binning will also result in artifacts in overdensities (some voxels may be empty due to their too small size). The right choice of voxel size for the evaluation of the \gls{remapping function} is the one giving the best evaluation of the \gls{density contrast}. This choice has to be made depending on the desired application of the remapped data.

The \gls{remapping function} describes how the \gls{pdf} for the \gls{density contrast} is affected by \glslink{non-linear evolution}{non-linear} \gls{structure formation}. For this reason, it depends on the nature of the gravitational interaction, as described by \gls{LPT} and by \glslink{full gravity}{full $N$-body dynamics}, but weakly depends on the detail of the \gls{cosmological parameters}. We checked the cosmology-dependence of the \gls{remapping function} in simulations with the \glslink{cosmological parameters}{dark matter and baryon density in the Universe}, $\Omega_\mathrm{m}$ and $\Omega_\mathrm{b}$, varying the \gls{WMAP-7} fiducial values (equation \eqref{eq:comological-parameters}) by $\pm$ $3 \sigma$ (still assuming a flat Universe):
\begin{equation}
\label{eq:cosmological-parameters-1}
\Omega_\Lambda = 0.750, \Omega_\mathrm{m} = 0.2494, \Omega_\mathrm{b} = 0.0428, \sigma_8 = 0.810, h = 0.704, n_{\mathrm{s}} = 0.967;
\end{equation}
\begin{equation}
\label{eq:cosmological-parameters-2}
\Omega_\Lambda = 0.700, \Omega_\mathrm{m} = 0.2992, \Omega_\mathrm{b} = 0.0488, \sigma_8 = 0.810, h = 0.704, n_{\mathrm{s}} = 0.967.
\end{equation}

\begin{figure*}
\begin{center}
\includegraphics[width=0.8\textwidth]{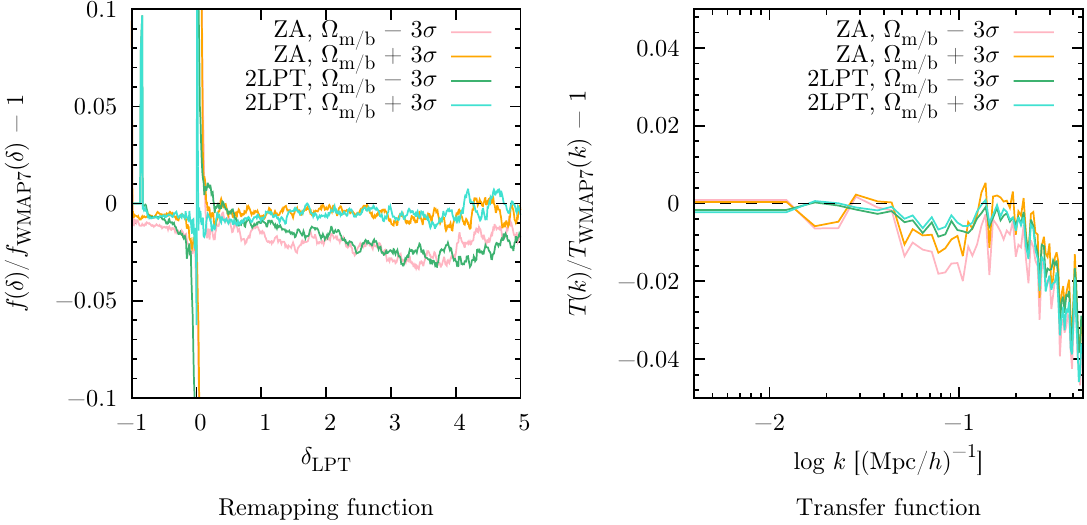}
\end{center}
\caption{Relative deviations of the \gls{remapping function} (left panel) and of the \gls{transfer function} (right panel) for varying \gls{cosmological parameters} (equations \eqref{eq:cosmological-parameters-1} and \eqref{eq:cosmological-parameters-2}), with respect to their behaviour in a fiducial cosmology (equation \eqref{eq:comological-parameters}).}
\label{fig:deviations_cosmology}
\end{figure*}

Even for these models notably far from the fiducial values, we found that the \gls{remapping function} almost perfectly overlaps that of our main analysis, for the density range $\delta \in [-1;5]$, containing typically 98 to 99\% of the voxels. We found a difference of less than 5\% for $\delta = 5$ (see the left panel of figure \ref{fig:deviations_cosmology}).

The \gls{transfer function} used in steps 1 and 3 of the improved procedure also exhibits very weak \gls{redshift}-dependence, with deviations limited to a few percents at the smallest scales of interest of this work ($k~\approx~0.4$ (Mpc/$h$)$^{-1}$, see the right panel of figure \ref{fig:deviations_cosmology}).

\section{Statistics of remapped fields}
\label{par:results}

In this section, we discuss the validity of the improved \gls{remapping} procedure described in section~\ref{par:improvement-remapping-procedure}, by studying the correlators of the remapped field in comparison to the input \gls{LPT} and \glslink{N-body simulation}{$N$-body} fields. The \gls{remapping} procedure based on the \glslink{EPT}{Eulerian} \gls{density contrast} essentially replaces the \gls{LPT} \glslink{one-point distribution}{one-point function} by that of the smoothed \glslink{N-body simulation}{$N$-body}-evolved field. Since the position and shape of structures is left unchanged, we expect the \glslink{high-order correlation function}{higher-order correlators} of the \gls{density field} to be respected by the \gls{remapping} procedure. Of particular interest is to check how \gls{remapping} affects \glslink{high-order correlation function}{higher-order statistics} and if possible improvements could be exploited in data analysis or \glslink{mock catalog}{artificial galaxy survey} applications.

We implemented a numerical algorithm that computes and analyzes a remapped \gls{density field}. The procedure can be divided in three steps:
\begin{enumerate}
\item We take as input two cosmological \glslink{density field}{density fields}, evolved from the same \gls{initial conditions} with \gls{LPT} (\gls{ZA} or \gls{2LPT}) and with \glslink{full gravity}{full $N$-body dynamics}, and estimate the \glslink{one-point distribution}{one-point statistics} (\gls{pdf} and \gls{cdf} for $\delta$) and the \gls{transfer function} for this particular realization. We repeat this step for the eight realizations used in our analysis.
\item We take as input all the \glslink{one-point distribution}{one-point statistics} computed with individual realizations, and we compute a precise \gls{remapping function} using the full statistics of all available realizations, as described in section~\ref{par:remapping-procedure}. The \gls{transfer function} used as a reference is the mean of all available realizations. At this point, the \gls{remapping function} and the \gls{transfer function} can be tabulated and stored for later use, and \glslink{N-body simulation}{$N$-body simulations} are no longer required.
\item For each realization, we \glslink{remapping}{remap} the \gls{density field} using the improved three-step procedure described in section~\ref{par:improvement-remapping-procedure} and we analyze its correlation functions.
\end{enumerate}

Our algorithm provides the \glslink{one-point distribution}{one-point} (section~\ref{par:eulerian-one-point-stats}) and \glslink{two-point correlation function}{two-point} (section~\ref{par:eulerian-two-point-stats}) statistics. We used the code described in \citet{Gil-Marin2011,Gil-Marin2012} to study the \glslink{three-point correlation function}{three-point statistics} (section~\ref{par:eulerian-three-point-stats}). The results are presented below.

\subsection{One-point statistics}
\label{par:eulerian-one-point-stats}

\begin{figure*}
\begin{center}
\includegraphics[width=0.85\textwidth]{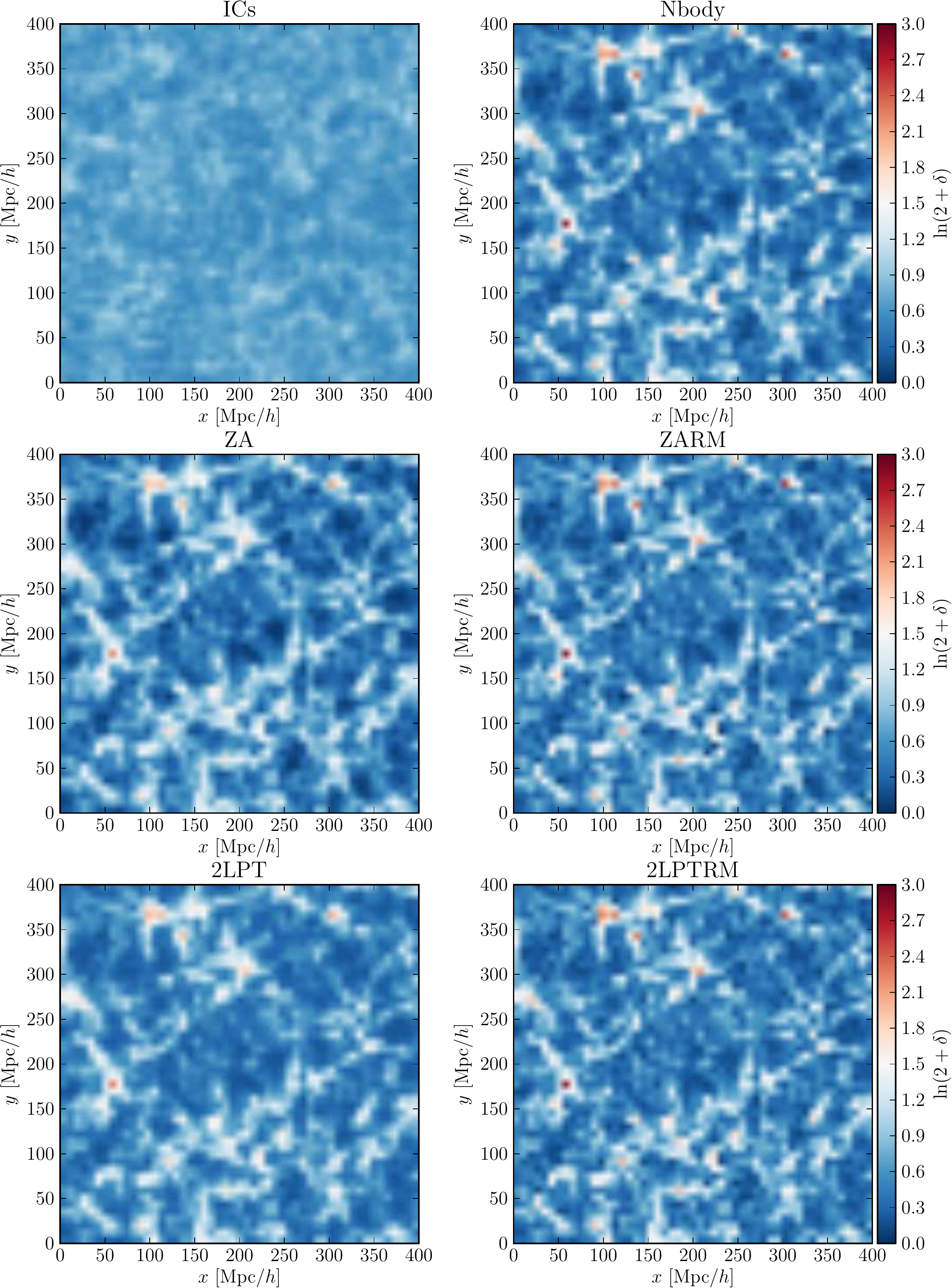}
\caption{\glslink{redshift}{Redshift}-zero \gls{density contrast} on a $128^2$-pixel slice of a $512^3$-\glslink{particle realization}{particles realization} in a 1024~Mpc/$h$ box with \gls{periodic boundary conditions}. For clarity, the slice is limited to a square of 400~Mpc/$h$ side, and the quantity shown is the log-density field, $\ln(2+\delta)$. For comparison with the \gls{initial conditions}, the \gls{density field} at high \gls{redshift} ($z=5$) is shown in the top left corner. The \gls{redshift}-zero \glslink{density field}{density fields} are determined using, from top to bottom and from left to right: a \glslink{full gravity}{full $N$-body simulation}, the \glslink{ZA}{Zel'dovich approximation}, alone (\gls{ZA}) and after \gls{remapping} (\gls{ZARM}), \glslink{2LPT}{second-order Lagrangian perturbation theory}, alone (\gls{2LPT}) and after \gls{remapping} (\gls{2LPTRM}). The \gls{remapping} and the \gls{transfer function} operations are performed on a $128^3$-voxel grid, corresponding to a mesh size of 8~Mpc/$h$.}
\label{fig:slices128}
\end{center}
\end{figure*}

The \gls{remapping} procedure, described in section~\ref{par:remapping}, is essentially a replacement of the cumulative distribution function of the \gls{density contrast} $\delta$ of the input \gls{LPT}-evolved field, $\mathpzc{C}_{\mathrm{LPT}}$, by that of the reference \glslink{N-body simulation}{$N$-body}-evolved field after smoothing, $\mathpzc{C}_{\mathrm{Nbody,S(LPT)}}$. After having applied the \gls{remapping} procedure, we recomputed the \gls{pdf} of the remapped field and verified that it matches that of the fiducial field as a sanity check.

Remapping and rescaling the density modes alters local density values but positions of structures remain unchanged. It is therefore important to check that \gls{remapping} visually alters the \gls{LPT}-evolved distribution in such a way that structures resemble more their \glslink{N-body simulation}{$N$-body} evolved counterparts. Figure \ref{fig:slices128} shows a slice of the \gls{density contrast} $\delta$, measured at \gls{redshift} zero, on a $128^2$-pixel sheet of a $512^3$-\glslink{particle realization}{particles realization} in a 1024~Mpc/$h$ box. The corresponding mesh size is 8~Mpc/$h$. Visually, remapped fields (\gls{ZARM} and \gls{2LPTRM}) are closer to the \glslink{full gravity}{full $N$-body} result than their originals (\gls{ZA} and \gls{2LPT}), with plausible particle distribution.

Since the improved \gls{remapping} procedure involves a rescaling of the density modes in Fourier space (step 3), the \gls{pdf} for the \gls{density contrast} of the remapped fields is not guaranteed to be correct by construction, as would be the case with a naive \gls{remapping} (section~\ref{par:remapping-procedure}). Therefore, the \gls{one-point distribution} has to be carefully checked at this point. In figure \ref{fig:pdf}, we plot the \gls{pdf} for the \gls{density contrast} at \gls{redshift} zero for \glslink{N-body simulation}{$N$-body simulations} and the approximately evolved fields: with the \gls{ZA} and \gls{2LPT} alone, and after \gls{remapping} (\gls{ZARM} and \gls{2LPTRM}). It can be observed that the peaks of the \glslink{pdf}{pdfs} get closer to the reference set by \glslink{N-body simulation}{$N$-body dynamics} and that the \gls{pdf} of remapped fields accurately follows that of \glslink{full gravity}{full gravitational dynamics} for $\delta > 0$. The procedure being successful on average for \glslink{one-point distribution}{one-point statistics} and accurate for the most common events in overdensities, we expect the \glslink{number function}{number count} of objects such as \glslink{cluster}{clusters} predicted by \gls{LPT} to be made more robust by our procedure.

\subsection{Two-point statistics}
\label{par:eulerian-two-point-stats}

\subsubsection{Power spectrum}
\label{par:eulerian-two-point-stats-power-spectrum}

In figure \ref{fig:PS}, we plot the \gls{redshift}-zero \gls{power spectrum} of the different \glslink{density field}{density fields}.\footnote{The reader is referred to section \ref{sec:eulerian-two-point-stats-power-spectrum} for practical details on the computation of these \glslink{power spectrum}{power spectra}.} The relative deviations of \glslink{power spectrum}{power spectra} with reference to the \gls{density field} computed with a \glslink{full gravity}{full $N$-body simulation} are presented in figures \ref{fig:PS-deviations-mesh} and \ref{fig:PS-deviations-redshift}.

At high \gls{redshift} ($z > 1$), we found no notable difference between the \gls{power spectrum} of matter evolved with \glslink{full gravity}{full $N$-body dynamics} and that of \gls{LPT}-remapped distributions. This indicates that our \gls{remapping} procedure is increasingly successful as we go backwards in time towards the \gls{initial conditions}, where \gls{LPT} gives accurate results.

At low \gls{redshift}, the \gls{power spectrum} shape of remapped \gls{LPT} fields is closer to the shape of the \glslink{full gravity}{full non-linear} \gls{power spectrum}, turning down at smaller scales than the \gls{LPT} \glslink{power spectrum}{power spectra}. In particular, \gls{LPT} fields exhibit more small-scale correlations after \gls{remapping}, illustrating the success of our procedure in the \gls{mildly non-linear regime} of large-scale \gls{structure formation}. 

Contrary to the \glslink{density field}{density fields} obtained via a naive \gls{remapping} approach, whose \glslink{power spectrum}{power spectra} exhibit a positive artificial offset at large scales as discussed in section~\ref{par:comparison-structure-types}, the fields obtained with the improved procedure have correct \glslink{two-point correlation function}{two-point statistics} at all scales for coarse grids (down to 8~Mpc/$h$). For finer grids, a negative large-scale bias appears in the \gls{power spectrum}, meaning that we have suppressed too much small-scale power in the \glslink{N-body simulation}{$N$-body} field in the first step of our procedure, which propagates to large scales with \gls{remapping}. Comparing the panels of figures \ref{fig:PS-deviations-mesh} and \ref{fig:PS-deviations-redshift}, it can be observed that this effect is suppressed at higher \gls{redshift} and for coarser binning. We found that a voxel size of 8~Mpc/$h$ is the best compromise, with a large-scale \gls{power spectrum} compatible with the error bars and clear improvement at small scales, as can be seen in figure \ref{fig:PS}. This mesh size corresponds to the target resolution for analyses in the \gls{mildly non-linear regime}, as discussed in the introduction of this chapter.

\subsubsection{Fourier-space cross-correlation coefficient}
\label{par:eulerian-two-point-stats-cross-correlation}

In figure \ref{fig:crosscorr}, we present the Fourier-space \gls{cross-correlation} coefficient $R \equiv P_{\delta \times \delta'}/\sqrt{P_\delta P_{\delta'}}$ between the \gls{redshift}-zero \gls{density field} in the \glslink{N-body simulation}{$N$-body simulation} and several other \glslink{density field}{density fields}. At \gls{redshift} zero and at small scales, the agreement is better with remapped \gls{LPT} fields than with \gls{LPT} alone, confirming the success of the \gls{remapping} procedure to explore the \gls{mildly non-linear regime}. In particular, the \gls{remapping} of \gls{2LPT} predicts more than 96\% level accuracy at $k = 0.4$ (Mpc/$h$)$^{-1}$ (corresponding to scales of 16~Mpc/$h$), where \gls{2LPT} gives only 93\%. The \gls{cross-correlation} coefficient indicates better agreements for the \gls{remapping} of \gls{2LPT} than for the \gls{remapping} of the \gls{ZA}, which is consistent with the better performance of \gls{2LPT} in predicting the \glslink{phase}{phases} of the \glslink{full gravity}{full $N$-body field} (see section \ref{sec:eulerian-two-point-stats-cross-correlation}).

\subsection{Three-point statistics}
\label{par:eulerian-three-point-stats}

We analyzed the accuracy of our method beyond second-order statistics by studying the \gls{bispectrum}, using the code described in \citet{Gil-Marin2011,Gil-Marin2012}.\footnote{Technical considerations concerning the computation of \glslink{bispectrum}{bispectra} are presented in section \ref{sec:eulerian-three-point-stats}.} Figure \ref{fig:BS} shows the \gls{redshift}-zero \gls{bispectrum} for equilateral triangles. The overall result is a clear improvement of the \gls{bispectrum} of \gls{LPT}-evolved fields with the \gls{remapping} procedure, especially on the small scales shown, probing the \gls{mildly non-linear regime}, $k \gtrsim 0.1$ (Mpc/$h$)$^{-1}$ corresponding to scales $\lesssim 62$ Mpc/$h$, where \gls{LPT} predicts less \glslink{three-point correlation function}{three-point correlation} than \gls{full gravity}. At large scales, the approximation error remains $\lesssim 1 \sigma$ of the estimated \gls{statistical uncertainty}, even for a resolution of 8~Mpc/$h$ and at late times ($z=0$).

The relative deviations of approximate \glslink{bispectrum}{bispectra} with reference to \glslink{full gravity}{full $N$-body simulations} are shown in figures \ref{fig:BS-deviations-mesh}, \ref{fig:BS-deviations-redshift}, \ref{fig:BS-deviations-triangle} and \ref{fig:BS-deviations-theta}. As expected, the success of our \gls{remapping} procedure in exploring small scales ($k \gtrsim 0.1$ (Mpc/$h$)$^{-1}$) is increased for more coarsely-binned \glslink{density field}{density fields} (see figure \ref{fig:BS-deviations-mesh}) and at higher \gls{redshift} (see figure \ref{fig:BS-deviations-redshift}). In figure \ref{fig:BS-deviations-triangle} we examine the scale-dependence of the \gls{bispectrum} for various triangle shapes. The precise dependence on the triangle shape at different scales is shown in figure \ref{fig:BS-deviations-theta}.

\section{Discussion and conclusion}
\label{par:conclusion}

The main subject of this chapter is the development of a method designed to improve the correspondence between approximate models for gravitational dynamics and \glslink{full gravity}{full numerical simulation} of large-scale \gls{structure formation}. Our methodology relies on a \gls{remapping} of the \gls{one-point distribution} of the \gls{density contrast} of the approximately evolved particle distribution using information extracted from \glslink{N-body simulation}{$N$-body simulations}.

Due to the differences in the precise structure of the \glslink{density field}{density fields} in \glslink{LPT}{Lagrangian perturbation theory} and in \gls{full gravity}, the naive implementation of this procedure, inspired by \citet{Weinberg1992}, gives a large-scale bias in the \gls{power spectrum}. This is not solved by a simple rescaling of Fourier modes, which leads to Gibbs ringing artifacts and an overall unphysical representation of the \glslink{density field}{density fields}. Smoothing \gls{LPT} and \glslink{N-body simulation}{$N$-body} \glslink{density field}{density fields} with the same kernel is also unsuccessful, as smoothed $N$-body fields will always keep a sharper structure than smoothed \gls{LPT} fields.

We figured out that the cause of this large-scale bias is not the different density predicted locally by \gls{LPT} and \glslink{N-body simulation}{$N$-body dynamics} on a point-by-point basis, but a problem of mismatch between the volume of extended objects. Our findings question the reliability of \gls{LPT} for \gls{LSS} data analysis and generation of \glslink{mock catalog}{mock catalogs} at low \gls{redshift} and high \gls{mass resolution}. They are also a likely explanation for the discrepancy between the \gls{power spectrum} of \gls{initial conditions} reconstructed via \gls{Gaussianization} and linear theory expectations, encountered by \citet{Weinberg1992}.

Considering these results, we improved \citeauthor{Weinberg1992}'s \gls{remapping} procedure for \glslink{final conditions}{present-day} \glslink{density field}{density fields} by the use of a \gls{transfer function}. In this fashion, we obtain a physically more accurate representation of the three-dimensional matter distribution in the \gls{mildly non-linear regime}, while \glslink{high-order correlation function}{improving higher-order statistics}. Since \gls{LPT} captures well the cosmological dependence and \gls{remapping} operates on small-scale aspects of the \gls{density field}, we found that our procedure that is nearly independent of \gls{cosmological parameters}.

The aim of this method is to develop a fast, flexible and efficient way to generate realizations of \gls{LSS} \glslink{density field}{density fields}, accurately representing the \gls{mildly non-linear regime}. Our procedure, therefore, responds to the increasing demand for numerically inexpensive models of three-dimensional \gls{LSS}, for applications to modern cosmological data analysis. At the level of statistical error in our numerical experiments, the approach provides a good method for producing \glslink{mock catalog}{mock halo catalogs} and low-\gls{redshift} \gls{initial conditions} for simulations, if desired. The resulting information can also be used in a variety of cosmological analyses of present and upcoming observations.

We showed that our approach allows fast generation of cosmological \glslink{density field}{density fields} that correlate with \glslink{N-body simulation}{$N$-body simulations} at better than 96\% down to scales of $k~\approx~0.4~(\mathrm{Mpc}/h)^{-1}$ at \gls{redshift} zero and are substantially better than standard \gls{LPT} results at higher \glslink{redshift}{redshifts} on the same \glslink{comoving coordinates}{comoving} scales. Remapping improves the fast \gls{LPT} \gls{bispectrum} predictions on small scales while the large scale \gls{bispectrum} remains accurate to within about 1$\sigma$ of the measurement in our \glslink{N-body simulation}{$N$-body simulations}. Since real observations will have larger statistical errors for the foreseeable future, our method provides an adequate fast model of the \glslink{non-linear evolution}{non-linear} \gls{density field} on scales down to $\sim~8$~Mpc/$h$. These results constitute a substantial improvement with respect to existing techniques, since \glslink{non-linear evolution}{non-linearities} begin to affect even large-scale measurements in \glslink{galaxy survey}{galaxy surveys}. Since the number of modes usable for cosmological exploitation scale as $k^3$, even minor improvements in the smallest scale $k$ allow access to much more knowledge from existing and upcoming observations. This work is a step further in the \gls{non-linear regime}, which contains a wealth of yet unexploited \glslink{information content}{cosmological information}. For possible applications, we provided a cosmographic and statistical characterization of approximation errors.

Our \gls{remapping} procedure predicts the \gls{two-point correlation function} at around $95$\% level accuracy and \gls{three-point correlation function} at around $80$\% level accuracy at \gls{redshift} $3$, for $k$ between $0.1$ and $0.4$ $(\mathrm{Mpc}/$h$)^{-1}$, illustrating the increasing success of our methods as we go backwards in time towards the \gls{initial conditions}, when \gls{LPT} is an accurate description of early \gls{structure formation}. This is of particular interest in several areas of high-\gls{redshift} cosmology, such as forecasting \glslink{21 cm surveys}{21~cm surveys} \citep{Lidz2007}, analyzing the properties of the intergalactic medium via the \glslink{Lyman-alpha forest}{Lyman-$\alpha$ forest} \citep{Kitaura2012b} or probing the \gls{reionization} epoch \citep{Mesinger2007}. This work might also add to methods of data analysis for the ongoing and upcoming high-\gls{redshift} \glslink{galaxy survey}{galaxy surveys} mentioned in the \hyperref[chap:intro]{introduction}.

However, the realization of \glslink{density field}{density fields} with these procedures stays approximate, since the \glslink{full gravity}{full non-linear gravitational physics} involves information contained in the shape of structures, which cannot be captured from a \glslink{one-point distribution}{one-point} modification of \gls{LPT}, especially after \gls{shell-crossing}. We studied the performance of \glslink{one-point distribution}{one-point} \gls{remapping} of \gls{LPT} and presented a statistical characterization of the errors, but additional refinements, such as a non-linear, density-dependent smoothing of the \glslink{N-body simulation}{$N$-body} field, could further improve on these approximations, for an increased computational cost. This is, however, beyond the scope and intent of this work. Generally, the complications at large scales that we encounter when applying a \gls{local} \gls{remapping} seem difficult to solve in a \glslink{EPT}{Eulerian} \gls{density field} approach and would favor a \glslink{LPT}{Lagrangian}, \glslink{dark matter particles}{particle}-based perspective.

As mentioned in section \ref{sec:Remapping_intro}, fast and accurate methods to model the \glslink{non-linear evolution}{non-linearly evolved} mass distribution in the Universe have the potential of profound influence on modern cosmological data analysis. Full Bayesian \gls{large-scale structure inference} methods such as the \textsc{borg} algorithm, which extract \glslink{information content}{information} on the matter distribution in the Universe from \glslink{galaxy survey}{galaxy redshift surveys}, rely on \gls{LPT} (see chapter \ref{chap:BORG}). The technique proposed in this chapter can be envisioned as a numerically efficient and flexible extension of these methods, permitting us to push dynamic analyses of the large scale structure further into the \gls{non-linear regime}.

%% file: Chapter7/Chapter7Content.tex
\chapter{Non-linear filtering of large-scale structure samples}
\label{chap:filtering}
\minitoc

\defcitealias{Hemans1826}{Felicia}
\begin{flushright}
\begin{minipage}[c]{0.6\textwidth}
\rule{\columnwidth}{0.4pt}

``While o'er him fast, through sail and shroud,\\
\hspace*{1cm} The wreathing fires made way.\\
They wrapt the ship in splendour wild,\\
\hspace*{1cm} They caught the flag on high,\\
And streamed above the gallant child,\\
\hspace*{1cm} Like banners in the sky.''\\
--- \citetalias{Hemans1826} \citet{Hemans1826}, \textit{Casabianca}

\vspace{-5pt}\rule{\columnwidth}{0.4pt}
\end{minipage}
\end{flushright}

\abstract{\section*{Abstract}
Due to the approximate \gls{2LPT} model implemented in the {\borg} algorithm, inferred large-scale structure \glslink{sample}{samples} are only correct in the \glslink{linear regime}{linear} and \gls{mildly non-linear regime} of \gls{structure formation}. This chapter describes subsequent improvement of such \glslink{sample}{samples} at non-linear scales, via an operation that we refer to as ``\gls{non-linear filtering}''. This process does not replace \glslink{full gravity}{fully non-linear} \gls{large-scale structure inference}, but rather fills small scales with physically reasonable information. Several approaches to \gls{non-linear filtering} are considered and discussed. 
}

This chapter discusses the generation of non-linear, constrained \glslink{particle realization}{realizations} of the late-time \glslink{LSS}{large-scale structure} via an operation that we call ``filtering'' of {\borg} \glslink{sample}{samples}. It is structured as follows. We give motivation for \gls{non-linear filtering} and describe two different approaches (direct improvement of \gls{final conditions}, and \glslink{constrained simulation}{constrained simulations}) in section \ref{sec:Introduction filtering}. For later use in chapter \ref{chap:dmvoids}, we describe a set of \glslink{sample}{samples} optimally filtered with \glslink{Gadget-2}{\textsc{Gadget}} in section \ref{sec:Filtering Gadget}. In section \ref{sec:Filtering COLA}, we describe the efficient {\cola} scheme for fast production of non-linear \glslink{LSS}{large-scale structure} \glslink{particle realization}{realizations}, and apply it to generate a large ensemble of \glslink{sample}{samples}, used in chapter \ref{chap:ts}.

\section{Introduction}
\label{sec:Introduction filtering}

\subsection{Motivation for non-linear filtering of large-scale structure samples}

As noted in section \ref{sec:Translating to the final density field}, the \glslink{large-scale structure likelihood}{likelihood} for Bayesian \gls{large-scale structure inference} involves a \gls{structure formation} model to translate from the \glslink{initial conditions}{initial} to the \glslink{final conditions}{final} \gls{density field}:
\begin{equation}
\Gray{\delta}^\mathrm{\BattleShipGrey{i}} \mapsto \BattleShipGrey{\delta}^\mathrm{\BattleShipGrey{f}} = \mathcal{G}(\BattleShipGrey{\delta}^\mathrm{\BattleShipGrey{i}},a) .
\end{equation}
Ideally, this step should involve a numerical model that \glslink{full gravity}{fully accounts} for the \glslink{non-linear evolution}{non-linearities} of the \gls{Vlasov-Poisson system}, which describes \gls{structure formation} (see chapter \ref{chap:theory}). Unfortunately, this is not currently computationally tractable. For this reason, {\borg} uses \gls{2LPT} as a proxy for \glslink{gravitational evolution}{gravitational dynamics}.\footnote{For the record, a {\borg} run, using \gls{2LPT}, takes of the order of a year (wall-clock time).} 

Nevertheless, the description of particular patterns of the \gls{cosmic web} (as presented in part \ref{part:IV} of this thesis) requires description of the \gls{LSS} not only correct at the scales correctly described by \gls{2LPT} ($k \lesssim 0.1$ $\mathrm{Mpc}/h$, see chapter \ref{chap:lpt}), but also physically reasonable at smaller scales, up to $k \sim 1$ $\mathrm{Mpc}/h$. At this point, it is also useful to recall that the number of Fourier modes usable for cosmology scales as the cube of the smallest accessible mode, $k^3$.

For these reasons, data-constrained, non-linear \glslink{particle realization}{realizations} of the \gls{LSS} have a large variety of applications. As noted before, constraining small, non-linear scales within the inference framework is not yet possible; therefore, such \glslink{particle realization}{realizations} will rely on fusing data-constrained large scales and unconstrained small scales that only reflect our theoretical understanding of \gls{structure formation}. Throughout this thesis, we refer to the production of data-constrained, non-linear \glslink{particle realization}{realizations}, on the basis of {\borg} \glslink{LSS}{large-scale structure} \glslink{sample}{samples}, as \textit{\gls{non-linear filtering}}.

\subsection{Filtering in the final conditions}

One possible way to perform \gls{non-linear filtering} is to directly improve the \gls{final conditions} produced as {\borg} outputs. The technique of \gls{remapping} \glslink{LPT}{Lagrangian perturbation theory} can be useful in this context: as demonstrated in chapter \ref{chap:remapping}, it cheaply yields improvements of \glslink{density field}{density fields} in the \gls{mildly non-linear regime}. A particular advantage of \gls{remapping} is its very low computational cost, which allows to process a large number of \glslink{sample}{samples}.\footnote{The computational cost for \gls{remapping} all the outputs of a {\borg} run, about $10,000$ \glslink{sample}{samples}, would be comparable to a few \glslink{full gravity}{full-gravity} dark matter simulations using \textsc{\gls{Gadget-2}}.} As seen in chapters \ref{chap:BORG} and \ref{chap:BORGSDSS}, this is crucial for adequate \gls{uncertainty quantification}.

\subsection{Filtering via constrained simulations}
\label{sec:Filtering via constrained simulations}

Another idea is to capitalize on the inference of the \gls{initial conditions} by {\borg}. Starting from inferred \glslink{density field}{density fields}, which contain the \gls{data} constraints (see in particular section \ref{sec:cosmic_history} for a discussion of \glslink{Lagrangian transport}{information transport}), it is possible to go \glslink{forward modeling}{forward in time} using an alternative \gls{structure formation} model, noted $\mathcal{G}_\textsc{nl}$, that improves upon $\mathcal{G}$ for the description of small scales structures:
\begin{equation}
\BattleShipGrey{\delta}^\mathrm{\BattleShipGrey{i}} \mapsto \BattleShipGrey{\delta}^\mathrm{\BattleShipGrey{f}}_\textsc{nl} = \mathcal{G}_\textsc{nl}(\BattleShipGrey{\delta}^\mathrm{\BattleShipGrey{i}},a) .
\end{equation}
This process is known in the literature as running \textit{\glslink{constrained simulation}{constrained simulations}}. \glslink{final conditions}{Final} \glslink{density field}{density fields} $\BattleShipGrey{\delta}^\mathrm{\BattleShipGrey{f}}_\textsc{nl}$ constructed in this way agree with corresponding {\borg} \gls{final conditions} $\BattleShipGrey{\delta}^\mathrm{\BattleShipGrey{f}}$ at large scales, but are also physically reasonable at smaller scales, up to the validity limit of $\mathcal{G}_\textsc{nl}$.

In this picture, interesting questions are the determination of the smallest scale influenced by the \gls{data} and the characterization of the reliability of structures extrapolated in unobserved regions, at high \gls{redshift} or near \glslink{survey geometry}{survey boundaries}. An upcoming publication will investigate the validity of \glslink{constrained simulation}{constrained simulations}, in particular the strength of \gls{data} constraints in domains or at scales that have not been considered in the inference scheme.

In the following, we examine two particular cases for $\mathcal{G}_\textsc{nl}$, corresponding to the \textsc{\gls{Gadget-2}} cosmological code (section \ref{sec:Filtering Gadget}) and to the fast \textsc{cola} scheme (section \ref{sec:Filtering COLA}).

\section{Fully non-linear filtering with Gadget}
\label{sec:Filtering Gadget}

\draw{This section draws from section II.B. in \citet{Leclercq2015DMVOIDS}.}

Optimal \gls{non-linear filtering} of {\borg} results is achieved when $\mathcal{G}_\textsc{nl}$ \glslink{full gravity}{fully accounts for non-linear} \glslink{gravitational evolution}{gravitational dynamics}. This is the case when a cosmological simulation code is used. For the purpose of this thesis, we consider that \gls{non-linear filtering} of {\borg} results with the \textsc{\gls{Gadget-2}} cosmological code \citep{Springel2001,Springel2005} is optimal.

For a variety of later uses, in particular for inference of \glslink{dark matter void}{dark matter voids} in the Sloan volume (chapter \ref{chap:dmvoids}), we generate a set of such optimally filtered, data-constrained \glslink{particle realization}{realizations} of the present \glslink{LSS}{large-scale structure}. To do so, we rely on a subset of statistically independent \gls{initial conditions} \glslink{particle realization}{realizations}, provided by \citet{Jasche2015BORGSDSS} (see chapter \ref{chap:BORGSDSS}). The \glslink{initial conditions}{initial} \gls{density field}, defined on a cubic equidistant grid with side length of 750~Mpc/$h$ and $256^3$ voxels, is populated by $512^3$ \gls{dark matter particles} placed on a regular Lagrangian grid. The \glslink{dark matter particles}{particles} are evolved with \gls{2LPT} to the \gls{redshift} of $z=69$, followed by a propagation with \textsc{\gls{Gadget-2}} from $z=69$ to $z=0$. In this fashion, we generate \glslink{full gravity}{fully non-linear}, data-constrained \glslink{reconstruction}{reconstructions} of the present-day large-scale dark matter distribution.

As discussed in section \ref{sec:Introduction filtering}, \gls{final conditions} inferred by {\borg} are accurate only at \glslink{linear regime}{linear} and \glslink{mildly non-linear regime}{mildly non-linear} scales. Application of \glslink{full gravity}{fully non-linear} dynamics to the corresponding \gls{initial conditions} acts as an additional filtering step, extrapolating predictions to unconstrained \glslink{non-linear regime}{non-linear regimes}. In a \glslink{Bayesian statistics}{Bayesian approach}, this new information can then be tested with complementary observations in the actual sky for updating our knowledge on the Universe.

\begin{figure} 
\begin{center}
\includegraphics[width=\columnwidth]{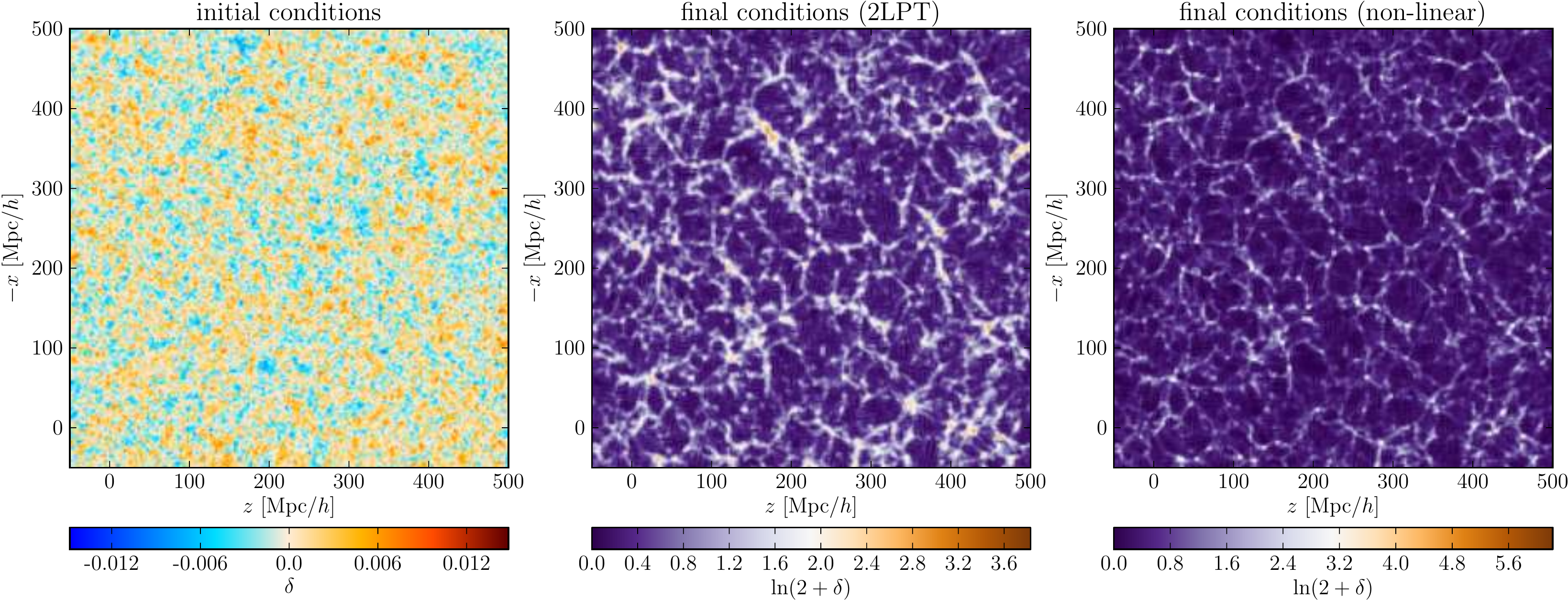}
\caption{Non-linear filtering of {\borg} results. Slices through one \gls{sample} of \glslink{initial conditions}{initial} (left panel) and \glslink{final conditions}{final} \glslink{density field}{density fields} (middle panel) inferred by {\borg}. The \glslink{final conditions}{final} \gls{density field} (middle panel) is a prediction of the \gls{2LPT} model used by {\borg}. On the right panel, a slice through the data-constrained \glslink{particle realization}{realization} obtained with the same \gls{sample} via \gls{non-linear filtering} (\glslink{full gravity}{fully non-linear} gravitational \gls{structure formation} starting from the same \gls{initial conditions}) is shown.\label{fig:filtering}}
\end{center}
\end{figure}

\begin{figure}
\begin{center}
\includegraphics[width=0.6\columnwidth]{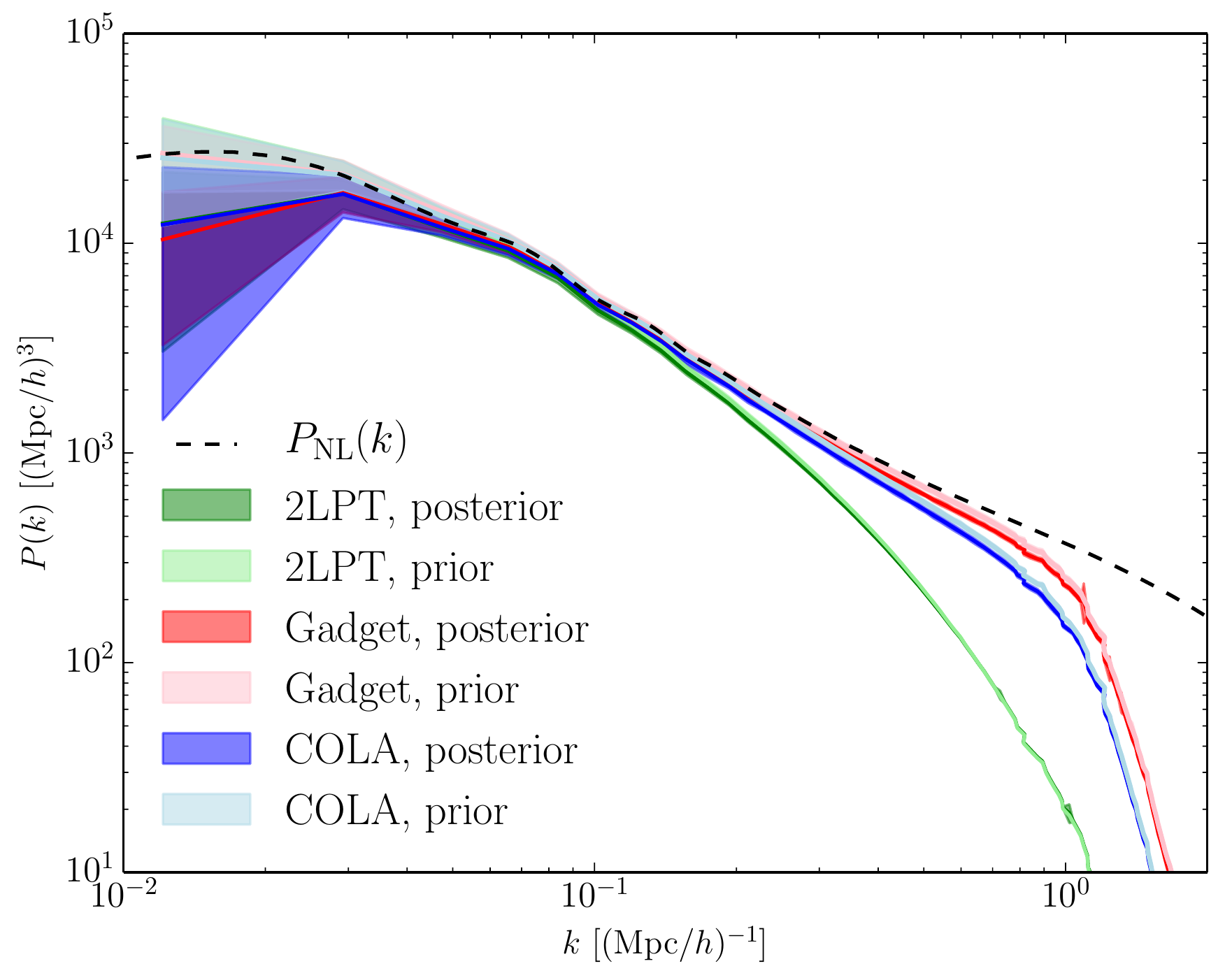}
\caption{\glslink{power spectrum}{Power spectra} of dark matter \glslink{density field}{density fields} at \gls{redshift} zero, computed with a mesh size of 3 Mpc/$h$. The particle distributions are determined using: $1,000$ unconstrained \gls{2LPT} \glslink{particle realization}{realizations} (``2LPT, prior''), $4,473$ constrained \gls{2LPT} \glslink{sample}{samples} inferred by {\borg} (``2LPT, posterior''), $11$ unconstrained \textsc{\gls{Gadget-2}} \glslink{particle realization}{realizations} (``Gadget, prior''), $11$ constrained \glslink{sample}{samples} inferred by {\borg} and filtered with \textsc{\gls{Gadget-2}} (``Gadget, posterior''), $1,000$ unconstrained {\cola} \glslink{particle realization}{realizations} (``COLA, prior''), $1,097$ constrained \glslink{sample}{samples} inferred by {\borg} and filtered with {\cola} (``COLA, posterior''). The solid lines correspond to the mean among all \glslink{particle realization}{realizations} used in this work, and the shaded regions correspond to the 2-$\sigma$ credible interval estimated from the standard error of the mean. The dashed black curve represents $P_\mathrm{NL}(k)$, the theoretical \gls{power spectrum} expected at $z=0$ from high-resolution \glslink{N-body simulation}{$N$-body simulations}.}
\label{fig:power_spectrum}
\end{center}
\end{figure}

An illustration of the \gls{non-linear filtering} procedure is presented in figure \ref{fig:filtering}.\footnote{In figure \ref{fig:filtering} and in all slice plots of the rest of this thesis, we keep the coordinate system of \citet{Jasche2015BORGSDSS}, also used in chapter \ref{chap:BORGSDSS}.} By comparing \glslink{initial conditions}{initial} and \glslink{final conditions}{final} \glslink{density field}{density fields}, one can see correspondences between structures in the present Universe and their origins. Comparing \glslink{final conditions}{final} \glslink{density field}{density fields} before and after filtering (middle and left panels), one can check the conformity of the \glslink{linear regime}{linear} and \glslink{mildly non-linear regime}{mildly non-linear} structures at large and intermediate scales, correctly predicted by \gls{2LPT}. Small-scale structures, corresponding to the deeply \gls{non-linear regime}, are much better represented after \gls{non-linear filtering} (resulting particularly in sharper \glslink{filament}{filaments} and \glslink{cluster}{clusters}). $N$-body dynamics also resolves much more finely the substructure of \glslink{void}{voids} -- known to suffer from spurious artifacts in \gls{2LPT}, namely the presence of peaky, overdense spots where there should be deep \glslink{void}{voids} \citetext{\citealp{Sahni1996,Neyrinck2013,Leclercq2013}; see also chapter \ref{chap:lpt}} -- which is of particular relevance for the purpose of inferring \glslink{dark matter void}{dark matter voids} (see chapter \ref{chap:dmvoids}).

The improvement introduced by \gls{non-linear filtering} at the level of \glslink{two-point correlation function}{two-point statistics} is presented in figure \ref{fig:power_spectrum}, where we plot the \glslink{power spectrum}{power spectra} of dark matter \glslink{density field}{density fields} at $z=0$. The agreement between unconstrained and constrained \glslink{particle realization}{realizations} at all scales can be checked. The plot also shows that our set of constrained \glslink{reconstruction}{reconstructions} contains the additional power expected in the \gls{non-linear regime}\footnote{Note that the lack of small scale power in \textsc{\glslink{Gadget-2}{Gadget}} and {\cola} with respect to theoretical predictions, for $k \gtrsim 0.5~(\mathrm{Mpc}/h)^{-1}$, is a gridding artifact due to the finite mesh size used for the analysis. This value corresponds to around one quarter of the \gls{Nyquist wavenumber}.}, up to $k \approx 0.4~(\mathrm{Mpc}/h)^{-1}$.

\section{Fast non-linear filtering with COLA}
\label{sec:Filtering COLA}

For means of \gls{uncertainty quantification} within \gls{large-scale structure inference}, it is necessary to process a large number of \glslink{sample}{samples}. Unfortunately, optimal \gls{non-linear filtering} with \textsc{\gls{Gadget-2}} is too expensive for the $\sim$ $10,000$ \glslink{sample}{samples} of a single {\borg} run. However, an approximate model for non-linear \gls{structure formation}, correct up to scales of a few Mpc/$h$, is enough for our purposes, as long as the approximation error is controlled and quantified.

\subsection{The COLA method}
\label{sec:The COLA method}

The {\cola} \citep[COmoving Lagrangian Acceleration,][]{Tassev2013,Tassev2015} technique offers a cheap way to perform \gls{non-linear filtering} of a large number of {\borg} \glslink{sample}{samples}. A particular advantage (in opposition to standard \glslink{PM}{particle-mesh codes}) is its flexibility in trading accuracy at small scales for computational speed, without sacrificing accuracy at the largest scales.

The general idea of {\cola} is to use our analytic understanding of \gls{structure formation} at large scales via \gls{LPT}, and to solve numerically only for a subdominant contribution describing small scales. Specifically, \citet{Tassev2012b} propose to expand the \glslink{displacement field}{Lagrangian displacement} of \glslink{dark matter particles}{particles} as
\begin{equation}
\boldsymbol{\Psi}(\textbf{x},\tau) = \boldsymbol{\Psi}_\mathrm{LPT}(\textbf{x},\tau) + \boldsymbol{\Psi}_\mathrm{MC}(\textbf{x},\tau)
\end{equation}
where $\boldsymbol{\Psi}_\mathrm{LPT}(\textbf{x},\tau)$ is the analytic displacement prescribed by \gls{LPT}\footnote{Following \citet{Tassev2012b}, this first term can be written more generally in Fourier space as $\boldsymbol{\Psi}_\bigstar(\textbf{k},\tau)~=~R_\mathrm{LPT}(k,\tau)~\boldsymbol{\Psi}_\mathrm{LPT}(\textbf{k},\tau)$, where $R_\mathrm{LPT}(k,\tau)$ is a \gls{transfer function} that we ignore here for simplicity.} (the \gls{ZA} or \gls{2LPT}, see chapter \ref{chap:lpt}) and $\boldsymbol{\Psi}_\mathrm{MC}(\textbf{x},\tau) \equiv \boldsymbol{\Psi}(\textbf{x},\tau) - \boldsymbol{\Psi}_\mathrm{LPT}(\textbf{x},\tau)$ is the ``\glslink{mode coupling}{mode-coupling residual}''. Using this Ansatz, the Eulerian position is $\textbf{x} = \textbf{q} + \boldsymbol{\Psi}_\mathrm{LPT} + \boldsymbol{\Psi}_\mathrm{MC}$, and the \gls{equation of motion}, which reads schematically (omitting constants and \glslink{Hubble flow}{Hubble expansion}; see equation \eqref{eq:equation-of-motion-p})
\begin{equation}
\label{eq:EoM-standard}
\deriv{^2 \textbf{x}}{\tau^2} = - \nabla_\textbf{x} \Phi ,
\end{equation}
can be rewritten in a frame comoving with ``\gls{LPT} observers'', whose trajectories are given by $\boldsymbol{\Psi}_\mathrm{LPT}$, as
\begin{equation}
\label{eq:EoM-COLA}
\deriv{^2 \boldsymbol{\Psi}_\mathrm{MC}}{\tau^2} = - \nabla_\textbf{x} \Phi - \deriv{^2 \boldsymbol{\Psi}_\mathrm{LPT}}{\tau^2} .
\end{equation}
In analogy with \gls{classical mechanics}, $\mathrm{d}^2 \boldsymbol{\Psi}_\mathrm{LPT}/\mathrm{d}\tau^2$ can be thought of as a fictitious force acting on \glslink{dark matter particles}{particles}, coming from the fact that we are working in a non-inertial frame of reference.

The standard approach in \gls{PM} codes (see appendix \ref{apx:simulations}) is to discretize the second-derivative time operator in equation \eqref{eq:EoM-standard}. At large scales, this is nothing more than solving for the \gls{linear growth factor}. Thereforce, if few timesteps are used in \gls{PM} codes, the large-scale structure will be miscalculated only because of a faulty estimation of the \glslink{linear growth factor}{growth factor}, the exact value of which being well-known.

In contrast, the {\cola} method uses a numerical discretization of the operator $\mathrm{d}^2/\mathrm{d}\tau^2$ only on the left-hand side of equation \eqref{eq:EoM-COLA} and exploits the exact analytic expression for the fictitious force, $\mathrm{d}^2 \boldsymbol{\Psi}_\mathrm{LPT}/\mathrm{d}\tau^2$. The equation solved by {\cola}, equation \eqref{eq:EoM-COLA}, is obviously equivalent to \eqref{eq:EoM-standard}. However, as demonstrated by \citet{Tassev2013}, using this framework requires a smaller number of timesteps to recover accurate particle \glslink{particle realization}{realizations}. In particular, they show that as few as 10 timesteps from $z=9$ to $z=0$ are sufficient to obtain an accurate description of \gls{halo} statistics up to \glslink{halo}{halos} of mass $10^{11}$ $\mathrm{M}_\odot/h$, without resolving the internal dynamics of \glslink{halo}{halos}. Concerning the description of the large-scale matter \gls{density field}, 10 {\cola} timesteps achieve better than $95$\% \gls{cross-correlation} with the true result up $k \sim 2$ Mpc/$h$.

\begin{figure}
\begin{center}
\includegraphics[width=\columnwidth]{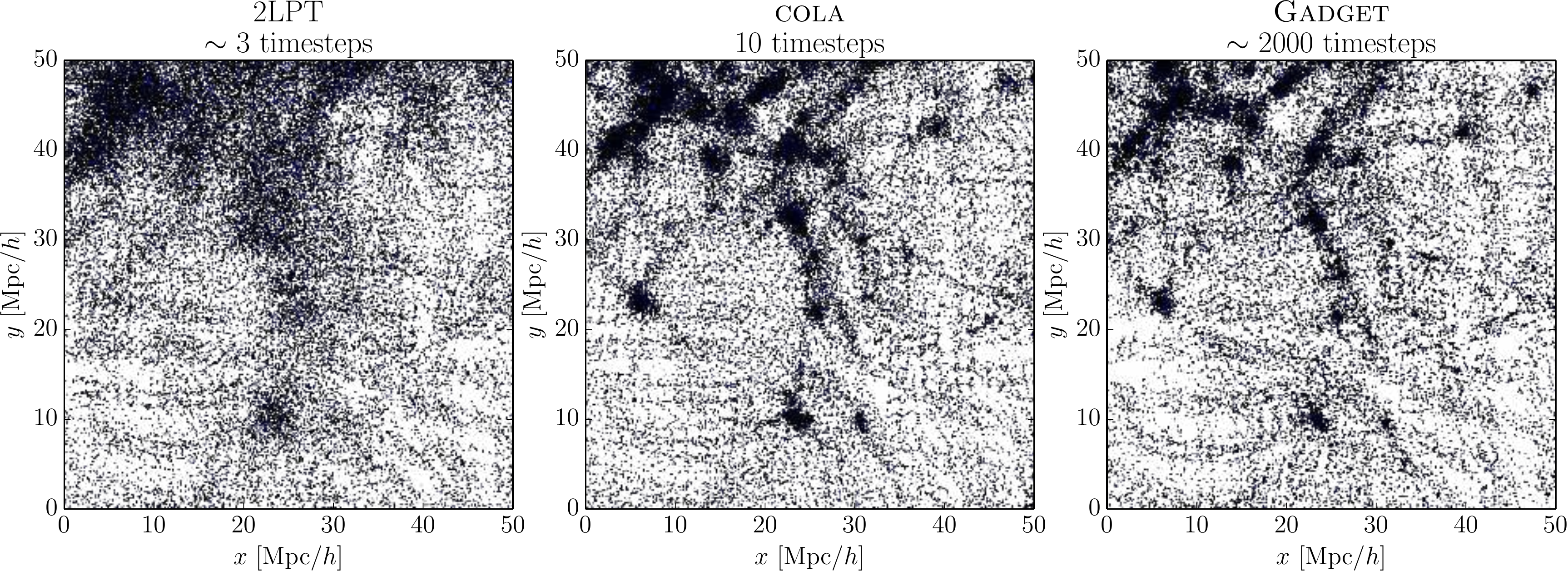}
\caption{Slices through three \glslink{particle realization}{particle realizations} evolved from the same \gls{initial conditions} up to $z=0$. The \glslink{dark matter particles}{particles} are shown as black points. Each slice is 20 Mpc/$h$ thick and 50 Mpc/$h$ on the side. The left panel shows the \gls{2LPT} approximation, of computational cost roughly equivalent to 3 timesteps of a $N$-body code. The right panel shows the reference result obtained from \textsc{\gls{Gadget-2}} after $\sim$~2000 timesteps, starting from \gls{2LPT} \gls{initial conditions} at $z=69$. The middle panel shows the result obtained with {\cola} with 10 timesteps, starting from \gls{2LPT} \gls{initial conditions} at $z=9$.}
\label{fig:particle_slice}
\end{center}
\end{figure}

As an illustration of the performance of {\cola}, we show slices through corresponding \gls{2LPT}, {\cola} and \glslink{Gadget-2}{\textsc{Gadget}} \glslink{particle realization}{particle realizations} in figure \ref{fig:particle_slice}. The simulations contain $512^3$ \glslink{dark matter particles}{particles} in a 750 Mpc/$h$ cubic box with \gls{periodic boundary conditions}. Forces are calculated on a \gls{PM} grid of $512^3$ cells. The \gls{initial conditions} are generated with \gls{2LPT} at a \gls{redshift} of $z=69$ for \glslink{Gadget-2}{\textsc{Gadget}} and $z=9$ for {\cola}.

\subsection{Non-linear BORG-COLA realizations}
\label{sec:Non-linear BORG-COLA realizations}

\draw{This section draws from section II.B. in \citet{Leclercq2015ST}.}

In chapter \ref{chap:ts}, we use an ensemble of $1,097$ large-scale structure \glslink{particle realization}{realizations} produced via \gls{non-linear filtering} of {\borg} \glslink{sample}{samples} with {\cola}. The \glslink{initial conditions}{initial} \gls{density field}, defined on a cubic equidistant grid with side length of 750~Mpc/$h$ and $256^3$ voxels, is populated by $512^3$ \glslink{dark matter particles}{particles} placed on a regular Lagrangian lattice. The \glslink{dark matter particles}{particles} are evolved with \gls{2LPT} to the \gls{redshift} of $z=69$ and with {\cola} from $z=69$ to $z=0$. The \glslink{final conditions}{final} \gls{density field} is constructed by binning the \glslink{dark matter particles}{particles} with a \gls{CiC} method on a $256^3$-voxel grid. This choice corresponds to a resolution of around 3~Mpc/$h$ for all the maps described in chapter \ref{chap:ts}. In this fashion, we generate a large set of data-constrained \glslink{reconstruction}{reconstructions} of the \glslink{final conditions}{present-day dark matter distribution} \citep[see also][]{Lavaux2010a,Kitaura2013,Hess2013,Nuza2014}. To ensure sufficient accuracy, $30$ timesteps logarithmically-spaced in the \gls{scale factor} are used for the evolution with \cola. 

\begin{figure}
\begin{center}
\includegraphics[width=0.6\columnwidth]{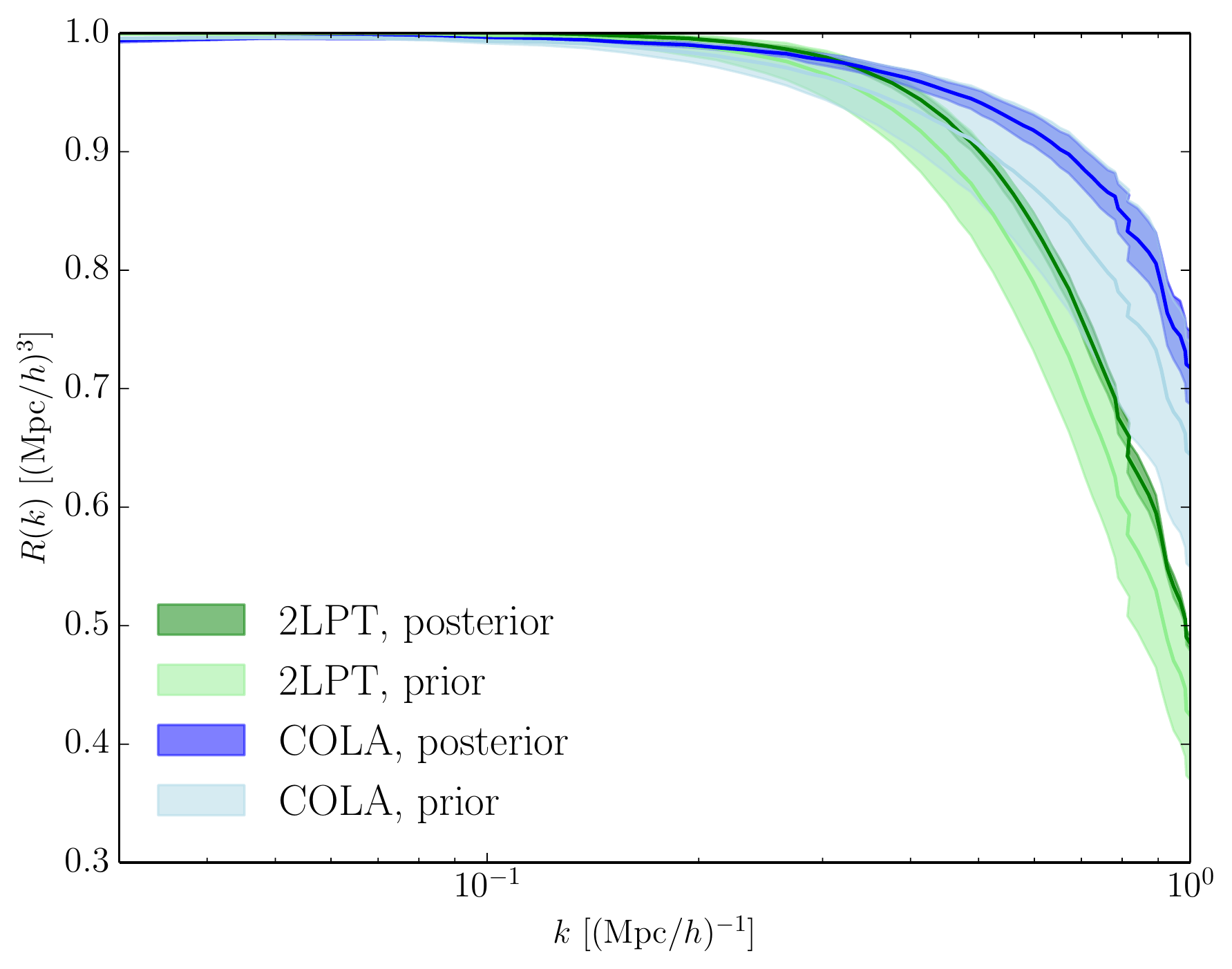}
\caption{\glslink{cross-correlation}{Cross-correlations} between \glslink{density field}{density fields} at \gls{redshift} zero, computed with a mesh size of $3$ Mpc/$h$. The reference fields are the result of \textsc{\gls{Gadget-2}}. The lines correspond to the cross-correlation between unconstrained \gls{2LPT} realizations and corresponding simulations (``2LPT, prior''), constrained \gls{2LPT} samples inferred by {\borg} and corresponding optimal filtering (``2LPT, posterior''), unconstrained {\cola} realizations and corresponding simulations (``COLA, prior''), constrained {\borg}-{\cola} samples and corresponding optimal filtering (``COLA, posterior''). In each case, we use $11$ constrained or unconstrained realizations. The solid lines correspond to the mean among all \glslink{particle realization}{realizations} used in this work, and the shaded regions correspond to the 2-$\sigma$ credible interval estimated from the standard error of the mean.}
\label{fig:cross_correlation_borgcola}
\end{center}
\end{figure}

{\cola} enables us to cheaply generate non-linear \glslink{density field}{density fields} at the required accuracy, as we now show. The power spectrum of non-linear {\borg}-{\cola} \glslink{particle realization}{realizations} is shown in figure \ref{fig:power_spectrum} in comparison to that of unconstrained \glslink{particle realization}{realizations} and to samples optimally filtered with \textsc{\gls{Gadget-2}}. In figure \ref{fig:cross_correlation_borgcola}, we plot the \gls{cross-correlation} between approximate density fields (predicted by \gls{2LPT} or by {\cola}) and the result of \textsc{\gls{Gadget-2}}, for both unconstrained and constrained realizations. On these plots, it can be checked that our constrained samples, inferred by {\borg} and filtered with {\cola}, contain the additional power expected in the \gls{non-linear regime} and \glslink{cross-correlation}{cross-correlate} at better that $95$\% accuracy with the corresponding \glslink{full gravity}{fully non-linear} \glslink{particle realization}{realizations}, up to $k \approx 0.4$ $\mathrm{Mpc}/h$. Therefore, as for unconstrained \glslink{N-body simulation}{simulations}, our setup yields vanishing difference between the representation of constrained \glslink{density field}{density fields} with {\cola} and with \textsc{\gls{Gadget-2}}, at the scales of interest of this work.

%% file: Chapter8/Chapter8Content.tex
\part{Cosmic web analysis}
\label{part:IV}

\chapter{Dark matter voids in the SDSS galaxy survey}
\label{chap:dmvoids}
\minitoc

\defcitealias{Tolkien1954}{John Ronald Reuel}
\begin{flushright}
\begin{minipage}[c]{0.6\textwidth}
\rule{\columnwidth}{0.4pt}

``Lost and forgotten be, darker than the darkness,\\
Where gates stand for ever shut, till the world is mended.''\\
--- \citetalias{Tolkien1954} \citet{Tolkien1954}, \textit{The Fellowship of the Ring}

\vspace{-5pt}\rule{\columnwidth}{0.4pt}
\end{minipage}
\end{flushright}

\abstract{\section*{Abstract}

What do we know about \glslink{void}{voids} in the \gls{dark matter} distribution given the \glslink{SDSS}{Sloan Digital Sky Survey} and assuming the {\LCDM} model? In chapter \ref{chap:BORGSDSS}, application of the Bayesian \glslink{large-scale structure inference}{inference} algorithm {\borg} to the \gls{SDSS} Data Release 7 main galaxy sample has generated detailed \glslink{final conditions}{Eulerian} and \glslink{initial conditions}{Lagrangian} representations of the \glslink{LSS}{large-scale structure} as well as the possibility to accurately \glslink{uncertainty quantification}{quantify corresponding uncertainties}. Building upon these results, we present constrained catalogs of \glslink{void}{voids} in the Sloan volume, aiming at a physical representation of \glslink{dark matter void}{dark matter underdensities} and at the alleviation of the problems due to \gls{sparsity} and \glslink{bias}{biasing} on \gls{galaxy void} catalogs. To do so, we generate \glslink{constrained simulation}{data-constrained reconstructions} of the presently observed \glslink{LSS}{large-scale structure} using a \glslink{full gravity}{fully non-linear gravitational model}. We then find and analyze \gls{void} candidates using the \textsc{vide} toolkit. Our methodology therefore predicts the properties of \glslink{void}{voids} based on fusing \gls{prior} information from \glslink{N-body simulation}{simulations} and \gls{data} constraints. For usual \gls{void} statistics (\gls{number function}, \gls{ellipticity distribution} and radial \gls{density profile}), all the results obtained are in agreement with \glslink{N-body simulation}{dark matter simulations}. Our \gls{dark matter void} candidates probe a deeper \gls{void hierarchy} than \glslink{void}{voids} directly based on the observed galaxies alone. The use of our catalogs therefore opens the way to high-precision \gls{void} cosmology at the level of the \gls{dark matter} field. We have made the \gls{void} catalogs used in this work available at \href{http://www.cosmicvoids.net}{http://www.cosmicvoids.net}.}

\draw{This chapter is adapted from its corresponding publication, \citet{Leclercq2015DMVOIDS}.}

\section{Introduction}

Observations of the cosmic \glslink{LSS}{large-scale structure} have revealed that galaxies tend to lie in thin \glslink{sheet}{wall}-like structures surrounding large underdense regions known as \glslink{void}{voids}, which constitute most of the volume of the Universe. Although the discovery of cosmic \glslink{void}{voids} dates back to some of the first \glslink{galaxy survey}{galaxy redshift surveys} \nopagebreak{\citep{Gregory1978,Kirshner1981,deLapparent1986}} and their significance was assessed in some early studies \citep{Martel1990,vandeWeygaert1993,Goldberg2004}, the systematic analysis of \gls{void} properties has only been considered seriously as a source of cosmological information in the last decade \citep[e.g.][and references therein]{Sheth2004,Colberg2005,Viel2008,BetancortRijo2009,Lavaux2010,Biswas2010,vandeWeygaert2011,Lavaux2012}. Like overdense tracers of the \gls{density field} such as \glslink{cluster}{clusters}, \glslink{void}{voids} can be studied by statistical methods in order to learn about their distribution and properties compared to theoretical predictions.

Generally, direct sensitivity of \gls{void} statistics to cosmology is only guaranteed for the underdense regions of the overall matter \gls{density field}, which includes a large fraction of \gls{dark matter}. These are the physical \glslink{void}{voids} in the \gls{LSS}, for which theoretical modeling is established. However, absent direct measurements of \glslink{dark matter void}{dark matter underdensities}, current \gls{void} catalogs are defined using the locations of galaxies in \glslink{galaxy survey}{large redshift surveys} \citep{Pan2012,Sutter2012DR7CATALOGS,Sutter2014DR9CATALOGS,Nadathur2014a}. Since galaxies trace the underlying mass distribution only \glslink{sparsity}{sparsely}, \gls{void} catalogs are subject to uncertainty and \gls{noise}. Additionally, numerical simulations show that there exists a population of particles in cosmic \glslink{void}{voids}. This is an indication of physical \glslink{bias}{biasing} in \gls{galaxy formation}: there is primordial \glslink{dark matter}{dark} and baryonic matter in \glslink{void}{voids}, but due to the low density, little \gls{galaxy formation} takes place there. Additionally, due to complex baryonic physics effects during their formation and evolution, galaxies are \glslink{bias}{biased} tracers of the underlying \gls{density field}, which gives rise to qualitatively different \gls{void} properties.

The sensitivity of \gls{void} properties to the \gls{sampling} density and \glslink{bias}{biasing} of the tracers has only been recently analyzed in depth on simulations, by using \glslink{mock catalog}{synthetic models} to mimic realistic \glslink{galaxy survey}{surveys}. \cite{Little1994,Benson2003,Tinker2009,Sutter2014SPARSITYBIAS} found that the statistical properties of \glslink{void}{voids} in \glslink{galaxy survey}{galaxy surveys} are not the same as those in dark matter distributions. At lower tracer density, small \glslink{void}{voids} disappear and the remaining \glslink{void}{voids} are larger and more spherical. Their \glslink{density profile}{density profiles} get slightly steeper, with a considerable increase of their \gls{compensation} scale, which potentially may serve as a static ruler to probe the \gls{expansion} history of the Universe \citep{Hamaus2014a}. \cite{Hamaus2014b} recently proposed a universal formula for the \glslink{density profile}{density profiles} of \glslink{void}{voids}, describing in particular \glslink{dark matter void}{dark matter voids} in simulations \citep[see also][]{Colberg2005,Paz2013,Ricciardelli2014,Nadathur2014b}. The connection between \glslink{galaxy void}{galaxy voids} and \glslink{dark matter void}{dark matter voids} on a one-by-one basis is difficult due to the complex internal \glslink{void hierarchy}{hierarchical structure} of \glslink{void}{voids} \citep{Dubinski1993,vandeWeygaert1993,Sahni1994,Sheth2004,Aragon-Calvo2013,Sutter2014DR9CATALOGS,Sutter2014DMGALAXYVOIDS}. However, the nature of this relationship determines the link between a \glslink{galaxy survey}{survey}, with its particular tracer density, and the portion of the \gls{cosmic web} that it represents. Understanding this connection is of particular importance in light of recent results that probe the \gls{LSS} via its effect on photons geodesics. These results include \cite{Melchior2014,Clampitt2014}, which probe the dark matter distribution via \gls{weak gravitational lensing}; \cite{Ilic2013,Planck2013ISW} for the detection of the \gls{integrated Sachs-Wolfe effect} in the \glslink{CMB}{cosmic microwave background}, sensitive to the properties of \gls{dark energy}. As a response to this demand, \cite{Sutter2014DMGALAXYVOIDS} found that \glslink{void}{voids} in \glslink{galaxy survey}{galaxy surveys} always correspond to \glslink{dark matter void}{underdensities in the dark matter}, but that their centers may be offset and their size can differ, in particular in \glslink{sparsity}{sparsely sampled} \glslink{galaxy survey}{surveys} where \gls{void} edges suffer fragmentation.

While previous authors offer broad prescriptions to assess the effects of \gls{sparsity} and \glslink{bias}{biasing} of the tracers on \glslink{void}{voids}, the connection between \glslink{galaxy void}{galaxy voids} of a particular \glslink{galaxy survey}{survey} and \glslink{dark matter void}{dark matter underdensities} remains complex. In particular, disentangling these effects from cosmological signals in presence of the uncertainty inherent to any cosmological observation (\gls{selection effects}, \glslink{galaxy survey}{survey} \gls{mask}, \gls{noise}, \gls{cosmic variance}) remains an open question. In this work, we propose a method designed to circumvent the issues due to the conjugate and intricate effects of \gls{sparsity} and \glslink{bias}{biasing} on \gls{galaxy void} catalogs. In doing so, we will show that \glslink{void}{voids} in the dark matter distribution can be constrained by the \textit{ab initio} analysis of \glslink{galaxy survey}{surveys} of tracers, such as galaxies. We will demonstrate the feasibility of our method and obtain catalogs of \glslink{dark matter void}{dark matter voids} candidates in the \glslink{SDSS}{Sloan Digital Sky Survey} Data Release 7.

Our method is based on the identification of \glslink{void}{voids} in the dark matter distribution \glslink{large-scale structure inference}{inferred} from \glslink{LSS}{large-scale structure} \glslink{galaxy survey}{surveys}. The constitution of such maps from galaxy positions, also known as ``\gls{reconstruction}'', is a field in which \glslink{Bayesian statistics}{Bayesian} methods have led to enormous progress over the last few years. Initial approaches typically relied on approximations such as a multivariate \glslink{grf}{Gaussian} or \gls{log-normal distribution} for \glslink{density field}{density fields}, with a prescription for the \gls{power spectrum} to account for the correct \glslink{two-point correlation function}{two-point statistics} \citep{Lahav1994,Zaroubi2002,Erdovgdu2004,Kitaura2008,Kitaura2009,Kitaura2010,JascheKitaura2010,Jasche2010a,Jasche2010b}. However, due to their potentially complex shapes, proper identification of structures such as \glslink{void}{voids} requires \glslink{reconstruction}{reconstructions} correct not only at the level of the \gls{power spectrum}, but also \glslink{high-order correlation function}{higher-order correlators}. \glslink{large-scale structure inference}{Inferences} of this kind from observational \gls{data} have only been made possible very recently by the introduction of physical models of \gls{structure formation} in the \glslink{large-scale structure likelihood}{likelihood}. This naturally moves the problem to the inference of the \gls{initial conditions} from which the \glslink{LSS}{large-scale structure} originates \citep{Jasche2013BORG,Kitaura2013,Wang2013}.

This work exploits the recent application of the {\borg} (Bayesian Origin Reconstruction from Galaxies, \citet{Jasche2013BORG}, see chapter \ref{chap:BORG}) algorithm to the \glslink{SDSS}{Sloan Digital Sky Survey} galaxies \citep[][see chapter \ref{chap:BORGSDSS}]{Jasche2015BORGSDSS}, and on the subsequent generation of constrained non-linear realizations of the present large-scale distribution of \gls{dark matter}. {\borg} is a full-scale Bayesian framework, permitting the four-dimensional physical inference of \glslink{density field}{density fields} in the \glslink{linear regime}{linear} and \gls{mildly non-linear regime}, evolving gravitationally from the \gls{initial conditions} to the presently observed \glslink{LSS}{large-scale structure}. By exploring a highly non-linear and \glslink{non-Gaussianity}{non-Gaussian} \gls{LSS} \gls{posterior} distribution via efficient \glslink{MCMC}{Markov Chain Monte Carlo} methods, it also provides naturally and fully self-consistently accurate \gls{uncertainty quantification} for all derived quantities. A straightforward use of reconstructed \gls{initial conditions} is to resimulate the considered volume \citep{Lavaux2010a,Kitaura2013,Hess2013}. In the same spirit, building upon the inference of the \gls{initial conditions} by {\borg}, one can generate a set of data-constrained realizations of the present \glslink{LSS}{large-scale structure} via \glslink{full gravity}{full $N$-body dynamics}. As we will show, we make use of \gls{initial conditions} reconstructed by {\borg} without any further post-processing, which demonstrates the high quality of inference results.

Due to the limited number of \glslink{phase space}{phase-space} foldings, the influence of \glslink{non-linear evolution}{non-linearity} in cosmic \glslink{void}{voids} is expected to be milder as compared to galaxies and dark matter \glslink{halo}{halos} (\citealp{Neyrinck2012,Neyrinck2013b,Leclercq2013}, see also \citealp{Abel2012,Falck2012,Shandarin2012}). For this reason, \glslink{void}{voids} are more closely related to the \gls{initial conditions} of the Universe, which makes them the ideal laboratories for physical application of Bayesian inference with {\borg}. In this work, we apply the \gls{void} finder algorithm {\vide} \citep{Sutter2015VIDE}, based on \textsc{\gls{zobov}} \citep{Neyrinck2008}, to data-constrained, non-linear \glslink{reconstruction}{reconstructions} of the \gls{LSS}. Each of them is a full physical realization of densely-sampled \glslink{dark matter particles}{particles tracing the dark matter} \gls{density field}. In this fashion, we construct catalogs of \glslink{dark matter void}{dark matter voids} in the \gls{SDSS} volume robust to \gls{sparsity} and \glslink{bias}{biasing} of galaxies. As we will show, this procedure drastically reduces \gls{statistical uncertainty} in \gls{void} catalogs. Additionally, the use of \glslink{constrained simulation}{data-constrained reconstructions} allows us to extrapolate the \gls{void} identification in existing \gls{data} (e.g. at very small or at the largest scales, at high \gls{redshift} or near the \glslink{survey geometry}{survey boundary}).

As described in chapters \ref{chap:BORG} and \ref{chap:BORGSDSS}, \citep[see][]{Jasche2013BORG,Jasche2015BORGSDSS}, the {\borg} inference framework possesses a high degree of control on observational \glslink{systematic uncertainty}{systematic} and \glslink{statistical uncertainty}{statistical uncertainties} such as \gls{noise}, \gls{survey geometry} and \gls{selection effects}. \glslink{uncertainty quantification}{Uncertainty quantification} is provided via efficient \gls{sampling} of the corresponding \glslink{large-scale structure likelihood}{LSS posterior} distribution. The resultant set of \glslink{initial conditions}{initial} and \glslink{final conditions}{final} \gls{density field} realizations yields a numerical representation of the full \gls{posterior} distribution, capturing all \gls{data} constraints and observational uncertainties. Building upon these results, in this work, we will extend our Bayesian reasoning to \gls{void} catalogs. Specifically, we apply \glslink{full gravity}{full non-linear $N$-body dynamics} to a set of data-constrained \gls{initial conditions} to  arrive at a set of non-linear dark matter \glslink{density field}{density fields} at the present epoch. As a result, we obtain a probabilistic description of non-linear \glslink{density field}{density fields} constrained by \gls{SDSS} observations. Applying the {\vide} \gls{void} finder to this set of \glslink{reconstruction}{reconstructions} yields $N$ data-constrained realizations of the catalog, representing the \gls{posterior} \glslink{pdf}{probability distribution} for \glslink{dark matter void}{dark matter voids} given observations. In this fashion, we have fully Bayesian access to \gls{uncertainty quantification} via the variation between different realizations. In particular, we are now able to devise improved estimators for any \gls{void} statistics by the use of \glslink{Blackwell-Rao estimator}{Blackwell-Rao estimators}. To assess the robustness of this technique for cosmological application, we focus on three key \gls{void} observables: \glslink{number function}{number functions}, \glslink{ellipticity distribution}{ellipticity distributions} and radial \glslink{density profile}{density profiles}. These are especially sensitive probes of non-standard cosmologies \citep{Bos2012} and are well understood in both \gls{data} and simulations \citep[e.g][]{Sutter2014DR9CATALOGS}.

As a general matter, we stress that these data-constrained realizations of \gls{dark matter void} catalogs were obtained assuming a {\LCDM} \gls{prior}. Using our products for \glslink{model comparison}{model testing} therefore requires care: in the absence of \gls{data} constraints, one will simply be dealing with realizations of the {\LCDM} \gls{prior}. Consequently, any departure from unconstrained {\LCDM} predictions are driven by the \gls{data}. Conversely, for \glslink{model comparison}{model tests} where the \gls{data} are not strongly informative, agreement with {\LCDM} is the default answer.

This chapter is organized as follows. In section \ref{sec:Methodology}, we describe our methodology: Bayesian \glslink{large-scale structure inference}{inference} with the {\borg} algorithm, \gls{non-linear filtering} of the results, \gls{void} identification technique and \glslink{Blackwell-Rao estimator}{Blackwell-Rao estimators} for \gls{void} statistics. In section \ref{sec:Properties of dark matter voids}, we examine the properties of the \glslink{dark matter void}{dark matter voids} in our catalogs. Finally, in section \ref{sec:Summary and conclusions} we summarize our results, discuss perspectives for existing and upcoming \glslink{galaxy survey}{galaxy surveys} and offer concluding comments. 

\section{Methodology}
\label{sec:Methodology}

In this section, we describe our methodology step by step:

\begin{enumerate}
\item {\glslink{large-scale structure inference}{inference} of the \gls{initial conditions} with {\borg} (section \ref{sec:Bayesian large-scale structure inference with the BORG algorithm}),}
\item {generation of data-constrained realizations of the \gls{SDSS} volume (section \ref{sec:Generation of data-constrained reconstructions}),}
\item {\gls{void} finding and processing (section \ref{sec:Void finding and processing}),}
\item {combination of different \gls{void} catalogs with \glslink{Blackwell-Rao estimator}{Blackwell-Rao estimators} (section \ref{sec:Blackwell-Rao estimators for dark matter void realizations}).}
\end{enumerate}

In section \ref{sec:Galaxy void catalog and dark matter simulation}, we describe the \gls{void} catalogs used as references for comparison with our results. These are \gls{galaxy void} catalogs directly based on \gls{SDSS} galaxies without use of our methodology, and catalogs of \glslink{void}{voids} in dark matter \glslink{N-body simulation}{simulations}.

\begin{figure} 
  \centering
  {\small Standard \gls{galaxy void} identification\normalsize\\
  \includegraphics[width=0.6\columnwidth]{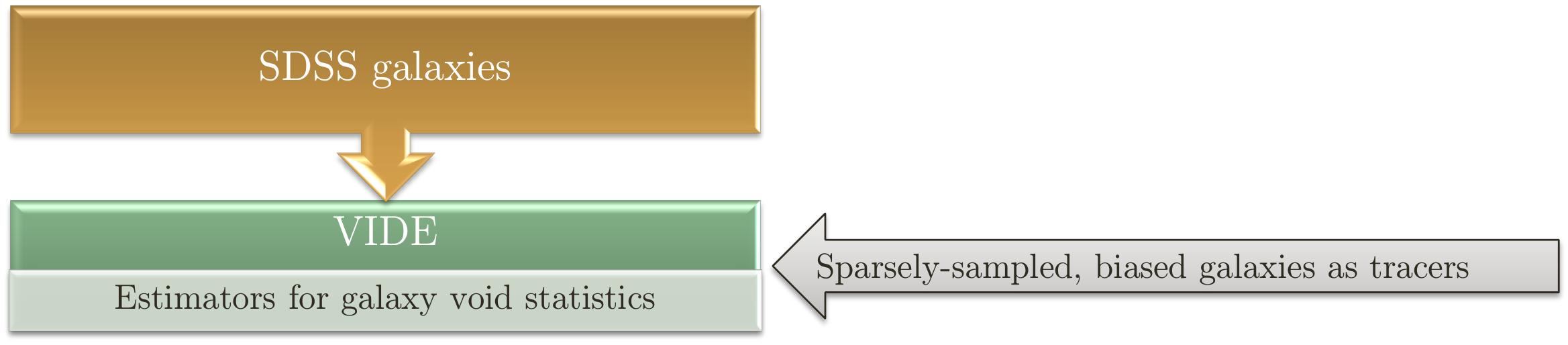}\vspace{15pt}\\
  \small Inference of \glslink{dark matter void}{dark matter voids}\normalsize\\\vspace{2pt}
  \includegraphics[width=0.6\columnwidth]{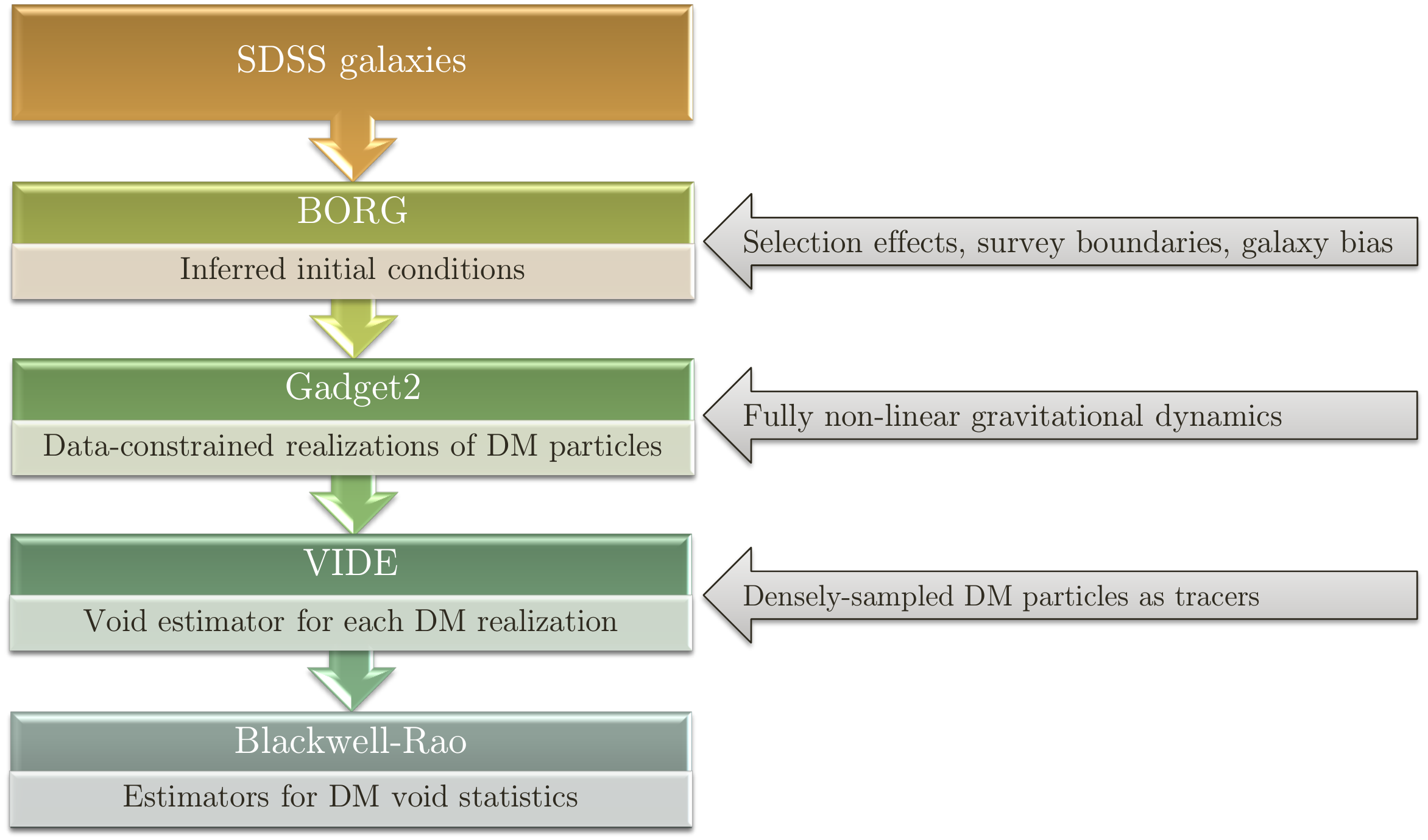}}
  \caption{
  Schematic representation of our methodology for the inference of \glslink{dark matter void}{dark matter voids} (lower panel) in comparison to the standard approach for the identification of \glslink{galaxy void}{galaxy voids} (upper panel).
           }
\label{fig:method}
\end{figure}

A schematic representation of our procedure is represented in figure \ref{fig:method}, in comparison to the standard approach of finding \glslink{void}{voids} using galaxies as tracers.

\subsection{Bayesian large-scale structure inference with the BORG algorithm}
\label{sec:Bayesian large-scale structure inference with the BORG algorithm}

This works builds upon previous results, obtained by the application of {\borg} \citep[Bayesian Origin Reconstruction from Galaxies,][]{Jasche2013BORG} to \gls{SDSS} main galaxy data \citep{Jasche2015BORGSDSS}. In the rest of this section, we summarize its most stringent features; the reader is referred to chapters \ref{chap:BORG} and \ref{chap:BORGSDSS} for all details. 

The {\borg} algorithm is a fully probabilistic inference machinery aiming at the analysis of \glslink{linear regime}{linear} and \glslink{mildly non-linear regime}{mildly-non-linear} \glslink{density field}{density} and \glslink{velocity field}{velocity fields} in galaxy observations. It incorporates a physical model of cosmological \gls{structure formation}, which translates the traditional task of reconstructing the non-linear three-dimensional \gls{density field} into the task of inferring corresponding \gls{initial conditions} from present cosmological observations. This approach yields a highly non-trivial Bayesian inference, requiring to explore very \glslink{high-dimensional parameter space}{high-dimensional} and non-linear spaces of possible solutions to the \gls{initial conditions} problem from incomplete observations. Typically, these parameter spaces comprise on the order of $10^6$ to $10^7$ parameters, corresponding to the elements of the discretized observational domain.

Specifically, the {\borg} algorithm explores a \gls{posterior} distribution consisting of a \glslink{grf}{Gaussian} \gls{prior}, describing the statistical behavior of the \glslink{initial conditions}{initial} \gls{density field} at a cosmic \gls{scale factor} of $a=10^{-3}$,  linked via \glslink{2LPT}{second-order Lagrangian perturbation theory} to a \glslink{Poisson likelihood}{Poissonian model} of \gls{galaxy formation} at the present epoch \citetext{for details see \citealp{Jasche2013BORG} and \citealp{Jasche2015BORGSDSS}}. As pointed out by previous authors \citep[see e.g.][and chapter \ref{chap:lpt}]{Moutarde1991,Buchert1994,Bouchet1995,Scoccimarro2000,Bernardeau2002,Scoccimarro2002}, \gls{2LPT} describes the \glslink{one-point distribution}{one-}, \glslink{two-point correlation function}{two-} and \glslink{three-point correlation function}{three-point statistics} correctly and represents \glslink{high-order correlation function}{higher-order statistics} very well. Consequently, the {\borg} algorithm naturally accounts for features of the \gls{cosmic web}, such as \glslink{filament}{filaments}, that are typically associated to \glslink{high-order correlation function}{high-order statistics} induced by \glslink{non-linear evolution}{non-linear} gravitational \gls{structure formation} processes.

Besides physical \gls{structure formation}, the \gls{posterior} distribution also accounts for \gls{survey geometry}, \gls{selection effects} and \gls{noise}, inherent to any cosmological observation (see section \ref{sec:The BORG data model}). Corresponding full Bayesian \gls{uncertainty quantification} is provided by exploring this highly \glslink{non-Gaussianity}{non-Gaussian} and non-linear \gls{posterior} distribution via an efficient \glslink{HMC}{Hamiltonian Markov Chain Monte Carlo} \gls{sampling} algorithm \citep[see][and sections \ref{sec:Hamiltonian Monte Carlo}, \ref{sec:Hamiltonian Monte Carlo and equations of motion for the LSS density}, for details]{Jasche2013BORG}. In order to account for \gls{luminosity} dependent galaxy \gls{bias} \citep{Jasche2013BIAS} and to make use of automatic \gls{noise} calibration, we further use modifications introduced to the original {\borg} algorithm by \citet{Jasche2015BORGSDSS} (see section \ref{sec:Calibration of the noise level}).

In this work, we make use of the $12,000$ \glslink{sample}{samples} of the \gls{posterior} distribution generated by \citet{Jasche2015BORGSDSS}, described in chapter \ref{chap:BORGSDSS}, which constitute highly-detailed and accurate \glslink{reconstruction}{reconstructions} of the \glslink{initial conditions}{initial} and present-day \glslink{density field}{density fields} constrained by \gls{SDSS} observations.

\subsection{Generation of data-constrained reconstructions}
\label{sec:Generation of data-constrained reconstructions}

Starting form 11 statistically \glslink{conditional independence}{independent} \gls{initial conditions} realizations from the {\borg} \gls{SDSS} analysis, we generated a set of \glslink{full gravity}{fully non-linear}, \glslink{constrained simulation}{constrained reconstructions} of the \gls{LSS}. This step is achieved via \glslink{non-linear filtering}{optimal filtering} of {\borg} results with the \textsc{\gls{Gadget-2}} \citep{Springel2001,Springel2005} cosmological code. For details on the \gls{non-linear filtering} procedure, see chapter \ref{chap:filtering}, in particular section \ref{sec:Filtering Gadget} for the description of the set of realizations used in this chapter.

\subsection{Void finding and processing}
\label{sec:Void finding and processing}

\subsubsection{Void finding}
\label{sec:Void finding}

We identify and post-process \glslink{void}{voids} with the {\vide} (Void IDentification and Examination) toolkit\footnote{\href{http://www.cosmicvoids.net}{http://www.cosmicvoids.net}}~\citep[][also described in section \ref{sec:apx-vide} of appendix \ref{apx:classification}]{Sutter2015VIDE}, which uses a highly modified version of \textsc{\gls{zobov}} \citep{Neyrinck2008,Lavaux2012,Sutter2012DR7CATALOGS} to create a \gls{Voronoi tessellation} of the tracer particle population and the \gls{watershed transform} to group \glslink{Voronoi tessellation}{Voronoi} cells into zones and voids \citep{Platen2007}. The \gls{watershed transform} identifies catchment basins as the cores of \glslink{void}{voids}, and ridgelines, which separate the flow of water, as the boundaries of \glslink{void}{voids}. It naturally builds a nested \glslink{void hierarchy}{hierarchy of voids} \citep{Lavaux2012,Bos2012}. For the purposes of this work, we examine all \glslink{void}{voids} regardless of their position in the \glslink{void hierarchy}{hierarchy}. The pipeline imposes a density-based threshold within the \gls{void} finding operation: \glslink{void}{voids} only include as additional members \glslink{Voronoi tessellation}{Voronoi} zones if the minimum ridge density between that zone and the \gls{void} is less than 0.2 times the mean particle density \citetext{\citealp{Platen2007}; see \citealp{Blumenthal1992,Sheth2004} for the role of the corresponding $\delta = -0.8$ \glslink{density contrast}{underdensity}}. If a \gls{void} consists of only a single zone (as they often do in \glslink{sparsity}{sparse} populations) then this restriction does not apply. 

{\vide} provides several useful definitions used in this work,  such as the effective radius,
\begin{equation}
R_\mathrm{v} \equiv \left( \frac{3}{4\pi} V \right)^{1/3} ,
\end{equation}
where $V$ is the total volume of the \glslink{Voronoi tessellation}{Voronoi} cells that contribute to the \gls{void}. We use this radius definition to ignore \glslink{void}{voids} with $R_\mathrm{v}$ below the mean particle spacing $\bar{n}^{-1/3}$ of the tracer population, as these are increasingly affected by Poisson fluctuations. {\vide} also reports the volume-weighted center, or macrocenter, as
\begin{equation}
\textbf{x}_\mathrm{v} \equiv \frac{1}{\sum_i V_i} \sum_i \textbf{x}_i V_i,
\end{equation}
where $\textbf{x}_i$ and $V_i$ are the positions and \glslink{Voronoi tessellation}{Voronoi} volumes of each tracer particle $i$, respectively.

In each tracer population, the {\vide} pipeline provides \gls{void} estimators ; in particular, the three statistics we will focus on in section \ref{sec:Properties of dark matter voids}: \glslink{number function}{number count}, \gls{ellipticity distribution} and radial \gls{density profile}. 

\begin{figure} 
  \centering
  {\includegraphics[width=0.49\columnwidth]{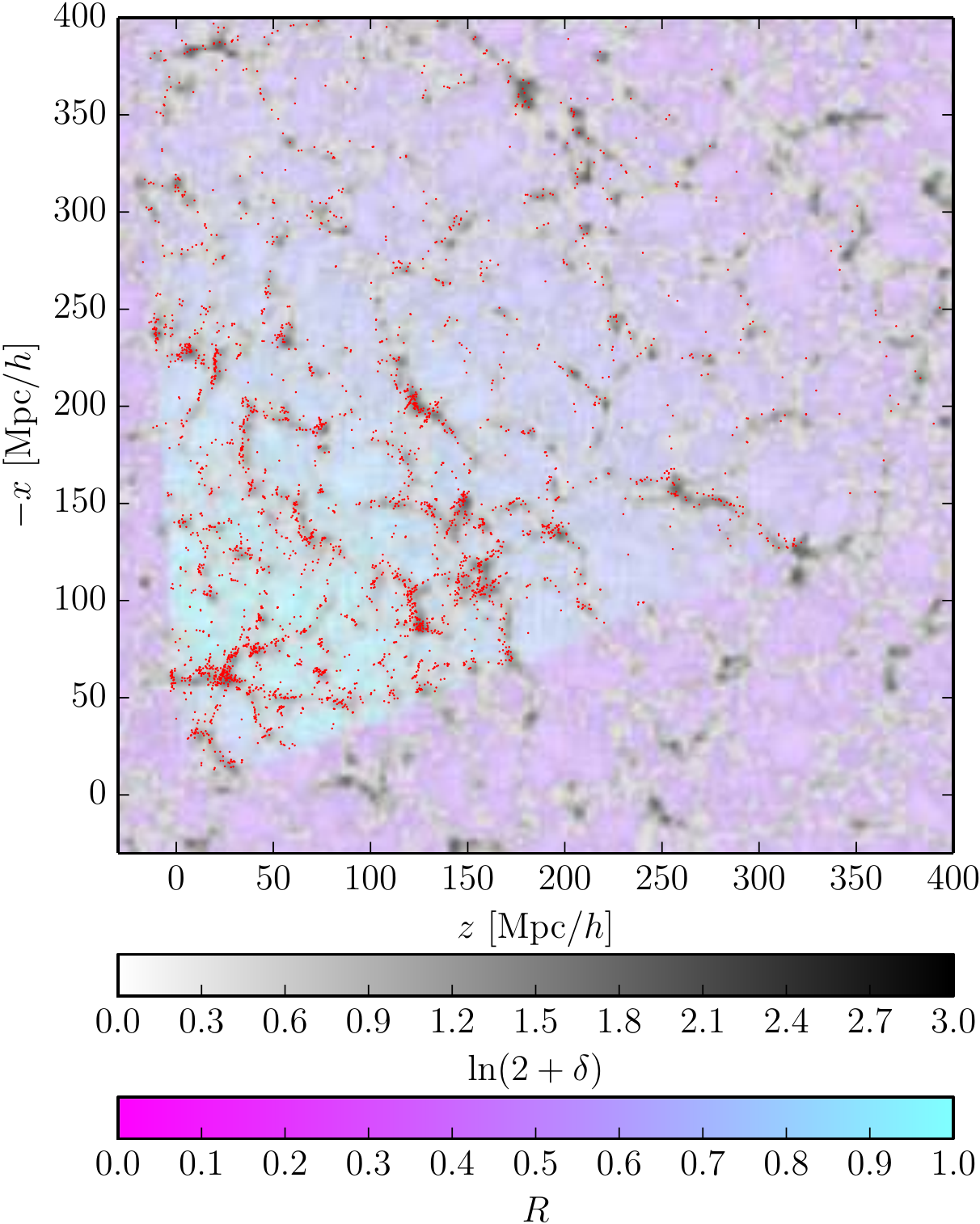}}
  {\includegraphics[width=0.49\columnwidth]{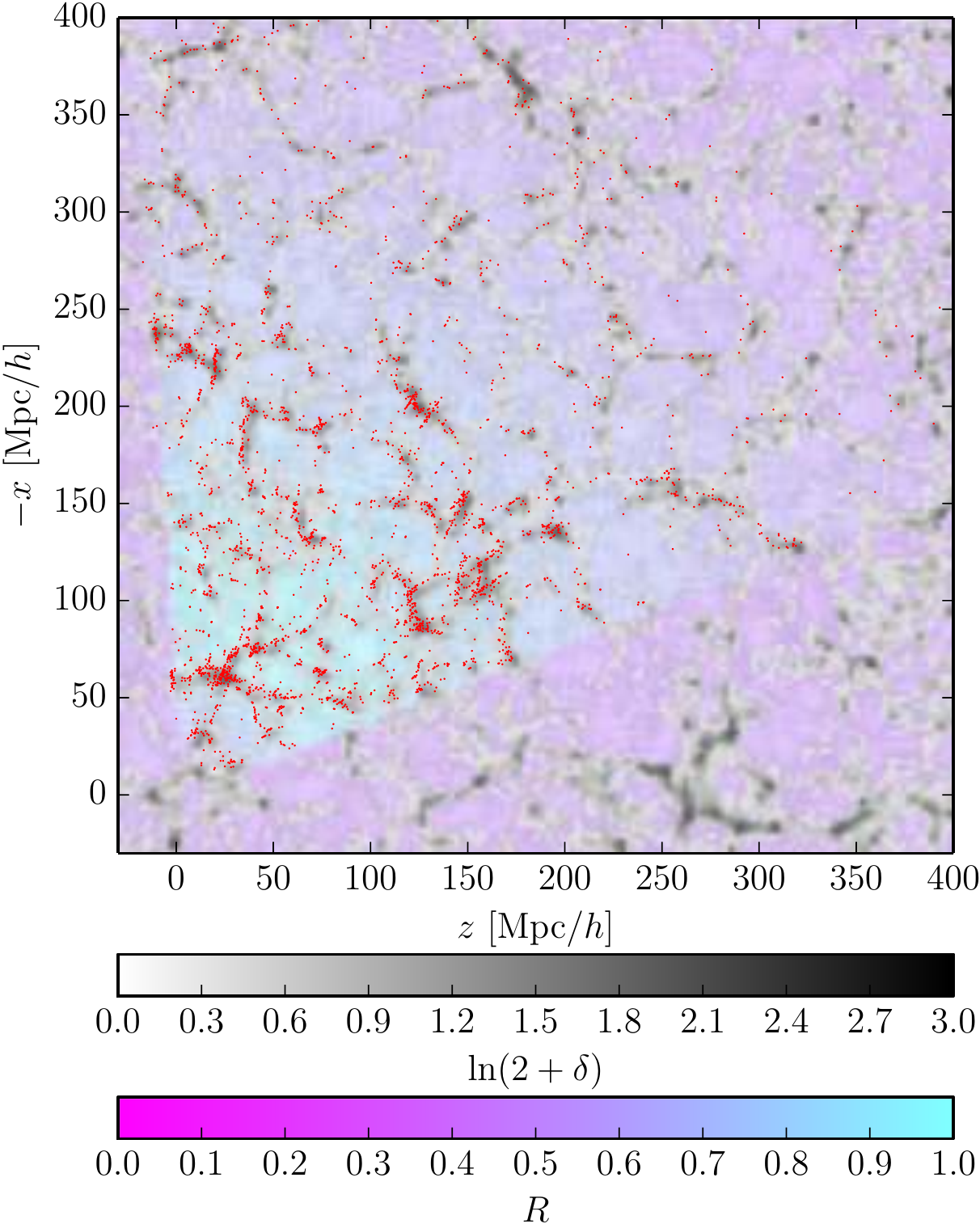}}
  {\includegraphics[width=0.49\columnwidth]{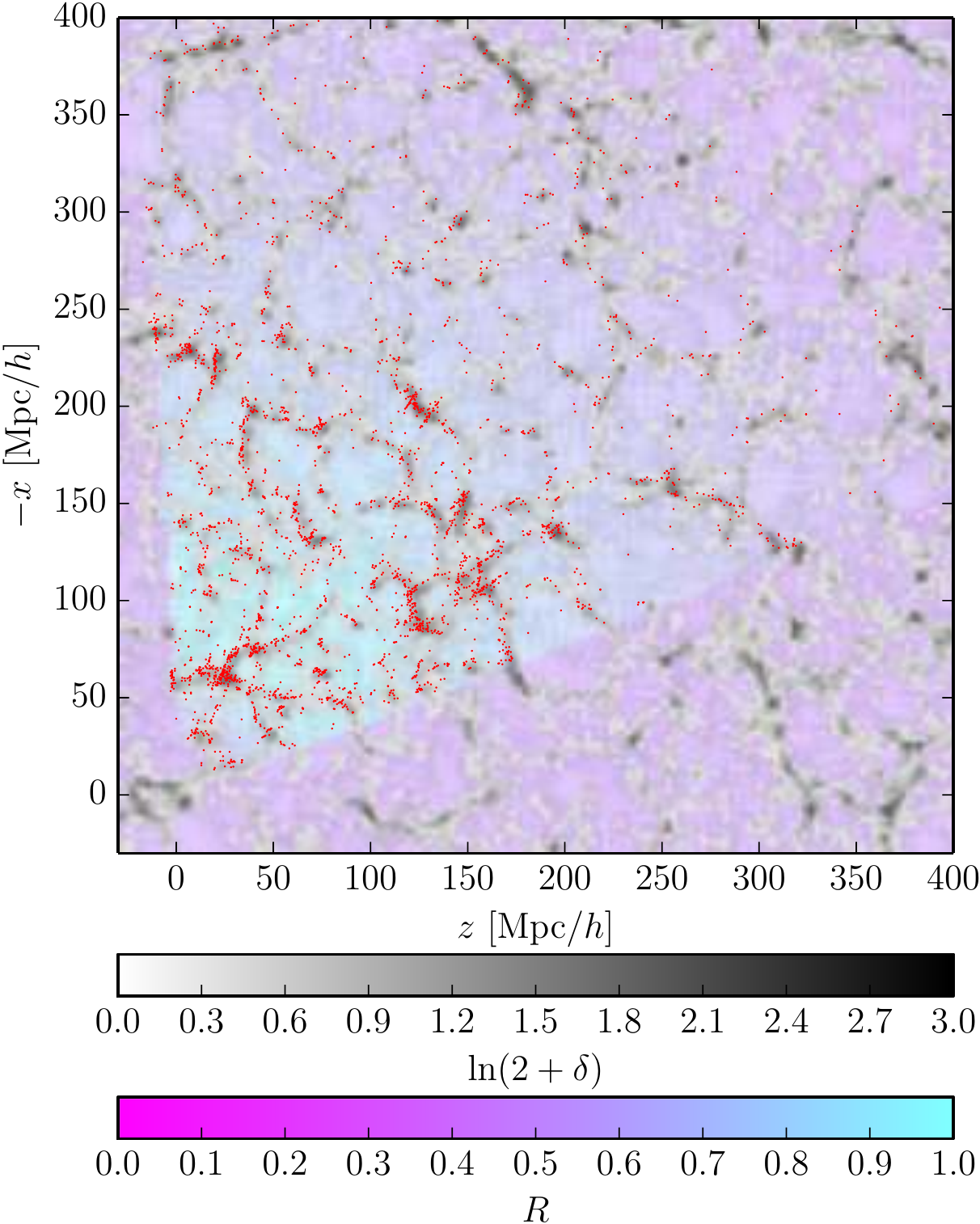}}
  {\includegraphics[width=0.49\columnwidth]{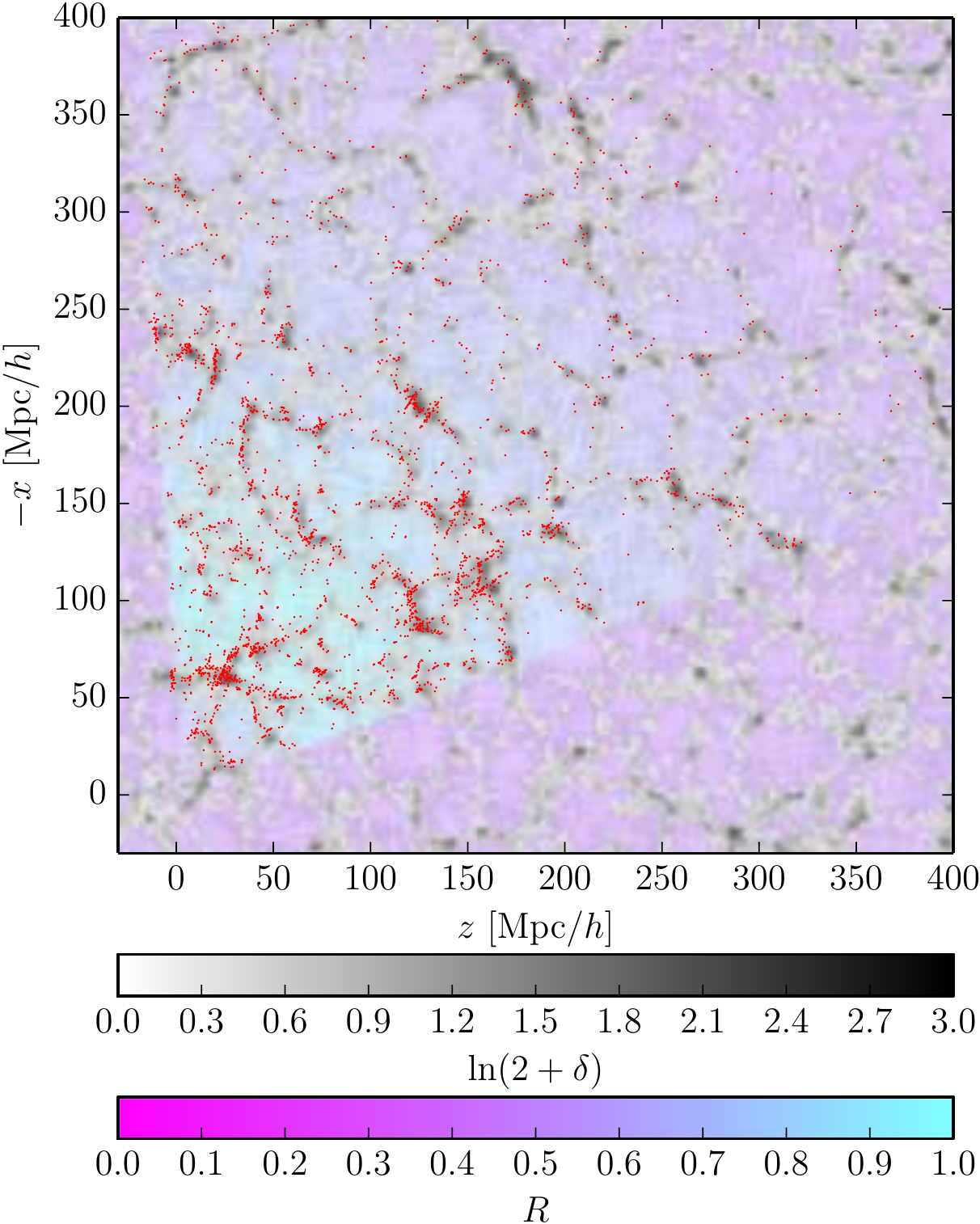}}
  \caption{
Slices through different data-constrained realizations used to build \glslink{sample}{samples} of the \gls{dark matter void} catalog. The \gls{SDSS} galaxies used for the inference with {\borg} are represented as red dots. The density of \gls{dark matter particles} identified by {\vide} as being part of a \gls{void} is shown in gray scale. In color, we show the \glslink{dark matter particles}{particles} that live in the core of \glslink{dark matter void}{dark matter voids} (in a density environment smaller than $-0.3$ times the average density). The \gls{survey response operator} $R$ shows how well the results are constrained by the \gls{data} (see text for details). In the observed region, the \gls{data} are strongly informative about the \gls{cosmic web} in general and \glslink{void}{voids} in particular; the \glslink{reconstruction}{reconstructions} are not \gls{prior}-dominated.
           }
\label{fig:slice_voids}
\end{figure}

In figure \ref{fig:slice_voids}, we show slices through different data-constrained realizations. The density of \gls{dark matter particles} identified by {\vide} as being part of a \gls{void} is represented in gray scale. Note that, since \textsc{\gls{zobov}} essentially performs a division of space in different \gls{void} regions with vanishingly-thin ridges, almost all \glslink{dark matter particles}{particles} initially present in the dark matter field are conserved. For clarity of the visualization, the quantity represented is $\ln(2+\delta)$ where $\delta$ is the \gls{density contrast} of \glslink{dark matter particles}{particles} in \glslink{void}{voids}. The \gls{SDSS} galaxies used for the {\borg} analysis are overplotted as red dots. The core of \glslink{dark matter void}{dark matter voids} (using a density threshold $\delta < -0.3$) is shown in color. As can be observed, \glslink{dark matter void}{dark matter voids} also correspond to \glslink{galaxy void}{underdensities in the field traced by galaxies}, which is in agreement with the results obtained by \cite{Sutter2014DMGALAXYVOIDS} in simulations.

\subsubsection{Selection of voids}
\label{sec:Selection of voids}

The {\vide} pipeline identifies all \glslink{dark matter void}{dark matter voids} in the non-linear data-constrained realizations described in section \ref{sec:Generation of data-constrained reconstructions}. These live in boxes of 750 Mpc/$h$ side length with \gls{periodic boundary conditions}. In order to select physically meaningful \gls{dark matter void} candidates, we have to select a subsample of \glslink{void}{voids} which intersect the volume of the box actually constrained by \gls{SDSS} galaxies.

As described in chapter \ref{chap:BORGSDSS}, unobserved and observed regions in the inferred \glslink{final conditions}{final} \glslink{density field}{density fields} do not appear visually distinct, a consequence of the excellent performance of the \gls{2LPT} model implemented in {\borg} as a physical description of \gls{structure formation}. In addition, due to the \gls{non-local} \glslink{Lagrangian transport}{transport} of observational information between \glslink{initial conditions}{initial} and \gls{final conditions}, the region influenced by \gls{data} extends beyond the \glslink{survey geometry}{survey boundaries} and the \glslink{LSS}{large-scale structure} appears continuous there. The fact that \gls{data} constraints can radiate out of the survey volume has been known since the first \glslink{constrained simulation}{constrained reconstructions} of the mass distribution \citep{Bertschinger1987,Hoffman1991,vandeWeygaert1996}, where a \gls{power spectrum} \gls{prior} was assumed to \glslink{sampling}{sample} constrained \glslink{grf}{Gaussian random fields}. Here, as detailed in chapters \ref{chap:BORG} and \ref{chap:BORGSDSS}, constraints are propagated by the \gls{structure formation} model assumed in the inference process (\gls{2LPT}), which accounts not only for \glslink{two-point correlation function}{two-point statistics}, but for the full hierarchy of correlators, in its regime of validity. Therefore, \glslink{dark matter void}{dark matter voids} candidates intersecting the \glslink{survey geometry}{survey boundaries} can be considered as physical if a significant fraction of their volume is influenced by the \gls{data}.

The \gls{survey response operator} $R$ is a voxel-wise function representing simultaneously the \gls{survey geometry} (observed and unobserved regions) and the \gls{selection effects} in galaxy catalogs. Here, we kept for $R$ the average over the six \gls{luminosity} bins used in the {\borg} \gls{SDSS} run (for details see chapter \ref{chap:BORGSDSS}). For the purpose of this work, we keep all \gls{void} candidates whose center is in a region where $R$ is strictly positive. This region represents $7.9 \times 10^7$ cubic Mpc/$h$, around $18.7$\% of the full box. In each of the $11$ realizations used in this work, we kept around $166,000$ data-constrained \glslink{void}{voids} out of $886,000$ \glslink{void}{voids} in the entire box.

In figure \ref{fig:slice_voids}, the \gls{survey response operator} is shown in color from purple (totally unobserved region) to blue (region fully constrained by the \gls{data}). One can see the correct propagation of information operated by {\borg}, as \glslink{void}{voids} appear continuous at the \glslink{survey geometry}{survey boundaries}.

\subsection{Blackwell-Rao estimators for dark matter void realizations}
\label{sec:Blackwell-Rao estimators for dark matter void realizations}

A particular advantage of our Bayesian methodology is the ability to provide accurate \gls{uncertainty quantification} for derived \gls{dark matter void} properties. In particular, the Markovian \glslink{sample}{samples} described in chapter \ref{chap:BORGSDSS} permit us to employ a \gls{Blackwell-Rao estimator} to describe the \gls{posterior} distribution for inferred \glslink{dark matter void}{dark matter voids}. Specifically, we are interested in deriving the \gls{posterior} distribution  $\p(x \vert d)$ of a \gls{dark matter void} property $x$ given observations $d$. Using the realizations of the \gls{initial conditions} $\delta^\mathrm{i}$ and the \gls{dark matter void} realizations $V$ generated by the approach described in sections \ref{sec:Generation of data-constrained reconstructions} and \ref{sec:Void finding and processing}, we obtain

\begin{eqnarray}
\p(x \vert d) &=& \int \p(x \vert V) \, \p(V ,\delta^{\mathrm{i}} \vert d) \, \mathrm{d}V \, \mathrm{d} \delta^{\mathrm{i}} \nonumber \\
&=& \int \p(x \vert V) \, \p(V \vert \delta^{\mathrm{i}},d) \, \p(\delta^{\mathrm{i}} \vert d) \, \mathrm{d}V \, \mathrm{d} \delta^{\mathrm{i}} \nonumber \\
&=& \int \p(x \vert V) \, \updelta_\mathrm{D} (V - \tilde{V}( \delta^{\mathrm{i}} )) \, \p(\delta^{\mathrm{i}} \vert d) \, \mathrm{d}V \, \mathrm{d} \delta^{\mathrm{i}} \nonumber \\
&=& \int \p(x \vert \tilde{V}( \delta^{\mathrm{i}} )) \,\p(\delta^{\mathrm{i}} \vert d)\, \mathrm{d} \delta^{\mathrm{i}} \nonumber \\
&\approx& \frac{1}{N} \sum_k \p\left(x \vert \tilde{V}( \delta^{\mathrm{i}}_k )\right) \nonumber \\
&=& \frac{1}{N} \sum_k \p\left(x \vert V_k\right) \, ,
\label{eq:bwrao}
\end{eqnarray}
where we assumed the \gls{dark matter void} templates $V$ to be \glslink{conditional independence}{conditionally independent} of the \gls{data} $d$ given the \gls{initial conditions} $\delta^\mathrm{i}$, and to derive uniquely from the \glslink{initial conditions}{initial} \gls{density field} via the procedure described in sections \ref{sec:Generation of data-constrained reconstructions} and \ref{sec:Void finding and processing}, yielding $\p(V \vert \delta^{\mathrm{i}}, d) = \p(V \vert \delta^{\mathrm{i}}) = \updelta_\mathrm{D} (V - \tilde{V}( \delta^{\mathrm{i}}))$. We also exploited the fact that we have a \glslink{sample}{sampled} representation of the \gls{initial conditions} \gls{posterior} distribution $\p(\delta^{\mathrm{i}} \vert d) \approx 1/N \sum_k \updelta_\mathrm{D}(\delta^{\mathrm{i}}-\delta^{\mathrm{i}}_k )$, where $k$ labels one of the $N$ \glslink{sample}{samples}. The last line of equation \eqref{eq:bwrao} represents the \gls{Blackwell-Rao estimator} for \gls{void} property $x$ to be inferred from our \gls{dark matter void} catalogs $V_k$, providing thorough Bayesian means to \glslink{uncertainty quantification}{quantify uncertainties}. It consists of a mixture distribution over different realizations of \gls{dark matter void} templates.

The {\vide} pipeline provides estimated means and variances for derived quantities \(x\), allowing us to model the distributions $\p(x \vert V_k)$ as Gaussians with mean $x_k$ and variance $\sigma_k^2$, for respective \gls{dark matter void} templates. The final expression for the \gls{posterior} distribution of $x$ given the \gls{data} is therefore
\begin{equation}
\label{eq:bwrao_estimator}
\p(x \vert d) \approx \frac{1}{N} \sum_k \frac{1}{\sqrt{2 \pi \sigma_k^2}} \, \exp \left(-\frac{1}{2} \frac{\left( x - x_k \right)^2}{\sigma_k^2} \right).
\end{equation}
Even though we have access to \glslink{non-Gaussianity}{non-Gaussian} \gls{uncertainty quantification} via the \gls{posterior} distribution given in equation \eqref{eq:bwrao_estimator}, for the presentation in this chapter we will be content with estimating means and variances. 
The mean for $x$ given $d$ is 
\begin{equation}
\label{eq:BRmean}
\langle x \vert d \rangle \approx \frac{1}{N} \sum_k x_k , 
\end{equation}
and the variance is
\begin{equation}
\label{eq:BRvar}
\langle \left( x - \langle x \rangle \right)^2 \vert d \rangle \approx \frac{1}{N} \sum_k (x_k^2 + \sigma_k^2) - \langle x \vert d \rangle^2 .
\end{equation}

As described in section \ref{sec:Selection of voids}, we select \glslink{void}{voids} in the data-constrained regions of \glslink{reconstruction}{reconstructions} of the dark matter \gls{density field}. Since these regions are the same in different \glslink{reconstruction}{reconstructions}, the different \gls{void} catalogs describe the same region of the actual Universe. For this reason, while estimating uncertainties, it is not possible to simply use all the \glslink{void}{voids} in our catalogs as if they were \glslink{conditional independence}{independent}.\footnote{We generally recommend special care for proper statistical treatment while working with the data-constrained realizations of our \gls{dark matter void} catalog, especially if one wants to use \glslink{frequentist statistics}{frequentist} estimators of \gls{void} properties.} However, using an increasing number of \glslink{reconstruction}{reconstructions}, we shall still see a decrease of \gls{statistical uncertainty}. Indeed, from \eqref{eq:BRmean} and \eqref{eq:BRvar} it follows that
\begin{equation}
\langle \left( x - \langle x \rangle \right)^2 \vert d \rangle \leq \frac{1}{N} \sum_k \sigma_k^2 ,
\end{equation}
which means that the combination of different realizations will generally yield an improved estimator for any original statistics.

Note that this procedure is completely general and applies to any estimator provided by the {\vide} pipeline.

\subsection{Void catalogs for comparison of our results}
\label{sec:Galaxy void catalog and dark matter simulation}

In section \ref{sec:Properties of dark matter voids}, we will compare our results for \glslink{dark matter void}{dark matter voids} to state-of-the-art results for \glslink{galaxy void}{galaxy voids}. To do so, we use the catalogs of \cite{Sutter2012DR7CATALOGS} based on the \gls{SDSS} DR7 galaxies, publicly available at \href{http://www.cosmicvoids.net}{http://www.cosmicvoids.net}. In particular, we compare to the \glslink{void}{voids} found in the \texttt{bright1} and \texttt{dim1} volume-limited galaxy catalogs, for which the mean galaxy separations are 8 and 3 Mpc/$h$, respectively \cite[for details, see][]{Sutter2012DR7CATALOGS}. 

Assessment of our results for \glslink{dark matter void}{dark matter voids} in \gls{SDSS} data also require systematic comparison to \glslink{dark matter void}{dark matter voids} found in cosmological simulations. We ran 11 such unconstrained simulations with the same setup as described in section \ref{sec:Generation of data-constrained reconstructions} for the generation of data-constrained realizations. We started from \glslink{grf}{Gaussian random fields} with an \cite{Eisenstein1998,Eisenstein1999} \gls{power spectrum} using the fiducial \gls{cosmological parameters} of the {\borg} analysis ($\Omega_\mathrm{m}~=~0.272$, $\Omega_\mathrm{\Lambda}~=~0.728$, $\Omega_\mathrm{b}~=~0.045$, $h~=~0.702$, $\sigma_8~=~0.807$, $n_\mathrm{s}~=~0.961$, see chapter \ref{chap:BORGSDSS}). These \glslink{initial conditions}{initial} \glslink{density field}{density fields}, defined in a 750 Mpc/$h$ cubic box of $256^3$ voxels, are occupied by a Lagrangian lattice of $512^3$ \gls{dark matter particles}. These are evolved to $z=69$ with \gls{2LPT} and from $z=69$ to $z=0$ with \textsc{\gls{Gadget-2}}. As for constrained realizations, in our simulations we selected the \glslink{void}{voids} located inside the observed \gls{SDSS} volume (see section \ref{sec:Selection of voids}) and combined properties using \glslink{Blackwell-Rao estimator}{Blackwell-Rao estimators} (see section \ref{sec:Blackwell-Rao estimators for dark matter void realizations}).

\section{Properties of dark matter voids}
\label{sec:Properties of dark matter voids}

In this section, we describe the statistical properties of the \glslink{dark matter void}{dark matter voids} found in the data-constrained parts of our \glslink{reconstruction}{reconstructions} of the \gls{SDSS} volume. We focus on three key statistical summaries abundantly described in the literature: \glslink{number function}{number count}, \gls{ellipticity distribution} and radial \glslink{density profile}{density profiles}.

\subsection{Number function}
\label{sec:Number function}

\begin{figure} 
  \centering 
  {\includegraphics[width=0.55\columnwidth]{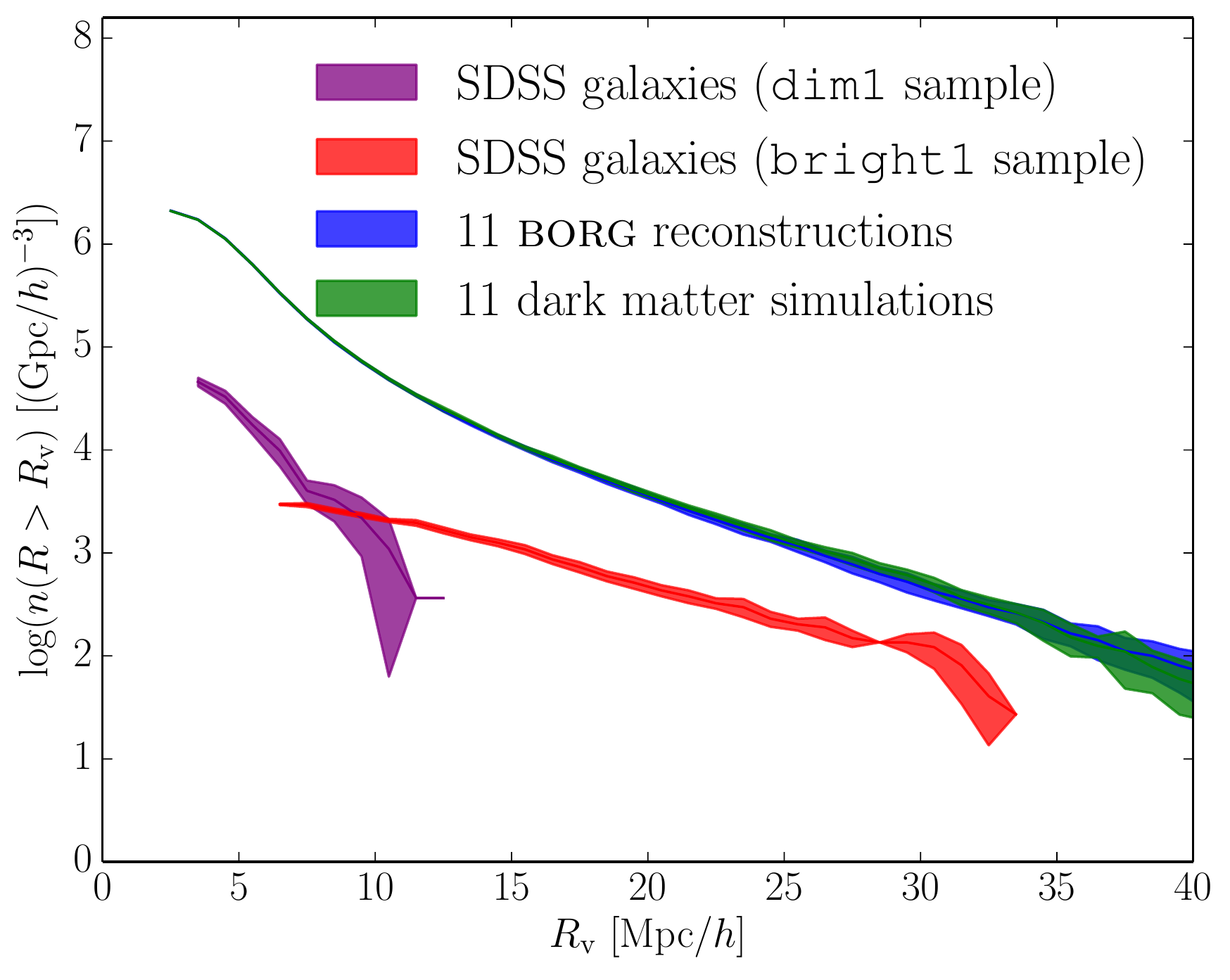}}
  \caption{
           Cumulative \gls{void} \glslink{number function}{number functions}. The results from 11 {\borg} \glslink{reconstruction}{reconstructions} (blue) are compared to a dark matter \glslink{N-body simulation}{$N$-body simulation} (green) and to the \glslink{galaxy void}{galaxy voids} directly found in two volume-limited sub-samples of the \gls{SDSS} DR7 (\texttt{dim1}, purple and \texttt{bright1}, red). The solid lines are the measured or predicted \glslink{number function}{number functions} and the shaded regions are the 2-$\sigma$ Poisson uncertainties. Fewer \glslink{void}{voids} are found in observations than in dark matter simulations, due to the \gls{sparsity} and \gls{bias} of tracers, as well as observational uncertainty coming from the \gls{survey geometry} and \gls{selection effects}. \glslink{number function}{Number functions} from {\borg} \glslink{reconstruction}{reconstructions} agree with simulations at all scales.}
\label{fig:numberfunc}
\end{figure}

The \gls{number function} of \glslink{void}{voids} provides a simple, easily accessible, and surprisingly sensitive cosmological probe. For example, the \gls{number function} has been shown to respond to coupled \gls{dark matter}-\gls{dark energy} \citep{Li2009,Sutter2015COUPLEDDEDM}, modified gravity \citep{Li2012,Clampitt2013}, and variations in fundamental \gls{cosmological parameters} \citep{Pisani2015}. While most studies of the \gls{number function} take place in \glslink{N-body simulation}{$N$-body simulations}, there has also been significant theoretical and analytical work, beginning with the excursion set formulation of~\cite{Sheth2004} and continuing through further enhancements to account for the complex nature of \gls{void} shapes \citep{Jennings2013}. As previous authors \citep{Muller2000,Sutter2012DR7CATALOGS,Sutter2014DR9CATALOGS,Nadathur2014a,Nadathur2014b} have noted, there tend to be fewer \glslink{void}{voids} in observations than in numerical simulations, especially for small \glslink{void}{voids}. This is due to the conjugate effects of \gls{sparsity} and \glslink{bias}{biasing} of tracers, which can modify the \gls{number function} in complex ways \citep{Furlanetto2006,Sutter2014SPARSITYBIAS,Sutter2014DR9CATALOGS}, as well as \glslink{survey geometry}{survey geometries} and \gls{selection effects}, which can non-trivially diminish the \gls{void} population. However, recently~\cite{Sutter2014DR9CATALOGS} showed a correspondence between observed and theoretical \glslink{number function}{number functions} once these factors are taken into account.

Figure \ref{fig:numberfunc} shows the cumulative \gls{void} \gls{number function} in {\borg} \glslink{reconstruction}{reconstructions} (blue) compared to dark matter simulations using the same setup (green) and to \glslink{galaxy void}{galaxy voids} in the \gls{SDSS} DR7 (red and purple). The confidence regions are 2-$\sigma$ Poisson uncertainties and the blue and green lines use \glslink{Blackwell-Rao estimator}{Blackwell-Rao estimators} to combine the results in 11 realizations.

We can immediately note the excellent agreement between simulations and \glslink{dark matter void}{dark matter voids} candidates in the \gls{SDSS} as found by our methodology. The two \gls{void} populations are almost indistinguishable at all scales, which demonstrates that the data-constrained \gls{number function} predicted by our methodology is exactly that of \glslink{dark matter void}{dark matter voids} in numerical simulations. In particular, this proves that our framework correctly permits to circumvent the effects of \gls{sparsity} and \glslink{bias}{biasing} of \gls{SDSS} galaxies on \gls{void} \glslink{number function}{number count}. Indeed, \glslink{dark matter void}{dark matter voids} in our \glslink{reconstruction}{reconstructions} are densely-sampled with the same number density as in simulations, $\bar{n}~=~0.318~(\mathrm{Mpc}/h)^{-3}$ ($512^3$ \glslink{dark matter particles}{particles} in $(750~\mathrm{Mpc}/h)^3$) compared to $\bar{n}~\approx~10^{-3}~(\mathrm{Mpc}/h)^{-3}$ for \gls{SDSS} galaxies \citep{Sutter2012DR7CATALOGS}. Furthermore, any incorrect treatment of galaxy \gls{bias} by the {\borg} algorithm would result in a residual \gls{bias} in our \glslink{reconstruction}{reconstructions} that would yield an erroneous \gls{void} \gls{number function} as compared to simulations \citep{Sutter2014SPARSITYBIAS}. The absence of any such feature confirms that galaxy \gls{bias} is correctly accounted for in our analysis and further validates the framework described in chapter \ref{chap:BORGSDSS} \citep{Jasche2015BORGSDSS}.

Additionally, due to the high density of tracer \glslink{dark matter particles}{particles}, we find at least around one order of magnitude more \glslink{void}{voids} at all scales than the \glslink{void}{voids} directly traced by the \gls{SDSS} galaxies, which \glslink{sampling}{sample} the underlying mass distribution only \glslink{sparsity}{sparsely}. This results in a drastic reduction of \gls{statistical uncertainty} in \gls{void} catalogs, as we demonstrate in sections \ref{sec:Ellipticity distribution} and \ref{sec:Radial density profiles}.

\subsection{Ellipticity distribution}
\label{sec:Ellipticity distribution}

\glslink{ellipticity distribution}{}The shape distribution of \glslink{void}{voids} is complementary to overdense probes of the dark matter \gls{density field} such as galaxy \glslink{cluster}{clusters}. Indeed, as matter collapses to form galaxies, \glslink{void}{voids} expand and can do so aspherically. While \citet{Icke1984} argued that \glslink{void}{voids} are expected to become more spherical as they expand, \citet{Platen2008} found that the shape distribution of \glslink{void}{voids} remains complex at late times and showed that the aspherical expansion of \glslink{void}{voids} is strongly linked to the external \glslink{tidal field}{tidal influence}.\footnote{\glslink{tidal effects}{Tidal effects} are taken into account in our analysis since {{\borg}} models \gls{gravitational evolution} up to \glslink{2LPT}{second order in Lagrangian perturbation theory}.} Therefore, the shapes of empty regions generally change during cosmic evolution and retain information on their \gls{formation history}. In particular, the \gls{void} shape distribution potentially serves as a powerful tracer of the \gls{equation of state} of \gls{dark energy} \citep{Lee2006,Park2007,Biswas2010,Lavaux2012,Bos2012}. In addition, the mean stretch of \glslink{void}{voids} along the line of sight may be used for an application of the \glslink{Alcock-Paczynski effect}{Alcock-Paczynski test} \citep{Alcock1979,Ryden1995,Lavaux2012,Sutter2012APSDSS,Sutter2014APSDSS,Hamaus2014c}.

For these applications, it is of crucial importance for the \gls{void} catalog to be unaffected by \glslink{systematic uncertainty}{systematics} due to baryonic physics. Furthermore, as pointed out by \cite{Bos2012}, in \glslink{sparsity}{sparse} populations such as galaxies it is very difficult to statistically separate {\LCDM} from alternative cosmologies using \gls{void} shapes. As we now show, our framework allows to access \gls{void} shapes at the level of the dark matter distribution, deeper than with the galaxies, and to reduce the \gls{statistical uncertainty} due to their \gls{sparsity}. Note that all the \gls{phase} information and spatial organization of the \gls{LSS} is unaffected by our \gls{prior} assumptions, which generally affect the density amplitudes via the cosmological \gls{power spectrum}. The geometry of \glslink{void}{voids} discussed here is therefore strongly constrained by the observations.

\begin{figure} 
  \centering 
  {\includegraphics[width=0.55\columnwidth]{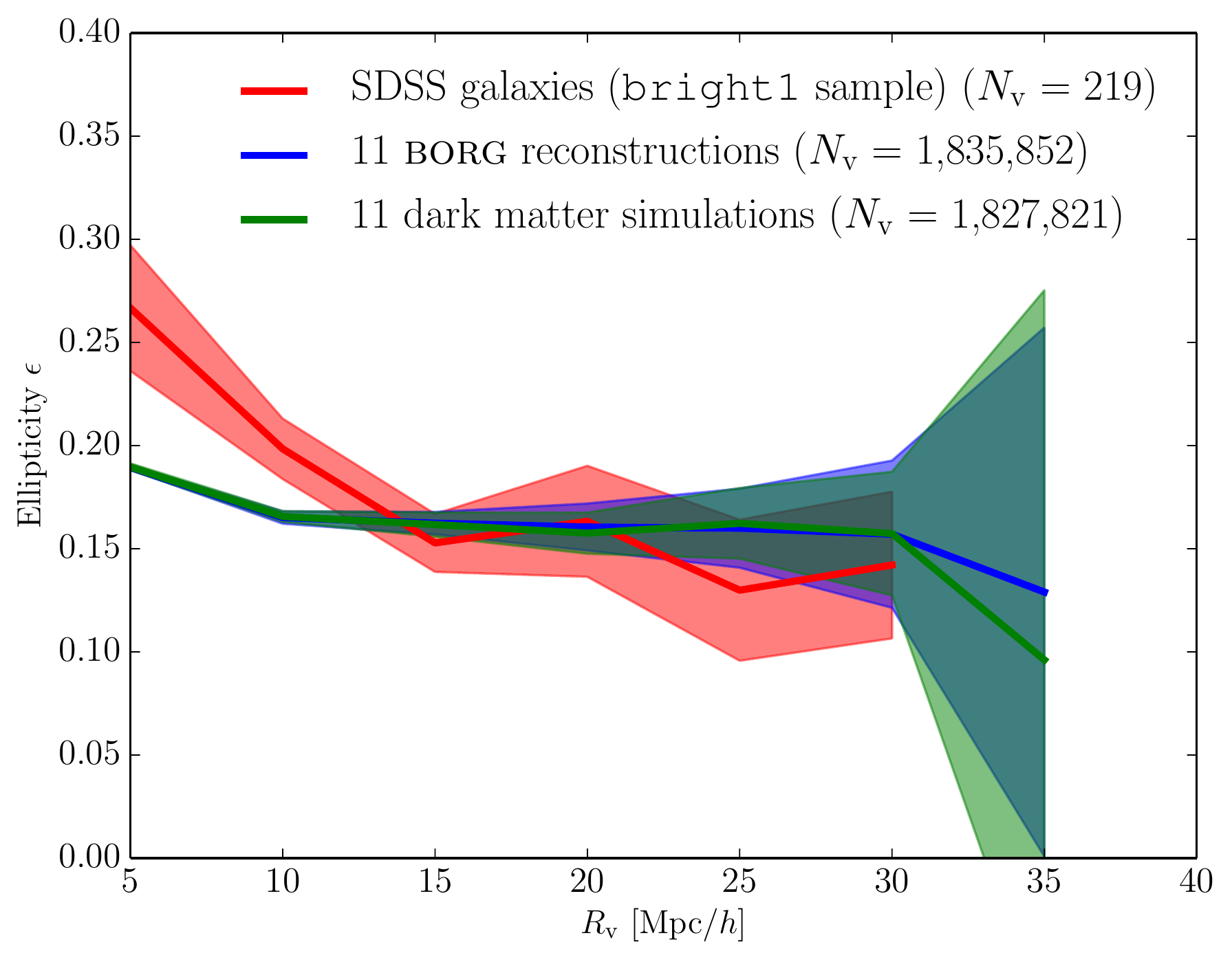}}
  \caption{
           \glslink{ellipticity distribution}{Distribution of ellipticities} $\epsilon$ versus effective radii of \glslink{void}{voids}. The solid line shows the mean, and the shaded region is the 2-$\sigma$ confidence region estimated from the standard error on the mean in each radial bin. Small \glslink{galaxy void}{galaxy voids} are found more elliptical than \glslink{dark matter void}{dark matter voids} because of important Poisson fluctuations below the mean galaxy separation ($8~\mathrm{Mpc}/h$). Ellipticities of \glslink{dark matter void}{dark matter voids} in {\borg} \glslink{reconstruction}{reconstructions} and simulations agree at all scales, and the \gls{statistical uncertainty} in their determination is drastically reduced in comparison to \gls{galaxy void} catalogs.
           }
\label{fig:ellipvsr}
\end{figure}

We simplify the discussion by focusing on the ellipticity, computed by the {\vide} toolkit using the eigenvalues of the inertia tensor \citep[for details, see section \ref{sec:apx-ellipticity} and][]{Sutter2015VIDE}. Figure \ref{fig:ellipvsr} shows the mean ellipticity and the standard error on the mean (i.e. $\sigma/\sqrt{N_\mathrm{v}}$, where $\sigma$ is the standard deviation and $N_\mathrm{v}$ is the number of voids) as a function of \gls{void} effective radius. The red line represents the \glslink{galaxy void}{galaxy voids} directly found in the \gls{SDSS} data, the blue line the \glslink{dark matter void}{dark matter voids} of our data-constrained catalogs, and the green line the \glslink{void}{voids} found in dark matter simulations prepared with the same setup. The blue and green lines use \glslink{Blackwell-Rao estimator}{Blackwell-Rao estimators} to combine the results of 11 realizations. For the interpretation of the ellipticity of small \glslink{galaxy void}{galaxy voids}, it is useful to recall that the mean galaxy separation in the \texttt{bright1} sample is 8~Mpc/$h$, meaning that Poisson fluctuations will be of importance for \glslink{void}{voids} whose effective radius is below this scale. 

The comparison between \glslink{dark matter void}{dark matter voids} of {\borg} \glslink{reconstruction}{reconstructions} and of simulations shows that the predicted ellipticities fully agree with the expectations at all scales. This further demonstrates that our candidates qualify as \glslink{dark matter void}{dark matter voids} as defined by numerical simulations, in particular alleviating the galaxy \gls{bias} problem. Furthermore, as already noted, our inference framework produces many more \glslink{void}{voids} than \glslink{sparsity}{sparse} galaxy catalogs, especially at small scales. This results in a radical reduction of \gls{statistical uncertainty} in the ellipticity prediction for small \glslink{dark matter void}{dark matter voids} as compared to \glslink{galaxy void}{galaxy voids}, as can be observed in figure \ref{fig:ellipvsr}.

\subsection{Radial density profiles}
\label{sec:Radial density profiles}

\begin{figure*} 
  \centering 
  {\includegraphics[width=0.49\columnwidth]{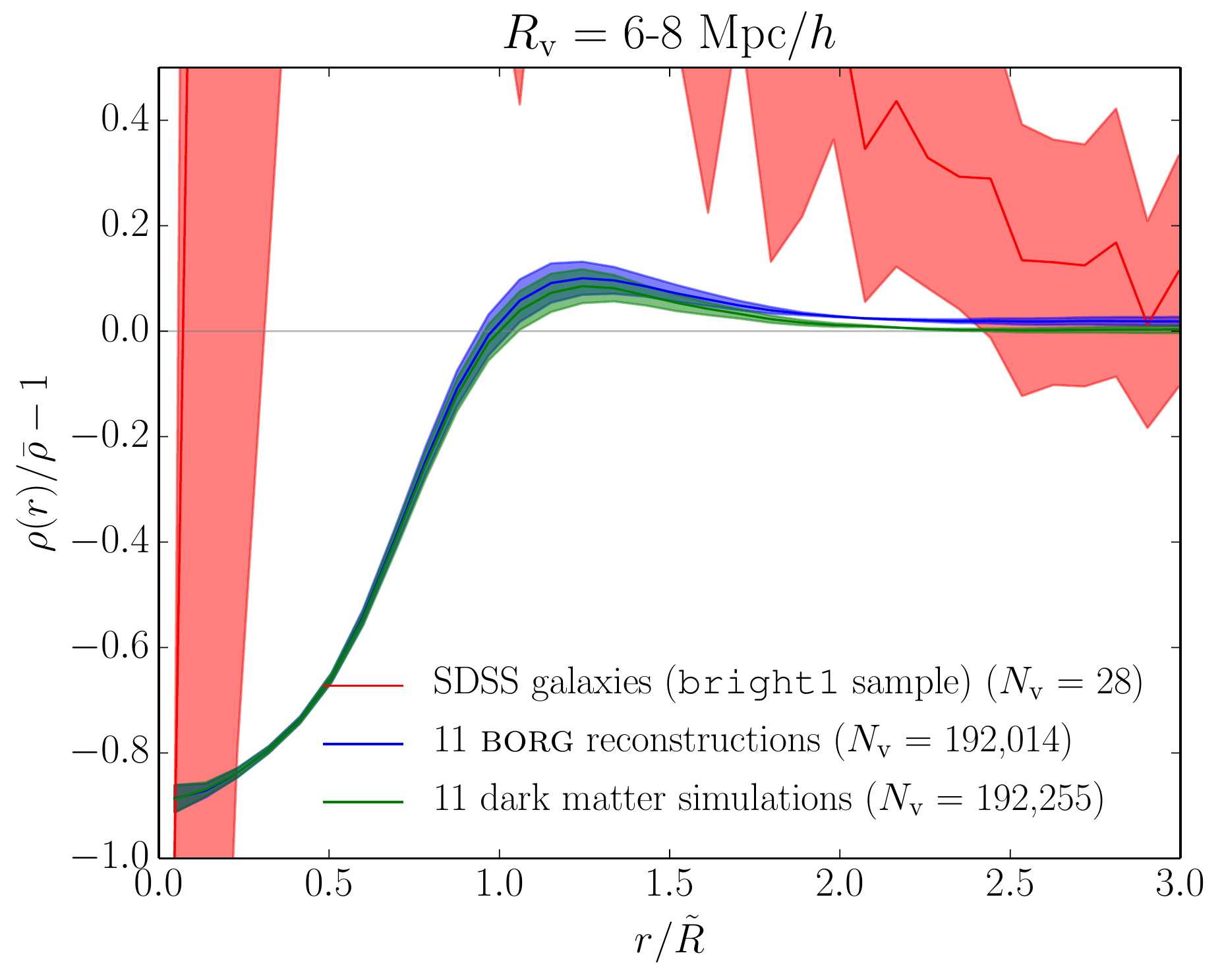}}
  {\includegraphics[width=0.49\columnwidth]{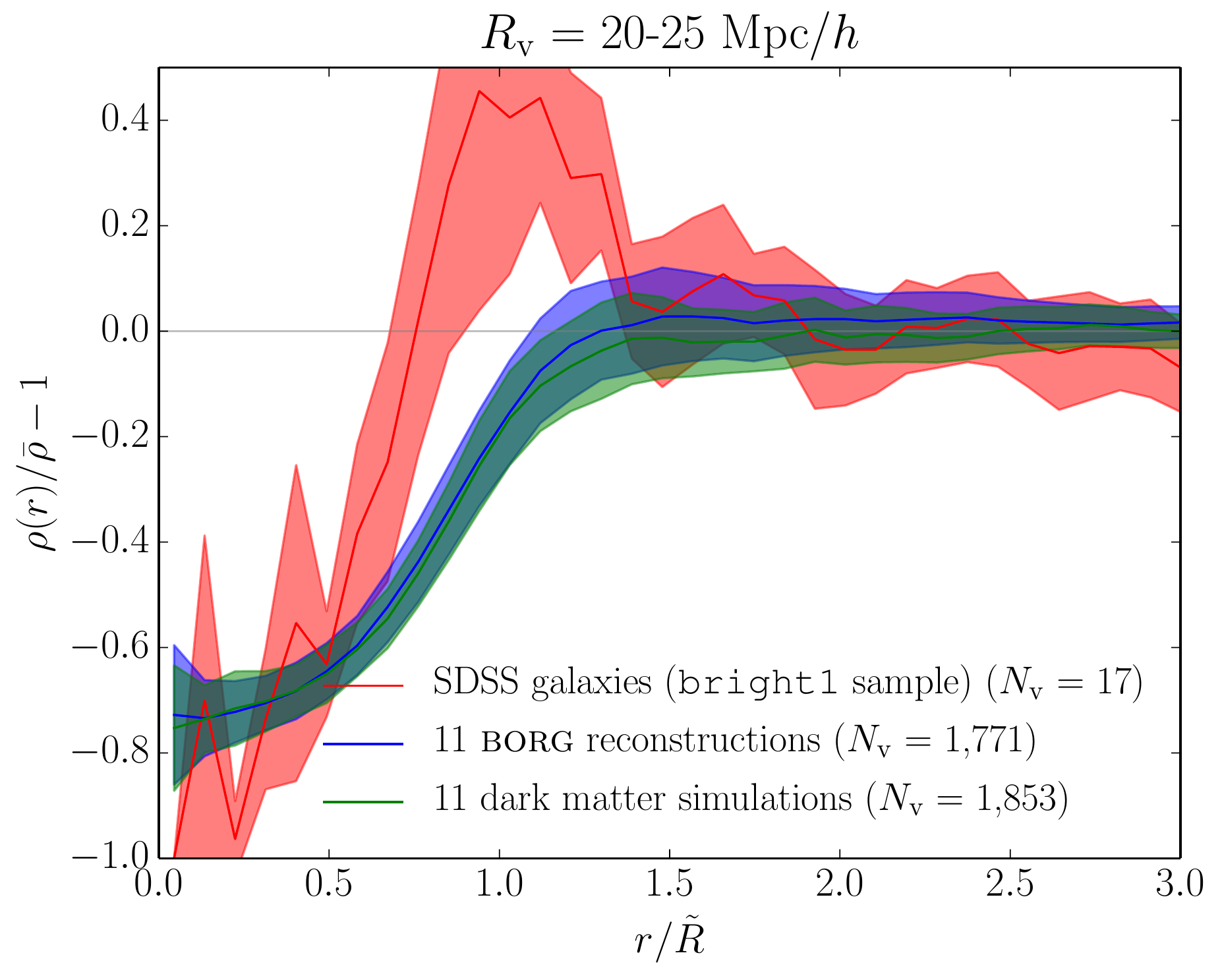}}
  \caption{
           One-dimensional radial \glslink{density profile}{density profiles} of stacked \glslink{void}{voids}, for \glslink{void}{voids} of effective radius in the range 6-8~Mpc/$h$ (left) and 20-25~Mpc/$h$ (right). $\tilde{R}$ corresponds to the median \gls{void} size in the stack. The solid line shows the mean, and the shaded region is the 2-$\sigma$ confidence region estimated from the standard error on the mean in each radial bin. \glslink{galaxy void}{Galaxy void} profiles are strongly \gls{noise}-dominated, contrary to \glslink{dark matter void}{dark matter voids}. The heights of \glslink{compensation}{compensation ridges} are different because \glslink{dark matter void}{dark matter voids} are identified in a higher density of tracers, which induces a deeper \gls{void hierarchy}.
           }
\label{fig:1dprofile}
\end{figure*}

The radial \gls{density profile} of \glslink{void}{voids}, reconstructed in real space using techniques such as those described in \cite{Pisani2014}, can be used to test \gls{general relativity} and constrain dynamical \gls{dark energy} models \citep{Shoji2012,Spolyar2013}. More generally, it shows a self-similar structure \citep{Colberg2005,Ricciardelli2014,Hamaus2014b,Nadathur2014b}, and characterizes the \gls{LSS} in a fundamental way \citep{vandeWeygaert1993}. All results presented in this section assume that \gls{dark matter particles} in {\borg} \glslink{reconstruction}{reconstructions} and in simulations live in physical space. The {\borg} algorithm automatically mitigates \gls{redshift-space distortions} by treating anisotropic features in the \gls{data} as \gls{noise} \citep{Jasche2015BORGSDSS}. Furthermore, as pointed out by \cite{Padilla2005}, \gls{redshift-space distortions} have very mild effects on \gls{void} \glslink{density profile}{density profiles}. We therefore expect our results to be robust under the transformation from real to redshift space.

Using {\vide}, we construct the one-dimensional radial \glslink{density profile}{density profiles} of stacked \glslink{void}{voids} for various \gls{void} sizes. Note that we do not apply any rescaling to the \gls{void} sizes as we stack. Figure \ref{fig:1dprofile} shows two such profiles, for \glslink{void}{voids} of effective radius in the range 6-8~Mpc/$h$ (left panel) and 20-25~Mpc/$h$ (right panel). The solid lines show the mean and the shaded regions are the 2-$\sigma$ confidence regions estimated from the standard error on the mean, using \glslink{Blackwell-Rao estimator}{Blackwell-Rao estimators} for {\borg} \glslink{reconstruction}{reconstructions} and dark matter simulations. At the level of statistical error in our results, our \glslink{reconstruction}{reconstructions} show radial \glslink{density profile}{density profiles} in agreement with simulations at all radii and for all \gls{void} sizes. Note that, if small \glslink{void}{voids} essentially reflect the \gls{prior} information used for the {\borg} analysis and $N$-body \glslink{non-linear filtering}{filtering}, bigger \glslink{void}{voids} are strongly constrained by the \gls{data}. The profile shapes agree nicely with the results of \cite{Sutter2014SPARSITYBIAS,Hamaus2014b} from dark matter simulations: higher ridges and lower central densities in smaller \glslink{void}{voids}. Specifically, our \glslink{reconstruction}{reconstructions} exhibit the same behaviour as simulations, with a transition scale between small overcompensated to large undercompensated \glslink{void}{voids} \citep{Ceccarelli2013,Paz2013,Cai2014,Hamaus2014a}.

In contrast, \gls{galaxy void} profiles at the same scales are strongly \gls{noise}-dominated. This is due to the \gls{sparsity} and \glslink{bias}{biasing} of galaxies, which are alleviated with the present approach. In particular, our methodology performs a meaningful compromise between \gls{data} and \gls{prior} information, which predicts corrected shapes and smaller variance for the profiles of \glslink{dark matter void}{dark matter voids} as compared to \glslink{galaxy void}{galaxy voids}. Note that at the same physical scales (e.g. $20~\mathrm{Mpc}/h$), \glslink{galaxy void}{galaxy voids} and \glslink{dark matter void}{dark matter voids} have different ridge heights. This is because a deeper \gls{void hierarchy} emerges in higher tracer \gls{sampling} densities, affecting the \gls{compensation} of \glslink{void}{voids} at a given size \citep{Sutter2014SPARSITYBIAS}. 

In addition to the location of all \gls{dark matter particles}, our inference framework also provides their individual \glslink{velocity field}{velocity vectors}, which are predicted from gravitational clustering. While the direct measurement of individual galaxy velocities is very difficult in most observations, our \gls{reconstruction} technique readily allows to infer the \gls{velocity profile} of \glslink{void}{voids}. This allows to make a connection between a static (based on the \glslink{density profile}{density profiles}) and a dynamic (based on the \glslink{velocity profile}{velocity profiles)} characterization of \glslink{void}{voids}. In particular, as mentioned before, our results agree with the existence of a transition scale between two regimes: undercompensated, inflowing \glslink{void}{voids} and overcompensated, outflowing \glslink{void}{voids}, respectively known as \gls{void-in-cloud} and \gls{void-in-void} in the terminology originally introduced by \cite{Sheth2004}.

\section{Summary and conclusions}
\label{sec:Summary and conclusions}

This chapter is an example of the rich variety of scientific results that have been produced by the recent application \citep{Jasche2015BORGSDSS} of the Bayesian inference framework {\borg} \citep{Jasche2013BORG} to the \glslink{SDSS}{Sloan Digital Sky Survey} main sample galaxies. We proposed a method designed to find \gls{dark matter void} candidates in the Sloan volume. In doing so, we proved that physical \glslink{void}{voids} in the dark matter distribution can be correctly identified by the \textit{ab initio} analysis of \glslink{galaxy survey}{galaxy surveys}. 

Our method relies on the physical inference of the \gls{initial conditions} for the entire \gls{LSS} \citep{Jasche2013BORG,Jasche2015BORGSDSS}. Starting from these, we generated realizations of the \gls{LSS} using a \glslink{full gravity}{fully non-linear cosmological code}. In this fashion, as described in section \ref{sec:Generation of data-constrained reconstructions}, we obtained a set of \glslink{constrained simulation}{data-constrained reconstructions} of the \glslink{final conditions}{present-day} dark matter distribution. The use of \glslink{full gravity}{fully non-linear dynamics} as a \glslink{non-linear filtering}{filter} allowed us to extrapolate the predictions of {\borg} to the unconstrained \glslink{non-linear regime}{non-linear regimes} and to obtain an accurate description of structures. As described in section \ref{sec:Void finding and processing}, we identified the \glslink{void}{voids} in these \glslink{reconstruction}{reconstructions} using the \gls{void} finder of the {\vide} pipeline \citep{Sutter2015VIDE} and applied an additional selection criterion to limit the final catalogs of \glslink{dark matter void}{dark matter voids} candidates to regions covered by observations. To check that these candidates qualify for physical \glslink{void}{voids}, we analyzed our catalogs in terms of a set of statistical diagnostics. We focused on three key \gls{void} statistics, well understood both in \gls{data} and in simulations, provided by the {\vide} toolkit: \gls{number function}, \gls{ellipticity distribution} and radial \gls{density profile}. As mentioned in section \ref{sec:Galaxy void catalog and dark matter simulation}, for comparison of our results, we used the \gls{void} catalog of \cite{Sutter2012DR7CATALOGS}, directly based on \gls{SDSS} main sample galaxies, and unconstrained dark matter simulations produced with the same setup as our \glslink{reconstruction}{reconstructions}.

For \glslink{uncertainty quantification}{quantifying the uncertainty}, we adopted the same Bayesian philosophy as in the \glslink{large-scale structure inference}{LSS inference} framework: several \gls{void} catalogs are produced, based on different \glslink{sample}{samples} of the {\borg} \gls{posterior} \glslink{pdf}{probability distribution function}. Each of them represents a realization of the actual \glslink{dark matter void}{dark matter voids} in the Sloan volume, and the variation between these catalogs \glslink{uncertainty quantification}{quantifies the remaining uncertainties} of various sources (in particular, \gls{survey geometry} and \gls{selection effects}, see chapter \ref{chap:BORGSDSS} for a complete discussion). In order to produce a statistically meaningful combination of our different \gls{dark matter void} catalogs, in section \ref{sec:Blackwell-Rao estimators for dark matter void realizations}, we introduced \glslink{Blackwell-Rao estimator}{Blackwell-Rao estimators}. We showed that the combination of different realizations generally yields an improved estimator for any original \gls{void} statistic.

For all usual \gls{void} statistics (\gls{number function} in section \ref{sec:Number function}, \gls{ellipticity distribution} in section \ref{sec:Ellipticity distribution} and radial \glslink{density profile}{density profiles} in section \ref{sec:Radial density profiles}), we found remarkably good agreement between predictions for \glslink{dark matter void}{dark matter voids} in our \glslink{reconstruction}{reconstructions} and expectations from numerical simulations. This validates our inference framework and qualifies the candidates to physically reasonable \glslink{dark matter void}{dark matter voids}, probing a level deeper in the mass distribution \glslink{void hierarchy}{hierarchy} than galaxies. Further, since \gls{sparsity} and \glslink{bias}{biasing} of tracers modify these statistics \citep{Sutter2014SPARSITYBIAS}, it means that these effects have been correctly accounted for in our analysis. Indeed, in chapter \ref{chap:BORGSDSS} we showed that {\borg} accurately accounts for \gls{luminosity}-dependent galaxy \gls{bias} and performs automatic calibration of the \glslink{noise parameter}{noise level} within a fully Bayesian approach. Building on the detailed representation of \glslink{initial conditions}{initial} \glslink{density field}{density fields}, our \glslink{reconstruction}{reconstructions} possess a high density of tracers, $\bar{n} = 0.318~(\mathrm{Mpc}/h)^{-3}$, contrary to galaxies, which \glslink{sampling}{sample} the underlying mass distribution only \glslink{sparsity}{sparsely} ($\bar{n} \approx 10^{-3}~(\mathrm{Mpc}/h)^{-3}$). 

Another important aspect of our methodology is that the use of full-scale physical \glslink{density field}{density fields} instead of a scarce population of galaxies allows to adjust the density of tracers to reduce \gls{shot noise} at the desired level. In our analysis, we found at least one order of magnitude more \glslink{void}{voids} at all scales. This yields a radical reduction of \gls{statistical uncertainty} in \gls{noise}-dominated \gls{void} catalogs, as we have shown for \glslink{ellipticity distribution}{ellipticity distributions} and \glslink{density profile}{density profiles}.

In summary, our methodology permits to alleviate the issues due to the conjugate and intricate effects of \gls{sparsity} and \glslink{bias}{biasing} on \gls{galaxy void} catalogs, to drastically reduce \gls{statistical uncertainty} in \gls{void} statistics, and yields new catalogs of \glslink{dark matter void}{dark matter voids} for a variety of cosmological applications. For example, these enhanced \gls{data} sets can be used for \gls{cross-correlation} with other cosmological probes such as the \glslink{CMB}{cosmic microwave background}, to study the \gls{integrated Sachs-Wolfe effect}, or \glslink{weak gravitational lensing}{gravitational lensing} shear maps. Along with the ensemble mean \gls{density field} and corresponding standard deviations inferred by {\borg}, published as supplementary material with \citet{Jasche2015BORGSDSS}, we believe that the catalogs of our \glslink{dark matter void}{dark matter voids} candidates in the Sloan volume can be of interest to the scientific community. For this reason, all the \gls{void} catalogs used to produce the results described in this chapter have been made publicly available at \href{http://www.cosmicvoids.net}{http://www.cosmicvoids.net}, along with the paper corresponding to this chapter \citep{Leclercq2015DMVOIDS}.

Our \glslink{Bayesian statistics}{Bayesian methodology}, based on \glslink{large-scale structure inference}{inference} with {\borg} and subsequent \gls{non-linear filtering} of the results, assumes some \gls{prior} information, namely the standard {\LCDM} cosmological framework and initially \glslink{grf}{Gaussian} density fluctuations. We want to emphasize that any analysis using our constrained catalogs will be \glslink{bias}{biased} toward the confirmation of these assumptions. Therefore, this method will be only applicable if the \gls{data} contain sufficient support for the presence of non-standard cosmology to overrule the preference for {\LCDM} and Gaussianity in our \gls{prior}. However, any significant departure from standard cosmology means that the \gls{prior} has been overridden by the \glslink{large-scale structure likelihood}{likelihood} and that such deviations really are supported by the \gls{data}.

While the recommendations of \cite{Sutter2014SPARSITYBIAS} for quantifying and disentangling the effects of \gls{sparsity} and \glslink{bias}{biasing} depend on specific \glslink{galaxy survey}{survey} details, our inference framework is extremely general. It allows to translate \gls{void} statistics from current and future \glslink{galaxy survey}{galaxy surveys} to theory-like, high-resolution \gls{dark matter} predictions. In this fashion, it is straightforward to decide if any particular \gls{void} statistic can be directly informative about cosmology. These results indicate a new promising path towards effective and precise \gls{void} cosmology at the level of the \gls{dark matter} field.

%% file: Chapter9/Chapter9Content.tex
\chapter{Bayesian analysis of the dynamic cosmic web in the SDSS galaxy survey}
\label{chap:ts}

\defcitealias{Shannon1948}{Claude}
\begin{flushright}
\begin{minipage}[c]{0.6\textwidth}
\rule{\columnwidth}{0.4pt}

``I just wonder how things were put together.''\\
--- \citetalias{Shannon1948} \citeauthor{Shannon1948}

\vspace{-5pt}\rule{\columnwidth}{0.4pt}
\end{minipage}
\end{flushright}

\minitoc

\abstract{\section*{Abstract}

Recent application of the Bayesian algorithm {\borg} to the \glslink{SDSS}{Sloan Digital Sky Survey} main sample galaxies resulted in the physical \glslink{large-scale structure inference}{inference} of the \gls{formation history} of the observed \glslink{LSS}{large-scale structure} from its origin to the present epoch. In this work, we use these inferences as inputs for a detailed probabilistic \glslink{structure type}{cosmic web-type} analysis. To do so, we generate a large set of data-\glslink{constrained simulation}{constrained realizations} of the \glslink{LSS}{large-scale structure} using a fast, \glslink{full gravity}{fully non-linear gravitational model}. We then perform a dynamic \glslink{cosmic web classification}{classification of the cosmic web} into four distinct components (\glslink{void}{voids}, \glslink{sheet}{sheets}, \glslink{filament}{filaments}, and \glslink{cluster}{clusters}) on the basis of the \gls{tidal field}. Our \glslink{large-scale structure inference}{inference} framework automatically and self-consistently propagates typical observational uncertainties to \glslink{cosmic web classification}{web-type classification}. As a result, this study produces accurate cosmographic \glslink{cosmic web classification}{classification of large-scale structure elements} in the \gls{SDSS} volume. By also providing the \glslink{formation history}{history} of these structure maps, the approach allows an analysis of the origin and growth of the early traces of the \gls{cosmic web} present in the \glslink{initial conditions}{initial} \gls{density field} and of the evolution of global quantities such as the \glslink{VFF}{volume} and \glslink{MFF}{mass filling fractions} of different structures. For the problem of \glslink{cosmic web classification}{web-type classification}, the results described in this chapter constitute the first connection between theory and observations at \glslink{non-linear regime}{non-linear} scales including a physical model of \gls{structure formation} and the demonstrated capability of \gls{uncertainty quantification}. A connection between cosmology and \gls{information theory} using real data also naturally emerges from our probabilistic approach. Our results constitute quantitative \gls{chrono-cosmography} of the complex web-like patterns underlying the observed galaxy distribution.}

\draw{This chapter is adapted from its corresponding publication, \citet{Leclercq2015ST}.}

\section{Introduction}
\label{sec:Tweb-introduction}

The large-scale distribution of matter in the Universe is known to form intricate, complex patterns traced by galaxies. The existence of this \glslink{LSS}{large-scale structure}, also known as the \textit{\gls{cosmic web}} \citep{Bond1996}, has been suggested by early observational projects aiming at mapping the Universe \citep{Gregory1978,Kirshner1981,deLapparent1986,Geller1989,Shectman1996}, and has been extensively analyzed since then by \glslink{galaxy survey}{massive surveys} such as the 2dFGRS \citep{Colless2003}, the \gls{SDSS} \citep[e.g.][]{Gott2005} or the 2MASS \glslink{galaxy survey}{redshift survey} \citep{Huchra2012}. The \gls{cosmic web} is usually segmented into different \glslink{structure type}{elements}: \glslink{void}{voids}, \glslink{sheet}{sheets}, \glslink{filament}{filaments}, and \glslink{cluster}{clusters}. At late times, low-density regions (\glslink{void}{voids}) occupy most of the volume of the Universe. They are surrounded by walls (or \glslink{sheet}{sheets}) from which departs a network of denser \glslink{filament}{filaments}. At the intersection of \glslink{filament}{filaments} lie the densest clumps of matter (\glslink{cluster}{clusters}). Dynamically, matter tends to flow out of the \glslink{void}{voids} to their \gls{compensation} walls, transits through \glslink{filament}{filaments} and finally accretes in the densest \glslink{halo}{halos}. 

Describing the \gls{cosmic web} morphology is an involved task due to the intrinsic complexity of individual structures, but also to their connectivity and the \glslink{void hierarchy}{hierarchical nature} of their global organization. First approaches \citep[e.g.][]{Barrow1985,Gott1986,Babul1992,Mecke1994,Sahni1998} often characterized the \gls{LSS} with a set of global and statistical diagnostics, without providing a way to locally identify \glslink{structure type}{cosmic web elements}. In the last decade, a variety of methods has been developed for segmenting the \gls{LSS} into its \glslink{structure type}{components} and applied to numerical simulations and observations. Among them, some focus on investigating one component at a time, in particular \glslink{filament}{filaments} (e.g. the Candy model -- \citealp{Stoica2005,Stoica2007,Stoica2010}, the skeleton analysis -- \citealp{Novikov2006,Sousbie2008}, and DisPerSE -- \citealp{Sousbie2011a,Sousbie2011b}) or \glslink{void}{voids} (e.g. \citealp{Plionis2002,Colberg2005,Shandarin2006,Platen2007,Neyrinck2008,Sutter2015VIDE,Elyiv2015}, see also \citealp{Colberg2008} for a \gls{void} finder comparison project). Unfortunately, this approach does not allow an analysis of the connections between \glslink{structure type}{cosmic web components}, identified in the same framework. Another important class of web classifiers dissects \glslink{cluster}{clusters}, \glslink{filament}{filaments}, walls, and \glslink{void}{voids} at the same time. In particular, several recent studies deserve special attention due to their methodological richness. The ``\gls{T-web}'' and ``\gls{V-web}'' \citep{Hahn2007a,Forero-Romero2009,Hoffman2012} characterize the \gls{cosmic web} based on the \glslink{tidal field}{tidal} and \glslink{velocity shear field}{velocity shear fields}. {\diva} \citep{Lavaux2010} rather uses the shear of the \glslink{displacement field}{Lagrangian displacement field}. {\origami} \citep{Falck2012} identifies single and multi-stream regions in the full six-dimensional \glslink{phase space}{phase-space} information \citep{Abel2012,Neyrinck2012,Shandarin2012}. The Multiscale Morphology Filter \citep{Aragon-Calvo2007} and later refinements \textsc{nexus}/\textsc{nexus+} \citep{Cautun2013} follow a multiscale approach which probes the \glslink{void hierarchy}{hierarchical nature} of the \gls{cosmic web}.

In the standard theoretical picture, the \gls{cosmic web} arises from the anisotropic nature of gravitational collapse, which drives the formation of structure in the Universe from \glslink{initial conditions}{primordial fluctuations} \citep{Peebles1980}. The capital importance of the large-scale \gls{tidal field} in the formation and evolution of the \gls{cosmic web} was first pointed out in the seminal work of \citet{Zeldovich1970}. In the \citeauthor{Zeldovich1970} approximation, the \glslink{final conditions}{late-time} morphology of structures is linked to the \glslink{eigenvalue}{eigenvalues} of the \gls{tidal tensor} in the \gls{initial conditions}. Gravitational collapse amplifies any anisotropy present in the primordial \gls{density field} to give rise to highly asymmetrical structures. This picture explains the segmented nature of the \gls{LSS}, but not its connectivity. The \gls{cosmic web} theory of \cite{Bond1996} asserted the deep connection between the \gls{tidal field} around rare density peaks in the \glslink{initial conditions}{initial fluctuations} and the final web pattern, in particular the \glslink{filament}{filamentary} \gls{cluster}-\gls{cluster} bridges. More generally, the shaping of the \gls{cosmic web} through gravitational clustering is essentially a deterministic process described by \gls{Einstein's equations} and the main source of stochasticity in the problem enters in the generation of \gls{initial conditions}, which are known, from \glslink{inflation}{inflationary theory}, to resemble a \glslink{grf}{Gaussian random field} to very high accuracy \citep{Guth1982,Hawking1982,Bardeen1983}. For these reasons, considerable effort has been devoted to a theoretical understanding of the \gls{LSS} in terms of perturbation theory in the \glslink{EPT}{Eulerian} and \glslink{LPT}{Lagrangian} frames \citep[for a review, see][]{Bernardeau2002}. While this approach offers important analytical insights, it only permits to describe \gls{structure formation} in the \glslink{linear regime}{linear} and \glslink{mildly non-linear regime}{mildly non-linear regimes} and it is usually limited to the first \glslink{high-order correlation function}{few correlation functions} of the \gls{density field}. The complete description of the connection between primordial fluctuations and the \glslink{final conditions}{late-time} \gls{LSS}, including a full \glslink{phase space}{phase-space} treatment and the entire \glslink{high-order correlation function}{hierarchy of correlators}, has to rely on a numerical treatment through \glslink{N-body simulation}{$N$-body simulations}. The characterization of \glslink{structure type}{cosmic web environments} in the \gls{non-linear regime} and the description of their time evolution has only been treated recently, following the application of web classifiers to state-of-the-art simulations. In particular, \citet{Hahn2007a,Aragon-Calvo2010} presented a \gls{local} description of \glslink{structure type}{structure types} in high-resolution cosmological simulations. \citet{Hahn2007b,Bond2010,Cautun2014} analyzed the time evolution of the \gls{cosmic web} in terms of the \glslink{MFF}{mass} and \glslink{VFF}{volume} content of \glslink{structure type}{web-type components}, their density distribution, and a set of new analysis tools especially designed for particular \glslink{structure type}{elements}.

To the best of our knowledge, neither the \glslink{cosmic web classification}{classification of cosmic environments} at non-linear scales in physical realizations of the \gls{LSS} nor the investigation of their genesis and growth, using real \gls{data} and with demonstrated capability of \gls{uncertainty quantification}, have been treated in the existing literature. In this work, we propose the first probabilistic \glslink{cosmic web classification}{web-type analysis} conducted with observational \gls{data} in the deeply \gls{non-linear regime} of \gls{LSS} formation. We build accurate maps of dynamic \glslink{structure type}{cosmic web components} with a resolution of around 3~Mpc/$h$, constrained by observations. In addition, our approach leads to the first quantitative \glslink{large-scale structure inference}{inference} of the \gls{formation history} of these environments and allows the construction of maps of the embryonic traces in the \glslink{initial conditions}{initial perturbations} of the \glslink{final conditions}{late-time} morphological features of the \gls{cosmic web}.

Cosmographic descriptions of the \gls{LSS} in terms of three-dimensional maps, and in particular  a dynamic \gls{structure type} cartography carry potential for a rich variety of applications. Such maps characterize the anisotropic nature of gravitational \gls{structure formation}, the clustering behavior of galaxies as a function of their \glslink{tidal field}{tidal environment} and permit to describe the traces of the \gls{cosmic web} already imprinted in the \gls{initial conditions}. So far, most investigations focused on understanding the physical properties of dark \glslink{halo}{halos} and galaxies in relation to the \gls{LSS}. \citet{Hahn2007a,Hahn2007b,Hahn2009,Hahn2014,Aragon-Calvo2010} found a systematic dependence of \gls{halo} properties such as morphological type, color, \gls{luminosity} and spin parameter on their cosmic environment (local \glslink{density field}{density}, \glslink{velocity field}{velocity} and \gls{tidal field}). In addition, a correlation between \gls{halo} shapes and spins and the orientations of nearby \glslink{filament}{filaments} and \glslink{sheet}{sheets}, predicted in simulations \citep{Altay2006,Hahn2007a,Hahn2007b,Hahn2009,Paz2008,Zhang2009,Codis2012,Libeskind2013,Welker2014,Aragon-Calvo2014,Laigle2015}, has been confirmed by observational galaxy \gls{data} \citep{Paz2008,Jones2010,Tempel2013,Zhang2013}. Cartographic descriptions of the \gls{cosmic web} also permit to study the environmental dependence of galaxy properties \cite[see e.g.][]{Lee2008a,Lee2008,Park2010,Yan2012,Kovavc2014} and to make the connection between the sophisticated predictions for galaxy properties in hydrodynamic simulations \citep[e.g.][]{Vogelsberger2014,Dubois2014,Codis2015} and observations. Another wide range of applications of \gls{structure type} \glslink{reconstruction}{reconstructions} is to probe the effect of the inhomogeneous \glslink{LSS}{large-scale structure} on photon properties and geodesics. For example, it is possible to interpret the \gls{weak gravitational lensing} effects of \glslink{void}{voids} \citep{Melchior2014,Clampitt2014}. Dynamic information can also be used to produce prediction templates for secondary effects expected in the \glslink{CMB}{cosmic microwave background} such as the kinetic \gls{Sunyaev-Zel'dovich effect} \citep{Li2014}, the \glslink{integrated Sachs-Wolfe effect}{integrated Sachs-Wolfe} and \glslink{Rees-Sciama effect}{Rees-Sciama} effects \citep[e.g.][]{Cai2010,Ilic2013,Planck2013ISW}. Lastly, as the \gls{cosmic web} morphology arises from \glslink{gravitational evolution}{gravitational instability}, it can be used to test \gls{general relativity} \citep{Falck2015a}.

Building such refined cosmographic descriptions of the Universe requires \glslink{high-dimensional parameter space}{high-dimensional}, non-linear \glslink{large-scale structure inference}{inferences}. In chapter \ref{chap:BORGSDSS} \citep{Jasche2015BORGSDSS}, we presented a \gls{chrono-cosmography} project, aiming at reconstructing simultaneously the \glslink{density field}{density} distribution, the \gls{velocity field} and the \gls{formation history} of the \gls{LSS} from galaxies. To do so, we used an advanced Bayesian \gls{inference} algorithm to assimilate the \glslink{SDSS}{Sloan Digital Sky Survey} DR7 data into the forecasts of a physical model of \gls{structure formation} (\glslink{2LPT}{second order Lagrangian perturbation theory}). Besides inferring the four-dimensional \glslink{formation history}{history} of the matter distribution, these results permit us an analysis of the genesis and growth of the complex web-like patterns that have been observed in our Universe. Therefore, this work constitutes a new \gls{chrono-cosmography} project, aiming at the analysis of the evolving \gls{cosmic web}.

Our investigations rely on the \glslink{large-scale structure inference}{inference} of the \gls{initial conditions} in the \gls{SDSS} volume (see chapter \ref{chap:BORGSDSS}). Starting from these, we generate a large set of \glslink{constrained simulation}{constrained realizations} of the Universe using the {\cola} method \citep[][see also section \ref{sec:The COLA method}]{Tassev2013}. This physical model allows us to perform the first description of the \gls{cosmic web} in the \gls{non-linear regime}, using real \gls{data}, and to follow the time evolution of its constituting \glslink{structure type}{elements}. Throughout this chapter, we adopt the \citet{Hahn2007a} dynamic ``\gls{T-web}'' classifier, which segments the \gls{LSS} into \glslink{void}{voids}, \glslink{sheet}{sheets}, \glslink{filament}{filaments}, and \glslink{cluster}{clusters}. This choice is motivated by the close relation between the equations that dictate the dynamics of the growth of structures in the \citeauthor{Zeldovich1970} formalism and the Lagrangian description of the \gls{LSS} which naturally emerges with {\borg}. As this procedure relies on the estimation of the \glslink{eigenvalue}{eigenvalues} of the \gls{tidal tensor} in Fourier space, it constitutes a non-linear and \gls{non-local} estimator of \glslink{structure type}{structure types}, requiring adequate means to propagate observational uncertainties to finally inferred products (web-type maps and all derived quantities), in order not to misinterpret results. The {\borg} algorithm naturally addresses this problem by providing a set of density realizations constrained by the \gls{data}. The variation between these \glslink{sample}{samples} constitute a thorough \glslink{uncertainty quantification}{quantification of uncertainty} coming from all observational effects (in particular the incompleteness of the \gls{data} because of the \glslink{galaxy survey}{survey} \gls{mask} and the radial \glslink{selection effects}{selection functions}, as well as \gls{luminosity}-dependent galaxy \glslink{bias}{biases}, see chapter \ref{chap:BORGSDSS} for details), not only with a point estimation but with a detailed treatment of the \gls{likelihood}. Hence, for all problems addressed in this work, we get a fully probabilistic answer in terms of a \gls{prior} and a \gls{posterior} distribution. Building upon the robustness of our \gls{uncertainty quantification} procedure, we are able to make the first observationally-supported link between cosmology and \gls{information theory} \citep[see][for theoretical considerations related to this question]{Neyrinck2015a} by looking at the \gls{entropy} and \gls{Kullback-Leibler divergence} of \glslink{pdf}{probability distribution functions}.

This chapter is organized as follows. In section \ref{sec:Methods}, we describe our methodology: Bayesian \gls{large-scale structure inference} with the {\borg} algorithm, \gls{non-linear filtering} of \glslink{sample}{samples} with {\cola} and \glslink{cosmic web classification}{web-type classification} using the \gls{T-web} procedure. In sections \ref{sec:The late-time large-scale structure} and \ref{sec:The primordial large-scale structure}, we describe the \gls{cosmic web} at \glslink{final conditions}{present} and \glslink{initial conditions}{primordial} times, respectively. In section \ref{sec:Evolution of the cosmic web}, we follow the time evolution of \glslink{structure type}{web-types} as structures form in the Universe. Finally, we summarize our results and offer concluding comments in section \ref{sec:Conclusion}.

\section{Methods}
\label{sec:Methods}

In this section, we describe our methodology step by step:

\begin{enumerate}
\item \glslink{large-scale structure inference}{inference} of the \gls{initial conditions} with {\borg} (section \ref{sec:Bayesian large-scale structure inference with BORG}),
\item generation of \glslink{constrained simulation}{data-constrained realizations} of the \gls{SDSS} volume via \gls{non-linear filtering} of {\borg} \glslink{sample}{samples} with {\cola} (section \ref{sec:Non-linear filtering of samples with COLA}),
\item \glslink{cosmic web classification}{classification of the cosmic web} in \glslink{void}{voids}, \glslink{sheet}{sheets}, \glslink{filament}{filaments}, and \glslink{cluster}{clusters}, using the \gls{T-web} algorithm (section \ref{sec:Classification of the cosmic web}).
\end{enumerate}

\subsection{Bayesian large-scale structure inference with BORG}
\label{sec:Bayesian large-scale structure inference with BORG}

This work builds upon results previously obtained by application of the {\borg} \citep[Bayesian Origin Reconstruction from Galaxies,][]{Jasche2013BORG} algorithm to the \glslink{SDSS}{Sloan Digital Sky Survey} data release 7 \citep{Jasche2015BORGSDSS}. {\borg} is a full-scale Bayesian \gls{inference} code which permits to simultaneously analyze morphology and \gls{formation history} of the \gls{cosmic web} (see chapters \ref{chap:BORG} and \ref{chap:BORGSDSS} for a complete description).

As discussed in \citet{Jasche2013BORG}, accurate and detailed cosmographic \glslink{large-scale structure inference}{inferences} from observations require modeling the \glslink{mildly non-linear regime}{mildly non-linear} and \gls{non-linear regime} of the presently observed matter distribution. The exact statistical behavior of the \gls{LSS} in terms of a full \glslink{pdf}{probability distribution function} for \glslink{non-linear evolution}{non-linearly evolved} \glslink{density field}{density fields} is not known. For this reason, the first full-scale \glslink{reconstruction}{reconstructions} relied on phenomenological approximations, such as multivariate \glslink{grf}{Gaussian} or \glslink{log-normal distribution}{log-normal distributions}, incorporating a cosmological \gls{power spectrum} to accurately represent correct \glslink{two-point correlation function}{two-point statistics} of \glslink{density field}{density fields} \cite[see e.g.][]{Lahav1994,Zaroubi2002,Erdovgdu2004,Kitaura2008,Kitaura2009,Kitaura2010,JascheKitaura2010,Jasche2010a,Jasche2010b}. However, these prescriptions only model the \glslink{one-point distribution}{one} and \glslink{two-point correlation function}{two-point statistics} of the matter distribution. Additional statistical complexity of the evolved \gls{density field} arises from the fact that gravitational \gls{structure formation} introduces \gls{mode coupling} and \gls{phase} correlations. This manifests itself not only in a sheer amplitude difference of \glslink{density field}{density} and \glslink{velocity field}{velocity fields} at different \glslink{redshift}{redshifts}, but also in a modification of their statistical behavior by the generation of \glslink{high-order correlation function}{higher-order correlation functions}. An accurate modeling of these \glslink{high-order correlation function}{high-order correlators} is of crucial importance for a precise description of the connectivity and \glslink{void hierarchy}{hierarchical nature} of the \gls{cosmic web}, which is the aim of this chapter.

While the statistical nature of the \glslink{final conditions}{late-time} \gls{density field} is poorly understood, the \gls{initial conditions} from which it formed are known to obey \glslink{grf}{Gaussian statistics} to very great accuracy \citep{Planck2015PNG}. Therefore, it is reasonable to account for the increasing statistical complexity of the evolving matter distribution by a dynamical model of \gls{structure formation} linking \glslink{initial conditions}{initial} and \gls{final conditions}. This naturally turns the problem of \gls{LSS} analysis to the task of inferring the \gls{initial conditions} from present cosmological observations \citep{Jasche2013BORG,Kitaura2013,Wang2013}. This approach yields a very \glslink{high-dimensional parameter space}{high-dimensional} and non-linear \gls{inference} problem. Typically, the parameter space to explore comprises on the order of $10^6$ to $10^7$ \glslink{structure type}{elements}, corresponding to the voxels of the map to be inferred. For reasons linked to computational cost, the {\borg} algorithm employs \gls{2LPT} as an approximation for the actual gravitational dynamics linking \glslink{initial conditions}{initial} three-dimensional \glslink{grf}{Gaussian} \glslink{density field}{density fields} to present, \glslink{non-Gaussianity}{non-Gaussian} \glslink{density field}{density fields}. As known from perturbation theory \citep[see e.g.][]{Bernardeau2002}, in the \glslink{linear regime}{linear} and \gls{mildly non-linear regime}, \gls{2LPT} correctly describes the \glslink{one-point distribution}{one-}, \glslink{two-point correlation function}{two-} and \glslink{three-point correlation function}{three-point statistics} of the matter distribution and also approximates very well \glslink{high-order correlation function}{higher-order correlators}. It accounts in particular for \gls{tidal effects} in its regime of validity. Consequently, the {\borg} algorithm correctly \glslink{Lagrangian transport}{transports} the observational \glslink{information content}{information} corresponding to complex web-like features from the \glslink{final conditions}{final} \gls{density field} to the corresponding \gls{initial conditions}. Note that such an explicit Bayesian \glslink{forward modeling}{forward-modeling} approach is always more powerful than constraining (part of) the \glslink{high-order correlation function}{sequence of correlation functions}, as it accounts for the entire \gls{dark matter} dynamics (in particular for the infinite \glslink{high-order correlation function}{hierarchy of correlators}), in its regime of validity. This is of particular importance, since the \glslink{high-order correlation function}{hierarchy of correlation functions} has been shown to be an insufficient description of \glslink{density field}{density fields} in the \gls{non-linear regime} \citep{Carron2012a,Carron2012b}.

As discussed in chapter \ref{chap:BORGSDSS} \citep{Jasche2015BORGSDSS}, our analysis comprehensively accounts for observational effects such as \glslink{selection effects}{selection functions}, \gls{survey geometry}, \gls{luminosity}-dependent galaxy \glslink{bias}{biases} and \gls{noise}. Corresponding \gls{uncertainty quantification} is provided by \gls{sampling} from the \glslink{high-dimensional parameter space}{high-dimensional} \gls{posterior} distribution via an efficient implementation of the \glslink{HMC}{Hamiltonian Markov Chain Monte Carlo} method \citep[see chapter \ref{chap:BORG} and][for details]{Jasche2013BORG}. In particular, \gls{luminosity}-dependent galaxy \glslink{bias}{biases} are explicitly part of the {\borg} \glslink{large-scale structure likelihood}{likelihood} and the \gls{bias} amplitudes are inferred self-consistently during the run. Though not explicitly modeled, \gls{redshift-space distortions} are automatically mitigated: due to the \gls{prior} preference for \glslink{statistical homogeneity}{homogeneity} and \glslink{statistical isotropy}{isotropy}, such anisotropic features are treated as \gls{noise} in the \gls{data}.

In the following, we make use of the $12,000$ \glslink{sample}{samples} of the \gls{posterior} distribution for \glslink{initial conditions}{primordial} \glslink{density field}{density fields}, obtained in chapter \ref{chap:BORGSDSS}. These \glslink{reconstruction}{reconstructions}, constrained by \gls{SDSS} observations, act as \gls{initial conditions} for the generation of constrained \glslink{LSS}{large-scale structure} realizations. It is important to note that we directly make use of {\borg} outputs without any further post-processing, which demonstrates the remarkable quality of our \glslink{large-scale structure inference}{inference} results. 

\subsection{Non-linear filtering of samples with COLA}
\label{sec:Non-linear filtering of samples with COLA}

In section \ref{sec:Comparison of structure types in LPT and $N$-body dynamics} \citep[][section 2.A]{Leclercq2013}, we performed a study of differences in the representation of \glslink{structure type}{structure types} in \glslink{density field}{density fields} predicted by \gls{LPT} and \glslink{N-body simulation}{$N$-body simulations}. To do so, we used the same \glslink{cosmic web classification}{web-type classification} procedure as in this work (see sections \ref{sec:Classification of the cosmic web} and \ref{sec:apx-tweb}). In spite of the visual similarity of \gls{LPT} and \glslink{N-body simulation}{$N$-body} \glslink{density field}{density fields} at large and intermediate scales (above a few Mpc/$h$), we found crucial differences in the representation of structures. Specifically, \gls{LPT} predicts fuzzier \glslink{halo}{halos} than \gls{full gravity}, and incorrectly assigns the surroundings of \glslink{void}{voids} as part of them. This manifests itself in an overprediction of the \glslink{VFF}{volume} occupied by \glslink{cluster}{clusters} and \glslink{void}{voids} at the detriment of \glslink{sheet}{sheets} and \glslink{filament}{filaments}. The substructure of \glslink{void}{voids} is also known to be incorrectly represented in \gls{2LPT} \citep{Sahni1996,Neyrinck2013,Leclercq2013}.

For these reasons, in this chapter we cannot directly make use of the \glslink{final conditions}{final} {\borg} density \glslink{sample}{samples}, which are a prediction of the \gls{2LPT} model. Instead, we rely on the inferred \gls{initial conditions}, which contain the \gls{data} constraints (as described in chapter \ref{chap:BORGSDSS}) and on a \gls{non-linear filtering} step (see chapter \ref{chap:filtering}) similar to the one described in chapter \ref{chap:dmvoids} \citep{Leclercq2015DMVOIDS}. Due to the large number of \glslink{sample}{samples} to be processed for this work, we do not use a \glslink{full gravity}{fully non-linear} \glslink{N-body simulation}{simulation} code as in chapter \ref{chap:dmvoids}, but the {\cola} method \cite[][see also section \ref{sec:The COLA method}]{Tassev2013}. 

The generation of the set of non-linear {\borg}-{\cola} \glslink{sample}{samples}, used in this chapter, is described in section \ref{sec:Non-linear BORG-COLA realizations}.

\subsection{Classification of the cosmic web}
\label{sec:Classification of the cosmic web}

The {\borg} filtered \glslink{reconstruction}{reconstructions} permit a variety of scientific analyses of the large scale structure in the observed Universe. In this work, we focus specifically on the possibility to characterize the \gls{cosmic web} by distinct \glslink{structure type}{structure types}. Generally, any of the methods cited in the introduction (section \ref{sec:Tweb-introduction}) can be employed for analysis of our density \glslink{sample}{samples}, however for the purpose of this chapter, we follow the ``\gls{T-web}'' \glslink{cosmic web classification}{classification} procedure as proposed by \cite{Hahn2007a}, described in section \ref{sec:apx-tweb}. 

The basic idea of this dynamical \glslink{cosmic web classification}{classification} approach is that the \glslink{eigenvalue}{eigenvalues} $\mu_1 \le \mu_2 \le \mu_3$ of the \gls{tidal tensor} $\mathscr{T}_{ij} \equiv \mathrm{H}(\tilde{\Phi})_{ij}$ (Hessian of the rescaled \gls{gravitational potential}) characterize the geometrical properties of each point in space. With these definitions, the three \glslink{eigenvalue}{eigenvalues} of the \gls{tidal tensor} form a decomposition of the \gls{density contrast} field, in the sense that the trace of $\mathscr{T}$ is $\mu_1+\mu_2+\mu_3 = \delta$. Each spatial point can then be classified as a specific \glslink{structure type}{web type} by considering the signs of $\mu_1$, $\mu_2$, $\mu_3$, according to the rules given in table \ref{tb:tweb-rules}.

Several extensions of this \glslink{cosmic web classification}{classification} procedure exist, that permit different \glslink{cosmic web classification}{classification} up to sub-megaparsec scales (see section \ref{sec:Extensions of the T-web}). In this work, we will probe scales down to $\sim$ 3~Mpc/$h$ (the voxel size in our \glslink{reconstruction}{reconstructions}). Therefore, we will be content with the original \glslink{cosmic web classification}{classification} procedure as proposed by \cite{Hahn2007a}.

It is important to note that the \gls{tidal tensor} and the rescaled \gls{gravitational potential} are both physical quantities, and hence their calculation requires the availability of a full physical \gls{density field} in contrast to a smoothed mean reconstruction of the \gls{density field}. As described in chapter \ref{chap:BORGSDSS}, density \glslink{sample}{samples} obtained by the {\borg} algorithm provide such required full physical \glslink{density field}{density fields}. The \gls{tidal tensor} can therefore easily be calculated in each density \gls{sample} from the Fourier space representations of equations \eqref{eq:reduced-Poisson} and \eqref{eq:tidal-tensor} \citep[see section \ref{sec:Tweb-implementation} and][for details on the technical implementation]{Hahn2007a,Forero-Romero2009}.

The web classifier provides four voxel-wise \glslink{scalar field}{scalar fields} that characterize the large scale structure. In a specific realization, the answer is unique, meaning that these fields obey the following conditions at each voxel position $\vec{x}_k$:
\begin{equation}
\mathrm{T}_i(\vec{x}_k) \in \{0,1\} \; \mathrm{for} \; i \in \llbracket 0,3 \rrbracket \quad \mathrm{and} \quad \sum_{i=0}^{3} \mathrm{T}_i(\vec{x}_k) = 1 
\end{equation}
where $\mathrm{T}_0=$ \gls{void}, $\mathrm{T}_1=$ \gls{sheet}, $\mathrm{T}_2=$ \gls{filament}, $\mathrm{T}_3=$ \gls{cluster}. In this work, we follow the Bayesian approach of \citet{Jasche2015BORGSDSS} and quantify the degree of belief in \gls{structure type} \glslink{cosmic web classification}{classification}. Specifically, our \glslink{cosmic web classification}{web classification} is given in terms of four voxel-wise \glslink{scalar field}{scalar fields} that obey the following conditions at each voxel position $\vec{x}_k$:
\begin{equation}
\mathpzc{T}_i(\vec{x}_k) \in [0,1] \; \mathrm{for} \; i \in \llbracket 0,3 \rrbracket \quad \mathrm{and} \quad \sum_{i=0}^{3} \mathpzc{T}_i(\vec{x}_k) = 1 .
\end{equation}
Here, $\mathpzc{T}_i(\vec{x}_k) \equiv \langle \mathrm{T}_i(\vec{x}_k) \rangle_{\p(\mathrm{T}_i(\vec{x}_k)|d)} = \p(\mathrm{T}_i(\vec{x}_k)|d)$ are the \gls{posterior} probabilities indicating the possibility to encounter specific \glslink{structure type}{structure types} at a given position in the observed volume, \glslink{conditional pdf}{conditional} on the \gls{data}. These are estimated by applying the \glslink{cosmic web classification}{web classification} to all density \glslink{sample}{samples} and counting the relative frequencies at each individual spatial coordinate within the set of \glslink{sample}{samples} \citep[see section 5 in][]{Jasche2010a}. With this definition, the cosmic \glslink{structure type}{web-type} \gls{posterior mean} is given by
\begin{equation}
\left\langle \p(\mathrm{T}_i(\vec{x}_k)|d) \right\rangle = \frac{1}{N}\sum_{n=1}^{N} \sum_{j=0}^{3} \updelta_{\mathrm{K}}^{\mathrm{T}_i(\vec{x}_k) \mathrm{T}^n_j(\vec{x}_k)} ,
\label{eq:pdf_mean}
\end{equation}
where $n$ labels one of the $N$ \glslink{sample}{samples}, $\mathrm{T}^n_j(\vec{x}_k)$ is the result of the web classifier on the $n$-th \gls{sample} (i.e. a unit four-vector at each voxel position $\vec{x}_k$ containing zeros except for one component, which indicates the \gls{structure type}), and $\updelta_{\mathrm{K}}^{ab}$ is a \gls{Kronecker symbol}.

\section{The late-time large-scale structure}
\label{sec:The late-time large-scale structure}

In this section, we discuss the results of our analysis of the \glslink{final conditions}{final} \gls{density field}, at $a=1$. For reasons of computational time with {\cola} \glslink{non-linear filtering}{filtering} (see section \ref{sec:Non-linear filtering of samples with COLA}), we kept around 10\% of the original set of \glslink{sample}{samples} obtained in chapter \ref{chap:BORGSDSS}. In order to mitigate as much as possible the effects of correlation among \glslink{sample}{samples}, we maximally separated the \glslink{sample}{samples} kept for the present analysis, keeping one out of ten consecutive \glslink{sample}{samples} of the original Markov Chain. Hence,  for all results discussed in this section, we used a total of $1,097$ \glslink{sample}{samples} inferred by {\borg} and filtered with {\cola}.

\subsection{Tidal environment}
\label{sec:Tidal environment final}

\begin{figure*}
\begin{center}
\includegraphics[width=\textwidth]{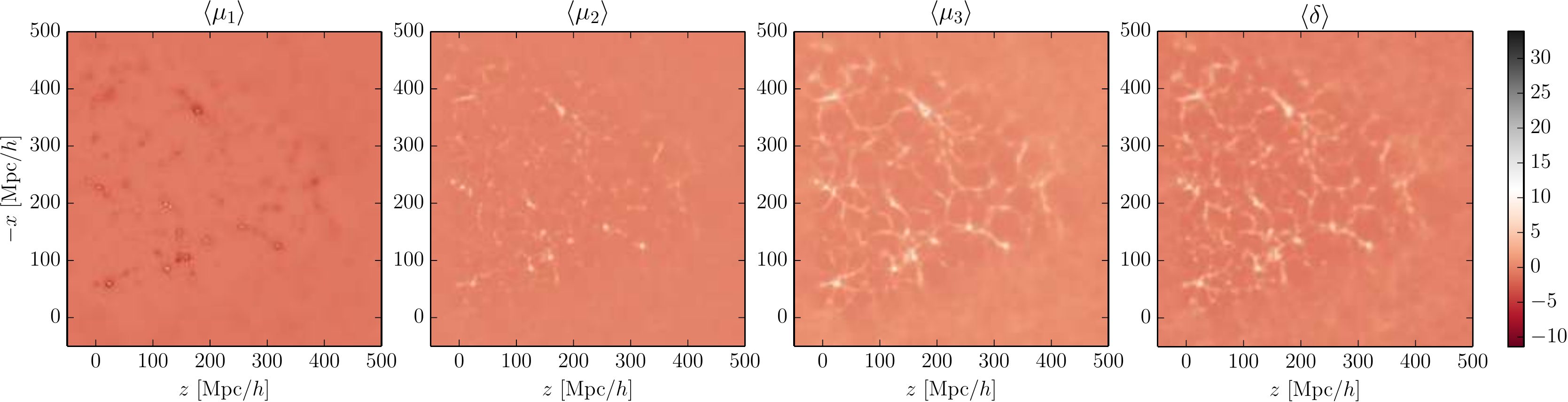}
\caption{Slices through the three-dimensional ensemble \gls{posterior mean} for the \glslink{eigenvalue}{eigenvalues} $\mu_1 \leq \mu_2 \leq \mu_3$ of the \glslink{tidal tensor}{tidal field tensor} in the \gls{final conditions}, estimated from $1,097$ \glslink{sample}{samples}. The rightmost panel shows the corresponding slice through the \gls{posterior mean} for the \glslink{final conditions}{final} \gls{density contrast} $\delta=\mu_1+\mu_2+\mu_3$, obtained in section \ref{sec:inf_density_field}. \label{fig:lambda_cola_final_mean}}
\end{center}
\end{figure*}

As a natural byproduct, the application of the \gls{T-web} classifier to density \glslink{sample}{samples} yields \glslink{sample}{samples} of the \glslink{pdf}{pdfs} for the three \glslink{eigenvalue}{eigenvalues} of the \glslink{tidal tensor}{tidal field tensor}. These \glslink{pdf}{pdfs} account for the assumed physical model of \gls{structure formation} and the \gls{data} constraints, and \glslink{uncertainty quantification}{quantify uncertainty} coming in particular from \gls{selection effects}, \glslink{survey geometry}{surveys geometries} and galaxy \glslink{bias}{biases}. In a similar fashion as described in section \ref{sec:inference_results}, the ensemble of \glslink{sample}{samples} permits us to provide any desired statistical summary such as \glslink{posterior mean}{mean} and \glslink{posterior standard deviation}{variance}.

\begin{figure*}
\begin{center}
\includegraphics[width=\textwidth]{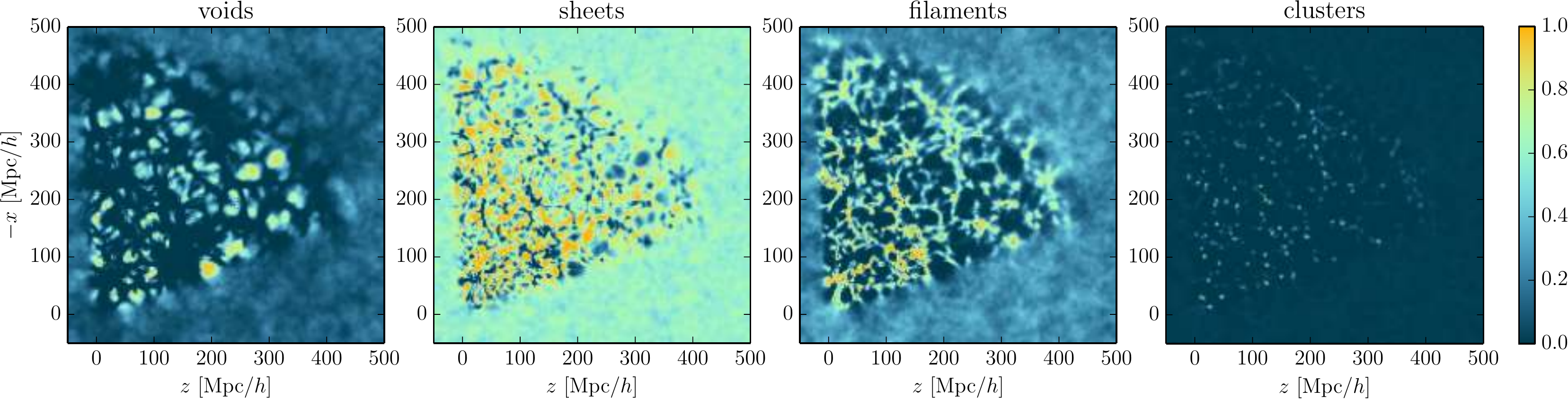}
\caption{Slices through the \gls{posterior mean} for different \glslink{structure type}{structure types} (from left to right: \gls{void}, \gls{sheet}, \gls{filament}, and \gls{cluster}) in the \glslink{final conditions}{late-time} \glslink{LSS}{large-scale structure} in the \glslink{SDSS}{Sloan} volume ($a=1$). These four three-dimensional voxel-wise \glslink{pdf}{pdfs} sum up to one on a voxel basis.\label{fig:pdf_final}}
\end{center}
\end{figure*}

In figure \ref{fig:lambda_cola_final_mean}, we show slices through the ensemble mean fields $\mu_1$, $\mu_2$ and $\mu_3$. For visual comparison, the rightmost panel of figure \ref{fig:lambda_cola_final_mean} shows the corresponding slice through the \gls{posterior mean} of the \glslink{final conditions}{final} \gls{density contrast}, $\delta = \mu_1 + \mu_2 + \mu_3$, obtained in section \ref{sec:inf_density_field}. Different morphologies can be observed in the data-constrained parts of these slices: $\mu_1$, $\mu_2$ and $\mu_3$ respectively trace well the \glslink{cluster}{clusters}, \glslink{filament}{filaments} and \glslink{sheet}{sheets}, as we now argue. The $\mu_1$ field is rather homogeneous, apart for small spots where all \glslink{eigenvalue}{eigenvalues} are largely positive, i.e. undergoing dramatic gravitational collapse along three axes. These correspond to the dynamic \glslink{cluster}{clusters}. Note that there exists a form of ``\glslink{tidal field}{tidal} \gls{compensation}'': these \glslink{cluster}{clusters} are surrounded by regions where $\mu_1$ is smaller than its cosmic mean. More patterns can be observed in the $\mu_2$ field: it also exhibits \glslink{filament}{filaments} (appearing as dots when piercing the slice). Finally, the $\mu_3$ field is highly-structured, as it also traces \glslink{sheet}{sheets} (which appear \glslink{filament}{filamentary} when sliced). Dynamic \glslink{void}{voids} can also be easily distinguished in this field, wherever $\mu_3$ is negative.

\subsection{Probabilistic web-type cartography}
\label{sec:Probabilistic web-type cartography final}

Building upon previous results and using the procedure described in section \ref{sec:Classification of the cosmic web}, we obtain probabilistic maps of structures. More precisely, we obtain four \glslink{pdf}{probability distributions} at each spatial position, $\p(\mathrm{T}_i(\vec{x_k})|d)$, indicating the possibility to encounter a specific \gls{structure type} (\gls{cluster}, \gls{filament}, \gls{sheet}, \gls{void}) at that position. As noted in section \ref{sec:Classification of the cosmic web}, these \glslink{pdf}{pdfs} take their values in the range $[0,1]$ and sum up to one on a voxel-basis. Figure \ref{fig:pdf_final} shows slices through their means (see equation \eqref{eq:pdf_mean}). The plot shows the anticipated behavior, with a high degree of structure and values close to certainty (i.e. zero or one) in regions covered by \gls{data}, while the unobserved regions approach a uniform value corresponding to the \gls{prior}. At this point, it is worth noting that the \gls{T-web} classifier has a \gls{prior} preference for some \glslink{structure type}{structure types}. Using unconstrained \glslink{LSS}{large-scale structure} realizations produced with the same setup,\footnote{By this, we specifically mean realizations obtained from \glslink{initial conditions}{initial} randomly-generated \glslink{grf}{Gaussian} \glslink{density field}{density fields} with an \cite{Eisenstein1998,Eisenstein1999} \gls{power spectrum} using the fiducial \gls{cosmological parameters} of the {\borg} analysis ($\Omega_\mathrm{m}~=~0.272$, $\Omega_\mathrm{\mu}~=~0.728$, $\Omega_\mathrm{b}~=~0.045$, $h~=~0.702$, $\sigma_8~=~0.807$, $n_\mathrm{s}~=~0.961$, see equation \eqref{eq:cosmo-BORGSDSS}\label{fn:setup}). The \gls{density field} is defined on a $750~\mathrm{Mpc}/h$ cubic grid of $256^3$-voxels and populated by $512^3$ \gls{dark matter particles}, which are evolved to $z=69$ with \gls{2LPT} and from $z=69$ to $z=0$ with {\cola}, using 30 timesteps logarithmically-spaced in the \gls{scale factor}. The particles are binned on a $256^3$-voxel grid with the \gls{CiC} scheme to get the \glslink{final conditions}{final} \gls{density field}.} we measured that these \gls{prior} probabilities, $\p(\mathrm{T}_i)$, can be well described by Gaussians whose mean and standard deviation are given in table \ref{tb:prior_final}. 

\begin{table}\centering
\begin{tabular}{lcc}
\hline\hline
Structure type & $\mu_{\p(\mathrm{T}_i)}$ & $\sigma_{\p(\mathrm{T}_i)}$ \\
\hline
\multicolumn{3}{c}{Late-time large-scale structure ($a=1$)} \\
Void & $0.14261$ & $6.1681 \times 10^{-4}$ \\
Sheet & $0.59561$ & $6.3275 \times 10^{-4}$ \\
Filament & $0.24980$ & $5.5637 \times 10^{-4}$ \\
Cluster & $0.01198$ & $5.8793 \times 10^{-5}$ \\
\hline\hline
\end{tabular}
\caption{\Gls{prior} probabilities assigned by the \gls{T-web} classifier to the different structures types, in the \glslink{final conditions}{late-time} \glslink{LSS}{large-scale structure} ($a=1$).}
\label{tb:prior_final}
\end{table}

In addition to their ensemble mean, the set of \glslink{sample}{samples} permits to propagate \gls{uncertainty quantification} to \glslink{cosmic web classification}{web-type classification}. In particular, it allows us to locally assess the strength of \gls{data} constraints. In \gls{information theory}, a convenient way to characterize the uncertainty content of a random source $\mathcal{S}$ is the Shannon \gls{entropy} \citep{Shannon1948}, defined by
\begin{equation}
H\left[\mathcal{S}\right] \equiv - \sum_i p_i \log_2(p_i) ,
\end{equation}
where the $p_i$ are the probabilities of possible events. This definition yields expected properties and accounts for the intuition that the more likely an event is, the less information it provides when it occurs (i.e. the more it contributes to the source \gls{entropy}). We follow this prescription and write the voxel-wise \gls{entropy} of the web-type \gls{posterior}, $\p(\mathrm{T}(\vec{x}_k)|d)$, as
\begin{equation}
H\left[ \p(\mathrm{T}(\vec{x}_k)|d) \right] \equiv - \sum_{i=0}^{3} \p(\mathrm{T}_i(\vec{x}_k)|d) \log_2(\p(\mathrm{T}_i(\vec{x}_k)|d)) .
\label{eq:definition_entropy}
\end{equation}
It is a number in the range $[0,2]$ and its natural unit is the shannon (Sh). $H=0$~Sh in the case of perfect certainty, i.e. when the \gls{data} constraints entirely determine the underlying \gls{structure type}: $\p(\mathrm{T}_{i_0}(\vec{x}_k)|d)$ is 1 for one $i_0$ and 0 for $i \neq i_0$. $H$ reaches its maximum value of $2$~Sh when all $\p(\mathrm{T}_i(\vec{x}_k)|d)$ are equal to $1/4$. This is the case of maximal randomness: all the events being equally likely, no information is gained when one occurs.

\begin{figure*}
\begin{center}
\includegraphics[width=0.49\textwidth]{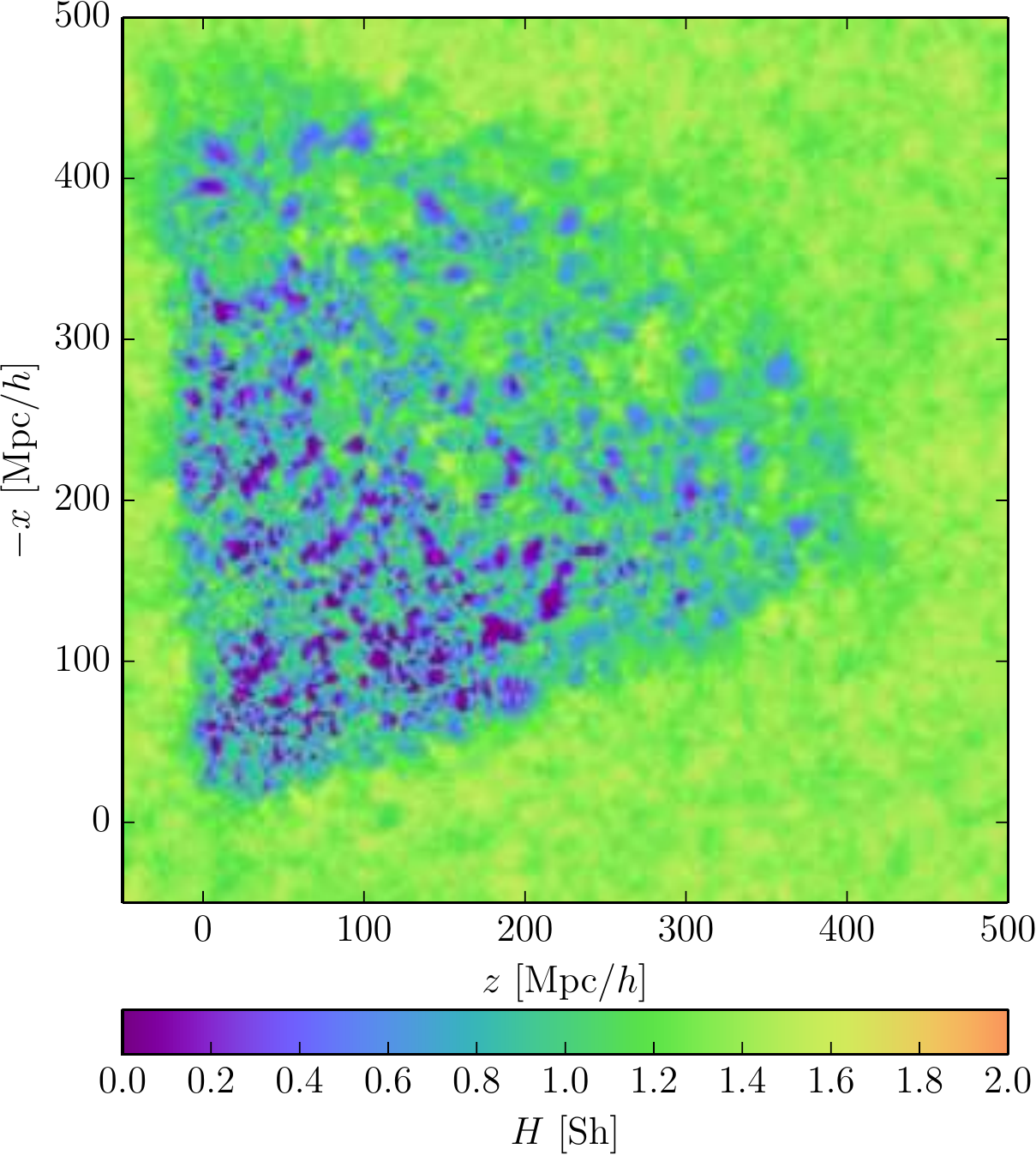}
\includegraphics[width=0.49\textwidth]{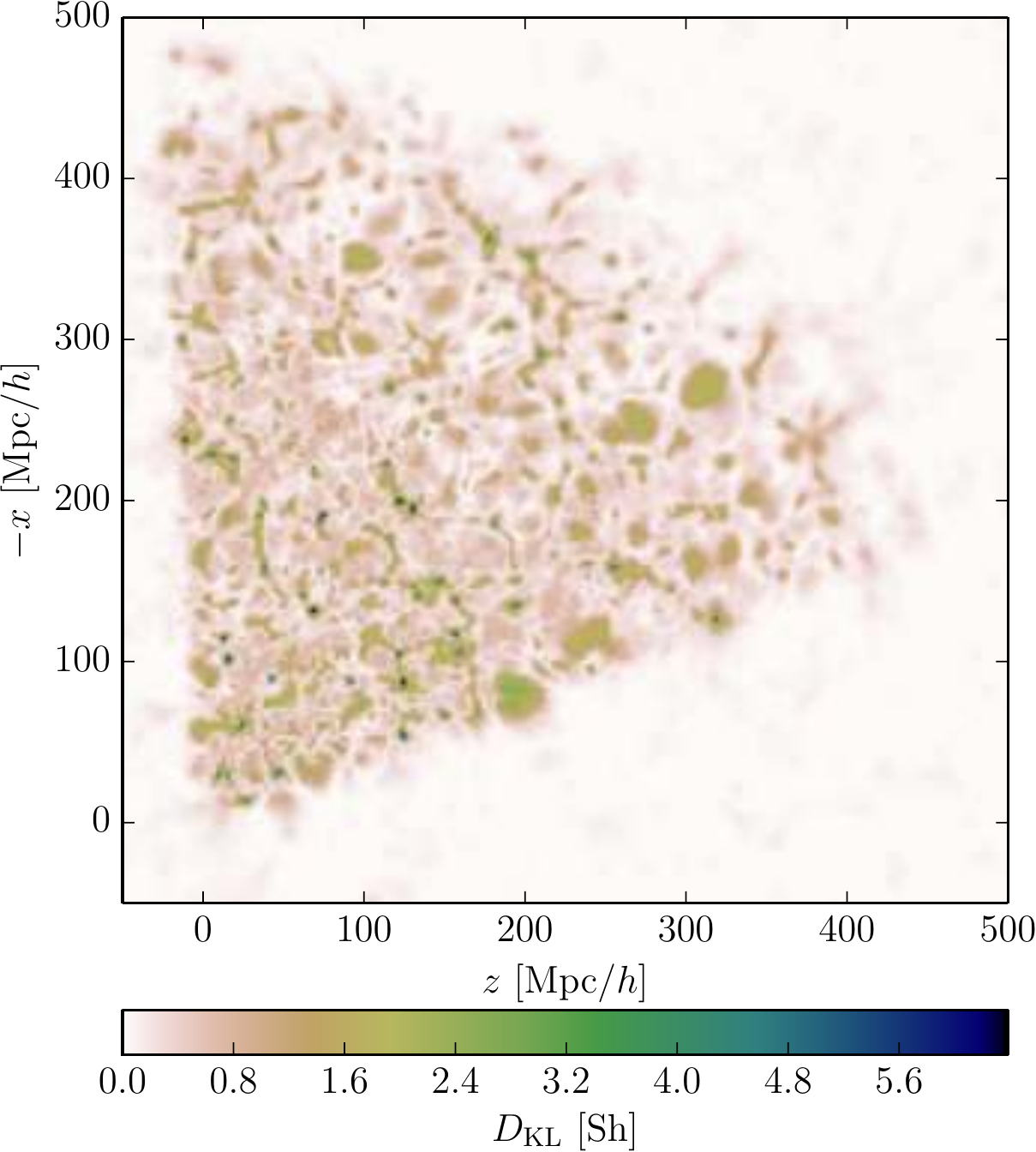}
\caption{Slices through the \gls{entropy} of the \glslink{structure type}{structure types} \gls{posterior} (left panel) and the \gls{Kullback-Leibler divergence} of the \gls{posterior} from the \gls{prior} (right panel), in the \gls{final conditions}. The \gls{entropy} $H$, defined by equation \eqref{eq:definition_entropy}, quantifies the \gls{information content} of the \gls{posterior} \gls{pdf} represented in figure \ref{fig:pdf_final}, which results from fusing the \gls{information content} of the \gls{prior} and the \gls{data} constraints. The \gls{Kullback-Leibler divergence} $D_\mathrm{KL}$, defined by equation \eqref{eq:KL_divergence}, represents the information gained in moving from the \gls{prior} to the \gls{posterior}. It quantifies the information that has been learned on \glslink{structure type}{structure types} by looking at \gls{SDSS} galaxies.\label{fig:pdf_final_entropy}}
\end{center}
\end{figure*}

A slice through the voxel-wise \gls{entropy} of the web-type \gls{posterior} is shown in the left panel of figure \ref{fig:pdf_final_entropy}. Generally, the \gls{entropy} map reflects the \gls{information content} of the \gls{posterior} \gls{pdf}, which comes from augmenting the \gls{information content} of the \gls{prior} \gls{pdf} with the \gls{data} constraints, in the \glslink{Bayesian statistics}{Bayesian way}.

The \gls{entropy} takes low values and shows a high degree of structure in the regions where \gls{data} constraints exist, and even reaches zero in some spots where the \gls{data} are perfectly informative. Comparing with figures \ref{fig:lambda_cola_final_mean} and \ref{fig:pdf_final}, one can note that this structure is highly non-trivial and does not follow any of the previously described patterns. This is due to the facts that in a \gls{Poisson process}, the signal (here the density, inferred in section \ref{sec:inf_density_field}) is correlated with the uncertainty and that \glslink{structure type}{structure types} \glslink{cosmic web classification}{classification} further is a non-linear function of the \gls{density field}. In the unobserved regions, the \gls{entropy} fluctuates around a constant value of about $1.4$~Sh, which characterizes the \gls{information content} of the \gls{prior}. This value is consistent with the expectation, which can be computed using equation \eqref{eq:definition_entropy} (unconditional on the \gls{data}) and the numbers given in table \ref{tb:prior_final}. 

The \glslink{information theory}{information-theoretic} quantity that measures the information gain (in shannons) due to the \gls{data} is the relative entropy or \gls{Kullback-Leibler divergence} \citep{Kullback1951} of the \gls{posterior} from the \gls{prior},
\begin{eqnarray}
D_\mathrm{KL}\left[ \p(\mathrm{T}(\vec{x}_k)|d) \middle\| \p(\mathrm{T}) \right]
& \equiv & \sum_{i=0}^{3} \p(\mathrm{T}_i(\vec{x}_k)|d) \log_2\left(\frac{\p(\mathrm{T}_i(\vec{x}_k)|d)}{\p(\mathrm{T}_i)}\right) \nonumber \\
& = & - H\left[ \p(\mathrm{T}(\vec{x}_k)|d) \right] - \sum_{i=0}^{3} \p(\mathrm{T}_i(\vec{x}_k)|d) \log_2(\p(\mathrm{T}_i)) .
\label{eq:KL_divergence}
\end{eqnarray}
It is a non-symmetric measure of the difference between the two \glslink{pdf}{probability distributions}.

A slice through the voxel-wise \gls{Kullback-Leibler divergence} of the web-type \gls{posterior} from the \gls{prior} is shown in the right panel of figure \ref{fig:pdf_final_entropy}. As expected, the information gain is zero out of the \glslink{survey geometry}{survey boundaries}. In the observed regions, \gls{SDSS} galaxies are informative on underlying \glslink{structure type}{structure types} at the level of at least $\sim$~1~Sh. This number can go to $\sim$~3~Sh in the interior of deep \glslink{void}{voids} and up to $\sim$~6~Sh in the densest \glslink{cluster}{clusters}. This map permits to visualize the regions where additional \gls{data} would be needed to improve \gls{structure type} \glslink{cosmic web classification}{classification}, e.g. in some high-\gls{redshift} regions where uncertainty remains due to \gls{selection effects}.

\subsection{Volume and mass filling fractions}
\label{sec:Volume and mass filling fractions final}

A characterization of large scale environments commonly found in literature involves evaluating global quantities such as the \glslink{VFF}{volume} and \glslink{MFF}{mass} content of these structures. In a particular realization, the \glslink{VFF}{volume filling fraction} (\gls{VFF}) for \gls{structure type} $\mathrm{T}_i$ is the number of voxels of type $\mathrm{T}_i$ divided by the total number of voxels in the considered volume,
\begin{equation}
\mathrm{VFF}(\mathrm{T}_i) \equiv \frac{\sum_{\vec{x}_k} \sum_{j=0}^{3} \updelta_{\mathrm{K}}^{\mathrm{T}_i(\vec{x}_k)\mathrm{T}^n_j(\vec{x}_k)}}{N_{\mathrm{vox}}} .
\label{eq:definition_VFF}
\end{equation}
The \glslink{MFF}{mass filling fraction} (\gls{MFF}) can be obtained in a similar manner by weighting the same sum by the local density $\rho(\vec{x}_k) = \bar{\rho}\,(1+\delta(\vec{x}_k))$,
\begin{equation}
\mathrm{MFF}(\mathrm{T}_i) \equiv \frac{\sum_{\vec{x}_k} \sum_{j=0}^{3} (1+\delta(\vec{x}_k)) \updelta_{\mathrm{K}}^{\mathrm{T}_i(\vec{x}_k)\mathrm{T}^n_j(\vec{x}_k)}}{\sum_{\vec{x}_k} (1+\delta(\vec{x}_k))} .
\label{eq:definition_MFF}
\end{equation}
To ensure that results are not \gls{prior}-dominated, we measured the \glslink{VFF}{VFFs} and \glslink{MFF}{MFFs} in the data-constrained parts of our realizations. More precisely, we limited ourselves to the voxels where the \gls{survey response operator} (representing simultaneously the \gls{survey geometry} and the \gls{selection effects}, see sections \ref{sec:The BORG data model} and \ref{sec:galaxy_sample}) is strictly positive. This amounts to $N_\mathrm{vox} = 3$,$148$,$504$ out of $256^3~=~16$,$777$,$216$ voxels, around $18.7$\% of the full box (see also section \ref{sec:Selection of voids} and figure \ref{fig:slice_voids}). In equations \eqref{eq:definition_VFF} and \eqref{eq:definition_MFF}, $\vec{x}_k$ labels one of these voxels. 

By measuring the \gls{VFF} and \gls{MFF} of different \glslink{structure type}{structure types} in each \glslink{constrained simulation}{constrained realization} of our ensemble, we obtained the \gls{posterior} \glslink{pdf}{pdfs}, $\p(\mathrm{VFF}(\mathrm{T}_i)|d)$ and $\p(\mathrm{MFF}(\mathrm{T}_i)|d)$, \glslink{conditional pdf}{conditional} on the \gls{data}. Similarly, we computed the \gls{prior} \glslink{pdf}{pdfs}, $\p(\mathrm{VFF}(\mathrm{T}_i))$ and $\p(\mathrm{MFF}(\mathrm{T}_i))$, using unconstrained realizations produced with the same setup. We found that all these \glslink{pdf}{pdfs} can be well described by Gaussians, the mean and variance of which are given in tables \ref{tb:final_vff} and \ref{tb:final_mff}. 

Previous studies on this topic \citep[e.g.][]{Doroshkevich1970b,Shen2006,Hahn2007a,Forero-Romero2009,Jasche2010a,Aragon-Calvo2010,Shandarin2012,Cautun2014} have found a wide range of values for the \gls{VFF} and \gls{MFF} of structures \citep[see e.g. table 3 in][]{Cautun2014}. For example, existing studies found that \glslink{cluster}{clusters} occupy at most a few percent of the \glslink{VFF}{volume} of the Universe but contribute significantly to the \glslink{MFF}{mass} content, with a \gls{MFF} ranging from $\sim 10\%$ \citep{Hahn2007a,Cautun2014} to $\sim 40\%$ \citep{Shandarin2012}. The \gls{void} \glslink{VFF}{volume} fraction can vary from $\sim 10\%$ \citep{Hahn2007a} to $\sim 80\%$ \citep{Aragon-Calvo2010,Shandarin2012,Cautun2014}; in the \cite{Forero-Romero2009} formalism (see section \ref{sec:Extensions of the T-web}), it is a very sensitive function of the threshold $\mu_\mathrm{th}$ \citep[figure 9 in][]{Jasche2010a}. These large disparities in the literature arise because different algorithms use various information and criteria for classifying the \gls{cosmic web}. For this reason, we believe that it is only relevant to make relative statements for the same setup, i.e. to compare our results to the corresponding \gls{prior} quantities, as done in tables \ref{tb:final_vff} and \ref{tb:final_mff}. In this purpose, the large number of \glslink{sample}{samples} used allowed a precise characterization of the \glslink{pdf}{pdfs} so that all digits quoted in the tables are significant. Note that all our analyses are repeatable for different setups, which allows in principle a comparison with any previous work.

As expected for a Bayesian update of the degree of belief, the \gls{posterior} quantities generally have smaller variance and a mean value displaced from the \gls{prior} mean. For the \gls{MFF}, the \glslink{posterior mean}{posterior means} are always within two standard deviations of the corresponding \gls{prior} means. The analysis shows that in the \gls{SDSS}, a larger \glslink{MFF}{mass fraction} is occupied by \glslink{cluster}{clusters}, \glslink{sheet}{sheets}, and \glslink{void}{voids}, at the detriment of \glslink{filament}{filaments}, in comparison to the \gls{prior} expectation. The \gls{data} also favor a smaller filling of the \glslink{SDSS}{Sloan} volume by \glslink{filament}{filaments} and \glslink{sheet}{sheets} and larger filling by \glslink{void}{voids} and \glslink{cluster}{clusters}. For the \gls{cluster} \gls{VFF}, the \gls{posterior mean}, $\mu_{\mathrm{VFF}(\mathrm{T}_3)|d} = 0.01499$ is at about $15$ standard deviations ($\sigma_{\mathrm{VFF}(\mathrm{T}_3)}= 1.9194 \times 10^{-4}$) of the \gls{prior} mean, $\mu_{\mathrm{VFF}(\mathrm{T}_3)} = 0.01198$. Given other results on the \gls{VFF} and \gls{MFF}, we believe that the \gls{data} truly favor a higher \glslink{VFF}{volume} content in \glslink{cluster}{clusters} as compared to the \gls{structure formation} model used as \gls{prior}. However, this surprising result should be treated with care; part of the discrepancy is likely due to the original \textsc{borg} analysis, which optimizes the \gls{initial conditions} for evolution with \gls{2LPT} (instead of the \gls{non-linear evolution} with \textsc{cola} used for the present work). \gls{LPT} predicts fuzzier \glslink{halo}{halos} than \glslink{N-body simulation}{$N$-body dynamics}, which results in the incorrect prediction of a high \gls{cluster} \gls{VFF} \citep[see section \ref{sec:Comparison of structure types in LPT and $N$-body dynamics};][]{Leclercq2013}. 

\begin{table}\centering
\begin{tabular}{lcccc}
\hline\hline
Structure type & $\mu_{\mathrm{VFF}}$ & $\sigma_\mathrm{VFF}$ & $\mu_{\mathrm{VFF}}$ & $\sigma_\mathrm{VFF}$ \\
\hline
\multicolumn{1}{c}{} & \multicolumn{4}{c}{Late-time large-scale structure ($a=1$)} \\
\multicolumn{1}{c}{} & \multicolumn{2}{c}{Posterior} & \multicolumn{2}{c}{Prior} \\
Void & $0.14897$ & $1.8256 \times 10^{-3}$ & $0.14254$ & $6.2930 \times 10^{-3}$ \\
Sheet & $0.58914$ & $1.3021 \times 10^{-3}$ & $0.59562$ & $2.2375 \times 10^{-3}$ \\
Filament & $0.24689$ & $1.1295 \times 10^{-3}$ & $0.24986$ & $4.4440 \times 10^{-3}$ \\
Cluster & $0.01499$ & $8.7274 \times 10^{-5}$ & $0.01198$ & $1.9194 \times 10^{-4}$ \\
\hline\hline
\end{tabular}
\caption{Mean and standard deviation of the \gls{prior} and \gls{posterior} \glslink{pdf}{pdfs} for the \glslink{VFF}{volume filling fraction} of different \glslink{structure type}{structure types} in the \glslink{final conditions}{late-time} \glslink{LSS}{large-scale structure} ($a=1$).}
\label{tb:final_vff}
\end{table}

\begin{table}\centering
\begin{tabular}{lcccc}
\hline\hline
Structure type & $\mu_{\mathrm{MFF}}$ & $\sigma_\mathrm{MFF}$ & $\mu_{\mathrm{MFF}}$ & $\sigma_\mathrm{MFF}$ \\
\hline
\multicolumn{1}{c}{} & \multicolumn{4}{c}{Late-time large-scale structure ($a=1$)} \\
\multicolumn{1}{c}{} & \multicolumn{2}{c}{Posterior} & \multicolumn{2}{c}{Prior} \\
Void & $0.04050$ & $8.3531 \times 10^{-4}$ & $0.03876$ & $2.3352 \times 10^{-3}$ \\
Sheet & $0.35605$ & $1.2723 \times 10^{-3}$ & $0.35286$ & $3.6854 \times 10^{-3}$ \\
Filament & $0.47356$ & $1.5661 \times 10^{-3}$ & $0.48170$ & $4.2215 \times 10^{-3}$ \\
Cluster & $0.12990$ & $6.4966 \times 10^{-4}$ & $0.12666$ & $1.8284 \times 10^{-3}$ \\
\hline\hline
\end{tabular}
\caption{Mean and standard deviation of the \gls{prior} and \gls{posterior} \glslink{pdf}{pdfs} for the \glslink{MFF}{mass filling fraction} of different \glslink{structure type}{structure types} in the \glslink{final conditions}{late-time} \glslink{LSS}{large-scale structure} ($a=1$).}
\label{tb:final_mff}
\end{table}

\section{The primordial large-scale structure}
\label{sec:The primordial large-scale structure}

In this section, we discuss the results of our analysis of the \glslink{initial conditions}{initial} \gls{density field}, at $a=10^{-3}$. Since the analysis of the primordial \glslink{LSS}{large-scale structure} does not involve an additional \glslink{non-linear filtering}{filtering} step, we have been able to keep a larger number of \glslink{sample}{samples} of the \gls{posterior} \gls{pdf} for \gls{initial conditions}, obtained in chapter \ref{chap:BORGSDSS}. Hence, for all results described in this section, we used a total of $4,473$ \glslink{sample}{samples}.

\subsection{Tidal environment}

\begin{figure*}
\begin{center}
\includegraphics[width=\textwidth]{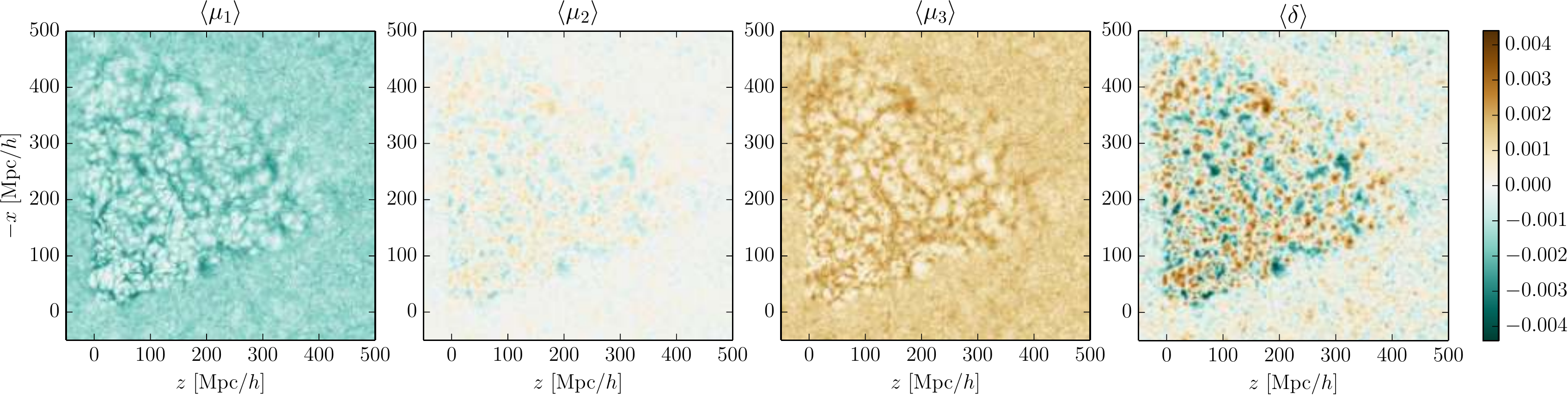}
\caption{Slices through the three-dimensional ensemble \gls{posterior mean} for the \glslink{eigenvalue}{eigenvalues} $\mu_1 \leq \mu_2 \leq \mu_3$ of the \glslink{tidal tensor}{tidal field tensor} in the \gls{initial conditions}, estimated from $4,473$ \glslink{sample}{samples}. The rightmost panel shows the corresponding slice through the \gls{posterior mean} for the \glslink{initial conditions}{initial} \gls{density contrast} $\delta=\mu_1+\mu_2+\mu_3$, obtained in section \ref{sec:inf_density_field}. \label{fig:lambda_initial_mean}}
\end{center}
\end{figure*}

\begin{figure*}
\begin{center}
\includegraphics[width=\textwidth]{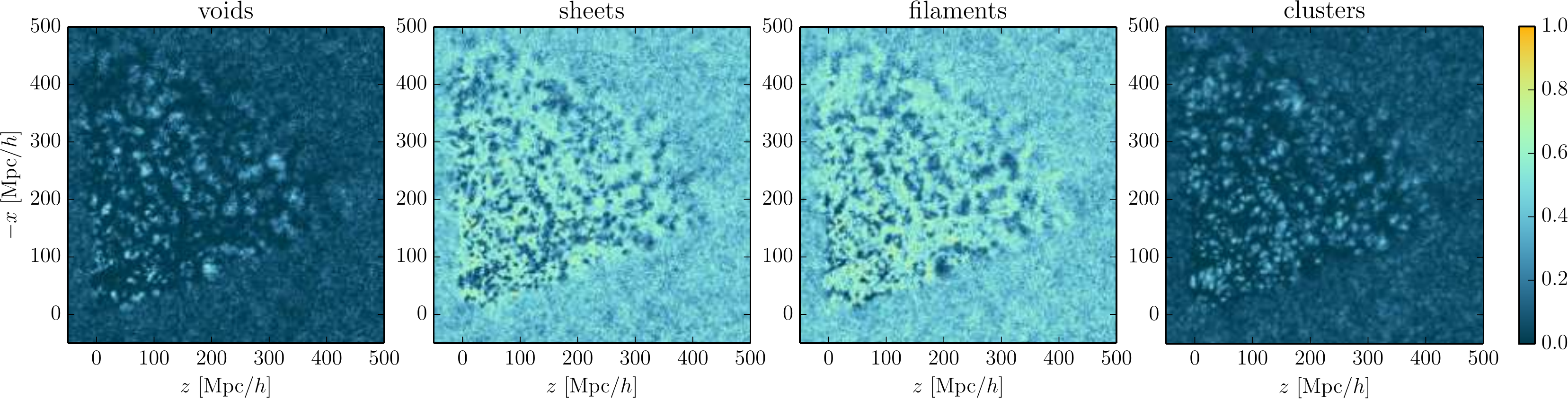}
\caption{Slices through the \gls{posterior mean} for different \glslink{structure type}{structure types} (from left to right: \gls{void}, \gls{sheet}, \gls{filament}, and \gls{cluster}) in the primordial \glslink{LSS}{large-scale structure} in the \glslink{SDSS}{Sloan} volume ($a=10^{-3}$). These four three-dimensional voxel-wise \glslink{pdf}{pdfs} sum up to one on a voxel basis.\label{fig:pdf_initial}}
\end{center}
\end{figure*}

In a similar fashion as in section \ref{sec:Tidal environment final}, the application of the \gls{T-web} classifier to \glslink{initial conditions}{initial} density \glslink{sample}{samples} yields the \gls{posterior} \gls{pdf} for the three \glslink{eigenvalue}{eigenvalues}, $\mu_1$, $\mu_2$ and $\mu_3$, of the \glslink{initial conditions}{initial} \glslink{tidal tensor}{tidal field tensor}. Figure \ref{fig:lambda_initial_mean} shows slices through their means. For visual comparison, the rightmost panel shows the corresponding slice through the \gls{posterior mean} of the \glslink{initial conditions}{initial} \gls{density contrast}, $\delta~=~\mu_1~+~\mu_2~+~\mu_3$, obtained in section \ref{sec:inf_density_field}.

In a \glslink{grf}{Gaussian random field}, $\mu_1$ is generally negative, $\mu_3$ is generally positive and $\mu_2$ close to zero (see the unobserved parts of the slices in figure \ref{fig:lambda_initial_mean}). In addition, $\mu_2$ closely resembles the total \gls{density contrast} $\delta$ up to a global scaling. In the constrained regions, the \glslink{eigenvalue}{eigenvalues} of the \glslink{initial conditions}{initial} \gls{tidal tensor} follow this behavior. The structure observed in their maps is visually consistent with the decomposition of Gaussian density fluctuations as shown by the right panel.

\subsection{Probabilistic web-type cartography}
\label{sec:Probabilistic web-type cartography initial}

Looking at the sign of the \glslink{eigenvalue}{eigenvalues} of the \glslink{initial conditions}{initial} \gls{tidal tensor} and following the procedure described in section \ref{sec:Classification of the cosmic web}, we obtain a probabilistic cartography of the primordial \glslink{LSS}{large-scale structure}. As before, we obtain four voxel-wise \glslink{pdf}{pdfs} $\p(\mathrm{T}_i(\vec{x_k})|d)$, taking their values in the range $[0,1]$ and summing up to one. Figure \ref{fig:pdf_initial} shows slices through their means, defined by equation \eqref{eq:pdf_mean}. As in the \gls{final conditions}, the maps exhibit structure in the data-constrained regions and approach uniform values in the unobserved parts, corresponding to the respective \glslink{prior}{priors}. Using unconstrained realizations of \glslink{grf}{Gaussian random fields} produced with the same setup,\footnote{We used the \gls{initial conditions} of our set of unconstrained simulations (see footnote \ref{fn:setup}).} we measured these \gls{prior} probabilities. Their means and standard deviations are given in table \ref{tb:prior_initial}.

At this point, it is worth mentioning that there exists an additional symmetry for \glslink{grf}{Gaussian random fields}. Since the definition of the \gls{tidal tensor} is linear in the \gls{density contrast} (see equations \eqref{eq:reduced-Poisson} and \eqref{eq:tidal-tensor}) and since positive and negative \glslink{density contrast}{density contrasts} are equally likely, a positive and negative value for a given $\mu_i$ have the same probabilities. Because of this sign symmetry, the \glslink{pdf}{pdfs} for \glslink{void}{voids} and \glslink{cluster}{clusters} (0 or 3 positive/negative \glslink{eigenvalue}{eigenvalues}) and the \glslink{pdf}{pdfs} for \glslink{sheet}{sheets} and \glslink{filament}{filaments} (1 or 2 positive/negative \glslink{eigenvalue}{eigenvalues}) are equal. This can be checked both in table \ref{tb:prior_initial} and in the unconstrained regions of the maps in figure \ref{fig:pdf_initial}. In the constrained regions, a qualitative complementarity between \glslink{pdf}{pdfs} for \glslink{void}{voids} and \glslink{cluster}{clusters} and for \glslink{sheet}{sheets} and \glslink{filament}{filaments} can be observed. This can be explained by the following. As $\sum_i \p(\mathrm{T}_i(\vec{x_k})|d)=1$ and assuming that $\p(\mathrm{T}_i(\vec{x_k})|d) \approx \p(\mathrm{T}_{3-i}(\vec{x_k})|d)$ for unlikely events, consistently with the previous remark, we get $\p(\mathrm{T}_0(\vec{x_k})|d) \approx 1 - \p(\mathrm{T}_3(\vec{x_k})|d)$ wherever $\p(\mathrm{T}_1(\vec{x_k})|d) \approx \p(\mathrm{T}_2(\vec{x_k})|d)$ is sufficiently small; and $\p(\mathrm{T}_1(\vec{x_k})|d) \approx 1 - \p(\mathrm{T}_2(\vec{x_k})|d)$ wherever $\p(\mathrm{T}_0(\vec{x_k})|d) \approx \p(\mathrm{T}_3(\vec{x_k})|d)$ is sufficiently small. These results are therefore consistent with expectations based on Gaussianity for the primordial \glslink{LSS}{large-scale structure} in the \glslink{SDSS}{Sloan} volume.

\begin{table}\centering
\begin{tabular}{lcc}
\hline\hline
Structure type & $\mu_{\p(\mathrm{T}_i)}$ & $\sigma_{\p(\mathrm{T}_i)}$ \\
\hline
\multicolumn{3}{c}{Primordial large-scale structure ($a=10^{-3}$)} \\
Void & $0.07979$ & $5.4875 \times 10^{-5}$ \\
Sheet & $0.42022$ & $1.0240 \times 10^{-4}$ \\
Filament & $0.42022$ & $1.0412 \times 10^{-4}$ \\
Cluster & $0.07978$ & $5.6337 \times 10^{-5}$ \\
\hline\hline
\end{tabular}
\caption{\Gls{prior} probabilities assigned by the \gls{T-web} classifier to the different structures types, in the primordial \glslink{LSS}{large-scale structure} ($a=10^{-3}$).}
\label{tb:prior_initial}
\end{table}

\begin{figure*}
\begin{center}
\includegraphics[width=0.49\textwidth]{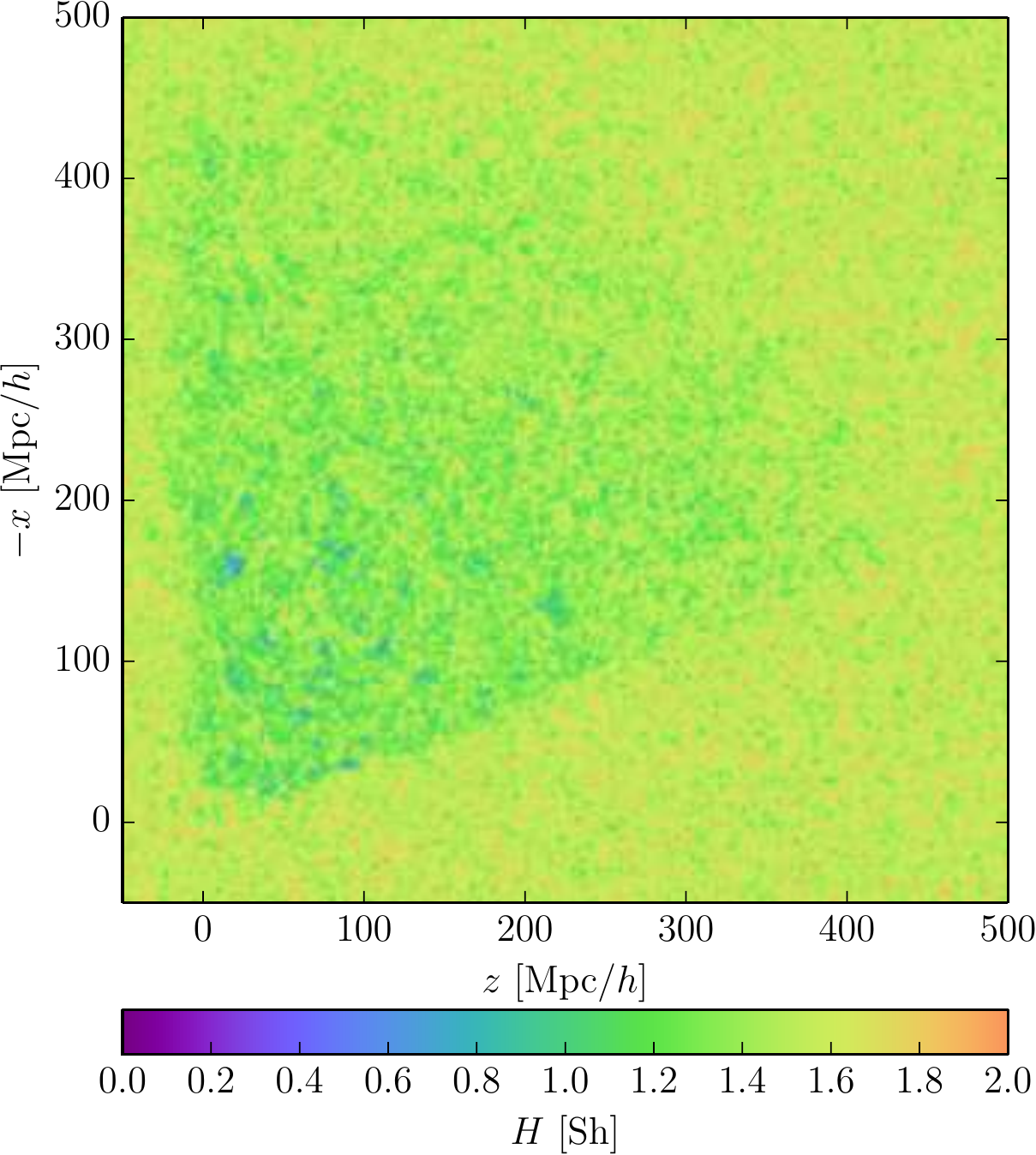}
\includegraphics[width=0.49\textwidth]{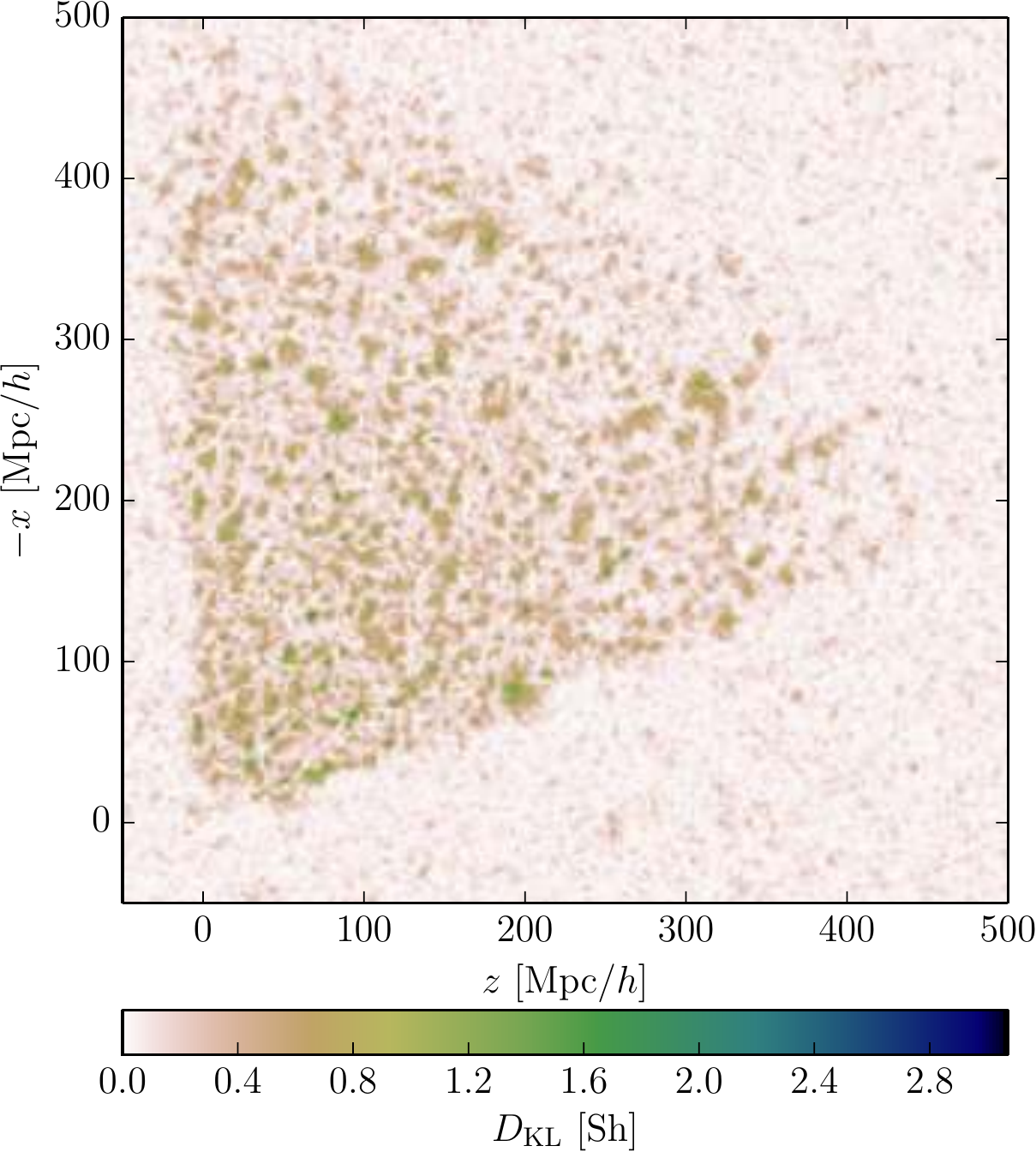}
\caption{Slices through the \gls{entropy} of the \glslink{structure type}{structure types} \gls{posterior} (left panel) and the \gls{Kullback-Leibler divergence} of the \gls{posterior} from the \gls{prior} (right panel), in the \gls{initial conditions}. The \gls{entropy} $H$, defined by equation \eqref{eq:definition_entropy}, quantifies the \gls{information content} of the \gls{posterior} \gls{pdf} represented in figure \ref{fig:pdf_initial}, which results from fusing the \gls{information content} of the \gls{prior} and the \gls{data} constraints. The \gls{Kullback-Leibler divergence} $D_\mathrm{KL}$, defined by equation \eqref{eq:KL_divergence}, represents the information gained in moving from the \gls{prior} to the \gls{posterior}. It quantifies the information that has been learned on \glslink{structure type}{structure types} by looking at \gls{SDSS} galaxies.\label{fig:pdf_initial_entropy}}
\end{center}
\end{figure*}

In a similar fashion as in section \ref{sec:Probabilistic web-type cartography final}, the ensemble of \glslink{sample}{samples} permits us to propagate uncertainties to \gls{structure type} \glslink{cosmic web classification}{classification} and to characterize the strength of \gls{data} constraints. In the left panel of figure \ref{fig:pdf_initial_entropy}, we show a slice through the voxel-wise \gls{entropy} of the web-type \gls{posterior} \gls{pdf} in the \gls{initial conditions}, defined by equation \eqref{eq:definition_entropy}. This function quantifies the \gls{information content} of the \gls{posterior}, which comes from both the \gls{prior} and the \gls{data} constraints. As in the \gls{final conditions}, the \gls{entropy} takes lower values inside the survey region. In the unobserved parts, the \gls{entropy} fluctuates around 1.6~Sh, value which characterizes the \gls{information content} of the \gls{prior}. Using equation \eqref{eq:definition_entropy} (unconditional on the \gls{data}) and the numbers given in table \ref{tb:prior_initial}, one can check that this number is consistent with the expectation. In the right panel of figure \ref{fig:pdf_initial_entropy}, we show a map of the \gls{Kullback-Leibler divergence} of the \gls{posterior} from the \gls{prior}, which represents the information gain due to the \gls{data}.

\subsection{Volume and mass filling fractions}
\label{sec:Volume and mass filling fractions initial}

We computed the \glslink{VFF}{volume} and \glslink{MFF}{mass filling fractions} (defined by equations \eqref{eq:definition_VFF} and \eqref{eq:definition_MFF}) of different \glslink{structure type}{structure types} in the primordial \glslink{LSS}{large-scale structure} in the \glslink{SDSS}{Sloan} volume. As for the \gls{final conditions}, we kept only the regions where the \gls{survey response operator} is strictly positive. Consequently, we obtained the \gls{posterior} \glslink{pdf}{pdfs} $\p(\mathrm{VFF}(\mathrm{T}_i)|d)$ and $\p(\mathrm{MFF}(\mathrm{T}_i)|d)$. Using a set of unconstrained \glslink{grf}{Gaussian random fields}, we also measured $\p(\mathrm{VFF}(\mathrm{T}_i))$ and $\p(\mathrm{MFF}(\mathrm{T}_i))$ and found that all these \glslink{pdf}{pdfs} are well described by Gaussians, the means and standard deviations of which are given in table \ref{tb:initial_vff} and \ref{tb:initial_mff}.

All \gls{posterior} quantities obtained are within two standard deviations of the corresponding \gls{prior} means, and show smaller variance, as expected. Hence, all results obtained are consistent with \glslink{grf}{Gaussian} \gls{initial conditions}.

\begin{table}\centering
\begin{tabular}{lcccc}
\hline\hline
Structure type & $\mu_{\mathrm{VFF}}$ & $\sigma_\mathrm{VFF}$ & $\mu_{\mathrm{VFF}}$ & $\sigma_\mathrm{VFF}$ \\
\hline
\multicolumn{1}{c}{} & \multicolumn{4}{c}{Primordial large-scale structure ($a=10^{-3}$)} \\
\multicolumn{1}{c}{} & \multicolumn{2}{c}{Posterior} & \multicolumn{2}{c}{Prior} \\
Void & $0.07994$ & $4.0221 \times 10^{-4}$ & $0.07977$ & $1.0200 \times 10^{-3}$ \\
Sheet & $0.41994$ & $6.1770 \times 10^{-4}$ & $0.42019$ & $1.7885 \times 10^{-3}$ \\
Filament & $0.42048$ & $6.3589 \times 10^{-4}$ & $0.42024$ & $1.7820 \times 10^{-3}$ \\
Cluster & $0.07964$ & $3.8043 \times 10^{-4}$ & $0.07980$ & $1.0260 \times 10^{-3}$ \\
\hline\hline
\end{tabular}
\caption{Mean and standard deviation of the \gls{prior} and \gls{posterior} \glslink{pdf}{pdfs} for the \glslink{VFF}{volume filling fraction} of different \glslink{structure type}{structure types} in the primordial \glslink{LSS}{large-scale structure} ($a=10^{-3}$).}
\label{tb:initial_vff}
\end{table}

\begin{table}\centering
\begin{tabular}{lcccc}
\hline\hline
Structure type & $\mu_{\mathrm{MFF}}$ & $\sigma_\mathrm{MFF}$ & $\mu_{\mathrm{MFF}}$ & $\sigma_\mathrm{MFF}$ \\
\hline
\multicolumn{1}{c}{} & \multicolumn{4}{c}{Primordial large-scale structure ($a=10^{-3}$)} \\
\multicolumn{1}{c}{} & \multicolumn{2}{c}{Posterior} & \multicolumn{2}{c}{Prior} \\
Void & $0.07958$ & $4.0122 \times 10^{-4}$ & $0.07941$ & $1.0163 \times 10^{-3}$ \\
Sheet & $0.41933$ & $6.1907 \times 10^{-4}$ & $0.41957$ & $1.7912 \times 10^{-3}$ \\
Filament & $0.42110$ & $6.3543 \times 10^{-4}$ & $0.42087$ & $1.7785 \times 10^{-3}$ \\
Cluster & $0.07999$ & $3.8206 \times 10^{-4}$ & $0.08015$ & $1.0293 \times 10^{-3}$ \\
\hline\hline
\end{tabular}
\caption{Mean and standard deviation of the \gls{prior} and \gls{posterior} \glslink{pdf}{pdfs} for the \glslink{MFF}{mass filling fraction} of different \glslink{structure type}{structure types} in the primordial \glslink{LSS}{large-scale structure} ($a=10^{-3}$).}
\label{tb:initial_mff}
\end{table}

\section{Evolution of the cosmic web}
\label{sec:Evolution of the cosmic web}

\begin{figure*}
\begin{center}
\includegraphics[width=0.49\textwidth]{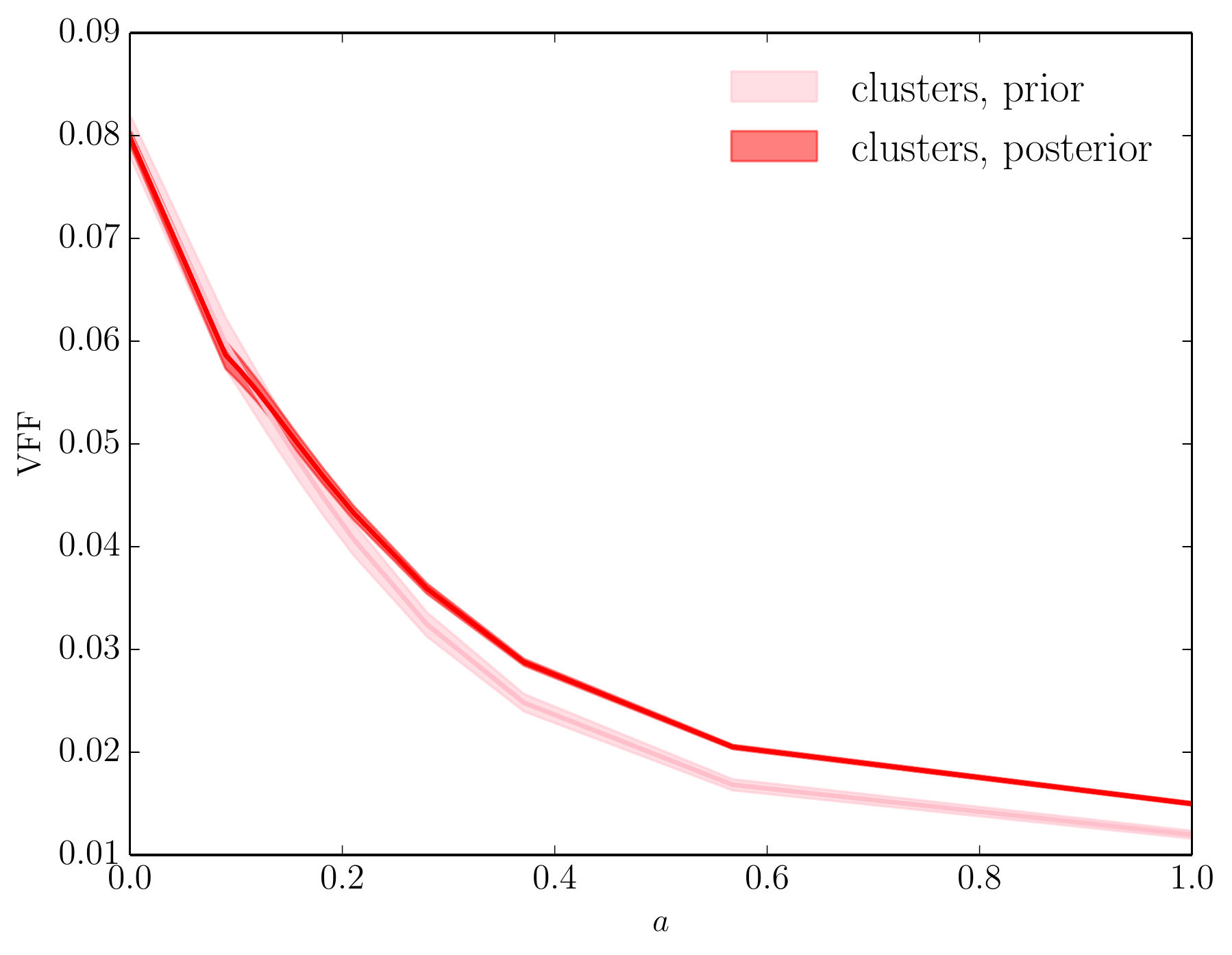}
\includegraphics[width=0.49\textwidth]{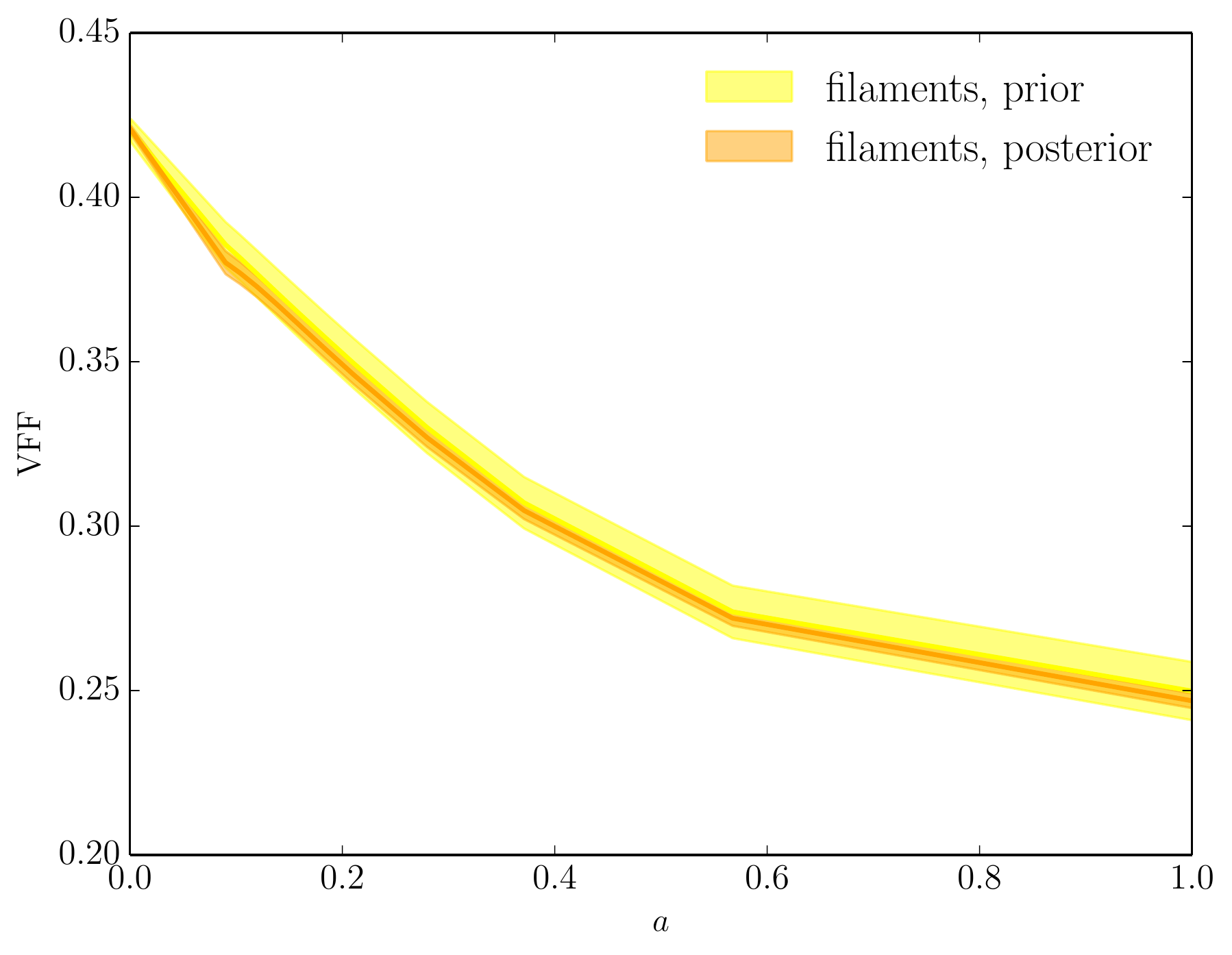} \\
\includegraphics[width=0.49\textwidth]{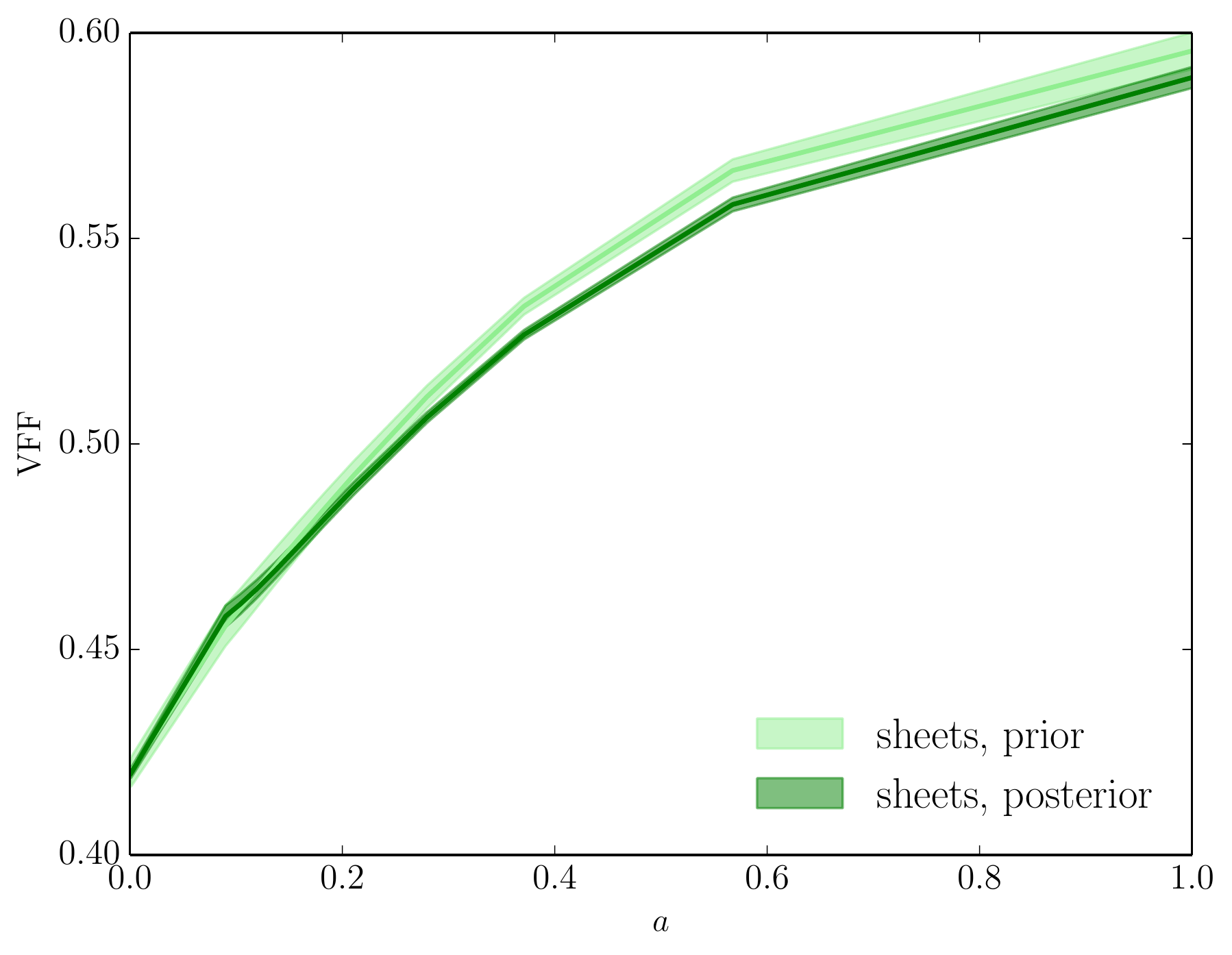}
\includegraphics[width=0.49\textwidth]{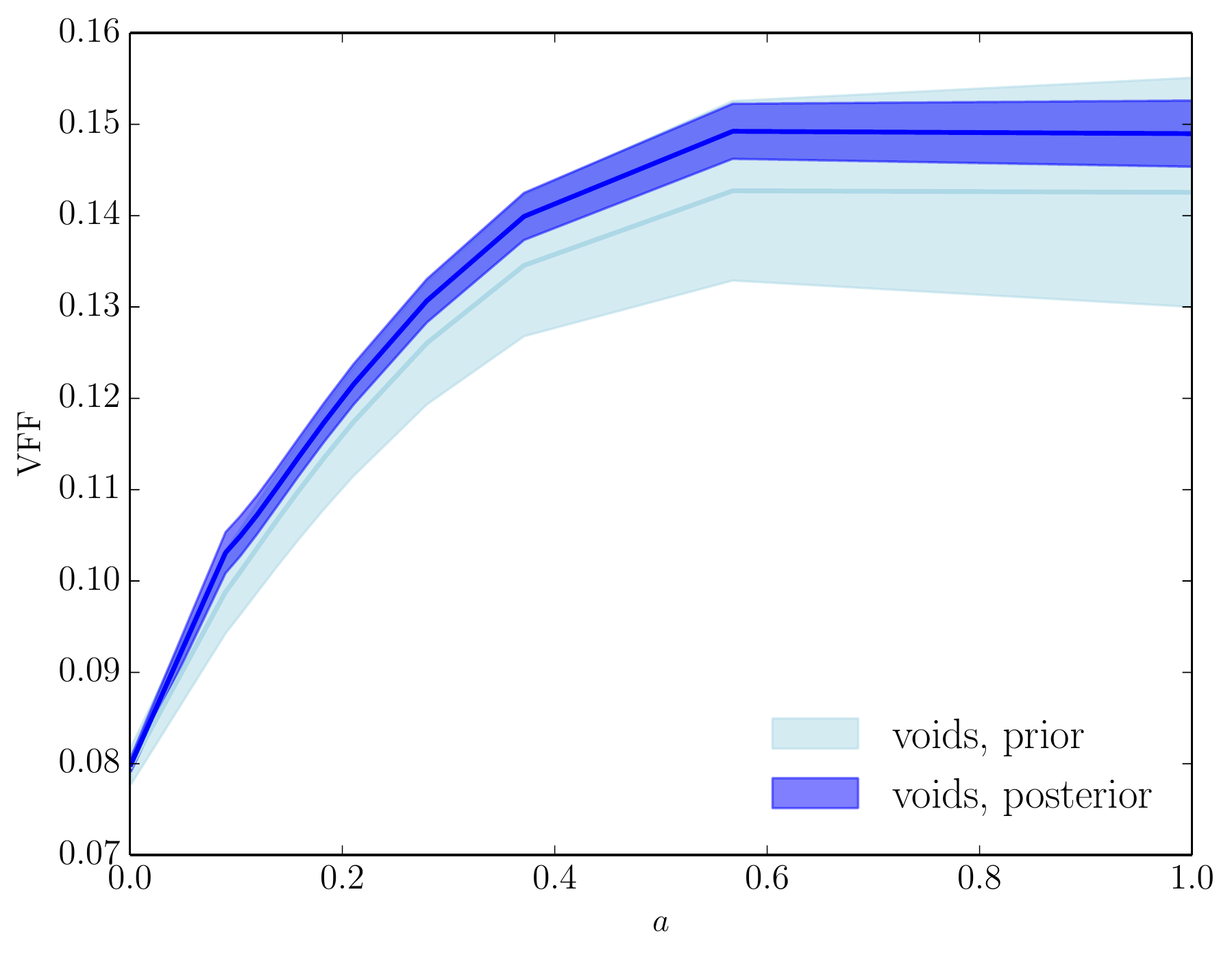} \\
\caption{Time evolution of the \glslink{VFF}{volume filling fractions} of different \glslink{structure type}{structure types} (from left to right and top to bottom: \glslink{cluster}{clusters}, \glslink{filament}{filaments}, \glslink{sheet}{sheets}, \glslink{void}{voids}). The solid lines show the \gls{pdf} means and the shaded regions are the 2-$\sigma$ credible intervals. Light colors are used for the \glslink{prior}{priors} and dark colors for the \glslink{posterior}{posteriors}.\label{fig:evolution_vff}}
\end{center}
\end{figure*}

\begin{figure*}
\begin{center}
\includegraphics[width=0.49\textwidth]{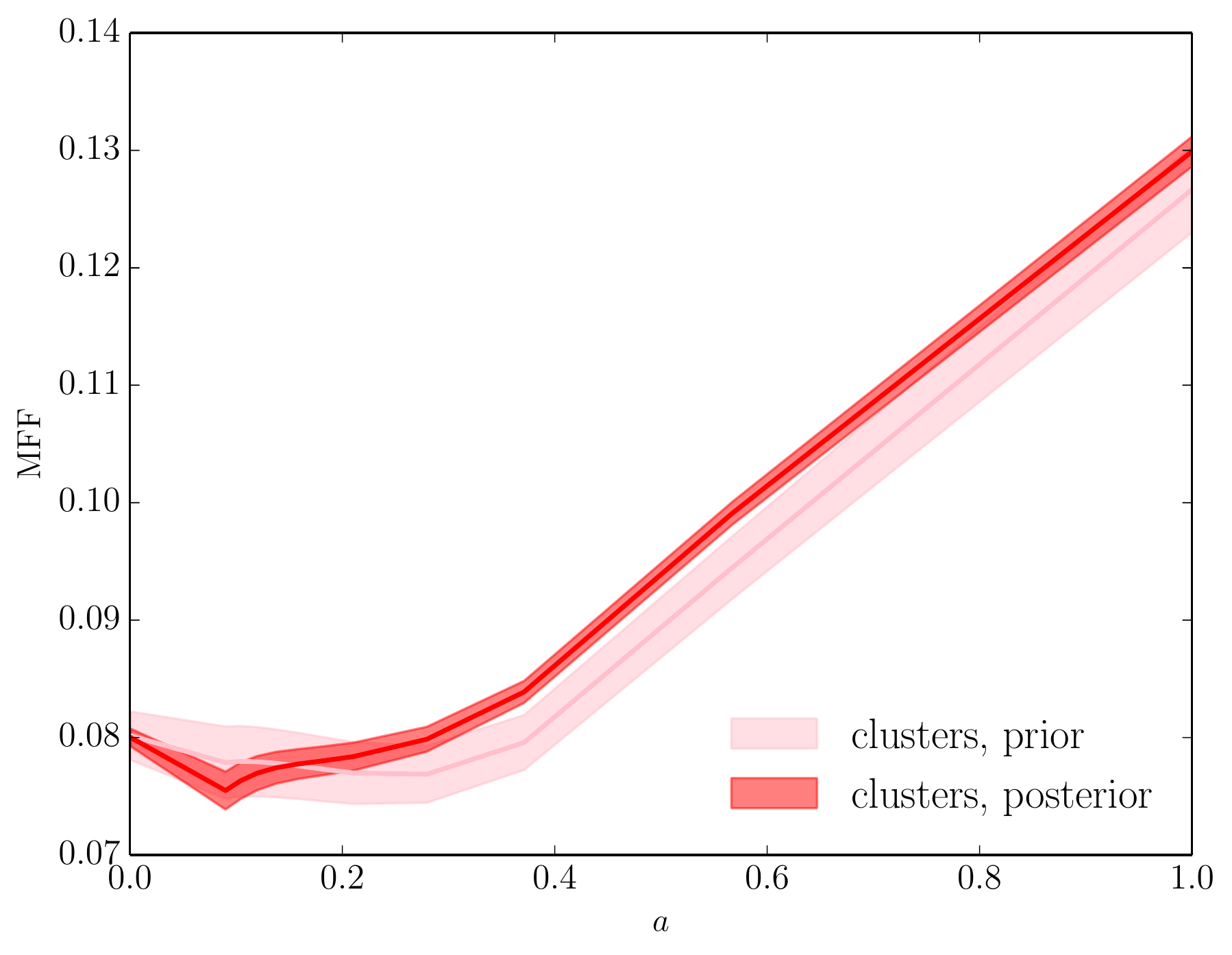}
\includegraphics[width=0.49\textwidth]{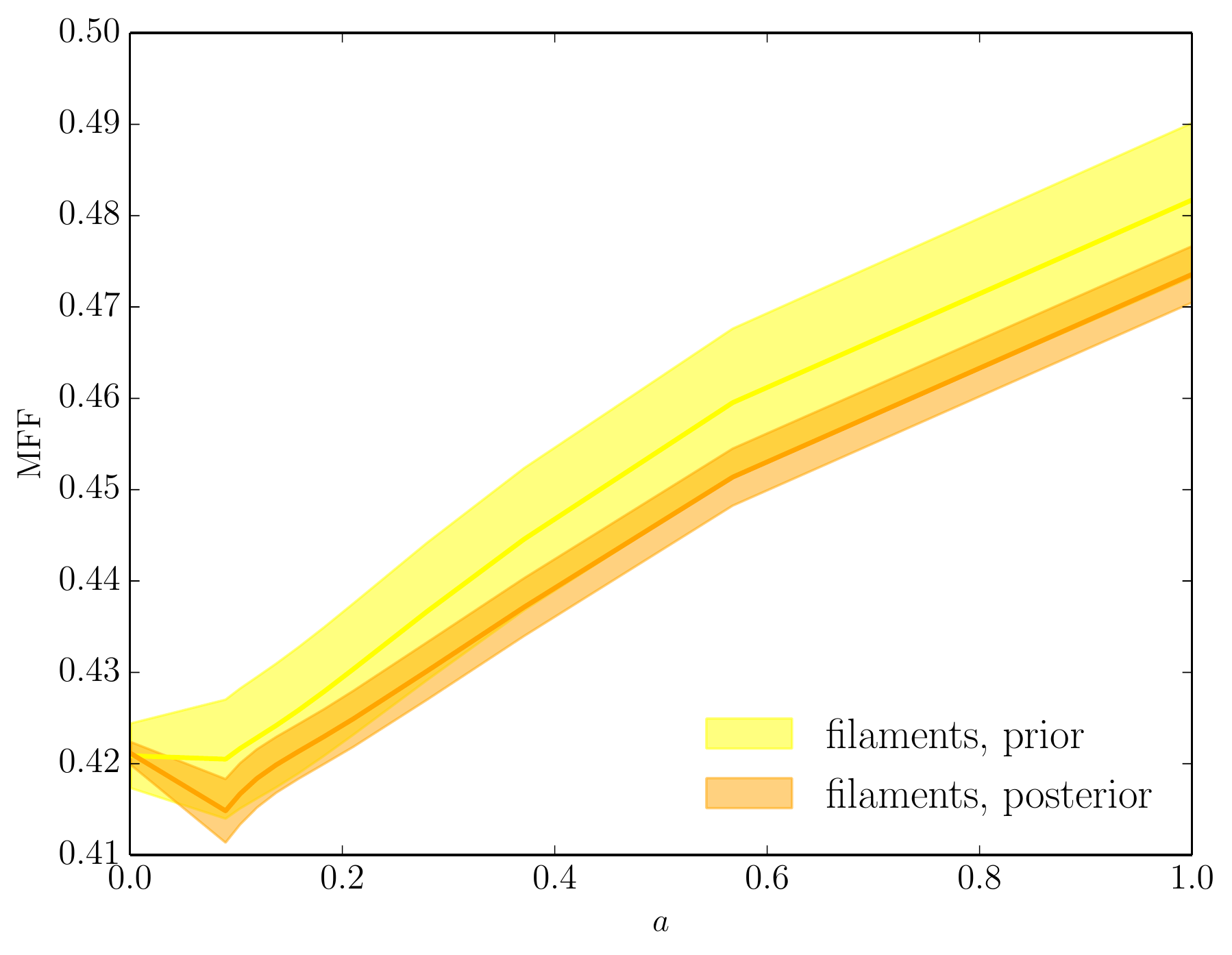} \\
\includegraphics[width=0.49\textwidth]{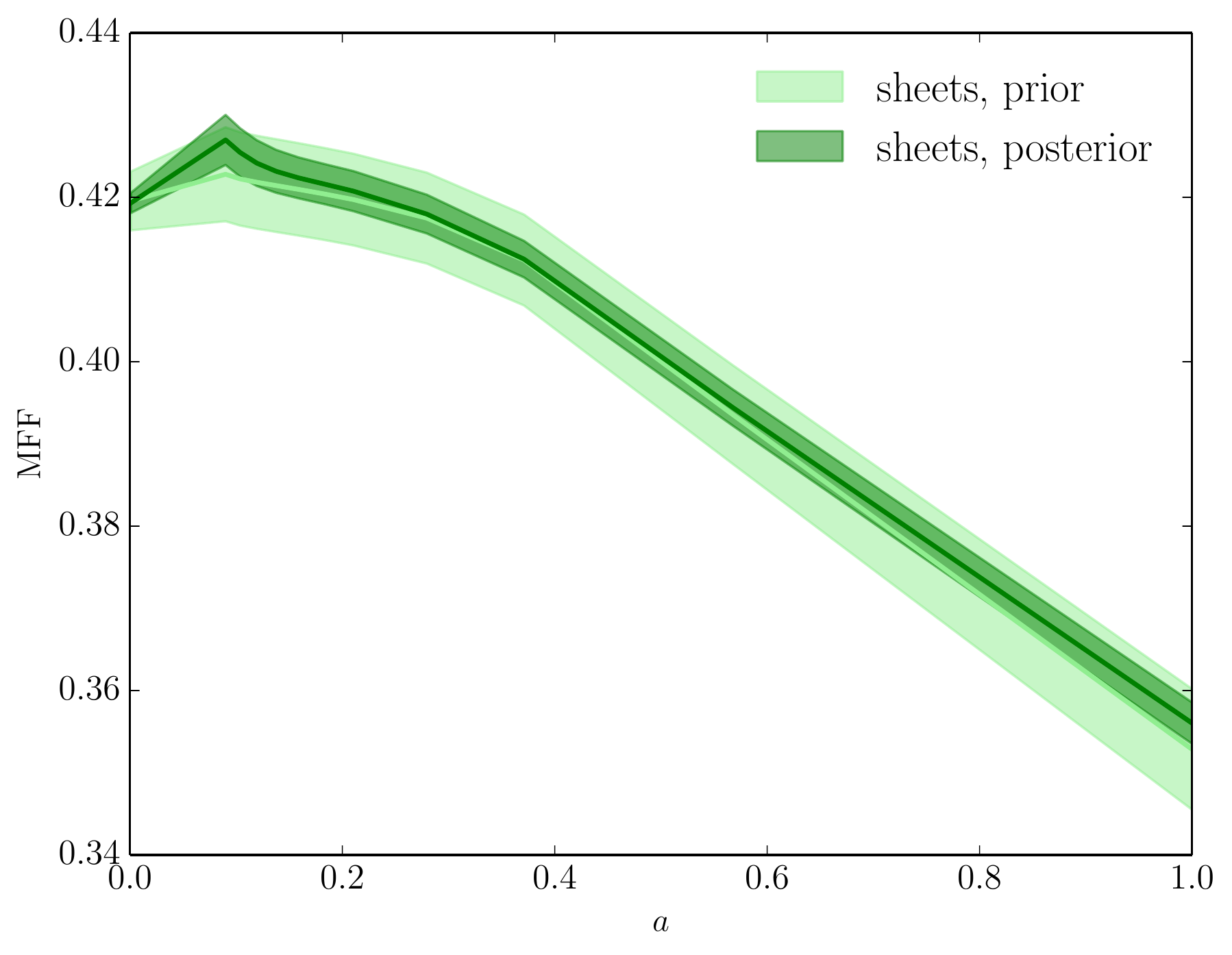}
\includegraphics[width=0.49\textwidth]{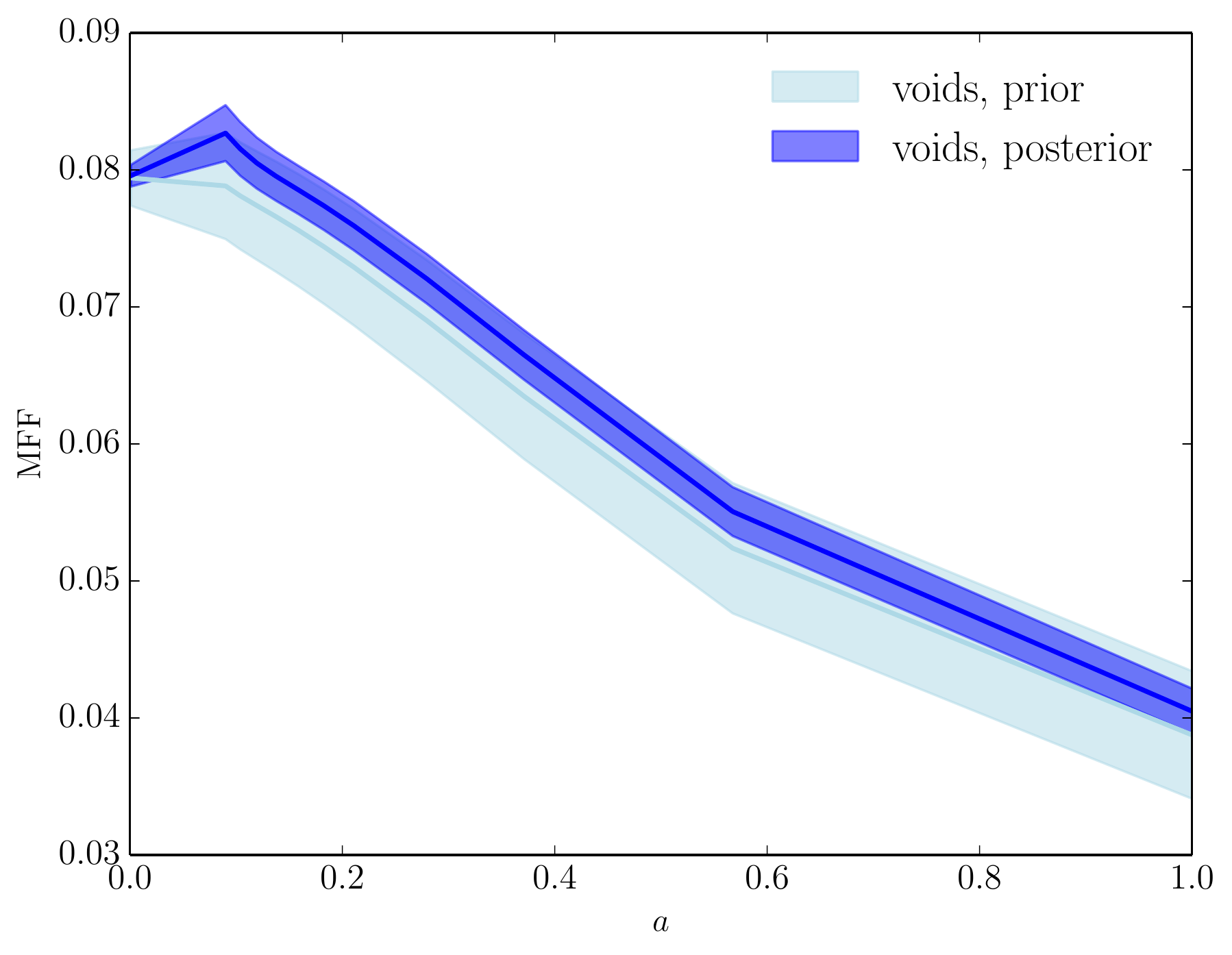} \\
\caption{Same as figure \ref{fig:evolution_vff} but for the \glslink{MFF}{mass filling fractions}.\label{fig:evolution_mff}}
\end{center}
\end{figure*}

In addition to the \gls{inference} of \glslink{initial conditions}{initial} and \glslink{final conditions}{final} \glslink{density field}{density fields}, {\borg} allows to simultaneously analyze the \gls{formation history} and morphology of the observed \glslink{LSS}{large-scale structure}, a subject that we refer to as \gls{chrono-cosmography}. In this section, we discuss the evolution of the \gls{cosmic web} from its origin ($a=10^{-3}$, analyzed in section \ref{sec:The primordial large-scale structure}) to the present epoch ($a=1$, analyzed in section \ref{sec:The late-time large-scale structure}). To do so, we use $11$ snapshots saved during the \textsc{cola} \glslink{non-linear filtering}{filtering} of our results (see section \ref{sec:Non-linear filtering of samples with COLA}). These are linearly separated in \gls{redshift} from $z=10$ to $z=0$. We perform this analysis in the $1,097$ \glslink{sample}{samples} filtered with \textsc{cola} considered in section \ref{sec:The late-time large-scale structure}. For each of these \glslink{sample}{samples} and for each \gls{redshift}, we follow the procedure described in sections \ref{sec:Non-linear filtering of samples with COLA} and \ref{sec:Classification of the cosmic web} to compute the \gls{density field} and to \glslink{cosmic web classification}{classify} the \glslink{structure type}{structure types}.

\subsection{Evolution of the probabilistic maps}
\label{sec:Evolution of the probabilistic maps}

We followed the time evolution of the probabilistic web-type maps from the primordial (figure \ref{fig:pdf_initial}) to the \glslink{final conditions}{late-time} \glslink{LSS}{large-scale structure} (figure \ref{fig:pdf_final}). In unconstrained regions, these maps show the evolution of the \gls{prior} preference for specific \glslink{structure type}{structure types} (see tables \ref{tb:prior_final} and \ref{tb:prior_initial}), in particular the breaking of the initial symmetry between \glslink{void}{voids} and \glslink{cluster}{clusters} and between \glslink{sheet}{sheets} and \glslink{filament}{filaments}, discussed in section \ref{sec:Probabilistic web-type cartography initial}.

In data-constrained regions, the time evolution of web-type maps permits to visually check the \gls{expansion} history of individual regions where the \gls{posterior} probability of one specific structure is high. In particular, it is easy to see that, as expected from their dynamical definition, \glslink{void}{voids} expand and \glslink{cluster}{clusters} shrink in \gls{comoving coordinates}, from $a=10^{-3}$ to $a=1$ (the reader is invited to compare the leftmost and rightmost panels of figures \ref{fig:pdf_final} and \ref{fig:pdf_initial}). Similarly, regions corresponding with high probability to \glslink{sheet}{sheets} and \glslink{filament}{filaments} expand along two and one axis, respectively, and shrink along the others. This phenomenon is more difficult to see in slices, however, as the slicing plane intersects randomly the eigendirections of the \gls{tidal tensor}. 

The time evolution of maps of the web-type \gls{posterior} \gls{entropy} (absolute and relative to the \gls{prior}) also exhibit some interesting features. There, it is possible to simultaneously check the increase of the \gls{information content} of the \gls{prior} (from $H \approx 1.6$~Sh to $H \approx 1.4$~Sh) and the \glslink{Lagrangian transport}{displacement of observational information} operated by the physical model. As the \glslink{LSS}{large-scale structure} forms in the \glslink{SDSS}{Sloan} volume, \gls{data} constraints are propagated and the complex structure of the \glslink{final conditions}{final} \gls{entropy} map (figure \ref{fig:pdf_final_entropy}), discussed in section \ref{sec:Probabilistic web-type cartography final}, takes shape.

\subsection{Volume filling fraction}

Our ensemble of snapshots allows us to check the time evolution of global characterizations of the \glslink{LSS}{large-scale structure} such as the \glslink{VFF}{volume} and \glslink{MFF}{mass filling fractions} of different structures. As in sections \ref{sec:Volume and mass filling fractions final} and \ref{sec:Volume and mass filling fractions initial}, we computed these quantities using only the volume where the \gls{survey response operator} is non-zero. In figure \ref{fig:evolution_vff}, we plot these \gls{VFF} as a function of the \gls{scale factor}. There, the solid lines correspond to the \gls{pdf} means and the shaded regions to the 2-$\sigma$ credible intervals, with light colors for the \glslink{prior}{priors} and dark colors for the \glslink{posterior}{posteriors}.

The time variation of the \gls{VFF} in figure \ref{fig:evolution_vff} is consistent with the expected dynamical behavior of structures. As \glslink{void}{voids} and \glslink{sheet}{sheets} expand along three and two axes, respectively, their \glslink{VFF}{volume fraction} increases. Here, the \gls{posterior} probabilities are mild updates of this prediction. Conversely, as \glslink{cluster}{clusters} and \glslink{filament}{filaments} shrink along three and two axes, respectively, their \glslink{VFF}{volume fraction} decreases. An explanation for the substantial displacement of the \gls{posterior} from the \gls{prior}, observed for \glslink{cluster}{clusters}, can be found in section \ref{sec:Volume and mass filling fractions final}.

As already noted, the \gls{VFF} is a very sensitive function of the precise definition of structures, grid size, \glslink{mesh assignment}{density assignment} scheme, smoothing scale, etc. For this reason, even for \gls{prior} probabilities, our results can be in qualitative disagreement with previous authors \citep[e.g. figure 23 in][]{Cautun2014}, due to their very different definitions of structures. Therefore, we only found relevant to compare our \gls{posterior} results with the \gls{prior} predictions based on unconstrained realizations. The same remark applies to the \gls{MFF} in the following section.

\subsection{Mass filling fraction}

In figure \ref{fig:evolution_mff}, we show the time evolution of the \glslink{MFF}{mass filling fractions} using the same plotting conventions. Results are consistent with an interpretation based on large scale flows of matter. According to this picture, \glslink{void}{voids} always loose mass while \glslink{cluster}{clusters} always become more massive. The behavior of \glslink{sheet}{sheets} and \glslink{filament}{filaments} can in principle be more complex, since these regions have both inflows and outflows of matter depending on the detail of their expansion profiles. In our setup, we found that the number of axes along which there is expansion dominates in the determination of the balance of inflow versus outflow, for global quantities such as the \gls{MFF}. Therefore, \glslink{filament}{filaments} always gain mass and \glslink{sheet}{sheets} always loose mass. Summing up our \gls{prior} predictions, as they expand along at least two axes, matter flows out of \glslink{void}{voids} and \glslink{sheet}{sheets} and streams towards \glslink{filament}{filaments} and \glslink{cluster}{clusters}.

The \gls{posterior} probabilities slightly update this picture. Observations support smaller outflowing of matter from \glslink{void}{voids}. For structures globally gaining matter, the \glslink{prior}{priors} are displaced towards less massive \glslink{filament}{filaments} and more massive \glslink{cluster}{clusters}. All \gls{posterior} predictions fall within the $\sim2$-$\sigma$ credible interval from corresponding \gls{prior} means.

\section{Summary and Conclusion}
\label{sec:Conclusion}

Along with chapter \ref{chap:dmvoids} \citep{Leclercq2015DMVOIDS}, this work exploits the high quality of \glslink{large-scale structure inference}{inference} results produced by the application of the Bayesian code {\borg} \citep[chapter \ref{chap:BORG};][]{Jasche2013BORG} to the \glslink{SDSS}{Sloan Digital Sky Survey} main galaxy sample \citep[chapter \ref{chap:BORGSDSS};][]{Jasche2015BORGSDSS}. We presented a Bayesian \gls{cosmic web} analysis of the nearby Universe probed by the northern cap of the \gls{SDSS} and its surrounding. In doing so, we produced the first probabilistic, four-dimensional maps of dynamic \glslink{structure type}{structure types} in real observations.

As described in section \ref{sec:Bayesian large-scale structure inference with BORG}, our method relies on the physical \glslink{large-scale structure inference}{inference} of the \glslink{initial conditions}{initial} \gls{density field} in the \gls{LSS} \citep{Jasche2013BORG,Jasche2015BORGSDSS}. Starting from these, we generated a large set of \glslink{constrained simulation}{data-constrained realizations} using the fast {\cola} method (section \ref{sec:Non-linear filtering of samples with COLA}). The use of \gls{2LPT} as a physical model in the \gls{inference} process and of the \glslink{full gravity}{fully non-linear gravitational dynamics}, provided by {\cola}, as a filter allowed us to describe structures at the required statistical accuracy, by very well representing the full \glslink{high-order correlation function}{hierarchy of correlation functions}. Even though \gls{initial conditions} were inferred with the approximate \gls{2LPT} model, we checked that the clustering statistics of \glslink{constrained simulation}{constrained non-linear model evaluations} agree with theoretical expectations up to scales considered in this work. As described in section \ref{sec:Classification of the cosmic web}, we used the dynamic \glslink{cosmic web classification}{web-type classification} algorithm \glslink{T-web}{}proposed by \citet{Hahn2007a} to dissect the \gls{cosmic web} into \glslink{void}{voids}, \glslink{sheet}{sheets}, \glslink{filament}{filaments}, and \glslink{cluster}{clusters}.

In sections \ref{sec:The late-time large-scale structure} and \ref{sec:The primordial large-scale structure}, we presented the resulting maps of structures in the final and \gls{initial conditions}, respectively, and studied the distribution of global quantities such as \glslink{VFF}{volume fraction} and \glslink{MFF}{mass filling fractions}. In section \ref{sec:Evolution of the cosmic web}, we further analyzed the time evolution of our results, in a rigorous \glslink{chrono-cosmography}{chrono-cosmographic} framework.

For all results presented in this chapter, we demonstrated a thorough capability of \gls{uncertainty quantification}. Specifically, for all inferred maps and derived quantities, we got a probabilistic answer in terms of a \gls{prior} and a \gls{posterior} distribution. The variation between \glslink{sample}{samples} of the \gls{posterior} distribution \glslink{uncertainty quantification}{quantifies the remaining uncertainties} of various origins (in particular \gls{noise}, \gls{selection effects}, \gls{survey geometry} and galaxy \gls{bias}, see chapters \ref{chap:BORG} and \ref{chap:BORGSDSS} for a detailed discussion). Building upon our accurate probabilistic treatment, we looked at the \gls{entropy} of the \gls{structure type} \gls{posterior} and at the \glslink{Kullback-Leibler divergence}{relative entropy} between \gls{posterior} and \gls{prior}. In doing so, we quantified the \glslink{information content}{information gain} due to \gls{SDSS} galaxy data with respect to the underlying dynamic \gls{cosmic web} and analyzed how this \glslink{Lagrangian transport}{information is propagated} during cosmic \glslink{formation history}{history}. This study constitutes the first link between cosmology and \gls{information theory} using real \gls{data}.

In summary, our methodology yields an accurate cosmographic description of \glslink{structure type}{web types} in the \gls{non-linear regime} of \gls{structure formation}, permits to analyze their time evolution and allows a precise \gls{uncertainty quantification} in a full-scale Bayesian framework. These \glslink{large-scale structure inference}{inference} results can be used for a rich variety of applications, ranging from studying galaxies inside their environment to \glslink{cross-correlation}{cross-correlating} with other cosmological probes. They count among the first steps towards accurate \gls{chrono-cosmography}, the subject of simultaneously analyzing the morphology and \gls{formation history} of the inhomogeneous Universe.

\textit{Note added:} As we were finalizing the paper corresponding to this chapter \citep{Leclercq2015ST} for submission, the works by \citet{Zhao2015} and \citet{Shi2015} appeared where the relationship between \glslink{halo}{halos} and the \gls{cosmic web} environment defined by the \gls{tidal tensor} is being studied.

%% file: Chapter10/Chapter10Content.tex
\chapter{Cosmic-web type classification using decision theory}
\label{chap:decision}
\minitoc

\defcitealias{Stover2003}{Matthew}
\begin{flushright}
\begin{minipage}[c]{0.6\textwidth}
\rule{\columnwidth}{0.4pt}

``If no mistake have you made, yet losing you are... a different game you should play.''\\Master Yoda, in recollections of Mace Windu, \\
--- \citetalias{Stover2003} \citet{Stover2003}, \textit{Star Wars: Shatterpoint}

\vspace{-5pt}\rule{\columnwidth}{0.4pt}
\end{minipage}
\end{flushright}

\abstract{\section*{Abstract}

We propose a \glslink{decision theory}{decision criterion} for segmenting the \gls{cosmic web} into different \glslink{structure type}{structure types} (\glslink{void}{voids}, \glslink{sheet}{sheets}, \glslink{filament}{filaments}, and \glslink{cluster}{clusters}) on the basis of their respective probabilities and the strength of \gls{data} constraints. Our approach is inspired by an analysis of games of chance where the gambler only plays if a positive expected net gain can be achieved based on some degree of privileged information. The result is a general solution for classification problems in the face of uncertainty, including the option of not committing to a class for a candidate object. As an illustration, we produce high-resolution maps of \glslink{structure type}{web-type} constituents in the nearby Universe as probed by the \glslink{SDSS}{Sloan Digital Sky Survey} main galaxy sample. Other possible applications include the selection and labeling of objects in catalogs derived from astronomical \glslink{galaxy survey}{survey} data.}

\draw{This chapter is adapted from its corresponding publication, \citet{Leclercq2015DT}.}
\vspace{-25pt}
\draw{Credit: Leclercq \textit{et al}. 2015, A\&A, 576, L17. Reproduced with permission $\copyright$ ESO.}

\section{Introduction}

Building accurate maps of the \gls{cosmic web} from \glslink{galaxy survey}{galaxy surveys} is one of the most challenging tasks in modern cosmology. Rapid progress in this field took place in the last few years with the introduction of \glslink{large-scale structure inference}{inference} techniques based on \glslink{Bayesian statistics}{Bayesian} \gls{probability theory} \citep{Kitaura2009,Jasche2010a,Nuza2014,Jasche2015BORGSDSS}. This facilitates the connection between the properties of the \gls{cosmic web}, thoroughly analyzed in \glslink{N-body simulation}{simulations} \citep[e.g.][]{Hahn2007a,Aragon-Calvo2010,Cautun2014}, and observations \citep[see chapter \ref{chap:stats} and][for a review on the interface between theory and \gls{data} in cosmology]{Leclercq2014Varenna}.

In chapter \ref{chap:ts} \citep{Leclercq2015ST}, we conducted a fully probabilistic analysis of \glslink{structure type}{structure types} in the \gls{cosmic web} as probed by the \glslink{SDSS}{Sloan Digital Sky Survey} main galaxy sample. This study capitalized on the \gls{large-scale structure inference} performed by \citet[][chapter \ref{chap:BORGSDSS}]{Jasche2015BORGSDSS} using the {\borg} \citep[Bayesian Origin Reconstruction from Galaxies,][chapter \ref{chap:BORG}]{Jasche2013BORG} algorithm. As the full gravitational model of \gls{structure formation} \textsc{cola} \citetext{COmoving Lagrangian Acceleration, \citealp{Tassev2013}; see also section \ref{sec:The COLA method}} was used, our approach resulted in the first probabilistic and time-dependent \glslink{cosmic web classification}{classification of cosmic environments} at \glslink{non-linear regime}{non-linear scales} in physical realizations of the \glslink{LSS}{large-scale structure} conducted with real \gls{data}. Using the \citet{Hahn2007a} \glslink{T-web}{definition} \citep[appendix \ref{sec:apx-tweb}, see also its \glslink{V-web}{extensions},][]{Forero-Romero2009,Hoffman2012}, we obtained three-dimensional, time-dependent maps of the \gls{posterior} probability for each voxel to belong to a \gls{void}, \gls{sheet}, \gls{filament} or \gls{cluster}.

These \gls{posterior} probabilities represent all the available \gls{structure type} information in the observational \gls{data} assuming the framework of \glslink{LCDM}{$\Lambda\mathrm{CDM}$} cosmology. Since the \glslink{LSS}{large-scale structure} cannot be uniquely determined from observations, uncertainty remains about how to assign each voxel to a particular \gls{structure type}. The question we address in this chapter is how to proceed from the \gls{posterior} probabilities to a particular choice of assigning a \gls{structure type} to each voxel. \glslink{decision theory}{Decision theory} \citep[see, for example,][]{Berger1985} offers a way forward, since it addresses the general problem of how to choose between different actions under uncertainty. A key ingredient beyond the \gls{posterior} is the \gls{utility function} that assigns a quantitative profit to different actions for all possible outcomes of the uncertain quantity. The optimal decision is that which maximizes the expected utility.

After setting up the problem using our example and briefly recalling the relevant notions of Bayesian \gls{decision theory}, we will discuss different \glslink{utility function}{utility functions} and explore the results based on a particular choice.

\section{Method}
\label{sec:Decision theory Method}

The \glslink{decision theory}{decision problem} for \glslink{cosmic web classification}{structure-type classification} can be stated as follows. We have four different \glslink{structure type}{web-types} that constitute the ``space of input features:'' \{$\mathrm{T}_0=$ \gls{void}, $\mathrm{T}_1=$ \gls{sheet}, $\mathrm{T}_2=$ \gls{filament}, $\mathrm{T}_3=$ \gls{cluster}\}. We want to either choose one of them, or remain undecided if the \gls{data} constraints are not sufficient. Therefore our ``space of actions'' consists of five different elements: \{$a_0=$ ``decide \gls{void},'' $a_1=$ ``decide \gls{sheet},'' $a_2=$ ``decide \gls{filament},'' $a_3=$ ``decide \gls{cluster},'' and $a_{-1}=$ ``do not decide.''\} The goal is to write down a \glslink{decision theory}{decision rule} prescribing which action to take based on the \gls{posterior} information.

Bayesian \gls{decision theory} states that the action $a_j$ that should be taken is that which maximizes the expected \gls{utility function} (\glslink{conditional pdf}{conditional} on the \gls{data} $d$), given in this example by
\begin{equation}
\label{eq:utility_function}
U(a_j(\vec{x}_k)|d) = \sum_{i=0}^3 G(a_j|\mathrm{T}_i) \, \p(\mathrm{T}_i(\vec{x}_k)|d) ,
\end{equation}
where $\vec{x}_k$ labels one voxel of the considered domain, $\p(\mathrm{T}_i(\vec{x}_k)|d)$ are the \gls{posterior} probabilities of the different \glslink{structure type}{structure types} given the \gls{data}, and $G(a_j|\mathrm{T}_i)$ are the \glslink{gain function}{gain functions} that state the profitability of each action, given the ``true'' underlying structure. Formally, $G$ is a mapping from the space of input features to the space of actions. For our particular problem, it can be thought of as a $5\times4$ matrix $\mathbf{G}$ such that $\mathbf{G}_{ij} \equiv G(a_j|\mathrm{T}_i)$, in which case eq. \eqref{eq:utility_function} can be rewritten as a linear algebra equation, $\mathbf{U}~=~\mathbf{G}.\mathbf{P}$ where the 5-vector $\mathbf{U}$ and the 4-vector $\mathbf{P}$ contain the elements $\mathbf{U}_j \equiv U(a_j(\vec{x}_k)|d)$ and $\mathbf{P}_i \equiv \p(\mathrm{T}_i(\vec{x}_k)|d)$, respectively.

Let us consider the choice of \glslink{gain function}{gain functions}. Several choices are possible. For example, the 0/1-\glslink{gain function}{gain functions} reward a correct \glslink{decision theory}{decision} with 1 for each voxel, while an incorrect \glslink{decision theory}{decision} yields 0. This leads to choosing the \gls{structure type} with the highest \gls{posterior} probability. While this seems like a reasonable choice, we need to consider that a \glslink{decision theory}{decision} is made in each voxel, whereas we are interested in identifying structures as objects that are made of many voxels. For instance, since \glslink{cluster}{clusters} are far smaller than \glslink{void}{voids}, the \textit{a priori} probability for a voxel to belong to a \gls{cluster} is much smaller than for the same voxel to belong to a \gls{void}. To treat different structures on an equal footing, it makes sense to reward the correct choice of \gls{structure type} $\mathrm{T}_i$ by an amount inversely proportional to the average volume $V_i$ of one such structure. In the following, we use the \gls{prior} probability as a proxy for the \glslink{VFF}{volume fractions},
\begin{equation}
\p(\mathrm{T}_i) \approx \frac{V_i}{V_0+V_1+V_2+V_3} .
\end{equation}
We further introduce an overall cost for choosing a structure with respect to remaining undecided, leading to the following specification of the utility,
\begin{equation}
\label{eq:gain_functions}
G(a_j|\mathrm{T}_i) = \left\{
\begin{array}{ll}
      \dfrac{1}{\p(\mathrm{T}_i)} - \alpha & \mathrm{if~} j \in \llbracket0,3\rrbracket \mathrm{~and~} i=j, \\
      -\alpha & \mathrm{if~} j \in \llbracket0,3\rrbracket \mathrm{~and~} i \ne j, \\
      0 & \mathrm{if~} j=-1.\\
\end{array} 
\right.
\end{equation}
This choice limits 20 free functions to only one free parameter, $\alpha$. With this set of \glslink{gain function}{gain functions}, making (or not) a \glslink{decision theory}{decision} between \glslink{structure type}{structure types} can be thought of as choosing to play or not to play a gambling game costing $\alpha$. Not playing the game, i.e. remaining undecided ($j =-1$), is always free ($G(a_{-1}|\mathrm{T}_i) = 0$ for all $i$). If the gambler decides to play the game, i.e. to make a \glslink{decision theory}{decision} ($j \in \llbracket0,3\rrbracket$), they pay $\alpha$ but may win a reward, $\frac{1}{\p(\mathrm{T}_i)}$, by betting on the correct underlying structure ($i=j$).

In the absence of \gls{data}, the \gls{posterior} probabilities in equation \eqref{eq:utility_function} are the \gls{prior} probabilities $\p(\mathrm{T}_i)$, which are independent of the position $\vec{x}_k$, and the \glslink{utility function}{utility functions} are, for $j \in \llbracket 0,3 \rrbracket$,
\begin{eqnarray}
U(a_j) & = & \sum_{i=0}^3 G(a_j|\mathrm{T}_i) \, \p(\mathrm{T}_i) \nonumber \\
& = & \left(\dfrac{1}{\p(\mathrm{T}_j)} - \alpha\right) \p(\mathrm{T}_j) - \sum_{\substack{i=0 \\ i\neq j}}^3 \alpha \, \p(\mathrm{T}_i) \nonumber \\
& = & 1 - \alpha \left( \p(\mathrm{T}_j) + \sum_{\substack{i=0 \\ i\neq j}}^3 \p(\mathrm{T}_i) \right) \nonumber \\
\label{eq:utility_function_prior_1}
& = & 1 - \alpha, \\
\mathrm{and} \quad U(a_{-1}) &=& 0.
\label{eq:utility_function_prior_2}
\end{eqnarray}
Equations \eqref{eq:utility_function_prior_1} and \eqref{eq:utility_function_prior_2} mean that, in the absence of \gls{data}, this reduces to the roulette game \gls{utility function}, where, if correctly guessed, \textit{a priori} unlikely outcomes receive a higher reward, inversely proportional to the fraction of the probability space they occupy. Betting on outcomes according to the \gls{prior} probability while paying $\alpha=1$ leads to a \textit{\gls{fair game}} with zero expected net gain. The gambler will always choose to play if the cost per game is $\alpha \leq 1$ and will never play if $\alpha > 1$.

The \gls{posterior} probabilities update the \gls{prior} information in light of the \gls{data}, providing an advantage to the gambler through privileged information about the outcome. In the presence of informative \gls{data}, betting on outcomes based on the \gls{posterior} probabilities will therefore ensure a positive expected net gain and the gambler will choose to play even if $\alpha>1$. Increasing the parameter $\alpha$ therefore represents a growing \textit{\glslink{risk aversion}{aversion for risk}} and limits the probability of losing. Indeed, for high $\alpha$, the gambler will only play in cases where the \gls{posterior} probabilities give sufficient confidence that the game will be won, i.e. that the \glslink{decision theory}{decision} will be correct.

\section{Maps of structure types in the SDSS}
\label{sec:Decision theory Maps}

\begin{figure*}
\begin{center}
\includegraphics[width=\textwidth]{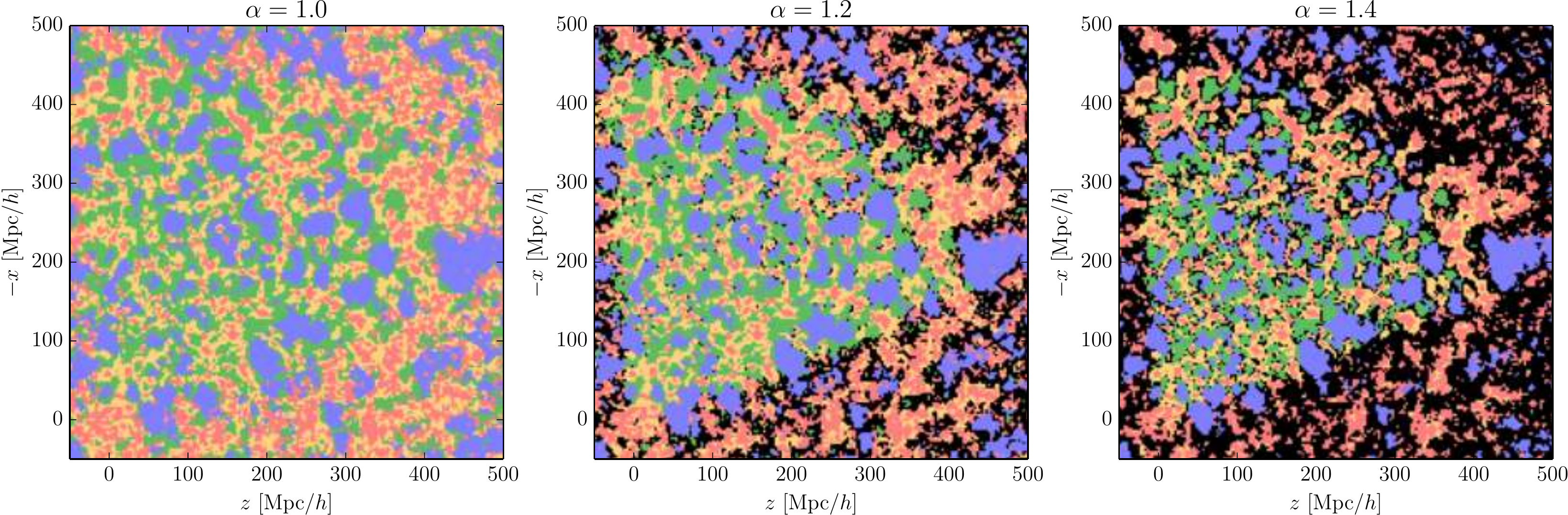}\\
\vspace{3pt}
\includegraphics[width=\textwidth]{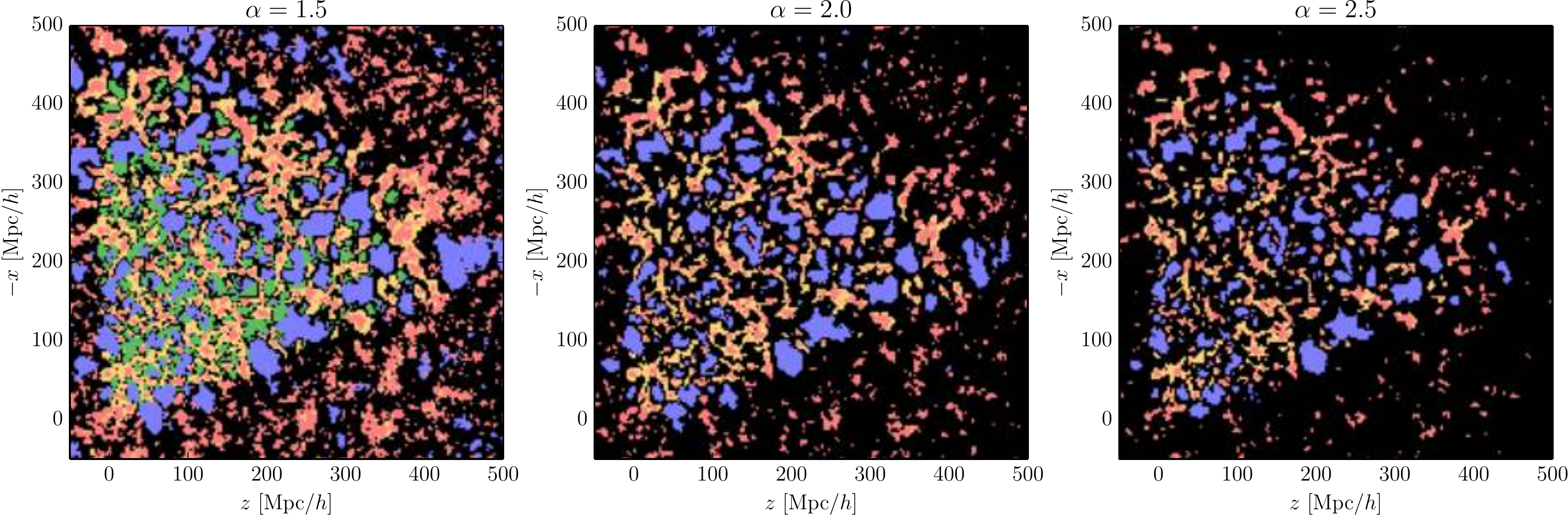}\\
\vspace{3pt}
\includegraphics[width=\textwidth]{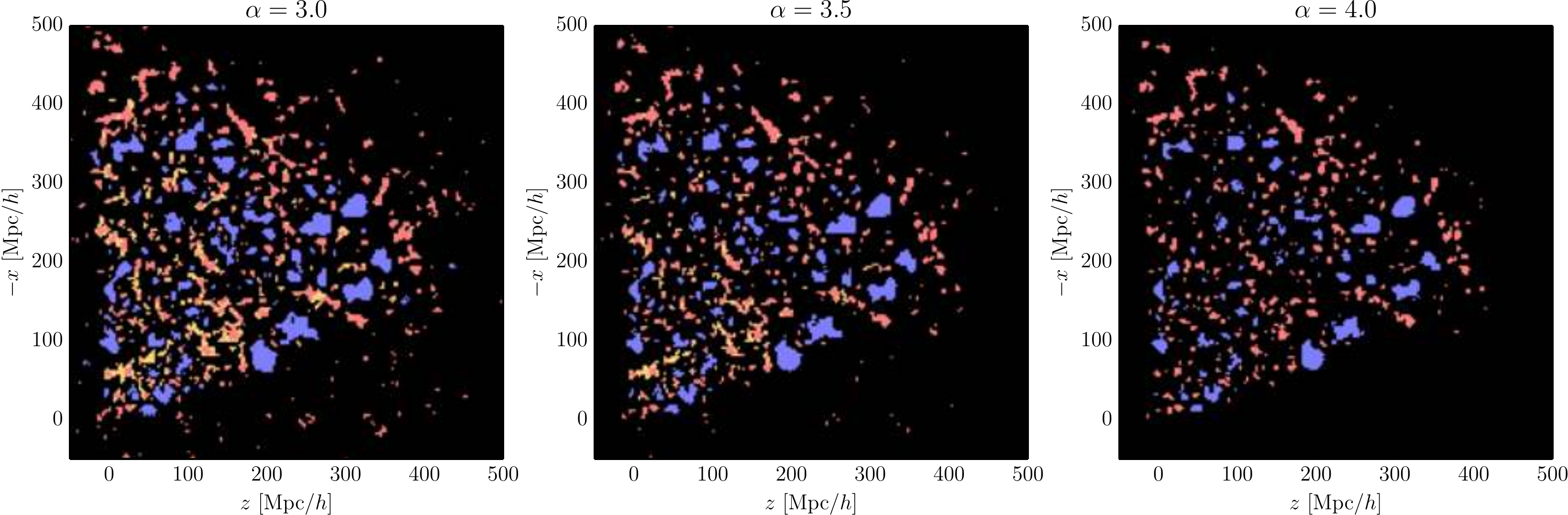}
\caption{Slices through maps of \glslink{structure type}{structure types} in the \glslink{final conditions}{late-time} \glslink{LSS}{large-scale structure}, at $a=1$. The color coding is blue for \glslink{void}{voids}, green for \glslink{sheet}{sheets}, yellow for \glslink{filament}{filaments}, and red for \glslink{cluster}{clusters}. Black corresponds to regions where \gls{data} constraints are insufficient to make a \glslink{decision theory}{decision}. The parameter $\alpha$, defined by equation \eqref{eq:gain_functions}, quantifies to the \gls{risk aversion} in the map: $\alpha=1.0$ corresponds to the most \gls{speculative map} of the \glslink{LSS}{large-scale structure}, and maps with $\alpha \geq 1$ are increasingly \glslink{conservative map}{conservative}. These maps are based on the \gls{posterior} probabilities inferred in chapter \ref{chap:ts} and on the Bayesian \glslink{decision theory}{decision rule} subject of the present chapter.\label{fig:decision_map_final}}
\end{center}
\end{figure*}

\begin{figure*}
\begin{center}
\includegraphics[width=\textwidth]{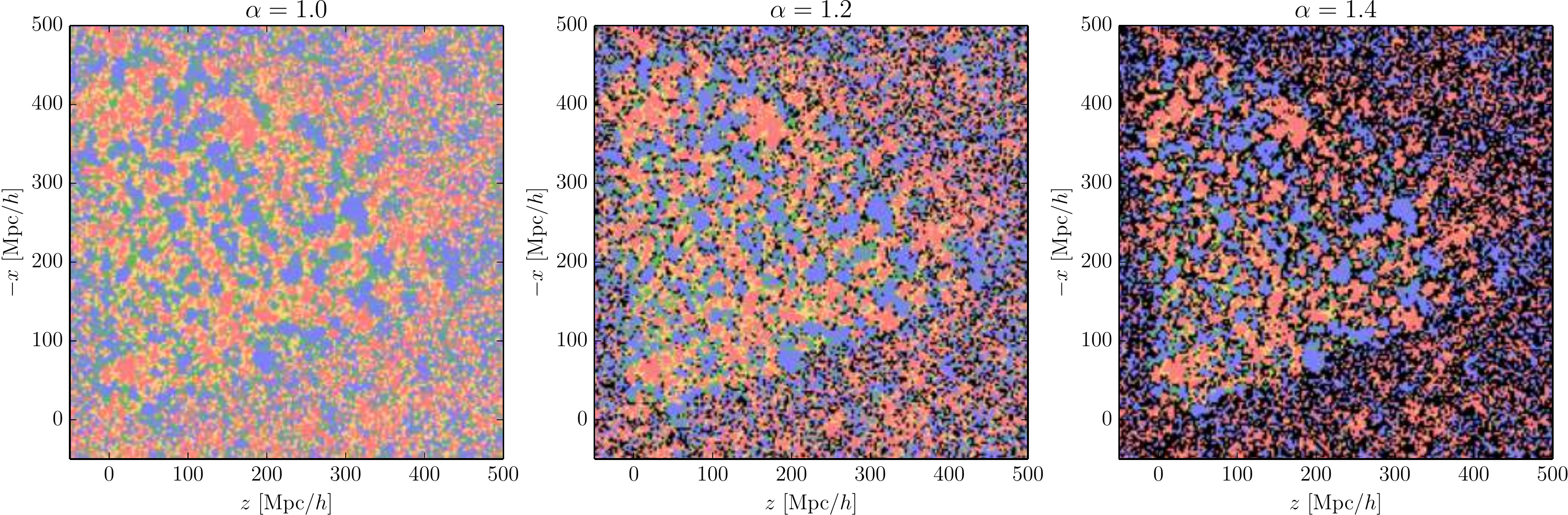}\\
\vspace{3pt}
\includegraphics[width=\textwidth]{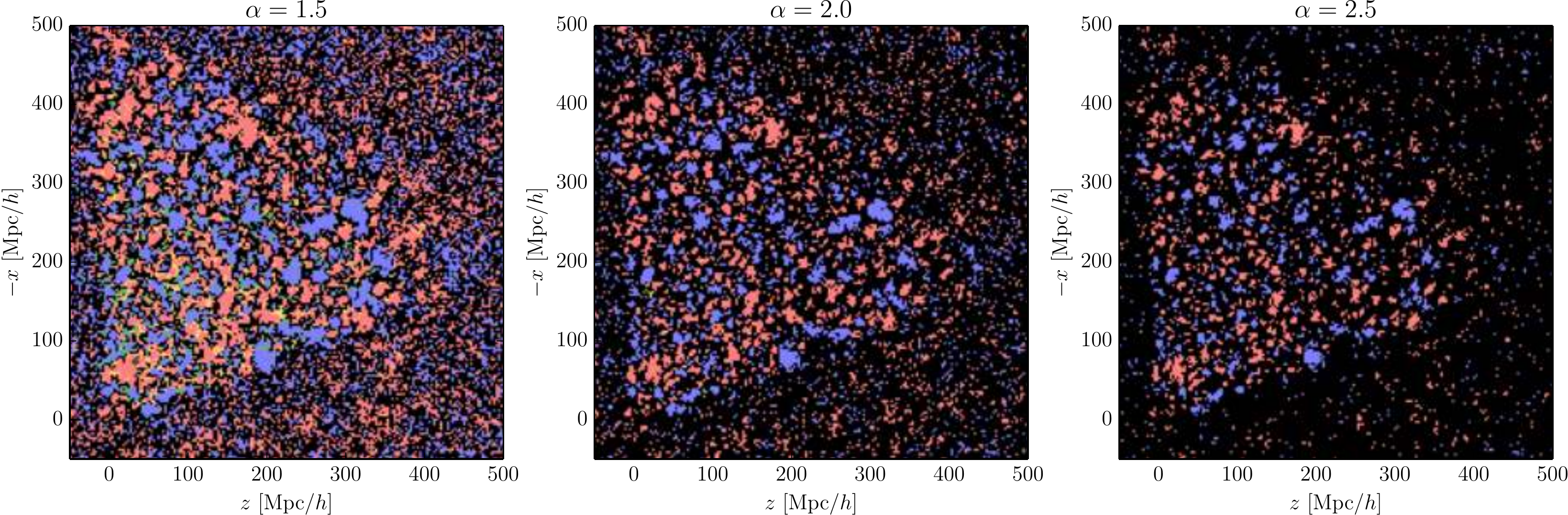}\\
\vspace{3pt}
\includegraphics[width=\textwidth]{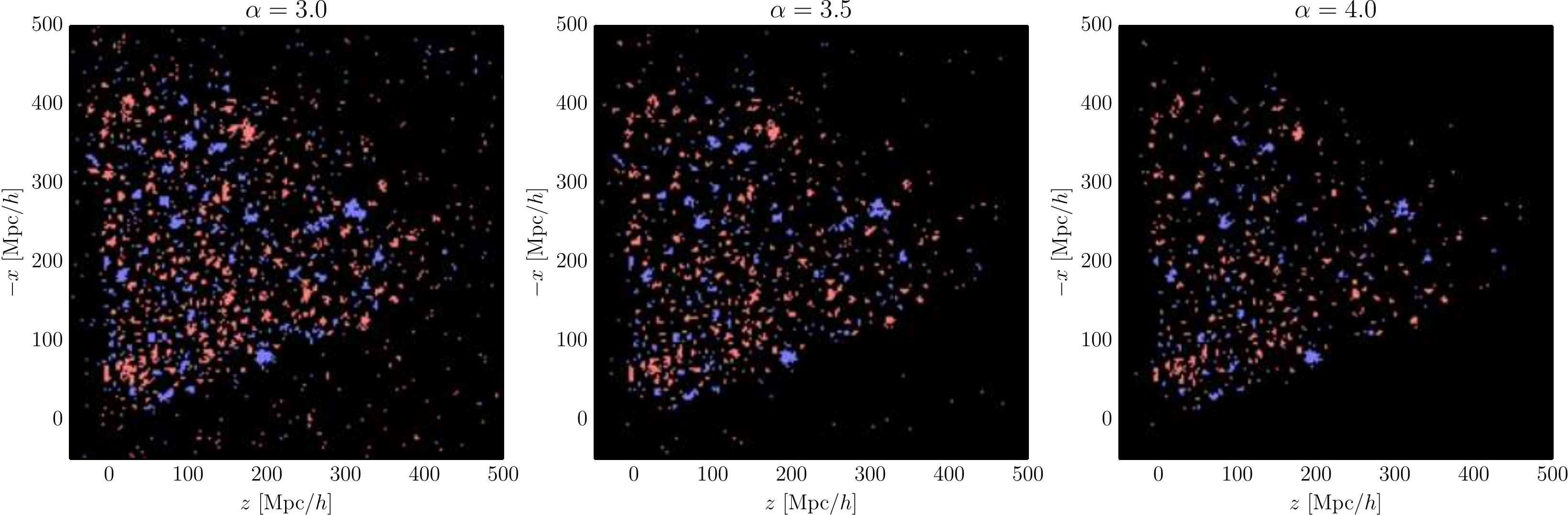}
\caption{Same as figure \ref{fig:decision_map_final} for the primordial \glslink{LSS}{large-scale structure}, at $a=10^{-3}$.
\label{fig:decision_map_initial}}
\end{center}
\end{figure*}

We applied the above \glslink{decision theory}{decision rule} to the \glslink{structure type}{web-type} \gls{posterior} probabilities presented in chapter \ref{chap:ts} \citep{Leclercq2015ST}, for different values of $\alpha \ge 1$ as defined by equation \eqref{eq:gain_functions}. In doing so, we produced various maps of the volume of interest, consisting of the northern Galactic cap of the \gls{SDSS} main galaxy sample and its surroundings. Slices through these three-dimensional maps are shown in figure \ref{fig:decision_map_final} for the \glslink{final conditions}{late-time} \glslink{LSS}{large-scale structure} (at $a=1$) and in figure \ref{fig:decision_map_initial} for the \glslink{initial conditions}{primordial} \glslink{LSS}{large-scale structure} (at $a=10^{-3}$).

When the game is \glslink{fair game}{fair} (namely when $\alpha =1$), it is always played, i.e. a \glslink{decision theory}{decision} between one of the four \glslink{structure type}{structure types} is always made. This results in the \textit{\gls{speculative map}} of \glslink{structure type}{structure types} (top left panel of figures \ref{fig:decision_map_final} and \ref{fig:decision_map_initial}). There, a \glslink{decision theory}{decision} is made even in regions that are not constrained by the \gls{data} (at high \gls{redshift} or outside of the \glslink{survey geometry}{survey boundaries}), based on \gls{prior} betting odds. 

By increasing the value of $\alpha >1$, we demand higher confidence in making the correct \glslink{decision theory}{decision}. This yields increasingly \textit{\glslink{conservative map}{conservative maps}} of the Sloan volume (see figures \ref{fig:decision_map_final} and \ref{fig:decision_map_initial}). In particular, at high values of $\alpha$, the algorithm makes \glslink{decision theory}{decisions} in the regions where \gls{data} constraints are strong (see figures \ref{fig:pdf_final_entropy} and \ref{fig:pdf_initial_entropy}), but often stays undecided in the unobserved regions. It can be observed that even at very high values, $\alpha \gtrsim 3$, a \glslink{decision theory}{decision} for one structure is made in some unconstrained voxels (typically in favor of the structure for which the reward is the highest: \glslink{cluster}{clusters} in the \gls{final conditions}, and \glslink{cluster}{clusters} or \glslink{void}{voids} in the \gls{initial conditions}). This effect is caused by the limited number of \glslink{sample}{samples} used in our analysis. Indeed, because of the finite length of the Markov Chain, the sampled representation of the \gls{posterior} has not yet fully converged to the true \gls{posterior}. For this reason, the numerical representation of the \gls{posterior} can be artificially displaced too much from the \gls{prior}, which results in an incorrect \glslink{structure type}{web-type} \glslink{decision theory}{decision}. This effect could be mitigated by obtaining more \glslink{sample}{samples} in the original {\borg} analysis (for an increased computational cost); or can be avoided by further increasing $\alpha$, at the expense of also degrading the map in the observed regions. We found the value of $\alpha=4$ (bottom right panel of figures \ref{fig:decision_map_final} and \ref{fig:decision_map_initial}) to be the best compromise between reducing the number of unobserved voxels in which a \glslink{decision theory}{decision} is made to a tiny fraction and keeping information in the volume covered by the \gls{data}.

As expected, structures for which the \gls{prior} probabilities are the highest disappear first from the map when one increases $\alpha$: betting on these structures being poorly rewarded, this choice is avoided in case of high \gls{risk aversion}. In the \gls{final conditions} (figure \ref{fig:decision_map_final}), we found that \glslink{sheet}{sheets} completely disappear for $\alpha \approx 1.68$ and \glslink{filament}{filaments} for $\alpha \approx 4.01$. In the \gls{initial conditions} (figure \ref{fig:decision_map_initial}), the critical value is around $\alpha \approx 2.36$ for both \glslink{sheet}{sheets} and \glslink{filament}{filaments}. In the most \glslink{conservative map}{conservative maps} displayed in figures \ref{fig:decision_map_final} and \ref{fig:decision_map_initial} ($\alpha = 4.0$), the \gls{SDSS} data provide extremely high evidence for the \glslink{void}{voids} and \glslink{cluster}{clusters} shown. In constrained parts, extended regions belonging to a given \gls{structure type} may not have the expected shape. This is true in particular for \glslink{filament}{filamentary} regions. Several factors can explain this: first, slicing through \glslink{filament}{filaments} make them appear as dots; second, with the dynamic \gls{T-web} definition, \gls{filament} regions often extend out into \glslink{sheet}{sheets} and \glslink{void}{voids}, and their static skeleton geometry is not the most prominent at the voxel scale (3 Mpc/$h$ in this work).

As detailed in chapters \ref{chap:BORG} and \ref{chap:BORGSDSS}, \gls{data} constraints are propagated by the \gls{structure formation} model assumed in the \gls{inference} process (\glslink{2LPT}{second-order Lagrangian perturbation theory}) and therefore radiate out of the \gls{SDSS} \glslink{survey geometry}{boundaries}. For this reason, for moderate values of $\alpha$, \glslink{cosmic web classification}{web-type classification} can be extended beyond the \glslink{survey geometry}{survey boundaries} to regions influenced by \gls{data}. This can be observed in figures \ref{fig:decision_map_final} and \ref{fig:decision_map_initial}, where one can see, for instance, that the shape of \glslink{void}{voids} that intersect the \gls{mask} is correctly recovered. Similarly, the \glslink{cosmic web classification}{classification} of high-\gls{redshift} structures confirms that the treatment of \gls{selection effects} by {\borg} is correctly propagated to \glslink{structure type}{web-type} analysis.

We finally comment on the required computational resources for the complete chain for running {\borg}, computing the \glslink{structure type}{web-type} \gls{posterior}, and making a \glslink{decision theory}{decision}. \glslink{large-scale structure inference}{Inference} with {\borg} is the most expensive part: on average, one \gls{sample} is generated in $1500$ seconds on $16$ cores \citep[chapter \ref{chap:BORGSDSS};][]{Jasche2015BORGSDSS}. Then, in each \gls{sample}, \glslink{tidal field}{tidal shear} analysis \citep[chapter \ref{chap:ts};][]{Leclercq2015ST} is a matter of a few seconds. Once the \glslink{structure type}{web-type} \gls{posterior} is known, making a \glslink{decision theory}{decision}, which is the subject of the present chapter, is almost instantaneous. Therefore, once the \gls{density field} has been inferred, which is useful for a much larger variety of applications, our method is substantially cheaper than several state-of-the-art techniques for \gls{cosmic web} analysis \citep[e.g. the method of ][for detecting \glslink{filament}{filaments}]{Tempel2013,Tempel2014}.

\section{Conclusions}
\label{sec:Conclusions}

In this chapter, we proposed a rule for optimal \glslink{decision theory}{decision making} in the context of \gls{cosmic web classification}. We described the problem set-up in Bayesian \gls{decision theory} and proposed a set of \glslink{gain function}{gain functions} that permit an interpretation of the problem in the context of game theory. This framework enables the dissection of the \gls{cosmic web} into different \glslink{structure type}{elements} (\glslink{void}{voids}, \glslink{sheet}{sheets}, \glslink{filament}{filaments}, and \glslink{cluster}{clusters}) given their \gls{prior} and \gls{posterior} probabilities and naturally accounts for the strength of \gls{data} constraints.

As an illustration, we produced three-dimensional templates of \glslink{structure type}{structure types} with various \gls{risk aversion}, describing a volume covered by the \gls{SDSS} main galaxy sample and its surrounding. These maps constitute an efficient statistical summary of the \gls{inference} results presented in chapter \ref{chap:ts} \citep{Leclercq2015ST} for \glslink{cross-correlation}{cross-use} with other astrophysical and cosmological \gls{data} sets.

Beyond this specific application, our approach is more generally relevant to the solution of classification problems in the face of uncertainty. For example, the construction of catalogs from astronomical surveys is directly analogous to the problem we describe here: it simultaneously involves a \glslink{decision theory}{decision} about whether or not to include a candidate object and which class label (e.g. star or galaxy) to assign to it.

%% file: Conclusion/ConclusionContent.tex
\chapter*{Summary, Conclusion and Outlook}
\label{chap:ccl}
\renewcommand{\leftmark}{Summary, Conclusion and Outlook}
\renewcommand{\rightmark}{Summary, Conclusion and Outlook}
\addstarredchapter{Summary, Conclusion and Outlook}

\defcitealias{Poincare1900}{Henri}
\begin{flushright}
\begin{minipage}[c]{0.6\textwidth}
\rule{\columnwidth}{0.4pt}

``Je crois qu'on obtiendra des r\'esultats \'etonnants. C'est justement pour cela que je ne puis rien vous en dire; car si je les pr\'evoyais, que leur resterait-il d'\'etonnant?''\\
--- \citetalias{Poincare1900} \citet{Poincare1900}

\vspace{-5pt}\rule{\columnwidth}{0.4pt}
\end{minipage}
\end{flushright}

\section*{Summary}

The main subject of this thesis is the process of \gls{data assimilation} for the analysis of the cosmological \glslink{LSS}{large-scale structure}. It aims at finding the best realizations of a physical model of \gls{structure formation} in light of the \gls{data}, while fully accounting for all uncertainties inherent to the \glslink{large-scale structure inference}{inference} problem. To this end, the {\borg} (Bayesian Origin Reconstruction from Galaxies) algorithm derives the \gls{initial conditions} and produces physical large-scale structure \glslink{reconstruction}{reconstructions}, by assimilating \glslink{galaxy survey}{survey} \gls{data} into a cosmological model. It is an inference engine that allows the simultaneous analysis of the morphology and the \gls{formation history} of the \gls{cosmic web}, a subject introduced in this work that we refer to as \gls{chrono-cosmography}. The present thesis contributed to the establishment of physical \gls{large-scale structure inference} as a functional and effective tool for the analysis of real \glslink{galaxy survey}{survey} \gls{data}.

We started by a review of the standard picture of \gls{LSS} formation, discussing the \glslink{gravitational evolution}{gravitational self-evolution} of the \gls{dark matter} \gls{fluid} and introducing cosmological perturbation theory (chapter \ref{chap:theory}). We then examined the accuracy of \glslink{LPT}{Lagrangian perturbation theory}, a tool widely applied in \gls{LSS} data analysis and also a key ingredient of the {\borg} algorithm (chapter \ref{chap:lpt}). We characterized the approximation error in \glslink{particle realization}{particle realizations} produced by \gls{LPT} at \glslink{ZA}{first} or \glslink{2LPT}{second order} instead of \glslink{full gravity}{fully non-linear gravity}. In particular, we analyzed the \glslink{one-point distribution}{one-}, \glslink{two-point correlation function}{two-} and \glslink{three-point correlation function}{three-point statistics} of the \gls{density field}, examined the \gls{displacement field} and compared the volume of different \glslink{structure type}{cosmic web elements}. In spite of visual similarities, we found that \gls{LSS} realizations produced by \gls{LPT} and by \glslink{N-body simulation}{$N$-body simulations} can drastically differ in some regimes, an effect of general interest for data analysis. 

Since unique recovery of signals from \gls{data} subject to observational effects (\glslink{survey geometry}{incomplete sky coverage}, \gls{selection effects}, \glslink{bias}{biases} and \gls{noise}) is not possible, {\borg} uses a \glslink{Bayesian statistics}{Bayesian approach} to \glslink{uncertainty quantification}{quantify uncertainties}. We discussed the fundamental concepts, mathematical framework and computer implementation of Bayesian \gls{probability theory} (chapter \ref{chap:stats}). Building upon these notions, we introduced physical \gls{large-scale structure inference} with the {\borg} algorithm (chapter \ref{chap:BORG}). We exposed the \gls{data model} and described the numerical \glslink{sampling}{sampler}, based on the \glslink{HMC}{Hamiltonian Monte Carlo} technique. Our approach allows the exploration of the \gls{posterior} distribution, which lives in an extremely \gls{high-dimensional parameter space} (usually of the order of $10$ million free parameters), in computationally reasonable times. As a result, it provides a \glslink{sample}{sampled} representation of the \glslink{LSS}{large-scale structure} inferred from the \gls{data}, in the form of four-dimensional cosmographic maps of the matter distribution. We applied the {\borg} algorithm to the \glslink{SDSS}{Sloan Digital Sky Survey} main galaxy sample data and presented our analysis (chapter \ref{chap:BORGSDSS}). Our results constitute accurate three-dimensional \glslink{reconstruction}{reconstructions} of the \glslink{final conditions}{present} \glslink{density field}{density} and \glslink{velocity field}{velocity fields}, but also of the \gls{initial conditions} and of the \gls{formation history} of the large-scale structures in the observed domain.

The full \glslink{data assimilation}{data-assimilation problem} is deeply \glslink{non-linear evolution}{non-linear}, implying that \glslink{pdf}{probability distributions} for observed cosmic fields are far from a \glslink{grf}{multivariate Gaussian}. We discussed the challenges associated with a fully \glslink{non-linear evolution}{non-linear description} of late-time \gls{structure formation}. We proposed a fast method to improve the correspondence between \glslink{density field}{density fields} in approximate models and in full \glslink{N-body simulation}{numerical simulations} (chapter \ref{chap:remapping}). The technique relies on remapping the one-point distribution of approximate fields using information extracted from \glslink{N-body simulation}{simulations}, and allows to extend the validity of \gls{LPT} beyond \gls{shell-crossing}, in the \gls{mildly non-linear regime}. We introduced the concept of \gls{non-linear filtering} of {\borg} \glslink{sample}{samples} (chapter \ref{chap:filtering}). This procedure improves \glslink{constrained simulation}{constrained realizations} by augmenting them with physically plausible information at small scales. We checked the accuracy of the fast {\cola} scheme as a non-linear filter versus \textsc{\gls{Gadget-2}}, and used it to produce a large ensemble of non-linear {\borg}-{\cola} samples for subsequent use. 

We finally made use of our results for \gls{cosmic web} analysis. With the {\vide} toolkit, we produced and analyzed constrained catalogs of cosmic \glslink{void}{voids} in the \glslink{SDSS}{Sloan} volume. In doing so, we showed that the inference of \glslink{dark matter void}{voids at the level of the dark matter field}, deeper than with the galaxies, is achievable, and that suitable inference technology is capable of tapping a mine of information even in existing \glslink{galaxy survey}{surveys}. In particular, we found at least one order of magnitude more \glslink{void}{voids} at all scales considered, between $5$ and $40$ Mpc/$h$. As a consequence, our method yields a drastic reduction of \gls{statistical uncertainty} for the determination of void properties and carries a vast potential for their use as cosmological probes. We presented a probabilistic analysis of the dynamic \gls{cosmic web}, \glslink{cosmic web classification}{dissected} into \glslink{void}{voids}, \glslink{sheet}{sheets}, \glslink{filament}{filaments}, and \glslink{cluster}{clusters}, on the basis of the \gls{tidal field} (chapter \ref{chap:ts}). We examined the \glslink{formation history}{history} and characterized the \gls{information content} of our web-type maps. This study demonstrated that our inference framework allows self-consistent propagation of observational uncertainties to \gls{cosmic web} analysis, and counts among the pioneering steps toward a data-supported connection between cosmology and \gls{information theory}. Eventually, we introduced a new framework for optimal \glslink{decision theory}{decision-making} based on the web-type \gls{posterior} probabilities and the strength of data constraints (chapter \ref{chap:decision}). We obtained efficient statistical summaries of our inference results and outlined more general applications to classification problems in the face of uncertainty.

\section*{Conclusion and Outlook}

\subsection*{What does chrono-cosmography predict about the Universe?}

The aim of physical \gls{large-scale structure inference} is to provide a cosmographic description of a subvolume of the observable Universe, as well as a probabilistic characterization of uncertainties. In this fashion, we have access to a wealth of information on the present and past of this region. This material can be used in a variety of subsequent astrophysical and cosmological analyses, many of which can be already conducted based on the results obtained during this PhD project.

A first class of possible projects build upon the \gls{inference} of the \gls{initial conditions}, in which gravitational \gls{non-Gaussianity} is largely suppressed.

\paragraph{Genesis and growth of the cosmic web.} As we have shown, Bayesian \gls{large-scale structure inference} paves the path toward a high-fidelity description of the complex web-like patterns in cosmic structure. A natural follow-up project is to exploit the richness of the information inferred by {\borg}, including quantities so far only accessible in simulations, to build and compare \glslink{cosmic web classification}{classifications} of the \gls{cosmic web}. These will use, for example: the Lagrangian \gls{displacement field} \citep[\textsc{diva},][]{Lavaux2010} or the stretchings and foldings of the \gls{dark matter} \glslink{phase space}{phase-space} sheet \citep[\textsc{origami},][]{Falck2012}. 

\paragraph{Primordial non-Gaussianity and inflation.} The \textsc{borg} algorithm has the potential to accurately characterize the statistics of primordial seeds. In particular, estimators of \glslink{bispectrum}{bispectra} \citep{Schmittfull2015} or based on the \glslink{phase}{phases} of inferred fields \citep{Obreschkow2013,Wolstenhulme2014} -- for which no \gls{prior} information is assumed -- may detect signatures of primordial \gls{non-Gaussianity} or constrain models of \gls{inflation}. While it will be extremely challenging for \gls{LSS} measurements to improve upon \gls{CMB} constraints on \gls{inflation}, the theoretically interesting threshold for many models involving \glslink{non-Gaussianity}{deviations from Gaussian} \gls{initial conditions} has not yet been reached. Hence, it is crucial to develop new methods that will extract this \glslink{information content}{information} from the \gls{LSS}.

\paragraph{Galaxies within the large-scale structure.} Physical properties of galaxies (\gls{luminosity}, color, spin, morphological type, etc.) are known to be correlated with their large-scale environment \citep[see e.g.][]{Lee2008a,Park2010,Eardley2015}. \glslink{cosmic web}{Cosmic web} analyses enabled by {\borg} straightforwardly allow to test models of \gls{galaxy formation} and evolution within their time-varying environment. The resulting information could then be used in future \gls{large-scale structure inference} procedures, to obtain more refined information on the matter \gls{density field} traced by these galaxies.

\vspace*{15pt}

{\borg} provides a complete description of the \glslink{gravitational evolution}{gravitational dynamics} of the volume of interest. A second class of projects is to use this information and see what it implies for other cosmological observables. In this regard, we use {\borg} as a forecast-generating machine, whose predictions can be tested with complementary observations in the actual sky.

\paragraph{Effects of the inhomogeneous large-scale structure on photons.} Inferred information permits to produce various prediction templates for \glslink{cross-correlation}{cross-correlations} with other cosmological \gls{data} sets. In particular, it is possible to predict the effects of inhomogeneities in the \gls{LSS} on photon properties and geodesics, given galaxy observations: deviations in \gls{redshift} (in the radial direction) and \gls{weak gravitational lensing} effects (in the angular directions). In a similar fashion, dynamic information such as \glslink{velocity field}{velocity fields} and evolution of the \gls{gravitational potential} can be used to enhance the detectability of secondary effects expected in the \glslink{CMB}{cosmic microwave background}, such as the kinetic \gls{Sunyaev-Zel'dovich effect}, the \gls{integrated Sachs-Wolfe effect}, and the non-linear \gls{Rees-Sciama effect}.

\paragraph{Cosmological parameters, baryon acoustic oscillations and dynamic dark energy.} The incorporation of a physical model in the likelihood provides a natural way to infer \gls{cosmological parameters} from observations. The work presented in this thesis is also expected to provide an alternative way to reconstruct the \glslink{BAO}{baryon acoustic oscillation} signal \citep{Padmanabhan2012} and to infer the \gls{equation of state} of a possible \gls{dark energy} component. This approach will yield a more precise picture of the \gls{expansion} history of the Universe and help to understand the origin of cosmic acceleration.

\subsection*{How do we include more aspects in the data model?}

Contrary to traditional approaches, which apply various cosmological tests to \gls{data} separately and combine constraints in a suboptimal fashion, the approach presented in this thesis automatically and fully self-consistently performs a joint analysis of all aspects. It models their interdependence and accounts for the ways in which different observables can mutually enhance one another. The joint analysis of all phenomena can be used to perform consistency tests of the \glslink{LCDM}{standard cosmological model} and has the potential to rule out some of its possible extensions.

However, many relevant aspects are still absent of the current {\borg} \gls{data model}: a fully \glslink{non-linear evolution}{non-linear} treatment of gravitational \gls{structure formation}, \gls{redshift-space distortions}, \gls{lightcone} effects, \glslink{photometric redshift}{photometric redshifts} uncertainty, density-dependent \gls{selection effects}, scale-dependent and stochastic galaxy \gls{bias} or predictions of non-standard cosmologies. The joint analysis of other probes of the \gls{LSS} (\gls{CMB lensing}, \glslink{weak gravitational lensing}{weak lensing} shear maps, etc.) should also be addressed in our framework, via a joint likelihood or sequential \gls{data assimilation}. The inclusion of these aspects in the \gls{LSS} \gls{data model} involves conceptual, but also technical challenges. Bayesian \gls{large-scale structure inference} is highly computationally expensive, to the degree that it touches the border of what is currently possible.

In the author's opinion, future progress will not only depend on adequate approximations, but also on the development of new methodological ways to implement \gls{sampling}. In comparison to the state-of-the-art \glslink{HMC}{Hamiltonian Monte Carlo} algorithm, efficient, advanced non-linear \gls{data assimilation} techniques will have to allow a much cheaper statistical \gls{inference} (presumably by several orders of magnitude), which will open the way for the inclusion of more physical effects.

\subsection*{What can ultimately be learned from the large-scale structure?}

How deterministic is the \glslink{structure formation}{formation of structure} in the Universe? In other words, at what scale does one-to-one mapping from \glslink{initial conditions}{initial} to \gls{final conditions} (valid at large, \glslink{linear regime}{linear} and \glslink{mildly non-linear regime}{weakly non-linear scales}) break down? Since state-of-the-art simulations are still very far from resolving all relevant physical processes, the issue of the scale at which \gls{structure formation} is non-deterministic is not yet considered crucial for numerical modeling. However, a theoretical understanding of this question would be of central interest in the context of an Effective Field Theory of the LSS \citep{Baumann2012,Carrasco2012,Senatore2014}. A quantification of the \gls{information content} of primordial patches that collapse to form structures (such as the Milky Way) characterizes the LSS in a fundamental way and contains a wealth of information on the properties of matter at the highest energies, far beyond the reach of particle colliders. 

High amount of primordial information contained in small-scale structures would for example disfavor \glslink{WDM}{warm dark matter} or any mechanism suppressing small-scale density fluctuations, and could even require further initial degrees of freedom such as \gls{isocurvature perturbations}. Issues related to the \gls{information content} of primordial patches have been recently speculatively examined by \cite{Neyrinck2015a}, who proposed, as a thought experiment, a test about the scale at which \gls{structure formation} is deterministic. Unfortunately, simulations do not provide much insight into the questions of where the \gls{information content} is, and how to optimally extract it from the \gls{data}.

Building upon the inference of \gls{initial conditions} from which the LSS originates, the first steps toward a practical implementation of a scale-dependent test of determinism in \gls{structure formation} can be taken. This task will involve the careful definition of a measure of complexity in Lagrangian patches inferred by \textsc{borg} and of a means to compare \glslink{initial conditions}{initial} and \glslink{final conditions}{final} \glslink{information content}{information}. It will also require careful analysis of information propagation via \gls{Lagrangian transport} within a fully probabilistic approach (see sections \ref{sec:cosmic_history} and \ref{sec:Evolution of the probabilistic maps}; figures \ref{fig:inside_out}, \ref{fig:pdf_final_entropy} and \ref{fig:pdf_initial_entropy} for a preparatory discussion). Further investigation will have to consider information sinks such as baryonic processes and black hole formation, and information sources that broadcast non-primordial randomness at large scales, such as supernovae, active galactic nuclei and unstable astrophysical phenomena. The link between astrophysical, thermodynamic entropy, as well as statistical, information-theoretic entropy will also have to be clarified.

\vspace*{50pt}

In the last few years, ESA’s \href{http://sci.esa.int/planck/}{Planck} mission confirmed our picture of the evolution of the \gls{homogeneous Universe} to spectacular accuracy and provided the highest precision probe to date of the physical origin of cosmic structure. Challenges for accurate cosmology now arise from studying the inhomogeneous cosmic structure. This research will provide an exceptionally detailed characterization of the \gls{cosmic web} underlying the observed galaxy distribution, extract information about the nature of \gls{dark energy}, and furnish unprecedentedly accurate information on the \gls{initial conditions} from which structure appeared in the Universe. Progress will not only depend on our ability to handle ever larger \gls{data} sets: crucial to the longer-term aims is developing efficient tools for \glslink{data assimilation}{assimilating data} into the forecasts of a physical model and quantifying the \gls{information content} uniquely encoded in the primordial large-scale structure. Only through such a quantitative statistical approach can we expand our understanding of the dynamic Universe and make significant progress on the age-old puzzles of cosmic beginning and ultimate fate of the Universe. I am confident that the methods and results described in this thesis, counting among the first steps towards precision \gls{chrono-cosmography}, will contribute to this endeavor.

%% file: AppendixA/AppendixAContent.tex
\part*{Appendices}
\renewcommand{\leftmark}{\leftmarkold}
\renewcommand{\rightmark}{\rightmarkold}
\addstarredpart{Appendices}
\adjustmtc[+1]

\chapter{Complements on Gaussian random fields}
\label{apx:complements GRFs}
\minitoc

\begin{flushright}
\begin{minipage}[c]{0.6\textwidth}
\rule{\columnwidth}{0.4pt}

``I have had my results for a long time: but I do not yet know how I am to arrive at them.''\\
--- Carl Friedrich Gau{\ss}\\
Quoted in \citet{Arber1954}, \textit{The Mind and the Eye}

\vspace{-5pt}\rule{\columnwidth}{0.4pt}
\end{minipage}
\end{flushright}

\abstract{\section*{Abstract} This appendix provides complements on \glslink{grf}{Gaussian random fields}. It offers a mathematical exposition of their definition and demonstrates well-known properties, used in particular in chapter \ref{chap:theory} and for the generation of \gls{initial conditions} for cosmological \glslink{N-body simulation}{simulations} (section \ref{sec:apx-Setting up initial conditions}).}

\section{Characteristic function}

\paragraph{Definition A.1.}For a random \glslink{scalar field}{scalar vector} $\lambda \in \mathbb{C}^n$ whose \gls{pdf} is $\p(\lambda)$, the \gls{characteristic function} $\varphi_\lambda$ is defined as the inverse \gls{Fourier transform} of $\p(\lambda)$. In other words, it is the expectation value of $\erm^{\i t^*\lambda}$, where $t \in \mathbb{C}^n$ is the argument of the \gls{characteristic function} \citep[e.g.][]{Manolakis2000}:
\begin{equation}
\varphi_\lambda(t) \equiv \left< \erm^{\i t^* \lambda} \right> = \int_\mathbb{C} \erm^{\i t^* \lambda} \, \p(\lambda) \, \drm \lambda .
\end{equation}

\glslink{characteristic function}{Characteristic functions} have well-known properties. In particular, an important theorem is the following.

\paragraph{Theorem A.2. (\gls{Kac's theorem}).\label{par:thm-A-2}} Let $\lambda_1, \lambda_2 \in \mathbb{C}^n$ be random vectors. The following statements are equivalent:
\begin{enumerate}
\item $\lambda_1$ and  $\lambda_2$ are \glslink{conditional independence}{independent} (we note $\lambda_1 \ci \lambda_2$),
\item the \gls{characteristic function} of the joint random vector $(\lambda_1, \lambda_2)$ is the product of the \glslink{characteristic function}{characteristic functions} of $\lambda_1$ and $\lambda_2$ i.e. $\varphi_{(\lambda_1, \lambda_2)} = \varphi_{\lambda_1}\varphi_{\lambda_2}$.
\end{enumerate}

\begin{proof} 1. $\Rightarrow$ 2. is straightforward using $\left< f(\lambda_1) g(\lambda_2) \right> = \left< f(\lambda_1) \right> \left< g(\lambda_2) \right>$.

2. $\Rightarrow$ 1. Let $\widetilde{\lambda_1}$ and $\widetilde{\lambda_2}$ be random vectors such that $\widetilde{\lambda_1}$ and $\lambda_1$ have the same \gls{pdf}, $\widetilde{\lambda_2}$ and $\lambda_2$ have the same \gls{pdf} and $\widetilde{\lambda_1} \ci \widetilde{\lambda_2}$. Then
\begin{eqnarray}
\varphi_{(\lambda_1, \lambda_2)} & = & \varphi_{\lambda_1}\varphi_{\lambda_2} \quad \mathrm{using~2.} \nonumber \\
& = & \varphi_{\widetilde{\lambda_1}}\varphi_{\widetilde{\lambda_2}} \quad \mathrm{using~the~pdfs} \nonumber \\
& = & \varphi_{(\widetilde{\lambda_1}, \widetilde{\lambda_2})} \quad \mathrm{using~1.} \Rightarrow \mathrm{2.} \nonumber
\end{eqnarray}
i.e. the \glslink{characteristic function}{characteristic functions} of $(\lambda_1, \lambda_2)$ and $(\widetilde{\lambda_1}, \widetilde{\lambda_2})$ coincide. From the uniqueness of the inverse \gls{Fourier transform} we conclude that $(\lambda_1, \lambda_2)$ and $(\widetilde{\lambda_1}, \widetilde{\lambda_2})$ are drawn from the same distribution, hence $\lambda_1 \ci \lambda_2$.
\end{proof}

\section{General definition of a Gaussian random vector}

\paragraph{Definition A.3.}A multivariate random \glslink{scalar field}{scalar vector} $\lambda \in \mathbb{C}^n$ is a \glslink{grf}{Gaussian random vector} if and only if there exists a vector $\mu \in \mathbb{C}^n$ and a Hermitian, positive semi-definite matrix $C \in \mathcal{M}_n(\mathbb{C})$ such that the \gls{characteristic function} of $\lambda$ is
\begin{equation}
\varphi_\lambda(t) = \exp\left( \i t^* \mu -\frac{1}{2} t^*Ct \right) .
\label{eq:characteristic-function-GRF}
\end{equation}
In this case, $\mu$ and $C$ are called the mean and covariance matrix of $\lambda$, respectively, and we note $\lambda \sim \mathcal{N}_n\left[ \mu, C \right]$. Here, the covariance matrix is allowed to be singular. This definition generalizes the one given in section \ref{sec:Gaussian random fields}, as we see from the following theorem.

\paragraph{Theorem A.4.}When $C$ is positive-definite (and therefore invertible), the distribution of $\lambda$ has a multivariate normal density
\begin{equation}
\p(\lambda|\mu,C) = \frac{1}{\sqrt{\vert 2\pi C\vert}} \exp\left(-\frac{1}{2}(\lambda-\mu)^* C^{-1}(\lambda-\mu) \right) .
\end{equation}

\begin{proof} By explicitly computing the inverse \gls{Fourier transform} of the multivariate normal distribution above (i.e. calculating the Gaussian integral), we can check that the \gls{characteristic function} of this distribution coincides with the value of equation \eqref{eq:characteristic-function-GRF}. From the uniqueness of the inverse \gls{Fourier transform}, we conclude that $\lambda$ is drawn from the distribution whose \gls{pdf} is given above.
\end{proof}

When this condition is fulfilled, we say that $\lambda$ is \textit{non-degenerate}.

\section{Some well-known properties of Gaussian random vectors}

\paragraph{Proposition A.5.\label{sec:prop-A-5}}Linear transformations preserve Gaussianity, i.e. for all $A \in \mathcal{M}_{m \times n}(\mathbb{C})$ and $b \in \mathbb{C}^m$, if $\lambda \sim \mathcal{N}_n\left[ \mu, C \right]$, then $A\lambda+b \sim \mathcal{N}_m \left[ A\mu+b, ACA^* \right]$.

\begin{proof} The \gls{characteristic function} of $A\lambda+b$ is, for all $s \in \mathbb{C}^m$,
\begin{eqnarray}
\varphi_{A\lambda+b}(s) & = & \left< \erm^{\i s^*(A\lambda+b) } \right> \nonumber\\
& = & \left< \erm^{\i (A^*s)^*\lambda } \right> \erm^{\i s^*b} \nonumber\\
& = & \varphi_\lambda(A^*s) \, \erm^{\i s^*b} \nonumber\\
& = & \exp\left( \i (A^*s)^*\mu - \frac{1}{2}(A^*s)^* C A^*s  \right) \exp\left( \i \, s^*b \right) \nonumber\\
& = & \exp\left( \i s^*(A\mu+b) - \frac{1}{2} s^*(ACA^*)s \right) . \nonumber 
\end{eqnarray}
\end{proof}

\paragraph{Proposition A.6.}Adding two \glslink{conditional independence}{independent} Gaussians yields a Gaussian, i.e. if $\lambda_1 \sim \mathcal{N}_n\left[\mu_1,C_1\right]$, $\lambda_2 \sim \mathcal{N}_n\left[\mu_2,C_2\right]$ and $\lambda_1 \ci \lambda_2$, then $\lambda_1+\lambda_2 \sim \mathcal{N}_n\left[\mu_1+\mu_2,C_1+C_2\right]$.

\begin{proof} The \glslink{conditional independence}{independence} of $\lambda_1$ and $\lambda_2$ implies the \glslink{conditional independence}{independence} of $\erm^{\i t^* \lambda_1}$ and $\erm^{\i t^* \lambda_2}$. Therefore,
\begin{equation}
\varphi_{\lambda_1+\lambda_2}(t) = \left< \erm^{\i t^* (\lambda_1+\lambda_2)} \right> = \left< \erm^{\i t^* \lambda_1} \erm^{\i t^* \lambda_2} \right> = \left< \erm^{\i t^* \lambda_1} \right> \left< \erm^{\i t^* \lambda_2} \right> = \varphi_{\lambda_1}(t) \varphi_{\lambda_2}(t) . \nonumber
\end{equation}
Using the \glslink{characteristic function}{characteristic functions} of $\lambda_1$ and $\lambda_2$ yields
\begin{equation}
\varphi_{\lambda_1+\lambda_2}(t) = \exp\left( \i t^* \mu_1 -\frac{1}{2} t^*C_1t \right) \exp\left( \i t^* \mu_2 -\frac{1}{2} t^*C_2t \right) = \exp\left( \i t^* (\mu_1+\mu_2) -\frac{1}{2} t^*(C_1+C_2)t \right) . \nonumber
\end{equation}
\end{proof}

\section{Marginal and conditionals of Gaussian random vectors}

To study the partition of Gaussian random vectors, let us define
\begin{equation}
\lambda=
\begin{pmatrix}
\lambda_{x} \\
\lambda_{y}
\end{pmatrix}, \quad
\mu=
\begin{pmatrix}
\mu_{x} \\
\mu_{y}
\end{pmatrix} \quad \mathrm{and} \quad C=
\begin{pmatrix}C_{xx}&C_{xy}\\ C_{yx}&C_{yy}
\end{pmatrix},
\end{equation}
where $\lambda_x, \mu_x \in \mathbb{C}^m$, $C_{xx} \in \mathcal{M}_m(\mathbb{C})$, $\lambda_y, \mu_y \in \mathbb{C}^{n-m}$, $C_{yy} \in \mathcal{M}_{n-m}(\mathbb{C})$, $C_{xy} \in \mathcal{M}_{m\times(n-m)}(\mathbb{C})$ and $C_{yx} = (C_{xy})^* \in \mathcal{M}_{(n-m) \times m}(\mathbb{C})$. We assume that $m<n$ and we want to prove that the \glslink{marginal pdf}{marginal} and \glslink{conditional pdf}{conditional distributions} of $\lambda_x$ and $\lambda_y$ are Gaussians with parameters given by equations \eqref{eq:GRF-marginals-1}--\eqref{eq:GRF-marginals-4} and \eqref{eq:GRF-conditionals-1}--\eqref{eq:GRF-conditionals-4}. By symmetry, we limit the discussion to $\lambda_x$ and $\lambda_x|\lambda_y$.

\paragraph{Proposition A.7.\label{sec:prop-A-7}}The \glslink{marginal pdf}{marginal distribution} of $\lambda_x$ is that of a Gaussian random vector with mean $\mu_x$ and variance $C_{xx}$.

\begin{proof} Consider $A=\begin{pmatrix}\textbf{1}_{xx}&\textbf{0}_{xy}\\ \textbf{0}_{yx}&\textbf{0}_{yy}\end{pmatrix}$. Proposition \hyperref[sec:prop-A-5]{A.5.} yields $A\lambda = \lambda_x \sim \mathcal{N}_m \left[ A\mu, ACA^* \right] = \mathcal{N}_m \left[ \mu_x, C_{xx} \right]$.
\end{proof}

Let us now consider the \glslink{conditional pdf}{conditionals}.

\paragraph{Lemma A.8.\label{sec:lemma-A-8}} $\lambda_x$ and $\lambda_y$ are \glslink{conditional independence}{independently distributed} if and only if $C_{xy} = \textbf{0}_{xy}$.

\begin{proof} This proposition follows by considering the \gls{characteristic function} of $\lambda$:
\begin{eqnarray}
\varphi_\lambda(t) & = & \varphi_{(\lambda_x,\lambda_y)}(t_x,t_y) \nonumber \\
& = & \exp \left( \i t^*\mu - \frac{1}{2}t^*Ct \right) \nonumber \\
& = & \exp \left( \i t_x^*\mu_x + \i t_y^*\mu_y - \frac{1}{2}t_x^*C_{xx}t_x - \frac{1}{2}t_x^*C_{xy}t_y - \frac{1}{2}t_y^*C_{yy}t_y - \frac{1}{2}t_y^*C_{yx}t_x \right) \nonumber \\
& = & \varphi_{\lambda_x}(t_x) \varphi_{\lambda_y}(t_y) \exp \left( - t_x^*C_{xy}t_y \right) \nonumber
\end{eqnarray}
and using \gls{Kac's theorem} (theorem \hyperref[sec:thm-A-2]{A.2.}), $\lambda_x \ci \lambda_y \Leftrightarrow \varphi_{(\lambda_x,\lambda_y)} = \varphi_{\lambda_x} \varphi_{\lambda_y} \Leftrightarrow C_{xy} =\textbf{0}_{xy}$.
\end{proof}

\paragraph{Definition A.9.}Let $C_{xx.y} \equiv C_{xx} - C_{xy}C_{yy}^{-1}C_{yx}$, the so-called generalized \gls{Schur-complement} of $C_{yy}$ in $C$.

\paragraph{Lemma A.10.\label{sec:lemma-A-10}}
\begin{equation}
\begin{pmatrix}
\lambda_x - C_{xy}C_{yy}^{-1}\lambda_y \\
\lambda_y
\end{pmatrix} \sim
\mathcal{N}_n
\left[\begin{pmatrix}
\mu_x - C_{xy}C_{yy}^{-1}\mu_y \\
\mu_y
\end{pmatrix},
\begin{pmatrix}
C_{xx.y} & \textbf{0}_{xy} \\
\textbf{0}_{yx} & C_{yy}
\end{pmatrix}
\right] .
\end{equation}

\begin{proof} Consider $A=\begin{pmatrix}\textbf{1}_{xx}&-C_{xy}C_{yy}^{-1}\\ \textbf{0}_{yx}&\textbf{1}_{yy}\end{pmatrix}$. The lemma follows by considering $A\lambda$ and using proposition \hyperref[sec:prop-A-5]{A.5.}
\end{proof}

\paragraph{Proposition A.11.} The \glslink{conditional pdf}{conditional distribution} of $\lambda_x$ given $\lambda_y$ is the Gaussian distribution given by 
\begin{equation}
\mathcal{N}_m \left[ \mu_x + C_{xy}C_{yy}^{-1}(\lambda_y - \mu_y), C_{xx.y} \right] . \nonumber
\end{equation}

\begin{proof}
Since $\lambda_x - C_{xy}C_{yy}^{-1}\lambda_y$ and $\lambda_y$ have zero covariance matrix (lemma \hyperref[sec:lemma-A-10]{A.10.}), they are \glslink{conditional independence}{independently distributed} according to lemma \hyperref[sec:lemma-A-8]{A.8.} Therefore, using also the result obtained for the \glslink{marginal pdf}{marginals} (proposition \hyperref[sec:prop-A-7]{A.7.}), we get
\begin{eqnarray}
(\lambda_x - C_{xy}C_{yy}^{-1}\lambda_y) | \lambda_y & \sim & \lambda_x - C_{xy}C_{yy}^{-1}\lambda_y \nonumber \\
& \sim & \mathcal{N}_m \left[ \mu_x - C_{xy}C_{yy}^{-1}\mu_y, C_{xx.y} \right] \nonumber
\end{eqnarray}
and hence
\begin{eqnarray}
\lambda_x|\lambda_y & \sim & (\lambda_x - C_{xy}C_{yy}^{-1}\lambda_y + C_{xy}C_{yy}^{-1}\lambda_y) | \lambda_y \nonumber \\
& \sim & \mathcal{N}_m \left[ \mu_x + C_{xy}C_{yy}^{-1}(\lambda_y - \mu_y), C_{xx.y} \right] \nonumber
\end{eqnarray}
by just translating the above normal density by the constant vector $C_{xy}C_{yy}^{-1}\lambda_y$.
\end{proof}

%% file: AppendixB/AppendixBContent.tex
\chapter{Simulating collisionless dark matter fluids}
\label{apx:simulations}
\minitoc

\begin{flushright}
\begin{minipage}[c]{0.6\textwidth}
\rule{\columnwidth}{0.4pt}

``Simulation:\\
\textbf{1.} \textbf{a.} The action or practice of simulating, with intent to deceive; false pretence, deceitful profession. (...)\\
\textbf{2.} A false assumption or display, a surface resemblance or imitation, of something. (...)''\\
--- The \href{http://www.oed.com/}{Oxford English Dictionary}\\
Quoted by \href{https://telescoper.wordpress.com/2014/05/08/illustris-cosmology-and-simulation/}{Peter Coles (2014)}

\vspace{-5pt}\rule{\columnwidth}{0.4pt}
\end{minipage}
\end{flushright}

\abstract{\section*{Abstract} This technical appendix describes the implementation of the \glslink{N-body simulation}{simulation} codes used in this thesis. It reviews the \glslink{PM}{particle-mesh} approach for simulating a collisionless \glslink{CDM}{cold dark matter} \gls{fluid}, as well as the {\cola} modification. The generation of \gls{initial conditions} using \glslink{LPT}{Lagrangian perturbation theory} is also discussed.}

Many of the projects described in this thesis rely on the \glslink{PM}{particle-mesh} (\gls{PM}) \glslink{N-body simulation}{simulation} technique. It has originally been introduced and applied in many different areas of physics, such as electromagnetism, hydrodynamics, magnetohydrodynamics, plasma physics and self-gravitating systems \citetext{see e.g. the books by \citealp{Hockney1981} and \citealp{Birdsall1985}}. In a cosmological context, the reference papers include \citet{Klypin1983,Efstathiou1985}.

This appendix reviews the \gls{PM} technique, the {\cola} modification, and the numerical implementation of \glslink{LPT}{Lagrangian perturbation theory}. More details on cosmological \gls{PM} codes can be found in the review by \citet{Bertschinger1998} or the lectures notes by \citet{Kravtsov2002,Springel2014,Teyssier2014}. The reader is also referred to the {\cola} papers, \citet{Tassev2013,Tassev2015}; and to \citet[][appendix D]{Scoccimarro1998b}, for the implementation of \gls{LPT}.

This appendix is organized as follows. In section \ref{sec:apx_Model_equations}, we write down the equations actually solved by \gls{PM}/{\cola} codes. We describe the main \gls{PM} steps and the required data structures in section \ref{sec:apx_Steps_data_structures}. Section \ref{sec:apx-Density assignments} reviews \glslink{mesh assignment}{mesh assignments} and \gls{interpolation} schemes; section \ref{sec:apx-Poisson equation and accelerations} discusses the resolution of the \gls{Poisson equation} and the computation of forces; and section \ref{sec:apx:KDK} examines how to update the positions and \glslink{momentum}{momenta} of \glslink{dark matter particles}{particles}. Finally, \ref{sec:apx-Setting up initial conditions} describes the generation of cosmological \gls{initial conditions} using \glslink{LPT}{Lagrangian perturbation theory}.

\section{Model equations}
\label{sec:apx_Model_equations}

\subsection{Model equations in the standard PM code}

A \gls{PM} codes solves the \gls{equation of motion} for \gls{dark matter particles} in \gls{comoving coordinates} (see equation \eqref{eq:equation-of-motion-p}; below the mass of \glslink{dark matter particles}{particles} $m$ is absorbed in the definition of the \gls{momentum} $\textbf{p}$):
\begin{eqnarray}
\textbf{p} = a \deriv{\textbf{x}}{\tau}, \\
\deriv{\textbf{p}}{\tau} = - a \nabla \Phi,
\end{eqnarray}
coupled with the \gls{Poisson equation} for the \gls{gravitational potential} (equation \eqref{eq:Poisson}),
\begin{equation}
\Delta \Phi = 4\pi \G a^2 \bar{\rho}(\tau) \delta = \frac{3}{2} \Omega_\mathrm{m}(\tau) \mathcal{H}^2(\tau) \delta.
\end{equation}
It is convenient to choose the \gls{scale factor} as time variable. Using $\partial_\tau = a' \, \partial_a = \dot{a}a \, \partial_a$ and $\bar{\rho}(\tau) = \rho^{(0)} a^{-3}$, the equations to solve are rewritten:
\begin{eqnarray}
\deriv{\textbf{x}}{a} & = & \frac{\textbf{p}}{a'a} = \frac{\textbf{p}}{\dot{a}a^2}, \\
\deriv{\textbf{p}}{a} & = & -\frac{a \nabla \Phi}{a'} = -\frac{\nabla \Phi}{\dot{a}} , \label{eq:kick-intermediate}\\
\Delta \Phi & = & 4\pi \G \rho^{(0)} a^{-1} \delta = \frac{3}{2} \Omega_\mathrm{m}^{(0)} a^{-1} \delta . \label{eq:Poisson-intermediate}
\end{eqnarray}
We will use the equivalent formulation
\begin{eqnarray}
\deriv{\textbf{x}}{a} & = & \mathpzc{D}(a) \textbf{p}, \label{eq:drift}\\
\deriv{\textbf{p}}{a} & = & \mathpzc{K}(a) \nabla \left( \Delta^{-1} \delta \right), \label{eq:kick}
\end{eqnarray}
where we have combined equations \eqref{eq:kick-intermediate} and \eqref{eq:Poisson-intermediate}, and defined $f(a) \equiv \dot{a}^{-1} = a/a' = \mathcal{H}^{-1}(a)$; $\mathpzc{D}(a) \equiv f(a)/a^2$ (the ``\gls{drift} prefactor'') and $\mathpzc{K}(a) \equiv -(3/2) \Omega_\mathrm{m}^{(0)} f(a)/a$ (the ``\gls{kick} prefactor'').

\subsection{Model equations with COLA}

If one desires to include the {\cola} scheme \citep[see][and section \ref{sec:The COLA method}]{Tassev2013}, then one works in a frame comoving with the Lagrangian \glslink{displacement field}{displacements}. Recall the \gls{LPT} position of a particle is given by (see section \ref{sec:LPT}),
\begin{equation}
\textbf{x}_\mathrm{LPT}(a) = \textbf{q} - D_1(a) \boldsymbol{\Psi}_1 + D_2(a) \boldsymbol{\Psi}_2 .
\label{eq:apx-mapping}
\end{equation}
Noting $\textbf{x}(a) = \textbf{x}_\mathrm{LPT}(a) + \textbf{x}_\mathrm{MC}(a)$ the real position of the same particle, including the \glslink{mode coupling}{mode-coupling residual} $\textbf{x}_\mathrm{MC}(a)$, one has (see equation \eqref{eq:apx-mapping}):
\begin{equation}
\deriv{\textbf{x}}{a} = \deriv{\textbf{x}_\mathrm{LPT}}{a} + \deriv{\textbf{x}_\mathrm{MC}}{a} ; \quad \mathrm{with} \quad \deriv{\textbf{x}_\mathrm{LPT}}{a} = - \deriv{D_1}{a} \boldsymbol{\Psi}_1 + \deriv{D_2}{a} \boldsymbol{\Psi}_2 \equiv \mathpzc{D}(a) \textbf{p}_\mathrm{LPT} .
\end{equation}
We also define $\textbf{p}_\mathrm{MC}$ such that $\drm \textbf{x}_\mathrm{MC}/\drm a \equiv \mathpzc{D}(a) \textbf{p}_\mathrm{MC}$. Then $\textbf{p} = \textbf{p}_\mathrm{LPT} + \textbf{p}_\mathrm{MC}$ (see equation \eqref{eq:drift}). Furthermore,
\begin{equation}
\deriv{\textbf{p}_\mathrm{LPT}}{a} = \frac{\drm}{\drm a} \left( \frac{1}{\mathpzc{D}(a)} \deriv{\textbf{x}_\mathrm{LPT}}{a} \right) \equiv - \mathpzc{K}(a) \mathpzc{V}[\textbf{x}_\mathrm{LPT}](a),
\end{equation}
where the differential operator $\mathpzc{V}[\cdot](a)$ is defined by
\begin{equation}
\mathpzc{V}[\cdot](a) \equiv - \frac{1}{\mathpzc{K}(a)} \frac{\drm}{\drm a} \left( \frac{1}{\mathpzc{D}(a)} \deriv{\,\cdot}{a} \right) .
\end{equation}
With these notations, equation \eqref{eq:kick} reads
\begin{equation}
\deriv{\textbf{p}}{a} = \deriv{\textbf{p}_\mathrm{LPT}}{a} + \deriv{\textbf{p}_\mathrm{MC}}{a} = - \mathpzc{K}(a) \mathpzc{V}[\textbf{x}_\mathrm{LPT}](a) + \deriv{\textbf{p}_\mathrm{MC}}{a} = \mathpzc{K}(a) \nabla \left( \Delta^{-1} \delta \right) .
\end{equation}
It is straightforward to check from equation \eqref{eq:apx-mapping} that $\mathpzc{V}[\textbf{x}_\mathrm{LPT}](a) = - \mathpzc{V}[D_1](a) \boldsymbol{\Psi}_1 + \mathpzc{V}[D_2](a) \boldsymbol{\Psi}_2$. Using the differential equation verified by \glslink{linear growth factor}{$D_1$} (equation \eqref{eq:evolution-growth-factor-a}) and the second \glslink{Friedmann's equations}{Friedmann equation} (equation \eqref{eq:Friedmann-2}), we get
\begin{equation}
\label{eq:VD1}
\mathpzc{V}[D_1](a) = D_1(a) .
\end{equation}
Similarly for the \gls{second-order growth factor}, using equation \eqref{eq:evolution-second-growth-factor-a},
\begin{equation}
\label{eq:VD2}
\mathpzc{V}[D_2](a) = D_2(a) - D_1^2(a) .
\end{equation}

In the {\cola} framework, the natural variables are therefore $\textbf{x}$ and $\textbf{p}_\mathrm{MC}$, and the equations of motion to solve (equivalents of equations \eqref{eq:drift} and \eqref{eq:kick}) are
\begin{eqnarray}
\deriv{\textbf{x}}{a} & = & \mathpzc{D}(a)\textbf{p}_\mathrm{MC} - \deriv{D_1}{a} \boldsymbol{\Psi}_1 + \deriv{D_2}{a} \boldsymbol{\Psi}_2, \label{eq:drift-cola}\\
\deriv{\textbf{p}_\mathrm{MC}}{a} & = & \mathpzc{K}(a) \left[ \nabla \left( \Delta^{-1} \delta \right) - \mathpzc{V}[D_1](a) \boldsymbol{\Psi}_1 + \mathpzc{V}[D_2](a) \boldsymbol{\Psi}_2 \right]. \label{eq:kick-cola}
\end{eqnarray}

In the \gls{initial conditions}, generated with \gls{LPT} (see section \ref{sec:apx-Setting up initial conditions}), we have $\textbf{p} = \textbf{p}_\mathrm{LPT}$; therefore the \glslink{mode coupling}{mode-coupling} \gls{momentum} residual in the rest frame of \gls{LPT} observers, $\textbf{p}_\mathrm{MC}$, should be initialized to zero \citep[this corresponds to the $\mathrm{L}_-$ operator in][appendix A]{Tassev2013}. At the end, the \gls{LPT} \gls{momentum} $\textbf{p}_\mathrm{LPT}$ has to be added to $\textbf{p}_\mathrm{MC}$ to recover the full \gls{momentum} of \glslink{dark matter particles}{particles}, $\textbf{p}$ \citep[this corresponds to the $\mathrm{L}_+$ operator in][appendix A]{Tassev2013}. In the following, wherever we do not make the explicit distinction between the standard \gls{PM} and the {\cola} approaches, we will drop the subscript ``MC'' for {\cola} \glslink{momentum}{momenta} and simply note $\textbf{p}$; however, one should keep in mind these two transformations at the beginning and at the end.

\section{Steps and data structures}
\label{sec:apx_Steps_data_structures}

\subsection{Main PM steps}
\label{sec:apx_PM_steps}

Equations \eqref{eq:drift} and \eqref{eq:kick} are solved iteratively in a \gls{PM} code, which consists of three main steps:
\begin{enumerate}
\item estimate the \gls{density field} on the grid from current particle positions; solve the \gls{Poisson equation} on the grid to get the \glslink{gravitational potential}{potential}; take the gradient of the \glslink{gravitational potential}{potential} to get the accelerations on the grid; and \glslink{interpolation}{interpolate} back to \glslink{dark matter particles}{particles} (see sections \ref{sec:apx-Density assignments} and \ref{sec:apx-Poisson equation and accelerations}),
\item advance particle \glslink{momentum}{momenta} using the new accelerations (equation \eqref{eq:kick}; see section \ref{sec:apx:KDK})
\item update particle positions using their new \glslink{momentum}{momenta} (equation \eqref{eq:drift}; see section \ref{sec:apx:KDK}).
\end{enumerate}

In the {\cola} scheme, steps 2 and 3 are replaced with the equivalents that come from equations \eqref{eq:kick-cola} and \eqref{eq:drift-cola}, respectively.

\subsection{Definitions and data structures}
\label{sec:apx-Definitions and data structures}

\paragraph{Grids and box size.}A \gls{PM} cosmological \glslink{N-body simulation}{simulation} is characterized by
\begin{itemize}
\item the number of \glslink{dark matter particles}{particles}, $N_\mathrm{p}$ (if \glslink{dark matter particles}{particles} start from a regular Lagrangian grid -- see section \ref{sec:apx-Setting up initial conditions} --, we note $N_{\mathrm{p}0}$, $N_{\mathrm{p}1}$, $N_{\mathrm{p}2}$ the number of \glslink{dark matter particles}{particles} along each direction, such that $N_\mathrm{p} \equiv N_{\mathrm{p}0} N_{\mathrm{p}1} N_{\mathrm{p}2}$);
\item the size of the \glslink{periodic boundary conditions}{periodic} box along each direction, $L_0$, $L_1$, $L_2$ (the total volume simulated is therefore $V \equiv L_0 L_1 L_2$);
\item and the number of cells of the \gls{PM} grid (i.e. the grid on which \glslink{density field}{density} and \glslink{gravitational potential}{potential} are defined) along each direction, $N_{\mathrm{g}0}$, $N_{\mathrm{g}1}$, $N_{\mathrm{g}2}$, with $N_\mathrm{g} \equiv N_{\mathrm{g}0} N_{\mathrm{g}1} N_{\mathrm{g}2}$.
\end{itemize}

In many cases we will assume that the box is cubic, and that the particle grid and the \gls{PM} grid are isotropic: $L_0 = L_1 = L_2 \equiv L$;  $N_{\mathrm{p}0} = N_{\mathrm{p}1} = N_{\mathrm{p}2}$; $N_{\mathrm{g}0} = N_{\mathrm{g}1} = N_{\mathrm{g}2}$. In the following, we denote the side lengths of cells by $\Delta x \equiv L_0/N_{\mathrm{g}0}$, $\Delta y \equiv L_1/N_{\mathrm{g}1}$, $\Delta z \equiv L_2/N_{\mathrm{g}2}$ and their volume by $V_\mathrm{c} \equiv \Delta x \Delta y \Delta z$. We have $V = N_\mathrm{g} V_\mathrm{c}$.

\paragraph{Particle variables.}Assuming that \glslink{dark matter particles}{particles} all have the same mass,\footnote{From the definition of $\Omega_\mathrm{m}^\mathrm{(0)}$, it is easy to see that the mass carried by each particle is $m = \dfrac{3 \Omega_\mathrm{m}^{(0)} H_0^2}{8\pi \G} \dfrac{V}{N_\mathrm{p}}$ (this number is called the \textit{\gls{mass resolution}}).\label{fn:particle_mass}} a \gls{PM} code needs a minimum of six real numbers ($\texttt{float}$ or $\texttt{double}$) for each particle: three coordinates and three \glslink{momentum}{momenta}. If the {\cola} modification is included (see section \ref{sec:The COLA method}), a minimum of nine (for \gls{LPT} at order one) or twelve  (for \gls{LPT} at order two) real numbers per particle is required (three additional real numbers per particle to store the \gls{LPT} \glslink{displacement field}{displacements} at each order).

We call these arrays $\texttt{x[mp]}$, $\texttt{y[mp]}$, $\texttt{z[mp]}$ (\glslink{dark matter particles}{particles}' positions); $\texttt{px[mp]}$, $\texttt{py[mp]}$, $\texttt{pz[mp]}$ (\glslink{dark matter particles}{particles}' \glslink{momentum}{momenta}); and if {\cola} is enabled, $\texttt{psix\_1[mp]}$, $\texttt{psiy\_1[mp]}$, $\texttt{psiz\_1[mp]}$ (for the \gls{ZA} \glslink{displacement field}{displacements}, $\boldsymbol{\Psi}_1$), $\texttt{psix\_2[mp]}$, $\texttt{psiy\_2[mp]}$, $\texttt{psiz\_2[mp]}$ (for the \gls{2LPT} \glslink{displacement field}{displacements}, $\boldsymbol{\Psi}_2$). Here $\texttt{mp}$ indexes a particle. It is interesting to note that the arrays containing the Lagrangian \glslink{displacement field}{displacements} are constants, i.e. that they are never updated within the code (their time-independence can be checked in equations \eqref{eq:drift-cola} and \eqref{eq:kick-cola}). Convenient data structures are 1D arrays of size $N_\mathrm{p}$ for particles' variables.

\paragraph{Grid variables.}In addition, the code needs real numbers ($\texttt{float}$ or $\texttt{double}$) for the \gls{density contrast} $\delta$ and the \glslink{gravitational potential}{potential} $\Phi$ at each grid cell. An array of size $N_\mathrm{g}$ is needed to store such grid variables. This array can be shared between \glslink{density field}{density} and \glslink{gravitational potential}{potential}: we first use it to store the \gls{density contrast} $\delta$, then replace its values with the \glslink{gravitational potential}{potential} when the \gls{Poisson equation} is solved.\footnote{The quantity stored is actually the reduced \gls{gravitational potential}, $\widetilde{\Phi} \equiv \Delta^{-1} \delta$, as the overall time-dependent coefficients needed to go from $\widetilde{\Phi}$ to $\Phi$ are factored out in $\mathpzc{K}(a)$ (see equations \eqref{eq:kick} and \eqref{eq:kick-cola}).\label{fn:apx-potential}}

We call this array $\texttt{density\_or\_Phi}$. A convenient data structure is a 3D array, such that the grid quantity at position $(i,j,k)$ is $\texttt{density\_or\_Phi[i,j,k]}$ (with $0\leq i < N_{\mathrm{g}0}$, $0\leq j < N_{\mathrm{g}1}$, $0\leq k < N_{\mathrm{g}2}$). Equivalently, we decided to implement  $\texttt{density\_or\_Phi}$ as a 1D array of size $N_\mathrm{g}$, such that the grid quantity at position $(i,j,k)$ is given by $\texttt{density\_or\_Phi[mc]}$ where the current cell is indexed by $\texttt{mc}=k+N_{\mathrm{g}2} \times (j+N_{\mathrm{g}1} \times i)$.

\paragraph{Accelerations.}It is also convenient to have three additional arrays of size $N_\mathrm{g}$ to store the components of the acceleration on the grid, and three arrays of size $N_\mathrm{p}$ to store the components of particles' acceleration.\footnote{Actually the reduced acceleration $\tilde{g} \equiv \nabla \left( \Delta^{-1} \delta \right)$ instead of the physical acceleration, see footnote \ref{fn:apx-potential}.} In the following, we note these arrays $\texttt{gx[mc]}$, $\texttt{gy[mc]}$, $\texttt{gz[mc]}$, $\texttt{gpx[mp]}$, $\texttt{gpy[mp]}$, $\texttt{gpz[mp]}$, where $0 \leq \texttt{mc} < N_\mathrm{g}$ indexes a grid cell and $0 \leq \texttt{mp} < N_\mathrm{p}$ indexes a particle.\footnote{These arrays are not absolutely required. Indeed, it is possible to get rid of them and to make the code more memory-efficient, if one performs in one step the \glslink{FDA}{finite difference} (to go from $\Delta^{-1} \delta$ to $\nabla (\Delta^{-1} \delta)$), the \gls{interpolation} (from the grid quantities to \glslink{dark matter particles}{particles}, see section \ref{sec:apx-Density assignments}) and the \gls{kick} operation (see section \ref{sec:apx:KDK}).} 

\section{Mesh assignments and interpolations}
\label{sec:apx-Density assignments}

This section describes how to assign to the grid a quantity carried by \glslink{dark matter particles}{particles} (the ``\gls{mesh assignment}'' operation, from \glslink{dark matter particles}{particles} to the grid), and how to distribute to \glslink{dark matter particles}{particles} a quantity that is known on the grid (the  ``\gls{interpolation}'' operation, from the grid to \glslink{dark matter particles}{particles}).

In a \gls{PM} code, the first operation is used to compute the \glslink{density field}{density} on the grid from particle positions; and the second operation is used to assign an acceleration to each particle from grid values. Both are used in step 1 of the main \gls{PM} steps (see section \ref{sec:apx_PM_steps}).

\subsection{The mesh assignment function}
\label{sec:apx-The mesh assignment function}

The general idea to assign \glslink{dark matter particles}{particles} to the grid is to assume that they have a ``shape'' $S$ that intersects the grid. Let us first describe the one-dimensional case, where $S(x)$ is the 1D particle shape. The fraction of the particle at $x_\mathrm{p}$ assigned to the cell at $x_\mathrm{c}$ is the \gls{shape function} averaged over this cell:
\begin{equation}
W(x_\mathrm{p}-x_\mathrm{c}) \equiv \int_{x_\mathrm{c}-\Delta x/2}^{x_\mathrm{c}+\Delta x/2} S(x' - x_\mathrm{p}) \, \drm x' = \int \Pi\left( \frac{x'-x_\mathrm{c}}{\Delta x} \right) S(x' - x_\mathrm{p}) \, \drm x'
\end{equation}
The \gls{assignment function} is hence the convolution:
\begin{equation}
W(x) = \Pi\left( \frac{x}{\Delta x} \right) * S(x) \quad \mathrm{where} \quad \Pi(s) =
\left\{\begin{array}{ll}
1 \quad \mathrm{if}~|s|\leq\frac{1}{2} \\
0 \quad \mathrm{otherwise}.
\end{array}\right.
\end{equation}
In 3D,
\begin{equation}
W(\textbf{x}_\mathrm{p}-\textbf{x}_\mathrm{c}) \equiv W(x_\mathrm{p}-x_\mathrm{c}) W(y_\mathrm{p}-y_\mathrm{c}) W(z_\mathrm{p}-z_\mathrm{c}).
\end{equation}
For some quantity $A$, if $A_\mathrm{p}$ are the values carried by the \glslink{dark matter particles}{particles} at positions $\textbf{x}_p$, the quantity $A$ at position $\textbf{x}_\mathrm{c}$ on the grid is 
\begin{equation}
\label{eq:A_on_grid}
A(\textbf{x}_\mathrm{c}) = \sum_{\{\textbf{x}_\mathrm{p}\}} A_\mathrm{p} W(\textbf{x}_\mathrm{p}-\textbf{x}_\mathrm{c}) .
\end{equation}
In particular, for gravitational \gls{PM} codes, the quantity carried by \glslink{dark matter particles}{particles} is their mass $m$. The \glslink{density field}{density} on the mesh is then a sum over the contributions of each particle as given by the \gls{assignment function},
\begin{equation}
\rho(\textbf{x}_\mathrm{c}) = \frac{1}{V_\mathrm{c}} \sum_{\{\textbf{x}_\mathrm{p}\}} m W(\textbf{x}_\mathrm{p}-\textbf{x}_\mathrm{c}) .
\end{equation}
The mean density is $\bar{\rho} = m N_\mathrm{p}/V$, from which we deduce the \gls{density contrast} $\delta \equiv \rho/\bar{\rho} - 1$ on the mesh,
\begin{equation}
\label{eq:delta_on_grid}
\delta(\textbf{x}_\mathrm{c}) = \left( \frac{N_\mathrm{g}}{N_\mathrm{p}} \sum_{\{\textbf{x}_\mathrm{p}\}}  W(\textbf{x}_\mathrm{p}-\textbf{x}_\mathrm{c}) \right) - 1.
\end{equation}

\subsection{Low-pass filtering}

The \gls{Nyquist-Shannon sampling theorem} \citep{Nyquist1928,Shannon1948,Shannon1949} states that the \gls{information content} of a sampled signal can be correctly recovered if two conditions hold: the signal must be band-limited, and the \gls{sampling} frequency must be greater than twice the maximum frequency present in the signal. If this is not the case, replicated spectra cannot be separated of the signal we seek to recover, a phenomenon known as \gls{aliasing} \citep[e.g.][]{Manolakis2000}. Natural signals, however, are generally not band-limited, so must be \glslink{low-pass filter}{low-pass filtered} before they are \glslink{sampling}{sampled}. Equivalently, the \gls{sampling} operation must include some form of \gls{local} averaging, reflecting the finite spatial resolution.

The Fourier representation\footnote{Here, we use the conventions for forward and inverse \glslink{Fourier transform}{Fourier transforms} as introduced in section \ref{sec:power-spectrum}.} of the ideal \gls{low-pass filter} that one should use as \gls{assignment function} is given as
\begin{equation}
W(k) = \frac{1}{\sqrt{2\pi}} \, \Pi\left( \frac{k}{k_{\mathrm{Nyq},x}} \right) = \frac{1}{\sqrt{2\pi}} \times
\left\{\begin{array}{ll}
1 \quad \mathrm{if}~|k| < k_\mathrm{max}\\
0 \quad \mathrm{if}~|k| \geq k_\mathrm{max},
\end{array}\right.
\end{equation}
where $k_\mathrm{max} \equiv k_{\mathrm{Nyq},x}/2$, and $k_{\mathrm{Nyq},x} \equiv 2\pi/\Delta x$ is the \gls{Nyquist wavenumber}. This filter is ideal in the sense that it has unity gain in the pass-band region and it perfectly suppresses all the power in the stop-band regions. The configuration space representation is 
\begin{equation}
W(x) = \frac{1}{\Delta x} \, \sinc\left( \frac{x}{\Delta x} \right),
\end{equation}
where $s \mapsto \sinc(s) \equiv \frac{\sin(\pi s)}{\pi s}$ is the cardinal sine function (using the signal processing convention). It is interesting to note that $W(x)$ is not always positive. Therefore, the physical property of a continuous \gls{density field} to be positive will not be reflected in its discretized representation, using ideal \glslink{low-pass filter}{low-pass filtering}. The loss of physicality is an expression of a fundamental problem of any data processing procedure: the loss of information due to discretizing the continuous signal.

Furthermore, due to the infinite support of the cardinal sine function in configuration space, the ideal \gls{sampling} method is generally not tractable, because computationally too expensive. For this reason, practical approaches often rely on approximating the ideal cardinal sine operator by less accurate, but faster calculable functions (often with compact support in configuration space). In Fourier space, this will generally introduce artificial attenuation of the pass-band modes and leakage of stop-band modes into the signal (i.e. incomplete suppression of the \gls{aliasing} power). The optimal choice of a \gls{low-pass filter} approximation is therefore always a choice between accuracy and computational speed \citep[see e.g.][for detailed studies]{Manolakis2000}. In the following section we discuss common approaches used in particle \glslink{N-body simulation}{simulations}.

\subsection{Common mesh assignment schemes}
\label{sec:apx-Common mesh assignment schemes}

Commonly used particle \glslink{shape function}{shape functions} and \glslink{mesh assignment}{assignment schemes} are often presented as a hierarchy \citep{Hockney1981}. The simplest scheme is to consider that \glslink{dark matter particles}{particles} are punctual and to assign each of them to the nearest grid point: $W(x_\mathrm{p}-x_\mathrm{c}) = 1$ if $x_\mathrm{c}- \frac{\Delta x}{2} \leq x_\mathrm{p} \leq x_\mathrm{c} + \frac{\Delta x}{2}$, $0$ otherwise. The \gls{shape function} is therefore
\begin{equation}
S_\mathrm{NGP}(x) \equiv \updelta_\mathrm{D}(x) \quad \mathrm{and} \quad S_\mathrm{NGP}(\textbf{x}) \equiv \updelta_\mathrm{D}(x) \updelta_\mathrm{D}(y) \updelta_\mathrm{D}(z).
\end{equation}
This is the \glslink{NGP}{Nearest Grid Point} (\gls{NGP}) \glslink{mesh assignment}{assignment scheme}.

The second particle \gls{shape function} in the hierarchy is a rectangular parallelepiped (a ``cloud'') of side length $\Delta x$, $\Delta y$, $\Delta z$. This scheme involves the 8 nearest cells for each particle and is called the \glslink{CiC}{Cloud-in-Cell} (\gls{CiC}) scheme. The \gls{shape function} is
\begin{equation}
S_\mathrm{CiC}(x) \equiv \frac{1}{\Delta x} \Pi\left( \frac{x}{\Delta x} \right) \quad \mathrm{and} \quad S_\mathrm{CiC}(\textbf{x}) \equiv  \frac{1}{\Delta x \Delta y \Delta z} \Pi\left( \frac{x}{\Delta x} \right) \Pi\left( \frac{y}{\Delta y} \right) \Pi\left( \frac{z}{\Delta z} \right).
\end{equation}

This \gls{shape function} can be seen as the convolution $\frac{1}{\Delta x} \Pi\left( \frac{x}{\Delta x} \right) * \updelta_\mathrm{D}(x)$. Higher-order \glslink{mesh assignment}{assignment schemes} are obtained by successively convolving with $\frac{1}{\Delta x} \Pi\left( \frac{x}{\Delta x} \right)$ along each direction. For example, the third-order scheme is called the \glslink{TSC}{Triangular Shaped Cloud} (\gls{TSC}) and involves the 27 neighboring cells for each particle. In one-dimension, the \gls{shape function} is
\begin{equation}
S_\mathrm{TSC}(x) \equiv \frac{1}{\Delta x} \Pi\left( \frac{x}{\Delta x} \right) * \frac{1}{\Delta x} \Pi\left( \frac{x}{\Delta x} \right) .
\end{equation}

\begin{figure}\centering
\includegraphics[width=\columnwidth]{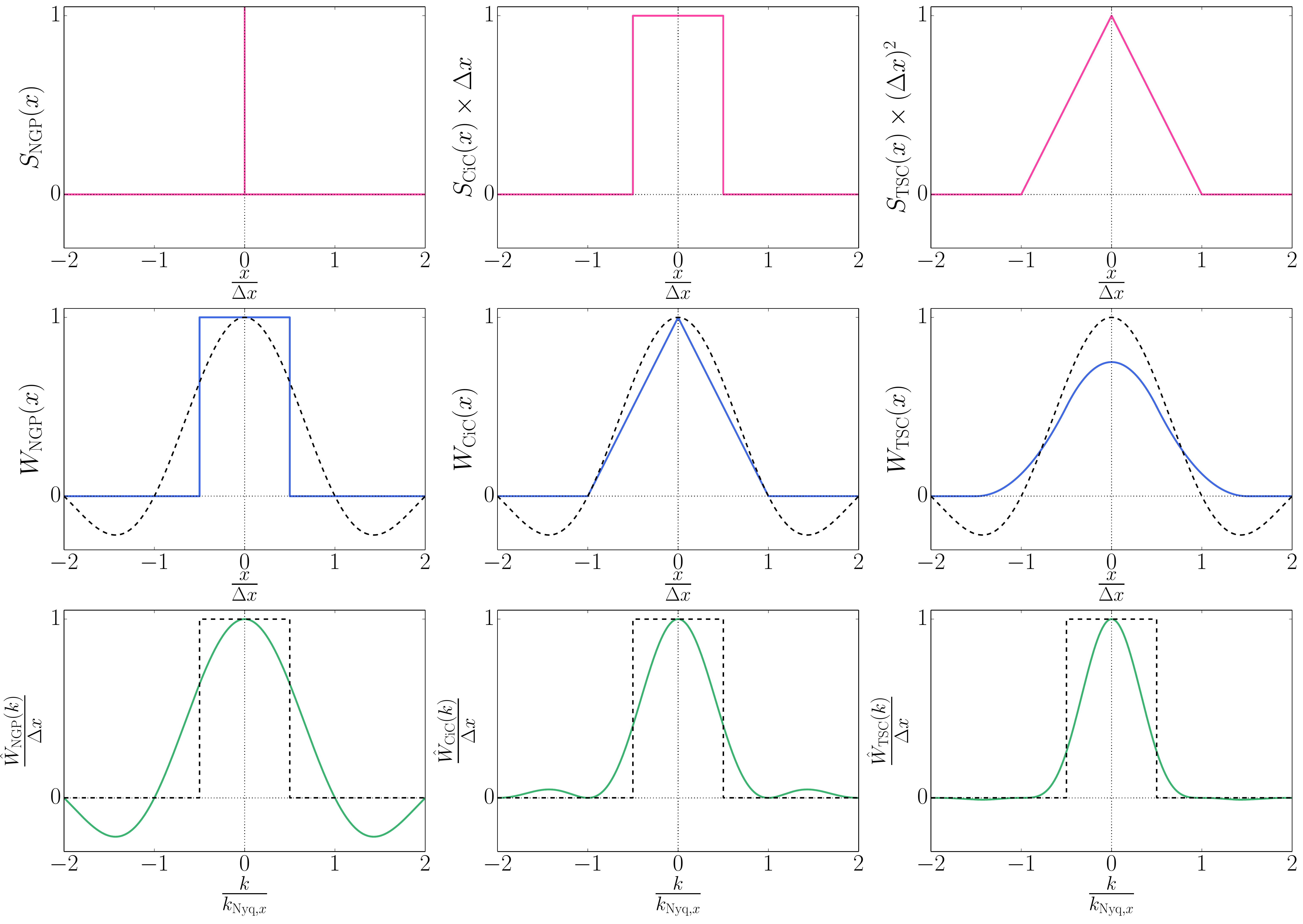}
\caption{Shape functions in configuration space for the first three schemes of the natural hierarchy of \glslink{mesh assignment}{mesh assignments} ($S$, first row); the corresponding \glslink{assignment function}{assignment functions} ($W$, second row) and their normalized \gls{Fourier transform} ($\hat{W}$, third row). From left to right, the schemes are: \glslink{NGP}{Nearest Grid Point} (\gls{NGP}), \glslink{CiC}{Cloud-in-Cell} (\gls{CiC}), \glslink{TSC}{Triangular Shaped Cloud} (\gls{TSC}). The \gls{Nyquist wavenumber} is defined by $k_{\mathrm{Nyq},x} \equiv 2 \pi/\Delta x$. For comparison, the dashed black lines show the configuration and Fourier space representations of the ideal \gls{low-pass filter} kernel.\label{fig:W_assignments}}
\end{figure}

The \gls{Fourier transform} of $x \mapsto \dfrac{1}{\Delta x} \Pi\left( \dfrac{x}{\Delta x} \right)$ is $k \mapsto \dfrac{1}{\sqrt{2\pi}} \, \sinc\left( \dfrac{k}{k_{\mathrm{Nyq},x}} \right)$. Therefore, in Fourier space, building the hierarchy is taking successive powers of $\dfrac{1}{\sqrt{2\pi}} \, \sinc\left( \dfrac{k}{k_{\mathrm{Nyq},x}} \right)$. The \gls{assignment function} $W$ is found by an additional convolution of $S$ with  $x \mapsto \Pi\left( \dfrac{x}{\Delta x} \right)$, which means, in Fourier space, an additional multiplication by $\dfrac{\Delta x}{\sqrt{2\pi}} \times \sinc\left( \dfrac{k}{k_{\mathrm{Nyq},x}} \right)$. In figure \ref{fig:W_assignments}, we show the \glslink{shape function}{shape functions} $S$ for the \gls{NGP}, \gls{CiC} and \gls{TSC} schemes (first row), the corresponding \glslink{assignment function}{assignment functions} $W$ (second row) and their normalized \glslink{Fourier transform}{Fourier transforms}, $\hat{W}/\Delta x$ (rescaled such that $\hat{W}(k=0)/\Delta x = 1$; third row).

High order schemes are obviously more expensive numerically, but they also give more precise results: from equation \eqref{eq:A_on_grid} and the \glslink{shape function}{shape functions}, we see that resulting quantities on the grid (\glslink{density field}{density}, forces) are piecewise constant in cells (\gls{NGP}); $C^0$ and piecewise linear (\gls{CiC}); $C^1$ with piecewise linear first derivative (\gls{TSC}), etc. (see figure \ref{fig:W_assignments}). The choice is a tradeoff between accuracy and computational expense. 

We summarize the results of this section in table \ref{tb:assignments}. In the following, we further comment on the well-known \gls{CiC} scheme, which is the prescription used to assign \glslink{dark matter particles}{particles} to the grid throughout this thesis, including in \gls{PM} and {\cola} implementations.

\begin{table}\centering
\begin{tabular}{llll}
\hline\hline
Name & \parbox[t]{10em}{Shape function\\ $S(x)$} & \parbox[t]{8em}{Number of\\ cells involved\\} & \parbox[t]{12em}{Properties of\\ grid-wise quantities}\\[-1.5ex]
\hline\\[-1.5ex]
NGP & $\delta(x)$ & $1^3=1$ & \parbox[t]{12em}{Piecewise constant in cells}\\[2ex]
CiC & $\dfrac{1}{\Delta x} \Pi \left(\dfrac{x}{\Delta x} \right)$ & $2^3=8$ & \parbox[t]{12em}{$C^0$, piecewise linear}\\[2ex]
TSC & $\dfrac{1}{\Delta x} \Pi \left(\dfrac{x}{\Delta x} \right) * \dfrac{1}{\Delta x} \Pi \left(\dfrac{x}{\Delta x} \right)$ & $3^3=27$ & \parbox[t]{12em}{$C^1$, differentiable with piecewise linear derivative}\\[3ex]
\hline\hline
\end{tabular}
\caption{Summary of the properties of commonly used particle \glslink{shape function}{shape functions}.}
\label{tb:assignments}
\end{table}

Let us consider the \gls{CiC} \glslink{mesh assignment}{density assignment} for a particle with coordinates $(x_\mathrm{p},y_\mathrm{p},z_\mathrm{p})$. The cell containing the particle has indexes given by
\begin{equation}
i = \left\lfloor \frac{x_\mathrm{p}}{\Delta x} \right\rfloor; \quad j = \left\lfloor \frac{y_\mathrm{p}}{\Delta y} \right\rfloor; \quad k = \left\lfloor \frac{z_\mathrm{p}}{\Delta z} \right\rfloor ,
\end{equation}
where $\left\lfloor \cdot \right\rfloor$ is the integer floor function. We consider that the cell center is at $(x_\mathrm{c},y_\mathrm{c},z_\mathrm{c}) = (i \times \Delta x, j \times \Delta y, k \times \Delta z)$.\footnote{The other common convention is to displace the cell center by half a voxel with respect to $(i \times \Delta x, j \times \Delta y, k \times \Delta z)$, i.e. $(x_\mathrm{c},y_\mathrm{c},z_\mathrm{c}) = (i \times \Delta x + \frac{\Delta x}{2}, j \times \Delta y + \frac{\Delta y}{2}, k \times \Delta z + \frac{\Delta z}{2})$.}

As noted before the particle may contribute to densities in the parent cell $(x_\mathrm{c},y_\mathrm{c},z_\mathrm{c})$ and the seven neighboring cells. Let us define
\begin{equation}
ii = \mathrm{mod}(i+1,N_{\mathrm{g}0}); \quad jj = \mathrm{mod}(j+1,N_{\mathrm{g}1}); \quad kk = \mathrm{mod}(k+1,N_{\mathrm{g}2}) .
\end{equation}
The modulo function enforces \gls{periodic boundary conditions}. The particle contributes to the eight cells indexed by $(i,j,k)$, $(ii,j,k)$, $(i,jj,k)$, $(i,j,kk)$, $(ii,jj,k)$, $(ii,j,kk)$, $(i,jj,kk)$ and $(ii,jj,kk)$. Let us define
\begin{eqnarray}
&d_x = \dfrac{x_\mathrm{p} - x_\mathrm{c}}{\Delta x} = \dfrac{x_\mathrm{p}}{\Delta x} - i; \quad d_y = \dfrac{y_\mathrm{p} - y_\mathrm{c}}{\Delta y} = \dfrac{y_\mathrm{p}}{\Delta y} - j; \quad d_z = \dfrac{z_\mathrm{p} - z_\mathrm{c}}{\Delta z} = \dfrac{z_\mathrm{p}}{\Delta z} - k; \\
&t_x = 1 - d_x; \quad t_y = 1 - d_y; \quad t_z = 1 - d_z .
\end{eqnarray}
Contributions to the eight cells are given by the formulae below, which also correspond to linear \glslink{interpolation}{interpolations} in 3D:
\begin{eqnarray}
W\!\left(\textbf{x}_\mathrm{p} - \textbf{x}_{(i,j,k)}\right) & = & t_x t_y t_z,\label{eq:CiC_1}\\
W\!\left(\textbf{x}_\mathrm{p} - \textbf{x}_{(ii,j,k)}\right) & = & d_x t_y t_z,\\
W\!\left(\textbf{x}_\mathrm{p} - \textbf{x}_{(i,jj,k)}\right) & = & t_x d_y t_z,\\
W\!\left(\textbf{x}_\mathrm{p} - \textbf{x}_{(i,j,kk)}\right) & = & t_x t_y d_z,\\
W\!\left(\textbf{x}_\mathrm{p} - \textbf{x}_{(ii,jj,k)}\right) & = & d_x d_y t_z,\\
W\!\left(\textbf{x}_\mathrm{p} - \textbf{x}_{(ii,j,kk)}\right) & = & d_x t_y d_z,\\
W\!\left(\textbf{x}_\mathrm{p} - \textbf{x}_{(i,jj,kk)}\right) & = & t_x d_y d_z,\\
W\!\left(\textbf{x}_\mathrm{p} - \textbf{x}_{(ii,jj,kk)}\right) & = & d_x d_y d_z. \label{eq:CiC_8}
\end{eqnarray}

Summing over all \glslink{dark matter particles}{particles} will result in the calculation of any quantity $A$ on the grid (equation \eqref{eq:A_on_grid}), in particular the \gls{density contrast} (equation \eqref{eq:delta_on_grid}).

\begin{figure}\centering
\includegraphics[width=0.5\columnwidth]{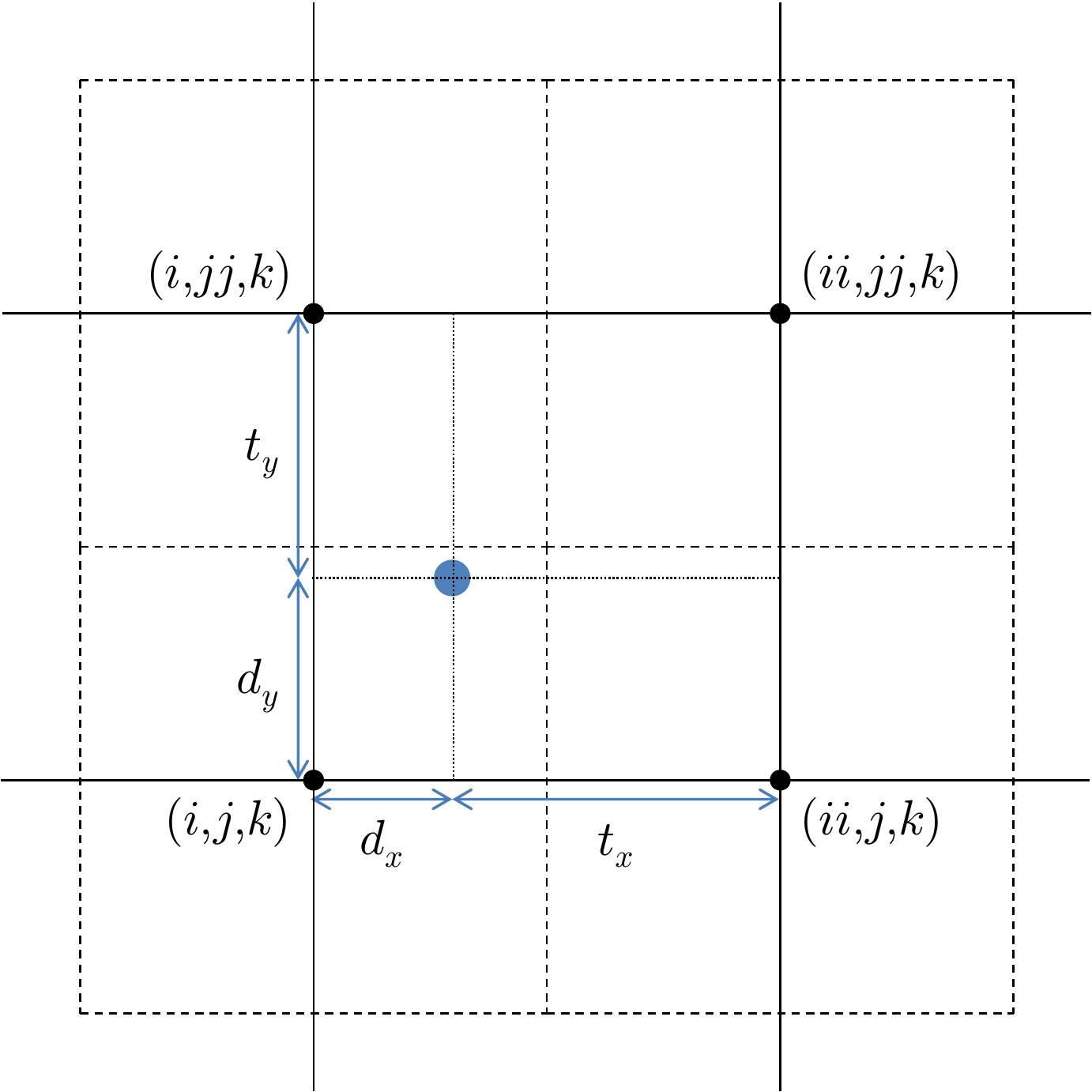}
\caption{Two-dimensional illustration of the \glslink{CiC}{Cloud-in-Cell} \glslink{mesh assignment}{assignment scheme}. Evertyhing is expressed in units of the cell size, $\Delta x$ along the $x$ direction and $\Delta y$ along the $y$ direction. In three dimensions, the particle is assigned to the eight neighboring cells with different weights given by equations \eqref{eq:CiC_1}--\eqref{eq:CiC_8}.\label{fig:CiC}}
\end{figure}

In figure \ref{fig:CiC}, we illustrate the \gls{CiC} scheme in two dimensions. The first step is to identify the cell indexes $i,ii,j,jj,k,kk$. Then, one computes the weight coefficients $d_x, t_x, d_y, t_y, d_z, t_z$ as shown on the figure, and assigns the particle to the neighboring cells using the formulae above.

\subsection{Interpolation}
\label{sec:apx-Interpolation}

Interpolation is used to distribute a grid-wise quantity to \glslink{dark matter particles}{particles}. For example, for \gls{PM} codes, accelerations are computed on the grid (see section \ref{sec:apx-Poisson equation and accelerations}), then \glslink{interpolation}{interpolated} back to each particle's position.

Using the same notations as before, for some quantity $A$, the problem is to compute $A_\mathrm{p}$ given the values of $A(\textbf{x}_\mathrm{c})$ in all the cells. This can be written in a similar fashion as equation \eqref{eq:A_on_grid}, but summing on grid cells instead of \glslink{dark matter particles}{particles}:
\begin{equation}
\label{eq:A_in_particles}
A_\mathrm{p} = A(\textbf{x}_\mathrm{p}) = \sum_{\{ \textbf{x}_\mathrm{c} \}} A(\textbf{x}_\mathrm{c}) W(\textbf{x}_\mathrm{p}-\textbf{x}_\mathrm{c}) .
\end{equation}
$W$ is the \gls{assignment function} defined in section \ref{sec:apx-The mesh assignment function}, which involves a \gls{shape function} $S$ as previously (\gls{NGP}, \gls{CiC}, \gls{TSC}, etc.). It is generally important to be consistent between the \gls{mesh assignment} scheme and the \gls{interpolation} scheme. In particular, for \gls{PM} codes, the same prescription should be used for \glslink{mesh assignment}{density assignment} and for \glslink{interpolation}{interpolating} accelerations at \glslink{dark matter particles}{particles}' positions. This ensures the absence of artificial self-forces (forces exerted by a particle on itself) and \gls{momentum} conservation \citep{Hockney1981}.

For the \gls{NGP} scheme, the value of $A_\mathrm{p}$ for a particle is just the value of $A(\textbf{x}_\mathrm{c})$ in its parent cell $(i,j,k)$. For the \gls{CiC} scheme, using equations \eqref{eq:A_in_particles} and \eqref{eq:CiC_1}--\eqref{eq:CiC_8}, we find:
\begin{eqnarray}
A_\mathrm{p} = & A_{(i,j,k)} t_x t_y t_z + A_{(ii,j,k)} d_x t_y t_z + A_{(i,jj,k)} t_x d_y t_z + A_{(i,j,k k)} t_x t_y d_z\nonumber\\
& + A_{(ii,jj,k)} d_x d_y t_z + A_{(ii,j,kk)} d_x t_y d_z + A_{(i,jj,kk)} t_x d_y d_z + A_{(ii,jj,kk)} d_x d_y d_z .
\end{eqnarray}
This is identical to trilinear \gls{interpolation}.

\section{Poisson equation and accelerations}
\label{sec:apx-Poisson equation and accelerations}

After \glslink{mesh assignment}{density assignment}, several steps are done on the mesh in \gls{PM} codes: solving the \gls{Poisson equation} to get the reduced \gls{gravitational potential} $\widetilde{\Phi} \equiv \Delta^{-1} \delta$ (section \ref{sec:apx-Solving the Poisson equation}), and then differentiating to get the reduced accelerations $\tilde{g} \equiv \nabla \left( \Delta^{-1} \delta \right)$ (section \ref{sec:apx-Computation of the accelerations}).

\subsection{Solving the Poisson equation}
\label{sec:apx-Solving the Poisson equation}

It is customary to solve the \gls{Poisson equation} in Fourier space:
\begin{enumerate}
\item the configuration-space \gls{density contrast} $\delta(\textbf{x})$ is \glslink{Fourier transform}{Fourier-transformed} to get $\delta(\textbf{k})$;
\item the reduced \gls{gravitational potential} is estimated by solving the \gls{Poisson equation} in Fourier space, $\widetilde{\Phi}(\textbf{k})~=~G(\textbf{k}) \delta(\textbf{k})$, where $G(\textbf{k})$ is a \glslink{Green function}{Green's function} for the Laplacian, discussed below;
\item the reduced \gls{gravitational potential} $\widetilde{\Phi}(\textbf{k})$ is transformed back to real space to get $\widetilde{\Phi}(\textbf{x})$.
\end{enumerate}

As noted in section \ref{sec:apx-Definitions and data structures}, the same array can be used to store $\delta$ and $\widetilde{\Phi}$, by doing in-place \glslink{Fourier transform}{Fourier transforms}.

\paragraph{Fourier transforms.}Steps 1 and 3 involve forward and backward discrete \glslink{Fourier transform}{Fourier transforms}. In the codes implemented for this thesis, we use the Fast Fourier Transform approach for discrete data, provided by the FFTW software library,\footnote{\href{http://www.fftw.org/}{http://www.fftw.org/}} defined and normalized as follows, for the forward and backward operations respectively:
\begin{eqnarray*}
\hat{f}_{\ell,m,n} & = & \Delta x \Delta y \Delta z \displaystyle \sum\limits_{i=0}^{N_{\mathrm{g}0}-1}\sum\limits_{j=0}^{N_{\mathrm{g}1}-1}\sum\limits_{k=0}^{N_{\mathrm{g}2}-1} f_{i,j,k} \, \erm^{- 2\i \pi (i\ell +jm + kn)/N_\mathrm{g} } ,\\
f_{i,j,k} & = & \frac{1}{L_0 L_1 L_2} \displaystyle \sum\limits_{i=0}^{N_{\mathrm{g}0}-1}\sum\limits_{j=0}^{N_{\mathrm{g}1}-1}\sum\limits_{k=0}^{N_{\mathrm{g}2}-1} \hat{f}_{\ell,m,n} \, \erm^{2\i \pi (i\ell +jm + kn)/N_\mathrm{g} } .
\end{eqnarray*}
In the following, we note the components of a Fourier mode $\textbf{k}$ as $k_x = \frac{2\pi}{L_0} \ell$, $k_y = \frac{2\pi}{L_1} m$, $k_z = \frac{2\pi}{L_2} n$. 

\paragraph{Green's function.}The choice for the \glslink{Green function}{Green's function} $G(\textbf{k})$ depends on how one wants to represent to Laplacian in configuration space. In Fourier space, the reduced \glslink{gravitational potential}{potential} obeys $-k^2 \widetilde{\Phi}(\textbf{k}) \equiv \delta(\textbf{k})$ where $k^2 \equiv |\textbf{k}|^2 = k_x^2 + k_y^2 + k_z^2$. It is therefore natural to simply use as \glslink{Green function}{Green's function} for the Laplacian $G(\textbf{k}) = -1/k^2$. This is the choice adopted in \textsc{\gls{Gadget-2}} \citep{Springel2001,Springel2005} and in the codes used in this thesis. Care should be taken however, as this choice corresponds to a highly \gls{non-local} function in configuration space \citep[see e.g. the discussion in][appendix E]{Birdsall1985}. Alternatively, we can discretize the Laplacian operator using the so-called 7-point template,
\begin{equation}
(\Delta \Phi)_{i,j,k} = \Phi_{i-1,j,k} + \Phi_{i+1,j,k} + \Phi_{i,j-1,k} + \Phi_{i,j+1,k} + \Phi_{i,j,k-1} + \Phi_{i,j,k+1} - 6 \Phi_{i,j,k} ,
\end{equation}
for which the \glslink{Green function}{Green's function} is given by
\begin{equation}
G(\textbf{k}) = -\frac{1}{4} \left[ \sin^2\left(\frac{k_x \Delta x}{2}\right) + \sin^2\left(\frac{k_y \Delta y}{2}\right) + \sin^2\left(\frac{k_z \Delta z}{2} \right) \right]^{-1} .
\end{equation}

\paragraph{Force smoothing.} Due to the finite resolution of the \gls{PM} grid, short-range forces cannot be accurately resolved, which can cause spurious effects in \glslink{N-body simulation}{simulations} \citep{Hockney1981}. For this reason, we smooth the short-range forces by multiplying by a \gls{Gaussian kernel} in Fourier space,
\begin{equation}
K_{k_\mathrm{s}}(k) = \exp\left( -\frac{1}{2} \frac{k^2}{k_\mathrm{s}^2} \right), \quad \mathrm{where} \quad k_\mathrm{s} \equiv \frac{2\pi}{L} A_\mathrm{s}.
\end{equation}
$A_\mathrm{s}$ is a free parameter that defines the split between long-range and short-range forces, in units of mesh cells. In our codes, we adopted $A_\mathrm{s} = 1.25$, the default value used in \textsc{\gls{Gadget-2}}.

\paragraph{Deconvolution of the CiC kernel.} We also correct for the convolution with the \gls{CiC} kernel, by dividing twice by (see section \ref{sec:apx-Common mesh assignment schemes})
\begin{equation}
K_\mathrm{CiC}(\textbf{k}) = \sinc^2\left( \frac{k_x}{k_{\mathrm{Nyq},x}} \right) \sinc^2\left( \frac{k_y}{k_{\mathrm{Nyq},y}} \right) \sinc^2\left( \frac{k_z}{k_{\mathrm{Nyq},z}} \right) .
\end{equation}
One deconvolution corrects for the smoothing effect of the \gls{CiC} in the \glslink{mesh assignment}{density assignment}, the other for the force \gls{interpolation} \citep{Springel2005}.

\paragraph{Overall factor in Fourier space.} Summing up our discussions in this section, the overall factor that we apply to $\delta$ in Fourier space (that we still note $G(\textbf{k})$ for convenience) is
\begin{equation}
G(\textbf{k}) = -\frac{1}{k^2} \times \frac{K_{k_\mathrm{s}}(k)}{K_\mathrm{CiC}(\textbf{k})^2} .
\end{equation}
After performing an inverse \gls{Fourier transform}, we obtain the reduced \gls{gravitational potential} on the mesh.

\subsection{Computation of the accelerations}
\label{sec:apx-Computation of the accelerations}

We get the reduced accelerations on the mesh by finite differencing the reduced \glslink{gravitational potential}{potential}. It would also be possible to take the gradient in Fourier space, by multiplying the \glslink{gravitational potential}{potential} by a factor $-\i \textbf{k}$ and obtaining directly the accelerations. However, this would require an inverse \gls{Fourier transform} for each coordinate (i.e. three instead of one), with little gain in accuracy compared to finite differences \citep{Springel2005}. 

We adopt central finite differences. Several schemes are possible depending on the desired accuracy. The two-point \glslink{FDA}{finite difference approximation} (\glslink{FDA}{FDA2}) is
\begin{equation}
\tilde{g}x_{(i,j,k)} \equiv \left. \pd{\widetilde{\Phi}}{x} \right|_{(i,j,k)} \approx \frac{1}{\Delta x} \left[ \frac{1}{2} \widetilde{\Phi}_{(i+1,j,k)} - \frac{1}{2} \widetilde{\Phi}_{(i-1,j,k)} \right]
\end{equation}
and similar formulae for the other coordinates $\tilde{g}y$ and $\tilde{g}z$. The accuracy is of order $\mathcal{O}(\Delta x^2)$.

In the codes implemented for this thesis, we adopted the four-point \glslink{FDA}{finite difference approximation} (\glslink{FDA}{FDA4}), as in \textsc{\gls{Gadget-2}},
\begin{equation}
\tilde{g}x_{(i,j,k)} \equiv \left. \pd{\widetilde{\Phi}}{x} \right|_{(i,j,k)} \approx \frac{1}{\Delta x} \left[ \frac{2}{3} \left( \widetilde{\Phi}_{(i+1,j,k)} - \widetilde{\Phi}_{(i-1,j,k)} \right) - \frac{1}{12} \left( \widetilde{\Phi}_{(i+2,j,k)} - \widetilde{\Phi}_{(i-2,j,k)} \right) \right]
\end{equation}
which offers order $\mathcal{O}(\Delta x^4)$ accuracy. In the two equations above, \gls{periodic boundary conditions} should always be enforced: $i+1$ is actually $\mathrm{mod}(i+1,N_{\mathrm{g}0})$, etc.

After having computed the three components of the accelerations on the grid, $\tilde{g}x(\textbf{x}_\mathrm{c})$, $\tilde{g}y(\textbf{x}_\mathrm{c})$, $\tilde{g}z(\textbf{x}_\mathrm{c})$, we \glslink{interpolation}{interpolate} with the \gls{CiC} scheme (see section \ref{sec:apx-Interpolation}) to get the accelerations at particles' positions, $\tilde{g}x(\textbf{x}_\mathrm{p})$, $\tilde{g}y(\textbf{x}_\mathrm{p})$, $\tilde{g}z(\textbf{x}_\mathrm{p})$.

\section{Update of positions and momenta}
\label{sec:apx:KDK}

Now that we have the accelerations for each particle from the grid-based Poisson solver (step 1 in section \ref{sec:apx_PM_steps}), we are able to update their \glslink{momentum}{momenta} (``\gls{kick}'') and their positions (``\gls{drift}''). This corresponds to steps 2 and 3 in section \ref{sec:apx_PM_steps}. At this point, we have to adopt a time integration scheme to update positions and \glslink{momentum}{momenta} from $a_i$ to $a_f$, and to define \Gls{kick} and \Gls{drift} operators. This is the object of sections \ref{sec:apx_Time_integrators} and \ref{sec:apx_K_D_operators}, respectively.

\subsection{Time integrators}
\label{sec:apx_Time_integrators}

Let us consider a Hamiltonian system, described in \gls{phase space} by the canonical coordinates $\textbf{z}=(q,p)$ and the Hamiltonian $\mathcal{H}(p,q) \equiv p^2/2 + \Phi(q)$. If we call $f(\textbf{z})=(p,-\partial \Phi/ \partial q)$, then \gls{Hamilton's equations} simply read $\dot{\textbf{z}} = f(\textbf{z})$. \gls{Hamilton's equations} are a \glslink{symplecticity}{symplectic} map, which means that the energy and the volume in \gls{phase space} are time-invariants:
\begin{equation}
\deriv{\mathcal{H}}{t} = 0 \quad \mathrm{and} \quad \nabla \cdot f = 0.  
\end{equation}

It is generally important to adopt a numerical integrator that respects these two conditions, at least approximately (see also the discussion in section \ref{sec:Hamiltonian Monte Carlo}). For a map $\textbf{z}(t) = \mathscr{F}(\textbf{z}_0)$, the volume in \gls{phase space} is conserved if $\det \pd{\mathscr{F}}{\textbf{z}} =1$. Classical first order time integrators use \gls{Euler's method}. In particular, the explicit \glslink{Euler's method}{Euler method},
\begin{equation}
\textbf{z}_{n+1} = \textbf{z}_{n} + f(\textbf{z}_n) \Delta t;  \quad \mathrm{for~which} \quad \det \pd{\mathscr{F}}{\textbf{z}} = 1 + \Delta t^2 \pd{^2 \Phi}{q^2} ,
\end{equation}
and the implicit \glslink{Euler's method}{Euler method},
\begin{equation}
\textbf{z}_{n+1} = \textbf{z}_{n} + f(\textbf{z}_{n+1}) \Delta t;  \quad \mathrm{for~which} \quad \det \pd{\mathscr{F}}{\textbf{z}} = \frac{1}{1 + \Delta t^2 \pd{^2 \Phi}{q^2}} ,
\end{equation}
are only approximately \glslink{symplecticity}{symplectic}. Using the \glslink{dark matter particles}{particles}' positions at time $t_n$ and \glslink{momentum}{momenta} at time $t_{n+1}$ makes the \glslink{Euler's method}{Euler integrator} \glslink{symplecticity}{symplectic}:
\begin{equation}
\textbf{z}_{n+1} = \textbf{z}_n + f(q_n, p_{n+1}) \Delta t; \quad \det \pd{\mathscr{F}}{\textbf{z}} = 1.
\end{equation}

For this thesis, we adopted the second-order \glslink{symplecticity}{symplectic} ``\glslink{KDK}{kick-drift-kick}'' algorithm, also known as the \gls{leapfrog} scheme \citep[e.g.][see also section \ref{sec:The leapfrog scheme integrator}]{Birdsall1985}:
\begin{eqnarray}
p_{n+1/2} & = & p_n - \left. \pd{\Phi}{q} \right|_n \frac{\Delta t}{2} ,\\
q_{n+1} & = & q_n + p_{n+1/2} \, \Delta t ,\\
p_{n+1} & = & p_{n+1/2} - \left. \pd{\Phi}{q} \right|_{n+1} \frac{\Delta t}{2} .
\end{eqnarray}
It is a straightforward exercise to check that this scheme exactly preserves volume in \gls{phase space}.

\begin{figure}\centering
\includegraphics[width=0.8\columnwidth]{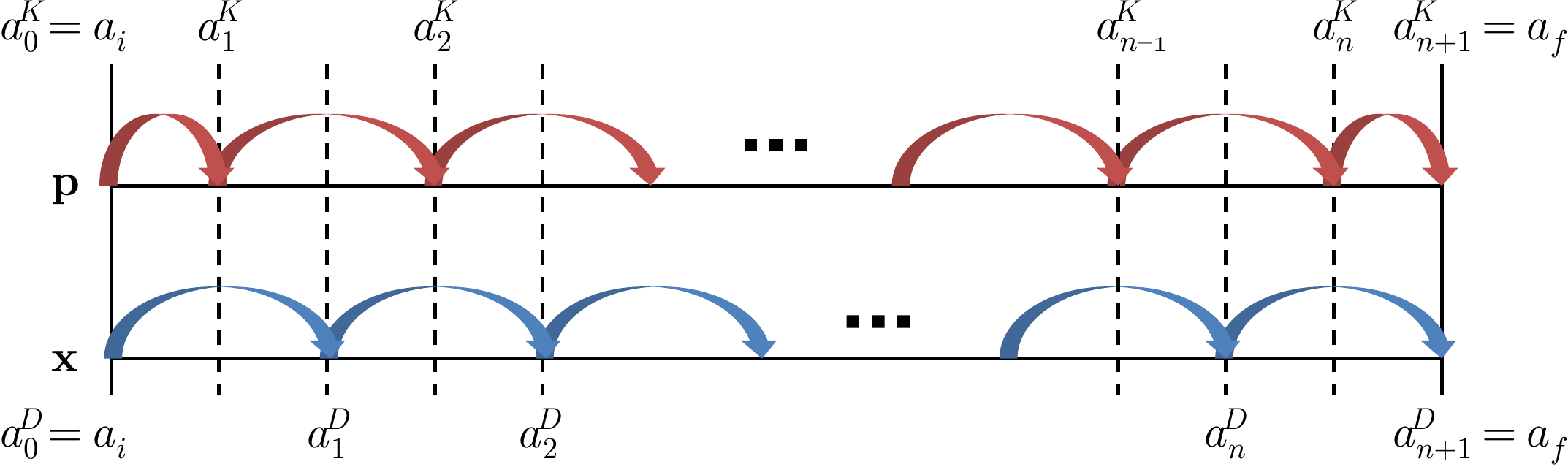}
\caption{Schematic illustration of the \gls{leapfrog} integrator. Particles' \glslink{momentum}{momenta} and positions are updated in turn, given the value of the other variable within the time interval.\label{fig:KDK}}
\end{figure}

For \gls{PM} and {\cola} codes, we assume a constant integration step $\Delta a \equiv \frac{a_f - a_i}{n}$, in such a way that the \glslink{initial conditions}{initial} \gls{scale factor} is $a_i = a_0$ and the \glslink{final conditions}{final} \gls{scale factor} is $a_f = a_{n+1} = a_i + n \Delta a$. A schematic view of the \gls{leapfrog} integration scheme is show in figure \ref{fig:KDK}. Note that during the evolution, positions and \glslink{momentum}{momenta} are not synchronized but displaced by half a timestep. For this reason during the first timestep, we give the \glslink{dark matter particles}{particles} only ``half a \gls{kick}'' using the accelerations computed at $a_i$; and during the last timestep, we give the \glslink{dark matter particles}{particles} an additional ``half a \gls{kick}'', to synchronize \glslink{momentum}{momenta} with positions at $a_f$. 

\subsection{Kick and Drift operators}
\label{sec:apx_K_D_operators}

In equations \eqref{eq:drift} and \eqref{eq:kick}, all the explicit dependence on the \gls{scale factor} is in the prefactors $\mathpzc{D}(a)$ and $\mathpzc{K}(a)$. The \gls{leapfrog} scheme algorithm relies on integrating the equations on a small timestep and approximating the \glslink{momentum}{momenta} or accelerations in the integrands by their value at some time within the interval. More precisely, for the ``\gls{drift} equation'':
\begin{equation}
\textbf{x}\!\left(a_f^\mathrm{D}\right) - \textbf{x}\!\left(a_i^\mathrm{D}\right) = \int_{a_i^\mathrm{D}}^{a_f^\mathrm{D}} \mathpzc{D}(\tilde{a}) \textbf{p}(\tilde{a}) \, \drm \tilde{a} \approx \left( \int_{a_i^\mathrm{D}}^{a_f^\mathrm{D}} \mathpzc{D}(\tilde{a}) \, \drm \tilde{a} \right) \textbf{p}\!\left(a^\mathrm{K}\right)
\end{equation}
and similarly for the ``\gls{kick} equation'':
\begin{equation}
\textbf{p}\!\left(a_f^\mathrm{K}\right) - \textbf{p}\!\left(a_i^\mathrm{K}\right) = \int_{a_i^\mathrm{K}}^{a_f^\mathrm{K}} \mathpzc{K}(\tilde{a}) \left[\nabla\left(\Delta^{-1}\delta\right)\right]\!\!(\tilde{a}) \, \drm \tilde{a} \approx \left( \int_{a_i^\mathrm{K}}^{a_f^\mathrm{K}} \mathpzc{K}(\tilde{a}) \, \drm \tilde{a} \right) \left[\nabla\left(\Delta^{-1}\delta\right)\right]\!\!(a^\mathrm{D})
\end{equation}
This defines the \Gls{drift} ($\mathrm{D}$) and \Gls{kick} ($\mathrm{K}$) operators:
\begin{eqnarray}
\label{eq:D_standard}
\mathrm{D}(a_i^\mathrm{D},a_f^\mathrm{D},a^\mathrm{K}): & \quad & \!\!\!\!\!\!\!\!\!\textbf{x}(a_i^\mathrm{D}) \mapsto \textbf{x}(a_f^\mathrm{D}) = \textbf{x}(a_i^\mathrm{D}) + \left( \int_{a_i^\mathrm{D}}^{a_f^\mathrm{D}} \mathpzc{D}(\tilde{a}) \, \drm \tilde{a} \right) \textbf{p}\!\left(a^\mathrm{K}\right) \\
\label{eq:K_standard}
\mathrm{K}(a_i^\mathrm{K},a_f^\mathrm{K},a^\mathrm{D}): & \quad & \!\!\!\!\!\!\!\!\!\textbf{p}(a_i^\mathrm{D}) \mapsto \textbf{p}(a_f^\mathrm{D}) = \textbf{p}(a_i^\mathrm{D}) + \left( \int_{a_i^\mathrm{K}}^{a_f^\mathrm{K}} \mathpzc{K}(\tilde{a}) \, \drm \tilde{a} \right) \left[\nabla\left(\Delta^{-1}\delta\right)\right]\!\!(a^\mathrm{D})
\end{eqnarray}
Consistently with the scheme described in section \ref{sec:apx_Time_integrators}, the time evolution between $a_0$ and $a_{n+1}$ is then achieved by applying the following operator, $\mathrm{E}(a_{n+1},a_0)$, to the \glslink{initial conditions}{initial state} $(\textbf{x}(a_0),\textbf{p}(a_0))$:
\begin{equation}
\label{eq:operator_E}
\mathrm{K}(a_{n+1/2},a_{n+1},a_{n+1}) \mathrm{D}(a_n,a_{n+1},a_{n+1/2}) \left[ \prod_{i=0}^n \mathrm{K}(a_{i+1/2},a_{i+3/2},a_{i+1}) \mathrm{D}(a_i,a_{i+1},a_{i+1/2}) \right] \mathrm{K}(a_0,a_{1/2},a_0) .
\end{equation}

If the {\cola} scheme is adopted, we obtain in a similar manner, from equations \eqref{eq:drift-cola} and \eqref{eq:kick-cola}:
\begin{eqnarray}
\textbf{x}\!\left(a_f^\mathrm{D}\right) - \textbf{x}\!\left(a_i^\mathrm{D}\right) & \!\! \approx & \!\!\left( \int_{a_i^\mathrm{D}}^{a_f^\mathrm{D}} \mathpzc{D}(\tilde{a}) \, \drm \tilde{a} \right) \textbf{p}_\mathrm{MC}\!\left(a^\mathrm{K}\right) - \left( \int_{a_i^\mathrm{D}}^{a_f^\mathrm{D}} \deriv{D_1(\tilde{a})}{\tilde{a}} \, \drm \tilde{a} \right) \boldsymbol{\Psi}_1 + \left( \int_{a_i^\mathrm{D}}^{a_f^\mathrm{D}} \deriv{D_2(\tilde{a})}{\tilde{a}} \, \drm \tilde{a} \right) \boldsymbol{\Psi}_2, \nonumber \\
& \!\! = & \!\!\left( \int_{a_i^\mathrm{D}}^{a_f^\mathrm{D}} \mathpzc{D}(\tilde{a}) \, \drm \tilde{a} \right) \textbf{p}_\mathrm{MC}\!\left(a^\mathrm{K}\right) - \left[ D_1 \right]_{a_i^\mathrm{D}}^{a_f^\mathrm{D}} \boldsymbol{\Psi}_1 + \left[ D_2 \right]_{a_i^\mathrm{D}}^{a_f^\mathrm{D}} \boldsymbol{\Psi}_2, \\
\textbf{p}_\mathrm{MC}\!\left(a_f^\mathrm{K}\right) - \textbf{p}_\mathrm{MC}\!\left(a_i^\mathrm{K}\right) & \!\! \approx & \!\!\left( \int_{a_i^\mathrm{K}}^{a_f^\mathrm{K}} \mathpzc{K}(\tilde{a}) \, \drm \tilde{a} \right) \left( \left[ \nabla\left(\Delta^{-1}\delta\right)\right]\!\!(a^\mathrm{D}) - \mathpzc{V}[D_1](a^\mathrm{D}) \boldsymbol{\Psi}_1 - \mathpzc{V}[D_2](a^\mathrm{D}) \boldsymbol{\Psi}_2 \right) \nonumber \\
& \!\! = & \!\!\left( \int_{a_i^\mathrm{K}}^{a_f^\mathrm{K}} \mathpzc{K}(\tilde{a}) \, \drm \tilde{a} \right) \left( \left[ \nabla\left(\Delta^{-1}\delta\right)\right]\!\!(a^\mathrm{D}) - D_1(a^\mathrm{D}) \boldsymbol{\Psi}_1 + \left( D_2(a^\mathrm{D}) - D_1^2(a^\mathrm{D}) \right) \boldsymbol{\Psi}_2 \right) \! .
\end{eqnarray}
In the last line we used equations \eqref{eq:VD1} and \eqref{eq:VD2}. This defines new \Gls{drift} ($\mathrm{\widetilde{D}}$) and \Gls{kick} ($\mathrm{\widetilde{K}}$) operators:
\begin{eqnarray}
\label{eq:D_COLA}
\mathrm{\widetilde{D}}(a_i^\mathrm{D},a_f^\mathrm{D},a^\mathrm{K}) : & \quad & \!\!\!\!\!\!\!\!\!\textbf{x}(a_i^\mathrm{D}) \mapsto \textbf{x}(a_f^\mathrm{D}) = \textbf{x}(a_i^\mathrm{D}) + \left( \int_{a_i^\mathrm{D}}^{a_f^\mathrm{D}} \mathpzc{D}(\tilde{a}) \, \drm \tilde{a} \right) \textbf{p}_\mathrm{MC}\!\left(a^\mathrm{K}\right) - \left[ D_1 \right]_{a_i^\mathrm{D}}^{a_f^\mathrm{D}} \boldsymbol{\Psi}_1 + \left[ D_2 \right]_{a_i^\mathrm{D}}^{a_f^\mathrm{D}} \boldsymbol{\Psi}_2 \\
\mathrm{\widetilde{K}}(a_i^\mathrm{K},a_f^\mathrm{K},a^\mathrm{D}) : & \quad & \!\!\!\!\!\!\!\!\!\textbf{p}_\mathrm{MC}(a_i^\mathrm{D}) \mapsto \textbf{p}_\mathrm{MC}(a_f^\mathrm{D}) = \textbf{p}_\mathrm{MC}(a_i^\mathrm{D}) + \left( \int_{a_i^\mathrm{K}}^{a_f^\mathrm{K}} \mathpzc{K}(\tilde{a}) \, \drm \tilde{a} \right) \times \nonumber \\
\label{eq:K_COLA}
& & \hfill \quad\quad\quad\quad\quad\quad\quad \left( \left[ \nabla\left(\Delta^{-1}\delta\right)\right]\!\!(a^\mathrm{D}) - D_1(a^\mathrm{D}) \boldsymbol{\Psi}_1 + \left( D_2(a^\mathrm{D}) - D_1^2(a^\mathrm{D}) \right) \boldsymbol{\Psi}_2 \right) .
\end{eqnarray}
With {\cola}, the time evolution between $a_0$ and $a_{n+1}$ is achieved by applying the following operator to the \glslink{initial conditions}{initial state} $(\textbf{x}(a_0),\textbf{p}(a_0))$:
\begin{equation}
\mathrm{L}_+(a_{n+1}) \widetilde{\mathrm{E}}(a_{n+1},a_0) \mathrm{L}_-(a_0),
\end{equation}
where $\widetilde{\mathrm{E}}(a_{n+1},a_0)$ is the operator given by equation \eqref{eq:operator_E}, replacing $\mathrm{D}$ by $\mathrm{\widetilde{D}}$ and $\mathrm{K}$ by $\mathrm{\widetilde{K}}$, and we where we use \citep[see][appendix A]{Tassev2013}:
\begin{equation}
\mathrm{L}_\pm(a) : \quad \textbf{p}(a) \mapsto \textbf{p}(a) \pm \textbf{p}_\mathrm{LPT}(a) = \textbf{p}(a) \pm \frac{1}{\mathpzc{D}(a)} \left( -\deriv{D_1}{a} \boldsymbol{\Psi}_1 +  \deriv{D_2}{a} \boldsymbol{\Psi}_2 \right) .
\end{equation}
$\mathrm{L}_-$ transforms the \gls{initial conditions} to the rest frame of \gls{LPT} observers (this is the same as initializing $\textbf{p}_\mathrm{MC}$ to zero), and $\mathrm{L}_+$ adds back the \gls{LPT} \glslink{momentum}{momenta} to $\textbf{p}_\mathrm{MC}$ at the end. 

In the codes implemented for this thesis, the integrals appearing in the \Gls{kick} and \Gls{drift} operators (equations \eqref{eq:D_standard}, \eqref{eq:K_standard}, \eqref{eq:D_COLA}, \eqref{eq:K_COLA}) are explicitly computed numerically. Another approach for the discretization of time operators is proposed by \citet[][section A.3.2.]{Tassev2013}. When needed, the \glslink{linear growth factor}{first order growth factor} $D_1$ and its logarithmic derivative $f_1$ are also computed numerically by explicit integration. For the \gls{second-order growth factor} and its logarithmic derivative, we use the fitting functions given by equations \eqref{eq:fitting_D2} and \eqref{eq:fitting_f2} \citep{Bouchet1995},
\begin{equation}
D_2(\tau) \approx -\frac{3}{7} D_1(\tau) \Omega_\mathrm{m}^{-1/143} \quad \mathrm{and} \quad f_2(\tau) \approx 2 f_1(\tau)^{54/55}.
\end{equation}

\section{Setting up initial conditions}
\label{sec:apx-Setting up initial conditions}

The last missing part for a full cosmological pipeline including the \gls{PM}/{\cola} codes described in previous sections is a way to set up \gls{initial conditions} at $a=a_i$. The first step (section \ref{sec:apx_GRF_ICs}) is to generate a realization of the random \gls{density field} describing the early Universe. As argued in chapter \ref{chap:theory}, it is physically relevant to describe this field as a \glslink{grf}{Gaussian random field}.

The second step (section \ref{sec:apx_highz_particle}) is to produce a high-\gls{redshift} \gls{particle realization} from this \glslink{initial conditions}{initial} \gls{density field}, to be given to the \gls{PM} code. The common approach is to use \glslink{LPT}{Lagrangian perturbation theory} (the \gls{ZA} or \gls{2LPT}). Several existing codes perform this task: among others, \textsc{Grafic} \citep{Bertschinger2001}, \gls{N-GenIC} \citep[][using the \gls{ZA}]{Springel2001,Springel2005} and its \gls{2LPT} extension, \textsc{\gls{2LPTic}} \citep{Crocce2006,Pueblas2009}, \textsc{MPGrafic} \citep{Prunet2008}, \textsc{music} \citep{Hahn2011}. However, for the purpose of this thesis, we implemented an independent \gls{ZA}/\gls{2LPT} \gls{initial conditions} generator. It is especially designed for full consistency with the {\borg} algorithm (see chapter \ref{chap:BORG}); in particular, it uses the same routine as {\borg} for the generation of \gls{LPT} \glslink{displacement field}{displacement fields}.

\subsection{The initial Gaussian random field}
\label{sec:apx_GRF_ICs}

There exists many software packages that allow generating normal random variates (i.e. single Gaussian random variates with mean $0$ and variance $1$), for example using the well-known \glslink{Box-Muller method}{Box-M\"uller method}. We choose the routines provided by the GNU scientific library \citep{GSL}. We generate one such normal random variate in each cell of the \glslink{initial conditions}{initial} grid, and call the resulting vector the ``\glslink{initial conditions}{initial} \gls{white noise} field'' $\xi$. It is a random signal with constant \gls{power spectrum} ($\left\langle \xi \xi^\intercal = \textbf{1} \right\rangle$). Alternatively, we can choose to import ``constrained \gls{white noise}'' that comes, for example, of \glslink{large-scale structure inference}{large-scale structure inferences} performed with {\borg}.

Generally, using a vector of normal variates $\xi$, one can generate a realization of a \gls{grf} with mean $\mu$ and covariance matrix $C$ by simply taking any matrix $\sqrt{C}$ that satisfies $\sqrt{C}\sqrt{C}^{\,\intercal} = C$ and computing $x=\sqrt{C} \xi + \mu$. One general way to generate $\sqrt{C}$ under the condition that $C$ has only positive definite \glslink{eigenvalue}{eigenvalues} is to use the so-called \gls{Cholesky decomposition}, implemented in many numerical packages.

For cosmological \gls{initial conditions}, however, the problem is generally much simpler. As we are generating a random realization of the \gls{density contrast} $\delta$, the mean is $\mu=0$ and, from \gls{statistical homogeneity} and \glslink{statistical isotropy}{isotropy}, the covariance matrix $C$ should be diagonal in Fourier space and contain the \gls{power spectrum} coefficients $P(k)/(2\pi)^{3/2}$ (see section \ref{sec:power-spectrum}). Hence, an obvious choice for the matrix $\sqrt{C}$ is the diagonal matrix containing the coefficients $\sqrt{P(k)/(2\pi)^{3/2}}$. Therefore, the procedure is to \glslink{Fourier transform}{Fourier-transform} $\xi$, to multiply each of its Fourier modes of norm $k$ by $\sqrt{P(k)/(2\pi)^{3/2}}$, and to perform an inverse \gls{Fourier transform} to get $\delta$ in configuration space.

Physical assumptions are needed for the \gls{power spectrum} coefficients $P(k)$. One possible approach is to use the outputs of Boltzmann codes that describe the early Universe \citetext{e.g. \textsc{cmbfast} -- \citealp{Seljak1996}, \textsc{camb} -- \citealp{Lewis2002a}, or \textsc{class} -- \citealp{Lesgourgues2011,Blas2011}}. However, in our implementation, we choose (as in {\borg}) to use the analytical \gls{power spectrum} from \citet{Eisenstein1998,Eisenstein1999} for the baryon-CDM \gls{fluid} (including \glslink{BAO}{baryonic wiggles}). It depends on the following \gls{cosmological parameters}, which have to be specified: $\Omega_\Lambda$, $\Omega_\mathrm{m}$, $\Omega_\mathrm{b}$, $n_\mathrm{s}$ and $\sigma_8$.

When performing \glslink{constrained simulation}{constrained simulations} (see section \ref{sec:Filtering via constrained simulations}), all the steps described in this section are bypassed, and we directly make use of the \glslink{initial conditions}{initial} \gls{density contrast} field inferred with {\borg}. 

\subsection{The high-redshift particle realization}
\label{sec:apx_highz_particle}

We start from ``grid-like'' \gls{initial conditions}, i.e. a \glslink{particle realization}{realization} of $N_\mathrm{p}$ \gls{dark matter particles}, placed on a regular lattice. More precisely, for $0 \leq \mathpzc{i} < N_{\mathrm{p}0}$, $0 \leq \mathpzc{j} < N_{\mathrm{p}1}$, $0 \leq \mathpzc{k} < N_{\mathrm{p}2}$, we place a particle at Lagrangian coordinates $\textbf{q} = (\mathpzc{i} L_0/N_{\mathrm{p}0}, \mathpzc{j} L_1/N_{\mathrm{p}1}, \mathpzc{k} L_2/N_{\mathrm{p}2})$. All the masses are set to the constant value given in footnote \ref{fn:particle_mass}, and at this point all the velocities are zero. Finally, each particle's id is set to $\texttt{mp} = \mathpzc{k} + N_{\mathrm{p}2} \times \left( \mathpzc{j} + N_{\mathrm{p}1} \times \mathpzc{i} \right)$. This allows to keep a memory of the \glslink{initial conditions}{initial} position of \glslink{dark matter particles}{particles} at any later time, even in the \gls{PM} code.

The following step is to compute the \gls{ZA} and \gls{2LPT} \glslink{displacement field}{displacements} for each particle, given the \glslink{initial conditions}{initial} \gls{density contrast} field $\delta(\textbf{q})$ generated in section \ref{sec:apx_GRF_ICs}. We proceed as follows. The \glslink{Lagrangian potential}{first-order potential} field, $\phi^{(1)}(\textbf{q})$, is evaluated on the Lagrangian grid by solving equation \eqref{eq:Poisson_phi1} in Fourier space,\footnote{We denote by $\boldsymbol{\upkappa}$ a Fourier mode on the Lagrangian grid, $\kappa$ its norm.}
\begin{equation}
\phi^{(1)}(\boldsymbol{\upkappa}) = -\delta(\boldsymbol{\upkappa})/\kappa^2.
\end{equation}
Each of its second order derivatives are also evaluated in Fourier space, using
\begin{equation}
\phi^{(1)}_{,ab}(\boldsymbol{\upkappa}) = -\phi^{(1)}(\boldsymbol{\upkappa}) \boldsymbol{\upkappa}_a \cdot \boldsymbol{\upkappa}_b.
\end{equation}
and inverse \glslink{Fourier transform}{Fourier-transformed}. From the configuration-space quantity
\begin{equation}
\boldsymbol{\phi}(\textbf{q}) \equiv \phi^{(1)}_{,xx}(\textbf{q})\phi^{(1)}_{,yy}(\textbf{q}) + \phi^{(1)}_{,xx}(\textbf{q})\phi^{(1)}_{,zz}(\textbf{q}) + \phi^{(1)}_{,yy}(\textbf{q})\phi^{(1)}_{,zz}(\textbf{q}) - \phi^{(1)}_{,xy}(\textbf{q})^2 - \phi^{(1)}_{,xz}(\textbf{q})^2 - \phi^{(1)}_{,yz}(\textbf{q})^2 ,
\end{equation}
we compute the \glslink{Lagrangian potential}{second-order potential} field, $\phi^{(2)}(\textbf{q})$, again in Fourier space, using (see equation \eqref{eq:Poisson_phi2})
\begin{equation}
\phi^{(2)}(\boldsymbol{\upkappa}) = - \boldsymbol{\phi}(\boldsymbol{\upkappa})/\kappa^2.
\end{equation}

Once $\phi^{(1)}(\textbf{q})$ and $\phi^{(2)}(\textbf{q})$ are known, we evaluate the first and second order \glslink{displacement field}{displacements} $\boldsymbol{\Psi}^{(1)}(\textbf{q}) \equiv \nabla_\textbf{q} \phi^{(1)}(\textbf{q})$ and $\boldsymbol{\Psi}^{(2)}(\textbf{q}) \equiv \nabla_\textbf{q} \phi^{(2)}(\textbf{q})$ on the \glslink{initial conditions}{initial} grid in configuration space, by using a \glslink{FDA}{finite difference approximation} scheme at order 2 (see section \ref{sec:apx-Computation of the accelerations}). Then, we \glslink{interpolation}{interpolate} from the grid to  \glslink{dark matter particles}{particles}' Lagrangian positions using a \gls{CiC} scheme (see section \ref{sec:apx-Interpolation}).

Finally, \glslink{dark matter particles}{particles} are displaced from their Lagrangian positions and their velocities are modified as prescribed by \gls{LPT} (equations \eqref{eq:mapping_LPT_code} and \eqref{eq:velocities_LPT_code}). More precisely, \glslink{dark matter particles}{particles} are given a zeroth ``\gls{kick}'',
\begin{equation}
\mathrm{K}_0(a_i): \quad \textbf{u} = 0 \mapsto \textbf{u}(a_i) = -f_1(a_i) D_1(a_i) \mathcal{H}(a_i) \boldsymbol{\Psi}_1(\textbf{q}) + f_2(a_i) D_2(a_i) \mathcal{H}(a_i) \boldsymbol{\Psi}_2(\textbf{q}) ,
\end{equation}
where $\textbf{u} \equiv \drm \textbf{x}/\drm \tau = a\mathcal{H} \drm \textbf{x}/\drm a$. From this we deduce the \glslink{initial conditions}{initial} \glslink{momentum}{momenta} in code units,
\begin{equation}
\textbf{p}(a_i) = \frac{1}{a_i \mathcal{H}(a_i) \mathpzc{D}(a_i)}\textbf{u}(a_i) .
\end{equation}
They also follow a zeroth ``\gls{drift}'':
\begin{equation}
\mathrm{D}_0(a_i): \quad \textbf{q} \mapsto \textbf{x}(a_i) = \textbf{q} - D_1(a_i) \boldsymbol{\Psi}_1(\textbf{q}) + D_2(a_i) \boldsymbol{\Psi}_2(\textbf{q}) .
\end{equation}
The required numerical prefactors are computed as described at the end of section \ref{sec:apx_K_D_operators}.

%% file: AppendixC/AppendixCContent.tex
\chapter{Cosmic structures identification and classification algorithms}
\label{apx:classification}
\minitoc

\defcitealias{Popper1972}{Karl}
\begin{flushright}
\begin{minipage}[c]{0.6\textwidth}
\rule{\columnwidth}{0.4pt}

``Whenever a theory appears to you as the only possible one, take this as a sign that you have neither understood the theory nor the problem which it was intended to solve.''\\
--- \citetalias{Popper1972} \citet{Popper1972}, \textit{Objective Knowledge: An Evolutionary Approach}

\vspace{-5pt}\rule{\columnwidth}{0.4pt}
\end{minipage}
\end{flushright}

\abstract{\section*{Abstract} This appendix discusses methods for identifying and \glslink{cosmic web classification}{classifying} \glslink{structure type}{structures} in the \gls{cosmic web}. As many approaches exist (see the introduction of chapter \ref{chap:ts}), in the following we only focus on the algorithms used in this thesis: the {\vide} toolkit for the identification of static \glslink{void}{voids} (section \ref{sec:apx-vide}), and the \gls{T-web} approach for dissecting the dynamic \gls{cosmic web} into \glslink{cluster}{clusters}, \glslink{filament}{filaments}, \glslink{sheet}{sheets}, and \glslink{void}{voids} (section \ref{sec:apx-tweb}).}

\section{VIDE: the Void IDentification and Examination toolkit}
\label{sec:apx-vide}

This section describes {\vide}, the Void IDentification and Examination toolkit. It is a static \gls{void} finder operating on \glslink{density field}{density fields}, used in chapter \ref{chap:dmvoids} of this thesis. The details behind {\vide} are described in its accompanying paper, \citet{Sutter2015VIDE}, and its website \href{http://www.cosmicvoids.net/}{http://www.cosmicvoids.net/}. {\vide} is based on \textsc{\gls{zobov}} \citep[ZOnes Bordering On Voidness,][]{Neyrinck2008} for the \gls{void} finding part (sections \ref{sec:apx-VTFE} and \ref{sec:apx-watershed}), and includes a set of additional features for pre- and post-processing \gls{void} catalogs (section \ref{sec:apx-voidcatalogs}).

\subsection{Voronoi Tessellation Density Estimation}
\label{sec:apx-VTFE}

The algorithm begins by building a \gls{Voronoi tessellation} of the tracer particle population \citep{Schaap2000,Schaap2007}. This provides a \gls{density field} \gls{estimator} (the Voronoi Tessellation Field Estimator, \textsc{vtfe}) based on the underlying particle positions. The \textsc{vtfe} (along with its dual, the Delaunay Tessellation Field Estimator, \textsc{dtfe}) is a \gls{local} density estimate that is especially suitable for astronomical data \citep{vandeWeygaert2009,Cautun2011}.

The \gls{Voronoi tessellation} is a partitioning of space into cells around each particle. For each particle $i$, the corresponding \glslink{Voronoi tessellation}{Voronoi cell} is the region consisting of all points closer to that particle than to any other. The density estimate at particle $i$ is $1/V(i)$, where $V(i)$ is the volume of the \glslink{Voronoi tessellation}{Voronoi cell} around particle $i$. It is further assumed constant density across the volume of each \glslink{Voronoi tessellation}{Voronoi cell}, which effectively sets a smoothing scale for the continuous \gls{density field}.

Finally, the \gls{Voronoi tessellation} also provides the adjacency measurement for each particle $i$ (i.e. the set of particles whose \glslink{Voronoi tessellation}{Voronoi cells} have a common boundary with $i$'s cell), which \textsc{\gls{zobov}} uses in the next step.

\subsection{The watershed algorithm}
\label{sec:apx-watershed}

\textsc{\gls{zobov}} then uses the \gls{watershed transform} \citep{Platen2007} to group \glslink{Voronoi tessellation}{Voronoi cells} into zones and subsequently \glslink{void}{voids}. Minima (also called cores or basins) are first identified as particles with lower density than any of their \glslink{Voronoi tessellation}{Voronoi} neighbors. Then, the algorithm merges nearby \glslink{Voronoi tessellation}{Voronoi cells} into zones (the set of cells for which density flows downward into the zone's core). Finally, the \gls{watershed transform} groups adjacent zones into \glslink{void}{voids} by finding minimum-density barriers between them and joining zones together. This can be thought of, for each zone $z$, as setting the ``water level'' to its minimum density and raising it gradually. Water may flow along lines joining adjacent \glslink{Voronoi tessellation}{Voronoi zones}, adding them to the \gls{void} defined around zone $z$. The process is stopped when water flows into a deeper zone (one with a lower core than $z$) or if $z$ is the deepest ``parent'' \gls{void}, when water floods the whole field. The \gls{void} corresponding to zone $z$ is defined as the set of zones filled with water just before this happens, and its boundary is the ridgeline which retains the flow of water. As can be understood from this description, the \gls{watershed transform} naturally builds a nested \glslink{void hierarchy}{hierarchy of voids} \citep{Lavaux2012,Bos2012}. 

\textsc{\gls{zobov}} imposes a density-based criterion within the \gls{void} finding operation: adjacent zones are only added to a \gls{void} if the density of the wall between them is less than $0.2$ times the mean particle density \citetext{\citealp{Platen2007}; see \citealp{Blumenthal1992,Sheth2004} for the role of the corresponding $\delta = -0.8$ \glslink{density contrast}{underdensity}}. This density threshold prevents \glslink{void}{voids} from expanding deeply into overdense structures and limits the depth of the \gls{void hierarchy} \citep{Neyrinck2008}. By default, {\vide} reports every identified basin as a \gls{void} (regardless of the density of the initial zone), but facilities exist for filtering the \gls{void} catalogs based on various criteria \citep{Sutter2015VIDE}. 

\subsection{Processing and analysis of void catalogs}
\label{sec:apx-voidcatalogs}

The {\vide} toolkit provides routines for performing many analysis tasks, such as manipulating, filtering, and comparing \gls{void} catalogs, plotting \gls{void} properties, stacking, computing clustering statistics and fitting \glslink{density profile}{density profiles} \citep{Sutter2015VIDE}. In this section, we briefly describe the details behind the three \gls{void} statistics used in chapter \ref{chap:dmvoids}: \glslink{number function}{number functions}, \glslink{ellipticity distribution}{ellipticity distributions}, and \glslink{density profile}{density profiles}.

\subsubsection{Number functions}

The effective radius of a \gls{void} is defined as
\begin{equation}
R_\mathrm{v} \equiv \left( \frac{3}{4\pi} V \right)^{1/3} ,
\end{equation}
where $V$ is the total volume of the \glslink{Voronoi tessellation}{Voronoi cells} that make up the \gls{void}. From this definition, \glslink{void}{voids} with effective radius smaller than $\bar{n}^{-1/3}$, where $\bar{n}$ is the mean number density of tracers, are excluded to prevent the effects of \gls{shot noise}. 

Based on this definition, {\vide} includes a built-in plotting routine for the cumulative \glslink{number function}{number functions} of multiple \gls{void} catalogs on a logarithmic scale (see figure \ref{fig:numberfunc}).

\subsubsection{Ellipticity distributions}
\label{sec:apx-ellipticity}

For each \gls{void} in the catalog, {\vide} also reports the volume-weighted center of all \glslink{Voronoi tessellation}{Voronoi cells} in the \gls{void}, or macrocenter:
\begin{equation}
\textbf{x}_\mathrm{v} \equiv \frac{1}{\sum_i V_i} \sum_i \textbf{x}_i V_i,
\end{equation}
where $\textbf{x}_i$ and $V_i$ are the positions and \glslink{Voronoi tessellation}{Voronoi volumes} of each tracer particle $i$, respectively.

\glslink{void}{Void} shapes are computed from \gls{void} member particles by constructing the inertia tensor:
\begin{eqnarray}
M_{xx} & = & \sum_{i=1}^{N_\mathrm{p}} \left( y_i^2 + z_i^2 \right) , \\
M_{xy} & = & - \sum_{i=1}^{N_\mathrm{p}} x_i y_i,
\end{eqnarray}
where $N_\mathrm{p}$ is the number of particles in the \gls{void}, and $(x_i,y_i,z_i)$ is the set of coordinates of particle $i$ relative to the \gls{void} macrocenter. The other components of the inertia tensor are obtained by cyclic permutation of coordinates. The \glslink{eigenvalue}{eigenstructure} of the inertia tensor gives the ellipticity of the \gls{void}:
\begin{equation}
\epsilon = 1 - \left( \frac{J_1}{J_3} \right)^{1/4} ,
\end{equation}
where $J_1$ and $J_3$ are the smallest and the largest \glslink{eigenvalue}{eigenvalues} of the inertia tensor, respectively. The \gls{ellipticity distribution} of \glslink{void}{voids} as a function of their effective radius follows from this definition (see figure \ref{fig:ellipvsr}).

\subsection{Radial density profiles}

{\vide} contains a routine to construct three-dimensional stacks of \glslink{void}{voids}, where \gls{void} macrocenters are superposed and particle positions are shifted to be expressed as relative to the stack center. This routine builds stacks of \glslink{void}{voids} whose effective radius is in some given range. From each of these three-dimensional stacks, {\vide} builds a spherically-averaged one-dimensional profile.

This is used in particular for building radial \glslink{density profile}{density profiles} of \glslink{void}{voids} at a given size (see figure \ref{fig:1dprofile}). 

\section{The T-web}
\label{sec:apx-tweb}

This section describes the ``\gls{T-web}'', a dynamic \glslink{cosmic web classification}{web classifier} which dissects the entire \glslink{LSS}{large-scale structure} into different \glslink{structure type}{structure types}: \glslink{void}{voids}, \glslink{sheet}{sheets}, \glslink{filament}{filaments}, and \glslink{cluster}{clusters}. It is used in section \ref{sec:Comparison of structure types in LPT and $N$-body dynamics}, chapters \ref{chap:ts} and \ref{chap:decision} of this thesis.

\subsection{The tidal tensor}

We start here from the \gls{Vlasov-Poisson system} in Eulerian coordinates, equations \eqref{eq:Poisson} and \eqref{eq:Vlasov}. It is always possible to rescale the cosmological \gls{gravitational potential} by defining $\tilde{\Phi} \equiv \Phi/(4\pi\G a^2 \bar{\rho})$ in such a way that $\tilde{\Phi}$ obeys a reduced \gls{Poisson equation},
\begin{equation}
\Delta \tilde{\Phi}(\textbf{x}) = \delta(\textbf{x}) .
\label{eq:reduced-Poisson}
\end{equation}
In this context, we define the \textit{\gls{tidal tensor}} $\mathscr{T}$ as the Hessian $\mathrm{H}(\tilde{\Phi})$ of the rescaled \gls{gravitational potential} $\tilde{\Phi}$,
\begin{equation}
\mathscr{T}_{ij} \equiv \mathrm{H}(\tilde{\Phi})_{ij} = \frac{\partial^2 \tilde{\Phi}}{\partial \textbf{x}_i \partial \textbf{x}_j} .
\label{eq:tidal-tensor}
\end{equation}
With this definition, the left-hand side of equation \eqref{eq:reduced-Poisson} can be seen as the application of the Laplace-Beltrami operator $\mathcal{LB}$ (or tensor Laplacian), trace of the Hessian, to $\tilde{\Phi}$:
\begin{equation}
\mathcal{LB}(\tilde{\Phi}) \equiv \mathrm{tr}(\mathrm{H}(\tilde{\Phi})) = \Delta \tilde{\Phi} .
\end{equation}
Let us denote by $\mu_1(\textbf{x}) \leq \mu_2(\textbf{x}) \leq \mu_3(\textbf{x})$ the three local \glslink{eigenvalue}{eigenvalues} of the \gls{tidal tensor}.\footnote{These \glslink{eigenvalue}{eigenvalues} are often noted $\lambda_i$ in the literature. We changed the notation in this thesis to avoid the confusion with the Zel'dovich formalism (see sections \ref{sec:Shell-crossing in the Zel'dovich approximation} and \ref{sec:Analogy with the Zel'dovich formalism}).} They are dimensionless and real (since $\mathscr{T}$ is symmetric). We have $\mathrm{tr}(\mathscr{T})(\textbf{x}) = \mu_1(\textbf{x})+\mu_2(\textbf{x})+\mu_3(\textbf{x})$, and the reduced \gls{Poisson equation} can therefore be seen as a decomposition of the Eulerian \gls{density contrast} field, in the sense that it reads 
\begin{equation}
\mu_1(\textbf{x})+\mu_2(\textbf{x})+\mu_3(\textbf{x}) = \delta(\textbf{x}) .
\label{eq:decomposition_tidal}
\end{equation}

At this point, it is useful to introduce some notations commonly found in the literature to characterize the \gls{tidal field}. Given equation \eqref{eq:decomposition_tidal}, the \glslink{eigenvalue}{eigenvalues} of the \gls{tidal tensor} define an ellipsoid with semi-axes \citep[e.g.][]{Peacock1985}
\begin{equation}
a_i(\textbf{x}) \equiv \sqrt{\frac{\delta(\textbf{x})}{\mu_i(\textbf{x})}} .
\end{equation}
The triaxiality parameters are defined by \citet{Bardeen1986} in terms of the \glslink{eigenvalue}{eigenvalues} as
\begin{equation}
\epsilon(\textbf{x}) = \frac{\mu_1(\textbf{x})-\mu_3(\textbf{x})}{2\delta(\textbf{x})} \quad \mathrm{and} \quad p(\textbf{x}) = \frac{\mu_1(\textbf{x})-2\mu_2(\textbf{x})+\mu_3(\textbf{x})}{2\delta(\textbf{x})} .
\end{equation}
$\epsilon$ is called the ellipticity (in the $\mu_1-\mu_3$ plane) and $p$ the prolateness (or oblateness). If $-\epsilon \leq p \leq 0$ then the ellipsoid is prolate-like, and if $0 \leq p \leq \epsilon$ it is oblate-like. The limiting cases are $p=-\epsilon$ for prolate spheroids and $p=\epsilon$ for oblate spheroids.

\subsection{Analogy with the Zel'dovich formalism}
\label{sec:Analogy with the Zel'dovich formalism}

The above equations have a strong similarity with that of the \glslink{ZA}{Zel'dovich formalism}. Indeed, we have seen that the first \gls{Lagrangian potential} $\phi^{(1)}$, defined by $\boldsymbol{\Psi}^{(1)}(\textbf{q},\tau) = -D_1(\tau) \nabla_\textbf{q} \phi^{(1)}(\textbf{q})$, satisfies a reduced \gls{Poisson equation} (equation \eqref{eq:Poisson_phi1}),
\begin{equation}
\Delta_\textbf{q} \phi^{(1)}(\textbf{q}) = \delta(\textbf{q}) .
\end{equation}
As discussed in section \ref{sec:Shell-crossing in the Zel'dovich approximation}, the shear of the displacement $\mathscr{R} \equiv \partial \boldsymbol{\Psi} /\partial \textbf{q}$ verifies
\begin{equation}
\mathscr{R}_{ij} = -D_1(\tau) \mathrm{H}(\phi^{(1)})_{ij} = -D_1(\tau) \frac{\partial^2 \phi^{(1)}}{\partial \textbf{q}_i \partial  \textbf{q}_j} .
\end{equation}
The local \glslink{eigenvalue}{eigenvalues} of Hessian of the first \gls{Lagrangian potential}, $\lambda_1(\textbf{q}) \leq \lambda_2(\textbf{q}) \leq \lambda_3(\textbf{q})$, permit to rewrite the reduced \gls{Poisson equation} as a decomposition of the \glslink{initial conditions}{initial} \gls{density contrast},
\begin{equation}
\lambda_1(\textbf{q}) + \lambda_2(\textbf{q}) + \lambda_3(\textbf{q}) = \delta(\textbf{q}) .
\end{equation}

\subsection{The T-web: original procedure}

In analogy with the Zel'dovich ``\gls{pancake}'' theory, where the sign of the $\lambda_i$ permit an interpretation of what happens at \gls{shell-crossing} in the \gls{ZA} in terms of \glslink{structure type}{structure types} (see section \ref{sec:Shell-crossing in the Zel'dovich approximation}), \citet{Hahn2007a} proposed to \glslink{cosmic web classification}{classify} structures using the sign of the $\mu_i$. Namely, a \gls{void} point corresponds to no positive \gls{eigenvalue}, a \gls{sheet} to one, a \gls{filament} to two, and a \gls{cluster} to three positive \glslink{eigenvalue}{eigenvalues} (see table \ref{tb:tweb-rules}). 

\begin{table}[h]\centering
\begin{tabular}{ll}
\hline\hline
Structure type & Rule\\
\hline
Void & $\mu_1,\mu_2,\mu_3 < 0$\\
Sheet & $\mu_1,\mu_2 < 0$ and $\mu_3 > 0$\\
Filament & $\mu_1 < 0$ and $\mu_2,\mu_3 > 0$\\
Cluster & $\mu_1,\mu_2,\mu_3 > 0$\\
\hline\hline
\end{tabular}
\caption{Rules for classification of \glslink{structure type}{structure types} according to the \gls{T-web} procedure \citep{Hahn2007a}.}
\label{tb:tweb-rules}
\end{table}

The interpretation of this rule is straightforward, as the sign of an \gls{eigenvalue} at a given position defines whether the gravitational force in the direction of the corresponding eigenvector is contracting (positive \glslink{eigenvalue}{eigenvalues}) or expanding (negative \glslink{eigenvalue}{eigenvalues}). Thus, the signature of the \gls{tidal tensor} characterizes the number of axes along which there is \glslink{gravitational evolution}{gravitational expansion or collapse}. This procedure is sometimes called the ``\gls{T-web}'', in reference to the \gls{tidal tensor}.

In \citet{Hahn2007a}, an interpretation of the above rule in terms of the orbit stability of test particles is also discussed. The \gls{equation of motion} in \gls{comoving coordinates} and in \gls{conformal time} reads (see equation \eqref{eq:equation-of-motion-p})
\begin{equation}
\frac{\drm \textbf{p}}{\drm \tau} = -ma \nabla \Phi \quad \mathrm{with} \quad \textbf{p} = ma \frac{\drm \textbf{x}}{\drm \tau}
\end{equation}
The local extrema of the \gls{gravitational potential} (i.e. points $\bar{\textbf{x}}$ such that $\nabla \Phi(\bar{\textbf{x}})=0$) are fixed points of the test particle \gls{equation of motion}. These can be, for example, the center of mass of \glslink{halo}{halos}. In their neighborhood, we can linearize the \gls{equation of motion} at the points $\bar{\textbf{x}}$, which yields the linear system
\begin{equation}
\frac{\drm}{\drm \tau} \left( ma \frac{\drm \textbf{x}}{\drm \tau}\right) \approx -ma \, \nabla^2 \Phi(\bar{\textbf{x}}) \cdot \left(  \textbf{x} - \bar{\textbf{x}}\right) ,
\end{equation}
or, in terms of coordinates,
\begin{equation}
\frac{\drm}{\drm \tau} \left( ma \frac{\drm \textbf{x}_i}{\drm \tau}\right) \approx -ma \, \sum_j \frac{\partial^2 \Phi}{\partial \textbf{x}_i \partial \textbf{x}_j}(\bar{\textbf{x}})  \left( \textbf{x}_j - \bar{\textbf{x}}_j \right) \propto -ma \, \sum_j \mathscr{T}_{ij}(\bar{\textbf{x}})  \left( \textbf{x}_j - \bar{\textbf{x}}_j \right).
\end{equation}
This equation means that the linear dynamics near local extrema of the \gls{gravitational potential} is fully governed by the \gls{tidal field}. The number of positive \glslink{eigenvalue}{eigenvalues} is equivalent to the dimension of the stable manifold at the fixed points:
\begin{itemize}
\item \glslink{void}{voids} are regions of space where the orbits of test particles are unstable (no positive \gls{eigenvalue});
\item \glslink{sheet}{sheets} correspond to one-dimensional stable manifolds (one positive, two negative \glslink{eigenvalue}{eigenvalues});
\item \glslink{filament}{filaments} correspond to two-dimensional stable manifolds (two positive, one negative \glslink{eigenvalue}{eigenvalues});
\item \glslink{cluster}{clusters} are attractive fixed points (three positive \glslink{eigenvalue}{eigenvalues}).
\end{itemize}
Dropping the assumption of local extrema of the \gls{gravitational potential} introduces a constant acceleration term to the linearized \gls{equation of motion}. This zeroth-order effect can be ignored by changing to free-falling coordinates. The behavior introduced by the first-order term, representing the \glslink{tidal effects}{tidal deformation of orbits}, and thus the \glslink{cosmic web classification}{web-type classification}, remain unchanged. 

\subsection{Extensions of the T-web}
\label{sec:Extensions of the T-web}

\subsubsection{Varying threshold}

Several extensions of this \glslink{cosmic web classification}{classification} procedure exist. \cite{Forero-Romero2009} pointed out that rather than using a threshold value $\mu_\mathrm{th}$ of zero, different positive values can be used. The corresponding set of rules is given by table \ref{tb:tweb-rules-extended}.

\begin{table}[h]\centering
\begin{tabular}{ll}
\hline\hline
Structure type & Rule\\
\hline
Void & $\mu_1,\mu_2,\mu_3 < \mu_\mathrm{th}$\\
Sheet & $\mu_1,\mu_2 < \mu_\mathrm{th}$ and $\mu_3 > \mu_\mathrm{th}$\\
Filament & $\mu_1 < \mu_\mathrm{th}$ and $\mu_2,\mu_3 > \mu_\mathrm{th}$\\
Cluster & $\mu_1,\mu_2,\mu_3 > \mu_\mathrm{th}$\\
\hline\hline
\end{tabular}
\caption{Rules for \glslink{cosmic web classification}{classification} of \glslink{structure type}{structure types} according to the extended \gls{T-web} procedure with varying threshold \citep{Forero-Romero2009}.}
\label{tb:tweb-rules-extended}
\end{table}

This introduces a new free parameter, which \textit{a priori} can take any value. However, \citet{Forero-Romero2009} argued that a natural threshold can be roughly estimated by equating the collapse time (determined by the \glslink{eigenvalue}{eigenvalues}) to the age of the Universe. For an \glslink{SC}{isotropic collapse}, they calculated explicitly $\mu_\mathrm{th}=3.21$ \citep[appendix A in][]{Forero-Romero2009}. As gravitational collapse is often highly anisotropic, they used an empirical approach to determine the threshold and argued that $\mu_\mathrm{th} \approx 1$ can yield better \glslink{cosmic web classification}{web classifications} than the original \gls{T-web}, down to the megaparsec scale.

The \gls{T-web} procedure and/or this extension have been used, for example, by \citet{Jasche2010a,Wang2012,Forero-Romero2014,Nuza2014,Alonso2015,Eardley2015,Forero-Romero2015,Leclercq2015ST,Zhao2015,AungCohn2015}.

\subsubsection{The V-web}

\cite{Hoffman2012} reformulated the extended \gls{T-web} procedure using the \glslink{velocity shear field}{velocity shear tensor} instead of the gravitational \gls{tidal tensor}. More precisely, they use the \glslink{eigenvalue}{eigenvalues} $\mu_i^V(\textbf{x})$ of the rescaled shear tensor defined by
\begin{equation}
\Sigma_{ij} \equiv -\frac{1}{2H(z)} \left( \frac{\partial \textbf{v}_i}{\partial \textbf{r}_j} + \frac{\partial \textbf{v}_j}{\partial \textbf{r}_i}\right) .
\end{equation}
This new scheme is generally referred to as the ``\gls{V-web}'' and the rules are given in table \ref{tb:vweb-rules}. \cite{Hoffman2012} showed that the two \glslink{cosmic web classification}{classifications} coincide at large scales (where the gravitational and \glslink{velocity field}{velocity fields} are proportional) and that the \gls{velocity field} resolves finer structure than the \gls{gravitational field} at the smallest scales (sub-megaparsec). They empirically determined the threshold value $\mu^V_\mathrm{th} = 0.44$ to yield the best visualization of the geometrical characteristics of the four environments at $z = 0$.

\begin{table}[h]\centering
\begin{tabular}{ll}
\hline\hline
Structure type & Rule\\
\hline
Void & $\mu_1^V,\mu_2^V,\mu_3^V < \mu^V_\mathrm{th}$\\
Sheet & $\mu_1^V,\mu_2^V < \mu_\mathrm{th}$ and $\mu_3^V > \mu^V_\mathrm{th}$\\
Filament & $\mu_1^V < \mu^V_\mathrm{th}$ and $\mu^V_2,\mu^V_3 > \mu^V_\mathrm{th}$\\
Cluster & $\mu_1^V,\mu_2^V,\mu_3^V > \mu^V_\mathrm{th}$\\
\hline\hline
\end{tabular}
\caption{Rules for \glslink{cosmic web classification}{classification} of \glslink{structure type}{structure types} according to the \gls{V-web} procedure \citep{Hoffman2012}.}
\label{tb:vweb-rules}
\end{table}

The \gls{V-web} has been used, for example, by \citet{Libeskind2013,Carlesi2014,Nuza2014,Lee2014,Libeskind2014}. In this thesis, we probe scales down to a few Mpc/$h$ (the voxel size in our \glslink{reconstruction}{reconstructions} or \glslink{N-body simulation}{simulations}). Therefore, we will be content with the original \gls{T-web} procedure as formulated by \citet{Hahn2007a}.

\subsection{Implementation}
\label{sec:Tweb-implementation}

This section gives details on how the \gls{T-web} procedure is implemented when used in this thesis. First, the \gls{density contrast} \glslink{density field}{field} is computed by \glslink{mesh assignment}{assigning} particles to the grid with a \gls{CiC} scheme (see section \ref{sec:apx-Density assignments}). It is transformed to Fourier space using a \gls{Fourier transform} on the grid. At this point, if desired, the \gls{density field} can be smoothed using a \gls{Gaussian kernel} $K_{k_\mathrm{s}}(k) \equiv \exp\left(-\frac{1}{2} \frac{k^2}{k_\mathrm{s}^2} \right)$ (usually this step is bypassed in the projects described in this thesis). This corresponds to a mass scale $M_\mathrm{s}$ which is linked to the smoothing length $R_\mathrm{s} \equiv \frac{2\pi}{k_\mathrm{s}}$ by
\begin{equation}
R_\mathrm{s} = \frac{1}{\sqrt{2\pi}} \left( \frac{M_\mathrm{s}}{\bar{\rho}} \right)^{1/3} .
\end{equation}
The reduced \gls{gravitational potential} is estimated by solving the \gls{Poisson equation} in Fourier space, $\tilde{\Phi}(\textbf{k}) = G(\textbf{k}) \delta(\textbf{k})$, where $G(\textbf{k})$ is the \gls{Green function} corresponding to the discretization adopted for the Laplacian. For the projects described in this thesis, we adopted the simple form $G(\textbf{k}) = -1/k^2$ (with also a smoothing of short-range forces and two deconvolutions of the \gls{CiC} kernel, see section \ref{sec:apx-Solving the Poisson equation}).
Hence, the \gls{gravitational potential} is given by the convolution
\begin{equation}
\tilde{\Phi}(\textbf{x}) = (G * \delta)(\textbf{x}),
\end{equation}
or, if the \gls{density field} had been smoothed, by
\begin{equation}
\tilde{\Phi}_{R_\mathrm{s}}(\textbf{x}) = (G * K_{k_\mathrm{s}} * \delta)(\textbf{x}) .
\end{equation}
We compute the components of the \gls{tidal tensor} in Fourier space using $\mathscr{T}_{ab} = -\tilde{\Phi}(\textbf{k}) \textbf{k}_a \textbf{k}_b$, and transform them back to configuration space by inverse \gls{Fourier transform}. In practice, only one \gls{Fourier transform} is required to go from $\delta$ to $\mathscr{T}_{ab} \propto -\delta(\textbf{k}) \textbf{k}_a \textbf{k}_b/k^2$ (or $\mathscr{T}_{ab} \propto -\delta(\textbf{k}) \textbf{k}_a \textbf{k}_b K_{k_\mathrm{s}}(k)/k^2$). Finally, we compute the \glslink{eigenvalue}{eigenvalues} of the \gls{tidal tensor} at each voxel of the grid and \glslink{cosmic web classification}{classify} \glslink{structure type}{structures} using the rules given in table \ref{tb:tweb-rules}. In this fashion, every voxel of the \gls{density field} gets assigned a flag corresponding to the \gls{structure type}: $\mathrm{T}_0$ for \glslink{void}{voids}, $\mathrm{T}_1$ for \glslink{sheet}{sheets}, $\mathrm{T}_2$ for \glslink{filament}{filaments}, $\mathrm{T}_3$ for \glslink{cluster}{clusters}.

The \gls{T-web} \glslink{cosmic web classification}{classification} takes a few seconds on 8 cores, for a typical \gls{density field} used in this thesis ($L=750$\nbsp Mpc/$h$, $N_\mathrm{v} = 256^3$).

\subsection{Example}

As an example, in this section, we show the results of the \gls{T-web} \glslink{cosmic web classification}{classification} for a simulated \gls{density field}.

The \glslink{N-body simulation}{simulation} contains $512^3$ \gls{dark matter particles} in a \glslink{comoving coordinates}{comoving} box of $750$~Mpc/$h$ with \gls{periodic boundary conditions}. The \gls{initial conditions} have been generated at $z=69$ using \glslink{2LPT}{second-order Lagrangian perturbation theory}. They obey \glslink{grf}{Gaussian statistics} with an \citet{Eisenstein1998,Eisenstein1999} \gls{power spectrum}. The \glslink{N-body simulation}{$N$-body simulation} has been run to $z=0$ with \textsc{\gls{Gadget-2}} \citep{Springel2001,Springel2005}. Particles are assigned to the grid using a \gls{CiC} method. The \gls{cosmological parameters} used are
\begin{equation}
\Omega_\Lambda = 0.728, \Omega_\mathrm{m}~=~0.272, \Omega_\mathrm{b}~=~0.045, \sigma_8 = 0.807, h = 0.702, n_{\mathrm{s}} = 0.961 ,
\end{equation}
which gives a \gls{mass resolution} of $2.37 \times 10^{11}~\mathrm{M}_\odot/h$.

\begin{figure}
\begin{center}
\includegraphics[width=\textwidth]{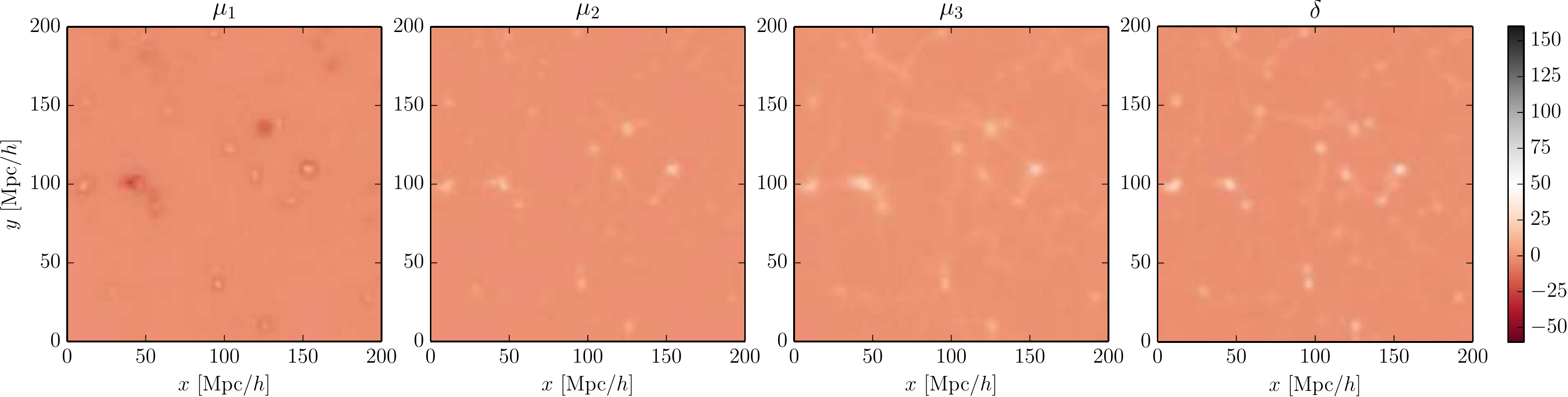} 
\caption{Slices through the voxel-wise \glslink{eigenvalue}{eigenvalues} $\mu_1 \leq \mu_2 \leq \mu_3$ of the \glslink{tidal tensor}{tidal field tensor} in the \gls{final conditions} of a dark matter \glslink{N-body simulation}{simulation}. The rightmost panel shows the corresponding slice through the \glslink{final conditions}{final} \gls{density contrast} $\delta=\mu_1+\mu_2+\mu_3$ (equation \eqref{eq:decomposition_tidal}). See also figure \ref{fig:pdf_final} for comparison.}
\label{fig:tweb_example_lambda}
\end{center}
\end{figure}

\begin{figure}
\begin{center}
\includegraphics[width=0.8\textwidth]{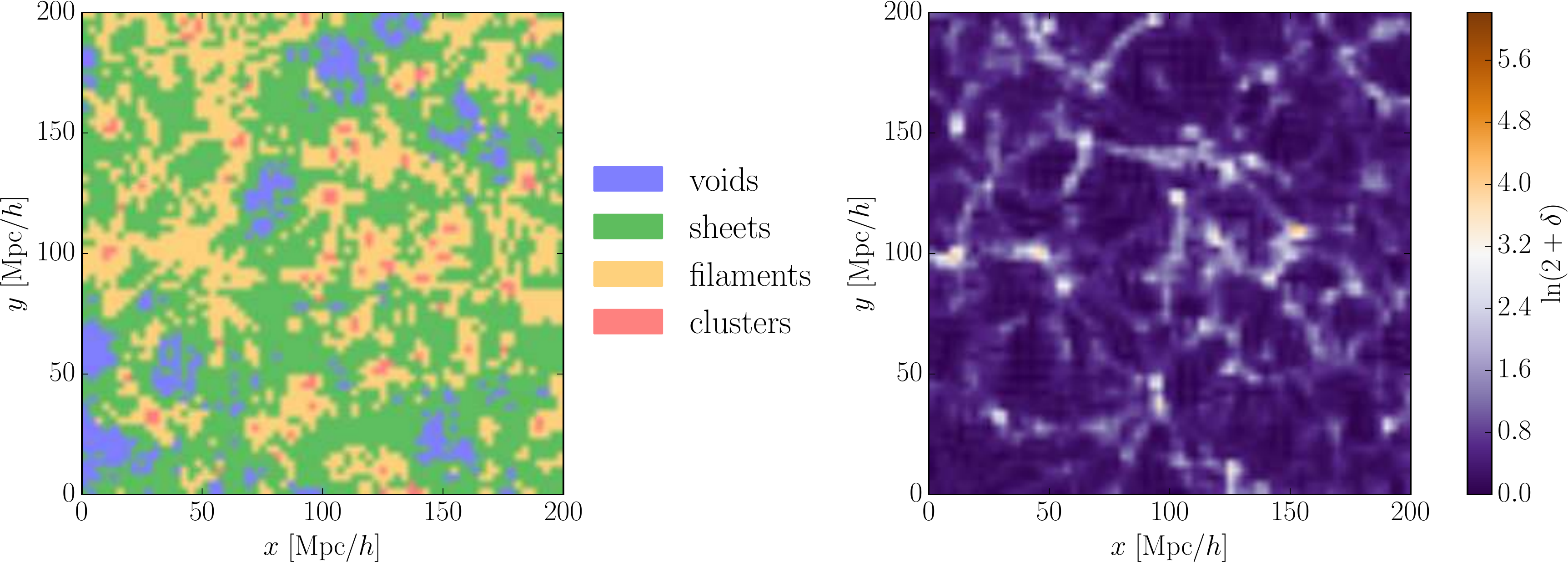} 
\caption{\textit{Left panel}. \glslink{cosmic web classification}{Classification of structures} with the \gls{T-web} procedure in the \gls{final conditions} of a dark matter simulation. The color coding is blue for \glslink{void}{voids}, green for \glslink{sheet}{sheets}, yellow for \glslink{filament}{filaments} and red for \glslink{cluster}{clusters}. \textit{Right panel}. Dark matter \glslink{density field}{density} in the corresponding slice (for convenience, the quantity shown in $\ln(2+\delta)$).}
\label{fig:tweb_example_objects}
\end{center}
\end{figure}

For clarity, we show slices through a $200$~Mpc/$h$ region of the \glslink{N-body simulation}{simulation}. Figure \ref{fig:tweb_example_lambda} shows the \glslink{eigenvalue}{eigenvalues} of the \gls{tidal tensor} and the \gls{density contrast}. A slice through the corresponding voxel-wise \glslink{cosmic web classification}{classification of structures} is shown in the left panel of figure \ref{fig:tweb_example_objects}.

